\documentclass[12pt]{book}

\usepackage{graphicx}
\usepackage{dcolumn}
\usepackage{color}
\usepackage{bm}
\usepackage{amsmath,amsfonts,amssymb}
\usepackage{accents}

\DeclareRobustCommand{\utilde}[1]{%
  \underaccent{\tilde}{\smash{#1}}%
}
\usepackage{multicol}

\usepackage{wrapfig}
\usepackage{epstopdf}
\usepackage{xkeyval}
\usepackage{cancel}
\usepackage{enumerate}
\usepackage{hyperref}

\hypersetup{
    colorlinks=true,
    urlcolor=magenta,
    filecolor=magenta,
    linktoc=page,
    citecolor=red,
    linkcolor=blue,
    bookmarks=true
}

\begin{document}

\title{\Huge {\color{red} A Treatise on Differential Geometry and its role in Relativity Theory}}
\author{\bf Subenoy Chakraborty}
\date{}
	
\maketitle

\tableofcontents


\chapter{Manifold and Differential Structure}


\section{Linear Space\,: Vectors}

~~An algebraic system is defined as a nonempty set $S$ with one or more binary operations on $S$.\\

For example, if we consider the set of all square matrices of same order as the set $S$ then ($S$,+,$\times$,$\cdot$), where `+' stands for the matrix addition, `$\times$' for the matrix multiplication and `$\cdot$' is the scalar multiplication, form an algebraic system. Note that a scalar multiplication is defined over a field $F$ (of real or complex numbers).\\

Let ($L$,+,$\cdot$) be an algebraic system defined over a field $F$. Here the binary operation addition (+) is defined as 
\begin{center}
$L$ $\times$ $L$ $\rightarrow$ $L$, \textit{i.e.}, for any $x$,$y$ $\in$ $L$, $\exists$ an element $z\in$ $L$ such that $z=x+y$.
\end{center}

Scalar multiplication `$\cdot$' is defined over the field $F$ as 
\begin{center}
$F$ $\times$ $L$ $\rightarrow$ $L$, \textit{i.e.}, for any $\alpha \in$ $F$, $x\in$ $L$, $\alpha x \in$ $L$.
\end{center}

An algebraic system with two binary operations ($L$,+,$\cdot$) is said to be a linear space or a vector space over a field $F$ if the binary operations satisfy the following axioms:
\begin{eqnarray}\label{1.1}
\left. \begin{array}{l}
\mbox{a)}~(L,+)~\mbox{is a commutative group}, \\
\mbox{b)~i)}~\alpha(x+y)=\alpha x + \alpha y \\
~~~~\mbox{ii)}~(\alpha +\beta)x=\alpha x+\beta x \\
~~~~\mbox{iii)}~(\alpha \beta)x=\alpha(\beta x) \\
~~~~\mbox{iv)}~1\cdot x=x~,
\end{array} \right.
\end{eqnarray}
for any $x$,$y$ $\in$ $L$ and $\alpha$, $\beta$ $\in$ $F$.\\

Here, elements of $F$ are called scalars. Usually, we say that the vector space is defined over the field $F$.\\

In vector algebra, a vector is just an arrow having forward point and end point in the space. Mathematically, a vector is an element of a vector space (\textit{i.e.}, linear space). A set of linearly independent vectors which span the whole vector space is called a basis of $L$. The number of vectors in the basis is called the dimension of the vector space.\\

Let $\lbrace e_i \rbrace$, $i=1,2,....n$ ($n$ is the dimension of the vector space) be a basis of the vector space. So any arbitrary vector $V$ $\in$ $L$ can be written as
\begin{equation} \label{1.2}
V=V^i e_i
\end{equation}
where the co-efficients $V^i$ are numbers and are called the components of the vector $V$ in the basis $\lbrace e_i \rbrace$. If we choose another basis $\lbrace e_i' \rbrace$ then $V$ can be written as 
\begin{equation} \label{1.3}
V=V'^{\,i} e_i'
\end{equation}
with $V'^{\,i}$ as the components of $V$ in the basis $\lbrace e_i' \rbrace$. Now considering $e_i'$ as an element of $L$ we can write it as a linear combination of the basis $\lbrace e_i \rbrace$, \textit{i.e.},
\begin{equation} \label{1.4}
e_i'=\Lambda _{i}^{~k} e_k.
\end{equation}

Similarly,
\begin{equation} \label{1.5}
e_i=\Lambda _{~i}^{k} e_k'.
\end{equation}

Combining these two, the co-efficients $\Lambda _{i}^{~k}$~'s will satisfy
\begin{eqnarray} \label{1.6}
&&\Lambda _{i}^{~k} \Lambda _{~k}^{l} = \delta _{i}^{~l} \nonumber \\
&\mbox{and~~~}&
\Lambda _{i}^{~k} \Lambda _{~l}^{i} = \delta _{~l}^{k}.
\end{eqnarray}

Now substituting (\ref{1.5}) in (\ref{1.2}) and equating with (\ref{1.3}), we have the transformation law for the components of $V$
\begin{equation} \label{1.7}
V'^k=\Lambda _{~i}^{k}V^i
\end{equation}
and similarly for the unprimed components
\begin{equation} \label{1.8}
V^k=\Lambda _{i}^{~k}V'^i.
\end{equation}

\section{Dual Space\,: Covectors}

~~~A linear map $w$: $L$ $\rightarrow$ $R$, is defined as if $V$ $\in$ $L$ then $w$($V$) is a real number. The collection of all such linear maps form a vector space having same dimension as $L$. This is called the dual vector space or simply dual space and is denoted by $L^*$. Any element of $L^*$ (\textit{i.e.}, linear map) is called a one-form (or a covector). We shall now show that, given a basis $\lbrace e_i \rbrace$ of $L$ we can find a class of one forms $\lbrace w^i \rbrace$ such that
\begin{equation} \label{1.9}
w^i (e_k)=\delta _{k}^{i},
\end{equation}
then this class of one forms constitute a basis for $L^*$.\\

Suppose the class of one-forms are linearly dependent, \textit{i.e.}, $\exists$ scalars $\lambda _i$ (not all zero) such that
\[
\begin{array}{l}
~~~~\lambda _i w^i = 0 \\
\Rightarrow \lambda _i w^i(e_k) = 0 \\
\Rightarrow \lambda _i \delta _{k}^{i} = 0 \\
\Rightarrow \lambda _k = 0,~~\forall ~~k=1,2,...n.
\end{array}\]

Hence this class of one-forms satisfying eq.\,(\ref{1.9}) are linearly independent.\\

Next we shall show that this class $\lbrace w^i \rbrace$ of one-forms generate $L^*$, \textit{i.e.}, any one-form can be expressed as a linear combination of this class of one-form. Let $B$ $\in$ $L^*$ and
$$B(e_k)=B_k~,~~k=1,2,\ldots ,n\,.$$

Also, ($B_{i}w^i$)($e_k$)=$B_{i}w^i(e_k)$=$B_{i}\delta _{k}^{i}=B_k$. Thus $B$ and $B_i w^i$ have the same actions on the basis vectors $\lbrace e_i \rbrace$ of $L$ and hence $B=B_i w^i$. Therefore, the class of one-forms $\lbrace w^i \rbrace$ satisfying eq.\,(\ref{1.9}) is a basis for L$^*$ and $B_i$'s are the components of $B$ in this basis. This basis is called the reciprocal or dual of the basis $\lbrace e_i \rbrace$. Similar to the original vector space $L$, the transformation laws for the dual basis and the components of a one-form are
\begin{equation} \label{1.10}
w'^k=\Lambda _{~i}^{k} w^i~,~~w^k=\Lambda _{i}^{~k} w'^i
\end{equation}
and
\begin{equation} \label{1.11}
B'_{i}=\Lambda _{i}^{~k} B_k~,~~B_i=\Lambda _{~i}^{k} B'_{k}.
\end{equation}

Often, vectors \textit{i.e.}, elements of the vector space $L$ are called contravariant vectors and elements of dual space $L^*$, \textit{i.e.}, the covectors are called covariant vectors. It is customary to write the components of a contravariant vector by an upper index (superscript) $v^a$ while a covariant vector by a lower index (subscript) $B_a$.\\

This convention of component notation shows the action of a dual vector on a vector in a simple way as
\begin{equation} \label{1.12}
B(V)=B_i w^i (V^k e_k)=B_i V^k w^i (e_k)=B_i V^k \delta _{k}^{i}=B_i V^i \in R.
\end{equation}

This suggests that it is sufficient to write only the components of the vectors (or covectors), there is no need of specifying the basis vectors. The formation of the number $B(V)$ is often called the contraction of $B$ with $V$. Further, the form of the action of covector $B$ on $V$ (given by Eq.\,(\ref{1.12})) can interpret vectors as linear maps on dual vectors as
\begin{equation} \label{1.13}
V(B)\equiv B(V)=B_i V^i.
\end{equation}

Therefore, the dual space to the dual vector space is the original vector space itself.\\

{\bf Examples:}
\begin{enumerate}[(a)]
	\item  The simplest example of a dual vector is the gradient of a scalar function.
\item Let the vector space $L$ be the space of $n$-component column vectors, \textit{i.e.}, if $V \in L$ then
$$V=\begin{bmatrix}
V^1 \\
V^2 \\
\vdots \\
V^n
\end{bmatrix}$$
So the dual space $L^*$ is that of $n$-component row vectors, \textit{i.e.}, for $w \in$ $L^*$,
$$w= \left[ w_1 , w_2 , \ldots , w_n \right]$$
The action of $w$ on $V$ is the ordinary matrix multiplication :
$$ w(V)=(w_1,w_2, \ldots , w_n)
 \begin{pmatrix}
V^1 \\
V^2 \\
\vdots \\
V^n
 \end{pmatrix} = w_{i}V^i$$
\item In quantum mechanics, the vectors are elements of the Hilbert space and are represented by kets $\vert \psi \rangle$. In this case the dual space is the space of bra\,s $\langle \phi \vert$ and the action gives the number $\langle \phi \vert \psi \rangle$.
\end{enumerate}

\section{Multilinear mapping of vectors and covectors\,: Tensors}

~~~Tensors can be considered as a generalization of the concept of vectors or covectors (in the sense that vectors (covectors) are one-index system of quantities while tensors are one or more index system of quantities). In Newtonian theory, one can write down the evolution equations in a compact form using the notion of vectors. Sometimes vectorial notations help us to solve problems and may help us to have geometrical as well as physical insight. Similarly, in relativity theory (and also in electro-magnetic theory) it is convenient to write the necessary equations in  a compact and elegant way using tensorial quantities. Basically, there are two distinct ways in which one can define tensors: the index free (or coordinate free) approach and the classical approach based on indices. Although the abstract index free approach shows deeper geometrical insight but it is not useful for practical calculations. In this chapter (i.e., Chapter One) index-free notion of tensor has been introduced for mathematical clarity but subsequently index base notion has been used for tensors. \\

The idea of linear map from vectors to real numbers can be extended by introducing a multilinear map ($T$) from a collection of covectors and vectors to $\mathbb{R}$:
\begin{center}
$T$: $L^* \times L^* \times \cdot \cdot \cdot L^* \times L\times L\times \cdot \cdot \cdot L\rightarrow \mathbb{R}$\\
($r$ times)~~~~~~~~~~~($s$ times)
\end{center}

Here multilinear mapping $T$ operates linearly on each vector and covector in the above Cartesian product. This multilinear mapping on a class of covectors and vectors to give a real number is called a tensor. In the above, $T$ is a ($r$, $s$)-type tensor or a tensor of rank ($r$, $s$). For example, if $T$ is a (2,2) tensor then the real number assigned by it with arguments $w$, $\eta$, $V$, $W$ is denoted by $T$($w$,$\eta$,$V$,$W$) and is called the value of the tensor with these arguments. Due to linearity of $T$ on its arguments, we have
\begin{equation} \label{1.14}
T(aw+b\eta,\sigma,cV+dW,U)=acT(w,\sigma,V,U)+adT(w,\sigma,W,U)+bcT(\eta,\sigma , V,U)+bdT(\eta,\sigma , W,U).
\end{equation}

A ($r$, $s$)-tensor field is a rule giving a ($r$, $s$)-tensor at each point. The linearity property of tensors is easily extendable to tensor fields, except that the numbers $a$, $b$, $c$ and $d$ in Eq. (\ref{1.14}) may have different values at each point.\\

In particular, a vector is a (1,0) tensor and a (0,1) tensor is a one-form. By convention, a (0,0)-tensor is termed as a scalar function. Note that for a (1,1) tensor $T$, $T$($w$,$V$) is a real number but for fixed $w$, $T$($w$;~$\cdot$) can be viewed as a one-form since it needs one vector as argument to give a real number. Similarly, $T$($\cdot$~;$V$) is a vector as operated on a one-form gives a scalar. Therefore, a (1,1)-tensor can be thought of as a linear vector valued function of vectors or a linear form-valued function of one-forms. This type of interpretation is possible for tensors of any order.\\

Moreover, the collection of all ($r$, $s$)-tensors at a point forms a vector space. In a particular basis, the addition of two ($r$, $s$)-tensors means the addition of the corresponding components and scalar multiplication is nothing but multiplication of the components by the scalars. But to construct a basis of the above vector space, we shall have to introduce a new operation known as the tensor product (or outer product). Suppose $A$ is a ($r$, $s$)-tensor and $B$ is a ($t$, $u$)-tensor, then their tensor product is denoted by $A\otimes B$ and is a ($r+t$, $s+u$)-tensor, defined as
$$A\otimes B\left(w^{(1)},w^{(2)}, \ldots , w^{(r+t)};V^{(1)},V^{(2)}, \ldots , V^{(s+u)}\right)$$
$$=A\left(w^{(1)},w^{(2)}, \ldots , w^{(r)};V^{(1)},V^{(2)}, \ldots , V^{(s)}\right)B\left(w^{(r+1)}, \ldots , w^{(r+t)};V^{(s+1)}, \ldots , V^{(s+u)}\right)$$

(Note that the index in $w^{(i)}$ or $V^{(k)}$ is to label the covector or the vector, but not their components). Thus the tensor product means the successive operation of the tensors $A$ and $B$ on the appropriate set of dual vectors and vectors and then multiplication of the corresponding numbers. So it is clear that in general this tensor product is not commutative, \textit{i.e.},
$$A\otimes B \neq B\otimes A$$
but it is associative, \textit{i.e.},
$$A\otimes (B \otimes C) = (A\otimes B)\otimes C .$$

Now the basis vectors for the vector space of all ($r$, $s$)-tensors can be constructed in a straightforward manner from the basis of $L$ and $L^*$. In fact, this set of basis is the tensor product of the form
$$e_{(\alpha _1)}\otimes e_{(\alpha _2)}\otimes \cdot \cdot \cdot \otimes e_{(\alpha _r)}\otimes w^{(\beta _1)}\otimes w^{(\beta _2)}\otimes \cdot \cdot \cdot \otimes w^{(\beta _s)}$$
and is denoted by $\theta _{\alpha _1 \alpha _2 \cdot \cdot \cdot \alpha _r}^{\beta _1 \beta _2 \cdot \cdot \cdot \beta _s}$.\\

So an arbitrary ($r$,$s$)-tensor $T$ can be written in compact form as
\begin{eqnarray} \label{1.15}
T &=& T _{\beta _1 \cdots \beta _s}^{\alpha _1 \cdots \alpha _r} \theta _{\alpha _1 \cdots \alpha _r}^{\beta _1 \cdots \beta _s} \nonumber \\
&=& T _{\beta _1 \cdots \beta _s}^{\alpha _1 \cdots \alpha _r} e_{(\alpha _1)}\otimes e_{(\alpha _2)}\otimes \cdots \otimes e_{(\alpha _r)}\otimes w^{(\beta _1)}\otimes w^{(\beta _2)}\otimes \cdots \otimes w^{(\beta _s)}.
\end{eqnarray}

In other words, the components can be obtained by operating the tensor on basis vectors and dual basis vectors
\begin{equation} \label{1.16}
T _{\beta _1 \beta _2 \cdots  \beta _s}^{\alpha _1 \alpha _2 \cdots \alpha _r}=T\left(w^{(\alpha_ 1)}, w^{(\alpha_ 2)}, \ldots , w^{(\alpha_ r)};e_{(\beta _1)}, \ldots , e_{(\beta _s)}\right).
\end{equation}

Further, the action of a ($r$, $s$)$-$tensor $T$ on an arbitrary vectors $V^{(1)},V^{(2)},\cdot \cdot \cdot V^{(s)}$ and covectors $B^{(1)},B^{(2)},\cdot \cdot \cdot B^{(r)}$ can be written as 
\begin{equation} \label{1.17}
T(B^{(1)},B^{(2)},\ldots , B^{(r)},V^{(1)},V^{(2)},\cdot \cdot \cdot , V^{(s)})=T _{\beta _1 \cdots \beta _s}^{\alpha _1 \cdots \alpha _r} B_{\alpha _1}^{(1)} B_{\alpha _2}^{(2)} \cdots B_{\alpha _r}^{(r)} V^{(1)\beta _1} V^{(2)\beta _2} \cdots V^{(s)\beta _s}.
\end{equation}

Note that the order of the indices is important as the tensor does not act on its various arguments in the same way. Using the transformation law for the basis vectors of $L$ (Eqs. (\ref{1.4}) and (\ref{1.5})) and of $L^*$ (Eq.\,(\ref{1.10})), the transformation law of the components of the tensor can be obtained from Eq.\,(\ref{1.15}) as
\begin{equation} \label{1.18}
{T'} _{r_1 \cdot \cdot \cdot r_s}^{\mu _1 \cdot \cdot \cdot \mu _r}=\Lambda _{~~\alpha _1}^{\mu _1}\Lambda _{~~\alpha _2}^{\mu _2} \cdot \cdot \cdot \Lambda _{~~\alpha _r}^{\mu _r} \Lambda _{\gamma _1}^{~~\beta _1}\Lambda _{\gamma _2}^{~~\beta _2} \cdot \cdot \cdot \Lambda _{\gamma _s}^{~~\beta _s} T _{\beta _1 \cdot \cdot \cdot \beta _s}^{\alpha _1 \cdot \cdot \cdot \alpha _r}.
\end{equation}

The transformation law shows that each superscript transforms like a vector and each subscript transforms like a dual vector. Similar to vectors, a tensor is conveniently described by its components.\\

We shall now introduce an important notion in tensor algebra called contraction. We illustrate it by examples. Suppose $A$ is a (1,1) tensor. Then it can act as a map on $L$ and the result will also be a vector, \textit{i.e.}, 
$$A: L\rightarrow L~\mbox{or}~A_{\beta}^{\alpha}: V^{\beta} \rightarrow A_{\beta}^{\alpha} V^{\beta} \in L~~\mbox{for any}~V^{\beta}\in L\,.$$
$$i.e.,~ A_{\beta}^{\alpha} V^{\beta}=W^{\alpha}\,.$$

Similarly, a tensor can act (fully or partly) on another tensor to give rise to a third tensor. For example,
$$A_{\gamma}^{\alpha \beta} B_{\beta \delta}^{\gamma}=C_{\delta}^{\alpha}~,~~\mbox{a (1,1) tensor}\,$$

Note that here outer product of two tensors $A$ and $B$ followed by contraction gives the tensor $C$.\\

In the first example, index $\beta$ is contracted, called the dummy index and $\alpha$ is the free index while in the second example both $\beta$ and $\gamma$ are dummy indices and $\alpha$, $\delta$ are free indices which will characterize the resulting tensor. By contraction operation on two tensors, if there does not remain any free index then the resulting tensor is a (0,0)$-$type, \textit{i.e.}, a scalar. Thus under the contraction operation, a tensor is usually reduced in one contravariant and one covariant order. It is easy to see that contraction is independent of the choice of basis.\\

{\bf Quotient Law:}\\
If the product $A_{\cdot \cdot \cdot \cdot}^{\cdot \cdot \cdot \cdot} B_{\cdot \cdot \cdot \cdot}^{\cdot \cdot \cdot \cdot}$ where dots represent indices which may involve contraction between indices of $A$ and $B$ and also $A$ is an indexed system of functions of the coordinate variables  and $B$ is an ordinary tensor of the type indicated by its indices, is a tensor of the type indicated by the free indices, then quotient law states that $A$ is a tensor of the type indicated by the indices.\\

{\bf Examples:}
\begin{enumerate}
	\item If $A(i,j,k)B^j=C_{i}^{k}$ then by quotient law $A(i,j,k)$ is a (1,2)$-$tensor with appropriate form $A_{ij}^{k}$.
\item Let $A(i,j)$ be a 2$-$index system of functions of the coordinate variables. If for any two arbitrary contravariant vectors $u^i$ and $v^i$ the expression $A(i,j)u^i v^j$ is a scalar then $A(i,j)$ is a (0,2)$-$tensor.
\end{enumerate}

\vspace{1cm}
 \textbf{Reciprocal Tensor:}\\

Let $a_{ij}$ be a non-singular (0,2)$-$tensor then a reciprocal tensor is defined as
\begin{equation} \label{1.19}
a^{ij}=\frac{\mbox{Cofactor of}~a_{ji}~\mbox{in}~|a_{ij}|}{|a_{ij}|}.
\end{equation}

In the following, we shall show that $a^{ij}$ is a non-singular (2,0)$-$tensor.\\

From the property of determinants,
\begin{eqnarray} \label{1.20}
a^{ik}a_{kj} &=& \delta _{j}^{i}\nonumber \\
\mbox{and} ~~~ a_{ik}a^{kj} &=& \delta _{i}^{j}.
\end{eqnarray}

Let $u_i$ be an arbitrary covariant vector. Then there exists a solution for the contravariant vector $v^i$ such that
$$a_{ij}v^j=u_i\,.$$

Now,
$$a^{ij}u_j=a^{ij} a_{jk}v^k=\delta^i_{~k}v^k=v^i$$

Thus $a^{ij} u_j$ is a contravariant vector for an arbitrary covariant vector $u_j$. Hence by the quotient law, it follows that $a^{ij}$ is a (2,0)$-$tensor. Further, from (\ref{1.20}), taking determinants we get
$$\left|a^{ik}\right|\left|a_{kj}\right|=\left|\delta _{j}^{i}\right|=1\,.$$

Hence $a^{ij}$ is a non-singular (2,0)$-$tensor.\\

{\bf Symmetric and Skew-symmetric tensor:}\\

For any (0,2)-tensor $A$, the symmetric part is denoted by $SA$ and is defined as
$$SA(V_1,V_2)=\frac{1}{2!}(A(V_1,V_2)+A(V_2,V_1))$$
for any $V_1$, $V_2$ $\in L$. In a particular basis if the components of $A$ are $A^{\alpha \beta}$ then components of its symmetric part is denoted by $A^{(\alpha \beta)}$ and is defined as
\begin{equation} \label{1.21}
A^{(\alpha \beta)}=\frac{1}{2!}(A^{\alpha \beta}+A^{\beta \alpha}).
\end{equation}

Similarly, the components of the skew-symmetric part is denoted by $A^{[\alpha \beta]}$ and is defined as
\begin{equation} \label{1.22}
A^{[\alpha \beta]}=\frac{1}{2!}\left(A^{\alpha \beta}-A^{\beta \alpha}\right).
\end{equation}

In general, a tensor of arbitrary order, say ($r$,$s$) having components $A_{\beta _1 \ldots \beta _s}^{\alpha _1 \ldots \alpha _r}$, the symmetric and anti-symmetric parts of $A$ are defined as
\begin{eqnarray}
A_{(\beta _1 \ldots \beta _s)}^{\alpha _1 \ldots \alpha _r} &=& \frac{1}{s!} \sum\limits_P A_{\beta _1 \ldots \beta _s}^{\alpha _1 \ldots \alpha _r} \nonumber \\
A_{[\beta _1 \ldots \beta _s]}^{\alpha _1 \ldots \alpha _r} &=& \frac{1}{s!} \sum\limits_P \delta _P A_{\beta _1 \ldots \beta _s}^{\alpha _1 \ldots \alpha _r}\,,  \label{1.23}
\end{eqnarray}
where $\Sigma _P$ stands for the sum over all permutations of the indices $(\beta _1,\beta _2,\ldots , \beta _s)$ and $\delta _P = +1$, for even permutation of ($1,2,\ldots , s$) and $\delta _P=-1$, for odd permutation of ($1,2,\ldots , s)$ and $=0$ if any two indices are equal. As an example,
$$A_{[\beta \gamma \delta]}^{\alpha}=\frac{1}{3!} \left [ A_{\beta \gamma \delta}^{\alpha}+A_{\gamma \delta \beta}^{\alpha}+A_{\delta \beta \gamma}^{\alpha}-A_{\beta \delta \gamma}^{\alpha}-A_{\gamma \beta \delta}^{\alpha}-A_{\delta \gamma \beta}^{\alpha} \right]\,.$$

Similar definition holds for symmetric (skew-symmetric) property of contravariant indices.\\

A tensor is said to be symmetric (skew-symmetric) in a given set of contravariant or covariant indices if it is identical to its symmetrized (skew-symmetrized) part on those indices. In particular, a (0,2)-tensor $B$ is symmetric if
$$T_{\alpha \beta}=T_{\beta \alpha}~,~~i.e.,~~T_{[\alpha,\beta]}=0$$
while it is anti-symmetric if 
$$T_{\beta \alpha}=-T_{\alpha \beta}~,~~i.e.,~~T_{(\alpha,\beta)}=0\,.$$

It should be noted that the above symmetric\,(skew-symmetric) property of a tensor is independent of the choice of basis.\\

{\bf Convention:} So far and henceforth we introduce an index convention due to Penrose and is called abstract index notation. According to this convention, a vector or tensor is identified by its components without mentioning the basis. For example, a (1,1)-tensor $T$ will be represented by $T_{\beta}^{\alpha}$.\\

We now introduce a special type of tensors namely the set of tensors of the type (0,$b$) and are anti-symmetric in all the $b$ indices. Such a tensor is called a \underline{$b-$form}. We define a new tensor product known as \underline{wedge product} (which is an anti-symmetric tensor product) as follows:\\

If $P$ is an $a-$form and $Q$ is a $b-$form then their wedge product is denoted by $P\wedge Q$, a $(a+b)-$form, having components
$$(P\wedge Q)_{\alpha \ldots \beta , \gamma \ldots \delta}=P_{[\alpha \ldots \beta} Q_{\gamma \ldots \delta ]}\,.$$

It is clear from the above definition that
$$P\wedge Q=(-1)^{ab} (Q\wedge P)\,.$$

Further, if ${e^\alpha}$ is a basis for covectors or one-froms then $e^{\alpha_1} \wedge e^{\alpha_2} \wedge \ldots \wedge e^{\alpha _s}$ form a basis for $s-$forms and any $s-$form $A$ can be written as
$$A=A_{\alpha _1 \alpha _2 \ldots \alpha _s} e^{\alpha_1}  \wedge e^{\alpha_2} \wedge \ldots \wedge e^{\alpha _s}~~\mbox{with}~~A_{\alpha _1 \alpha _2 \ldots \alpha _s}=A_{[\alpha _1 \alpha _2 \ldots \alpha _s]}\,$$

{\bf Note:} The space of all $s$-forms for all $s$ (including scalars as zero form) constitutes the Grassmann algebra of forms.\\

{\bf Examples:}\\

We shall give some examples of tensors. Although the definition of a tensor is rather abstract but there are some very common examples of tensors.
\begin{enumerate}[(a)]
	\item Previously we have shown that a column matrix is a vector while a row matrix is a dual vector. Then the matrix is a (1,1)-tensor (by quotient law) as multiplication of a matrix by a vector (column matrix) or a dual vector (row matrix) gives a vector (or a dual vector), {\it \textit{i.e.},}
$$A(i,j)v^j=T^i~,~~w_i A(i,j)=S_j\,.$$
Also, if a matrix is operated (multiplied) by a vector and a dual vector (in the usual way) then the result will be a scalar.
\item In continuum mechanics, for a given stress material, if we imagine a plane passing through the material then the force per unit area exerted by the material on one side of the plane upon that on the other is characterized by $\tau ^{\mu \nu}$, a (2,0)-symmetric tensor and is known as stress tensor. The force is termed as stress vector. As a plane, {\it i.e.,} a surface is represented by a one-form so the stress tensor can be thought of as a linear vector valued function of one-forms.
\end{enumerate}

\section{Metric tensor and inner product}

~~~In a vector space, an inner product between two vectors is a bilinear function $g$ which assigns a real number with them:
$$g: L\times L\rightarrow \mathbb{R}\,,$$
{\it i.e.,} $g(i,j)v^i w^j=\lambda$, a scalar ($v^i$, $w^j$ are components of $V$ and $W$ $\in L$ in a particular basis $\{e_i\}$)\,.\\

From quotient law, it is clear that $g$ is a (0,2)-tensor called the metric tensor. So in a particular basis, we write
\begin{equation} \label{1.24}
g(V,W)=g(W,V) \equiv g_{\alpha \beta} V^{\alpha} W^{\beta}.
\end{equation}

It is evident that $g$ is a symmetric (0,2)-tensor and the components of $g$ are defined as 
\begin{equation} \label{1.25}
g_{\alpha \beta}=g(e_{\alpha},e_{\beta}).
\end{equation}

If the vector space is of dimension $n$ then the components of the metric tensor can be written as a $n\times n$ symmetric matrix. Normally, the metric is chosen to be non-degenerate, {\it i.e.,} $g(U,V)\neq 0$ for non-zero $U$, $V$ $\in L$. In terms of components this implies the matrix representation of the metric tensor to be non-singular. If the vector space is of dimension `$n$' then the components of the metric tensor can be written as a $n\times n$ symmetric matrix. The inverse matrix are the components of a (2,0)-tensor $g^{\alpha \beta}$ (such that $g^{\alpha \beta}g_{\beta \gamma}=\delta _{\gamma}^{\alpha}$) which is also symmetric and is called the reciprocal metric tensor. The transformation law of the components of the metric tensor for the choice of a new basis $\{e'_{\alpha}\}$ are given by
\begin{equation} \label{1.26}
g'_{\mu \nu}=\Lambda _{\mu}^{~~\alpha} \Lambda _{\nu}^{~~\beta} g_{\alpha \beta} ,
\end{equation}
or in matrix notation
\begin{equation} \label{1.27}
g'=\Lambda ^{T}g\Lambda .
\end{equation}

{\bf Reduction to canonical form:}\\

We shall now prove the following theorem:\\

{\bf Theorem:} In a vector space with a metric tensor, there always exists a basis in which the metric tensor has the canonical form $diag \lbrace -1,-1,\cdot \cdot \cdot -1,+1,\cdot \cdot \cdot +1 \rbrace$.\\

{\bf Proof}: Suppose we choose the arbitrary matrix $\Lambda$ in Eq. (1.26) as the product of an orthogonal matrix $U$, {\it i.e., $U^{-1}=U^{T}$} and a diagonal matrix $D$ (self transpose), {\it i.e.,} $\Lambda =UD$ then Eq. (1.27) can be written as 
$$g'=DU^{-1}gUD\,.$$

From the property of similarity transformation of symmetric matrix, we can reduce $g_{u}=U^{-1}gU$ in the diagonal form and hence $g'$ becomes diagonal. In particular, if 
$$g_u=\mbox{diag}(g_{\alpha _1}, g_{\alpha _2},\ldots, g_{\alpha _n})~~\mbox{and}~~D=\mbox{diag}(d_{\alpha _1}, d_{\alpha _2},\ldots, d_{\alpha _n})$$
then
$$g'=\mbox{diag}(g_{\alpha _1} d_{\alpha _1}^{2}, g_{\alpha _2} d_{\alpha _2}^{2},\ldots , g_{\alpha _n} d_{\alpha _n}^{2})\,.$$

Thus choosing $d_{\alpha _k}={\lbrace |g_{\alpha _k}|\rbrace }^{-\frac{1}{2}}$, we have the elements of the diagonal matrix $g'$ as $+1$ or $-1$. In fact, the elements $g_{\alpha _1}$ of the diagonal matrix $g_u$ are the eigen values of the metric tensor $g$ and hence they are unique except for the order. Further, due to existence of the inverse matrix, the eigen values are non-zero. Thus by choosing the orthogonal matrix $U$ appropriately it is possible to arrange the metric tensor in the canonical form $\mbox{diag} \lbrace -1,-1,\cdot \cdot \cdot -1,+1,\cdot \cdot \cdot +1 \rbrace$.\\

\underline{{\bf Note I:}} The basis for which the metric tensor is reduced to canonical form is known as orthonormal basis. The trace of the canonical form, {\it i.e.,} the sum of the diagonal elements is called the \underline{signature} of the metric.\\

\underline{{\bf Note II:}} If the absolute value of the signature of the metric is equal to the dimension ($n$) of the vector space then it is said to be the Euclidean space ($R^n$). In this case the canonical form of the metric is either $\mbox{diag} \lbrace +1, +1, \ldots , +1 \rbrace $ or $\mbox{diag} \lbrace -1, -1, \ldots , -1\rbrace$. In Euclidean space, the orthonormal basis is called Cartesian and for which $g_{ij}=\delta _{ij}$, {\it i.e.,} in matrix form $g=I$. For a transformation matrix $\Lambda _E$ from one such basis to another, we have
$$I=\Lambda _{E}^{T}I\Lambda _{E},~{\it i.e.,}~~\Lambda _{E}^{T}\Lambda _{E}=I\,,$$

which shows that the transformation matrices are orthogonal. This class of orthogonal matrices forms a group, called Euclidean symmetry group or simply the \underline{orthogonal group} $O(n)$.\\

\begin{wrapfigure}[11]{r}{0.4\textwidth}\vspace{-\intextsep}
	\includegraphics[height=5.85 cm , width=7 cm ]{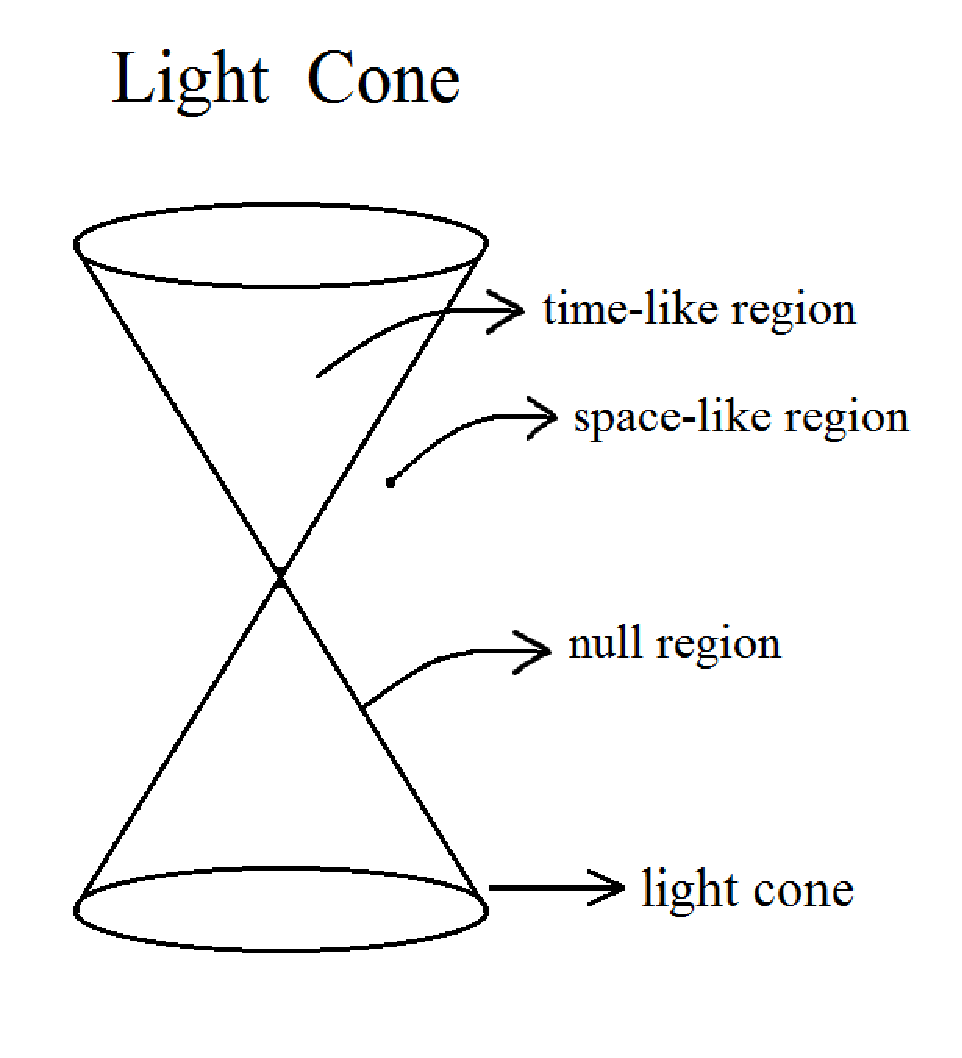}\vspace{-\intextsep}
	\begin{center}
		Fig. 1.1
	\end{center}\vspace{-\intextsep}
\end{wrapfigure}
\underline{{\bf Note III:}} If the absolute value of the signature of the metric is less than the dimension of the vector space then the metric is called indefinite. In particular, for a $n$ dimensional vector space, if the signature of the metric is $n-2$ (in absolute value) then it is called a \underline{Lorentzian metric}. So the canonical form of the metric is either $diag \lbrace -1, +1, +1,\ldots , +1 \rbrace $ or $diag \lbrace +1, -1, -1, \ldots , -1\rbrace$. A four dimensional Lorentzian metric is called a \underline{Minkowski metric} and the corresponding vector space is called the Minkowski space$-$the space of Einstein's special theory of relativity.\\

If we denote the matrix corresponding to the Lorentz metric by $\eta$, {\it i.e.,}
$$\eta=\mbox{diag}\lbrace -1, +1, +1,\ldots , +1 \rbrace \mbox{~or~diag}\lbrace +1, -1, -1, \ldots ,-1\rbrace$$
then the transformation matrix $\Lambda _L$ from one Lorentz basis to another satisfies
$$\eta = \Lambda _{L}^{T}\eta \Lambda _L\,.$$

These transformation matrices correspond to Lorentz transformation in special theory of relativity and the group formed by them is called the \underline{Lorentz group} $L(n)$.\\

\underline{{\bf Note IV:}} In Minkowski space (having Lorentz metric), the non-zero vectors at any point can be classified into three cases: time-like, null (light-like) and space-like. A non-zero vector $\textit{\textbf{v}}$ can be classified as follows:
$$g(\textit{\textbf{v}},\textit{\textbf{v}})<0~~(\mbox{time-like})~,~~~=0~~(\mbox{null})~,~~~>0~~(\mbox{space-like})\,.$$

The set of all light-like vectors form the surface of a double cone and is termed as \underline{null cone} or \underline{light cone}. This light cone separates the time-like and the space-like vectors. All vectors inside the light cone are time-like while space-like vectors lie outside the light cone (see figure 1.1).\\

\section{Manifolds}

~~~Given a set\,(topological space) $M$, if there exists a one-one mapping $f$ from an open subset $U$ of $M$ onto an open subset of $\mathbb{R}^n$ (the mapping is a homeomorphism), then $M$ is said to be a \underline{manifold} of dimension $n$. This definition of a manifold suggests that the set $M$ looks locally like $\mathbb{R}^n$ but globally they are quite distinct. From the above definition, it is clear that there will be other open subsets of $M$ with their own maps and any point of $M$ must lie in at least one such open subsets. The pair namely an open subset and a mapping, {\it i.e.,} ($U$,$f$) is called a \underline{chart}. A collection (class) of charts is called an \underline{atlas}, provided every point of $M$ is in at least one open subset of the class and for any two overlapping open subsets in the class, there exists functional relation between the corresponding mappings. In fact, if any two overlapping charts in an atlas are $C^k-$related then the manifold is said to be a \underline{$C^k-$manifold}. A manifold of class $C^1$ is called a \underline{differentiable manifold}.\\

For a more rigorous mathematical definition of a manifold let us start with a topological space having the following properties:\\

a)~\underline{Hausdorff property:} A topological space is said to be a Hausdorff space if any pair of distinct points in it has disjoint neighbourhoods.\\

b)~\underline{Second countable property:} A class of open sets of a topological space is said to form an open base of the topological space if any open set of the space is a union of elements of this class. A topological space with a countable open base is called a second countable space.\\

A Hausdorff, second countable topological space in which every point has a neighbourhood homeomorphic to an open set in $\mathbb{R}^n$, is called a manifold of dimension $n$.\\

\begin{figure}[h]
\begin{center}
\includegraphics[height=5 cm , width=7 cm ]{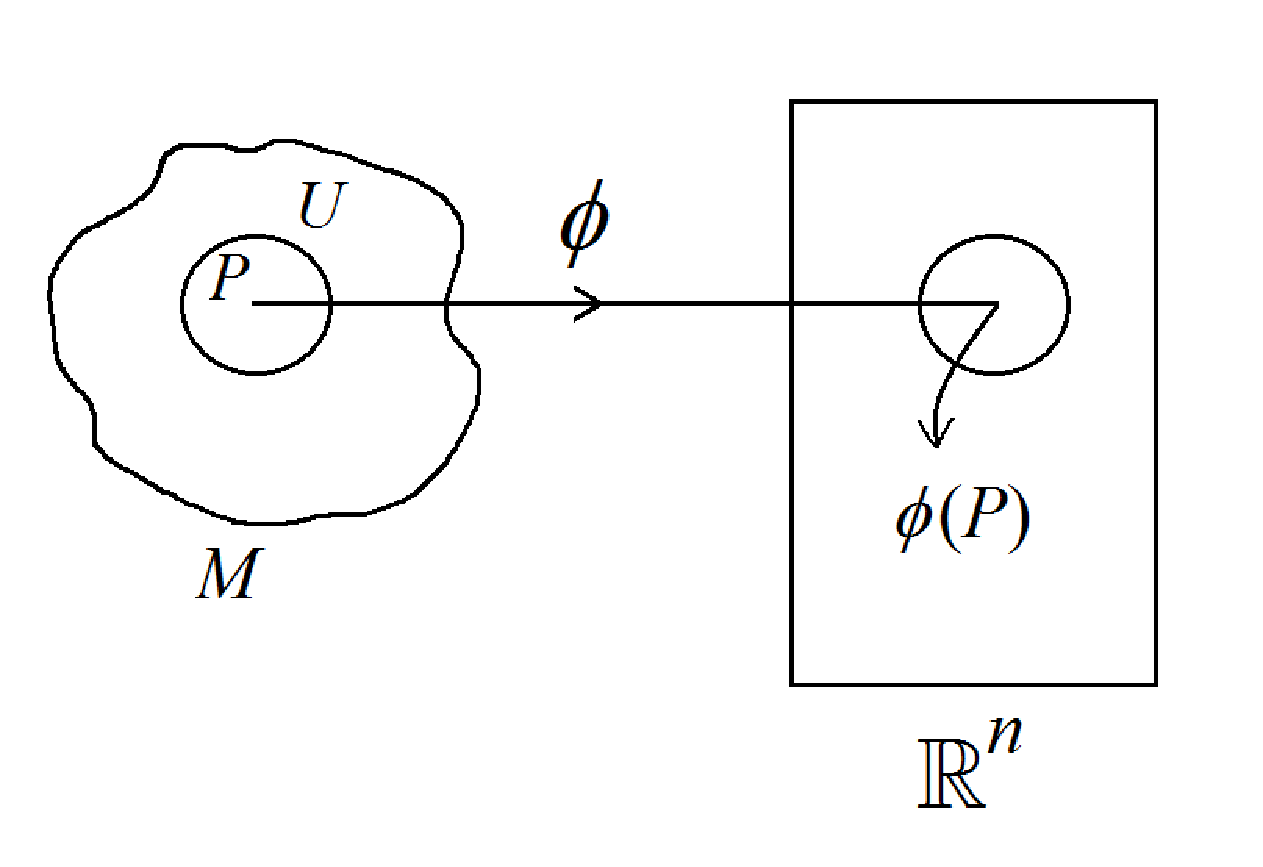}~~~~~~
\includegraphics[height=6 cm , width=8.5 cm ]{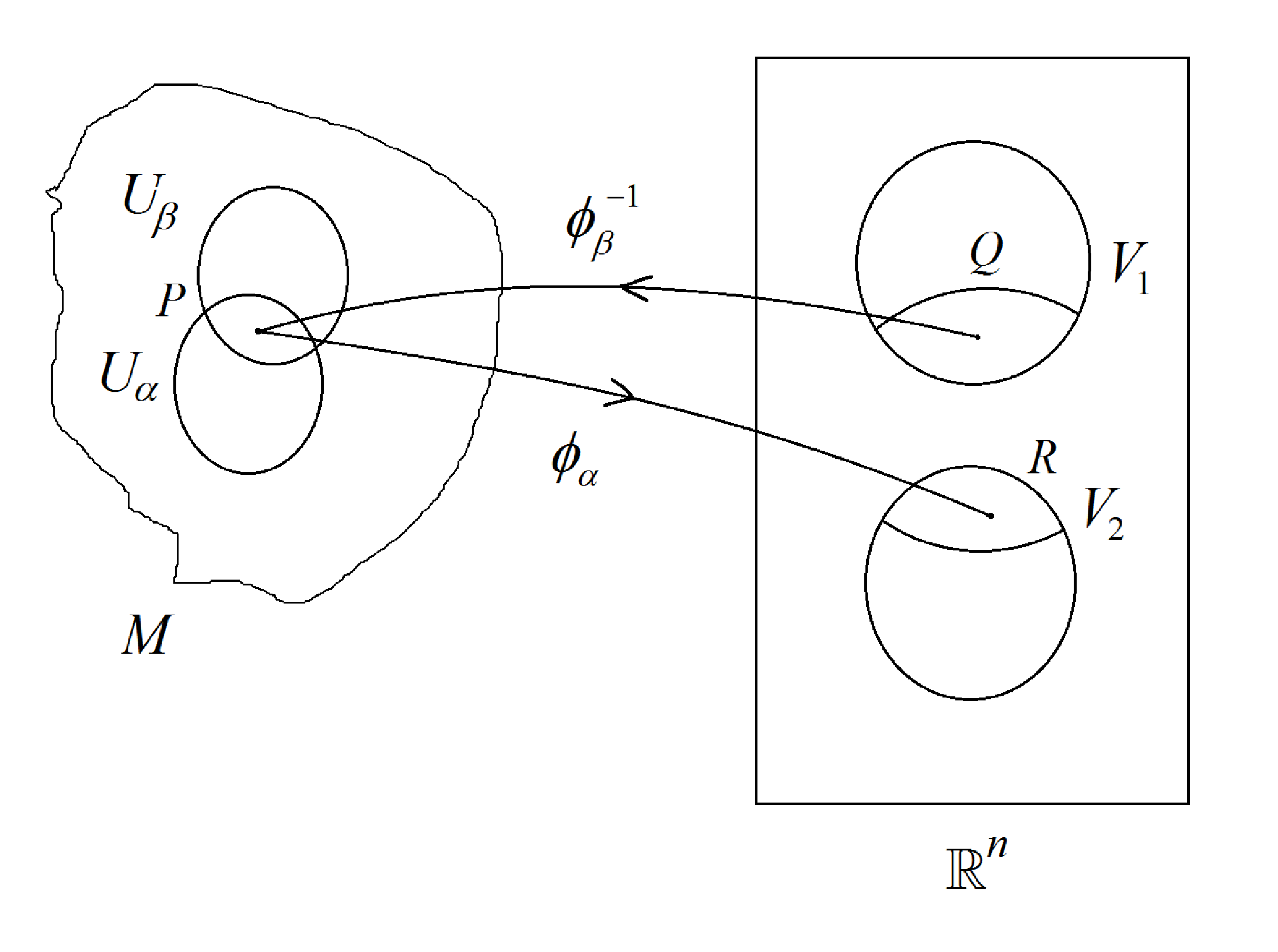}
\end{center}\vspace{-\intextsep}
\begin{center}
Fig. 1.2~~~~~~~~~~~~~~~~~~~~~~~~~~~~~~~~~~~~~~~~~~~~~~~~~~Fig 1.3
\end{center}\vspace{-\intextsep}
\end{figure}

In  figure1.2 $U$ is a neighbourhood\,(nhb) of any point $P$ of the manifold $M$ and $\phi$ is a homeomorphism from $U$ to an open set in $\mathbb{R}^n$. So the pair $(U,\phi)$ is called a chart at $P$.\\

Mathematically, a collection of charts $\{u_\alpha,\phi _\alpha\}$ is said to form a $C^r-$atlas if i) the class of subsets $\{u_\alpha\}$ cover $M$, {\it i.e.,} $M=\bigcup _\alpha U_\alpha$ and ii) for $U_\alpha \bigcap U_\beta \neq \phi$, the map $\phi _\alpha \circ \phi _{\beta}^{-1}: \phi _\beta \left( U_\alpha \bigcap U_\beta \right) \longrightarrow \phi _\alpha \left( U_\alpha \bigcap U_\beta \right)$ is a $C^r-$map of an open subset of $\mathbb{R}^n$ to an open subset of $\mathbb{R}^n$.\\

The figure clearly shows how two different regions of $\mathbb{R}^n$ are related by the mapping in the manifold $M$. Let $Q$ and $\mathbb{R}$ are the image points of $P$ under the mappings $\phi _\beta$ and $\phi _\alpha$ respectively. Being a point in $\mathbb{R}^n$ let the points $Q$ and $\mathbb{R}$ are represented by ($x^1$, $x^2$,$\cdot$ $\cdot$ $\cdot$ $x^n$) and ($y^1$, $y^2$,$\cdot$ $\cdot$ $\cdot$ $y^n$) respectively. Then the above mapping demands that
$$y^i=y^i(x^j)~~\mbox{or}~~x^k=x^k(y^l)~,~~~(i,j,k,l=1,2,\ldots ,n)\,.$$

For a $C^r-$atlas these functions have partial derivative upto order $r$. A manifold having a $C^r-$atlas is called a \underline{$C^r-$manifold}.\\

An atlas containing every possible compatible chart is called a \underline{maximal atlas}. A manifold can be defined as a set with a maximal atlas.\\

\underline{\textbf{Note 1:}} The necessity of the atlas to be maximal is that two equivalent spaces equipped with different atlases do not count as different manifolds.\\

\underline{\textbf{Note 2:}} A Hausdorff, second countable topological space admits partitions of unity and thereby Riemannian metric can be defined\,(which is our main interest). However, for studying manifold it is enough to consider only a topological space.\\

\underline{\textbf{Note 3:}} Two manifolds $M$ and $N$ are said to be equivalent if their local geometry is same. But globally the two manifolds are not necessarily identical. For example, the manifolds $S^2$ and $\mathbb{R}^2$ are locally equivalent but globally they are distinct.\\

\underline{\textbf{Note 4:}} A manifold can be considered as a set $M$ that can be parameterized continuously and the dimension of the manifold is the number of independent parameters involved.\\

\underline{\textbf{Note 5:}} We shall mostly deal with local geometry depending on the differential structure of the manifold. The global properties of the manifold are needed in studying fiber bundles and integration of functions.\\

\underline{\textbf{Note 6:}} The differentiability of a manifold gives an enormous structure in it. Subsequently, we shall study some of these differential structure. It should be mentioned that we have not introduced any notion of `distance' on the manifold nor the notion of shape or `curvature' of the manifold$-$only ingredient is the locally smooth nature.\\

{\bf Examples of manifold:}
\begin{enumerate}[(a)]
	\item The surface of a sphere (known as $S^2$) is a manifold of dimension two. There is always a homeomorphism from any open set of $S^2$ to an open set of $\mathbb{R}^2$ i.e., any point in $S^2$ has a sufficiently small neighbourhood which has a one-one correspondence onto a disc in $\mathbb{R}^2$. Although $S^2$ and $\mathbb{R}^2$ are clearly different (global properties) but neighbourhood of a point in $S^2$ looks very much like a neighbourhood of a point in $\mathbb{R}^2$. 

\item A vector space $V$ of dimension $n$ over the real numbers is a manifold. Here in a particular basis $\{e_\alpha\}$ any vector $v$ can be written as $v=a^\alpha e_\alpha$. Thus there is a mapping from $V-\mathbb{R}^n$: $v\rightarrow$($a^1,\cdot \cdot \cdot a^n$).

\item For an algebraic (or differential) equation with one dependent variable $y$ and an independent variable $x$, the set of all ($y$,$x$) satisfying the above equation forms a manifold. Here a particular solution is a curve in the manifold.

\item For a $n$ particle system, the phase space, consisting of positions and momentum\,(velocities) is a $6n-$dimensional manifold.

\item The set of all pure Lorentz transformations\,(boost) is a 3D manifold with components of the velocity of the boost as the parameters.
\end{enumerate}

\section{Differentiable mapping}

~~~Let $M_1$ and $M_2$ be two differentiable manifolds of dimensions $n_1$ and $n_2$. A mapping $f: M_1 \rightarrow M_2$ is said to be a \underline{differentiable mapping} of class $C^k$ if for every chart ($U_1$,$\phi _1$) containing any point $P$ of $M_1$ and every chart ($U_2$,$\phi _2$) containing the corresponding point $f(P)$ of $M_2$ and such that\\
$~~~~~~~~$i) $f(U_1)\subset U_2$\\
and~~ii) the mapping $\phi _2 \circ f \circ \phi _{1}^{-1}$: $\phi _1(U_1)\subset \mathbb{R}^{n_1}\rightarrow \phi _2(U_2)\subset \mathbb{R}^{n_2}$, is of class $C^k$.\\

A mapping $f$: $M\rightarrow N$ is called \underline{diffeomorphism} if i) $f$ is a bijection and ii) both $f$ and $f^{-1}$ are differentiable mapping. Then $M$ and $N$ are said to be \underline{diffeomorphic} to each other.\\

{\bf Note:} A diffeomorphism $f$ of $M$ onto itself is called a transformation of $M$.\\

A real-valued function on $M$, {\it i.e.,} $f$: $M\rightarrow \mathbb{R}$ is said to be a \underline{differentiable function} (of class $C^k$) if for every chart ($U$,$\phi$) containing $P \in M$, the function $f \circ \phi ^{-1}$: $\phi (U) \subset \mathbb{R}^n \rightarrow \mathbb{R}$ is of class $C^k$.\\

\underline{\bf{Note:}}
\begin{enumerate}[\bfseries a.]
	\item  A $C^\infty$ differentiable function is also called a smooth function.
	\item Two diffeomorphic manifolds can be considered as two distinct copies of a single abstract manifold and hence they are said to be equivalent. It is similar to the notion of isomorphism  in groups$-$two groups can be regarded as the same group if they are isomorphic to each other.
\item The set of all diffeomorphisms of a manifold $M$ onto itself ({\it i.e.,} transformations of $M$) forms a group denoted as Diff($M$). This group plays an important role in various branches of modern theoretical physics ({\it e.g.} loop quantum gravity).
\item The set of all differentiable functions on $M$ (denoted by $F(M)$) from i) an algebra over $\mathbb{R}$, ii) a ring over $\mathbb{R}$.
\end{enumerate}

\section{Curves on a manifold\,: Tangent spaces}

~~~A curve $\gamma$ through a point $P$ in $M$ is a differentiable mapping $\sigma : [\mu,\lambda]\subset \mathbb{R}\rightarrow M$ such that $\sigma (t_0)=P$ ($\mu < t_0 < \lambda$).\\

\underline{Note:} Two distinct differentiable mappings from $[\mu,\lambda]\rightarrow M$ gives two distinct curves in $M$.\\

The tangent vector to the curve $\gamma$ at $P$ is the function 
$$X_P: F(P)\rightarrow R$$
which is defined as
$$X_P f=\left[\frac{d}{dt}f(\sigma (t))\right]_{t=t_0}=\left\lbrace \lim_{\delta \rightarrow 0} \left[\frac{f(\sigma (t+\delta))-f(\sigma (t))}{\delta}\right]\right\rbrace _{t=t_0}~.$$
\begin{wrapfigure}[9]{r}{0.3\textwidth}\vspace{-\intextsep}
\includegraphics[height=4.3 cm , width=5 cm ]{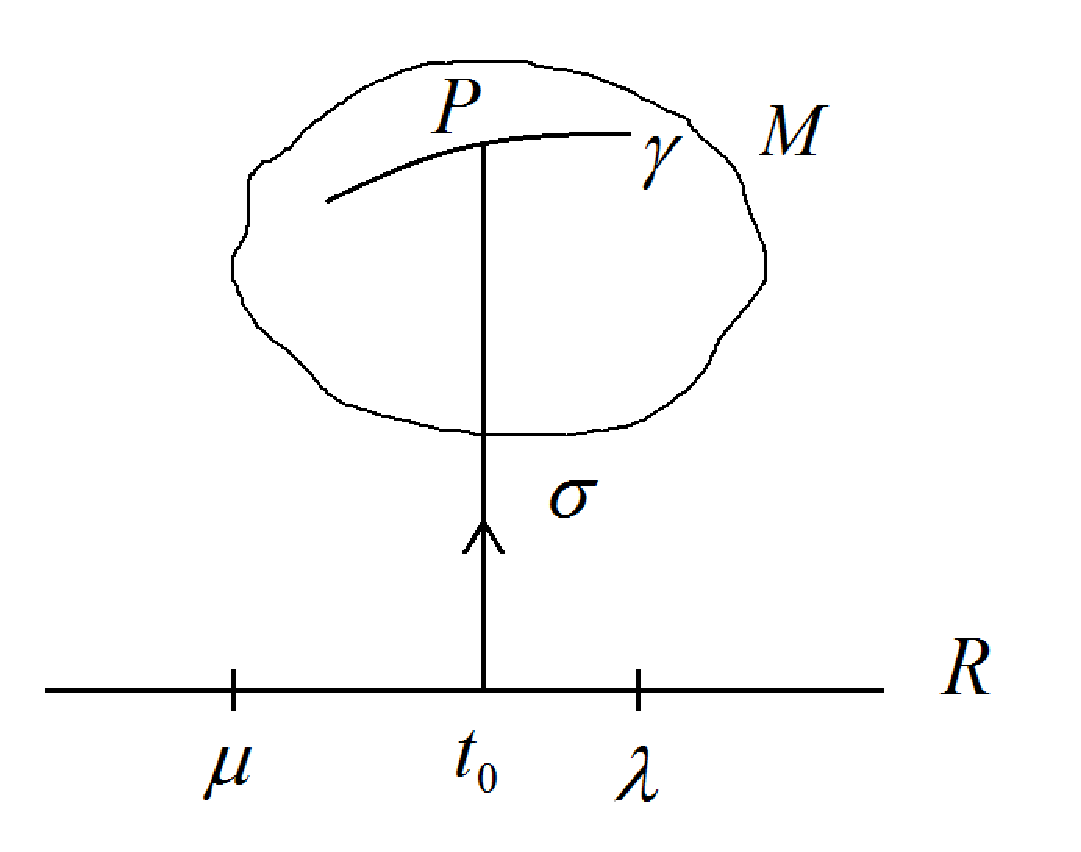}\vspace{-\intextsep}
\begin{center}
Fig. 1.4
\end{center}
\end{wrapfigure}

Here $f \in F(P)$ is the set of all differentiable functions on $M$ at $P$. From the property of the usual derivative operator (on the real line), the tangent vector $X_P$ is a linear function and obey Leibnitz product rule, {\it i.e.,}\\

i) $X_P(af+bg)=a(X_P f)+b(X_P g)$~~(linearity)\\

and
ii) $X_P (fg)=g(P)X_P f + f(P)X_P g$~(Leibnitzian property)\\
for all $f$, $g$ $\in F(P)$ and $a$, $b$ $\in R$.\\

\underline{Note:} Each function $X_P: F(P)\rightarrow \mathbb{R}$, defined above cannot be a tangent vector to some curve at $P$ unless it is a linear function and satisfies Leibnitz product rule.\\

The set of all tangent vectors to $M$ at $P$ is a vector space over $\mathbb{R}$. This vector space is called the tangent space and is denoted by $T_P(M)$. The dimension of the tangent space is same as that of the manifold itself.\\

A \underline{vector field} $X$ on $M$ is a rule that associates to each point $P \in M$, a vector $X_P \in T_P(M)$. Thus, if $f \in F(M)$, the set of all differentiable functions on $M$, then $Xf$ is defined to be a real-valued function on $M$ as
$$(Xf)(P)=X_P f.$$

The vector field $X$ is called differentiable if $Xf$ is differentiable for every $f \in F(M)$. If $\chi (M)$ denotes the set of all differentiable vector fields on $M$ then i) $\chi (M)$ is a vector space over $\mathbb{R}$ and ii) for every $f \in F(M)$, $fX$ is defined to be a vector field on $M$, defined as $(fX)(P)=f(P)X_P$.\\

A curve  $\sigma$ is called an \underline{integral curve} of a vector field if the tangent to the curve at every point is the vector corresponding to the vector field.\\

\underline{\bf{Note I:}} If ($x^1$,$x^2$, $\cdot$ $\cdot$ $\cdot$ $x^n$) be a local coordinate system in a nhb $U$ of $P \in M$, then the basis of the tangent space $T_P(M)$ is given by $\left\lbrace \left(\dfrac{\partial}{\partial x^1}\right)_P, \left(\dfrac{\partial}{\partial x^2}\right)_P,\cdot \cdot \cdot \left(\dfrac{\partial}{\partial x^n}\right)_P\right\rbrace$. Thus the tangent vectors of the coordinate curves through $P$ form a basis of $T_P$ and are denoted by the partial derivative operators at $P$.\\

\underline{\bf{Note II:}} The tangent spaces at different points of the manifold are distinct vector spaces and their elements are unrelated.\\

\underline{\bf{Note III:}} The collection of all tangent spaces over the manifold $M$ is called the \underline{tangent bundle}, denoted by $TM$. Thus $TM=\bigcup\limits _{P\in M}T_P(M)$. One can define a natural projection map $\Pi : TM\rightarrow M$ which relate each tangent vector to the point on the manifold at which it is the tangent. The inverse mapping associates to every point $P \in M$, the set of all tangents at $P$, {\it i.e.,} $T_P(M)$. This inverse mapping is called the fibre over $P$ in the context of fibre bundle.\\

\textbf{Cotangent space: Covector and dual vector}\\

We remind that $\chi (M)$, the set of all differentiable vector fields on $M$, is a vector space over $R$. Also, the set of all differentiable functions on $M$ is a vector space over $R$ and is denoted by $F(M)$.\\

Let us consider a map $w: \chi (M)\rightarrow F(M)$, that satisfies\\
~i) $w(X+Y)=w(X)+w(Y)$\\
~ii) $w(bX)=bw(X)$~~~$\forall b \in R, X,Y \in \chi (M)$\\
~iii) $(w_1+w_2)X=w_1(X)+w_2(X)$\\
then $w$ is called a linear mapping over $R$. Usually, a linear mapping $w: \chi (M)\rightarrow F(M)$ denoted by $w: X\rightarrow w(X)$ is called a \underline{1-form} on $M$.\\

The set of all one-forms on $M$ denoted by $D_1(M)$, is a vector space over $R$, called the \underline{dual} of $\chi (M)$. As $w(X)\in F(M)$, {\it i.e.,} $w(X): M\rightarrow R$, so for any point $P$ of $M$, we have
$$\lbrace w(X) \rbrace(P)=w_P(X_P),~~i.e.,~~w_P: T_P(M)\rightarrow R.$$

The collection of all $w_P$, {\it i.e.,} the collection of all one-form\,(dual vectors) at $P$, is a vector space known as \underline{cotangent space} or dual of the tangent space and is denoted by $T_{P}^{*}M$. So elements of 
$T_{P}^{*}M$, {\it i.e.,} covectors at $P$ are linear functionals of $T_P(M)$.\\

For any function $f \in F(M)$, we denote the total differential of $f$ by $df$ and is defined as
$$(df)_P(X_P)=(Xf)(P)=X_P f~,~~\forall P,$$
$$i.e.,~~df(X)=Xf.~~~~~~~~~~~~~~~~~~~~~~~~~~~~~~~~~~~~~$$

\underline{\bf{Note I:}} $df$ is a one-form on $M$.\\

\underline{\bf{Note II:}} If ($x^1$,$x^2$, $\cdot$ $\cdot$ $\cdot$ $x^n$) are coordinate functions defined in a nhb of $P$ (in $M$) then $dx^i$, $i=1,2, \ldots , n$ are 1-form on $M$\,(for each $i$) and they form a basis of $T_{P}^{*}M$.\\

\underline{\bf{Note III:}} The linearity property of the action of covectors on vectors enables us to regard vectors and covectors as dual of each other. Their action (or value) on one another is notationally represented by
$$w(V)\equiv V(w)\equiv \left\langle w,V \right\rangle.$$
The action $w(V)$ is also called the contraction of $w$ with $V$.\\

\section{Metric tensor on a manifold\,: Metric tensor field}

~~~A metric tensor field $g$ on a manifold $M$ is a symmetric tensor of type (0,2) such that it behaves as a metric on the tangent space $T_P$ at every point $P$ of $M$. We term this (0,2) tensor on the manifold as the metric of the manifold so that it is possible to define the notion of the distance between two points on the manifold and the curvature of the space\,(will be discussed in Sec. 1.11.). The differentiability of the metric tensor is essential for defining more structure on the manifold. So the metric tensor must be at least continuous. As a result, the canonical form of the metric tensor is same throughout the manifold and hence the signature is fixed. This is called the signature of the manifold.\\

Suppose $\gamma$ is a curve on the manifold with parameter $\lambda$, {\it i.e.,} $\gamma : x^\alpha =x^{\alpha} (\lambda)$. So the tangent vector at any point on the curve is $t^\alpha =\dfrac{dx^\alpha}{d\lambda}$. Then distance between two infinitesimal points on the curve is defined as
$$dS^2=g_{\alpha \beta}t^\alpha t^\beta (d\lambda)^2 =g(t,t)(d\lambda)^2.$$

If the metric is positive definite ({\it i.e.,} signature=dimension of the manifold) then $g(t,t)$ is positive and hence
$$dS={\lbrace g(t,t) \rbrace}^{\frac{1}{2}}d\lambda,$$
is the length of an element of the curve. However, for indefinite metric $dS^2$ is not of definite sign. For space-like curve (curve having tangent vector is a space-like vector) $dS^2$ is positive while it will be negative for time-like curve and the magnitude
$$dS=\left|g(t,t)\right|^{\frac{1}{2}}d\lambda$$
defines the proper distance for space-like curves and proper time for time-like curves. Also $dS=0$ for null curves.\\

\textbf{Note:} In case of indefinite metric, a null vector and a zero vector must be handled carefully. A null vector $n^\alpha$ has a zero norm ({\it i.e.},~ $g(n,n)=g_{ab}n^a n^b=0$) while a zero vector has all its components identically zero.\\

\section{Differential structure on the manifold}

~~~We shall now discuss three differential operators on manifolds namely a) Lie derivative, b) Exterior differentiation and c) Covariant differentiation. For the first two types of differential operators, manifold structure is sufficient while for covariant differentiation we need extra structure on the manifold\,(known as connection).\\

\subsection{Lie Derivative}
~~~Before defining Lie derivative we first introduce the idea of Lie dragging. We recapitulate that a vector field is a rule by which we get a vector at every point of the manifold. Given a vector field, an integral curve is a curve on the manifold such that the tangent vector at every point on it is the vector corresponding to the vector field. A family of integral curves which fill the whole (or a part of the) manifold is called a \underline{congruence}. So it is clear that integral curves do not intersect each other, {\it i.e.,} through each point of the manifold there exists one and only one integral curve of the congruence. Also, it is evident that congruence generates a natural mapping of the manifold into itself.\\

Suppose we consider a typical integral curve of the congruence which is parametrized by $\lambda$. For an infinitesimal small number $\Delta \lambda$ we can imagine a mapping along the integral curve so that each point on it is shifted to another point (on it) a parameter distance $\Delta \lambda$. Clearly, this mapping is a one-one mapping and can be termed as diffeomorphism provided the vector is differentiable ({\it i.e.,} $C^\infty$). If this mapping is possible for all $\Delta \lambda$ then we have a family of differentiable mappings, known as dragging along the congruence or a Lie dragging. We shall show below the conditions for Lie dragging of a scalar function and a vector field.\\
\begin{wrapfigure}[9]{r}{0.27\textwidth}\vspace{-2\intextsep}
	\includegraphics[height=4 cm , width=5 cm ]{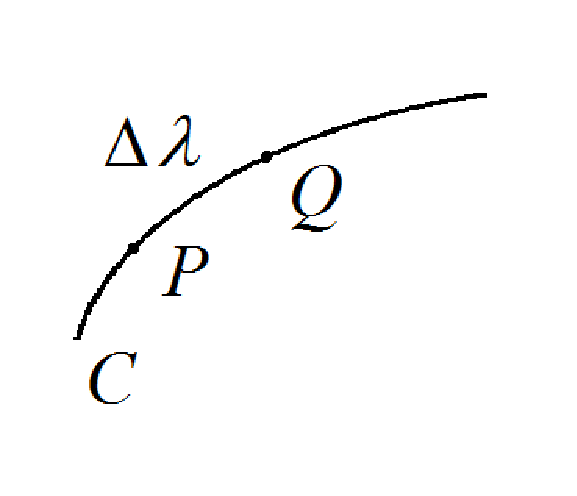}\vspace{-\intextsep}
	\begin{center}
		Fig. 1.5
	\end{center}\vspace{-\intextsep}
\end{wrapfigure}
{\bf~~ Scalar function:} Let $f$ be a function defined on the manifold $M$ and suppose $C$ is a typical integral curve of a congruence, parameterized by $\lambda$. Let $P$ and $Q$ be two points on $C$ separated by an infinitesimal parameter distance $\Delta \lambda$. Now due to Lie dragging, the function is carried along $C$ and as a result a new function ($f_d$) is defined such that its value at $Q$ is same as the value of $f$ at $P$, {\it i.e.,} $f_d(Q)=f(P)$. But if it so happens that the dragged function has the same value at $Q$ as the old one and it is true for all $Q$ along $C$ then we say that the function is invariant under this dragging or simply the function is \underline{Lie dragged}. So if a function is Lie dragged along any congruence ({\it i.e.,} an integral curve of the congruence) then it must be constant along it, {\it i.e.,} $\dfrac{df}{d\lambda}=0$.\\

{\bf Vector field:} Let $V$ be a vector field in a manifold $M$ and it generates the congruence having integral curves $\alpha$, $\beta$, $\gamma$, $\delta$, $\cdot \cdot \cdot$. Suppose $W$ be another vector field which we want to Lie dragged and $a$, $b$,$\cdot \cdot \cdot$ are the integral curves corresponding to $W$. As before let $\lambda$ be the parameter along the integral curve of $V$. The points $P_1$, $P_2$, $P_3$, $P_4$, $\cdot \cdot \cdot$ are the points of intersection of the integral curve $a$ (of $W$) with the integral curves $\alpha$, $\beta$, $\gamma$, $\delta$ (of $V$). These points are mapped to the points $Q_1$, $Q_2$, $Q_3$, $Q_4$, respectively by dragging through an infinitesimal parameter $\Delta \lambda$ (see fig. 1.6). Thus the integral curve $a$ is dragged to $a'$, an integral curve of another vector field (say $W'$). If $a'$ coincide with an integral curve of $W$ (say $b$ in the figure) and it is true for all $\Delta \lambda$ then we say that the vector field $W$ (and its congruence) is Lie dragged by the vector field $V$.
\begin{wrapfigure}[12]{r}{0.53\textwidth}\vspace{-\intextsep}
	\includegraphics[height=5 cm , width=9 cm ]{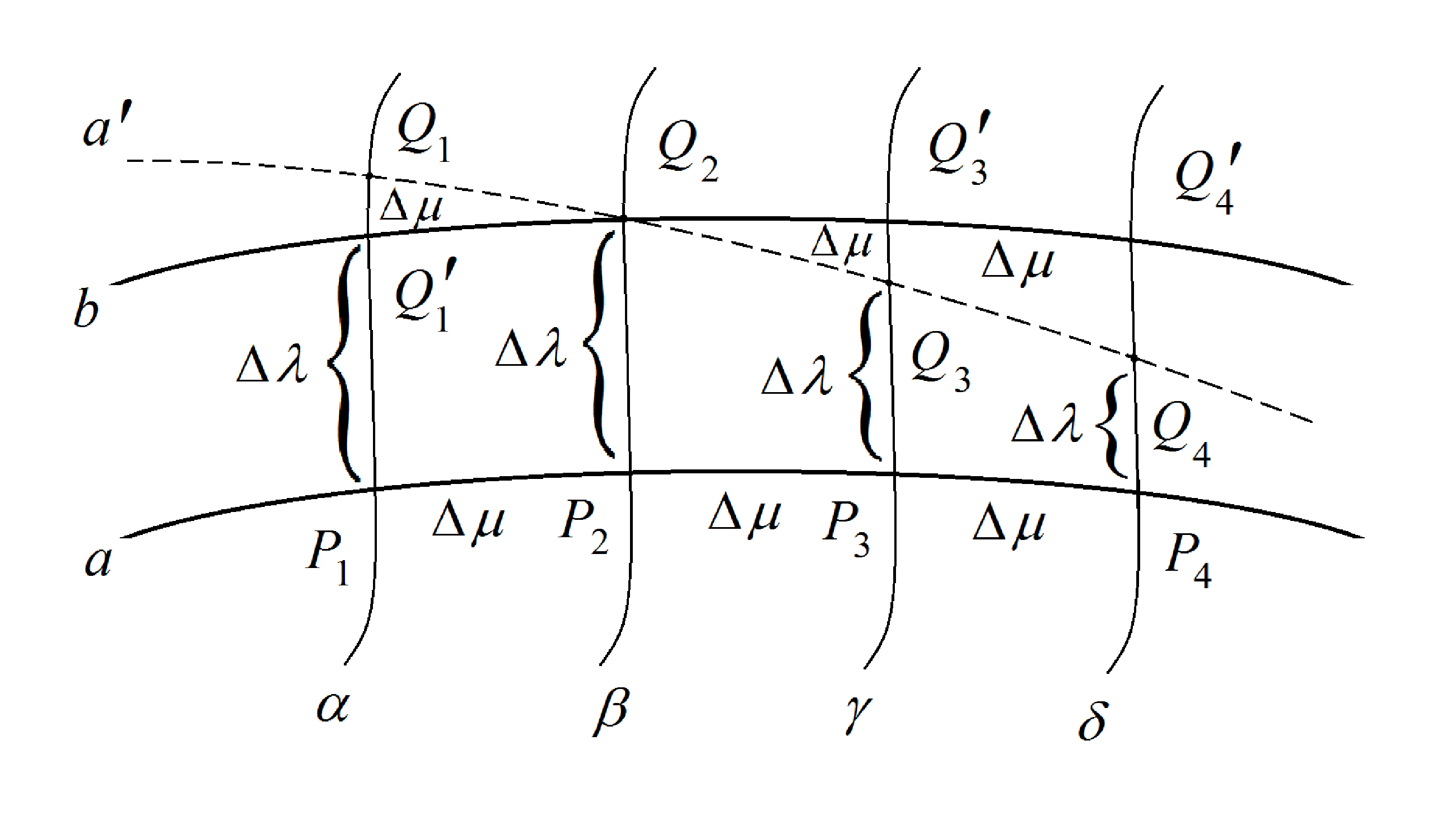}\vspace{-\intextsep}
	\begin{center}
		Fig. 1.6
	\end{center}\vspace{-\intextsep}
\end{wrapfigure}

There is a nice geometrical picture of Lie dragging of a vector field with respect to another vector field. Let $\lambda$ and $\mu$ be the parameters corresponding to the vector fields $V$ and $W$ respectively. Suppose the point $P_2$ is Lie dragged along the congruence of $V$ to the point $Q_2$ by a parameter distance $\Delta \lambda$ and then $Q_2$ is Lie dragged to the point $Q'_3$ along the integral curve $b$ of $W$ through a parameter distance $\Delta \mu$. On the other hand, if we first Lie dragged the point $P_2$ along $a$, the integral curve of $W$ through a parameter distance $\Delta \mu$ and reach to the point $P_3$ and subsequently $P_3$ is Lie dragged to $Q_3$ along the congruence of $V$ through the parameter distance $\Delta \lambda$. In general, $Q_3$ and $Q'_3$ are distinct points but if the vector field $W$ is Lie dragged by $V$ then the dragged integral curve $a'$ coincides with the integral curve $b$ and consequently $Q_3$ and $Q'_3$ are same point. Therefore, successive Lie dragging of two vector fields with respect to each other does not depend on the order of the vector field. So mathematically, we write
$$[V,W]=0\,.$$

Using this idea of Lie dragging, we now introduce the notion of Lie derivative which is essentially a derivative operator along a given congruence. In general, there are two inherent difficulties to define derivatives of vectors and tensor fields. The primary difficulty is that there is no mechanism to compare vectors (or tensors) at different points on the manifold (in Euclidean space, one set of basis vectors cover the whole space, so comparison of vectors at different points has no problem). This difficulty is resolved using the idea of Lie dragging along the congruence of a given vector field. The second problem is to define distance between points (in Euclidean space or manifold with metric distance is well defined). However, we consider the congruence of the given vector field and derivative is defined only along the congruence by defining the distance as the difference between the parameter values at the points on the congruence. Therefore, it is possible to define the Lie derivative only along the congruence of a given vector field as follows:\\

Let $\textit{\textbf{V}}$ be a given vector field whose congruence is parameterized by $\lambda$. For a scalar function having values $f(\lambda)$ and $f(\lambda +\Delta \lambda)$ at parameter points $P(\lambda)$ and $Q(\lambda +\Delta \lambda)$, we define a new function $f_d$ by dragging $f$ from $Q$ to $P$ so that $f_d(P)=f(Q)$, {\it i.e.}, $f_d(\lambda)=f(\lambda +\Delta \lambda)$. Then the Lie derivative of $f$ at $P$ is denoted by $L_{_V}f$ and is defined as
$$\lim_{\Delta \lambda \rightarrow 0}\frac{f_d(\lambda)-f(\lambda)}{\Delta \lambda}=\lim_{\Delta \lambda \rightarrow 0}\frac{f(\lambda +\Delta \lambda)-f(\lambda)}{\Delta \lambda}=\frac{df}{d\lambda}~,$$
which is the ordinary derivative as expected because a scalar function is frame independent. In particular, if $f$ is Lie dragged by the vector field then $\dfrac{df}{d\lambda}=0$ and hence $L_{\textit{\textbf{V}}}f=0$.\\

Suppose $\textit{\textbf{W}}$ be another vector field whose congruence is characterized by arbitrary parameter $\mu$. On the integral curve $\beta$ of $V$ (see the figure 1.6) let the vectors of the vector field $\textit{\textbf{W}}$ at $P_2$ and $Q_2$ be $\textit{\textbf{W}}(\lambda)$ and $\textit{\textbf{W}}(\lambda +\Delta \lambda)$ respectively. By Lie dragging of the vector $\textit{\textbf{W}}(\lambda +\Delta \lambda)$ at $Q_2$ to the point $P_2$ we introduce a new vector field $\textit{\textbf{W}}_d$ whose vector at $Q_2$ coincide with $\textit{\textbf{W}}(\lambda +\Delta \lambda)$, {\it i.e.,} $\textit{\textbf{W}}_d(\lambda +\Delta \lambda)=\textit{\textbf{W}}(\lambda +\Delta \lambda)$ and the commutator $[\textit{\textbf{W}}_d,\textit{\textbf{V}}]=0$.\\

By Taylor's expansion,
\begin{eqnarray}
\textit{\textbf{W}}_d(\lambda +\Delta \lambda) &=& \textit{\textbf{W}}_d(\lambda)+\Delta \lambda \left[ \frac{d}{d\lambda}\textit{\textbf{W}}_d(\lambda)\right] +O(\Delta \lambda ^2) \nonumber \\
i.e.,~~\textit{\textbf{W}}_d(\lambda) &=& \textit{\textbf{W}}_d(\lambda +\Delta \lambda)-\Delta \lambda \left[ \frac{d}{d\lambda}\textit{\textbf{W}}_d(\lambda)\right] +O(\Delta \lambda ^2) \nonumber \\
&=& \textit{\textbf{W}}(\lambda +\Delta \lambda)-\Delta \lambda \left[\frac{d}{d\lambda}\textit{\textbf{W}}_d(\lambda)\right] +O(\Delta \lambda ^2) \nonumber \\
&=& \textit{\textbf{W}}(\lambda)+\Delta \lambda \frac{d}{d\lambda}\textit{\textbf{W}}(\lambda)-\Delta \lambda \frac{d}{d\lambda}\textit{\textbf{W}}_d(\lambda) +O(\Delta \lambda ^2) \nonumber \\
&=& \textit{\textbf{W}}(\lambda)+\Delta \lambda \textit{\textbf{V}} \cdot \textit{\textbf{W}}(\lambda)-\Delta \lambda \textit{\textbf{V}} \cdot \textit{\textbf{W}}_d(\lambda) +O(\Delta \lambda ^2) \nonumber \\
&=& \textit{\textbf{W}}(\lambda)+\Delta \lambda \textit{\textbf{V}} \cdot \textit{\textbf{W}}(\lambda)- \Delta \lambda \textit{\textbf{W}}_d \cdot \textit{\textbf{V}}(\lambda) +O(\Delta \lambda ^2),~~\left(\mbox{as} \left[\textit{\textbf{W}}_d\,,\textit{\textbf{V}} \right]=0\right). \nonumber
\end{eqnarray}

Now the Lie derivative of the vector field $\textit{\textbf{W}}$ with respect to the vector field $\textit{\textbf{V}}$ is denoted by $L_{\textit{\textbf{V}}} \textit{\textbf{W}}$ and is defined as
\begin{eqnarray}
L_{\textit{\textbf{V}}} \textit{\textbf{W}} &=& \lim_{\Delta \lambda \rightarrow 0} \frac{\textit{\textbf{W}}_d(\lambda)-\textit{\textbf{W}}(\lambda)}{\Delta \lambda}\nonumber\\
&=&\lim_{\Delta \lambda \rightarrow 0} \textit{\textbf{V}}\cdot \textit{\textbf{W}}-\textit{\textbf{W}}_d\cdot \textit{\textbf{V}} \nonumber \\
&=& \textit{\textbf{V}}\cdot \textit{\textbf{W}}-\textit{\textbf{W}}\cdot \textit{\textbf{V}}\nonumber\\
&=&[\textit{\textbf{V}},\textit{\textbf{W}}]. \nonumber
\end{eqnarray}

(Note that the difference between $\textit{\textbf{W}}_d(\lambda)$ and $\textit{\textbf{W}}(\lambda)$ is a term of first order in $\Delta \lambda$ and hence in the limit they are equal.) In particular, if the vector field $\textit{\textbf{W}}$ is Lie dragged by the vector field $\textit{\textbf{V}}$ then $[\textit{\textbf{V}},\textit{\textbf{W}}]=0$ and hence $L_{\textit{\textbf{V}}} \textit{\textbf{W}}=0$.\\

Also the antisymmetric nature of the commutator bracket shows 
$$L_{\textit{\textbf{V}}} \textit{\textbf{W}}=-L_{\textit{\textbf{W}}} \textit{\textbf{V}}~.$$

 Alternatively, one can define Lie derivative using the idea of one parameter local group of diffeomorphisms. Let us consider a congruence of curves through each point of the manifold. Suppose 
	$$x^a=x^a(u),$$
	be a member of the congruence. Then the vector field $X^a=\dfrac{dx^a}{du}$ be the tangent vector to the curve and it can be extended over the entire manifold.\\
	
	On the otherway, given a non-zero vector field $X^a(u)$ over the manifold, it is possible to define a congruence of curves in the manifold termed as orbits (or trajectories) of $X^a$. In principle, these orbits (integral curves) are obtained by solving the ordinary differential equations:
	$$\frac{dx^a}{du}=X^a(x(u)).$$
	
	Now due to existence and uniqueness theorem for ordinary differential equations there always have solution at least in some neighbourhood (i.e., locally) of the initial point $P$. For any $q \in M,~\exists$ an open neighbourhood $B$ of $q$ and an $\epsilon >0$ so that one can define a family of diffeomorphisms $\phi_t : B \rightarrow M$ with $|u|< \epsilon$ by taking each point $P \in B$ a parameter distance $u$ along the integral curves of $\overrightarrow{X}$. Due to this diffeomorphism each tensor field $T$ at $P$ of type $(a, b)$ maps into $\phi_{_{u^{*}}}T_{_{|_{_{\phi_u (p)}}}}$\\
	
	Now the Lie derivative $\mathcal{L_{_{\overrightarrow{X}}}} T$ of a tensorfield $T$ with respect to $\overrightarrow{X}$ is defined as
	$$\mathcal{L_{_{\overrightarrow{X}}}} T=\lim_{u \rightarrow 0} \frac{1}{u}\bigg\{T_{_{|_{_{p}}}}-\phi_{_{u^{*}}}T_{_{|_{_{p}}}}\bigg\}.$$
	
	As under the map $\phi_{_{u}}$, the point $q=\phi_{_{-u}}(p)$ is mapped into $p$ so $\phi_{_{u^{*}}}$ is a map from $T_q$ to $T_p$. Using local coordinates $\{x^i\}$ in a neighbourhood of $p$ the coordinate components of $\phi_{_{u^{*}}} Y$ at $p$ are ($Y$ is a vector field)
	\begin{eqnarray}
		\bigg(\phi_{_{u^*}} Y\bigg)^i \Bigg|_p&=&\phi_{_{u^*}} Y\Bigg|_p x^i=Y^l\Bigg|_q \frac{\partial}{\partial x^l(q)}(x^i(p))\nonumber\\
		&=&\frac{\partial x^i (\phi_{_{u}}(q))}{\partial x^l(q)} Y^l\Bigg|_q\nonumber\\
		\mbox{i.e.,}~\frac{dx^i (\phi_t(q))}{dt}&=&X^i\Bigg|_{\phi_t(q)}\nonumber\\
		\mbox{Hence}~~~\frac{d}{dt}\Bigg(\frac{\partial x^i (\phi_{_{u}}(q))}{\partial x^l(q)}\Bigg)\Bigg|_{u=0}&=&\frac{\partial x^i}{\partial x^l}\Bigg|_p\nonumber
	\end{eqnarray}
	and one gets
	\begin{eqnarray}
		\bigg(L_{_{X}} Y\bigg)^i&=&-\frac{d}{dt}\bigg(\phi_{_{u^{*}}} Y\bigg)^i \Bigg|_{_{u=0}}\nonumber\\
		&=&\frac{\partial Y^i}{\partial x^l} X^l-\frac{\partial Y^i}{\partial x^l} Y^l\nonumber\\
		\mbox{i.e.,~~}\bigg(L_{_{X}} Y\bigg)f &=& X(Yf)-Y(Xf)\nonumber
	\end{eqnarray}
	where $f$ is a $C^2$ function. Thus $L_{_{X}} Y=[X, Y]=-L_{_{Y}} X$ . \\

We shall now deduce the Leibnitz rule for the Lie derivative. Let $f$ be a function and $\textit{\textbf{W}}$ be a vector field defined on a manifold $M$. $\textit{\textbf{V}}$ is the vector field with respect to which we determine the Lie derivative. We note that $f\textit{\textbf{W}}$ is also a vector field on $M$. Let us denote this vector field by $\textit{\textbf{W}}_f$. The dragged field of $\textit{\textbf{W}}_f$ is denoted by $\textit{\textbf{W}}_{fd}$. Then as before
\begin{eqnarray}
\textit{\textbf{W}}_{fd}(\lambda) &=& \textit{\textbf{W}}_f(\lambda)+\Delta \lambda \frac{d}{d\lambda}\textit{\textbf{W}}_f(\lambda)-\Delta \lambda \textit{\textbf{W}}_{fd}(\lambda)\cdot \textit{\textbf{V}} \nonumber \\
&=& \textit{\textbf{W}}_f(\lambda)+\Delta \lambda \frac{d}{d\lambda}\lbrace f(\lambda)\textit{\textbf{W}}(\lambda)\rbrace -\Delta \lambda f\textit{\textbf{W}}(\lambda)\cdot \textit{\textbf{V}}+O(\Delta \lambda ^2) \nonumber \\
&=& \textit{\textbf{W}}_f(\lambda)+\Delta \lambda \left\lbrace \frac{df}{d\lambda}\textit{\textbf{W}}(\lambda) +f\frac{d\textit{\textbf{W}}}{d\lambda} -f\textit{\textbf{W}}\cdot \textit{\textbf{V}}\right\rbrace +O(\Delta \lambda ^2) \nonumber \\
&=& \textit{\textbf{W}}_f(\lambda)+\Delta \lambda \left\lbrace \frac{df}{d\lambda}\textit{\textbf{W}}(\lambda) +f(\textit{\textbf{V}}\cdot \textit{\textbf{W}}-\textit{\textbf{W}}\cdot \textit{\textbf{V}})\right\rbrace +O(\Delta \lambda ^2) \nonumber \\
&=& \textit{\textbf{W}}_f(\lambda)+\Delta \lambda \left\lbrace \frac{df}{d\lambda}\textit{\textbf{W}}(\lambda) +f[\textit{\textbf{V}},\textit{\textbf{W}}]\right\rbrace +O(\Delta \lambda ^2) \nonumber
\end{eqnarray}

Thus,
\begin{eqnarray}
L_{\textit{\textbf{V}}}\textit{\textbf{W}}_f(\lambda) &=& \lim_{\Delta \lambda \rightarrow 0}\frac{\textit{\textbf{W}}_{fd}(\lambda)-\textit{\textbf{W}}_f(\lambda)}{\Delta \lambda}=\frac{df}{d\lambda}\textit{\textbf{W}}+f[\textit{\textbf{V}}, \textit{\textbf{W}}] \nonumber \\
\mbox{or},~~L_{\textit{\textbf{V}}}(f\textit{\textbf{W}}) &=& (L_{\textit{\textbf{V}}}f)\textit{\textbf{W}}+ f L_{\textit{\textbf{V}}}\textit{\textbf{W}} \nonumber,
\end{eqnarray}
the Leibnitz rule for differential operator.\\

We shall now introduce the Lie derivative of one-form using the above results for Lie derivative of vectors and scalars and by the application of Leibnitz rule. Let $\underline{w}$ be a one-form and $\textit{\textbf{W}}$ be an arbitrary vector field then by Leibnitz rule,
\begin{eqnarray}
L_{\textit{\textbf{V}}} \utilde{w}(\textit{\textbf{W}}) &=& (L_{\textit{\textbf{V}}}\utilde{w})(\textit{\textbf{W}})+\utilde{w}\cdot L_{\textit{\textbf{V}}}\textit{\textbf{W}} \nonumber \\
i.e.,~~(L_{\textit{\textbf{V}}}\utilde{w})(\textit{\textbf{W}}) &=& L_{\textit{\textbf{V}}} \utilde{w}(\textit{\textbf{W}})-\utilde{w}\cdot L_{\textit{\textbf{V}}}\textit{\textbf{W}}=\frac{df}{d\lambda}-\utilde{w}[\textit{\textbf{V}},\textit{\textbf{W}}], \label{1.28}
\end{eqnarray}
where $f$ is the inner product $\utilde{w}(\textit{\textbf{W}})$.\\

Using the above definitions of Lie derivative of scalar, vector and one-form, we shall now extend this definition of Lie derivative for an arbitrary tensor $T$ as follows:
\begin{center}
$L_{\textit{\textbf{V}}}T(\utilde{w}, \cdot \cdot \cdot; \textit{\textbf{W}}, \cdot \cdot \cdot)=(L_{\textit{\textbf{V}}}T)(\utilde{w}, \cdot \cdot \cdot; \textit{\textbf{W}}, \cdot \cdot \cdot)+T(L_{\textit{\textbf{V}}} \utilde{w}, \cdot \cdot \cdot; \textit{\textbf{W}}, \cdot \cdot \cdot)+ \cdot \cdot \cdot +T(\utilde{w}, \cdot \cdot \cdot; L_{\textit{\textbf{V}}} \textit{\textbf{W}}, \cdot \cdot \cdot)+ \cdot \cdot \cdot$
\end{center}
where $\utilde{w}, \cdot \cdot \cdot$ and $\textit{\textbf{W}}, \cdot \cdot \cdot$ are arbitrary one-form and vectors respectively.\\

Also, for arbitrary tensors $S$ and $T$
 $$L_{\textit{\textbf{V}}}(S \otimes T)=(L_{\textit{\textbf{V}}} S) \otimes T+S \otimes (L_{\textit{\textbf{V}}}T).$$
 
The next step is to find components of Lie derivative ({\it i.e.,} $L_{\textit{\textbf{V}}} \textit{\textbf{W}}$) in a coordinate basis. Given a coordinate system $\left\lbrace x^i \right\rbrace$, the set $\left\lbrace \dfrac{\partial}{\partial x^i} \right\rbrace$ is chosen as the basis (coordinate basis) for the vector fields. Suppose $\textit{\textbf{V}}=\dfrac{d}{d\lambda}$ and $\textit{\textbf{W}}=\dfrac{d}{d\mu}$ be two arbitrary vector fields. So in the coordinate basis, we have
$$\textit{\textbf{V}}=\frac{d}{d\lambda}=\frac{dx^i}{d\lambda}\frac{\partial}{\partial x^i}=V^i \frac{\partial}{\partial x^i}$$
and
$$\textit{\textbf{W}}=\frac{d}{d\mu}=\frac{dx^i}{d\mu}\frac{\partial}{\partial x^i}=W^i \frac{\partial}{\partial x^i}~.$$

Now,
\begin{eqnarray}
L_{\textit{\textbf{V}}} \textit{\textbf{W}}=[\textit{\textbf{V}}, \textit{\textbf{W}}]=\textit{\textbf{V}}\textit{\textbf{W}}-\textit{\textbf{W}}
\textit{\textbf{V}} &=& V^i \frac{\partial}{\partial x^i}\left(W^j\frac{\partial}{\partial x^j}\right)-W^j \frac{\partial}{\partial x^j}\left(V^i \frac{\partial}{\partial x^i}\right) \nonumber \\
&=& V^i W^j \left(\frac{\partial}{\partial x^i}\frac{\partial}{\partial x^j}-\frac{\partial}{\partial x^j}\frac{\partial}{\partial x^i}\right)+V^i \frac{\partial W^j}{\partial x^i}\frac{\partial}{\partial x^j}-W^j \frac{\partial V^i}{\partial x^j}\frac{\partial}{\partial x^i} \nonumber \\
&=& V^i \frac{\partial W^j}{\partial x^i}\frac{\partial}{\partial x^j}-W^j \frac{\partial V^i}{\partial x^j}\frac{\partial}{\partial x^i} \nonumber \\
&=& \left(V^j\frac{\partial W^i}{\partial x^i}-W^j\frac{\partial V^i}{\partial x^j}\right)\frac{\partial}{\partial x^i}. \nonumber
\end{eqnarray}

Hence,
$$\left(L_{\textit{\textbf{V}}} \textit{\textbf{W}}\right)^i=\left(V^j\frac{\partial W^i}{\partial x^j}-W^j\frac{\partial V^i}{\partial x^j}\right)~.$$

However, in an arbitrary basis $\left\lbrace e_{\alpha}\right\rbrace$, we have
$$\textit{\textbf{V}}=V^{\alpha} e_{\alpha}~~~\mbox{and}~~~\textit{\textbf{W}}=W^{\alpha} e_{\alpha}~.$$

So,
\begin{eqnarray}
L_{\textit{\textbf{V}}} \textit{\textbf{W}}=\textit{\textbf{V}}\textit{\textbf{W}}-\textit{\textbf{W}}
\textit{\textbf{V}} &=& V^{\alpha} e_{\alpha} W^{\beta} e_{\beta}-W^{\beta} e_{\beta} V^{\alpha} e_{\alpha} \nonumber \\
&=& V^{\alpha} W^{\beta}(e_{\alpha}e_{\beta}-e_{\beta}e_{\alpha})+V^{\alpha}(e_{\alpha}W^{\beta})e_{\beta}-W^{\beta}(e_{\beta}V^{\alpha})e_{\alpha} \nonumber \\
&=& V^{\alpha} W^{\beta}[e_{\alpha}, e_{\beta}]+\left[V^{\beta}(e_{\beta}W^{\alpha})-W^{\beta}(e_{\beta}V^{\alpha})\right]e_{\alpha} \nonumber \\
&=& V^{\alpha} W^{\beta} L_{e_{\alpha}}^{e_{\beta}}+\left[V^{\beta}(e_{\beta}W^{\alpha})-W^{\beta}(e_{\beta}V^{\alpha})\right]e_{\alpha}. \nonumber
\end{eqnarray}

Therefore
$$\left(L_{\textit{\textbf{V}}} \textit{\textbf{W}}\right)^\mu =V^{\alpha} W^{\beta} \left(L_{e_{\alpha}}^{e_{\beta}}\right)^\mu+\left[V^{\beta}(e_{\beta}W^{\mu})-W^{\beta}(e_{\beta}V^{\mu})\right].$$

In particular, if $\textit{\textbf{V}}$ is along a coordinate basis, say $\dfrac{\partial}{\partial x^l}$, then 
$$\left(L_{\textit{\textbf{V}}} \textit{\textbf{W}}\right)^i=V^l\frac{\partial W^i}{\partial x^l}~.$$

In the coordinate basis, the one-form $\utilde{w}$ can be written as $\utilde{w}=w_i dx^i$ and we have $\textit{\textbf{W}}=W^i \dfrac{\partial}{\partial x^i}$, so we get
\begin{eqnarray}
\left(L_{\textit{\textbf{V}}} \utilde{w}\right)(\textit{\textbf{W}}) &=& L_{\textit{\textbf{V}}} \utilde{w}(\textit{\textbf{W}})-\utilde{w}L_{\textit{\textbf{V}}} \textit{\textbf{W}} \nonumber \\
&=& \frac{d}{d\lambda}(w_i W^i)-w_i \left(V^j\frac{\partial W^i}{\partial x^j}-W^j\frac{\partial V^i}{\partial x^j}\right) \nonumber \\
&=& \frac{\partial}{\partial x^k}(w_i W^i)V^k - w_i \left(V^j\frac{\partial W^i}{\partial x^j}-W^j\frac{\partial V^i}{\partial x^j}\right) \nonumber \\
&=& \left(\frac{\partial w_i}{\partial x^k}\right)W^i V^k +w_i \frac{\partial W^i}{\partial x^k}V^k -w_i V^k\frac{\partial W^i}{\partial x^k}+w_i W^j \frac{\partial V^i}{\partial x^j} \nonumber \\
&=& \frac{\partial w_i}{\partial x^k} W^i V^k+w_k W^i \frac{\partial V^k}{\partial x^i} \nonumber \\
&=& \left(\frac{\partial w_i}{\partial x^k} V^k+w_k \frac{\partial V^k}{\partial x^i}\right)W^i. \nonumber
\end{eqnarray}

Thus,
$$\left(L_{\textit{\textbf{V}}} \utilde{w}\right)_i W^i=\left(\frac{\partial w_i}{\partial x^k} V^k+w_k \frac{\partial V^k}{\partial x^i}\right)W^i~.$$

As $\textit{\textbf{W}}$ is an arbitrary vector field, so
$$\left(L_{\textit{\textbf{V}}} \utilde{w}\right)_i =V^k \frac{\partial w_i}{\partial x^k}+w_k \frac{\partial V^k}{\partial x^i}~.$$

\textbf{Note}:\\

\textbf{I.} Lie derivative preserves the order of the tensor \textit{i.e.}, if T be a (k,l) tensor, then $L_{\textit{\textbf{V}}}T$ will also be a $(k,l)$ tensor. In particular, for any vector field $\textit{\textbf{W}}$, $L_{\textit{\textbf{V}}} \textit{\textbf{W}}$ is also a vector field distinct from $\textit{\textbf{W}}$.\\

\textbf{II.} Lie derivative is the co-ordinate independent form of the partial derivative. In particular, it commutes with the partial derivative.\\

\textbf{III.} Lie derivative obeys Leibnitz rule as in ordinary calculus.\\

\textbf{IV.} Lie derivative preserves contraction of tensor indices (though Lie derivative of metric tensor does not vanish ) and maps tensors linearly.\\

\textbf{V.} Lie derivative can be applied to arbitrary linear geometrical objects (for example Christoffel symbols discussed later).\\

 \textbf{VI.} Two vector fields are said to commute if the Lie derivative of one of them with respect to the other vanishes. Geometrically, this commutation means the following:\\

In figure 1.6, if we start from $P_2$ moves a parameter distance $\triangle \lambda$ along the integral $\beta$ of $\textit{\textbf{V}}$ and then moves a parameter distance $\triangle \mu$ along the integral curve `$b$' of $\textit{\textbf{W}}$ we reach to the point $Q_3'$. However, if we move in the reverse order \textit{i.e.}, at first we move along the integral curve `$a$' of $\textit{\textbf{W}}$ through a parameter distance $\triangle \mu$ to reach the point $P_3$ and then go along the integral curve $\gamma$ of $\textit{\textbf{V}}$ to a parameter distance $\triangle \lambda$ to obtain the point $Q_3$. If $Q_3$ is distinct from $Q_3'$ then the vector fields $\textit{\textbf{V}}$ and $\textit{\textbf{W}}$ are not commutative while if $Q_3$ coincides with $Q_3'$ then [$\textit{\textbf{V}}, \textit{\textbf{W}}$]$=0$. Thus for non-commutating vector fields\,(having non-zero Lie derivative) the end point will not be same if starting from the same initial point we go along the above two distinct paths.\\

\textbf{VII.} If $\textit{\textbf{A}}$ and $\textit{\textbf{B}}$ are any two twice-differentiable vector fields, then the operators [$L_{\textit{\textbf{A}}},L_{\textit{\textbf{B}}}$] and $L_{[{\textit{\textbf{A}}},{\textit{\textbf{B}}}]}$ are equivalent with respect to functions and vector fields over the manifold.\\

\textbf{Proof.}  For any function `$f$' on the manifold we have\\

$L_{[{\textit{\textbf{A}}},{\textit{\textbf{B}}}]}f=[\textit{\textbf{A}}, \textit{\textbf{B}}]f~~\left(\because L_{\textit{\textbf{V}}}f=\textit{\textbf{V}}(f)\right)$\\

Also $[L_{\textit{\textbf{A}}}, L_{\textit{\textbf{B}}}]f=(L_{\textit{\textbf{A}}} L_{\textit{\textbf{B}}}-L_{\textit{\textbf{B}}} L_{\textit{\textbf{A}}})f=L_{\textit{\textbf{A}}}(L_{\textit{\textbf{B}}}f)-L_{\textit{\textbf{B}}}(L_{\textit{\textbf{A}}}f)$\\
$~~~~~~~~~~~~~~~~~~~~~~~~~=L_{\textit{\textbf{A}}}(\textit{\textbf{B}}f)-L_{\textit{\textbf{B}}}(\textit{\textbf{A}}f)=\textit{\textbf{A}}\textit{\textbf{B}}(f)-\textit{\textbf{B}}\textit{\textbf{A}}(f)$\\
$~~~~~~~~~~~~~~~~~~~~~~~~~=(\textit{\textbf{A}}\textit{\textbf{B}}-\textit{\textbf{B}}
\textit{\textbf{A}})(f)=[\textit{\textbf{A}}, \textit{\textbf{B}}](f)=L_{[\textit{\textbf{A}}, \textit{\textbf{B}}]}(f)$\\

As `$f$' is arbitrary so, $[L_{\textit{\textbf{A}}}, L_{\textit{\textbf{B}}}]=L_{[\textit{\textbf{A}}, \textit{\textbf{B}}]}$.\\

Similarly, if $\textit{\textbf{X}}$ be any vector field then \\
$L_{[\textit{\textbf{A}},\textit{\textbf{B}}]}\textit{\textbf{X}}=\left[[\textit{\textbf{A}},\textit{\textbf{B}}],\textit{\textbf{X}}\right]$\\

Now, $[L_{\textit{\textbf{A}}}, L_{\textit{\textbf{B}}}]\textit{\textbf{X}}=(L_{\textit{\textbf{A}}}
L_{\textit{\textbf{B}}}-L_{\textit{\textbf{B}}}L_{\textit{\textbf{A}}})\textit{\textbf{X}}$\\
$~~~~~~~~~=L_{\textit{\textbf{A}}}(L_{\textit{\textbf{B}}}\textit{\textbf{X}})-L_{\textit{\textbf{B}}}(L_{\textit{\textbf{A}}}\textit{\textbf{X}})$\\
$~~~~~~~~~=L_{\textit{\textbf{A}}}[\textit{\textbf{B}}, \textit{\textbf{X}}]-L_{\textit{\textbf{B}}}[\textit{\textbf{A}}, \textit{\textbf{X}}]$\\
$~~~~~~~~~=[\textit{\textbf{A}},[\textit{\textbf{B}},\textit{\textbf{X}}]]-[\textit{\textbf{B}},[\textit{\textbf{A}},\textit{\textbf{X}}]]$\\
$~~~~~~~~~=[\textit{\textbf{A}},[\textit{\textbf{B}},\textit{\textbf{X}}]]+[\textit{\textbf{B}},[\textit{\textbf{X}},\textit{\textbf{A}}]]$\\
$~~~~~~~~~=-[\textit{\textbf{X}},[\textit{\textbf{A}},\textit{\textbf{B}}]]$
\begin{flushright}
	(by Jacobi's identity for vector fields $\textit{\textbf{A}}$, $\textit{\textbf{B}}$ and $\textit{\textbf{X}}$, assuming every one as $c^2$ functions)
\end{flushright}
$~~~~~~~=[[\textit{\textbf{A}}, \textit{\textbf{B}}], \textit{\textbf{X}}]$\\

Hence~~ $[L_{\textit{\textbf{A}}}, L_{\textit{\textbf{B}}}] \textit{\textbf{X}}=L_{[\textit{\textbf{A}}, \textit{\textbf{B}}]}\textit{\textbf{X}}$\\
$~~~~~~~\mbox{\it i.e.},~~ [L_{\textit{\textbf{A}}}, L_{\textit{\textbf{B}}}] =L_{[\textit{\textbf{A}}, \textit{\textbf{B}}]}$~~~~~~(as $\textit{\textbf{X}}$ is arbitrary)\\

\textbf{VIII.} For any three $c^3$-vector fields $\textit{\textbf{A}}$, $\textit{\textbf{B}}$ and $\textit{\textbf{C}}$ we have Jacobi identity for Lie derivatives \textit{i.e.},\\
$$[[L_{\textit{\textbf{A}}}, L_{\textit{\textbf{B}}}], L_{\textit{\textbf{C}}}]+[[L_{\textit{\textbf{B}}}, L_{\textit{\textbf{C}}}], L_{\textit{\textbf{A}}}]+[[L_{\textit{\textbf{C}}}, L_{\textit{\textbf{A}}}], L_{\textit{\textbf{B}}}]=0$$
which operates on functions and vector fields on the manifold.\\

\textbf{Proof.}~~ We have seen above $$[L_{\textit{\textbf{A}}}, L_{\textit{\textbf{B}}}] f=L_{[\textit{\textbf{A}}, \textit{\textbf{B}}]}f$$

so, $[[L_{\textit{\textbf{A}}}, L_{\textit{\textbf{B}}}], L_{\textit{\textbf{C}}}] f=[L_{[\textit{\textbf{A}}, \textit{\textbf{B}}]}, L_{\textit{\textbf{C}}}]f= L_{[[\textit{\textbf{A}}, \textit{\textbf{B}}], \textit{\textbf{C}}]} f=[[\textit{\textbf{A}}, \textit{\textbf{B}}], \textit{\textbf{C}}] f$\\

Similarly, $[[L_{\textit{\textbf{B}}}, L_{\textit{\textbf{C}}}], L_{\textit{\textbf{A}}}] f=[[\textit{\textbf{B}}, \textit{\textbf{C}}], \textit{\textbf{A}}] f$ ~~and~~$[[L_{\textit{\textbf{C}}}, L_{\textit{\textbf{A}}}], L_{\textit{\textbf{B}}}] f=[[\textit{\textbf{C}}, \textit{\textbf{A}}], \textit{\textbf{B}}] f$\\

Now, $$[[L_{\textit{\textbf{A}}}, L_{\textit{\textbf{B}}}], L_{\textit{\textbf{C}}}] f+[[L_{\textit{\textbf{B}}}, L_{\textit{\textbf{C}}}], L_{\textit{\textbf{A}}}] f+[[L_{\textit{\textbf{C}}}, L_{\textit{\textbf{A}}}], L_{\textit{\textbf{B}}}] f=[[\textit{\textbf{A}}, \textit{\textbf{B}}], \textit{\textbf{C}}] f+[[\textit{\textbf{B}}, \textit{\textbf{C}}], \textit{\textbf{A}}] f+[[\textit{\textbf{C}}, \textit{\textbf{A}}], \textit{\textbf{B}}] f=0$$

Hence we have the Jacobi's identity for Lie derivatives.\\

 {\bf IX. Invariance}:\\

A tensor field is said to be \underline{Lie~transported} along a curve $\gamma$ if its Lie derivative along the curve vanishes. Further, if the Lie derivative of a tensor field with respect to a vector field vanishes then we say that the tensor field is \underline{invariant} along the congruence of the vector field. In particular, if the vector field $\textit{\textbf{V}}$ is chosen along a co-ordinate basis vector (say $\dfrac{\partial}{\partial x^l}$) in a co-ordinate system then
$$L_{\textit{\textbf{V}}} T=0 \Rightarrow  \frac{\partial T}{\partial x^l}=0~~\textrm{( T is any tensor field)}$$
Thus, if a tensor field is independent of a particular co-ordinate then its Lie derivative along the corresponding co-ordinate curve vanishes.\\

This notion of invariance of a tensor field under a vector field is of importance in physics for analyzing the symmetries of tensor fields ( \textit{e. g.} metric tensor in GR, scalar field describing potential of a particle or a vector field of force etc. )\\

 {\bf X. Killing vector fields}:\\

The idea of Killing vector field is of importance for manifolds with a metric structure. A vector field $\textit{\textbf{V}}$ is said to be a Killing vector if
\begin{equation} \label{1.29}
L_{\textit{\textbf{V}}} g=0
\end{equation}
where g is the metric tensor of the manifold.\\

In components, the above Killing equation can be written as 
\begin{equation} \label{1.30}
(L_{\textit{\textbf{V}}} g)_{\mu \gamma}=0~i.e.,~V^\delta \frac{\partial }{\partial x^\delta} g_{\mu \gamma}+g_{\mu \delta}\frac{\partial }{\partial x^\gamma}V^\delta+g_{\delta \gamma}\frac{\partial }{\partial x^\mu}V^\delta =0
\end{equation}

For simplicity, if $\textit{\textbf{V}}$ is a co-ordinate basis (say $\dfrac{\partial}{\partial x^{\alpha}}$) \textit{i.e.}, the integral curves of $\textit{\textbf{V}}$ are family of co-ordinate lines for $x^\alpha$ then we have 
$$\frac{\partial}{\partial x^{\alpha}} g_{\mu \gamma}=0\,,$$
which implies that the components of the metric tensor are independent of the co-ordinates $x^{\alpha}$. In other words, if there exists a co-ordinate system in which the components of the metric are independent of a particular co-ordinate then the corresponding basis vector is a Killing vector\,(for details see \S 1.13).\\

As an example, we consider the metric in a 3 dimensional Euclidean manifold. In Cartesian system the metric components are $g_{\mu \gamma}=\delta_{\mu \gamma}$ \textit{i.e.}, independent of the co-ordinates $x$, $y$ and $z$ so $\dfrac{\partial}{\partial x}$, $\dfrac{\partial}{\partial y}$ and $\dfrac{\partial}{\partial z}$ are the Killing vectors. Further, writing the metric in polar co-ordinates $(r, \theta, \phi)$ \textit{i.e}., $ds^2=dr^2+r^2d\theta^2+r^2(\sin \theta)^2d\phi^2$, it is clear that
$$\frac{\partial}{\partial \phi}=x\frac{\partial}{\partial y}-y\frac{\partial}{\partial x}=\textit{\textbf{l}}_{\bm z}$$
is a Killing vector. From symmetry $\textit{\textbf{l}}_{\bm x}$ and $\textit{\textbf{l}}_{\bm y}$ are also Killing vector fields. Therefore, 3-D Euclidean manifold has six Killing vectors\,(of which three correspond to translational invariance and other three correspond to rotational invariance).\\

In general, a manifold of dimension `$n$' has at most $\dfrac{n(n+1)}{2}$ Killing vectors. A space with maximal Killing vectors is called a \underline{maximally symmetric space}. A maximally symmetric space is both homogeneous and isotropic. In the above example the 3-D Euclidean manifold is maximally symmetric space. We shall extensively discuss it again in \S 1.13.\\

\textbf{XI.} The set of all vector fields under which a tensor field or a class of tensor fields are invariant forms a Lie algebra. (A Lie algebra of vector fields is a vector space under addition and is closed under Lie-bracket\,(commutation) operation). This follows from the facts that\\\\
i) if a tensor field ($T$) is invariant under both $\textit{\textbf{V}}$ and $\textit{\textbf{W}}$ then it will also be invariant under $c_1\textit{\textbf{V}}+c_2\textit{\textbf{W}}$ ($c_1, c_2$ are scalers).\\\\
ii) $L_{\textit{\textbf{V}}}T=0=L_{\textit{\textbf{W}}}T\Rightarrow L_{[\textit{\textbf{V}}, \textit{\textbf{W}}]} T=0$.\\

\subsection{Exterior Differentiation}

~~~The exterior differential operator introduces another differentiation on manifold but it acts only on forms and preserve its character as forms - it raises the degree of the form by unity. So we can define exterior differential operator 'd' as a mapping which transform a form of arbitrary degree `$r$' to a $(r+1)$-form. Suppose in a co-ordinate system $\{x^\alpha\}$ a $r$-form $B$ can be written as
$$B=B_{\alpha \beta \ldots \lambda}dx^\alpha \wedge dx^\beta \wedge \ldots dx^\lambda$$

Then under exterior differentiation it becomes 
$$dB=dB_{\alpha \beta \ldots \lambda}\wedge  dx^\alpha \wedge dx^\beta \wedge \ldots \wedge dx^\lambda,$$
which is a $(r+1)$-form.\\

If we now make a co-ordinate transformation $\{x^\alpha\}\rightarrow \{x^{\alpha'}\}$ then 
$$B=B_{\alpha' \beta' \ldots \lambda'}dx^{\alpha'} \wedge dx^{\beta'} \wedge \ldots dx^{\lambda'}$$
where $$B_{\alpha' \beta' \ldots \lambda'}=\frac{\partial x^\alpha}{\partial x^{\alpha'}}\frac{\partial x^\beta}{\partial x^{\beta'}} \ldots \frac{\partial x^\lambda}{\partial x^{\lambda'}}B_{\alpha \beta \ldots \lambda}$$

Now
\begin{eqnarray}
dB&=&dB_{\alpha' \beta' .....\lambda'}\wedge dx^{\alpha'} \wedge dx^{\beta'} \wedge \ldots \wedge dx^{\lambda'} \nonumber \\
&=&d\left(\frac{\partial x^\alpha}{\partial x^{\alpha'}}\frac{\partial x^\beta}{\partial x^{\beta'}}......\frac{\partial x^\lambda}{\partial x^{\lambda'}}B_{\alpha \beta .....\lambda} \right)\Lambda dx^{\alpha'} \wedge dx^{\beta'} \wedge .....\wedge dx^{\lambda'} \nonumber \\
&=&dB_{\alpha \beta \ldots \lambda} \wedge (\frac{\partial x^\alpha}{\partial x^{\alpha'}} dx^{\alpha'})\wedge(\frac{\partial x^\beta}{\partial x^{\beta'}}dx^{\beta'})\wedge \ldots (\frac{\partial x^\lambda}{\partial x^{\lambda'}}dx^{\lambda'}) \nonumber \\
&+&\frac{\partial^2 x^\alpha}{\partial x^{\alpha'} \partial x^{\delta'}} \frac{\partial x^\beta}{\partial x^{\beta'}}\ldots \frac{\partial x^\lambda}{\partial x^{\lambda'}}B_{\alpha \beta \ldots \lambda}dx^{\delta'}\Lambda dx^{\alpha'}\wedge dx^{\beta'}\wedge \ldots \wedge dx^{\lambda'}+ \ldots + \ldots \nonumber \\
&=&dB_{\alpha \beta \ldots \lambda}\Lambda dx^\alpha \wedge dx^\beta \wedge \ldots \wedge dx^\lambda \,, \nonumber
\end{eqnarray}
where other terms vanish due to the product of symmetric and antisymmetric parts namely $\dfrac{\partial^2 x^\alpha }{\partial x^{\alpha'} \partial x^{\delta'}}$ and $dx^{\delta'} \wedge dx^{\alpha'}$\,.\\

This shows that the resulting $(r+1)-$form field is independent of the co-ordinate system. In particular, if a  scalar function `$f$' is termed as zero-form then the exterior differentiation of `$f$' is an one-form `$df$' defined by\,(in a co-ordinate system) 
$$df=\frac{\partial f}{\partial x^\alpha}dx^\alpha$$
In general $\left\langle df, \textit{\textbf{X}} \right\rangle=\textit{\textbf{X}}f$\,, for any vector field $\textit{\textbf{X}}$.\\

Further, from the above
$$d(dB)=\left\{\frac{\partial^2 B_{\alpha \beta .....\lambda}}{\partial x^\mu \partial x^\gamma}dx^\mu \wedge dx^\gamma\right\}\wedge dx^\alpha \wedge ....\wedge dx^\wedge=0$$
as the second order partial derivative is symmetric with respect to interchange of indices $\mu ,\gamma$ while $dx^\mu \wedge dx^\gamma$ is antisymmetric with respect to this interchange.\\

Thus we can summarize the result of exterior differentiation:\\

i) The exterior differentiation acts linearly on forms \textit{i.e.}, for any two $r$-forms $A$ and $B$
$$d(A+B)=dA+dB$$

ii) For any $r$-form field `$A$', $dA$ is a ($r+1$)-form field independent of the choice of co-ordinate. However, co-ordinate independence will be lost if tensor product is used instead of Wedge product.\\

iii) \underline{Leibnitz rule}: If A is a r-form and C is a form then 
$$d(A \wedge C)=dA \wedge C +(-1)^r A \wedge dC$$

iv) $d(dA)=0$ for any form $A$.

A form $A$ for which $dA=0$ is said to be closed while if $A=dB$ for some form $B$ then $A$ is said to be exact.\\

\textbf{Note}: From the above definition every exact form is closed but the converse is true only for a sufficiently small neighbourhood of the point under consideration \textit{i.e.}, if $A$ is a closed form then $\exists$ a form $B$ such that $A=dB$. It should be noted that this choice of $B$ is not unique because we can replace $B$ by $B+dC$ for arbitrary form $C$.\\

v) \textbf{Commutativity of a Lie derivative and exterior derivative}:\\
The Lie derivative of a $r$-form `$B$' with respect to a vector field $\textit{\textbf{V}}$ can be obtained by mathematical induction starting from a zero-form\,(a scalar) and using the result for one-form.\\

Thus we have (for detail proof see the appendix-I):\\
$$L_{\textit{\textbf{V}}}B=d[B(\textit{\textbf{V}})]+dB(\textit{\textbf{V}})$$
In particular, if $B=dA$, $A$ is an ($r-1$)-form, then
$$L_{\textit{\textbf{V}}}dA=d[dA(\textit{\textbf{V}})]~~(\because ~ d(dB)=0).$$
But from the above derivative formula\\
$$dA(\textit{\textbf{V}})=L_{\textit{\textbf{V}}}A-d[A(\textit{\textbf{V}})].$$
$\therefore L_{\textit{\textbf{V}}}dA=dL_{\textit{\textbf{V}}}A$\\
Hence Lie derivative and exterior differentiation commutes with each other.\\

\subsection{Covariant Differentiation}

~~~On a differentiable manifold, one can not identify vectors at different points to be parallel to each other. There is no well defined prescription of the intrinsic notion of parallelism on the manifold. The affine connection is a rule of introducing the idea of parallelism (\textit{i.e.}, parallel transport) of vectors at different points.\\

\begin{wrapfigure}[10]{r}{0.55\textwidth}\vspace{-1.5\intextsep}
\includegraphics[height=4.5 cm , width=10 cm ]{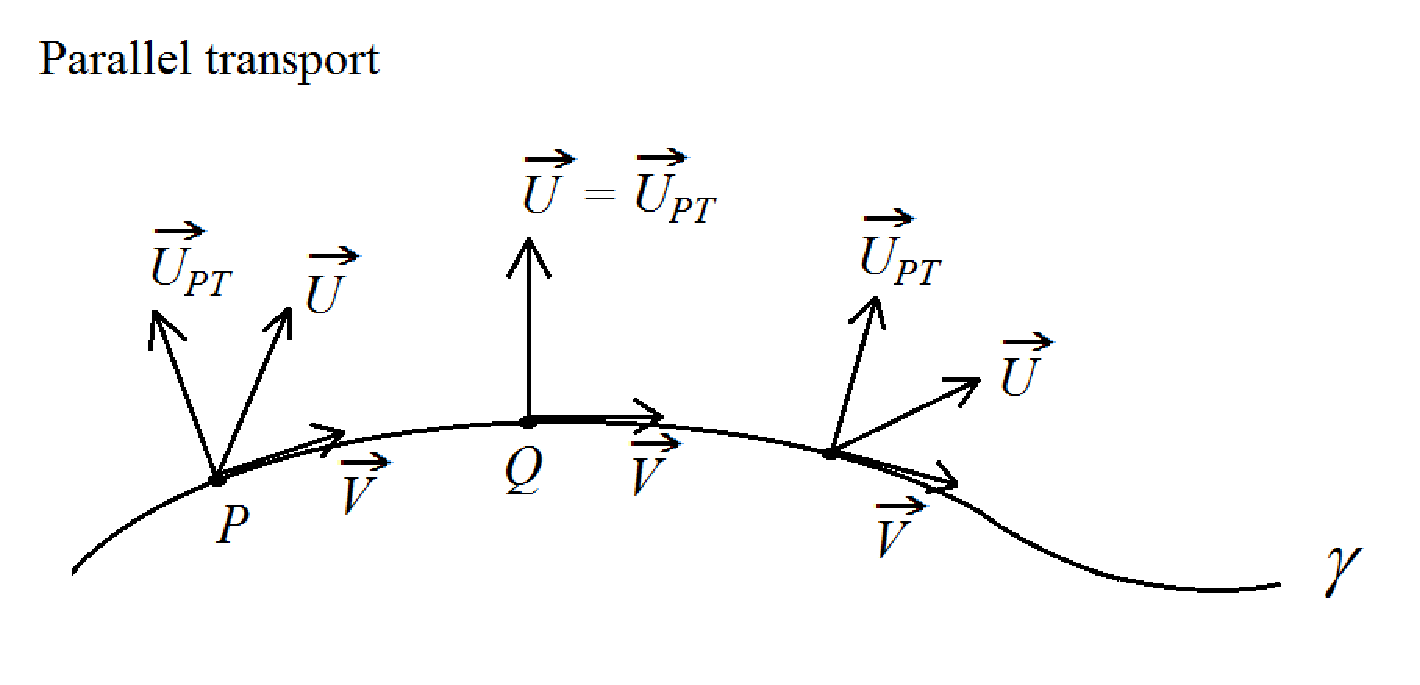}\vspace{-\intextsep}
\begin{center}
Fig. 1.7
\end{center}\vspace{-\intextsep}
\end{wrapfigure}

Let `$\gamma$' be a curve on the manifold and we denote the tangent vector to `$\gamma$' by $\textit{\textbf{V}} \left(=\dfrac{d}{d \lambda}\right)$ . The rule of connection then introduces a vector field $\textit{\textbf{U}}$ along $\gamma$ by the notion of parallel transport. So we can say that $\textit{\textbf{U}}$ does not change along $\gamma$ and hence a differential operator can be defined along `$\gamma$' such that $\textit{\textbf{U}}$ has zero differential. This differentiation is called covariant derivative with respect to $\textit{\textbf{V}}$ and is denoted by $\nabla_{\textit{\textbf{V}}}\textit{\textbf{U}}$ and we have $\nabla_{\textit{\textbf{V}}}\textit{\textbf{U}}=0$. It implies $\textit{\textbf{U}}$ to be parallely-transported along the curve $\gamma$ for which $\textit{\textbf{V}}$ is the tangent. The converse is also true. Thus a connection $\nabla$ at a point P on M is a rule which maps an arbitrary vector field $\textit{\textbf{U}}$ into another vector field $\nabla_{\textit{\textbf{V}}}\textit{\textbf{U}}$. Using this idea of parallel transport, we shall now define the covariant derivative of any vector field $\textit{\textbf{U}}$ defined over $\gamma$ as follows:\\

For convenience, let us express the vector field $\textit{\textbf{U}}$ as function of the parameter $\lambda$. So we have $\textit{\textbf{U}}(\lambda)$ and $\textit{\textbf{U}}(\lambda +\triangle \lambda)$ be the members of the vector field at $P$ and the neighbouring point $Q$ (at a parameter distance $\triangle \lambda$ from $P$ along the curve $\gamma$ ). We then define a new vector $\textit{\textbf{U}}_{PT}$ which equals $\textit{\textbf{U}}$ at $Q$ and is parallel-transported along $\gamma$ \textit{i.e.}, $\textit{\textbf{U}}_{PT}(\lambda +\triangle \lambda)=\textit{\textbf{U}}(\lambda +\triangle \lambda)$ and $\nabla_{\textit{\textbf{V}}}\textit{\textbf{U}}_{PT}=0$. Then the covariant derivative of $\textit{\textbf{U}}$ at $P$ is defined to be 
$$\nabla_{\textit{\textbf{V}}}\textit{\textbf{U}}=\lim_{\triangle \lambda \rightarrow0} \frac{\textit{\textbf{U}}_{PT}(\lambda)-\textit{\textbf{U}}(\lambda)}{\triangle \lambda}$$
\textbf{Note:}\\

{\bf I.} The derivative is evaluated entirely in the vector space $T_{P}$.\\

{\bf II.} Though there is similarity with Lie derivative but the significant difference between these two derivatives is that the notion of `dragging back' in Lie derivative needs the entire congruence \textit{i.e.}, the vectors are to be defined not only in $\gamma$ but also in the neighbourhood of $\gamma$. On the other hand, for covariant derivative we require the vector fields only on $\gamma$ but with an extra structure namely the connection on the curve.\\

{\bf III.} The covariant derivative $\nabla_{\textit{\textbf{v}}}$ at P depends only on the direction of $\textit{\textbf{v}}$ at P. Thus for any two functions $\alpha$ and $\beta$ 
$$\nabla_{\alpha\textit{\textbf{v}}_{\bm 1} +\beta\textit{\textbf{v}}_{\bm 2}}\textit{\textbf{U}}=\alpha \nabla_{\textit{\textbf{v}}_{\bm 1}} \textit{\textbf{U}}+\beta\nabla_{\textit{\textbf{v}}_{\bm 2}}\textit{\textbf{U}}$$  

{\bf IV.} From the definition $\nabla_{\textit{\textbf{V}}}\textit{\textbf{U}}$ is linear in $\textit{\textbf{U}}$ \textit{i.e.},
$$\nabla_{\textit{\textbf{V}}}(c_1\textit{\textbf{U}}_{\bm 1}+c_2\textit{\textbf{U}}_{\bm 2}) =c_1\nabla_{\textit{\textbf{V}}}\textit{\textbf{U}}_{\bm 1}+
c_2\nabla_{\textit{\textbf{V}}}\textit{\textbf{U}}_{\bm 2},~~c_1, c_2 ~\textrm{are constants.}$$

{\bf V.} For any scalar function `$f$' we have 
$$\nabla_{\textit{\textbf{V}}}f=\textit{\textbf{V}}(f)$$
$$\textrm{and}~\nabla_{\textit{\textbf{V}}}(f\textit{\textbf{U}})=\textit{\textbf{V}}(f).\textit{\textbf{U}}+f\nabla_{\textit{\textbf{V}}}\textit{\textbf{U}}$$

{\bf VI.} As $\nabla_{\textit{\textbf{V}}}\textit{\textbf{U}}$ is the covariant derivative of $\textit{\textbf{U}}$ (for a given connection) in the direction of $\textit{\textbf{V}}$ at P, so one can define $\nabla \textit{\textbf{U}}$ as a (1, 1)-type tensor field, contracting with $\textit{\textbf{V}}$ gives the vector $\nabla_{\textit{\textbf{V}}}\textit{\textbf{U}}$ \textit{i.e.}, 
$$\nabla \textit{\textbf{U}}(~ ; \textit{\textbf{V}})=\nabla_{\textit{\textbf{V}}}\textit{\textbf{U}}.$$
Also
$$\nabla(f\textit{\textbf{U}})=df\otimes \textit{\textbf{U}}+f\nabla \textit{\textbf{U}}.$$

{\bf VII.} Given basis $\{\textit{\textbf{e}}_{\bm \alpha}\}$ for vectors and $\{\utilde{e^\alpha}\}$ for one-form, the component of $\nabla U$ are denoted by $U^\alpha_{;\beta}$ and we write
$$\nabla \textit{\textbf{U}}=U^\alpha_{;\beta}\utilde{e^\beta}\otimes \textit{\textbf{e}}_{\bm \alpha}$$

{\bf VIII.} A connection $\nabla$ is a rule which maps a vector field $\textit{\textbf{U}}$ to a (1,1)-tensor field $\nabla \textit{\textbf{U}}$ (without any reference to a curve).\\

{\bf IX.} Though for a (1, 0)- tensor $\textit{\textbf{U}}$ (vector field), $\nabla \textit{\textbf{U}}$ is a (1,1)- tensor field but $\nabla$ is not a (0,1) tensor field as $\nabla (f\textit{\textbf{U}}) \neq f\nabla \textit{\textbf{U}}$ \textit{i.e.}, connection is not a tensor field. Here the tensor $\nabla \textit{\textbf{U}}$ is called the gradient of $\textit{\textbf{U}}$ and we write $$\nabla \textit{\textbf{U}}(\utilde{w} ; \textit{\textbf{V}})=\langle \utilde{w}, \nabla_{\textit{\textbf{V}}}\textit{\textbf{U}} \rangle$$

{\bf X.} The Leibnitz rule enable us to generalize the covariant derivative to tensors of arbitrary type:
$$\nabla (\textit{\textbf{W}}_{\bm 1} \otimes \textit{\textbf{W}}_{\bm 2})=(\nabla \textit{\textbf{W}}_{\bm 1})\otimes \textit{\textbf{W}}_{\bm 2}+\textit{\textbf{W}}_{\bm 1} \otimes \nabla \textit{\textbf{W}}_{\bm 2}$$
and
$$\nabla \langle \utilde{w}, \textit{\textbf{A}} \rangle = \langle \nabla \utilde{w}, \textit{\textbf{A}} \rangle +\langle \utilde{w}, \nabla \textit{\textbf{A}}\rangle ~~\textrm{(commutativity with contraction)}$$

Hence, as a consequence,
$$\nabla (A \otimes B)= \nabla A \otimes B+A\otimes \nabla B,$$
for any arbitrary tensors $A$ and $B$.\\

We shall now determine the components of covariant derivative with reference to some basis $\{\textit{\textbf{e}}_{\bm \alpha}\}$ of vectors and $\{\utilde{e^\alpha}\}$ of one forms. First of all let us consider $\nabla \textit{\textbf{e}}_{\bm \alpha}$, a (1,1) tensor. As tensors can be written as a linear combination of basis tensors which are the exterior product of basis vectors and basis one-forms so we write 
$$\nabla \textit{\textbf{e}}_{\bm \alpha}=\Gamma^\gamma_{\alpha \beta} \utilde{e^\beta} \otimes \textit{\textbf{e}}_{\bm \gamma}~~~i.e., \nabla_{\textit{\textbf{e}}_{\bm \delta}}\textit{\textbf{e}}_{\bm \alpha}=\Gamma_{\alpha \delta}^\gamma \textit{\textbf{e}}_{\bm \gamma}$$

Thus the co-efficients $\Gamma^\gamma_{\alpha \beta}$ can be written as the inner product 
$$\Gamma^\gamma_{\alpha \beta} =\langle \utilde{e^\gamma}, \nabla_{\textit{\textbf{e}}_{\bm \beta}}\textit{\textbf{e}}_{\bm \alpha} \rangle$$

These $n^3$ functions $\Gamma^\gamma_{\alpha \beta}$ are called christoffel symbols and they completely determine the affine connection. To have a clear idea about the mathematical objects $\Gamma^\gamma_{\alpha \beta}$ let us determine the transformation law for christoffel symbols from the transformation of basis vectors namely,
$$\textit{\textbf{e}}_{\bm \alpha'}=\Lambda^\beta _{~ \alpha'}\textit{\textbf{e}}_{\bm \beta}~\mbox{and}~~ \utilde{e}^{\lambda'}=\Lambda_\delta^{\lambda'} \utilde{e^\delta}$$
\begin{eqnarray}
\textrm{So,}~~\Gamma_{\alpha' \beta'}^{\gamma'}=\langle \utilde{e}^{\gamma'}, \nabla_{\textit{\textbf{e}}_{\bm \beta'}}^{\textit{\textbf{e}}_{\bm \alpha'}}\rangle
&=& \langle \Lambda_\gamma^{~ \gamma'}\utilde{e^\gamma}, \nabla_{(\Lambda_{\beta'}^\beta \textit{\textbf{e}}_{\bm \beta})}^{~ \Lambda^\alpha_{\alpha'}\textit{\textbf{e}}_{\bm \alpha}}\rangle\nonumber\\
&=&\Lambda_\gamma^{\gamma'} \Lambda_{\beta'}^\beta \langle \utilde{e^\gamma}, \nabla_{\textit{\textbf{e}}_{\bm \beta}}(\Lambda^\alpha_{\alpha'} \textit{\textbf{e}}_{\bm \alpha}) \rangle \nonumber \\
&=& \Lambda_\gamma^{\gamma'} \Lambda_{\beta'}^\beta \langle \utilde{e^\gamma}, \Lambda_{\alpha'}^\alpha \nabla_{\textit{\textbf{e}}_{\bm \beta}} \textit{\textbf{e}}_{\bm \alpha}+\textit{\textbf{e}}_{\bm \alpha}\nabla_{\textit{\textbf{e}}_{\bm \beta}} \Lambda_{\alpha'}^\alpha \rangle \nonumber \\
&=& \Lambda_{\alpha'}^\alpha \Lambda_{\beta'}^\beta \Lambda_\gamma^{\gamma'} \Gamma^\gamma_{\alpha \beta}+ \Lambda_\gamma^{\gamma'} \Lambda_{\beta'}^\beta \nabla_{\textit{\textbf{e}}_{\bm \beta}} \Lambda_{\alpha'}^\alpha \delta^\gamma_\alpha \nonumber \\
&=& \Lambda_{\alpha'}^\alpha \Lambda_{\beta'}^\beta \Lambda_\gamma^{\gamma'} \Gamma^\gamma_{\alpha \beta}+\Lambda^{~ \gamma'}_\alpha \nabla_{\textit{\textbf{e}}_{\bm \beta'}} \Lambda_{~ \alpha'}^\alpha \nonumber
\end{eqnarray}

In the above transformation law, the presence of the second term on the R.H.S shows that christoffel symbols are not component of a tensor- they are simply 3-index functions. However, for fixed `$\alpha$', $\Gamma^\gamma_{\alpha \beta}$ are components of a (1, 1) tensor \textit{i.e.}, $\{\Gamma^{\gamma}_{\alpha \beta} \utilde{e^\beta} \otimes \textit{\textbf{e}}_{\bm \gamma}\}$ is a collection of n (1, 1)- tensors.\\

In the above we have defined the covariant derivative of the basis vectors $\{\textit{\textbf{e}}_{\bm \alpha}\}$. Now we shall introduce in the following the covariant derivative of the dual basis vectors $\{\utilde{e^\alpha}\}$:
\begin{eqnarray}
\nabla_{\textit{\textbf{e}}_{\bm \alpha}}\langle \utilde{e^\gamma}, \textit{\textbf{e}}_{\bm \delta} \rangle &=&\langle \nabla_{\textit{\textbf{e}}_{\bm \alpha}} \utilde{e^\gamma}, \textit{\textbf{e}}_{\bm \delta} \rangle + \langle \utilde{e^\gamma}, \nabla_{\textit{\textbf{e}}_{\bm \alpha}}\textit{\textbf{e}}_{\bm \delta} \rangle\nonumber \\
\Rightarrow ~~~~\nabla_{\textit{\textbf{e}}_{\bm \alpha}} \delta^\gamma _\delta~~~& = &\langle \nabla_{\textit{\textbf{e}}_{\bm \alpha}} \utilde{e^\gamma}, \textit{\textbf{e}}_{\bm \delta} \rangle +\langle \utilde{e^\gamma}, \Gamma^\lambda_{\alpha \delta} \textit{\textbf{e}}_{\bm \lambda} \rangle \nonumber\\
\Rightarrow~~~~~~ 0~~~~~~~&=& \langle \nabla_{\textit{\textbf{e}}_{\bm \alpha}} \utilde{e^\gamma}, \textit{\textbf{e}}_{\bm \delta} \rangle +\Gamma^\lambda_{\alpha \delta} \delta^\gamma_\lambda\nonumber\\
 \Rightarrow \langle \nabla_{\textit{\textbf{e}}_{\bm \alpha}} \utilde{e^\gamma}, \textit{\textbf{e}}_{\bm \delta} \rangle& =& -\Gamma^\gamma_{\alpha \delta},~~i.e.,~~\nabla_{\textit{\textbf{e}}_{\bm \alpha}}\utilde{e^\gamma}=-\Gamma^\gamma_{\alpha \beta} \utilde{e^\delta}\nonumber
\end{eqnarray}

For arbitrary vectors $\textit{\textbf{U}}=U^\alpha \textit{\textbf{e}}_{\bm \alpha}$ and $\textit{\textbf{V}}=V^\alpha \textit{\textbf{e}}_{\bm \alpha}$ ,
\begin{eqnarray}
\nabla_{\textit{\textbf{V}}}\textit{\textbf{U}}=V^\alpha \nabla_{\textit{\textbf{e}}_{\bm \alpha}} U^\beta \textit{\textbf{e}}_{\bm \beta}
&=& V^\alpha \left(\nabla_{\textit{\textbf{e}}_{\bm \alpha}}U^\beta\right)\textit{\textbf{e}}_{\bm \beta} +V^\alpha U^\beta \nabla_{\textit{\textbf{e}}^{\bm \alpha}} \textit{\textbf{e}}_{\bm \beta} \nonumber \\
&=& V^\alpha\left(\nabla_{\textit{\textbf{e}}_{\bm \alpha}}U^\beta\right)\textit{\textbf{e}}_{\bm \beta}+V^\alpha U^\beta \Gamma^\gamma_{\alpha \beta} \textit{\textbf{e}}_{\bm \gamma} \nonumber \\
&=& V^\alpha \left(\nabla_{\textit{\textbf{e}}_{\bm \alpha}}U^\beta + U^\delta \Gamma^\beta_{\alpha \delta}\right)\textit{\textbf{e}}_{\bm \beta} \nonumber
\end{eqnarray}

However, if we choose $\textit{\textbf{V}}=\dfrac{d}{d\lambda}$ then we obtain\\
$$\nabla_{\textit{\textbf{V}}}\textit{\textbf{U}}=\left(\frac{dU^\beta}{d \lambda}+\Gamma^\beta_{\alpha \gamma}U^\gamma V^\alpha\right)\textit{\textbf{e}}_{\bm \beta}$$

So $\nabla \textit{\textbf{U}}=\nabla (U^\alpha \textit{\textbf{e}}_{\bm \alpha})=dU^\alpha \otimes \textit{\textbf{e}}_{\bm \alpha}+U^\alpha \Gamma^\gamma_{\alpha \beta} \utilde{e^\beta} \otimes \textit{\textbf{e}}_{\bm \gamma}$.\\

Similarly for covectors we have 
\begin{eqnarray}
\nabla_{\textit{\textbf{V}}}\utilde{w}=\nabla_{V^\alpha \textit{\textbf{e}}_{\bm \alpha}}\left(w_\beta \utilde{e^\beta}\right)
&=& V^\alpha \left[\left(\nabla_{\textit{\textbf{e}}_{\bm \alpha}}\utilde{e^\beta}\right)w_\beta+\left(
\nabla_{\textit{\textbf{e}}_{\bm \alpha}}w_\beta\right)\utilde{e^\beta}\right] \nonumber \\
&=& V^\alpha\left[\left(\nabla_{\textit{\textbf{e}}_{\bm \alpha}}w_\beta\right)\utilde{e^\beta}-
\Gamma^\beta_{\alpha \delta}\utilde{e^\delta}w_\beta\right] \nonumber \\
&=& V^\alpha \left[\left(\nabla_{\textit{\textbf{e}}_{\bm \alpha}}w_\beta \right)-\Gamma^\delta_{\alpha \beta} w_\delta\right]\utilde{e^\beta} \label{1.31}
\end{eqnarray}

Again for $\textit{\textbf{V}}=\dfrac{d}{d\lambda}$,
$$\nabla_{\textit{\textbf{V}}}\utilde{w}=\left[\frac{dw_\beta}{d\lambda}-\Gamma^\delta_{\alpha \beta} w_\delta V^{\alpha}\right]\utilde{e^\beta}$$

Also $\nabla \utilde{w}=\nabla (w_\alpha \utilde{e^\alpha})=dw_\alpha \otimes \utilde{e^\alpha}-w_\alpha \Gamma^\alpha_{\beta \gamma} \utilde{e^\beta} \otimes \utilde{e^\gamma}$\\

In particular, if we choose the co-ordinate basis namely $\{\frac{\partial}{\partial x^\alpha}\}$ and $\{dx^\alpha\}$ then the components of $\nabla \textit{\textbf{U}}$ are denoted by $U^\alpha;_{\beta}$ and are given by
\begin{equation} \label{1.32}
\left(\nabla U\right)^\alpha_\beta \equiv U^\alpha;_\beta = \frac{\partial U^\alpha}{\partial x^\beta}+ \Gamma^\alpha_{\beta \gamma} U^\gamma
\end{equation}

Note that  neither the first term nor the second term on the r.h.s are components of a tensor but their sum is a component of a tensor. \\

Similarly, the components of the covariant derivative of one-form $\utilde{\omega}$ are denoted by $\omega_{\alpha ; \beta}$ and has the expression 
$$(\nabla \omega)_{\alpha \beta} \equiv \omega_{\alpha ; \beta}=\frac{\partial \omega_\alpha}{\partial x^\beta} - \Gamma^\gamma_{\alpha \beta}\omega_\gamma$$

For two co-ordinate systems (\textit{i.e.}, two co-ordinate bases) the transformation law for Christoffel symbols are 
\begin{equation} \label{1.33}
\Gamma^{\gamma'}_{\alpha' \beta'}= \frac{\partial x^\alpha}{\partial x^{\alpha'}}\frac{\partial x^\beta}{\partial x^{\beta'}}\frac{\partial x^{\gamma'}}{\partial x^{\gamma}}\Gamma^{\gamma}_{\alpha \beta}+\frac{\partial x^{\gamma'}}{\partial x^{\delta}}\frac{\partial^2 x^\delta}{\partial x^{\alpha'} \partial x^{\beta'}}
\end{equation}

The above transformation law shows that if we have two different connections and $\Gamma$ and $\widehat{\Gamma}$ be the corresponding Christoffel symbols then, in co-ordinate transformation 
$$\Gamma^{\gamma'}_{\alpha' \beta'}-\widehat{\Gamma}^{\gamma'}_{\alpha' \beta'}=\frac{\partial x^\alpha}{\partial x^{\alpha'}}\frac{\partial x^\beta}{\partial x^{\beta'}}\frac{\partial x^{\gamma'}}{\partial x^{\gamma}}(\Gamma^{\gamma}_{\alpha \beta}-\widehat{\Gamma}^{\gamma}_{\alpha \beta})$$
\textit{i.e.}, the difference of two Christoffel symbols are components of a (1,2)-tensor.\\

For any scalar `$f$', the components of covariant derivative is simply the partial derivative \textit{i.e.},
$$(\nabla f)_\beta \equiv f_{; \beta}=\frac{\partial f}{\partial x^\beta}=f_{, \beta}$$

Suppose $A$ is a (1,1)-tensor having components $A^\alpha_\beta$ in a given basis \textit{i.e.}, 
$$A=A^\alpha_\beta \utilde{e^\beta} \otimes \textit{\textbf{e}}_{\bm \alpha}$$
then 
$$\nabla A=(\nabla A^\alpha_\beta)\utilde{e^\beta}\otimes \textit{\textbf{e}}_{\bm \alpha}+A^\alpha_\beta \nabla \utilde{e^\beta} \otimes \textit{\textbf{e}}_{\bm \alpha}+A^\alpha_\beta \utilde{e^\beta}\otimes \nabla \textit{\textbf{e}}_{\bm \alpha}$$

So in a co-ordinate basis we have
$$(\nabla A)^\alpha_{\beta \gamma}=A^\alpha_{\beta; \gamma}=\frac{\partial A^\alpha_\beta}{\partial x^\gamma}-\Gamma^\delta_{\beta \gamma}A^\alpha_\delta+\Gamma^\alpha_{\delta \gamma}A^\delta_\beta$$

In general for any ($r,s$)-tensor $B$ we have
\begin{equation} \label{1.34}
\left(\nabla B \right)^{\alpha_1 \ldots \alpha_r}_{\beta_1 \ldots \beta_s \gamma}\equiv B^{\alpha_1 \ldots \alpha_r}_{\beta_1 \ldots \beta_s ;\gamma}=\frac{\partial B^{\alpha_1 \ldots \alpha_r}_{\beta_1 \ldots \beta_s }}{\partial x^\gamma}+\Gamma^{\alpha_1}_{\delta_1 \gamma}B^{\delta_1 \alpha_2 \ldots \alpha_r}_{\beta_1 \ldots \beta_s}+\ldots -\Gamma^{\delta_1}_{\beta_1 \gamma}B^{\alpha_1 \ldots \alpha_r}_{\delta_1 \beta_2 \ldots \beta_s}- \cdots \cdots
\end{equation}

Note that covariant derivative of a ($r,s$)-tensor is a ($r,s+1$)- tensor.\\

For any arbitrary vector fields $\textit{\textbf{U}}$ and $\textit{\textbf{V}}$, a (1,2)-tensor field $T$ can be defined as 
\begin{equation} \label{1.35}
T(\textit{\textbf{U}}, \textit{\textbf{V}})=\nabla_{\textit{\textbf{U}}}\textit{\textbf{V}}-
\nabla_{\textit{\textbf{V}}}\textit{\textbf{U}}-[\textit{\textbf{U}}, \textit{\textbf{V}}]
\end{equation}

In a co-ordinate basis, the components of $T$ are
$$T^\alpha_{\beta \gamma}=\Gamma^\alpha_{\beta \gamma}-\Gamma^\alpha_{\gamma \beta}$$

This tensor is called the torsion tensor. It is an antisymmetric tensor. A connection is said to be symmetric (or torsion free) if the torsion tensor is identically zero and we have $\Gamma^\alpha_{\beta \gamma}=\Gamma^\alpha_{\gamma \beta}$. Further, for any scalar function `$f$' if $f_{;\alpha \beta}=f_{;\beta \alpha}$ then the corresponding connection is torsion-free.

For a torsion free connection, we have a relation between Lie derivative and covariant derivative as follows:
$$L_{\textit{\textbf{V}}}\textit{\textbf{U}}=[\textit{\textbf{V}}, \textit{\textbf{U}}]=\nabla_{\textit{\textbf{V}}}\textit{\textbf{U}}-
\nabla_{\textit{\textbf{U}}}\textit{\textbf{V}}$$
\textit{i.e.}, in components
$$\left(L_{\textit{\textbf{V}}}\textit{\textbf{U}}\right)^\alpha=U^\alpha_{;\beta} V^\beta-V^\alpha_{;\beta }U^\beta$$

Thus for any arbitrary tensor $T$ of type ($r,s$)
$$\left(L_{\textit{\textbf{V}}}T\right)^{\alpha_1 \ldots \alpha_r}_{\beta_1 \ldots \beta_s}=T^{\alpha_1 \ldots \alpha_r}_{\beta_1 \ldots \beta_s; \gamma}V^\gamma-T^{\gamma \alpha_2 \ldots \alpha_r}_{\beta_1 \ldots \beta_s}V^{\alpha_1}_{;\gamma}\ldots +T^{\alpha_1 \ldots \alpha_r}_{\gamma \beta_1 \ldots \beta_s}V^\gamma_{;\beta_1}+\ldots $$

Similarly, one can relate the covariant derivative with exterior derivative by the following relation:\\

For any $p$-form $B$, the ($p+1$)-form $dB$ can be written as 
$$dB=B_{\alpha_1 \alpha_2 \ldots \alpha_p ; \delta}dx^\delta \Lambda dx^{\alpha_1} \Lambda \ldots dx^{\alpha_p}$$
or in component form
$$(dB)_{\alpha_1 \alpha_2 \ldots \alpha_p \delta}= (-1)^p B_{[\alpha_1 \ldots \alpha_p; \delta]}$$

Thus we have seen that Lie derivative and exterior derivative are related to the covariant derivative for symmetric connection. But it should be remember that Lie derivative and exterior derivative do not need extra structure on the manifold \textit{i.e.}, independent of the connection, so the above relations between Lie derivative (or exterior derivative) and covariant derivative do not depend on connection (\textit{i.e.}, semicolons may be replaced by comma).\\

\subsection{Intrinsic Differentiation}

~~~We now extend the covariant differentiation demanding that differentiation does not change the order of the tensor. Such differentiation is called intrinsic differentiation. Let $\gamma$ be any curve in the manifold parametrized by $\lambda$ then intrinsic derivative of any tensor A of type $(r,s)$ is denoted by $\dfrac{\delta A}{d \lambda}$ and is also a (r,s)- tensor. If $\textit{\textbf{v}}$ be the tangent vector to $\gamma$ then the components of $\dfrac{\delta A}{d \lambda}$ is defined as
\begin{equation} \label{1.36}
\frac{\delta A^{\alpha_1 \ldots \alpha_r}_{\beta_1 \ldots \beta_s}}{d \lambda}=A^{\alpha_1 \ldots \alpha_r}_{\beta_1 \ldots \beta_s; \delta}v^\delta
\end{equation}

If we choose a local co-ordinate basis such that $v^\delta=\dfrac{dx^\delta}{d\lambda}$ then the intrinsic derivative of a vector $\textit{\textbf{W}}$ is expressed as 
\begin{equation} \label{1.37}
\frac{\delta W^\alpha}{d\lambda}=\frac{\partial W^\alpha}{\partial \lambda}+\Gamma^\alpha_{\beta \delta}W^\beta \frac{dx^\delta}{d \lambda}
\end{equation}

We shall now introduce the notion of parallel transport using intrinsic derivative. A tensor $A$ is said to be parallely transported along a curve $\gamma$ (parametrized by $\lambda$) if $\dfrac{\delta A}{d \lambda}=0$. In local co-ordinate system this gives a system of first order linear differential equations for the components of the parallely transported tensor. The uniqueness of the solution of such ordinary differential equation shows that we obtain a unique tensor at each point of $\gamma$ by parallely transporting a tensor along $\gamma$. So we can consider this idea of parallel transfer as a linear map from $T^r_s(P)$ to $T^r_s(Q)$ ($A$ is a ($r,s$)-tensor and $P$ and $Q$ are points on $\gamma$). It is clear that this linear map preserves all tensor products and tensor contractions. In particular, if we parallely transported the basis vectors of $T_P$ to $T_Q$ along $\gamma$ then the transported vectors at $Q$ forms a basis for $T_Q$ and there will be isomorphism between $T_P$ and $T_Q$.\\

\section{Geodesics}

~~~In this section we consider as  a particular case the parallel transport of the tangent vector along the curve itself. A curve $\gamma$ is said to be a geodesic if it parallely transported its own tangent vector. So for the geodesic curve
\begin{equation} \label{1.38}
\frac{\delta \textit{\textbf{v}}}{d \lambda}=0~~i.e., v^\alpha_{;\beta} v^\beta=0
\end{equation}
where $\lambda$ is the parameter along the curve $\gamma$ and $\textit{\textbf{v}}$ is the tangent vector to $\gamma$. In a local co-ordinate system $\{x^\alpha\}$ the explicit form of the differential equation of the geodesic equation is 
\begin{eqnarray}\frac{d v^\alpha}{d \lambda}~+~\Gamma^\alpha_{\beta \gamma}v^\beta v^\gamma~=~0~~\nonumber\\
i.e., \frac{d^2 x^\alpha}{d \lambda^2}+\Gamma^\alpha_{\beta \gamma}\frac{d x^\beta}{d \lambda}\frac{dx^\gamma}{d\lambda}=0\label{1.39}
\end{eqnarray}

This system of second order quasi-linear differential equation for $x^\alpha(\lambda)$ determine the geodesic curve. Here $\lambda$ is termed as the affine parameter. It is clear that $\lambda$ is unique upto an additive and multiplicative constant \textit{i.e.}, if $\mu=a\lambda+b$ ($a,~b$ are constants) then $\mu$ is also an affine parameter of the geodesic. The arbitrary constant `$b$' gives the freedom of choosing the initial point and the freedom in the choice of `$a$' suggest that we can scale the tangent vector by any constant (renormalization).\\

Consider the above quasi-linear differential equation for the geodesic, by the standard existence theorems for ordinary differential equations. It is possible to have a geodesic through any point P of the manifold such that the tangent to the geodesic at $P$ is a given vector from $T_P$. This geodesic $\gamma (\lambda)$ is unique and depends continuously both on the point $P$ and the direction at $P$. Such a geodesic is called a maximal geodesic.\\

We now introduce the idea of exponential mapping from $T_P$ to $M$ as follows. Given any $\textit{\textbf{v}}\in T_P$, exp($\textit{\textbf{v}}$) maps to the point in $M$ at a unit parameter distance from $P$ along  $\gamma_{\textit{\textbf{v}}}(\lambda)$, the maximal geodesic through $P$ in the direction of $\textit{\textbf{v}}$. This map may not be defined for all $\textit{\textbf{v}}\in T_P$ as $\gamma_{\textit{\textbf{v}}}(\lambda)$ may have restriction on $\lambda$. A maximal geodesic is said to be \underline{complete} if it is defined for all parametric values of $\lambda$. A manifold is said to be \underline{geodesically complete} if all geodesic on M are complete \textit{i.e.}, exponential mapping can be defined for all elements of $T_P$ and also for all points $P$ of $M$.

The idea of exponential mapping enable us to obtain a neighbourhood $N_P$ of a point $P$ of the manifold such that each point of $N_P$ are at unit parameter distance from $P$ along some maximal geodesic through $P$. By the implicit function theorem there exists an open neighbourhood of the origin of $T_P$ which maps to $N_P$ by exponential map which is also a diffeomorphism. Such open neighbourhood $N_P$ is called a normal neighbourhood of $P$. If any two points $Q$ and $R$ in the normal neighbourhood are such that they can be joined by a unique geodesic which is completely within $N_P$ then $N_P$ is called a \underline{convex normal neighbourhood}. Suppose 
$\gamma_{\textit{\textbf{v}}}(\lambda)$ be a maximal geodesic through $P$. By \underline{exponential map} exp($\textit{\textbf{v}}$) we obtain a point $Q$ in $N_P$ along $\gamma_{\textit{\textbf{v}}}(\lambda)$. The co-ordinates of Q can be written as $x^\alpha=\lambda v^\alpha (P)~(\lambda=0~\mbox{at}~P)$. So $\dfrac{d^2 x^\alpha}{d \lambda^2}=0$ and by the geodesic equation $\Gamma^\alpha_{\beta \gamma}v^\beta_{(P)}v^\gamma_{(P)}=0$. Due to arbitrary choice of $\textit{\textbf{v}}$ at $P$ one must have $\Gamma^\alpha_{\beta \gamma}=0 ~\mbox{at}~P$. Thus it is possible to have a co-ordinate system in $N_P$ such that the components of christoffel symbols vanish at $P$ (not necessarily at other points of $N_P$) but not the derivatives of it. Such a co-ordinate system is called a \underline{normal co-ordinates} in $N_P$. This co-ordinate system is useful (simplification to a great extend) for proving some properties of the manifold.\\
\textbf{Note:}\\\\
{\bf I.} If two vectors in $T_P$ are parallel then their geodesic curves will be identical but the affine parameters are different. So by exp. map we obtain different points on the geodesic path.\\\\
{\bf II.} The geodesic equation enables us to give some geometrical picture of torsion. Let $\gamma$ be a geodesic through $P$ having tangent vector $\textit{\textbf{v}}$. Suppose ${\bm {\xi}}$ be another vector linearly independent to $\textit{\textbf{v}}$ and $C$ be the geodesic with ${\bm {\xi}}$ as the tangent vector. We then parallel transport $\textit{\textbf{v}}$ along the geodesic `$C$' through a small affine parameter distance $\epsilon$ and construct a new geodesic $\gamma_P$ (having tangent $\textit{\textbf{v}}$) through this new point. Thus we obtain a congruence of geodesic in the neighbourhood of $P$. Finally, we parallely transport the linking vector ${\bm {\xi}}$ along this congruence of geodesics.\\
Then we have
$$\left(\mathcal{L}_{\textit{\textbf{v}}}{\bm {\xi}}\right)^\alpha=-T^\alpha_{\beta \delta} \xi^\beta v^\delta ~~~~~~\left(\because ~~\nabla _{\textit{\textbf{v}}}{\bm {\xi}} = 0 = \nabla _{\bm {\xi}}\textit{\textbf{v}}\right)$$
Thus if the torsion is non-zero \textit{i.e.}, the connection is not symmetric then the vector ${\bm {\xi}}$ does not remain fixed in this congruence. We say that ${\bm {\xi}}$ is rotated relatively to nearby geodesic by the effect of torsion. Conversely, if we take parallely transported vector ${\bm {\xi}}$ as the standard fixed direction then the congruence of parallel geodesic twists with respect to the geodesic of ${\bm {\xi}}$.\\\\
{\bf III.} The geodesic equation shows that only the symmetric part of a connection contributes to the geodesic equation.\\

\section{Riemann Curvature Tensor}

~~~In this section we start with parallel transport of a vector along a closed curve. In general, if we parallely transport a vector $\textit{\textbf{W}}$ starting from a point $P$ on a closed curve $\gamma$ and return to the same point along $\gamma$ then the transported vector $\textit{\textbf{W}}_{TP}$ is different from $\textit{\textbf{W}}$. Further, if we parallely transport the same vector $\textit{\textbf{W}}$ along a different closed path $\gamma^{~\prime}$ through $P$ then the resulting vector will be $\textit{\textbf{W}}_{TP}'$ which is in general different from $\textit{\textbf{W}}_{TP}$ (and $\textit{\textbf{W}}$). This non-integrability of parallel transfer is related to the non-commutativity of covariant derivatives as follows:\\

\begin{wrapfigure}{r}{0.45\textwidth}\vspace{-1.9\intextsep}
\includegraphics[height=4 cm , width=8 cm ]{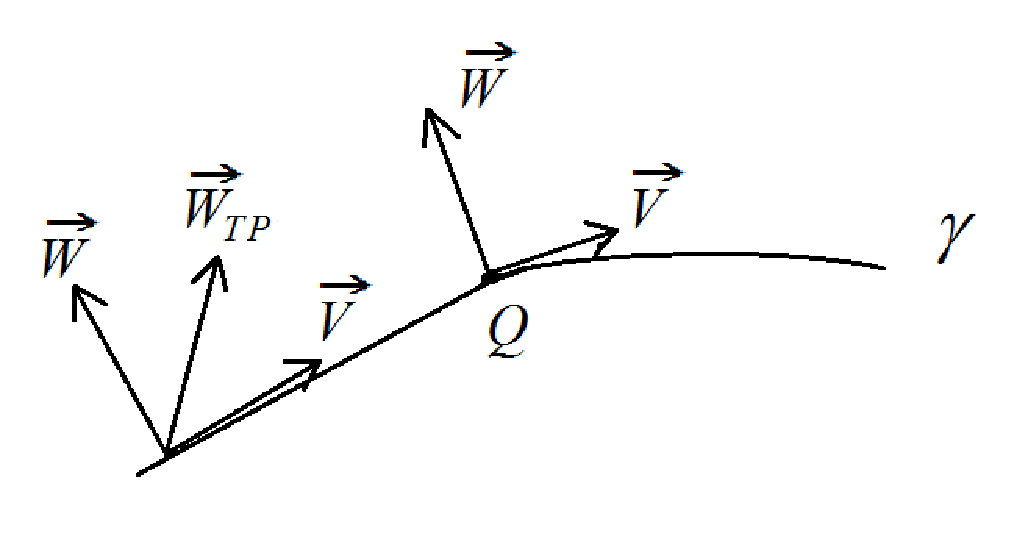}\vspace{-\intextsep}
\begin{center}
Fig. 1.8
\end{center}\vspace{-\intextsep}
\end{wrapfigure}
Let $\textit{\textbf{W}}$ be a vector field defined along a curve $\gamma$ for which the tangent vector is $\textit{\textbf{V}}$. $P$ and $Q$ are two neighbouring points on $\gamma$ at a parameter distance $\triangle \lambda$. If $\textit{\textbf{W}}$ at $Q$ is parallely transported at $P$ and we denote the transported vector at $P$ by $\textit{\textbf{W}}_{TP}$, called the images of $\textit{\textbf{W}}(Q)$ at $P$. Then from the definition of covariant derivative we write
$$\textit{\textbf{W}}_{TP}(P)=\textit{\textbf{W}}_{P}(P)+\Delta \lambda \nabla _{\textit{\textbf{V}}}\textit{\textbf{W}}_{P}(P)+ O(\Delta \lambda^2)$$

We now consider two family of integral curves (congruences) for which the tangent vectors are $\textit{\textbf{U}}$ and $\textit{\textbf{V}}$ and let $\mu$ and $\lambda$ be the parameters along these congruences.\\

Let $RQ,SP, \ldots $ are the integral curves for $\textit{\textbf{V}}$ and $RS,QP, \ldots $ are the integral curves for $\textit{\textbf{U}}$ (see fig. 1.9). We assume $[\textit{\textbf{V}},\textit{\textbf{U}}]=0$ \textit{i.e.}, $\textit{\textbf{U}}$ is Lie dragged by $\textit{\textbf{V}}$, so that it is possible to form a closed loop by their interaction. Suppose $\textit{\textbf{W}}$ be a vector field defined over these congruences .We parallely transport the vector field $\textit{\textbf{W}}$ from $R$ to $P$ along two paths- First we transport $\textit{\textbf{W}}$ from $R$ to $Q$ along the integral curve of $\textit{\textbf{V}}$ and then from $Q$ to $P$ along the integral curve of $\textit{\textbf{U}}$. In the second path, we first parallely transport from $R$ to $S$ along the integral curve of $\textit{\textbf{U}}$ and then from $S$ to $P$ along the integral curve of $\textit{\textbf{V}}$. Thus
$$\textit{\textbf{W}}_{TP}(R\rightarrow Q\rightarrow P)=\textit{\textbf{W}}_{P}+\Delta \mu \Delta \lambda \nabla_{\textit{\textbf{U}}}\nabla_{\textit{\textbf{V}}}\textit{\textbf{W}}_{P}+O(\Delta \lambda^n \Delta \mu ^m)~~~~~~\left((n+m) \geq 3.\right)$$

Similarly
$$\textit{\textbf{W}}_{TP}(R\rightarrow S\rightarrow P)=\textit{\textbf{W}}_{P}+ \Delta \lambda \Delta \mu \nabla_{\textit{\textbf{V}}} \nabla_{\textit{\textbf{U}}}\textit{\textbf{W}}_{P}+O(\Delta \lambda^n \Delta \mu ^m)$$

Hence,
$$\Delta \textit{\textbf{W}}(P)=\textit{\textbf{W}}_{TP}(R\rightarrow Q\rightarrow P)-\textit{\textbf{W}}_{TP}(R\rightarrow S\rightarrow P)=\Delta \mu \Delta \lambda [\nabla_{\textit{\textbf{U}}}, \nabla_{\textit{\textbf{V}}}]\textit{\textbf{W}}(P)$$

We shall now introduce the notion of geodesic deviation - another geometrical aspect of non-commutativity of covariant derivatives.\\

\begin{figure}[h]\vspace{-\intextsep}
\begin{center}
\includegraphics[height=5 cm , width=7 cm ]{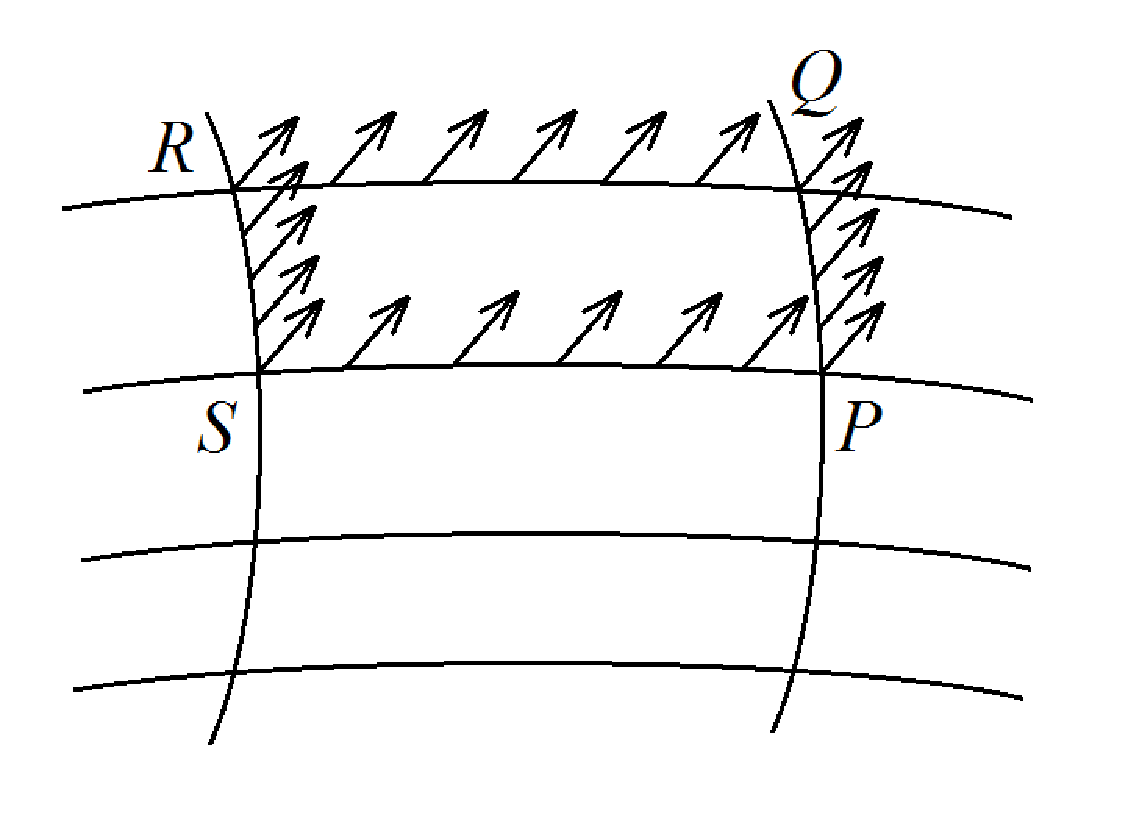}~~~~~~
\includegraphics[height=4 cm , width=7 cm ]{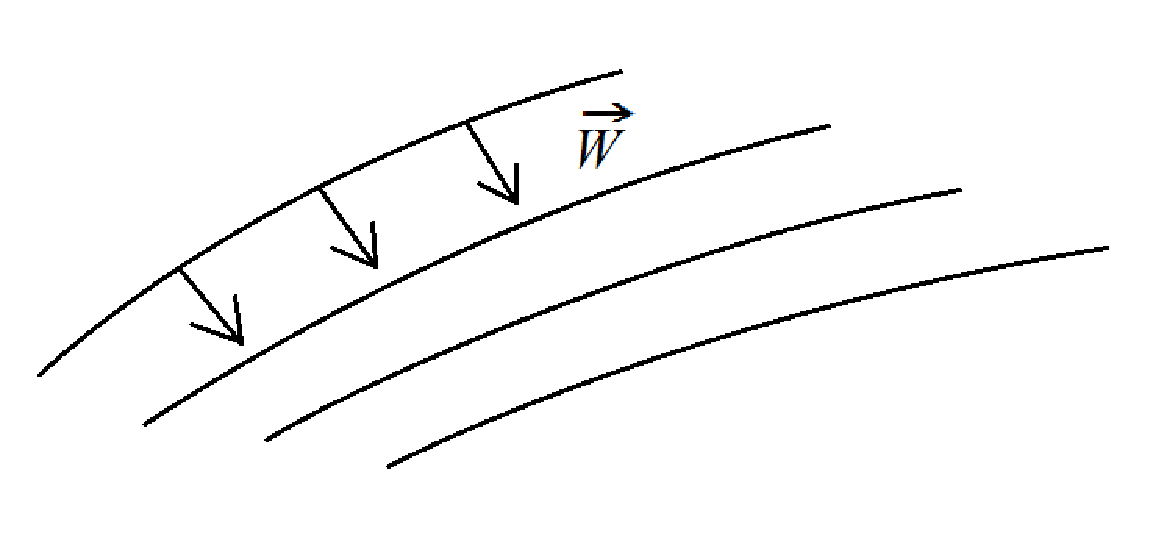}
\end{center}\vspace{-\intextsep}
\begin{center}
Fig. 1.9~~~~~~~~~~~~~~~~~~~~~~~~~~~~~~~~~~~~~~~~~~~~~~~~~~Fig 1.10
\end{center}\vspace{-\intextsep}
\end{figure}
In general geodesics which were parallel at the beginning do not remain parallel throughout. Suppose we consider a congruence of geodesics with tangent vector $\textit{\textbf{V}}~(i.e., ~ \nabla_{\textit{\textbf{V}}}\textit{\textbf{V}}=0)$ and $\textit{\textbf{W}}$ be a connecting vector, lie dragged by the congruence (\textit{i.e.}, $L_{\textit{\textbf{V}}}\textit{\textbf{W}}=0$). The change of $\textit{\textbf{W}}$ along the geodesics give the measure of geodesic deviation. We can say that $\nabla_{\textit{\textbf{V}}}\textit{\textbf{W}}$ depends on whether the geodesics are parallel or not at the starting point, while $\nabla_{\textit{\textbf{V}}}\nabla_{\textit{\textbf{V}}}\textit{\textbf{W}}$ gives the initial rate of separation of the geodesic changes. Using the relation between Lie derivative and co-variant derivative for torsionless connection\,(\textit{i.e.}, symmetric connection) we have
\begin{eqnarray}
\nabla_{\textit{\textbf{V}}}\nabla_{\textit{\textbf{V}}}\textit{\textbf{W}}
&=& \nabla_{\textit{\textbf{V}}}(\nabla_{\textit{\textbf{W}}}\textit{\textbf{V}}+L_{\textit{\textbf{V}}}
\textit{\textbf{W}})=
\nabla_{\textit{\textbf{V}}}\nabla_{\textit{\textbf{W}}}\textit{\textbf{V}} \nonumber \\
&=& [\nabla_{\textit{\textbf{V}}},\nabla_{\textit{\textbf{W}}}] \textit{\textbf{V}}+\nabla_{\textit{\textbf{W}}}\nabla_{\textit{\textbf{V}}}
\textit{\textbf{V}}=[\nabla_{\textit{\textbf{V}}},\nabla_{\textit{\textbf{W}}}] \textit{\textbf{V}}, \nonumber
\end{eqnarray}
which shows that the measure of geodesic deviation is also related to the non-commutativity of the covariant derivatives.\\

We shall define this non-commutativity of covariant derivative as the measure of curvature of the manifold. We define an operator R as
\begin{equation} \label{1.40}
R(\textit{\textbf{U}}, \textit{\textbf{V}})\equiv [\nabla_{\textit{\textbf{U}}},\nabla_{\textit{\textbf{V}}}]-\nabla_{[\textit{\textbf{U}}, \textit{\textbf{V}}]}
\end{equation}
\textit{i.e.},
$$R(\textit{\textbf{U}}, \textit{\textbf{V}})\textit{\textbf{W}}=\nabla_{\textit{\textbf{U}}}(\nabla_{\textit{\textbf{V}}}\textit{\textbf{W}})-\nabla_{\textit{\textbf{V}}}(\nabla_{\textit{\textbf{U}}}\textit{\textbf{W}})-\nabla_{[\textit{\textbf{U}}, \textit{\textbf{V}}]}\textit{\textbf{W}}$$

Then it is easy to see that
\[
\begin{array}{l}
\mbox{i)}~R(\textit{\textbf{U}}, \textit{\textbf{V}})(f\textit{\textbf{W}})=fR(\textit{\textbf{U}}, \textit{\textbf{V}})(\textit{\textbf{W}}), \\
\mbox{ii)}~R(g\textit{\textbf{U}}, \textit{\textbf{V}})(\textit{\textbf{W}})=gR(\textit{\textbf{U}}, \textit{\textbf{V}})(\textit{\textbf{W}})~.
\end{array}
\]
for any arbitrary functions $f$ and $g$.\\

Hence $R$ is simply a multiplicative operator, does not depend on derivatives of $\textit{\textbf{U}}$ and $\textit{\textbf{V}}$. In fact $R(\textit{\textbf{U}}, \textit{\textbf{V}})\textit{\textbf{W}}$ is linear in $\textit{\textbf{U}}, \textit{\textbf{V}}~\mbox{and}~ \textit{\textbf{W}}$ and it depends only on the values of $\textit{\textbf{U}}, \textit{\textbf{V}}~\mbox{and}~ \textit{\textbf{W}}$ at the given point. So it is a (1,3)-tensor. This is called the Riemann curvature tensor. Now with respect to dual bases $\{\textit{\textbf{e}}_{\bm \alpha}\}$ and $\{\utilde{e^\alpha}\}$, the components of the Riemann curvature tensor $R^\alpha_{\beta \gamma \delta}$ are defined as 
$$R^\alpha_{\beta \gamma \delta}\textit{\textbf{e}}_{\bm \alpha}=[\nabla_\gamma, \nabla_\delta] \textit{\textbf{e}}_{\bm \beta}-\nabla_{[\textit{\textbf{e}}_{\bm \gamma}, \textit{\textbf{e}}_{\bm \delta}]}\textit{\textbf{e}}_{\bm \beta}$$
\textit{i.e.}, $R^\alpha_{\beta \gamma \delta}=\langle \utilde{e^\alpha}, R(\textit{\textbf{e}}_{\bm \gamma}, \textit{\textbf{e}}_{\bm \delta})\textit{\textbf{e}}_{\bm \beta}\rangle$\\

Thus for arbitrary vector fields $\textit{\textbf{U}}~, \textit{\textbf{V}}~\mbox{and}~ \textit{\textbf{W}}$\\
$$R^\alpha_{\beta \gamma \delta} U^\gamma V^\delta W^\beta=(W^\alpha_{;\delta}V^\delta)_{;\gamma}U^\gamma-(W^\alpha_{;\delta}U^\delta)_{;\gamma}V^\gamma- W^\alpha_{;\delta}(V^\delta_{;\gamma}U^\gamma-U^\delta_{;\gamma}V^\gamma)$$

So for arbitrary vector fields $\textit{\textbf{U}}~\mbox{and}~ \textit{\textbf{V}}$ we have
$$W^\alpha_{;\delta \gamma}-W^\alpha_{; \gamma \delta}=R^\alpha_{\beta \gamma \delta}W^\beta,$$
which shows the non-commutation of second covariant derivatives in terms of the Riemann tensor. We now find an explicit expression of the components of Riemann curvature tensor in a co-ordinate basis. As\\
\[
\begin{array}{l}
R^\alpha_{\beta \gamma \delta}=\langle \utilde{e^\alpha}, R(\textit{\textbf{e}}_{\bm \gamma}, \textit{\textbf{e}}_{\bm \delta})\textit{\textbf{e}}_{\bm \beta}\rangle \\
~~~~~~=\langle \utilde{e^\alpha}, \nabla_{\textit{\textbf{e}}_{\bm \gamma}}(\nabla_{\textit{\textbf{e}}_{\bm \delta}}\textit{\textbf{e}}_{\bm \beta})-\nabla_{\textit{\textbf{e}}_{\bm \delta}}(\nabla_{\textit{\textbf{e}}_{\bm \gamma}}\textit{\textbf{e}}_{\bm \beta})-\nabla_{\left[\textit{\textbf{e}}_{\bm \gamma},\textit{\textbf{e}}_{\bm \delta} \right]}\textit{\textbf{e}}_{\bm \beta} \rangle \\
~~~~~~=\langle \utilde{e^\alpha}, \nabla_{\textit{\textbf{e}}_{\bm \gamma}}\left(\Gamma^\mu_{\delta \beta}\textit{\textbf{e}}_{\bm \mu}\right)  \rangle-\langle \utilde{e^\alpha},  \nabla_{\textit{\textbf{e}}_{\bm \delta}}\left(\Gamma^\mu_{\gamma \beta}\textit{\textbf{e}}_{\bm \mu}\right) \rangle-\langle \utilde{e^\alpha},  \nabla_{\left[\textit{\textbf{e}}_{\bm \gamma},\textit{\textbf{e}}_{\bm \delta} \right]}\textit{\textbf{e}}_{\bm \beta}  \rangle
\end{array}
\]

As in co-ordinate basis $[\textit{\textbf{e}}_{\bm \gamma},\textit{\textbf{e}}_{\bm \delta} ]=0$, so we get
\begin{eqnarray}
R^\alpha_{\beta \gamma \delta} &=& \langle \utilde{e^\alpha}, \left(\frac{\partial}{\partial x^\gamma}\Gamma^\mu_{\delta \beta}\right)\textit{\textbf{e}}_{\bm \mu}+\Gamma^\mu_{\delta \beta}\nabla_{\textit{\textbf{e}}_{\bm \gamma}}\textit{\textbf{e}}_{\bm \mu} \rangle-\langle \utilde{e^\alpha}, \left(\frac{\partial}{\partial x^\delta}\Gamma^\mu_{\gamma \beta}\right)\textit{\textbf{e}}_{\bm \mu}+\Gamma^\mu_{\gamma \beta}\nabla_{\textit{\textbf{e}}_{\bm \delta}}\textit{\textbf{e}}_{\bm \mu} \rangle \nonumber \\
&=& \left(\frac{\partial}{\partial x^\gamma}\Gamma^\mu_{\delta \beta}\right)\langle\utilde{e^\alpha},\textit{\textbf{e}}_{\bm \mu} \rangle+ \Gamma^\mu_{\delta \beta}\langle\utilde{e^\alpha},\nabla_{\textit{\textbf{e}}_{\bm \gamma}}\textit{\textbf{e}}_{\bm \mu} \rangle - \left(\frac{\partial}{\partial x^\delta}\Gamma^\mu_{\gamma \beta}\right)\langle\utilde{e^\alpha},\textit{\textbf{e}}_{\bm \mu} \rangle -\Gamma^\mu_{\gamma \beta}\langle\utilde{e^\alpha},\nabla_{\textit{\textbf{e}}_{\bm \delta}}\textit{\textbf{e}}_{\bm \mu} \rangle \nonumber \\
&=& \frac{\partial}{\partial x^\gamma}\Gamma^\alpha_{\delta \beta}-\frac{\partial}{\partial x^\delta}\Gamma^\alpha_{\gamma \beta}+\Gamma^\mu _{\delta \beta} \Gamma^\alpha_{\mu \gamma}-\Gamma^\mu_{\gamma \beta}\Gamma^\alpha_{\delta \mu} \label{1.41}
\end{eqnarray}

This gives the co-ordinate components of Riemann curvature tensor in terms of co-ordinate components of the connection. It is clear from the above expression for $R^\alpha_{\beta \gamma \delta}$ that
\begin{eqnarray}\label{1.42}
	&\mbox{i.)}&~R^\alpha_{\beta \gamma \delta }=-R^\alpha_{\beta \delta \gamma}~i.e.,~R^\alpha_{\beta (\gamma \delta)}=0=R_{\alpha \beta (\gamma \delta)}\nonumber\\
	&\mbox{ii)}& R_{(\alpha \beta) \gamma \delta}=0
\end{eqnarray}

Also we have the Bianchi's identities:
\begin{eqnarray}
	&\mbox{I.}&~R^\alpha_{[\beta \gamma \delta]} = 0~~i.e.,~~R^\alpha_{\beta \gamma \delta}+R^\alpha_{\gamma \delta \beta}+R^\alpha_{\delta \beta \gamma}=0 \nonumber \\
	&\mbox{II.}&~R^\alpha_{\beta [\gamma \delta ; \mu]} = 0~~i.e.,~~R^\alpha_{\beta \gamma \delta; \mu}+R^\alpha_{\beta \delta \mu ; \gamma}+R^\alpha_{\beta \mu \gamma; \delta}=0~. \label{1.43}
\end{eqnarray}

From Bianchi's first identity one has 
\begin{eqnarray}\label{1.44}
	R_{\alpha \beta \gamma \delta}=R_{\gamma \delta \alpha \beta}
\end{eqnarray}

(For detail proof of these results see section 2.19).\\

In an $n$-dimensional manifold, the number of linearly independent components of curvature tensor are $\frac{1}{3}n^2(n^2-1)$. From the geometric interpretation of Riemann curvature tensor, $R^\alpha_{\beta \gamma \delta}=0$ at all points of the manifold means that if a vector is parallely transported along a closed path then we get back the original vector at the starting point and we say that the connection is flat. This flatness property of a manifold is the global concept of parallelism. So in this case two vectors at two different points $P$ and $Q$ can be said to be parallel as vectors can be parallely transported from $P$ to $Q$ independent of the path along which we approaches $Q$ from $P$. Thus tangent spaces at all points of the manifold can be considered to be identical and the manifold may be identified with its tangent space.\\

\textbf{Note:} For a flat space, the Riemann curvature tensor vanishes but the Christoffel symbols are not necessarily zero.\\

By contracting the curvature tensor one can define a (0,2) symmetric tensor, known as Ricci tensor having components
$$R_{\beta \delta}=R^\alpha_{\beta \alpha \delta}$$

Also contracting further the scalar so obtained is known as Ricci scalar
$$R=g^{\alpha \beta}R_{\alpha \beta}$$

We then define another symmetric (0,2)-tensor $G_{\mu \nu}$ known as Einstein tensor
\begin{equation} \label{1.45}
	G_{\mu \nu}=R_{\mu \nu}-\frac{1}{2}Rg_{\mu \nu}
\end{equation}

From the contracted Bianchi identities namely
$$R^\alpha_{\beta [\alpha \gamma; \delta]}=0~~\mbox{and}~~g^{\beta \gamma}R^\alpha_{\beta [\alpha \gamma; \delta]}=0$$
a straight forward calculation results 
$$G^{\alpha \beta}_{;\beta}=0$$
\textit{i.e.}, Einstein tensor is divergence free (a distinct result from Ricci tensor). \\

\textbf{Note:}~~ The field equations for the Einstein's theory of gravitation are given by\\
$$G_{\alpha \beta}=\frac{8\pi G}{c^4}T_{\alpha \beta}$$

Space-time is represented by a four-dimensional manifold with metric (a generalization of flat Minkowski metric) obtained by solving the above Einstein field equations. $G^{\alpha \beta}_{; \beta}=0$ results the energy-momentum conservation relation $T^{\alpha \beta}_{; \beta}=0$. Due to symmetry of $G_{\alpha \beta}$ there are ten field equations to determine ten unknown metric co-efficients $g_{\alpha \beta}$. But due to the above divergence relations there are only six independent field equations. Hence metric components are determined only upto the four functional degrees of freedom to characterize coordinate transformations of $g_{\alpha \beta}$.\\

We now define a (0, 4)-tensor in `$n$'-dimension ($n \geq 3$) as 
\begin{equation} \label{1.46}
	C_{\rho \sigma \mu \nu}=R_{\rho \sigma \mu \nu}-\frac{2}{n-2}\left(g_{\rho [ \mu}~ R_{\nu ] \sigma}-g_{\sigma [ \mu}~ R_{\nu ] \rho}\right )+ \frac{2}{(n-1)(n-2)}R g_{\rho [ \mu} ~g_{\nu ] \sigma}.
\end{equation}

It is a linear combination of curvature tensor, Ricci tensor and Ricci scalar and is known as Weyl tensor.\\

In four dimension the Weyl tensor has the explicit form
\begin{equation} \label{1.47}
	C_{\rho \sigma \mu \nu}=R_{\rho \sigma \mu \nu}-\frac{1}{2}\bigg(g_{_{\rho \mu}}~ R_{_{\nu \sigma}}+g_{_{\sigma \nu}}~ R_{_{\mu \rho}}-g_{_{\rho \nu}}~ R_{_{\mu \sigma}}-g_{_{\sigma  \mu}}~ R_{_{\nu \rho}}\bigg)+ \frac{R}{6}\bigg( g_{_{\rho \mu}} ~g_{_{\nu  \sigma}}- g_{_{\rho \nu}} ~g_{_{\mu  \sigma}}\bigg).
\end{equation}

Now due to properties (\ref{1.42})--(\ref{1.44}) of the Riemannian curvature tensor the Weyl tensor has similar properties (note that Ricci tensor and Ricci scalar are contraction of curvature tensor) namely
\begin{eqnarray}
	&\mbox{i)}&~C_{\rho \sigma \mu \nu}=-C_{\rho \sigma \nu \mu},\nonumber\\
	&\mbox{ii)}&~C_{\rho \sigma \mu \nu}=-C_{\sigma \rho \mu \nu },\nonumber\\
	&\mbox{iii)}&~C_{\rho \sigma \mu \nu}+C_{\rho \mu \nu \sigma }+C_{\rho \nu \sigma \mu }=0,\nonumber\\
	\mbox{and}
	&\mbox{iv)}&~C_{\rho \sigma \mu \nu}=C_{\mu  \nu \rho \sigma }.\nonumber
\end{eqnarray}

Now we shall calculate
\begin{eqnarray}
	g^{\sigma \mu} C_{\rho \sigma \mu \nu}&=&g^{\sigma \mu} R_{_{\rho \sigma \mu \nu}}-\frac{1}{n-2}\bigg(g^{\sigma \mu} g_{_{\rho \mu}} R_{_{\nu \sigma}}+g^{\sigma \mu} g_{_{\sigma \nu}} R_{_{\mu \rho}}-g^{\sigma \mu} g_{_{\rho \nu}} R_{_{\mu \sigma}}-g^{\sigma \mu} g_{_{\sigma \mu}} R_{_{\nu \rho}}\bigg)\nonumber\\
	&&+\frac{R}{(n-1)(n-2)}\bigg(g^{\sigma \mu} g_{_{\rho \mu}} g_{_{\nu \sigma}}-g^{\sigma \mu} g_{_{\rho \nu}} g_{_{\mu \sigma}}\bigg)\nonumber\\
	&=&-g^{\sigma \mu} R_{_{\sigma \rho \mu \nu}}-\frac{1}{n-2}\bigg(\delta^{\sigma}_{_{\rho}} R_{_{\nu \sigma}}+\delta^{\mu}_{_{\nu}} R_{_{\mu \rho}}-g_{_{\rho \nu }} R-n R_{_{\nu \rho}}\bigg)+\frac{R}{(n-1)(n-2)}\bigg(\delta^{\sigma}_{_{\rho}} g_{_{\nu \sigma}}-n g_{_{\rho \nu}}\bigg)\nonumber\\
	&=&-R_{_{\rho \nu}}-\frac{1}{n-2}\bigg(R_{_{\rho \nu}}+R_{_{\rho \nu}}-g_{_{\rho \nu}} R-n R_{_{ \nu \rho}}\bigg)+\frac{R}{(n-1)(n-2)}\bigg(g_{_{\rho \nu}}-n g_{_{\rho \nu}}\bigg)\nonumber\\
	&=&-R_{_{\rho \nu}}+R_{_{\rho \nu}}+\frac{R}{(n-2)}g_{_{\rho \nu}}-\frac{R}{(n-2)}g_{_{\rho \nu}}=0\nonumber
\end{eqnarray}

Hence Weyl tensor is traceless.\\

So one can have $C^{\rho}_{_{\sigma \rho \nu}}=0$.\\

Thus Weyl tensor vanishes for any pair of contracted indices. In other words, Weyl tensor can be considered as that part of the curvature tensor for which all contractions vanish.\\

Further, for three dimension the number of independent components of the curvature tensor is six--the components of the Ricci tensor (see problem 1.15). So a simple algebra shows that $C_{_{\rho \sigma \mu \nu}}\equiv 0$ for three dimension.\\

Moreover, an important property of the Weyl tensor is its conformal invariance. Two metrices $g_{_{\alpha \beta}}$ and$\bar{g}_{_{\alpha \beta}}$ in a manifold are said to be conformally related (or simply conformal) if
$$\bar{g}_{_{\alpha \beta}}=\phi^2 g_{_{\alpha \beta}},$$
where $\phi$ is a non zero differentiable function. Then the ratio of the lengths of the vectors, angle between two vectors and null geodesics remain unaltered for the conformal metrices. Two conformally related metrices have same Weyl tensor i.e.,
$$\bar{C}_{_{\rho \sigma \mu \nu}}=C_{_{\rho \sigma \mu \nu}}.$$

A metric is said to be conformally flat if $\exists$ some scalar function $\psi$ such that $g_{_{\alpha \beta}}=\psi^2 \eta_{_{\alpha \beta}}$ ($\eta_{_{\alpha \beta}}$ is the flat metric). As curvature tensor identically vanishes for flat metric so Weyl tensor also vanishes. Hence conformally flat metric has zero Weyl tensor.
\section{Concept of Compatibility of the Connection}

~~Usually a manifold may have a connection, a volume form and a metric. So it is natural to introduce some compatibility relations among them. We shall first find a compatibility relation between the connection and the volume form and then between connection and the metric.

For any vector field $\textit{\textbf{V}}$, the covariant divergence is defined as
$$V^\alpha_{; \alpha}={\bm {\nabla}} \cdot \textit{\textbf{V}}=\nabla_\alpha V^\alpha,$$
while for volume-form $\utilde{\omega}$, divergence can be defined as 
$$\mathcal{L}_{\textit{\textbf{V}}}\utilde{\omega}= \left(\mbox{div}_{\utilde{\omega}}\textit{\textbf{V}}\right)\utilde{\omega}$$

Thus, we can say that connection and volume form are compatible if the above two divergences are equal for any $\textit{\textbf{V}}$ \textit{i.e.}, 
$$\mbox{div}_{\utilde{\omega}}\textit{\textbf{V}}=V^\alpha_{; \alpha}$$

Now, $(\mathcal{L}_{\textit{\textbf{V}}}\utilde{\omega})_{\alpha \ldots \delta}=\utilde{\omega}_{\alpha \ldots \delta ; \lambda} V^\lambda+\omega_{\alpha \ldots \delta }V^\lambda_{~; \lambda}$\\

Hence for compatibility of the connection and volume from we have
$$\nabla \utilde{\omega}=0$$

Let g be the metric tensor of the manifold. For any two vectors $\textit{\textbf{v}}$ and ${\bm {\omega}}$, the inner product is defined as $g(\textit{\textbf{v}}, {\bm {\omega}})$.\\

The connection $\nabla$ and the metric tensor `$g$' are said to be compatible if this inner product remains invariant for parallel transport of the vectors $\textit{\textbf{v}}$ and $\textit{\textbf{w}}$ along any curve $\gamma$ (also for any vector field $\textit{\textbf{v}}$ and $\textit{\textbf{w}}$). Suppose $\textit{\textbf{u}}$ be the tangent vector to $\gamma$ then invariance of the inner product demands
$$\nabla_{\textit{\textbf{u}}}g(\textit{\textbf{v}}, \textit{\textbf{w}})=0$$
$$i.e.,~(\nabla_{\textit{\textbf{u}}}g)(\textit{\textbf{v}},\textit{\textbf{w}})+g(\nabla_{\textit{\textbf{u}}}\textit{\textbf{v}}, \textit{\textbf{w}})+g(\textit{\textbf{v}}, \nabla_{\textit{\textbf{u}}}\textit{\textbf{w}})=0$$

As $\textit{\textbf{v}}$ and $\textit{\textbf{w}}$ are parallely transported along $\gamma$ so $\nabla_{\textit{\textbf{u}}}\textit{\textbf{v}}=0=\nabla_{\textit{\textbf{u}}}
\textit{\textbf{w}}$.\\

Hence we have
$$(\nabla_{\textit{\textbf{u}}}g)(\textit{\textbf{v}}, \textit{\textbf{w}})=0,~\mbox{for any}~\textit{\textbf{u}},~\textit{\textbf{v}}~\mbox{and}~\textit{\textbf{w}}.$$

Thus we have the compatibility relation $\nabla g=0$.\\

In particular, for a co-ordinate system $\{x^\alpha\}$ the above compatibility relation becomes
$$g_{\alpha \beta ; \gamma}=0~i.e.,~\frac{\partial g_{\alpha \beta}}{\partial x^\gamma}=\Gamma^\delta_{\alpha \gamma}g_{\delta \beta}+\Gamma^\delta_{\beta \gamma}g_{\alpha \delta}=\Gamma_{\beta \alpha \gamma}+\Gamma_{\alpha \beta \gamma}$$

Hence we have
$$\Gamma_{\alpha \beta \gamma}=\frac{1}{2}\left[\frac{\partial g_{\alpha \gamma}}{\partial x^\beta}+\frac{\partial g_{\beta \gamma}}{\partial x^\alpha}-\frac{\partial g_{\alpha \beta}}{\partial x^\gamma}\right]$$

So symmetric connection compatible with metric is unique and is termed as metric connection.\\

Further, if we use normal co-ordinates at any point $P$ then we have $\Gamma^\gamma_{\alpha \beta}=0$ at $P$ and hence $\dfrac{\partial g_{\alpha \beta}}{\partial x^\gamma }=0$ at $P$.\\

Thus at P the components of the Riemann curvature tensor can be written as 
$$R_{\alpha \beta \gamma \delta}=g_{\alpha \lambda}R^\lambda_{\beta \gamma \delta}=\frac{1}{2}\left[g_{\alpha \delta , \beta \gamma}-g_{\alpha \gamma, \beta \delta}+g_{\beta \gamma, \alpha \delta}-g_{\beta \delta ,\alpha \gamma}\right]$$

It is easy to see that $R_{\alpha \beta \gamma \delta}=R_{\gamma \delta \alpha \beta}$.\\

\section{Isometries\,: Killing vectors}

~~~In tensor calculus, objects which do not change under co-ordinate transformations are called invariants. A co-ordinate transformation which keeps metric to be invariant is called an isometry. It is of importance in Riemannian manifold as it carries information about the symmetries of the manifold.\\

For a co-ordinate transformation $x^\mu \longrightarrow x'^\mu$ the form invariance of $g_{\mu \nu}$ implies $g'_{\mu \nu}(x')=g_{\mu \nu}(x')$. As $g_{\alpha \beta}$ is a (0, 2)\,-tensor so under co-ordinate transformation the transformed metric is related to the original metric by the relation 
$$g_{\mu \nu}(x)=\frac{\partial x'^\alpha}{\partial x^\mu}\frac{\partial x'^\beta}{\partial x^\nu}g'_{\alpha \beta }(x')$$

But due to isometry the above transformation equation becomes
$$g_{\mu \nu}(x)=\frac{\partial x'^\alpha}{\partial x^\mu}\frac{\partial x'^\beta}{\partial x^\nu}g_{\alpha \beta }(x')$$

Now using the transformation of co-ordinates \textit{i.e.}, $x'^\mu =x'^\mu (x^\alpha)$ it is possible to write the above equation in terms of the old co-ordinates. But in general, it will be a very complicated equation. However, a lot of simplification is possible for infinitesimal co-ordinate transformation. Moreover, any finite transformtion (with non-zero Jacobian) can be constructed (by an integration process) from an infinite sequence of infinitesimal transformations.\\

Suppose the infinitesimal co-ordinate transformation is given by 
$$x'^\mu =x^\mu +\delta x^\mu~~\mbox{with}~~\delta x^\mu=\varepsilon \theta^\mu (x)$$
where $\varepsilon$ is an arbitrary infinitesimal parameter and $\theta^\mu$ is a vector field. Thus we have
$$\frac{\partial x'^\mu}{\partial x^\nu}=\delta^\mu_\nu+\varepsilon \theta^\mu_{,\nu}$$

Substituting into the transformation equation for metric tensor we get (by Taylor's expansion)\\
\[
\begin{array}{l}
g_{\mu \nu} (x)=(\delta^\alpha_\mu+\varepsilon \theta^\alpha_{, \mu})(\delta^\beta_\nu+\varepsilon \theta^\beta_{, \nu})g_{\alpha \beta }(x^\lambda+\varepsilon \theta ^\lambda) \\
~~~~~~~~=\left[\delta^\alpha_\mu \delta^\beta_\nu+\varepsilon \delta^\beta_\nu \theta^\alpha_{, \mu}+ \varepsilon \delta^\alpha_\mu  \theta^\beta_{, \nu}+O(\varepsilon^2)\right]\left[g_{\alpha \beta }(x)+ \varepsilon \theta ^\lambda \partial _\lambda g_{\alpha \beta}(x)+O(\varepsilon^2)\right] \\
~~~~~~~~=g_{\mu \nu}(x)+\varepsilon \left[g_{\alpha \nu}\partial_\mu \theta ^\alpha+g_{\mu \beta}\partial_\nu \theta^\beta+\theta^\lambda\partial_\lambda g_{\mu \nu}\right]+O(\varepsilon^2)~. \\\\
i.e.,~~\theta^\lambda\partial_\lambda g_{\mu \nu}+g_{\alpha \nu}\partial_\mu \theta^\alpha+g_{\mu \beta}\partial _\nu \theta ^\beta=0 \\
i.e.,~~\mathcal{L}_\theta g_{\mu \nu}=0~.
\end{array}
\]

As in the Lie derivative any partial derivative can be replaced by covariant differentiation so the above condition for infinitesimal isometry can be written as 
$$\mathcal{L}_\theta g_{\mu \nu}=0~~i.e.,~~\theta_{\mu ; \nu}+\theta_{\nu ; \mu}=0~~i.e.,~~\theta_{(\mu ;\nu)}=0\,.$$
The vector field $\theta^\mu$ is called the killing vector field (see $\S 1.9.1$) and the above equation is called the Killing equation. In the notion of Lie derivative, any vector field by which the metric tensor is Lie dragged is called a Killing vector field. The symmetry properties of Riemannian space are characterized by the Killing vectors.\\

In particular, for a co-ordinate system if the vector field $\theta^\mu$ is along any co-ordinate direction (say $x^\lambda$) \textit{i.e.}, $\theta^\mu=(0,0, \ldots 0,1,0,\ldots)$ then the above Killing equation simplifies to $\dfrac{\partial g_{\mu \nu}}{\partial x^\lambda}=0$. Then metric tensor does not depend on the particular co-ordinate $x^\lambda$. On the other way, if all the metric coefficients are independent of any particular co-ordinate (say $x^l$) then $\dfrac{\partial}{\partial x^l}$ will be a killing vector field of the space. So we can say that the Killing equation is the generalized version of the symmetry `independence of a co-ordinate'.\\

The above Killing equations are first order linear differential equations in the Killing vector $\theta^\mu$ (for 4D they are ten in number) and they depend on the metric tensor.
The integral curves for the Killing vector field are characterized by the differential equation $\dfrac{dx^\mu}{d \lambda}=\theta^\mu (x)$.\\

We shall now determine the symmetry of three simple spaces namely the Euclidean space, the Minkowskian space and the surface of a sphere by calculating the Killing vectors for these spaces.\\

\textbf{a) $3D$ Euclidean Space:}\\

The line element in Cartesian coordinates takes the form:
$$ds^2=dx^2+dy^2+dz^2.$$

As all the metric coefficients are independent of the co-ordinates so $\dfrac {\partial}{\partial x}, \dfrac {\partial}{\partial y}$ and $\dfrac {\partial}{\partial z}$ are Killing vectors. Now writing the line element in polar co-ordinates i.e., 
$$ds^2=dr^2+r^2 d \theta^2+r^2 \sin^2 \theta d\phi^2.$$

We see that all the metric coefficients are independent of the angular co-ordinate $\phi$ so
$$l_z=\frac {\partial}{\partial \phi}=x \frac {\partial}{\partial y}-y\frac {\partial}{\partial x}$$
is a Killing vector. Now the symmetry of the three Cartesian co-ordinates shows
$$l_x=y \frac {\partial}{\partial z}-z\frac {\partial}{\partial y}~\mbox{and}~l_y=z \frac {\partial}{\partial x}-x\frac {\partial}{\partial z}$$
are also Killing vectors. Hence the $3D$ Euclidean space has six Killing Vectors.\\\\

\textbf{b) Minkowskian Space:}\\

The metric tensor for this 4D space has the simple diagonal form 
$$g_{\mu \nu}=\mbox{diag}(-1,1,1,1)$$
and consequently all christoffel symbols vanish identically. So in Killing equation all covariant derivatives simplify to partial derivatives and we get
$$\theta_{\mu , \nu}+\theta_{\nu, \mu}=0~i.e., ~\theta_{(\mu, \nu)}=0.$$

Now differentiating once more we have the relations
$$\theta_{\mu, \nu \lambda}+\theta_{\nu, \mu \lambda}=0;~\theta_{\nu, \lambda \mu}+\theta_{\lambda, \nu \mu}=0; \theta_{\lambda,\mu \nu}+\theta_{\mu, \lambda \nu}=0,$$
which on combination gives
$$\theta_{\mu, \nu \lambda}=0$$

The general solution of this equation can be written as
$$\theta_\mu=\alpha_\mu+\tau_{\mu \nu} x^\nu$$
where $\tau_{\mu \nu}$ is antisymmetric in its indices to satisfy the Killing equation. Thus in 4D Minkowski flat space, we have ten linearly independent killing vectors which are characterized by the parameters ($\alpha_\mu, \tau_{\mu \nu}$). Four possible values of $\alpha_\mu$ correspond to translation along the four space-time axes while six independent values of $\tau_{\mu \nu}$ corresponds to rotation of axes in 4D (they are 3 usual spatial rotations and three spatial Lorentz transformation).\\

\textbf{c) Spherical Surface:}\\

The line element on the spherical surface (known as 2-sphere) can be written as 
$$ds^2=d\theta^2+\sin^2 \theta~ d\phi^2$$

So the metric tensor has the form $g_{\mu \nu}=\mbox{diag}(1, \sin^2 \theta)$.
Then the explicit form of the Killing equations are 
$$\odot^1_{,1}=0,~\odot^1_{,2}+\sin^2 \theta \odot^2_{,1}=0,~\odot^1 \cos \theta +\sin \theta \odot^2_{,2}=0$$
where the indices 1 and 2 corresponds to the angular co-ordinates $\theta$ and $\phi$ respectively. The solution of these differential equations are
$$\odot^1=\alpha \sin (\phi +\beta);~\odot^2=\alpha\cos (\phi +\beta) \cot \theta +\odot_0$$

The three independent parameters ($\alpha, \beta ,\odot_0$) show that there are three linearly independent Killing vectors namely $\odot^\mu_{(1)}=(\sin \phi, \cos \phi \cot \theta),~\odot^\mu_{(2)}=(\cos \phi, - \sin \phi \cot \theta),~\odot^\mu_{(3)}=(0,1)$.
Thus the number of Killing vectors on spherical surface is same as the plane. However, on the plane the Killing vectors correspond to two translation and one rotation but here we have no such geometrical picture of Killing vectors.\\

We shall now address  questions that naturally arise namely ``what is the maximum number of killing vectors possible in a Riemannian space? What is the nature of such space?"\\

For any Killing vector $\theta^\mu$, we have from the definition of Riemann curvature tensor
$$\theta_{\mu; \nu; \lambda}-\theta_{\mu; \lambda;\nu}=-R^\delta_{\mu \nu \lambda}\theta^\delta$$

Then from the Bianchi's first identity
$$R^\delta_{(\mu \nu \lambda)}=0$$

We have the following identity
$$(\theta_{\mu; \nu}-\theta_{\nu; \mu})_{;\lambda}+(\theta_{\nu; \lambda}-\theta_{\lambda; \nu})_{;\mu}+(\theta_{\lambda; \mu}-\theta_{\mu; \lambda})_{;\nu}=0$$

Using the Killing equation $\theta_{(\mu; \nu)}=0$, the above identity simplifies to
$$\theta_{\mu;\nu;\lambda}+\theta_{\nu;\lambda;\mu}+\theta_{\lambda;\mu;\nu}=0$$

Now again using the Killing equation and the definition of curvature tensor yields
$$\theta_{\lambda;\nu;\mu}=-R^\delta_{\mu \nu \lambda}\theta_ \delta\,.$$

The above equation tells us that if the Killing vector $\theta_\mu$ and its first derivative $\theta_{\mu;\nu}$ are known at any point $P$ of the manifold then second and higher derivatives of $\theta_\mu$ can be determined at $P$ and consequently the Killing vector can be evaluated in the nhb of $P$ by Taylor expansion. Thus in an `$n$' dimensional Riemannian space, we have at most `$n$' number of $\theta_\mu$ and $\dfrac{n(n-1)}{2}$ number of $\theta_{\mu;\nu}$ (as the Killing equations are antisymmetric) at $P$. Hence we have $n+\dfrac{n(n-1)}{2}=\dfrac{n(n+1)}{2}$ number of initial data at $P$ and consequently the maximum number of independent Killing vectors possible at any point is $\dfrac{n(n+1)}{2}$ (for 4D the maximum no. of killing vectors is ten).\\

If in a Riemannian space, the maximum number of Killing vectors exists then such space is called a maximally symmetric space. It can be shown easily that for such space the curvature scalar $R$ must be constant and the curvature tensor can be written as 
$$R_{\mu \nu \lambda \sigma}=\frac{R(g_{\nu \lambda}g_{\mu \sigma}-g_{\mu \lambda}g_{\nu \sigma})}{n(n-1)}$$

The space of maximum symmetry is also called the space of constant curvature with $\kappa=\sqrt{\dfrac{n(n-1)}{|R|}}$ as the radius of the curvature. Flat space with vanishing curvature is a particular example of maximally symmetric space. Euclidean spaces are maximally symmetric spaces (see example (a) above). For such spaces every point and every direction are equivalent. Hence such spaces are homogeneous and isotropic in nature.\\

Let us next consider symmetries along a geodesic in a Riemannian space. The equation of the geodesic can be written as
$$\frac{\delta u^\mu}{ds}=0~~i.e.,~~u^\mu_{;\nu}u^\nu=0$$
where $u^\nu$ is the particle 4-velocity. We now multiply the geodesic equation by the Killing vector $\theta_\mu$ and on simplification we have\\
\[
\begin{array}{l}
~~~~~~~\theta_\mu u^\mu_{;\nu}u^\nu=0 \\
i.e.,~~(\theta_\mu u^\mu)_{;\nu}u^\nu-(u^\mu u^\nu)(\theta_{\mu; \nu})=0 \\
i.e.,~~(\theta_\mu u^\mu)_{;\nu}=0 \\
i.e.,~~\theta_\mu u^\mu= \mbox{constant}~.
\end{array}
\]

Thus throughout the motion of the particle the product $\theta_\mu u^\mu$ remains constant \textit{i.e.}, the quantity is a constant of motion. Therefore, in particle mechanics conservation laws are associated with Killing vector fields of the space. For example, in Minkowskian space having ten Killing vector there are ten conservation laws\,: a) conservation of four momentum associated with four translational Killing vectors, b) conservation of angular momentum associated with three special rotations, c) conservation of the motion of C. G. associated with three special Lorentz transformation.\\\\
\textbf{Note\,:}\\
{\bf I.} It is possible to have more conservation laws than the number of Killing vectors \textit{i.e.}, there may have conservation laws which can not reflect any symmetry.\\\\
{\bf II.} If in a Riemannian space $T^{\mu \nu}$ is the energy momentum tensor for an arbitrary field satisfying the conservation law $T^{\mu \nu}_{; \nu}=0$, then for any killing vector $\theta_\mu$ we have
$$\theta_\mu T^{\mu \nu}_{; \nu}=0~~i.e.,~~(\theta_\mu T^{\mu \nu})_{; \nu}-T^{\mu \nu}\theta_{\mu; \nu}=0 $$
$$i.e.,~~(\theta _\mu T^{\mu \nu})_{; \nu}=0~~i.e.,~~\theta_\mu T^{\mu \nu}~~\textrm{is conserved quantity}.$$
{\bf III.} In general, a space without any symmetry does not have any Killing vector. In particular, if a space contains `$l$' linearly independent Killing vectors $\left(0 \leq l \leq \dfrac{n(n+1)}{2}\right)$ then they form a Lie algebra of dimension `$l$' over $R$ with Lie bracket as the algebra product.
\begin{center}
	\underline{-----------------------------------------------------------------------------------} 
\end{center}
\newpage
\vspace{5mm}
\begin{center}
	\underline{\bf Exercise} 
\end{center}
\vspace{3mm}
{\bf 1.1.} Suppose a $(0,2)$\,-tensor $A$ is such that $A\left(\textit{\textbf{U}},\textit{\textbf{U}}\right)=0$ for any contravariant vector $\textit{\textbf{U}}$ , then show that $A$ is antisymmetric in its arguments.\\\\
{\bf 1.2.} Suppose $A$ is an antisymmetric $(0,2)$\,-tensor and $B$ is an arbitrary $(2,0)$\,-tensor, then show that the contraction of $A$ with $B$ involves only the antisymmetric part of $B$\,.\\\\
{\bf 1.3.} Show that under a general co-ordinate transformation, partial derivative of a vector does not transform as a tensor while the commutator components $\left[\textit{\textbf{u}},\textit{\textbf{v}}\right]^{\alpha}$ transform as a $(1,0)$\,-tensor.\\\\
{\bf 1.4.} If the components of a $(0,p)$\,-tensor $\utilde{A}$ are antisymmetric with respect to interchange of any two indices, then show that $\utilde{A}$ is a completely antisymmetric tensor.\\\\
{\bf 1.5.} Suppose $\left\{P_{\alpha \beta \gamma}\right\}$ are the components of a completely antisymmetric $(0,3)$\,-tensor $P$ then
$$P_{[\alpha \beta \gamma]} = P_{\alpha \beta \gamma}\,.$$
{\bf 1.6.} Show that a completely antisymmetric $(0,p)$\,-tensor defined on an $n$-dimensional vector space vanishes identically if $p>n$\,.\\\\
{\bf 1.7.} In an $n$-dimensional vector space, show that the set of all $q$-forms\,$(q<n)$ for fixed $q$ is a vector space. Also show that it is a subspace of all $(0,q)$\,-tensors and has the dimension ${}^{n}C_q$\,.\\\\
{\bf 1.8.} If $\utilde{A}$ and $\utilde{B}$ are one-forms then show that $\utilde{A}\wedge \utilde{B}$ is a two-form. Also show that $\utilde{A}\wedge \utilde{A}= \utilde{0}$\\\\
{\bf 1.9.} If $\left\{\textit{\textbf{e}}_i\,,\,i=1,2, \ldots , n\right\}$ and $\left\{\utilde{w^i}\,,\,i=1,2, \ldots , n\right\}$ be the basis of a vector space and the corresponding basis of the dual vector space respectively, then show that $\left\{\utilde{w^i}\wedge \utilde{w^k}\,,\,i,k=\right.$\\
$\left.1,2, \ldots , n\right\}$ is a basis for the vector space of all 2-forms. Also show that for an arbitrary two-form $\utilde{A}$ we have
$$\utilde{A} = \frac{1}{2!}A_{lm}\,\utilde{w^l}\wedge \utilde{w}^m$$
where $A_{lm} = \utilde{A}(\textit{\textbf{e}}_l,\textit{\textbf{e}}_m)$\,.\\\\
{\bf 1.10.} If $\utilde{A}$ is a $r$-form and $\utilde{B}$ is a $s$-form the show that
$$(\utilde{A}\wedge \utilde{B})_{i_1 \ldots i_r\,j_1 \ldots j_s} = {}^{r+s}C_{r}\,A_{[i_1 \ldots i_r}\,B_{j_1 \ldots j_s]}\,.$$
{\bf 1.11.} Prove the following :\\

$(i)$ $\utilde{d}(f\utilde{d}g) = \utilde{d}f\wedge \utilde{d}g$ , for any scalar $f$\\

$(ii)$ $\utilde{d}\utilde{A} = \dfrac{1}{p!}\dfrac{\partial}{\partial x^k}(A_{i\ldots l})\utilde{d}x^{k}\wedge \utilde{d}x^{i}\wedge \cdots \utilde{d}x^{l}$\\

where $\utilde{A} = \dfrac{1}{p!}A_{i\ldots l}\utilde{d}x^{i}\wedge \cdots \utilde{d}x^{l}$ is a $p$-form expressed in a co-ordinate basis. Also show that
$$(\utilde{d}\utilde{A})_{mn \ldots r} = (p+1)\frac{\partial}{\partial x^{[m}}A_{n\ldots r]}\,.$$\\
{\bf 1.12.} Show that for an arbitrary vector $\textit{\textbf{V}}$
$$\left(\mathcal{L}_{\textit{\textbf{V}}}g\right)_{ij}= V_{(i;j)}\,.$$
{\bf 1.13.} Show that in a co-ordinate basis the components of $\mathcal{L}_{\textit{\textbf{V}}}\utilde{w}$ are given by
$$\left(\mathcal{L}_{\textit{\textbf{V}}}\utilde{w}\right)_{i}= V^j \frac{\partial}{\partial x^j}w_i + w_j \frac{\partial}{\partial x^i}V^j = V^j w_{i,j} + w_j V_{,i}^j\,.$$
{\bf 1.14.} The shear of a velocity field $\textit{\textbf{u}}$ is defined in Cartesian co-ordinates by the equation
$$\sigma _{\mu \nu}= V_{\mu , \nu}+ V_{\nu , \mu} - \frac{1}{3}\delta _{\mu \nu}\theta$$
where the expansion scalar $\theta$ is given by
$$\theta = \nabla \cdot \textit{\textbf{u}}\,.$$
Show that in an arbitrary co-ordinate system
$$\theta = \frac{1}{2}g^{ij}\mathcal{L}_{\textit{\textbf{u}}}g_{ij}~~~\mbox{and}~~~\sigma _{ij} = \mathcal{L}_{\textit{\textbf{u}}}g_{ij} - \frac{1}{3}\theta g_{ij}\,.$$
{\bf 1.15.} Show that the no. of linearly independent components of the curvature tensor in an $n$-dimensional manifold is\\
({\it i}) $\dfrac{1}{3}n^2 (n^2 - 1)$ for the mixed form of the curvature tensor $R_{\nu \delta \lambda}^{\mu}$\\
and ({\it ii}) $\dfrac{1}{12}n^2 (n^2 - 1)$ for fully covariant form of the curvature tensor $R_{\mu \nu \delta \lambda}$\,.\\\\

\begin{center}
	\underline{\bf Solution and Hints} 
\end{center}
\vspace{3mm}

{\bf Solution 1.1:} Let us write $\textit{\textbf{U}}$ as a vector sum of two contravariant vectors $\textit{\textbf{V}}$ and $\textit{\textbf{W}}$ {\it i.e.} $\textit{\textbf{U}} = \textit{\textbf{V}} + \textit{\textbf{W}}$ , then
\begin{eqnarray}
A\left(\textit{\textbf{U}},\textit{\textbf{U}}\right) &=& A\left(\textit{\textbf{V}}+\textit{\textbf{W}},\textit{\textbf{V}}+\textit{\textbf{W}}\right) = A\left(\textit{\textbf{V}},\textit{\textbf{V}}\right) + A\left(\textit{\textbf{V}},\textit{\textbf{W}}\right) + A\left(\textit{\textbf{W}},\textit{\textbf{V}}\right) + A\left(\textit{\textbf{W}},\textit{\textbf{W}}\right) \nonumber \\
&=& A\left(\textit{\textbf{V}},\textit{\textbf{W}}\right) + A\left(\textit{\textbf{W}},\textit{\textbf{V}}\right) \nonumber
\end{eqnarray}
Hence $A\left(\textit{\textbf{U}},\textit{\textbf{U}}\right)=0$ implies that $A\left(\textit{\textbf{V}},\textit{\textbf{W}}\right)= -A\left(\textit{\textbf{W}},\textit{\textbf{V}}\right)$ {\it i.e.} $A$ is antisymmetric in its arguments.\\

{\bf Solution 1.2:} We decompose the arbitrary $(2,0)$\,-tensor $B$ as the sum of its symmetric part and antisymmetric part {\it i.e.}
$$B^{\alpha \beta} = B^{(\alpha \beta)} + B^{[\alpha \beta]}$$
$$\mbox{so}~~~~A_{\alpha \beta}\,B^{\alpha \beta} = A_{\alpha \beta}\,B^{(\alpha \beta)} + A_{\alpha \beta}\,B^{[\alpha \beta]} = A_{\alpha \beta}\,B^{[\alpha \beta]}$$
(the first term in the r.h.s. vanishes due to product of a symmetric and antisymmetric part)\\

\underline{Note}\,: If $B$ is a symmetric $(2,0)$\,-tensor then
$$A_{\alpha \beta}\,B^{\alpha \beta} = 0\,.$$\\

{\bf Solution 1.3:} Let the general co-ordinate transformation be denoted by
$$x^{\alpha}= \Lambda _{\mu}^{~\alpha}\,x^{\prime \mu}$$
So we have
$$\frac{\partial}{\partial x^{\prime \mu}} = \Lambda _{\mu}^{~\alpha}\,\frac{\partial}{\partial x^{\alpha}}$$
For any vector $\textit{\textbf{u}}$ the transformation of its components are given by
$$u' = \Lambda _{~\beta}^{\nu}u^{\beta}$$
\begin{eqnarray}
\mbox{So}~~~~~~\frac{\partial u^{\prime \nu}}{\partial x^{\prime \mu}} &=& \frac{\partial}{\partial x^{\prime \mu}}\left(\Lambda _{~\beta}^{\nu}u^{\beta}\right) = \Lambda _{\mu}^{~\alpha}\,\frac{\partial}{\partial x^{\alpha}}\left(\Lambda _{~\beta}^{\nu}u^{\beta}\right) \nonumber \\
&=& \Lambda _{\mu}^{~\alpha}\,\Lambda _{~\beta}^{\nu}\,\frac{\partial u^{\beta}}{\partial x^{\alpha}} + \Lambda _{\mu}^{~\alpha}\,\frac{\partial \Lambda _{~\beta}^{\nu}}{\partial x^{\alpha}}\,u^{\beta} \nonumber
\end{eqnarray}
Thus due to non-vanishing of the 2nd term on the r.h.s. partial derivative of a vector is not a tensor. Now by definition, the commutator components in primed frame can be written as
\begin{eqnarray}
\left[\textit{\textbf{u}}',\textit{\textbf{v}}'\right]^{\nu} &=& u^{\prime \mu}\,\frac{\partial v^{\prime \nu}}{\partial x^{\prime \mu}} - v^{\prime \mu}\,\frac{\partial u^{\prime \nu}}{\partial x^{\prime \mu}} \nonumber \\
&=& \Lambda _{~\alpha}^{\mu}u^{\alpha}\left\{\Lambda _{\mu}^{~\delta}\Lambda _{~\beta}^{\nu}\,\frac{\partial v^{\beta}}{\partial x^{\delta}} + \Lambda _{\mu}^{~\delta}\,\frac{\partial \Lambda _{~\beta}^{\nu}}{\partial x^{\delta}}v^{\beta}\right\} - \Lambda _{~\alpha}^{\mu}v^{\alpha}\left\{\Lambda _{\mu}^{~\delta}\Lambda _{~\beta}^{\nu}\,\frac{\partial u^{\beta}}{\partial x^{\delta}} + \Lambda _{\mu}^{~\delta}\,\frac{\partial \Lambda _{~\beta}^{\nu}}{\partial x^{\delta}}u^{\beta}\right\} \nonumber
\end{eqnarray}
\begin{eqnarray}
&=& \left(\Lambda _{~\alpha}^{\mu}\Lambda _{\mu}^{~\delta}\right)\Lambda _{~\beta}^{\nu}u^{\alpha}\,\frac{\partial v^{\beta}}{\partial x^{\delta}} - \left(\Lambda _{~\alpha}^{\mu}\Lambda _{\mu}^{~\delta}\right)\Lambda _{~\beta}^{\nu}
v^{\alpha}\,\frac{\partial u^{\beta}}{\partial x^{\delta}} + \left(\Lambda _{~\alpha}^{\mu}\Lambda _{\mu}^{~\delta}\right)\frac{\partial \Lambda _{~\beta}^{\nu}}{\partial x^{\delta}}u^{\alpha}v^{\beta} - \left(\Lambda _{~\alpha}^{\mu}\Lambda _{\mu}^{~\delta}\right)\frac{\partial \Lambda _{~\beta}^{\nu}}{\partial x^{\delta}}v^{\alpha}u^{\beta} \nonumber \\
&=& \delta _{\alpha}^{~\delta}\,\Lambda _{~\beta}^{\nu}u^{\alpha}\frac{\partial v^{\beta}}{\partial x^{\delta}} - \delta _{\alpha}^{~\delta}\,\Lambda _{~\beta}^{\nu}v^{\alpha}\frac{\partial u^{\beta}}{\partial x^{\delta}} + \delta _{\alpha}^{~\delta}\,\frac{\partial \Lambda _{~\beta}^{\nu}}{\partial x^{\delta}}u^{\alpha}v^{\beta} - \delta _{\alpha}^{~\delta}\,\frac{\partial \Lambda _{~\beta}^{\nu}}{\partial x^{\delta}}v^{\alpha}u^{\beta} \nonumber \\
&=& \Lambda _{~\beta}^{\nu}\left(u^{\alpha}\frac{\partial v^{\beta}}{\partial x^{\alpha}} - v^{\alpha}\frac{\partial u^{\beta}}{\partial x^{\delta}}\right) + \frac{\partial \Lambda _{~\beta}^{\nu}}{\partial x^{\alpha}}u^{\alpha}v^{\beta} - \frac{\partial \Lambda _{~\beta}^{\nu}}{\partial x^{\alpha}}v^{\alpha}u^{\beta} \nonumber \\
&=& \Lambda _{~\beta}^{\nu}\left[\textit{\textbf{u}},\textit{\textbf{v}}\right]^{\beta} + \frac{\partial ^2 x^{\prime \nu}}{\partial x^{\alpha}\partial x^{\beta}}u^{\alpha}v^{\beta} - \frac{\partial ^2 x^{\prime \nu}}{\partial x^{\beta}\partial x^{\alpha}}u^{\beta}v^{\alpha} \nonumber \\
&=& \Lambda _{~\beta}^{\nu}\left[\textit{\textbf{u}},\textit{\textbf{v}}\right]^{\beta} \nonumber
\end{eqnarray}
Hence the commutator components transform as a $(1,0)$\,-tensor.\\

{\bf Solution 1.4:} In a particular basis we write
\begin{eqnarray}
\utilde{A}\left(\ldots , \textit{\textbf{U}} \ldots  \textit{\textbf{V}} \ldots \right) &=& A_{\ldots \mu \ldots \nu \ldots}\ldots U^{\mu} \ldots  V^{\beta} \ldots \nonumber \\
&=& -A_{\ldots \beta \ldots \alpha \ldots}\ldots U^{\alpha} \ldots  V^{\beta} \ldots \nonumber \\
&=& -A\left(\ldots \textit{\textbf{V}} \ldots  \textit{\textbf{U}} \ldots\right) \nonumber
\end{eqnarray}
Thus $\utilde{A}$ is a completely antisymmetric $(0,p)$\,-tensor.\\

{\bf Solution 1.5:} Since $P$ is a completely antisymmetric $(0,3)$\,-tensor so by interchange of indices, its components will satisfy
$$P_{\alpha \beta \gamma} = P_{\beta \gamma \alpha} = P_{\gamma \alpha \beta} = -P_{\alpha \gamma \beta} = -P_{\beta \alpha \gamma} = -P_{\gamma \beta \alpha}$$
Hence
\begin{eqnarray}
P_{[\alpha \beta \gamma]} &=& \frac{1}{3!}\left(P_{\alpha \beta \gamma} + P_{\beta \gamma \alpha} + P_{\gamma \alpha \beta} - P_{\alpha \gamma \beta} - P_{\beta \alpha \gamma} - P_{\gamma \beta \alpha}\right) \nonumber \\
&=& P_{\alpha \beta \gamma} \nonumber
\end{eqnarray}\\
{\bf Solution 1.6:} The vector space is spanned by $n$ basis vectors. In order to write down the components of a $(0,p)$\,-tensor we need a set of $p$ basis vectors of which at least two vectors $(p\geq n+1)$ are identical. So interchange of these two identical basis vectors make no change but due to antisymmetric property, the components have a sign change. Hence the components are all identically zero.\\

{\bf Solution 1.7:} For any two $q$-forms $A$ and $B$\,, their components will have the totally antisymmetric property namely
$$A_{i_1 i_2 \ldots \ldots i_q} = A_{[i_1 i_2 \ldots \ldots i_q]}$$
and
$$B_{i_1 i_2 \ldots \ldots i_q} = B_{[i_1 i_2 \ldots \ldots i_q]}$$
So
$$C_{i_1 i_2 \ldots \ldots i_q} = A_{i_1 i_2 \ldots \ldots i_q} + B_{i_1 i_2 \ldots \ldots i_q} = A_{[i_1 i_2 \ldots \ldots i_q]} + B_{[i_1 i_2 \ldots \ldots i_q]}$$
$$~~~~~~~~~~~~~~~~~~~~~~~~~~= C_{[i_1 i_2 \ldots \ldots i_q]}$$
Hence $C$ is also a $q$-form.\\

Similarly, $\lambda A$ (for some scalar $\lambda$) is also a $q$-form. Thus the set of all $q$-forms form a vector space. It is clear that this vector space is a subspace of the vector space of all $(0,q)$\,-tensors. Now the number of independent components of a $(0,q)$ -form in an $n$-dimensional vector space is ${}^{n}C_q$ so the dimension of this vector space is ${}^{n}C_q$\,.\\

{\bf Solution 1.8:} By definition,
$$\utilde{A}\wedge \utilde{B} = \utilde{A}\otimes \utilde{B} - \utilde{B}\otimes \utilde{A}$$
So for any two contravariant vectors $\textit{\textbf{U}}$ and $\textit{\textbf{V}}$ we have
$$\utilde{A}\wedge \utilde{B}(\textit{\textbf{U}},\textit{\textbf{V}}) = \utilde{A}(\textit{\textbf{U}})\utilde{B}(\textit{\textbf{V}}) - \utilde{B}(\textit{\textbf{U}})\utilde{A}(\textit{\textbf{V}})$$
$$\mbox{Hence,}~~~\utilde{A}\wedge \utilde{B}(\textit{\textbf{U}},\textit{\textbf{V}}) = -\utilde{A}\wedge \utilde{B}(\textit{\textbf{V}},\textit{\textbf{U}})\,,$$
It shows that the components of $\utilde{A}\wedge \utilde{B}$ in any basis are completely antisymmetric. So $\utilde{A}\wedge \utilde{B}$ is a two-form. Also from the above
$$\utilde{A}\wedge \utilde{A}(\textit{\textbf{U}},\textit{\textbf{V}}) = \utilde{A}(\textit{\textbf{U}})\utilde{A}(\textit{\textbf{V}}) - \utilde{A}(\textit{\textbf{V}})\utilde{A}(\textit{\textbf{U}}) = \utilde{0}\,,$$
a null two-form.\\\\
{\bf Solution 1.9:} By definition, for any two arbitrary elements $\textit{\textbf{u}}$ , $\textit{\textbf{v}}$ of the vector space we have
\begin{eqnarray}
\utilde{A}(\textit{\textbf{u}},\textit{\textbf{v}}) &=& \frac{1}{2}A_{lm}\left[\utilde{w^l}(\textit{\textbf{u}})\utilde{w}^m(\textit{\textbf{v}}) - \utilde{w}^m(\textit{\textbf{u}})\utilde{w^l}(\textit{\textbf{v}})\right] \nonumber \\
&=& \frac{1}{2}A_{lm}\left[u^l v^m - u^m v^l\right] \nonumber \\
&=& \frac{1}{2}A_{lm}\,u^l v^m - \frac{1}{2}A_{ml}\,u^l v^m \nonumber \\
&=& \frac{1}{2}A_{lm}\,u^l v^m + \frac{1}{2}A_{lm}\,u^l v^m \nonumber \\
&=& A_{lm}\,u^l v^m = \frac{1}{2!}A_{lm}\,\utilde{w^l}\wedge \utilde{w}^m(\textit{\textbf{u}},\textit{\textbf{v}}) \nonumber 
\end{eqnarray}
which shows that
$$\utilde{A} = \frac{1}{2!}A_{lm}\,\utilde{w^l}\wedge \utilde{w}^m$$
Hence $\left\{\utilde{w^l}\wedge \utilde{w}^m\right\}$ is a basis for the vector space of two-forms.\\\\
{\bf Note I :} The no. of independent components $A_{ij}$ are
$$\frac{1}{2}(n^2 - n) = \frac{n(n-1)}{2} = {}^n C_2\,.$$
So the dimension of the vector space of 2-forms is ${}^n C_2$\,.\\

{\bf Note II :} The dimension of the vector space consists of all $r$-forms\,$(r \leq n)$ in a vector space of dimension $n$ is ${}^n C_r$\,. So the sum of the dimensions of all possible forms\,$(\leq n)$ in the vector space is $\sum\limits _{r=0}^{n} {}^n C_r = 2^n$\,.\\

{\bf Solution 1.10:} We start with $r=2\,,s=1$ {\it i.e.} $\utilde{A}$ is a 2-form and $\utilde{B}$ is a one-form. By the previous example we write
$$\utilde{A} = \frac{1}{2!}A_{ij}\,\utilde{w^i}\wedge \utilde{w^j}~~,~~\utilde{B} = B_l \utilde{w^l}$$
$$\mbox{So}~~~~~~~~~~~~~\utilde{A}\wedge \utilde{B} = \frac{1}{2!}A_{ij}B_l\,\utilde{w^i}\wedge \utilde{w^j}\wedge \utilde{w^l}\,.~~~~~~~~~~~~~~~~~$$
As $\left\{\utilde{w^i}\wedge \utilde{w^j}\wedge \utilde{w^l}\right\}$ form a basis for the three-forms so we write
\begin{eqnarray}
(\utilde{A}\wedge \utilde{B})_{lmn} &=& \frac{3!}{2!}\,A_{[lm}B_{n]} \nonumber \\
&=& \frac{3!}{2!}\cdot \frac{1}{3!}\left[A_{lm}B_{n} + A_{mn}B_{l} + A_{nl}B_{m} - A_{ml}B_{n} - A_{nm}B_{l} - A_{ln}B_{m}\right] \nonumber \\
&=& A_{lm}B_{n} + A_{mn}B_{l} + A_{nl}B_{m} \nonumber
\end{eqnarray}
Thus we write,
$$(\utilde{A}\wedge \utilde{B})_{lmn} = \frac{3!}{2!}\,A_{[lm}B_{n]} = {}^3 C_{2}\,A_{[lm}B_{n]}$$
Now extending this result to any $r$-form $\utilde{A}$ and $s$-form $\utilde{B}$ we have
$$(\utilde{A}\wedge \utilde{B})_{i_1 \ldots i_r\,j_1 \ldots j_s} = {}^{r+s}C_{r}\,A_{[i_1 \ldots i_r}\,B_{j_1 \ldots j_s]}\,.$$
{\bf Note :} As ${}^{r+s}C_{r} = {}^{r+s}C_{s}$ so the above result is also true if $\utilde{A}$ is any $s$-form and $\utilde{B}$ is any $r$-form.\\

{\bf Solution 1.11:} For exterior differentiation of the Wedge product of two forms $A$ and $B$ we have
$$\utilde{d}(\utilde{A}\wedge \utilde{B}) = (\utilde{d}\utilde{A})\wedge \utilde{B} + (-1)^{l}\utilde{A}\wedge \utilde{d} \utilde{B}$$
where $\utilde{A}$ is a $l$-form.\\

Thus
\begin{eqnarray}
(i)~~\utilde{d}(f\utilde{d}g) &=& \utilde{d}f \wedge \utilde{d}g + (-1)^{0}f \wedge \utilde{d}(\utilde{d}\utilde{g}) \nonumber \\
&=& \utilde{d}f \wedge \utilde{d}g~~~\left(\because\utilde{d}(\utilde{d}\utilde{g}) = 0\right) \nonumber
\end{eqnarray}
({\it ii}) As $\utilde{A}= \frac{1}{p!}A_{i\ldots l}\utilde{d}x^{k}\wedge \utilde{d}x^{i}\wedge \cdots \utilde{d}x^{l}$\\\\
So using the above result
\begin{eqnarray}
\utilde{d}\utilde{A} &=& \frac{1}{p!}A_{i\ldots l}\frac{\partial}{\partial x^k}(A_{i\ldots l})\utilde{d}x^{k}\wedge \utilde{d}x^{i}\wedge \cdots \utilde{d}x^{l} \nonumber \\
&=& \frac{1}{p!}\frac{\partial A_{i\ldots l}}{\partial x^k}\utilde{d}x^{k}\wedge \utilde{d}x^{i}\wedge \cdots \utilde{d}x^{l} \nonumber
\end{eqnarray}
Hence
\begin{eqnarray}
(\utilde{d}\utilde{A})_{k\,i \ldots l} &=& \frac{(p+1)!}{p!}\frac{\partial}{\partial x^{[k}}A_{i \ldots l]} \nonumber \\
&=& (p+1)\frac{\partial}{\partial x^{[k}}A_{i \ldots l]} \nonumber
\end{eqnarray}\\
{\bf Solution 1.12:} From the formula for Lie derivative
$$\left(\mathcal{L}_{\textit{\textbf{V}}}g\right)_{ij}= V^k \frac{\partial}{\partial x^j}g_{ij} + g_{ik}\frac{\partial}{\partial x^j}V^k + g_{kj}\frac{\partial}{\partial x^i}V^k$$
For symmetric connection
$$\frac{\partial g_{ij}}{\partial x^k}= \Gamma _{ik}^{l}g_{lj} + \Gamma _{jk}^{l}g_{il}$$
\begin{eqnarray}
\therefore ~~\left(\mathcal{L}_{\textit{\textbf{V}}}g\right)_{ij} &=& V^k\left(\Gamma _{ik}^{l}g_{ij} + \Gamma _{jk}^{l}g_{il}\right) + g_{ik}\frac{\partial}{\partial x^j}V^k + g_{jk}\frac{\partial}{\partial x^i}V^k \nonumber \\
&=& \left(g_{ik}\frac{\partial}{\partial x^j}V^k + g_{ik}\Gamma _{jl}^{k}V^l\right) + \left(g_{jk}\frac{\partial}{\partial x^i}V^k + g_{kj}\Gamma _{il}^{k}V^l\right) \nonumber\\
&&{\tiny ~~~~~~~~~~~~~~~~~~~~~~~~~~~~~~~(l \rightleftharpoons k)~~~~~~~~~~~~~~~~~~~~~~~~~~~~~~~~~~~~~~~~(k \rightleftharpoons l)}\nonumber\\
&=& g_{ik}\left(\frac{\partial}{\partial x^j}V^k + \Gamma _{jl}^{k}V^l\right) + g_{jk}\left(\frac{\partial V^k}{\partial x^i} + \Gamma _{il}^{k}V^l\right) \nonumber \\
&=& g_{ik}\nabla _{j} V^k + g_{jk}\nabla _{i} V^k \nonumber \\
&=& V_{i;j} + V_{j;i} = V_{(i;j)} \nonumber
\end{eqnarray}
{\bf Note:} If $\textit{\textbf{V}}$ is a Killing vector field then $\left(\mathcal{L}_{\textit{\textbf{V}}}g\right)_{ij}=0$ and we have $V_{(i;j)}=0$ .\\

{\bf Solution 1.13:} Let $\utilde{w}$ be an one-form and $\textit{\textbf{W}}$ be an arbitrary vector field. Then by Leibnitz rule we have
$$\mathcal{L}_{\textit{\textbf{V}}}\left[\utilde{w}\left(\textit{\textbf{W}}\right)\right]= \left(\mathcal{L}_{\textit{\textbf{V}}}\utilde{w}\right)\left(\textit{\textbf{W}}\right) + \utilde{w}\left(\mathcal{L}_{\textit{\textbf{V}}}\textit{\textbf{W}}\right)$$
But we know
$$\mathcal{L}_{\textit{\textbf{V}}}\textit{\textbf{W}} = \left[\textit{\textbf{V}},\textit{\textbf{W}}\right]$$
$$\therefore ~~\mathcal{L}_{\textit{\textbf{V}}}\left[\utilde{w}\left(\textit{\textbf{W}}\right)\right]= \left(\mathcal{L}_{\textit{\textbf{V}}}\utilde{w}\right)\left(\textit{\textbf{W}}\right) + \utilde{w}\left[\textit{\textbf{V}},\textit{\textbf{W}}\right]$$
$$\mbox{So}~~~~~~~~\left(\mathcal{L}_{\textit{\textbf{V}}}\utilde{w}\right)\left(
\textit{\textbf{W}}\right)= \mathcal{L}_{\textit{\textbf{V}}}\left[\utilde{w}\left(\textit{\textbf{W}}\right)\right] - \utilde{w}\left[\textit{\textbf{V}},\textit{\textbf{W}}\right]$$
\begin{eqnarray}
\mbox{Thus}~~~ \left(\mathcal{L}_{\textit{\textbf{V}}}\utilde{w}\right)_{i}W^i &=& V^j \frac{\partial}{\partial x^j}\left(w_i W^i\right) - w_i \left[V^j \frac{\partial}{\partial x^j}W^i - W^j \frac{\partial}{\partial x^j}V^i\right] \nonumber \\
&=& V^j W^i \frac{\partial}{\partial x^j}w_i + w_i W^j \frac{\partial}{\partial x^j}V^i \nonumber \\
&=& \left(V^j \frac{\partial}{\partial x^j}w_i + w_j \frac{\partial}{\partial x^i}V^j\right)W^i ~~~~~~(i \rightleftharpoons j~~\mbox{in the 2nd term}) \nonumber
\end{eqnarray}
Hence in a co-ordinate basis
\begin{eqnarray}
\mbox{Thus}~~~ \left(\mathcal{L}_{\textit{\textbf{V}}}\utilde{w}\right)_{i} &=& \left(V^j \frac{\partial}{\partial x^j}w_i + w_j \frac{\partial}{\partial x^i}V^j\right) \nonumber \\
&=& V^j w_{i,j} + w_j V_{,i}^j \nonumber
\end{eqnarray}
{\bf Note:} For a symmetric connection we can replace the comma\,s in the above expression for $\left(\mathcal{L}_{\textit{\textbf{V}}}\utilde{w}\right)_{i}$ by semicolons {\it i.e.} we can write $\left(\mathcal{L}_{\textit{\textbf{V}}}\utilde{w}\right)_{i}= V^j w_{i;j} + w_j V_{;i}^j$\,.\\

{\bf Solution 1.14:} As
$$\left(\mathcal{L}_{\textit{\textbf{V}}}g\right)_{ij} = u^k \frac{\partial}{\partial x^k}g_{ij} + g_{ik}\frac{\partial}{\partial x^j}u^k + g_{kj}\frac{\partial}{\partial x^i}u^k$$
\begin{eqnarray}
\mbox{So}~~~ g^{ij}\mathcal{L}_{\textit{\textbf{u}}}g_{ij} &=& u^k g^{ij}\frac{\partial}{\partial x^k}g_{ij} + g^{ij}g_{ik}\frac{\partial}{\partial x^j}u^k + g^{ij}g_{kj}\frac{\partial}{\partial x^i}u^k \nonumber \\
&=& \frac{1}{2}u^k \frac{\partial}{\partial x^k}\left(g^{ij}g_{ij}\right) + \delta _{k}^{j} \frac{\partial}{\partial x^j}u^k + \delta _{k}^{i} \frac{\partial}{\partial x^i}u^k \nonumber \\
&=& \frac{1}{2}u^k \frac{\partial}{\partial x^k}(n) + \frac{\partial}{\partial x^k}u^k + \frac{\partial}{\partial x^k}u^k = 2\,\nabla \cdot \textit{\textbf{u}} \nonumber
\end{eqnarray}
Hence $\theta = \dfrac{1}{2}g^{ij}\mathcal{L}_{\textit{\textbf{u}}}g_{ij}$ .\\\\
$$\mbox{Again}~~~~~~\mathcal{L}_{\textit{\textbf{u}}}g_{ij} = u^k \frac{\partial}{\partial x^k}g_{ij} + g_{ik}\frac{\partial}{\partial x^j}u^k + g_{kj}\frac{\partial}{\partial x^i}u^k$$
As $g_{ij}$'s are constants in Cartesian co-ordinate system so
\begin{eqnarray}
\mathcal{L}_{\textit{\textbf{u}}}g_{ij} &=& 0 + \frac{\partial}{\partial x^j}\left(g_{ik}u^k\right) + \frac{\partial}{\partial x^i}\left(g_{kj}u^k\right) \nonumber \\
&=& \frac{\partial}{\partial x^j}u_i + \frac{\partial}{\partial x^i}u_j = u_{i,j} + u_{j,i} \nonumber
\end{eqnarray}
$$\therefore ~~~~~~~\sigma _{ij} = \mathcal{L}_{\textit{\textbf{u}}}g_{ij} - \frac{1}{3}\theta g_{ij}$$
As these equations are tensorial equations so they hold in any arbitrary co-ordinate system.\\

{\bf Solution 1.15:} The Riemannian curvature tensor in mixed form $R_{\nu \delta \lambda}^{\mu}$ has the following algebraic properties\,:\\

({\it i}) ~~$R_{\nu (\delta \lambda)}^{\mu}= 0$ ~~~~{\it i.e.}~~ curvature tensor is antisymmetric in the last two lower indices.\\

({\it ii}) ~~$R_{[\nu \delta \lambda]}^{\mu}= 0$ ~~~~~~~~~ the Bianchi identity\\
$~~~~~~~~~~~~~~~~~~~~~~~~~~~~~~~~~~\mbox{(the cyclic sum of the three lower indices to be zero)}$.\\

No. of constraint due to 1st algebraic property are
$$n^2 \, \frac{n(n+1)}{2}$$
No. of constraint due to Bianchi identity are
$$n\cdot {}^{n}C_3 = n\frac{n(n-1)(n-2)}{3!}$$
So the number of independent components of the Riemann curvature tensor in mixed form
$$=n^4 - \frac{n^3(n+1)}{2} - \frac{n^2(n-1)(n-2)}{6} = \frac{1}{3}n^2(n^2 - 1)$$
For the fully covariant form of the curvature tensor the algebraic properties are the following\,:\\

({\it i}) ~~$R_{\mu \nu (\delta \lambda)}= 0$ , ({\it ii}) ~~$R_{(\mu \nu) \delta \lambda}= 0$ , ({\it iii}) ~~$R_{\mu \nu \delta \lambda}= R_{\delta \lambda \mu \nu}$\\

({\it iv}) ~~$R_{\mu [\nu \delta \lambda]}= 0~~~i.e.~~~~R_{\mu \nu \delta \lambda} + R_{\mu \delta \lambda \nu} + R_{\mu \lambda \nu \delta}= 0$\\

Due to antisymmetric property of the curvature tensor $R_{\mu \nu \delta \lambda}$ in the first pair and in the last pair there are $N= \frac{1}{2}n(n-1)$ ways of choosing independent pairs among them. As the tensor is symmetric due to exchange of these pairs so there are $\frac{1}{2}N(N+1)$ independent ways of choosing the combination $\mu \nu \delta \lambda$\,. For the cyclic identity it should be noted that due to pair symmetry property it is trivially satisfied unless $\mu , \nu , \delta$ and $\lambda$ are all distinct. So the number of distinct constraints due to this cyclic identity is ${}^{n}C_4$\,. Hence the number of independent components for fully covariant curvature tensor is
\begin{eqnarray}
\frac{1}{2}N(N+1) - {}^{n}C_4 &=& \frac{1}{2}\frac{n(n-1)}{2}\left\{\frac{n(n-1)}{2}+1\right\} - \frac{n(n-1)(n-2)(n-3)}{4!} \nonumber \\
&=& \frac{n(n-1)(n^2 -n +2)}{8} - \frac{n(n-1)(n-2)(n-3)}{24} \nonumber \\
&=& \frac{n^2}{12}(n^2 - 1) \nonumber
\end{eqnarray}
{\bf Note:} For $n<4$ , the Bianchi 1st identity is trivially satisfied.\\


\chapter{Differential geometry in Local Co-ordinate Basis}


\section{Euclidean Space}

~~~The set of all real numbers, {\it i.e.,} the whole real line is denoted by $R^1$ (or, simply by $R$). The Cartesian product of $R^1$ with itself $n$ times ($n$ is a +ve integer) is denoted by $R^n$ and it is the set of all ordered $n$-tuples ($x^1, x^2, ..., x^n$) of real numbers. The addition and scalar multiplication in $R^n$ can be defined as follows:\\

Let, $a=(a^1, a^2, ..., a^n)$ and $b=(b^1, b^2, ..., b^n)$ be two elements of $R^n$, then, their sum, $a+b$ and scalar multiplication $\alpha a$ ($\alpha$ is a real number) are defined as\\

$a+b=(a^1+b^1, a^2+b^2, ..., a^n+b^n)$,\\

and\\

$\alpha a=(\alpha a^1, \alpha a^2, \ldots , \alpha a^n)$\\

This co-ordinate wise algebraic operations make $R^n$ a real linear space (a vector space) having identity element (or, the zero element) with respect to addition is $\theta$= (0, 0, ..., 0), and the inverse of $a$ is $-a$= ($-a^1, -a^2, ..., -a^n$). Here, inverse is also called the negative of the corresponding element. Thus, each element, $x=(x^1, x^2, ..., x^n)$ of the $n$ dimensional linear space $R^n$ is ordered array of its $n$ co-ordinates $x^1, x^2, ..., x^n$.\\

Further, any element ({\it i.e.,} a point) in $R^n$ can be represented as a vector from the origin to that point and the above definition of addition and scalar multiplication can be considered as the vector addition and scalar multiplication of vector. Thus, elements of $R^n$ can be viewed either as points or as vectors from the origin to those points. Moreover, any element $x=(x^1, x^2, ..., x^n)$ can be thought of as a real function

$$f: (1, 2, ..., n) \rightarrow (x^1, x^2, ..., x^n)$$\\

such that, $f(i)=x^i$, {\it i.e.,} $R^n$ can be considered as the space of all real functions defined on the set of first $n$ positive integers. Therefore, elements of $R^n$ can be viewed as points, as vectors and as functions. The norm in $R^n$ can be defined suitably as follows:\\

For any element $x=(x^1, x^2, ..., x^n)$ of $R^n$, the norm is denoted by $\left\|x\right\|$ and is given by 

$$\left\|x\right\|= \sqrt{|x^1|^2+|x^2|^2+...+|x^n|^2} = (\Sigma_{i=1} ^n |x^i|^2)^{\frac{1}{2}}$$

Thus, if x is considered as a point then $\left\|x\right\|$ is the distance from the origin, while if the elements of $R^n$ are viewed as vectors, then norm is the magnitude of the vector. On the other hand, considering $R^n$ as composed of real functions $f$ defined on $\lbrace 1, 2, ..., n \rbrace$, the above norm can be written as $$\left\|f\right\|= (\Sigma_{i=1} ^n |f(i)|^2)^{\frac{1}{2}}$$

This is called the Euclidean norm on $R^n$ and the real linear space $R^n$ with the Euclidean norm is called $n$-dimensional Euclidean space.\\

The Euclidean space $R^n$ has the usual non-compact, metric topology and can be given a differential structure with a globally-defined co-ordinate chart.\\

In general, any finite dimensional vector space $M$ can be considered as a differentiable manifold in the sense that any basis set of vectors from $M$ can be considered to map $M$ isomorphically onto $R^n$. Then, `$n$' is called the dimension of the manifold $M$. Suppose, ($x^1, x^2, ..., x^n$) and ($y^1, y^2, ..., y^n$) be two co-ordinate systems corresponding to overlapping charts. Since, there is one-one correspondence between these co-ordinate charts, so the Jacobian $\bigg|\dfrac{\partial y^i}{\partial x^j}\bigg|$ or its inverse $\bigg|\dfrac{\partial x^i}{\partial y^j}\bigg|$ is non-zero throughout the overlap. If the Jacobian is positive definite, then the two co-ordinate systems are said to have the same orientation.\\

A manifold is said to be orientable if it admits of an atlas, such that, any two co-ordinate systems in it having an overlap have the same orientation. We shall deal with only orientable manifolds.\\

{\bf Example:} In 2D Euclidean plane $R^2$, let ($x^1, x^2$) and ($y^1, y^2$) be two rectangular Cartesian co-ordinate systems related as $$y^1= x^1 \cos \theta+x^2 \sin \theta$$
$$y^2=x^1 \sin \theta-x^2 \cos \theta$$

Then the Jacobian

\[
J= \left|\frac{\partial y^i}{\partial x^i}\right|=
\left| \begin{array}{cc}
\cos \theta & \sin \theta \\
\sin \theta & - \cos \theta
\end{array} \right|
= -1
\]

On the otherhand, if ($z^1, z^2$) be another co-ordinate system related to ($x^1, x^2$) as $z^1= x^1$, $z^2= -x^2$. Then, we have 
$$y^1= z^1 \cos \theta-z^2 \sin \theta$$
$$y^2= z^1 \sin \theta+z^2\cos \theta$$ so, the Jacobian is $$J= \left|\frac{\partial y^i}{\partial z^j}\right|= 1$$
This shows that, $R^2$ is an orientable manifold.\\\\

\section{Tangent vector and Tangent space}

~~~Let $\gamma$: $x^i$= $g^i (\lambda)~,~~i=1, 2, 3, \ldots , n$ be a differentiable curve passing through a point $P$ in $M$. Suppose, $\left\lbrace  x^i _0 = g^i (0); i=1, 2, 3, ..., n\right\rbrace$    be the co-ordinate of $P$ and a neighboring point $Q$ is identified by the parameter $\Delta \lambda$, {\it i.e.,} $Q$ has co-ordinates $\left\lbrace g^i (\Delta \lambda), i=1, 2, 3, \ldots , n \right\rbrace$. Then, the tangent vector to the curve $\gamma$ at $P$ is defined as
\begin{equation} \label{2.1}
t^i= \frac{dx^i}{d\lambda}{\bigg|_{\lambda=0}}
\end{equation}  

It should be noted that for any $n$-tuples of real numbers ($\alpha^1, \alpha^2,....,\alpha^n$), $\exists$ a curve in $M$ through $P$ such that the tangent vector has the components ($\alpha^1, \alpha^2,....,\alpha^n$) in some $x-$co-ordinate system. It is clear that the equation to the curve can be written as 
$$ x^i=x^i _0 +\alpha^i \lambda~,~~i.e.,~~\textit{\textbf{r}}= \textit{\textbf{r}}_{\bm 0}+\lambda {\bm {\alpha}}$$

Thus, the set of all tangent vectors corresponding to all curves through $P$ forms a vector space of dimension `$n$'. This vector space is called the tangent space at $P$ to $M$ and is denoted by $T_P (M)$ or simply by $T_P$.\\\

\subsection{Basis in $\mathbf{T_P}$}

~~~The `$n$' dimensional vector space $T_P$ has a natural choice of basis $\left\lbrace \textit{\textbf{e}}_{\bm i} \right\rbrace$, where $\textit{\textbf{e}}_{\bm i}$=  $\left\lbrace 0, 0,\ldots,0,1,\right.$\\
$\left.0,\ldots,0 \right\rbrace$ (`1' is in $i^{\mbox{th}}$ position), $i=1, 2, 3, \ldots , n$. This is also called the co-ordinate basis. Thus, any element of $T_P$ ({\it i.e.,} a tangent vector to the curve $\gamma$ in $M$) can be written as 

\begin{equation}\label{2.2}
\textit{\textbf{v}}= \frac{dx^i}{d\lambda} \textit{\textbf{e}}_{\bm i},
\end{equation}
being evaluated at $P$ (The summation convention due to Einstein has been introduced).\\

Suppose ($\overline{x}^1, \overline{x}^2, ..., \overline{x}^n$) be another co-ordinate system and $\left\lbrace \overline{\bm e}_{\bm 1}, \overline{\bm e}_{\bm 2},..., \overline{\bm e}_{\bm n}\right\rbrace$ be the natural basis for it, then for any $\textit{\textbf{v}}$ $\in$ $T_P$, we have

\begin{equation}\label{2.3}
\textit{\textbf{v}}= \frac{d\overline{x}^i}{d \lambda} \overline{\bm e}_{\bm i}~~~~~~(\mbox{evaluated~at}~P)
\end{equation}

So comparing Equations (\ref{2.2}) and (\ref{2.3}) we have

$$\frac{dx^i}{d\lambda} \textit{\textbf{e}}_{\bm i}= \frac{d \overline{x}^i}{d\lambda} \overline{\bm e}_{\bm i}= \frac{d \overline{x}^k}{d\lambda} \overline{\bm e}_{\bm k}$$

\textit{i.e.}, $$(\overline{\bm e}_{\bm k}-\frac{\partial x^i}{\partial \overline{x}^k}\bm e_{\bm i})\frac{d \overline{x}^k}{d\lambda}= 0$$

This relation is true for tangent vector to arbitrary curve $\gamma$ in $M$ {\it i.e.} for arbitrary values of $\dfrac{d \overline{x}^k}{d \lambda}$. Hence we have

\begin{equation}\label{2.4}
\overline{\bm e}_{\bm k}= \frac{\partial x^i}{\partial \overline{x}^k} \textit{\textbf{e}}_{\bm i}
\end{equation}

In a similar way, we have, comparing (\ref{2.2}) and (\ref{2.3})

\begin{equation}\label{2.5}
\bm e_{\bm i}= \frac{\partial \overline{x}^k}{\partial x^i} \overline{\bm e}_{\bm k}
\end{equation}

Now any arbitrary element $\textit{\textbf{A}}$ of $T_P$ can be written as 
\[
\textit{\textbf{A}}=
\left\{
\begin{array}{ll}
A^i \textit{\textbf{e}}_{\bm i} & (\mbox{in}~x~\mbox{co-ordinate~system}) \\
\overline{A}^i\overline{\bm e}_{\bm i} & (\mbox{in}~\overline{x}~\mbox{co-ordinate~system})
\end{array}
\right.
\]

where $A^i$'s and $\overline{A}^i$'s are called the components of $\textit{\textbf{A}}$ in $x$ and $\overline{x}$ co-ordinate systems respectively. So we write

$$\overline{A}^i \overline{\bm e}_{\bm i}= A^i \textit{\textbf{e}}_{\bm i}$$

{\it i.e.,} $$\overline{A}^k \overline{\bm e}_{\bm k}= A^i \textit{\textbf{e}}_{\bm i}$$

or using the transformation (\ref{2.5}) we get

$$(\overline{A}^k-\frac{\partial \overline{x}^k}{\partial x^i} A^i) \overline{\bm e}_{\bm k}= 0$$

As the basis vectors $\lbrace \overline{\bm e}_{\bm k} \rbrace$ are linearly independent, so the co-efficients vanish identically, {\it i.e.,} 

\begin{equation}\label{2.6}
\overline{A}^k= \frac{\partial \overline{x}^k}{\partial x^i} A^i
\end{equation}

Similarly using (\ref{2.4}) we have 

\begin{equation}\label{2.7}
A^i= \frac{\partial x^i}{\partial \overline{x}^k} \overline{A}^k
\end{equation}

So we have the following definition:\\

If a mathematical object is represented by a one-index system of functions $A^i$ of the co-ordinate variables $x^i$ of any one $x-$co-ordinate system, and by the functions $\overline{A}^i$ of the co-ordinate variables $\overline{x}^i$ of an other $\overline{x}-$co-ordinate system, and the two representations are connected by the transformation law $$\overline{A}^i= \frac{\partial \overline{x}^i}{\partial x^k} A^k$$
then $A^i$ are called the contravariant components of a vector (in $x-$co-ordinates). We also briefly say that $A^i$ is a contravariant vector.\\\\

\section{Covectors}

~~~A linear map $\utilde{L}: T_P \longrightarrow R$ (the real line $R$ can be considered as one dimensional vector space over itself), {\it i.e.,} a linear functional on $T_P$ is called a covector or a dual vector at $P$. As a linear map from a vector space $V$ into another vector space is completely determined by its action on the basis vectors of $V$, so $\utilde{L}$ is completely determined by its action on the basis vectors $\lbrace \overline{\bm e}_{\bm 1}, \overline{\bm e}_{\bm 2}, \ldots , \overline{\bm e}_{\bm n}\rbrace$ of $T_P$. Suppose $\utilde{L}(e_k)$ = $l_k~,~~k= 1, 2, \ldots , n$. We call $l_k$ the component of the covector $\utilde{L}$ in $x-$co-ordinates.\\

Let $\overline{l_k}$ be the components of $\utilde{L}$ in $\overline{x}$ co-ordinates. Then $\utilde{L}(\overline{\bm e}_{\bm k})$= $\overline{l_k}$, where as before $\lbrace \overline{\bm e}_{\bm 1}, \overline{\bm e}_{\bm 2}, \ldots , \overline{\bm e}_{\bm n}\rbrace$ is the natural basis of $\overline{x}$ co-ordinates. We now determine the transformation law for the components of covectors as follows:

$$\overline{l}_k= \utilde{L}(\overline{\bm e}_{\bm k})= \utilde{L}(\frac{\partial x^i}{\partial \overline{x}^k} \textit{\textbf{e}}_{\bm i})= \frac{\partial x^i}{\partial \overline{x}^k} \utilde{L}(\textit{\textbf{e}}_{\bm i})= \frac{\partial x^i}{\partial \overline{x}^k} l_i$$

\textit{i.e.},

\begin{equation}\label{2.8}
\overline{l}_k= \frac{\partial x^i}{\partial \overline{x}^k} l_i
\end{equation}

In a similar way, we get

\begin{equation}\label{2.9}
l_i= \frac{\partial \overline{x}^k}{\partial x^i}\overline{l_k}
\end{equation}

{\bf Definition:}\\

If a mathematical object is represented by an one index system of functions $l_i$ of the co-ordinate variables $\lbrace x^i \rbrace$ of some $x-$co-ordinate system and by the functions $\overline{l_i}$ of the co-ordinate variables $\lbrace \overline{x}^i\rbrace$ of any other $\overline{x}$ co-ordinate system and the two representations are connected by the transformation law $$\overline{l}_i= \frac{\partial \overline{x}^k}{\partial x^i}\overline{l_k}$$ 
 
then $\lbrace l_i \rbrace$ are called the covariant components of a vector and we simply write $l_i$ as covariant vector.\\

{\bf Reciprocal natural basis:}\\

The set of all linear mappings $T_P\longrightarrow R$, form a vector space ($T_P^\ast$) under usual addition of linear mappings and scalar multiplication of linear mappings.\\

Let ($\textit{\textbf{e}}_{\bm 1}, \textit{\textbf{e}}_{\bm 2}, \ldots , \textit{\textbf{e}}_{\bm n}$) be the natural basis of $T_P$ and  ($\utilde{e}^1, \utilde{e}^2, \ldots , \utilde{e}^n$) be elements of $T_P^\ast$ such that

$$\utilde{e}^i (\textit{\textbf{e}}_{\bm k})= \delta^i _k$$

where $\delta^i _k$= 1, if $i=k$ and 0, if $i \neq k$ is the usual Kronecker delta {\it i.e.}
\[
\delta _k ^i =
\left\{
\begin{array}{ll}
1 & \mbox{if}~i = k \\
0 & \mbox{if}~i \neq k~.
\end{array}
\right.
\]

We first prove that $\lbrace \utilde{e}^i\rbrace$ are linearly independent.\\

If possible, let
\[
\begin{array}{l}
~~~~~~a_i \cdot \utilde{e^i}= 0 \\
\Rightarrow (a_i \cdot \utilde{e^i}) \textit{\textbf{e}}_{\bm k}= 0 \\
\Rightarrow a_i \cdot \utilde{e^i} (\textit{\textbf{e}}_{\bm k})= 0 \\
\Rightarrow a_i \cdot \delta_k ^i = 0 \\
{\it i.e.,}~~ a_k= 0,~~k=1, 2, \ldots , n~.
\end{array}
\]
\\

Hence $\lbrace \utilde{e}^i \rbrace$ are linearly independent.\\

Now we show that $\lbrace \utilde{e}^i \rbrace$ generate $T_P^\ast$, {\it i.e.,} any element of $T_P^\ast$ can be expressed as a linear combination of $\lbrace \utilde{e}^i \rbrace$.\\

Suppose $\utilde{B}\in T_P^\ast$ and $\underline{B}(\textit{\textbf{e}}_{\bm k})$= $B_k$, k=1, 2, \ldots , n.\\

Also $$ \left( B_i \cdot \utilde{e^i} \right) (\textit{\textbf{e}}_{\bm k})= B_i \cdot \utilde{e^i} (\textit{\textbf{e}}_{\bm k})= B_i \delta_k ^i= B_k $$

Thus $\utilde{B}$ and $B_i\utilde{e^i}$ have the same action on the basis vectors $\lbrace \textit{\textbf{e}}_{\bm 1}, \textit{\textbf{e}}_{\bm 2}, \ldots , \textit{\textbf{e}}_{\bm n}\rbrace$ of $T_P$. Hence $\utilde{B}$= $B_i \cdot \utilde{e^i}$. Therefore, $\lbrace e^i\rbrace$ is a basis of $T_P^\ast$ and dim $T_P^\ast=$ dim $T_P$= n. Here the basis $\lbrace \utilde{e^1}, \utilde{e^2}, \ldots , \utilde{e^n}  \rbrace$ is called the reciprocal or dual of the basis $\lbrace \textit{\textbf{e}}_{\bm 1}, \textit{\textbf{e}}_{\bm 2}, \ldots , \textit{\textbf{e}}_{\bm n}\rbrace$.\\

{\bf Transformation law for reciprocal basis:}\\

Let $\lbrace \utilde{e}^i\rbrace$ and $\lbrace \overline{\utilde{e}}^i\rbrace$ be the natural reciprocal bases in $x$ and $\overline{x}$ co-ordinate systems. For any covector $\utilde{L} \in T_P^\ast$, let $l_i$ and $\overline{l}_i$ be the components with respect to the above choices of basis for $T_P^\ast$. Then
\begin{eqnarray}
\utilde{L}= l_i \utilde{e^i}&\mbox{and}&\utilde{L}= \overline{l}_i \overline{\utilde{e}}^i\nonumber\\
i.e.,~~l_i \cdot \utilde{e^i}&=& \overline{l}_i \cdot \overline{\utilde{e^i}}\nonumber
\end{eqnarray}

Using the transformation laws (\ref{2.8}) and (\ref{2.9}) for the components of the covector, we have the transformation laws for the dual basis as
\begin{equation}\label{2.10}
\overline{\utilde{e^i}}= \frac{\partial \overline{x}^i}{\partial x^j}\utilde{e}^j
\end{equation}

and 
\begin{equation}\label{2.11}
\utilde{e^i}= \frac{\partial x^i}{\partial \overline{x}^k} \utilde{e}^k
\end{equation}
\\

\section{Multilinear mapping of vectors and covectors: Tensors}

~~~~Suppose $T$ be a multilinear mapping that maps $p$ arbitrarily co-vectors and $q$ arbitrary vectors into a scalar such that it is linear in every argument. Thus
\begin{equation}\label{2.12}
T(\utilde{e^{i_1}}, \utilde{e^{i_2}},\ldots,\utilde{e^{i_p}}, \textit{\textbf{e}}_{k_1}, \textit{\textbf{e}}_{k_2}, \ldots , \textit{\textbf{e}}_{k_q})= T^{i_1, i_2, \ldots , i_p}_{k_1, k_2,\ldots , k_p},
\end{equation}

are called the components of the mapping $T$ in $x-$co-ordinate system.\\

Similarly, in $\bar{x}$ co-ordinate system we have 
$$T(\utilde{\overline{e}^{i_1}}, \utilde{\overline{e}^{i_2}},\ldots,\utilde{\overline{e}^{i_p}}, \overline{\bm e}_{\bm {k_1}}, \overline{\bm e}_{\bm {k_2}},\ldots, \overline{\bm e}_{\bm {k_q}})= \overline{T}^{i_1, i_2,\ldots,i_p}_{k_1, k_2,\ldots , k_p},$$

as the components of T. Using the transformation laws (\ref{2.4}), (\ref{2.5}), (\ref{2.10}), (\ref{2.11}) for the basis of $T_P$ and $T_P^\ast$ we have the relation between the components of the multilinear mapping in different co-ordinate system as 

\begin{equation}\label{2.13}
\overline{T}^{i_1, i_2,\ldots , i_p}_{k_1, k_2,\ldots , k_p}= \frac{\partial \overline{x}^{i_1}}{\partial x^{u_1}} \frac{\partial \overline{x}^{i_2}}{\partial x^{u_2}} \cdots \frac{\partial \overline{x}^{i_p}}{\partial x^{u_p}} \frac{\partial x^{v_1}}{\partial \overline{x}^{k_1}} \frac{\partial x^{v_2}}{\partial \overline{x}^{k_2}} \cdots  \frac{\partial x^{v_q}}{\partial \overline{x}^{k_q}} T^{u_1, u_2,\ldots , u_p}_{v_1, v_2,\ldots , v_p}.
\end{equation}

In modern terminology, a \underline{tensor} is defined as a multilinear mapping of the above form. The classical definition is as follows: \\

{\bf Definition:} If a mathematical object is represented by an ($p+ q$)--indexed system of functions $T^{i_1, i_2,...i_p}_{k_1, k_2,...k_q}$ of the co-ordinate variables $x^i$ of any $x-$co-ordinate system and by the functions $\overline{T}^{i_1, i_2,...i_p}_{k_1, k_2,...k_q}$ of the co-ordinate variables $\overline{x}^i$ of any other $\overline{x}$ co-ordinate system and the two representations are connected by the transformation law (\ref{2.13}), then $T^{i_1, i_2,...i_p}_{k_1, k_2,...k_q}$ are called the components in $x-$co-ordinate system of a tensor of contravariant order `$p$' and covariant order `$q$' or simply component of an $(p,~q)$ tensor. we also briefly say that $T^{i_1, i_2,...i_p}_{k_1, k_2,...k_q}$ is a $(p,~q)$ tensor.\\

If the contravariant order is zero but not the covariant order, then the tensor is called a\,(fully) covariant tensor. Similarly, we have a\,(fully) contravariant tensor if the covariant order is zero but not the contravariant order. If both the orders are different from zero, then the tensor is called a mixed tensor.\\

{\bf Note-1 :} A contravariant vector is a contravariant tensor of order one \textit{i.e.}, a $(1,0)$ tensor; a covariant vector is a covariant tensor of order one, \textit{i.e.}, a $(0,~1)$- tensor . A scalar function is called a tensor of order zero, \textit{i.e.}, a $(0,~0)$\,-tensor.\\

The negative of a tensor is defined by component wise negatives and is a tensor of the same order-type as that of the given tensor. The sum or difference of two tensors of the same order-type are defined by component wise sum or difference and are tensors of the same order-type.The scalar multiplication of a tensor is defined by component wise multiplication by the scalar and is a tensor of the same order-type as that of the given tensor. From these facts it follows that the set of all tensors of a particular order-type forms a vector space. The space formed by all tensors of some particular order-type $(p,~q)$ is called the  $(p,~q)$\,-tensor space at P and is denoted as $(T_p)_q^p$. The spaces $(T_p)_0^1$ , $(T_p)_1^0$ are denoted as $T_p$ and $T_p^\ast$ respectively.\\

{\bf Note-2 :} The homogeneous nature of the transformation laws of the tensors (eq.\,(\ref{2.13})) shows that if an $(p,~q)$ tensor has all its components equal to zero in one co-ordinate system at $P$ then it has all components zero at $P$ in any other co-ordinate system. If this happens at every point of the region under consideration then the tensor is called a zero tensor of $(p,~q)$-type. Finally, we can also conclude that if a tensorial equation is valid in one coordinate system then it is valid in any other coordinate system.\\

The {\underline rank} of a tensor is defined as the total no of real indices per component. So a $(r,~s)$\,-tensor is of rank $(r+s)$.\\

\section{Product of tensors}

~~~Let $R_{j_1 \ldots j_q}^{i_1 \ldots i_p}$ and $T_{l_1 \ldots l_s}^{k_1 \ldots k_r}$ are components of two tensors R and T of orders $(p,~q)$ and $(r, ~s)$ respectively. Then 
$$S^{i_1 \ldots i_pk_1 \ldots k_r}_{j_1 \ldots j_ql_1 \ldots l_s}=R_{j_1 \ldots j_q}^{i_1 \ldots i_p}T_{l_1 \ldots l_s}^{k_1 \ldots k_r}$$
is defined as the components of an $(p+r,~q+s)$ tensor and is called the outer product of the tensors $R$ and $T$. It is denoted as before by $R\otimes T$.\\

{\bf Note-I.} In general , this product is not commutative.\\

{\bf II.} The outer product of two tensors is a tensor where order is the sum of the orders of the two tensors.\\

{\bf III.} If in the outer product of two tensors at least one contravariant and one covariant index are identical then the outer product is called an inner product.\\

\section{Kronecker delta}

~~~From the point of view of tensor algebra, Kronecker delta is a $(1,~1)$-tensor. So its appropriate form should be $\delta_j^i$, The explicit form of the components of this mixed tensor is
\begin{eqnarray}\label{2.24}
\delta_j^i &=& 1~~~\mbox{if}~~i=j\nonumber \\
&=& 0~~~\mbox{if}~~i\neq j
\end{eqnarray}

{\bf Note:} If $A_j^i=\delta_j^i$ , then
$\bar{A}_j^i=\dfrac{\partial \bar{x}^i}{\partial x^p}\dfrac{\partial x^q}{\partial \bar{x}^j}A_q^p=\dfrac{\partial \bar{x}^i}{\partial x^p}\dfrac{\partial x^q}{\partial \bar{x}^j}\delta_q^p=\dfrac{\partial \bar{x}^i}{\partial x^p}\dfrac{\partial x^p}{\partial \bar{x}^j}=\delta_j^i$\\\\

\section{Contraction}

~~~Contraction is an operation on tensor under which the tensor is reduced in one contravariant order and one covariant order. The repeated index is called a dummy index which has no contribution in defining the order-type of the tensor while the free indices of the resulting tensor give the order-type of the resulting tensor.\\

{\bf Note:} In each process of contraction, the rank of tensor is reduced by two.\\


\section{Symmetry and Skew-Symmetry}

~~~let $T$ be a $(p,~q)$ tensor having components $T^{i_1 \ldots i_p}_{k_1 \ldots k_ik_m \ldots k_q}$ in some $x$- coordinate system. $T$ is said to be symmetric or skew symmetric with respect to the covariant indices $k_l$ and $k_m$ if
\begin{equation}\label{2.25}
T_{k_1 \ldots k_l \ldots k_m \ldots k_q}^{i_1 \ldots i_p}=T_{k_1 \ldots k_m \ldots k_l \ldots k_q}^{i_1 \ldots i_p}~~\mbox{or}~~-T_{k_1 \ldots k_m \ldots k_l \ldots k_q}^{i_1 \ldots i_p}
\end{equation}
for all possible values of the other indices.\\

Symmetry or anti-symmetry with respect to contravariant indices can be defined in a similar way.\\

{\bf Note-I :} Symmetry property can\,not be defined for mixed indices {\it i.e.} for one contravariant and one covariant index.\\

{\bf II :} The symmetry (or skew-symmetry) property for a tensor is independent of any particular co-ordinate system.\\

{\bf Proof :} Suppose in $\bar{x}$- co-ordinate system the components of a symmetric (or anti-symmetric) tensor $T$ can be written as :
\begin{eqnarray}\bar{T}_{v_1....v_l..v_m...v_q}^{u_1...u_p}&=&\frac{\partial \bar{x}^{u_1}}{\partial x^{i_1}}.....\frac{\partial \bar{x}^{u_p}}{\partial x^{i_p}} \cdot \frac{\partial x^{k_1}}{\partial \bar{x}^{v_1}}.......\frac{\partial x^{k_l}}{\partial \bar{x}^{v_l}}.....\frac{\partial x^{k_m}}{\partial \bar{x}^{v_m}}....\frac{\partial x^{k_q}}{\partial \bar{x}^{v_q}}T_{k_1....k_l..k_m...k_q}^{i_1...i_p}\nonumber\\
&=&\frac{\partial \bar{x}^{u_1}}{\partial x^{i_1}}.....\frac{\partial \bar{x}^{u_p}}{\partial x^{i_p}} \cdot \frac{\partial x^{k_1}}{\partial \bar{x}^{v_1}}.......\frac{\partial x^{k_m}}{\partial \bar{x}^{v_m}}....\frac{\partial x^{k_l}}{\partial \bar{x}^{v_l}}.....\frac{\partial x^{k_q}}{\partial \bar{x}^{v_q}}T_{k_1....k_m..k_l...k_q}^{i_1...i_p}~~~\mbox{(by~the~given~symmetry)}\nonumber\\
&=&\bar{T}_{v_1....v_m..v_l...v_q}^{u_1...u_p}\nonumber
\end{eqnarray}

Hence the components of $T$ in $\bar{x}$ co-ordinate system is also symmetric.\\

{\bf III :} Any (2, 0) or (0, 2) tensor can be expressed as a sum of a symmetric and a skew-symmetric tensor.\\\\

\section{Quotient Law}

~~~If the contraction of an indexed system of functions of the co-ordinate variables with an arbitrary tensor results another tensor then quotient law states that the indexed system of functions is also a tensor. The order-type is indicated by the free indices. For example, consider the product $A^{....}_{....}B^{....}_{....}$, where dots represent indices which may involve contraction between indices of $A$ and $B$. If $A$ is an indexed system of functions of the co-ordinate variables, $B$ is an arbitrary tensor of the type indicated by its indices and the product is a tensor of the type indicated by the free indices then $A$ is also a tensor.\\

\section{Relative Tensor}

~~~  If a set of mathematical quantities $A_{ij}$ satisfy the following transformation law:
\begin{equation}\label{2.26}
A_{ij}'=A_{pq}\frac{\partial \bar{x}^{p}}{\partial x^{i}}\frac{\partial \bar{x}^{q}}{\partial x^{j}}\left|\frac{\partial x}{\partial x'}\right|^{\omega}
\end{equation}
  then $A_{ij}$ is called a \underline{relative-tensor} of weight $w$. A relative tensor of weight one is called tensor density. If $w=0$ then it is the usual tensor. A relative tensor of order zero is called relative scalar. A relative scalar of weight one is called a scalar density.\\\\

\section{Riemannian space\,: Metric Tensor}

 ~~~  In this chapter we have so far developed local co-ordinate basis in Euclidean space where the co-ordinate systems are orthogonal frame of references. Now we shall introduce Riemannian space having only curvilinear co-ordinate system.\\
	
   A Riemannian space of dimension `$n$' is a n-dimensional differentiable manifold in which any pair of neighbouring points $P(x^i)$ and $Q(x^i+dx^i)$ belonging to a co-ordinate neighbourhood `$x$' is associated with an elementary scalar $dS$, called the distance between $P$ and $Q$, given by positive definite elementary quadratic form
\begin{equation}\label{2.27}
ds^2=g_{ij}dx^idx^j
\end{equation}
called the metric form or the first fundamental form or the ground form. (If the metric is non-singular but indefinite then the space is called semi-Riemannian.If in particular, the metric is of signature $(+,+, \ldots , +,-)$ then the space is called Lorentzian space. A semi-Riemannian space is also called a pseudo-Riemannian space.)\\
	
From the quotient law it follows that $g_{ij}$ is a covariant (symmetric) tensor of order two.\\

{\bf Note:} For Euclidean space, the first fundamental form has the same expression as equation (\ref{2.27}) but all the components of the metric tensor are constant in a preferred co-ordinate system.\\

\section{Algebraic operations of vectors in Riemannian space}

~~~   For any two vectors $\textit{\textbf{A}}=(A^i)$ and $\textit{\textbf{B}}=(B^i)$ in some $x$- co-ordinate system, the scalar product is defined as
$$\textit{\textbf{A}} \cdot \textit{\textbf{B}}=(\textit{\textbf{A}} \textit{\textbf{B}})=g^{ij}A^iB^j$$  
 
   So magnitude of a vector is given by
	   \begin{equation}\label{2.28}
		\left\|\textit{\textbf{A}}\right\|=(\textit{\textbf{A}},\textit{\textbf{A}})^{1/2}=(g_{ij}A^iA^j)^{1/2}
    \end{equation}
    
   Thus if $\theta$ be the angle between any two vectors $\textit{\textbf{A}}$ and $\textit{\textbf{B}}$ then
   \begin{equation}\label{2.29}
	\cos \theta=\frac{\textit{\textbf{A}} \cdot \textit{\textbf{B}}}{\left\|\textit{\textbf{A}}\right\| \left\|\textit{\textbf{B}}\right\|}=\frac{g_{ij}A^iB^j}{(g_{pq}A^pA^q)^{1/2}(g_{lm}B^lB^m)^{1/2}}
    \end{equation}
    
   Further, if $A_i$ and $B_i$ are the covariant components of any two vectors in some co-ordinate system then
  $$\textit{\textbf{A}} \cdot \textit{\textbf{B}}=g^{ij}A_iB_j~~\mbox{and}~~\left\|\textit{\textbf{A}}\right\| = (g^{ij}A_iA_j)^{1/2}$$
  where $g^{ij}$ is the reciprocal tensor (defined in \S 1.3) of the metric tensor.\\
  
   The covariant and contravariant components of a vector are connected by the metric tensor(or it's reciprocal) as follows :
 $$A_i=g_{ij}A^j~~~,~~~A^i=g^{ij}A_j$$
 
   The reciprocal tensor contracts with the metric tensor as
  $$g_{ik}g^{kj}=\delta^j_i~~~,~~~g^{kl}g_{li}=\delta^k_i$$
  
  Here the contravariant vector $A^i$ is said to be associated to $A_i$ .\\
  
  Thus the scalar product can be written as
  $$\textit{\textbf{A}} \cdot \textit{\textbf{B}}=g_{ij}A^iB^j=A^iB_i=A_iB^i$$
  or equivalently,
  $$\Vert{A}\Vert=(A_iA^i)^{1/2}.$$\\

\section{Length of a curve}

~~~  Let $\Gamma$ be a curve in a Riemannian space $V_n$. Suppose $P_0$ is a fixed point on $\Gamma$ and `$s$' denotes the are length of the curve measured from $P_0$ to any point $P$, `$t$' is assumed to be the affine parameter along the curve $\Gamma$. If $l$ be the length along the curve between two variable points $P_1$ and $P_2$ having affine parameters $t_1$ and $t_2$ then
  $$l=\int\limits _{P_2}^{P_1} dS =\int\limits _{t_2}^{t_1} \left(g_{ij}\frac{dx^i}{dt}\frac{dx^j}{dt}\right)^{1/2} dt.$$
	
	If $g_{ij}\dfrac{dx^i}{dt}\dfrac{dx^j}{dt} = 0$ along the curve then $l = 0$ {\it i.e.} the points $P_1$ and $P_2$ are at zero distance though they are not coincident. Then the curve is called a minimal or null curve. In Minkowskian space (the space-time continuum of special relativity) these curves are called the world lines of light and they lie on the surface of the light cone.\\
	
	Let $x^i\,(i=1,2, \ldots ,n)$ be the co-ordinates in a Riemannian space $V_n$ . The co-ordinate curve of parameter $x^l$ is defined as $x^i = c^i~,~\forall i$ except $i= l~,~c^i$'s are constants. Hence $dx^i = 0~~\forall i \neq l$ and $x^l \neq 0$. So the tangent vector along this co-ordinate curve is denoted by
	$$\textit{\textbf{t}}_{\bm l}= (0,0, \ldots ,0,1,0, \ldots ,0)~~~~~\mbox{(`1' in the}~l\mbox{-th position)}$$
	
Similarly, the tangent vector $\textit{\textbf{t}}_{\bm m}$ to the co-ordinate curve having parameter $x^m$ is taken to be
	$$\textit{\textbf{t}}_{\bm m}= (0,0, \ldots ,0,0,1,0, \ldots ,0)~~~~~\mbox{(here `1' in the}~m\mbox{-th position)}.$$
	
So the angle between these two co-ordinate curves is given by
\begin{eqnarray}
\cos \theta _{lm} &=& \frac{\textit{\textbf{t}}_{\bm l} \cdot \textit{\textbf{t}}_{\bm m}}{\left\|\textit{\textbf{t}}_{\bm l}\right\| \left\|\textit{\textbf{t}}_{\bm m}\right\|}= \frac{g_{ij}t_l ^i t_m ^j}{\sqrt{\left(g_{\alpha \beta} t_l ^{\alpha} t_l ^{\beta}\right)\left(g_{\mu \nu} t_m ^{\mu} t_m ^{\nu}\right)}} \nonumber \\
&=& \frac{g_{lm}}{\sqrt{g_{ll} \, g_{mm}}}~. \nonumber
\end{eqnarray}

Hence $\theta _{lm} = \dfrac{\pi}{2}$ implies $g_{lm} = 0$ {\it i.e.} the two co-ordinate curves will be orthogonal to each other if $g_{lm} = 0$.\\

\section{Angle between two co-ordinate hypersurfaces}

Let $\phi(x^i)$= constant be a hypersurface to a Riemannian manifold $V$. Then
\begin{equation}\label{2.30}
d\phi=\frac{\partial\phi}{\partial x^i}dx^i=0.
\end{equation}

This shows that $dx^i$ is orthogonal to $\dfrac{\partial\phi}{\partial x^i}$. But $dx^i$ is along the tangential direction to the hypersurfsce, so $\dfrac{\partial\phi}{\partial x^i}$ is normal to the hypersurface. Thus if $\theta$ be the angle between two hypersurfsces, $\phi(x^i)$= constant and $\psi(x^i)$= constant then 

\begin{equation}\label{2.31}
\cos \theta = \frac{g^{\alpha\beta}\frac{\partial\phi}{\partial x^\alpha}\frac{\partial\psi}{\partial x^\beta}}{\sqrt{\left(g^{\alpha\beta}\frac{\partial\phi}{\partial x^\alpha}\frac{\partial\phi}{\partial x^\beta}\right)\left(g^{\mu\nu}\frac{\partial\phi}{\partial x^\mu}\frac{\partial\phi}{\partial x^\nu}\right)}}.
\end{equation}

In particular, if we choose $\phi(x^i)=x^l=$ constant and $\psi(x^i)=x^m=$ constant as the co-ordinate hypersurfaces then

\begin{equation}\label{2.32}
\cos \theta = \frac{g^{lm}}{\sqrt{g^{ll}g^{mm}}}~.
\end{equation} 
\\\\

\section{Covariant Differentiation}

~~~Let $V_n$ be a $n$ dimensional submanifold of a manifold $M$ of dimension $m(>n)$. Let ${\stackrel{\rightarrow}{e_i}}$ be a natural basis (in some $x-$co-ordinate system) of the sub-manifold $V_n$. So for any vector $\stackrel{\rightarrow}{v}\in V_n$ we write
$$\stackrel{\rightarrow}{v}=v^i \stackrel{\rightarrow}{e_i}~.~~~~~~~~$$
\begin{equation}\label{2.36}
\partial _k \stackrel{\rightarrow}{v}=\left(\frac{\partial v^i}{\partial x^k}\right)\stackrel{\rightarrow}{e_i} + v^i\frac{\partial \stackrel{\rightarrow}{e_i}}{\partial x^k}
\end{equation}

The second term on the R.H.S. of the above equation shows that $\partial_k\stackrel{\rightarrow}{v}$ may not be a sub-manifold vector. Let us denote the projection of $\partial_k\stackrel{\rightarrow}{v}$ on $V_n$ by $(\partial_k\stackrel{\rightarrow}{v})_{||V_n}$ and that perpendicular to $V_n$ by $(\partial_k\stackrel{\rightarrow}{v})_{\bot V_n}$. Thus the above equation\,(\ref{2.36}) on $V_n$ can be written as
\begin{equation}\label{2.37}
(\partial_k\stackrel{\rightarrow}{v})_{||V_n}=(\partial_k v^i)\stackrel{\rightarrow}{e_i}+v^i\left(\frac{\partial \stackrel{\rightarrow}{e_i}}{\partial x^k}\right)_{||V_n}~.
\end{equation}

Now, $\left(\dfrac{\partial \stackrel{\rightarrow}{e_i}}{\partial x^k}\right)_{||V_n}$ can be written as a linear combination of basis vectors {\it i.e,}
\begin{equation}\label{2.38}
(\partial_k \stackrel{\rightarrow}{e_i})_{||V_n}=\Gamma^l_{ki}\stackrel{\rightarrow}{e_l}
\end{equation}

Then
\begin{eqnarray}
(\partial_k\stackrel{\rightarrow}{v})_{||V_n} &=& (\partial_K v^i)\stackrel{\rightarrow}{e_i}+ v^i\Gamma^l_{ki}\stackrel{\rightarrow}{e_l} \nonumber \\
&=& (\partial_K v^i)\stackrel{\rightarrow}{e_i}+v^i\Gamma^i_{ks}\stackrel{\rightarrow}{e_i}~~(\mbox{changing~the~repeated~indices}~`i\mbox{'}~\mbox{and}~`l\mbox{'}~
\mbox{to} \nonumber
\\&&~~~~~~~~~~~~~~~~~~~~~~~~~~~~~`s\mbox{'}~\mbox{and}~`i\mbox{'}~\mbox{respectively~}~i.e.~i\rightarrow s~,~l\rightarrow i)\nonumber\\
&=& \left[\partial_Kv^i+\Gamma^i_{ks} v^s\right]\stackrel{\rightarrow}{e_i}=\left(\nabla_k v^i\right) \stackrel{\rightarrow}{e_i}\nonumber
\end{eqnarray}
where, $\nabla_K v^i=\partial_Kv^i+\Gamma^i_{ks} v^s$ , are called the covariant differentiation w.r.t. $x^k$ of the contravariant components of $\stackrel{\rightarrow}{v}$. The scalar co-efficients $\Gamma^i_{ks}$ in the above equation are called the Riemann Christoffel symbols of the second kind.\\

Now, $$\partial_k\stackrel{\rightarrow}{e_i}=\left(\partial_k \stackrel{\rightarrow}{e_i}\right)_{||V_n}+\left(\partial_k \stackrel{\rightarrow}{e_i}\right)_{\bot V_n}~.$$

So,
\begin{eqnarray}\label{2.39}
\left(\partial_k\stackrel{\rightarrow}{e_i}\right) \cdot \stackrel{\rightarrow}{e_l} &=& \left(\partial_k \stackrel{\rightarrow}{e_i}\right)_{||V_n} \cdot \stackrel{\rightarrow}{e_l}+0 \nonumber \\
&=& \left(\Gamma^j_{ki}\stackrel{\rightarrow}{e_j}\right) \cdot \stackrel{\rightarrow}{e_l} = g_{jl}\Gamma^j_{ki} = \Gamma_{kil}
\end{eqnarray}
where $\Gamma_{kil}=g_{jl}\Gamma^j_{ki}$ , is called the Riemann-Christoffel symbol of first kind. From the above definitions of the Christoffel symbols we write
\begin{equation}\label{2.40}
g^{pl}\Gamma_{kil}=g^{pl}g_{jl}\Gamma^j_{ki}=\delta^p_j\Gamma^j_{ki}=\Gamma^p_{ki}~.
\end{equation}

Thus Christoffel symbols of 1st and 2nd kind are convertable by lowering and raising the indices using metric tensor or its reciprocal {\it i.e;}
\begin{equation}\label{2.41}
\Gamma_{ijk}=g{kp}\Gamma^p_{ij}~~\mbox{and}~~\Gamma^k_{ij}=g^{pk}\Gamma _{ijp}~.
\end{equation}
	
Let $\stackrel{\rightarrow}{X}=x^1\stackrel{\rightarrow}{e_1}+x^2\stackrel{\rightarrow}{e_2}+ \ldots \ldots + x^n\stackrel{\rightarrow}{e_n}$ , be any vector in $V_n$ then $$\frac{\partial \stackrel{\rightarrow}{X}}{\partial x^k} = \stackrel{\rightarrow}{e_k}~.$$

So, $$\frac{\partial}{\partial x^i}\stackrel{\rightarrow}{e_k}=\frac{\partial}{\partial x^i}(\frac{\partial\stackrel{\rightarrow}{X}}{\partial x^k})=\frac{\partial^2\stackrel{\rightarrow}{X}}{\partial x^i \partial x^k}=\frac{\partial^2\stackrel{\rightarrow}{X}}{\partial x^k \partial x^i}=\frac{\partial}{\partial x^k}(\frac{\partial \stackrel{\rightarrow}{X}}{\partial x^i})=\frac{\partial \stackrel{\rightarrow}{e_i}}{\partial x^k}$$
$$i.e.,~~~~\partial_i \stackrel{\rightarrow}{e_k}=\partial_k \stackrel{\rightarrow}{e_i}~.$$

Then from equation (\ref{2.39})
$$\Gamma_{kil}= (\partial_k \stackrel{\rightarrow}{e_i}) \cdot \stackrel{\rightarrow}{e_l} = (\partial_i \stackrel{\rightarrow}{e_k}) \cdot \stackrel{\rightarrow}{e_l}=\Gamma_{ikl}$$

Hence $\Gamma_{kil}$ is symmetric in $i$ and $k$ {\it i.e.} in the first two indices. Consequently, the Christoffel symbol of 2nd kind is also symmetric in the two lower indices.\\

Now,
\begin{eqnarray}\label{2.42}
\partial_i g_{kl}=\partial_i (\stackrel{\rightarrow}{e_k} \cdot \stackrel{\rightarrow}{e_l}) &=& (\partial_i \cdot \stackrel{\rightarrow}{e_k}) \cdot \stackrel{\rightarrow}{e_l}+(\partial_i \stackrel{\rightarrow}{e_l}) \cdot \stackrel{\rightarrow}{e_k} \nonumber \\
&=& \Gamma_{ikl}+\Gamma_{ilk}~.
\end{eqnarray}

Similarly
\begin{equation}\label{2.43}
\partial_l g_{ik}=\Gamma_{lik}+\Gamma_{lki}
\end{equation}
\begin{equation}\label{2.44}
\partial_k g_{li}=\Gamma_{kli}+\Gamma_{kil}~.
\end{equation}

Using the symmetry property of $\Gamma_{ijk}$ , we have from $\lbrace$(\ref{2.42})-(\ref{2.43})+(\ref{2.44})$\rbrace$,
\begin{equation}\label{2.45}
\Gamma_{ikl}=\frac{1}{2}(\partial_i g_{kl}+\partial_k g_{il}-\partial_l g_{ik})~.
\end{equation}

Further, if $w_i$ be the components of a covector $\utilde{w}$ {\it i.e.,} $$\utilde{w}=w_i \utilde{e^i}$$
then, $$\partial_k \utilde{w}=(\partial_k w_i)\utilde{e^i}+w_i(\partial_k \utilde{e^i})~.$$
Hence, $$\left(\partial_k \utilde{w}\right)_{||V_n}=(\partial_k w_i)\utilde{e^i}+w_i\left(\partial_k \utilde{e^i}\right)_{||V_n}$$

As before let us write
\begin{equation}\label{2.46}
\left(\partial_k \utilde{e^i}\right)_{||V_n}=\Gamma^{*i}_{kl} \utilde{e^l}~,
\end{equation}
then
\begin{eqnarray}\label{2.47}
\left(\partial_k \utilde{w}\right)_{||V_n} &=& (\partial_k w_i)\utilde{e^i}+\Gamma^{*i}_{kl} \utilde{e^l}w_i \nonumber \\
&=& (\partial_k w_i)\utilde{e^i}+\Gamma^{*l}_{ki} \utilde{e^i}w_l~~~~(i \rightleftharpoons l) \nonumber \\
&=& (\partial_k w_i+\Gamma^{*l}_{ki} w_l)\utilde{e^i}~.
\end{eqnarray}

Again, $$(\partial_k \utilde{e^i})=(\partial_k \utilde{e^i})_{||V_n}+(\partial_k \utilde{e^i})_{\bot V_n}~.$$

Taking dot product with $\stackrel{\rightarrow}{e_l}$ , we get
$$(\partial_k \utilde{e^i}) \cdot \stackrel{\rightarrow}{e_l} = \Gamma^{*i}_{kj}\utilde{e^j}\stackrel{\rightarrow}{e_l} = \Gamma^{*i}_{kj}\delta^j_l = \Gamma^{*i}_{kl}~.$$

But, $$~~~\utilde{e^i}\stackrel{\rightarrow}{e_l}=\delta^i_l~.$$

So, $$(\partial_k \utilde{e^i})\stackrel{\rightarrow}{e_l}+\utilde{e^i}(\partial_k \stackrel{\rightarrow}{e_l})=0$$
or, $$\Gamma^{*i}_{ks}\utilde{e^s}\stackrel{\rightarrow}{e_l}+\utilde{e^i}\Gamma^u_{kl}\stackrel{\rightarrow}{e_u}=0~~(\mbox{using~equation~(\ref{2.38})~and~(\ref{2.46})})$$
or, $$\Gamma^{*l}_{ks} \delta^s_l+\Gamma^u_{kl} \delta^i_u=0$$
or, $$\Gamma^{*i}_{kl} + \Gamma^i_{kl}=0$$
{\it i.e.,}
\begin{equation}\label{2.48}
\Gamma^{*i}_{kl}=-\Gamma^i_{kl}.
\end{equation}

Hence from equation (\ref{2.47})
\begin{equation}\label{2.49}
(\partial_k \utilde{w})_{||V_n}=(\partial_k w_i-\Gamma^l_{ki}w_l)\utilde{e^i}=(\nabla_k w_i)\utilde{e^i}~.
\end{equation}

Here,
\begin{equation}\label{2.50}
\nabla_k w_i=(\partial_k w_i-\Gamma^l_{ki}w_l)
\end{equation}
is called the covariant derivative w.r.t. $x^k$ of the covariant components of $\utilde{w}$.\\

{\bf Note:} In $n$ dimensional Riemannian space the no. of independent Christoffel symbols are $\dfrac{n^2(n+1)}{2}$.\\

\section{Transformation Laws for Christoffel symbols}

~~~Let $\Gamma_{ijk}$ and $\overline{\Gamma}_{ijk}$ be the components of the christoffel symbols of first kind in some $x-$co-ordinate and $\overline{x}-$co-ordinates respectively.\\

Then,
\begin{eqnarray}\label{2.51}
\overline{\Gamma}_{lmp} &=& \frac{1}{2}\left(\frac{\partial \overline{g}_{mp}}{\partial \overline{x}^l}+\frac{\partial \overline{g}_{lp}}{\partial \overline{x}^m}-\frac{\partial \overline{g}_{lm}}{\partial \overline{x}^p}\right)
\nonumber \\
&=&\frac{1}{2}\left[\frac{\partial}{\partial \overline{x}^l}\left(\frac{\partial x^k}{\partial \overline{x}^m}\frac{\partial x^s}{\partial \overline{x}^p}g_{ks}\right)+\frac{\partial}{\partial \overline{x}^m}\left(\frac{\partial x^i}{\partial \overline{x}^l}\frac{\partial x^s}{\partial \overline{x}^p}g_{is}\right)-\frac{\partial}{\partial \overline{x}^p}\left(\frac{\partial x^i}{\partial \overline{x}^l}\frac{\partial x^k}{\partial \overline{x}^m}g_{ik}\right)\right]
\nonumber \\
&=& \frac{1}{2}\left[\left(\frac{\partial^2 x^k}{\partial \overline{x}^l \partial \overline{x}^m}\frac{\partial x^s}{\partial \overline{x}^p}g_{ks}+\frac{\partial x^k}{\partial \overline{x}^m}\frac{\partial^2 x^s}{\partial \overline{x}^l \partial \overline{x}^p}g_{ks}+\frac{\partial x^k}{\partial \overline{x}^m}\frac{\partial x^s}{\partial \overline{x}^p}\frac{\partial x^i}{\partial \overline{x}^l}\frac{\partial g_{ks}}{\partial x^i}\right)\right. \nonumber \\
&&+ \left(\frac{\partial^2 x^i}{\partial \overline{x}^m \partial \overline{x}^l}\frac{\partial x^s}{\partial \overline{x}^p}g_{is}+\frac{\partial x^i}{\partial \overline{x}^l}\frac{\partial^2 x^s}{\partial \overline{x}^m \partial \overline{x}^p}g_{is}+\frac{\partial x^i}{\partial \overline{x}^l}\frac{\partial x^s}{\partial \overline{x}^p}\frac{\partial x^k}{\partial \overline{x}^m}\frac{\partial g_{is}}{\partial x^k}\right) \nonumber \\
&&- \left. \left(\frac{\partial^2 x^i}{\partial \overline{x}^p \partial \overline{x}^l}\frac{\partial x^k}{\partial \overline{x}^m}g_{ik}+\frac{\partial x^i}{\partial \overline{x}^l}\frac{\partial^2 x^k}{\partial \overline{x}^m \partial \overline{x}^p}g_{ik}-\frac{\partial x^i}{\partial \overline{x}^l}\frac{\partial x^k}{\partial \overline{x}^m}\frac{\partial x^s}{\partial \overline{x}^p}\frac{\partial g_{ik}}{\partial x^s}\right)\right] \nonumber \\
&=& \frac{\partial^2 x^k}{\partial \overline{x}^l \partial \overline{x}^m}\frac{\partial x^s}{\partial \overline{x}^p}g_{ks}+\frac{\partial x^i}{\partial \overline{x}^l}\frac{\partial x^k}{\partial \overline{x}^m}\frac{\partial x^s}{\partial \overline{x}^p}\left[\frac{1}{2}(\partial_i g_{ks}+\partial_k g_{is}-\partial_s g_{ik})\right] \nonumber \\
\overline{\Gamma}_{lmp} &=& \frac{\partial^2 x^k}{\partial \overline{x}^l \partial \overline{x}^m}\frac{\partial x^s}{\partial \overline{x}^p}g_{ks}+\frac{\partial x^i}{\partial \overline{x}^l}\frac{\partial x^k}{\partial \overline{x}^m}\frac{\partial x^s}{\partial \overline{x}^p} \Gamma_{iks}~.
\end{eqnarray}

This is the transformation law for $\Gamma_{ijk}$. The presence of the 1st term in the R.H.S. (containing second order partial derivatives of the co-ordinate variables) shows that $\Gamma_{ijk}$ is not a tensor, it is simply a three index symbol.\\

We shall now deduce the transformation law of $\Gamma^k_{ij}$, the Christoffel symbols of second kind.\\

The transformation law of the reciprocal metric tensor (a , $(2,~0)-$tensor) is
\begin{equation}\label{2.52}
\overline{g}^{pq}=\frac{\partial \overline{x}^p}{\partial x^f} \frac{\partial \overline{x}^q}{\partial x^h}g^{fh}~.
\end{equation}

Now contracting the L.H.S. of equations (\ref{2.51}) using (\ref{2.52}) and accordingly the R.H.S. we have 
\begin{eqnarray}\label{2.53}
\overline{\Gamma}_{lmp} \overline{g}^{pq} &=& \frac{\partial^2 x^k}{\partial \overline{x}^l \partial \overline{x}^m} \left[\frac{\partial x^s}{\partial \overline{x}^p} \cdot \frac{\partial \overline{x}^p}{\partial x^f}\right] \frac{\partial \overline{x}^q}{\partial x^h} g^{fh} g_{ks}+\frac{\partial x^i}{\partial \overline{x}^l}\frac{\partial x^k}{\partial \overline{x}^m}\left[\frac{\partial x^s}{\partial \overline{x}^p} \cdot \frac{\partial \overline{x}^p}{\partial x^f}\right]\frac{\partial \overline{x}^q}{\partial x^h}g^{fh}\Gamma_{iks} \nonumber \\
\mbox{or,}~~~~ \overline{\Gamma}^q_{lm} &=& \frac{\partial^2 x^k}{\partial \overline{x}^l \partial \overline{x}^m}\frac{\partial \overline{x}^q}{\partial x^h}\delta^s_f g^{fh} g_{ks}+\frac{\partial x^i}{\partial \overline{x}^l}\frac{\partial x^k}{\partial \overline{x}^m}\frac{\partial \overline{x}^q}{\partial x^h}\delta^s_f g^{fh} \Gamma_{iks} \nonumber \\
i.e,~~~~ \overline{\Gamma}^q_{lm} &=& \frac{\partial^2 x^k}{\partial \overline{x}^l \partial \overline{x}^m} \cdot \frac{\partial \overline{x}^q}{\partial x^k}+\frac{\partial x^i}{\partial \overline{x}^l}\frac{\partial x^k}{\partial \overline{x}^m}\frac{\partial \overline{x}^q}{\partial x^h} \Gamma^h_{ik}~.
\end{eqnarray}

This transformation law of Christoffel symbols of second kind shows that it is also not a tensor, rather a 3-index system of functions.\\

Further, contracting equations (\ref{2.53}) with $\dfrac{\partial x^u}{\partial \overline{x}^q}$ we have
\begin{eqnarray}\label{2.54}
\overline{\Gamma}^q_{lm}\frac{\partial x^u}{\partial \overline{x}^q}&=&\frac{\partial^2 x^k}{\partial \overline{x}^l \partial \overline{x}^m}\delta^u_k+\frac{\partial x^i}{\partial \overline{x}^l}\frac{\partial x^k}{\partial \overline{x}^m}\delta^u_h \Gamma^h_{ik} \nonumber \\
&=&\frac{\partial^2 x^u}{\partial \overline{x}^l \partial \overline{x}^m}+\frac{\partial x^i}{\partial \overline{x}^l}\frac{\partial x^k}{\partial \overline{x}^m}\Gamma^u_{ik}~.
\end{eqnarray}

Hence,
\begin{equation}\label{2.55}
\frac{\partial^2 x^u}{\partial \overline{x}^l \partial \overline{x}^m}=\overline{\Gamma}^q_{lm}\frac{\partial x^u}{\partial \overline{x}^q}-\frac{\partial x^i}{\partial \overline{x}^l}\frac{\partial x^k}{\partial \overline{x}^m}\Gamma^u_{ik}~.
\end{equation}
\\

\section{Tensorial property of covariant derivative}

~~~We have defined the covariant $x^k-$derivative of a contravarinat vector $v^i$ as
$$\nabla _kv^i=\partial_k v^i+\Gamma^i_{kl}v^l~,$$
which we shall show later to be a (1, 1)-tensor. Similarly, the covariant $x^k-$derivative of a covariant vector $w_i$ is defined as
$$\nabla _kw_i=\partial_kw_i-\Gamma^l_{ki}w_l~,$$
which will be shown to be a (0, 2) tensor.\\

We shall now extend this covariant differentiation to arbitrary tensors assuming linearity and Leibnitzian property of the covariant derivative.\\

First of all the covariant derivative of a scalar is defined as the partial derivative w.r.t. the corresponding co-ordinate $i.e.$,
$$\nabla _kf=\frac{\partial f}{\partial x^k}=\partial_kf~,$$
which is clearly a (0, 1) tensor.\\

We shall now deduce the covariant derivative of a (1, 1) tensor having components $A^i_j$ in $x-$co-ordinate system.\\

For any arbitrary covariant and contravariant vectors $u_i$ and $v^i$ , the expression $A_j^i u_i v^j$ is a scalar\,(by contraction), so we have
\begin{eqnarray}
\nabla _k(A^i_j u_i v^j) &=& \partial_k(A^i_j u_i v^j) \nonumber \\
&=& (\partial_k A^i_j)u_iv^j+A^i_j(\partial_k u_i)v^j+A^i_j u_i(\partial_k v^j) \nonumber \\
&=& (\partial_k A^i_j)u_iv^j+A^i_j(\nabla _k u_i+\Gamma^s_{ki} u_s)v^j+A^i_j u_i(\nabla _k v^j-\Gamma^j_{ks}v^s) \nonumber \\
&=& (\partial_k A^i_j)u_iv^j+\Gamma^s_{ki} A^i_j u_s v^i-\Gamma^j_{ks} A^i_j u_i v^s+A^i_j v^j(\nabla _ku_i)+A^i_j u_i(\nabla _v v^j)~. \nonumber
\end{eqnarray}

Due to Leibnitzian property , the L.H.S can be written as :
$$(\nabla _k A^i_j)u_i v^j+A^i_j v^j(\bigtriangledown_k u_i)+A^i_j u_i(\bigtriangledown_kv^j).$$

So comparing with the R.H.S we have for any arbitrary vectors $\utilde{u}$ and $\stackrel{\rightarrow}{v}$
$$\nabla _k A^i_j = \partial_k A^i_j+\Gamma^i_{ks} A^s_j-\Gamma^s_{kj} A^i_s~.$$

Similarly,
$$\nabla _kA_{ij}=\partial_k A_{ij}-\Gamma^s_{ki}A_{sj}-\Gamma^u_{kj} A_{iu}$$
and
$$\nabla _kA^{ij}=\partial_k A^{ij}+\Gamma^i_{ks} A^{sj}+\Gamma^j_{ks}A^{is}~.$$

In general, for an arbitrary $(r,~s)$ tensor
$$\nabla _kA^{i_1,i_2,\ldots \ldots ,i_r}_{j_1,j_2,\ldots \ldots ,j_s}=\partial_k A^{i_1,i_2,\ldots \ldots ,i_r}_{j_1,j_2,\ldots \ldots ,j_s}+\sum^r_{p=1} \Gamma^{i_p}_{ks} A^{i_1,i_2,\ldots ,i_{p-1},s,i_{p+1}\ldots ,i_r}_{j_1,j_2,\ldots \ldots ,j_s}-\sum^s_{q=1} \Gamma^u_{kj_q} A^{i_1,i_2,\ldots \ldots ,i_r}_{j_1,j_2,\ldots ,j_{q-1},u,j_{q+1}\ldots ,j_s}~.$$
\\


\section{Intrinsic Derivative}

~~~Let $A^{i_1,i_2,\ldots ,i_r}_{j_1,j_2,\ldots ,j_s}$ be the components of an $(r,~s)$ tensor A in some $x-$co-ordinate system defined in a region $D$ of a Riemannian space $M$. Suppose $\gamma:x^i=x^i(u)$ be a curve in this region. Just as an ordinary derivative $\dfrac{d}{du}$ will satisfy the relation
$$\frac{d}{du}A^{i_1,i_2,\ldots , i_r}_{j_1,j_2,\ldots , j_s}=\left(\partial_k A^{i_1,i_2,\ldots , i_r}_{j_1,j_2,\ldots , j_s}\right)\frac{dx^k}{du}.$$

We now define an operator $\dfrac{\delta}{du}$ as
\begin{equation}\label{2.59}
\frac{\delta}{du}A^{i_1,i_2,\ldots , i_r}_{j_1,j_2,\ldots , j_s}=\left(\nabla _k \cdot A^{i_1,i_2,\ldots , i_r}_{j_1,j_2,\ldots , j_s}\right)\frac{dx^k}{du}
\end{equation}
and call it the {\bf intrinsic} $u-$derivative of $A$ or the intrinsic derivative of $A$ along the curve $\gamma$.\\

Using quotient law, it follows that the intrinsic derivative of an $(r,~s)$ tensor is also an $(r,~s)$ tensor $i.e.,$ the intrinsic derivative does not alter the order-type of the tensor. Thus
\begin{eqnarray}\label{2.60}
\frac{\delta A^i}{du}=(\nabla _k A^i)\frac{dx^k}{du} &=& \left(\partial_k A^i+\Gamma^i_{ks} A^s\right)\frac{dx^k}{du} \nonumber \\
&=& \frac{dA^i}{du}+\Gamma^i_{ks}\frac{dx^k}{du}A^s~.
\end{eqnarray}

Similarly;
\begin{equation}\label{2.61}
\frac{\delta A_i}{du}=\frac{dA_i}{du}-\Gamma^s_{ik} A_s\frac{dx^k}{du}.
\end{equation}

For any scalar, $\sigma$
\begin{equation}\label{2.62}
\frac{\delta \sigma}{du}=(\nabla _k \sigma) \cdot \frac{dx^k}{du}=(\partial_k \sigma)\frac{dx^k}{du}=\frac{d \sigma}{du}~.
\end{equation}

\underline{\bf Note:} If the covariant derivative of a tensor is zero then obviously its intrinsic derivative is also zero.

\begin{center}
\underline{\bf Directional derivative}
\end{center}
~~~If $\stackrel{\rightarrow}{X}=(X^i)$ be any vector field defined in the region $D$, then
$$X^k\nabla _k A^{i_1,i_2, \ldots , i_r}_{j_1,j_2, \ldots , j_s}$$
is called the directional (tensor) derivative of the tensor $A$ in the direction of the vector $\stackrel{\rightarrow}{X}$ and is often denoted by $\nabla _{\stackrel{\rightarrow}{X}} A^{i_1,i_2,\ldots ,i_r}_{j_1,j_2,\ldots ,j_s}$.\\

\underline{\bf Divergence of a vector:} The divergence of a contravariant vector is defined as contraction of its covariant derivative $i.e.,$\\
$$\mbox{div}A^i=A^i_{;i}~.$$

Similarly, the divergence of a covariant vector $A_i$ is denoted by $div A_i$ and is defined as 
$$\mbox{div}A_i=g^{ik} A_{i;k}~.$$

\underline{\bf Note:}
$$\mbox{div}A_i = \mbox{div}A^i$$

\underline{\bf Curl of a vector:} $$\mbox{curl}A_i=A_{i;j}-A_{j;i}=\frac{\partial A_i}{\partial x^j}-\frac{\partial A_j}{\partial x^i}$$

\underline{\bf Note-I:} If the covariant derivative of a covariant vector is symmetrical then the vector must be gradient of some scalar function.\\

As, $A_{i;j}=A_{j;i}$ , so $\mbox{curl}A_i=0 ~~~~ \Rightarrow A_i=\nabla \phi$.\\

\underline{\bf II:} If $\nabla ^2$ be the Laplacian operator then for any scalar function $\phi$
\begin{eqnarray}
\nabla ^2 \phi= \mbox{div~grad} \phi &=& \frac{1}{\sqrt {g}}\frac{\partial}{\partial x^i}\left[\sqrt{g}g^{ij}\phi_{,j}\right] \nonumber \\
&=& \frac{1}{\sqrt {g}} \partial _i \left[\sqrt{g} g^{ij} \phi _{,j} \right] \nonumber
\end{eqnarray}\\

\section{Riemann Curvature Tensor}

We know that the commutator $\left[\partial_k,\partial_j\right] \equiv \partial_k \partial_j-\partial_j \partial_k$ of partial derivatives is zero, when acting on functions of class $C^r(r\geq 2).$ However, similar is not the case with covariant derivatives.The study of commutator of the covariant derivatives leads to the notion of what is called the Riemann Curvature Tensor.\\

Let us calculate the value of 
$$\left[\nabla_k,\nabla_j\right]w_i\equiv\left(\nabla_k \nabla_j-\nabla_j\nabla_k\right)w_i~.$$

As, $\nabla_j w_i=\partial_jw_i-\Gamma^s_{ji} w_s$ ~~~ so,
\begin{eqnarray}
\nabla_k\nabla_j w_i &=& \partial_k (\nabla_j w_i)-\Gamma^{\sigma}_{kj}\nabla_{\sigma}w_i-\Gamma^{\sigma}_{ki}\nabla_j w_{\sigma} \nonumber \\
&=& \partial_k \left(\partial_j w_i-\Gamma^s_{ji} w_s\right)-\Gamma^{\sigma}_{kj}\left(\partial_{\sigma} w_i-\Gamma^s_{\sigma i}w_s\right)-\Gamma^{\sigma}_{ki}\left(\partial_j w_{\sigma}-\Gamma^s_{j \sigma}w_s\right) \nonumber \\
&=& \partial_k \partial_j w_i-\left(\partial_k \Gamma^s_{ji}\right)w_s-\Gamma^s_{ji}(\partial_k w_s)-\Gamma^{\sigma}_{kj}\partial_{\sigma}w_i+\Gamma^{\sigma}_{kj}\Gamma^s_{\sigma i}w_s-\Gamma^{\sigma}_{ki}(\partial_j w_{\sigma})+\Gamma^{\sigma}_{ki}\Gamma^s_{j {\sigma}}w_s~. \nonumber
\end{eqnarray}

Now performing $k\rightleftharpoons j$ we have
$$\nabla_j \nabla_k w_i=\partial_j \partial_k w_i-(\partial_j \Gamma^s _{ki})w_s-\Gamma^s_{ki}(\partial_j w_s)-\Gamma^{\sigma}_{jk}\partial_{\sigma} w_i+\Gamma^{\sigma}_{jk}\Gamma^s_{\sigma i}w_s-\Gamma^{\sigma}_{ki}(\partial_k w_{\sigma})+\Gamma^{\sigma}_{ji}\Gamma^s_{k \sigma}w_s$$

Thus,
\begin{eqnarray}
[\nabla_k,\nabla_j]w_i &=& -\left(\partial _k \Gamma ^s_{ji}-\partial_j \Gamma ^s_{ki}-\Gamma ^{\sigma}_{ki}\Gamma^s_{j \sigma}+\Gamma^{\sigma}_{ji}\Gamma ^s_{k \sigma}\right)A_s \nonumber \\
&=& -\left(\partial _k \Gamma ^s_{ji}-\partial_j \Gamma ^s_{ki}+\Gamma ^s_{\sigma k}\Gamma^{\sigma}_{ji} - \Gamma ^s_{\sigma j}\Gamma ^{\sigma}_{ki}\right)A_s~. \nonumber
\end{eqnarray}

We now define,
\begin{equation}\label{2.63}
[\nabla_k,\nabla_j]w_i=R^s_{ijk} w_s
\end{equation}
where,
\begin{equation}\label{2.64}
R^s_{ijk}=-\left(\partial_k\Gamma^s_{ji}-\partial_j \Gamma^s_{ki}+\Gamma^s_{\sigma k}\Gamma^{\sigma}_{ji}-\Gamma^s_{\sigma j}\Gamma^{\sigma}_{ki}\right)~.
\end{equation}

From quotient law we see that $R^s_{ijk}$ is a (1, 3)\,-tensor and is called the Riemann curvature tensor (in mixed form).\\

We next calculate the commutator $[\nabla_k,\nabla_j]v^i.$\\

As,$$\nabla_j v^i=\partial_jv^i+\Gamma^i_{js}v^s$$
so,\begin{eqnarray}\nabla_k \nabla_j v^i&=&\partial_k(\nabla_j v^i)-\Gamma^{\sigma}_{kj} \nabla_{\sigma} v^i+\Gamma^i_{k \sigma} \nabla_j v^{\sigma}\nonumber\\
&=&\partial_k \partial_j v^i+(\partial_k \Gamma^i_{js})v^s+\Gamma^i_{js}(\partial_k v^s)-\Gamma^{\sigma}_{kj}(\partial_{\sigma}v^i)-\Gamma^{\sigma}_{kj} \Gamma^i_{\sigma s}v^s+\Gamma^i_{k \sigma}(\partial_j v^{\sigma})+\Gamma^i_{k \sigma}\nonumber \Gamma^{\sigma}_{js} v^s.\end{eqnarray}

Commuting on $k$ and $j$ we get
\begin{eqnarray}\label{2.65}
[\nabla_k,\nabla_j]v^i &=& \left(\partial_k \Gamma^i_{js}-\partial_j \Gamma^i_{ks}+\Gamma^i_{\sigma k} \Gamma^{\sigma}_{js}-\Gamma^i_{\sigma j} \Gamma^{\sigma}_{ks}\right)v^s \nonumber \\
&=& -R^i_{sjk} v^s.
\end{eqnarray}

In general,
\begin{equation}\label{2.66}
[\nabla_k,\nabla_j]A^{i_1,i_2,\ldots , i_r}_{p_1,p_2,\ldots , p_m}=\sum ^m_{u=1}A^{i_1,i_2,\ldots , i_r}_{p_1,p_2,\ldots ,p_{u-1},s, p_{u+1},\ldots , p_m}R^s_{p_u jk}-\sum^r_{u=1}A^{i_1,i_2,\ldots , i_{u-1},s, i_{u+1},\ldots ,i_r}_{p_1,p_2,\ldots , p_m}R^{i_u}_{sjk}~.
\end{equation}

\begin{center}
\underline{\bf Curvature Tensor in Fully Covariant Form:}
\end{center}

We define
\begin{equation}\label{2.70}
R_{sijk}=g_{sh}R^h_{ijk}~,
\end{equation}

as the Riemann curvature tensor in fully covariant form.\\

Then
$$g^{ps} R_{sijk}=g^{ps}g_{sh}R^h_{ijk}=R^p_{ijk}$$
\begin{equation}\label{2.71}
R^p_{ijk}=g^{ps}R_{sijk}~.
\end{equation}

Now using equation (\ref{2.64}) in equation (\ref{2.70}) , we get
$$R_{sijk}=-\left[g_{sh}\partial_k\Gamma^h_{ji}-g_{sh}\partial_s\Gamma^h_{ki}+\Gamma_{rks}\Gamma^r_{ji}-\Gamma_{rjs}\Gamma^r_{ki}\right].$$

But $$g_{sh}\partial_k\Gamma^h_{ji}=\partial_k(g_{sh}\Gamma^h_{ji})-(\partial_k g_{sh})\Gamma^h_{ji}~.$$

Hence
\begin{eqnarray}
R_{sijk} &=& -\partial_k \left(\Gamma_{jis}\right)+\left(\partial_k g_{sh}\right)\Gamma^h_{ji}+\partial_j \Gamma_{kis}-\left(\partial_j g_{sh}\right)\Gamma^h_{ki}-\Gamma_{rks}\Gamma^r_{ji}+\Gamma_{rjs}\Gamma^r_{ki} \nonumber\\
&=& -\partial_k \Gamma_{jis}+\left(\Gamma_{ksh}+\Gamma_{khs}\right)\Gamma^h_{ji}+\partial_j \Gamma_{kis}-\left(\Gamma_{jsh}+\Gamma_{jhs}\right)\Gamma^h_{ki}-\Gamma_{rks}\Gamma^r_{ji}+\Gamma_{rjs}\Gamma^r_{ki} \nonumber\\
&=& -\left[\partial_k \Gamma_{jis}-\partial_j \Gamma_{kis}-\Gamma^h_{ks}\Gamma_{jih}+\Gamma^h_{js} \Gamma_{kih}\right].\nonumber
\end{eqnarray}
$$\left( \mbox{\textbf{Note~:}}~~\Gamma_{ksh}\Gamma^h_{ji} = g_{hp}\Gamma^p_{ks}\Gamma^h_{ji} = \Gamma^p_{ks}\Gamma_{jip} = \Gamma^h_{ks}\Gamma_{jih} \right)$$

Thus
\begin{equation}\label{2.72}
R_{sijk}=-\left[\partial_k \Gamma_{jis}-\partial_j \Gamma_{kis}-\Gamma^h_{sk}\Gamma_{jih}+\Gamma^h_{sj}\Gamma_{kih}\right]. 
\end{equation}

\underline{\bf Note:} The number of independent components (not necessarily vanishing) of curvature tensor does not exceed $\dfrac{n^2(n^2-1)}{12}$\\

\begin{center}
\underline{\bf Properties of Curvature Tensor}
\end{center}

{\bf I.} Both $R^h_{ijk}$ and $R_{hijk}$ are skew-symmetric in the last two indices $i.e.,$
$$R^h_{i(jk)}=0=R_{hi(jk)}~.$$

It follows directly from the defining relation for $R^h_{ijk}$ or $R_{hijk}~.$\\

{\bf II.} $R_{hijk}$ is skew symmetric also in the 1st two indices.
$$i.e.,~~~~ R_{(hi)jk}=0$$

{\bf \underline{Proof:}}  As $g_{pq}$ is covariant constant so we have 
$$\left(\nabla_k \nabla_j-\nabla_k \nabla_j\right)g_{pq}=0.$$

Using Equation (\ref{2.68}) we get
$$g_{sq} R^s_{pjk}+g_{ps} R^s_{qjk}=0$$
$$i.e.,~~ R_{qpjk}+R_{pqjk}=0~~~~~~~~~~~~$$
$$i.e.,~~ R_{(pq)jk}=0.~~~~~~~~~~~~~~~~~~~$$

{\bf III. \underline{Bianchi's first identity:}}
\begin{equation}\label{2.73}
\mbox{a)}~~ R^h_{ijk}+R^h_{jki}+R^h_{kij}=0,~~i.e;~ R^h_{(ijk)}=0.
\end{equation}
\begin{equation}\label{2.74}
\mbox{b)}~~ R_{hijk}+R_{hjki}+R_{hkij}=0,~~i.e;~R_{h(ijk)}=0
\end{equation}

{\bf \underline{Note:}} The above two relations are not independent. In fact 2nd relation (\ref{2.74}) can be obtained from the first one (\ref{2.73}) by multiplying $g_{ph}$ and using the definition (\ref{2.72}).\\

{\bf \underline{Proof:}} From the definition of the curvature tensor in equation (\ref{2.64})
$$R^h_{ijk}=-\left(\partial_k \Gamma^h_{ji}-\partial_j \Gamma^h_{ki}+\Gamma^h_{pk}\Gamma^p_{ji}-\Gamma^h_{pj}\Gamma^p_{ki}\right)$$

By cyclic rotation of $i,j,k$ we get
$$R^h_{jki}=-\left(\partial_i \Gamma^h_{kj}-\partial_k \Gamma^h_{ij}+\Gamma^h_{pi}\Gamma^p_{kj}-\Gamma^h_{pk}\Gamma^p_{ij}\right)$$
$$R^h_{kij}=-\left(\partial_j \Gamma^h_{ik}-\partial_i \Gamma^h_{jk}+\Gamma^h_{pj}\Gamma^p_{ik}-\Gamma^h_{pi}\Gamma^p_{jk}\right)$$

Adding these three equations and using symmetry of the Christoffel symbol of 2nd kind in the lower two indices, we get
$$R^h_{ijk}+R^h_{jki}+R^h_{kij}=0.$$

{\bf IV.} $R_{hijk}=R_{jkhi}$ $i.e.,$ for Riemann curvature tensor in fully covariant form, the first and last pair can be interchanged without changing the value of the tensor.\\

{\bf \underline{Proof:}} The Bianchi 1st identity in fully covariant form gives 
$$R_{hijk}+R_{hjki}+R_{hkij}=0$$

By cyclic rotation of the indices $(h,i,j,k)$ we get
$$R_{ijkh}+R_{ikhj}+R_{ihjk}=0$$
$$R_{jkhi}+R_{jhik}+R_{jikh}=0$$
$$R_{khij}+R_{kijh}+R_{kjhi}=0$$

Adding these four relations and using the skew symmetric properties I and II we get
$$2R_{hjki}+2R_{ikhj}=0$$
$$or,~ R_{hjki}+R_{ikhj}=0~~~~~$$
$$i.e;~ R_{hjki}=-R_{ikhj}=R_{kihj}.$$

Hence the property.\\

{\bf V. \underline{Bianchi's second identity:}}
\begin{equation}\label{2.75}
\mbox{a)}~~ \nabla_h R^i_{sjk}+\nabla_j R^i_{skh}+\nabla_k R^i_{shj}=0,
\end{equation}
\begin{equation}\label{2.76}
\mbox{b)}~~ \nabla_h R_{isjk}+\nabla_j R_{iskh}+\nabla_k R_{ishj}=0,
\end{equation}

{\bf Note:} As in Bianchi's 1st identity, it will be enough to prove the 1st result only, the second one can be obtained from the 1st by lowering the index $'i'$ with the help of the metric tensor.\\

{\bf Proof:}  For an arbitrary vector field $\stackrel{\rightarrow}{v},$ we have
$$\left[\nabla_k, \nabla_j\right]v^i = -R^i_{sjk} v^s~.$$
Now taking covariant $x^h$ derivative we get
\begin{equation}\label{2.77}
\left(\nabla_h \nabla_k \nabla_j - \nabla_h \nabla_j \nabla_k\right)v^i = -\left(\nabla_h v^s\right) R^i_{sjk}-v^s \nabla_h R^i_{sjk}~.
\end{equation}

Cyclic rotations of $j, k, h $ gives the following two relations
\begin{equation}\label{2.78}
\left(\nabla_j \nabla_h \nabla_k - \nabla_j \nabla_k \nabla_h\right)v^i=-\left(\nabla_j v^s\right) R^i_{skh}-v^s \nabla_j R^i_{skh}~.
\end{equation}
\begin{equation}\label{2.79}
\left(\nabla_k \nabla_j \nabla_h - \nabla_k \nabla_h \nabla_j\right)v^i=-\left(\nabla_k v^s\right) R^i_{shj}-v^s \nabla_k R^i_{shj}~.
\end{equation}

The sum of the left hand side of (\ref{2.77}), (\ref{2.78}) and (\ref{2.79}) may be written as
\begin{eqnarray}
\left(\nabla_h \nabla_k- \nabla_k \nabla_h\right) \left(\nabla_j v^i\right) &+&
\left(\nabla_j \nabla_h- \nabla_h \nabla_j\right) \left(\nabla_k v^i\right) + \left(\nabla_h \nabla_k- \nabla_k \nabla_h\right) \left(\nabla_j v^i\right) \nonumber \\
&=& \left(\nabla_s v^i\right) R^s_{jkh}- \left(\nabla_j v^s\right) R^i_{skh} +\left(\nabla_s v^i\right) R^s_{khj} \nonumber \\
&&- \left(\nabla_k v^s\right) R^i_{shj}+ \left(\nabla_s v^i\right) R^s_{hjk}-\left(\nabla_h v^s\right) R^i_{sjk} \nonumber \\
&=& \left(\nabla_s v^i\right) \left[R^s_{jkh} +R^s_{khj} +R^s_{hjk}\right]- \left(\nabla_j v^s\right) R^i_{skh} \nonumber \\
&&- \left(\nabla_k v^s\right) R^i_{shj} -\left(\nabla_h v^s\right) R^i_{sjk}~.\nonumber
\end{eqnarray}

By, virtue of Bianchi's first identity the above expression simplifies to
$$ -\left[\left(\nabla_j v^s\right) R^i_{skh}+ \left(\nabla_k v^s\right) R^i_{shj}+  \left(\nabla_h v^s\right) R^i_{sjk}\right].$$

Now, equating this with the sum of the right hand sides of (\ref{2.77}), (\ref{2.78}), and (\ref{2.79}) we get
$$-v^s\left[\nabla_h R^i_{sjk}+ \nabla_j R^i_{skh}+ \nabla_k R^i_{shj}\right]=0.$$

Since this holds for arbitrary $v^s$ , so we obtain
$$  \nabla_h R^i_{sjk}+\nabla_j R^i_{skh}+\nabla_k R^i_{shj}=0.$$

Hence the proof.\\

\section{Ricci Tensor and Scalar Curvature}

~~~The curvature tensor $R^h_{ijk}$ can have three types of contractions namely of $h$ with $i, j~\mbox{or}~k.$ Now
$$R^h_{ijk}=-\left(\partial_k \Gamma^h_{ji}- \partial_j \Gamma^h_{ki}+\Gamma^h_{\rho k} \Gamma^{\rho}_{ji}-\Gamma^h_{\rho j}\Gamma^{\rho}_{ki}\right)$$
$$R^h_{hjk}=-\left(\partial_k \Gamma^h_{jh}- \partial_j \Gamma^h_{kh}+\Gamma^h_{\rho k} \Gamma^{\rho}_{jh}-\Gamma^h_{\rho j}\Gamma^{\rho}_{kh}\right).$$

But, $$\Gamma^h_{\rho k} \Gamma^{\rho}_{jh}= \Gamma^{\rho}_{jh} \Gamma^h_{\rho k} = \Gamma^{\rho}_{hj} \Gamma^h_{k \rho}= \Gamma^h_{\rho j} \Gamma^{\rho}_{kh}~~(h \rightleftharpoons \rho)$$

Hence, $$R^h_{hjk}=-\left(\partial_k \Gamma^h_{jh}- \partial_j \Gamma^h_{kh}\right)=-\left(\partial_k~\partial_j \log \sqrt{g}- \partial_j~\partial_k \log \sqrt{g}\right)=0$$

Thus, $R^h_{hjk}$ is a zero tensor.\\

Next, we consider
\begin{eqnarray}\label{2.80}
R^h_{ijh} &=& -\left(\partial_h \Gamma^h_{ji}-\partial_j \Gamma^h_{hi}-\Gamma^h_{\rho  h} \Gamma^{\rho}_{ji}-\Gamma^h_{\rho j} \Gamma^{\rho}_{hi}\right) \nonumber \\
&=& -\left[\partial_h \Gamma^h_{ji}- \partial_j \partial_i \log \sqrt{g}+\left(\partial_{\rho} \log \sqrt{g}\right)\Gamma^{\rho}_{ji}-\Gamma^h_{\rho j} \Gamma^{\rho}_{hi}\right]. 
\end{eqnarray}

This is a $(0,~2)$ tensor denoted by $R_{ij}$ and is called the {\bf Ricci} tensor or the contracted curvature tensor. Further,
$$\Gamma^h_{\rho j} \Gamma^{\rho}_{hi}=\Gamma^{\rho}_{hi} \Gamma^h_{\rho j}=\Gamma^h_{\rho i} \Gamma^{\rho}_{hj} ~~~~(h \rightleftharpoons \rho)$$

The first three terms in the right hand sides of eq.\,(\ref{2.80}) are symmetric in $i$ and $j$ and the above result shows that the fourth term is also symmetric in $i$ band $j$. Hence the Ricci tensor is symmetric in its indices \textit{i.e.}, Ricci tensor is a symmetric (0, 2) tensor.\\

Lastly considering $R^h_{ihk}$, we see that
$$R^h_{ihk}=-R^h_{ikh}=-R_{ik}$$ {\it i.e.,} no new tensor.\\

{\bf \underline{Note-I.}} ~~~~$g^{hk}~R_{hijk}=R^k_{~ijk}=R_{ij}~.$\\

{\bf \underline{II.}} ~~~~$g^{ij}~ R_{hijk}=g^{ij}~R_{ihkj}=R^j_{~hkj}=R_{hk}~.$\\

The scalar $g^{ij}R_{ij}$ is denoted by $R$ and is called the scalar curvature.\\

The tensor $g^{ki}~R_{ij}=R^k_{~j}$ is called the Ricci tensor in mixed form.\\

\section{Space of Constant Curvature}

~~~~~If in a Riemannian space
\begin{equation}\label{2.81}
R_{hijk}=k\left(g_{hj} g_{ik}-g_{hk} g_{ij}\right).
\end{equation}

where $k$ is a scalar then it can be proved that $k$ is a constant. Such a space is called a space of constant curvature.\\

A Riemannian space with $R_{hijk}=0$ is called a {\bf flat space}. A flat space is obviously a space of constant curvature.\\

\begin{center}
	\underline{-----------------------------------------------------------------------------------} 
\end{center}
\newpage
\vspace{5mm}
\begin{center}
	\underline{\bf Exercise} 
\end{center}
\vspace{3mm}
{\bf 2.1.} In Euclidean 3-space $E_3$, show that any basis $\lbrace \textit{\textbf{e}}_{\bm 1}, \textit{\textbf{e}}_{\bm 2}, \textit{\textbf{e}}_{\bm 3}\rbrace$ and its reciprocal basis $\lbrace \utilde{e^1}, \utilde{e^2}, \utilde{e^3}\rbrace$ are connected as
$$\utilde{e^1}= \frac{\textit{\textbf{e}}_{\bm 2}\times \textit{\textbf{e}}_{\bm 3}}{[\textit{\textbf{e}}_{\bm 1} ~ \textit{\textbf{e}}_{\bm 2} ~ \textit{\textbf{e}}_{\bm 3}]},~~~ \utilde{e^2}= \frac{\textit{\textbf{e}}_{\bm 3}\times \textit{\textbf{e}}_{\bm 1}}{[\textit{\textbf{e}}_{\bm 1} ~ \textit{\textbf{e}}_{\bm 2} ~ \textit{\textbf{e}}_{\bm 3}]},~~~ \utilde{e^3}= \frac{\textit{\textbf{e}}_{\bm 1}\times \textit{\textbf{e}}_{\bm 2}}{[\textit{\textbf{e}}_{\bm 1} ~ \textit{\textbf{e}}_{\bm 2} ~ \textit{\textbf{e}}_{\bm 3}]}$$
and similarly
$$\textit{\textbf{e}}_{\bm 1}= \frac{\utilde{e^2} \times \utilde{e^3}}{[\utilde{e^1} ~ \utilde{e^2} ~ \utilde{e^3}]},~~~ \textit{\textbf{e}}_{\bm 2}= \frac{\utilde{e^3}\times \utilde{e^1}}{[\utilde{e^1} ~ \utilde{e^2} ~ \utilde{e^3}]},~~~ \textit{\textbf{e}}_{\bm 3}= \frac{\utilde{e^1}\times \utilde{e^2}}{[\utilde{e^1} ~ \utilde{e^2} ~ \utilde{e^3}]}$$
Also show that the two box products are inverse of each other.\\\\
{\bf 2.2.} How many components does a tensor of rank 3 have in a space of dimension 4 ?\\\\
{\bf 2.3.} If a contravariant vector $\textit{\textbf{v}}$ has components $v^1=\lambda$, $v^k=0$, $(k=2,3,\ldots,n)$ in $x-$coordinate system then show that its components $\bar{v}^i$ in another $\bar{x}$-coordinate system are given by $$\bar{v}^i=\lambda \frac{d \bar{x}^i}{dx^1}\,.$$\\
{\bf 2.4.} Show that there is no distinction between contravariant and covariant components of a vector, when rectangular Cartesian co-ordinates are used.\\\\
{\bf 2.5.} Show that the equations of transformation of a mixed tensor possesses the group property (or equivalently the transformations of a mixed tensor is transitive).\\\\
{\bf 2.6.} Let $A^i_{jk}$ is a $(1,~2)$\,-tensor and $B^{lm}$  is a $(2,~0)$\,-tensor. Then $A^i_{jk}B^{jm}$ is a $(2,~1)$\,-tensor of the form $C^{im}_k$.~~~(Here $j$ is the dummy index and $i,~k~~\mbox{and}~~m$ are the free indices characterized by the order-type of the resulting tensor $C$).\\\\
{\bf 2.7.} Show that $A_{ij}B^{jk}$ is a $(1,~1)$ tensor while $A_{ij}B^{ji}$ is a scalar.\\\\
{\bf 2.8.} If $S^{ij}$ is a symmetric tensor and $a_{ij}$ is an alternating tensor (\textit{i.e.} skew-symmetric) then the inner product $S^{ij}$ $a_{ij}$ vanish identically.\\\\
{\bf 2.9.} Show that if the quadratic form $a_{ij}x^ix^j$ is identically zero then $a_{ij}$ is skew-symmetric.\\\\
{\bf 2.10.} Show that in an n-dimensional space the no. of independent components of a symmetric tensor $S_{ij}$ is $\dfrac{1}{2} n(n+1)$ and that of a skew-symmetric tensor $a_{ij}$ is  $\dfrac{1}{2} n(n-1)$.\\\\
{\bf 2.11} Let $A(i,j,k)$ be a 3-index system of functions of the co-ordinate variables. If for an arbitrary (1, 0)\,-tensor $B^u$,the system :
$$\sum_j A(i,~j,~k)B^j$$
is a (1, 1)\,-tensor then show that $A(i,j,k)$ is a (1, 2)\,-tensor. Also write the appropriate expression for $A$.\\\\
{\bf 2.12.} Let $A(i,j,k,)$ be a 3 index system of functions of the coordinate variables. If for an arbitrary (1,1)\,-tensor $B_v^u$, the expression
$$\sum_k A(i,~j,~k)B^k_m~,$$
is a (2, 1)\,-tensor then show that $A(i,j,k)$ is also a (2, 1)\,-tensor.\\\\
{\bf 2.13.} Let $A(i,j)$ be a 2-index system of functions of the co-ordinate variables. If for two arbitrary contravariant vectors $u^i$ and $v^i$, the expression  $A(i,j)u^iv^j$ is a scalar then show that $A(i,j)$ is a (0, 2)\,-tensor.\\\\
{\bf 2.14.} Let $A(i,j)$ be a 2-index system of functions of the co-ordinate variables. If for an arbitrary contravariant vector $u^i$, the expression
$$\sum_i\sum_j A(i,j)u^iu^j,$$
is a scalar,then show that $A(i,j)+A(i,j)$ is a $(0,~2)$\,-tensor. Further, if $A(i,j)$ is symmetric then $A(i,j)$ itself is a $(0,~2)$\,-tensor.\\\\
{\bf 2.15.} Show that Kronecker's delta is a $(1,~1)$\,-tensor by using the quotient law.\\
\textit{Hints:} Use $\delta_j^iu^j=u^i$ ,$u^i$ is an arbitrary vector.\\\\
{\bf 2.16.} If $a_{ijk}\lambda^i \lambda^j \lambda^k$ is a scalar for arbitrary contravariant vector $\lambda$, then show that\\
~~~~~~~~$a_{ijk}+a_{ikj}+a_{jki}+a_{jik}+a_{kij}+a_{kji}$ is a $(0,3)$\,-tensor.\\\\
{\bf 2.17.} Prove that the equations of transformation of a relative tensor possess the group property.\\\\
{\bf 2.18.} Show that there is no distinction between contravariant and co-variant vectors when we restrict ourselves to transformations of the type
$$x'^{\alpha}=a_{\beta}^{\alpha}x^{\beta}+b^{\alpha}$$
where $a'$s and $b'$s are constants such that  
$$\sum_{\alpha=1}^{3}a_\beta^\alpha a_{\alpha}^{\gamma}=\delta_{\beta}^{\gamma}$$\\
{\bf 2.19.} If the tensor $a_{ij}$ and $g_{ij}$ are symmetric and $u^i$ and $v^i$ are components of contravariant vectors satisfying the equation
\[
\left.
\begin{array}{l}
\left(a_{ij}-kg_{ij}\right)u^i=0 \\
\left(a_{ij}-k'g_{ij}\right)v^i=0
\end{array}
\right\}
~~(i,~j)=1,2, \ldots ,n~;~~ k\neq k'
\]
prove that, $g_{ij}u^iu^j=0$ , $a_{ij}u^iu^j=0$.\\\\
{\bf 2.20.} If $a_{mn}x^mx^n=b_{mn}x^mx^n$ for arbitrary values of $x^r$, show that $a_{(mn)}=b_{(mn)}$.\\\\
{\bf 2.21.} If $a_{hijk}\lambda^h\mu^i\lambda^j\mu^k=0$~, where $\lambda^i,~\mu^i$ are components of two arbitrary vectors, then show that
$$a_{hijk}+a_{hkji}+a_{jihk}+a_{jkhi}=0.$$\\
{\bf 2.22.} If $A^i$ is an arbitrary contravariant vector and $C_{ij}A^iA^j$ is an invariant, then show that $C_{ij}+C_{ji}$ is a covariant tensors of the $2^{nd}$ order.\\\\
{\bf 2.23.} If $a^{ij}_k \lambda_i\mu_j\gamma^k$ is a scalar invariant, $\lambda_i,~\mu_j,~\gamma^k$ are arbitrary vectors, then show that $a^{ij}_k$ is a $(2,~1)$\,-tensor.\\\\
{\bf 2.24.} If in a Riemannian space the co-ordinate curves are orthogonal to each other then the co-ordinate hypersurfaces are also orthogonal to each other and conversely.\\\\
{\bf 2.25.} Express the fundamental tensors $g_{ij}$ and $g^{ij}$ in terms of the components of the unit tangents $e^i_h$ $(h=1,2, \ldots ,n)$ to an orthogonal ennuple.\\\\
{\bf 2.26.} If $g=\left|g_{ij}\right|$ where $g_{ij}$ is a non-singular tensor and if $g^{ij}$ is its reciprocal tensor then prove that
$$\frac{\partial}{\partial x^k}\log g = g^{ji}\frac{\partial}{\partial x^k}g_{ij}=-g_{ij}\frac{\partial}{\partial x^k}g^{ji}.$$\\
{\bf 2.27.} Prove that $\sqrt{g}\,dx^1dx^2 \ldots \ldots dx^n$ is an invariant volume element.\\\\
{\bf 2.28.} Show that $\dfrac{\partial}{\partial x^k} \ln \left|\det A\right|=\mbox{Tr}A^{-1}\dfrac{\partial A}{\partial x^k}$ , for any square matrix $A$.\\\\
{\bf 2.29.} Show that the laws of transformations of Christoffel symbols possess transitive property.\\\\
{\bf 2.30.} For any scalar $\phi$, show that $\nabla _k \phi$ is a (0, 1) tensor.\\\\
{\bf 2.31.} Show that $\nabla _k A_i=\partial_k A_i-\Gamma^s_{ki} A_s$ is a (0, 2) tensor.\\\\
{\bf 2.32.} Show that $\partial_m g_{ij}=\Gamma_{mij}+\Gamma_{mji}$.\\\\
{\bf 2.33.} If $\theta_{ij}$ is the angle between the $i$-th and $j$-th co-ordinate hypersurfaces then show that 
$$\cos \theta_{ij}=\frac{g^{ij}}{\sqrt{g^{ii} g^{jj}}}\,.$$
\\
{\bf 2.34.} Prove that
$$\Gamma^s_{ms}=\partial_m \log {\sqrt{g}}=\frac{1}{\sqrt{g}}\partial_m \sqrt{g}=\frac{1}{2g}\partial_mg.$$\\
{\bf 2.35.} Show that $\delta^i_j~,~g_{ij}$ and $g^{ij}$ are covariant constants.\\\\
{\bf 2.36.} Find the commutation formulas for covariant derivatives of the tensor $A^p_q$ , $A_{pq}$ and $A^{pq}.$\\\\
{\bf 2.37.} In a Riemannian $n-$space show that $R_{ij}=\lambda g_{ij}$ implies $\lambda=\dfrac{R}{n}$.\\\\
{\bf 2.38.} If in a Riemannian space the relation $$g_{ij}R_{kl}-g_{il}R_{jk}+g_{jk}R_{il}-g_{kl}R_{ij}=0$$ holds then show that the space is Einstein.\\\\
{\bf 2.39.} Show that a space of constant curvature is an Einstein space.\\\\
{\bf 2.40.} If $R_{ij}$ is the Ricci tensor and $R^h_{~j}=g^{hi}R_{ij}$ then prove that
$$\nabla_h R^h_{~j}=\frac{1}{2} \frac{\partial R}{\partial x^j}\,.$$\\
{\bf 2.41.} Show that in an Einstein space of dimension $n>2$ , the scalar curvature $R$ is a constant.\\\\
{\bf 2.42} Show that if in a $V_3$ the coordinates can be chosen so that the components of a tensor $g_{ij}$ are zero for $i \neq j$ , then
$$(i)~~R_{hj}=\frac{1}{g_{ii}} R_{hiij}~~,~~(ii)~~R_{hh}=\frac{1}{g_{ii}}R_{hiij}+\frac{1}{g_{jj}}R_{hjjh}~.$$\\\\

\begin{center}
	\underline{\bf Solution and Hints} 
\end{center}
\vspace{3mm}
{\bf Solution 2.1:} Since the action of the dual basis on the basis of $T_P$ are given by $$\utilde{e^i}(\textit{\textbf{e}}_{\bm j})= \delta^i _j$$
So we have in 3D 
$$\utilde{e^1}(\textit{\textbf{e}}_{\bm 1})= 1,~~\utilde{e^1}(\textit{\textbf{e}}_{\bm 2})= 0,~~\utilde{e^1}(\textit{\textbf{e}}_{\bm 3})= 0$$
{\it i.e.,}~~ $\utilde{e^1}$ is orthogonal to $\textit{\textbf{e}}_{\bm 2}$ and $\textit{\textbf{e}}_{\bm 3}$. So we write
$$\utilde{e^1}= \lambda(\textit{\textbf{e}}_{\bm 2}\times \textit{\textbf{e}}_{\bm 3})$$
Now, $$1= \utilde{e^1}(\textit{\textbf{e}}_{\bm 1})= \lambda [\textit{\textbf{e}}_{\bm 1} ~ \textit{\textbf{e}}_{\bm 2} ~ \textit{\textbf{e}}_{\bm 3}]$$
\textit{i.e.}, $$\lambda = \frac{1}{[\textit{\textbf{e}}_{\bm 1} ~ \textit{\textbf{e}}_{\bm 2} ~ \textit{\textbf{e}}_{\bm 3}]}.$$
So, $$\utilde{e^1}= \frac{\textit{\textbf{e}}_{\bm 2}\times \textit{\textbf{e}}_{\bm 3}}{[\textit{\textbf{e}}_{\bm 1} ~ \textit{\textbf{e}}_{\bm 2} ~ \textit{\textbf{e}}_{\bm 3}]}$$
Similarly for $\utilde{e^2}$ and $\utilde{e^3}.$\\
Again, $$\utilde{e^1}(\textit{\textbf{e}}_{\bm 1})= 1~~~,~~~ \utilde{e^2}(\textit{\textbf{e}}_{\bm 1})= 0~~~,~~~ \utilde{e^3}(\textit{\textbf{e}}_{\bm 1})= 0~,$$
So,  $\textit{\textbf{e}}_{\bm 1}$ is orthogonal to $\textit{\textbf{e}}_{\bm 2}$ and $\textit{\textbf{e}}_{\bm 3}$, \textit{i.e.}, $\textit{\textbf{e}}_{\bm 1}$= $\mu$ ($\utilde{e^2} \times \utilde{e^3}$)\\
Hence, $$1= \utilde{e^1}(\textit{\textbf{e}}_{\bm 1})= \mu \utilde{e^1} \cdot (\utilde{e^2} \times \utilde{e^3})= \mu [\utilde{e_1} ~ \utilde{e_2} ~ \utilde{e_3}]$$
\textit{i.e.}, $$\textit{\textbf{e}}_{\bm 1}= \frac{\utilde{e^2} \times \utilde{e^3}}{[\utilde{e_1} ~ \utilde{e_2} ~ \utilde{e_3}]}$$
Now, 
\begin{eqnarray}
[\utilde{e^1} ~ \utilde{e^2} ~ \utilde{e^3}] &=& \utilde{e^1} \cdot (\utilde{e^2} \times \utilde{e^3}) \nonumber \\
&=& \frac{\textit{\textbf{e}}_{\bm 2} \times \textit{\textbf{e}}_{\bm 3}}{[\textit{\textbf{e}}_{\bm 1} ~ \textit{\textbf{e}}_{\bm 2} ~ \textit{\textbf{e}}_{\bm 3}]} \left\{ \frac{(\textit{\textbf{e}}_{\bm 3} \times \textit{\textbf{e}}_{\bm 1})\times(\textit{\textbf{e}}_{\bm 2} \times \textit{\textbf{e}}_{\bm 3})}{[\textit{\textbf{e}}_{\bm 1} ~ \textit{\textbf{e}}_{\bm 2} ~ \textit{\textbf{e}}_{\bm 3}]^2} \right\} \nonumber \\
&=& \frac{\textit{\textbf{e}}_{\bm 2} \times \textit{\textbf{e}}_{\bm 3}}{[\textit{\textbf{e}}_{\bm 1} ~ \textit{\textbf{e}}_{\bm 2} ~ \textit{\textbf{e}}_{\bm 3}]^3} \lbrace(\textit{\textbf{e}}_{\bm 3} \times \textit{\textbf{e}}_{\bm 1})\times(\textit{\textbf{e}}_{\bm 1} \times \textit{\textbf{e}}_{\bm 2})\rbrace \nonumber \\
&=& \frac{\textit{\textbf{e}}_{\bm 2} \times \textit{\textbf{e}}_{\bm 3}}{[\textit{\textbf{e}}_{\bm 1} ~ \textit{\textbf{e}}_{\bm 2} ~ \textit{\textbf{e}}_{\bm 3}]^3} \lbrace ((\textit{\textbf{e}}_{\bm 1} \times \textit{\textbf{e}}_{\bm 2}) \cdot \textit{\textbf{e}}_{\bm 3})\textit{\textbf{e}}_{\bm 1}-((\textit{\textbf{e}}_{\bm 1} \times \textit{\textbf{e}}_{\bm 2}) \cdot \textit{\textbf{e}}_{\bm 1})\textit{\textbf{e}}_{\bm 3}\rbrace \nonumber \\
&=& \frac{(\textit{\textbf{e}}_{\bm 2} \times \textit{\textbf{e}}_{\bm 3})[\textit{\textbf{e}}_{\bm 1} ~ \textit{\textbf{e}}_{\bm 2} ~ \textit{\textbf{e}}_{\bm 3}] \textit{\textbf{e}}_{\bm 1}}{[\textit{\textbf{e}}_{\bm 1} ~ \textit{\textbf{e}}_{\bm 2} ~ \textit{\textbf{e}}_{\bm 3}]^3} \nonumber \\
&=& \frac{1}{[\textit{\textbf{e}}_{\bm 1} ~ \textit{\textbf{e}}_{\bm 2} ~ \textit{\textbf{e}}_{\bm 3}]}~.~~~\mbox{(proved)} \nonumber
\end{eqnarray}\\
{\bf Solution 2.2:} Total no. of components of a tensor of rank $m$ in an $n$-dimensional space $=n^m$. Thus no. of components $=4^3=64$.\\\\
{\bf Solution 2.4:} Without any loss of generality, we choose for simplicity the space dimension to be two. Let $(X,~Y)$ and $(X',~Y')$ be two sets of rectangular Cartesian coordinate systems. So if $(x^1,~x^2)$ and $(x^{1'},~x^{2'})$ be the coordinates of a point $P$ in these two coordinate systems then we have,

\begin{equation}\label{2.14}
x^{1'}=l_1x^1+m_1x^2~~,~~x^{2'}=l_2x^1+m_2x^2
\end{equation}
or equivalently,
\begin{equation}\label{2.15}
x^{1}=l_1x^{1'}+l_2x^{2'}~~,~~x^{2}=m_1x^{1'}+m_2x^{2'}
\end{equation}
suppose $A^i$ be the contravariant components of a vector $\textit{\textbf{A}}$, then the transformation law equations (\ref{2.6}) and (\ref{2.7}) give
$$A^{i'}=\frac{\partial x^{i'}}{\partial x^k} A^k~~~~,~~~~i=1,2~~;~k=1,2$$
or explicitly,
\begin{equation}\label{2.16}
A^{1'}=\frac{\partial x^{1'}}{\partial x^1} A^1   +\frac{\partial x^{1'}}{\partial x^2} A^2=l_1A^1+m_1A^2     
\end{equation}
\begin{equation}\label{2.17}
A^{2'}=\frac{\partial x^{2'}}{\partial x^1} A^1   +\frac{\partial x^{2'}}{\partial x^2} A^2=l_2A^1+m_2A^2     
\end{equation}
Similarly, if $A_u$ be the covariant components of the vector $\textit{\textbf{A}}$, then the transformation laws equations (\ref{2.8}) and (\ref{2.9}) give
$$A_{u}^{'}=\frac{\partial x^v}{\partial x^{u'}} A_v  ~~, ~~(u,~v)=(1,~2) $$

\begin{equation}\label{2.18}
i.e.~~~~~~A_{1}^{'}=\frac{\partial x^{1'}}{\partial x^1} A_1  + \frac{\partial x^{1'}}{\partial x^2} A_2=l_1A_1+m_1A_2
\end{equation}

\begin{equation}\label{2.19}
A_{2}^{'}=\frac{\partial x^{2'}}{\partial x^1} A_1   +\frac{\partial x^{2'}}{\partial x^2} A_2=l_2A_1+m_2A_2     
\end{equation}
The transformation equations (\ref{2.16})$-$(\ref{2.19}) show that there is no distinction between covariant and contravariant components of a vector under rectangular Cartesian coordinates.\\\\
{\bf Solution 2.5:} Let $A_j^i$ and $\bar{A}_q^p$ be the components of a $(1,~1)$ tensor in some $x$ and $\bar{x}-$co-ordinate systems. So the components are related by the transformation laws
\begin{equation}\label{2.20}
\bar{A}_q^p=   \frac{\partial \bar{x}^{p}}{\partial x^i}  \frac{\partial x^j}{\partial \bar{x}^q}  A_j^i
\end{equation} 
If we choose another $\bar{\bar{x}}$-coordinate system in which the components of the $(1,~1)$-tensor are $\bar{\bar{A}}_v^u$, then $\bar{\bar{A}}_v^u$ and $\bar{A}_q^p$ are related by the transformation laws

\begin{equation}\label{2.21}
\bar{\bar{A}}_v^u= \frac{\partial \bar{\bar{x}}^{u}}{\partial \bar{x}^p} \frac{\partial \bar{x}^q}{\partial \bar{\bar{x}}^{v}} \bar{A}_q^p
\end{equation}
Now substituting $\bar{A}^p_q$ from (\ref{2.20}) into (\ref{2.21}) we have
\begin{eqnarray}\label{2.22}
\bar{\bar{A}}_v^u &=& \frac{\partial \bar{\bar{x}}^{u}}{\partial \bar{x}^p} \frac{\partial \bar{x}^q}{\partial \bar{\bar{x}}^{v}} \frac{\partial \bar{x}^{p}}{\partial x^i}  \frac{\partial x^j}{\partial \bar{x}^q}  A_j^i \nonumber \\
&=& \left(\frac{\partial \bar{\bar{x}}^{u}}{\partial \bar{x}^p}\frac{\partial \bar{x}^{p}}{\partial x^i}\right)\left(\frac{\partial \bar{x}^q}{\partial \bar{\bar{x}}^{v}}  \frac{\partial x^j}{\partial \bar{x}^q}  \right)A_j^i \nonumber \\
&=& \frac{\partial \bar{\bar{x}}^{u}}{\partial x^i}\frac{\partial x^j}{\partial \bar{\bar{x}}^{v}}A_j^i~.
\end{eqnarray}

Equation\,(\ref{2.22}) is nothing but the transformation laws of the components of the $(1,~1)$-tensor when there is a transformation of coordinate from $x$-system to $\bar{x}$-system. If the transformation equations (\ref{2.20}) and (\ref{2.21}) are denoted by $T(A)$ and $\bar{T}(A)$ then Eq.\,(\ref{2.22}) tells us
\begin{equation}\label{2.23}
\bar{\bar{T}}(A)=T(A) ~o~\bar{T}(A)
\end{equation}
\textit{i.e.}, the transformation laws follow transitive property. If the co-ordinates $x^i$ transforms to $x^i$ itself then the components of the tensor remain same and is called the identity transformation. The transformation from $\bar{x}$-co-ordinate system to $x$-co-ordinate system is the inverse of that from $x$-co-ordinate system to $\bar{x}$-co-ordinate system and equation (\ref{2.23}) by the combination give the identity transformation. So the set of all transformation equations of a tensor form a group. In fact it is an abelian group.\\\\
{\bf Solution 2.6:} From the transformation law for tensor
\begin{eqnarray}
A^i_{jk}B^{jm} &=& \frac{\partial x^i}{\partial \bar{x}^p}\frac{\partial \bar{x}^q}{\partial x^j}\frac{\partial \bar{x}^r}{\partial x^k}\frac{\partial x^j}{\partial \bar{x}^s}\frac{\partial x^m}{\partial \bar{x}^t}\bar{A}^p_{qr}\bar{B}^{st} \nonumber \\
&=& \frac{\partial x^i}{\partial \bar{x}^p}\frac{\partial x^m}{\partial \bar{x}^t}\frac{\partial \bar{x}^r}{\partial x^k}\left(\frac{\partial \bar{x}^q}{\partial x^j}\frac{\partial x^j}{\partial \bar{x}^s} \right)\bar{A}^p_{qr}\bar{B}^{st} \nonumber \\
&=& \frac{\partial x^i}{\partial \bar{x}^p}\frac{\partial x^m}{\partial \bar{x}^t}\frac{\partial \bar{x}^r}{\partial x^k}\delta_s^q\bar{A}^p_{qr}\bar{B}^{st} \nonumber \\
&=& \frac{\partial x^i}{\partial \bar{x}^p}\frac{\partial x^m}{\partial \bar{x}^t}\frac{\partial \bar{x}^r}{\partial x^k}\bar{A}^p_{qr}\bar{B}^{qt} \nonumber
\end{eqnarray}
Which clearly shows that $A_{jk}^iB^{jm}$ is a $(2,~1)$-tensor.\\\\
{\bf Solution 2.11:} Let the given components be in some $x$-co-ordinate system and let those under any other $\bar{x}$-co-ordinate system be denoted by bar signs over the main letters. Suppose summation convention is used for any repeated index.\\

Since in the product $A(i,j,k) B^j$,the free indices are $i$ and $k$ while the expression is given to be a (1, 1)-tensor, then it must be a tensor of the form $C^i_k$ or $C^k_i$.\\\\
{\bf Case-I :}~~~~Let $A(i,~j,~k)B^j=C_k^i$. Then $\bar{A}(p,~q,~r)\bar{B}^q=\bar{C}_r^p$.\\\\
But from the transformation law of tensor
\begin{eqnarray}
\bar{C}_r^p &=& \frac{\partial \bar{x}^{p}}{\partial x^{i}}\frac{\partial x^{k}}{\partial \bar{x}^{r}}C_k^i \nonumber \\
\mbox{or,}~~~~\bar{A}(p,~q,~r)\bar{B}^q &=& \frac{\partial \bar{x}^{p}}{\partial x^{i}}\frac{\partial x^{k}}{\partial \bar{x}^{r}}A(i,~j,~k)B^j \nonumber
\end{eqnarray}
$$~~~~~~~~~~~~~~~~~~~~~~~~~~~~~~~~~~~~~~~~~~~~~~~~~~~~~~~~~~~~~~~~~= \frac{\partial \bar{x}^{p}}{\partial x^{i}}\frac{\partial x^{k}}{\partial \bar{x}^{r}}A(i,~j,~k)\frac{\partial x^{j}}{\partial \bar{x}^{q}}\bar{B}^q~~~~~~\mbox{(as}~B~\mbox{is~(1,~0)~tensor)}$$
$$\mbox{or,}~~~~ \left[\bar{A}(p,~q,~r) -\frac{\partial \bar{x}^{p}}{\partial x^{i}}\frac{\partial x^{k}}{\partial \bar{x}^{r}}\frac{\partial x^{kj}}{\partial \bar{x}^{q}}A(i,~j,~k)\right]\bar{B}^q = 0~.$$
Since it holds for arbitrary $B^u$ and hence for arbitrary $\bar{B}^q$, so we have
$$\bar{A}(p,~q,~r)=\frac{\partial \bar{x}^{p}}{\partial x^{i}}\frac{\partial x^{k}}{\partial \bar{x}^{r}}\frac{\partial x^{j}}{\partial \bar{x}^{q}}A(i,~j,~k)$$
This shows that $A(i,~j,~k)$ is a $(1,~2)$-tensor and its appropriate form is $A^i_{~jk}$.\\\\
{\bf Case-II :}~~~~If $A(i,~j,~k)B^j=C_i^k$, then $A(i,~j,~k)$ is again a $(1,~2)$ tensor , but its appropriate form will be $A^k_{ij}$.\\\\
{\bf Hints 2.15:} Use $\delta_j^iu^j=u^i$ ,$u^i$ is an arbitrary vector.\\\\
{\bf Solution 2.16:} As $a_{ijk}\lambda^i \lambda^j \lambda^k$ is a scalar so,$$a_{ijk}\lambda^i \lambda^j \lambda^k=\bar{a}_{pqr}\bar{\lambda}^p\bar{\lambda}^q\bar{\lambda}^r=\bar{a}_{pqr}
\frac{\partial \bar{x}^{p}}{\partial x^{i}}\frac{\partial \bar{x}^{q}}{\partial x^{j}}\frac{\partial \bar{x}^{r}}{\partial x^{k}}\lambda^i \lambda^j \lambda^k$$
$$\mbox{or,} ~\left(a_{ijk}-\bar{a}_{pqr}
\frac{\partial \bar{x}^{p}}{\partial x^{i}}\frac{\partial \bar{x}^{q}}{\partial x^{j}}\frac{\partial \bar{x}^{r}}{\partial x^{k}}\right)\lambda^i \lambda^j \lambda^k=0~~i.e.,~~C_{ijk}\lambda^i \lambda^j \lambda^k=0$$
Since this holds for arbitrary contravariant vector $\lambda$, so we have
$$C_{ijk}+C_{ikj}+C_{jki}+C_{jik}+C_{kij}+C_{kji}=0.$$
Now putting the values of the $C$-system and adjusting the dummy indices we get,
$$a_{ijk}+a_{ikj}+a_{jki}+a_{jik}+a_{kij}+a_{kji}=\frac{\partial \bar{x}^{p}}{\partial x^{i}}\frac{\partial \bar{x}^{q}}{\partial x^{j}}\frac{\partial \bar{x}^{r}}{\partial x^{k}}(\bar{a}_{pqr}+\bar{a}_{prq}+\bar{a}_{qrp}+\bar{a}_{qpr}+\bar{a}_{rpq}+\bar{a}_{rqp}).$$
Here  $a_{ijk}+a_{ikj}+a_{jki}+a_{jik}+a_{kij}+a_{kji}$  is a (0, 3)-tensor.\\\\
{\bf Solution 2.18:}~~~~From the transformation law
$$x'^{\alpha}=a_{\beta}^{\alpha}x^{\beta}+b^{\alpha}$$
we have $$\frac{\partial x^{'\alpha}}{\partial x^{\delta}}=a^{\alpha}_{\delta}$$
Again from the transformation law
$$\sum_{\alpha=1}^{3}a_\alpha^\gamma x'^\alpha=\sum_{\alpha=1}^{3}
a_\alpha^\gamma a_\beta^\alpha x^\beta+\sum_{\alpha=1}^{3}a_\alpha^\gamma b^{\alpha}
=\delta_\beta^{\gamma}x^\beta+\sum_{\alpha=1}^{3}a_\alpha^\gamma b^{\alpha}$$
$$\Rightarrow x^\gamma=\sum_{\alpha=1}^{3}a_\alpha^\gamma x'^\alpha - \sum_{\alpha=1}^{3}a_\alpha^\gamma b^\alpha$$
$$\mbox{so,}~~\frac{\partial x^{\gamma}}{\partial x^{'m}}=a_m^{\gamma}~~,~i.e.,~~~~\frac{\partial x^{\delta}}{\partial x^{'\gamma}}= a^{\delta}\gamma$$
$$\mbox{Now,}~~A^{'\alpha}=\frac{\partial x^{'\alpha}}{\partial x^{\delta}}A^{\delta}=a^{\alpha}_{\delta}A^{\delta}$$
$$\mbox{Similarly,}~~A'_{\gamma}=\frac{\partial x^{\delta}}{\partial x^{'\gamma}}A_{\delta}=a_{\gamma}^{\delta}A_{\delta} $$
Hence the contravariant and covariant components transform in the same way.\\\\
{\bf Solution 2.19:}~~~~We have
$$(a_{ij}-kg{ij})u^i=0$$
$$\mbox{and}~~~~(a_{ij}-k'g{ij})v^i=0~~~~~$$
Now multiply the first equation by $v^j$ and second one by $u^j$ and then subtracting we have
$$a_{ij}u^iv^j-a_{ij}v^iu^j-kg_{ij}u^iv^j+k'g_{ij}v^iu^j=0$$
Now, interchanging $i$ and $j$ in the second and forth term and noting that
$$a_{ij}=a_{ji}~~\mbox{and}~~g_{ij}=g_{ji}.$$
We have,
$$(k'-k)g_{ij}u^iv^j=0$$
$$i.e.~~~~g_{ij}u^iv^j=0.$$
Now multiplying the first equation by $v^j$ and using this result we obtain,
$$a_{ij}u^iv^j=0\,.$$\\
{\bf Solution 2.20:}~~~~As $a_{mn}x^mx^n=b_{mn}x^mx^n$, so we have
$$A=(a_{mn}-b{mn})x^mx^n=0.$$
So, $$\frac{\partial{A}}{\partial{x^i}}=(a_{in}-b_{in})x^n+(a_{mi}-b_{mi})x^m$$
and so, $$\frac{\partial^2{A}}{\partial{x^i}\partial{x^j}}=(a_{ij}-b_{ij})+(a_{ji}-b_{ji})=0$$
$$\Rightarrow a_{ij}+a_{ji}=b_{ij}+b_{ji}$$
$$i.e.~~~~a_{(ij)}=b_{(ij)}.$$\\
{\bf Solution 2.21:}~~~~Let $$A=a_{hijk}\lambda^h\mu^i\lambda^j\mu^k=0$$
$$\frac{\partial{A}}{\partial{\lambda^h}}=a_{hijk}\mu^i\lambda^j\mu^k+a_{pihk}\lambda^p\mu^i\mu^k=0$$
$$\frac{\partial^2 {A}}{\partial{\lambda^{\mu}}\partial\lambda^{j}}=a_{hijk}\mu^i\mu^k+a_{jihk}\mu^i\mu^k=0$$ 
$$\frac{\partial^3 {A}}{\partial{\lambda^{\mu}}\partial\lambda^{j}\partial\mu^{i}}=
a_{hijk}\mu^k+a_{hkji}\mu^k+ a_{jihk}\mu^k+a_{jkhi}\mu^k=0$$   
$$\frac{\partial^4 {A}}{\partial{\lambda^{\mu}}\partial\lambda^{j}\partial\mu^{i}\partial\mu^{k}}=
a_{hijk}+a_{hkji} +a_{jihk}+a_{jkhi}=0$$
Hence the result.\\\\
{\bf Solution 2.22:}~~~~As $C_{ij}A^iA^j$ is an invariant for arbitrary contravariant vector $A^i$, so
$$C_{ij}A^iA^j=C_{ij}^{'}A^{'i}B^{'j}$$
using tensor law of transformation
$$C_{ij}A^iA^j=C^{'}_{ij}\frac{\partial x^{'i}}{\partial x^{\alpha}}A^{\alpha}\frac{\partial x^{'j}}{\partial x^{\beta}}A^{\beta}.$$
Now interchanging the suffixes $i$ and $j$
$$C_{ji}A^jA^i=C^{'}_{ji}\frac{\partial x^{'j}}{\partial x^{\alpha}}\frac{\partial x^{'i}}{\partial x^{\beta}}A^{\alpha}A^{\beta}=C^{'}_{ji}\frac{\partial x^{'i}}{\partial x^{\alpha}}\frac{\partial x^{'j}}{\partial x^{\beta}}A^{\alpha}A^{\beta}$$
(interchanging the dummy suffixes $\alpha$ and $\beta$).\\\\
Thus,
$$\left(C_{ji}+C_{ij}\right)A^iA^j=\left(C'_{ji}+C'_{ij}\right)\frac{\partial x^{'i}}{\partial x^{\alpha}}\frac{\partial x^{'j}}{\partial x^{\beta}}A^{\alpha}A^{\beta}$$   
$$\Rightarrow \left(C_{\alpha \beta}+C_{\beta \alpha}\right)A^{\alpha}A^{\beta}=\left(C'_{ji}+C'_{ij}\right)\frac{\partial x^{'i}}{\partial x^{\alpha}}\frac{\partial x^{'j}}{\partial x^{\beta}}A^{\alpha}A^{\beta}$$
$$\Rightarrow \left[\left(C_{\alpha \beta}+C_{\beta \alpha}\right)-\left(C'_{ij}+C'_{ji}\right)\frac{\partial x^{'i}}{\partial x^{\alpha}}\frac{\partial x^{'j}}{\partial x^{\beta}}\right]A^{\alpha}A^{\beta} = 0$$
As $A^\alpha$ is arbitrary so the expression within the square bracket vanishes. Hence $C_{\alpha\beta}+C_{\beta\alpha}$ is a $(0,~2)$-tensor.\\\\
{\bf Solution 2.24:} The vectors $\lbrace \stackrel{\rightarrow}{e_1},\stackrel{\rightarrow}{e_2}, \ldots \ldots ,\stackrel{\rightarrow}{e_n} \rbrace$, the natural basis in some $x-$co-ordinate system are respectively tangential to the 1st, 2nd,$\ldots n\mbox{-th}$ co-ordinate curves. We first suppose that the co-ordinate curves are orthogonal to each other $i.e,~~ \stackrel{\rightarrow}{e_i} \cdot \stackrel{\rightarrow}{e_j}=0~~(i\neq j)$. As $g_{ij} = \stackrel{\rightarrow}{e_i} \cdot \stackrel{\rightarrow}{e_j}$ , so we have $g_{ij}=0~~(i\neq j)$. Thus $g_{ij}$ as a square matrix is diagonal. Since $g_{ij}$ is non-singular so all the diagonal elements are nonzero $i.e.,~~ g_{ij} \neq 0$ if $i\neq j$. So similar result will hold for reciprocal metric tensor $g^{ij}$ $i.e.,$ $g^{ij}=0$ if $i \neq j~~; \neq 0$ if $i=j$. But $g^{ij}=\utilde{e^i} \cdot \utilde{e^j}$ and hence $\utilde{e^i} \cdot \utilde{e^j}=0$~~$\mbox{for}~~i \neq j$. As $\utilde{e^i}$ is normal to the $i^{th}$ co-ordinate hypersurface so the co-ordinate hypersurfaces are orthogonal to each other.\\

A family of curves such that through each point of $V_n$ only one of the curves of the family passes, is called a congruence of curves. A congruence of curves is defined by a vector field. An orthogonal ennuple in a Riemannian space of dim $n$ is a set of $n$ mutually orthogonal congruences of curves.\\\\
{\bf Solution 2.25:} Let $e^i_\alpha~(\alpha=1,2, \ldots ,n)$ be the $n$ unit tangent vectors to an orthogonal ennuple in $V_n$. As these congruences are orthogonal to each other so
$$g_{ij}e^i_\alpha e^j_\beta=\delta_{\alpha\beta}$$
If we define, $$e^i_\alpha=\frac{\mbox{co-factor~of}~ e_{(\alpha)^i}~\mbox{in~determinant}~\left|e_{(\alpha)^i}\right|}{\left|e_{(\alpha)^i}\right|}~,$$
then from the property of determinant
\begin{equation}\label{2.33}
\sum_{i=1}^n e_{(\alpha)j} e^i_(\alpha)=\delta^i_j~.
\end{equation}
Now multiply both side by $g^{jk}$ we get\\
$$\sum_{i=1}^n e^i_{(\alpha)} e_{(\alpha)j} g^{jk}=\delta^i_j g^{jk}$$
\begin{equation}\label{2.34}
i.e,~~~~\sum_{i=1}^n e^i_\alpha e^k_\alpha=g^{ik}~.
\end{equation}
Again multiplying equation(\ref{2.33}) by $g_{ik}$ we have
$$\sum_{i=1}^n e^i_{(\alpha)} e_{(\alpha)j} g_{ik}=\delta^i_j g_{ik}$$
\begin{equation}\label{2.35}
i.e,~~~~ \sum_{i=1}^n e_{(\alpha)k} e_{(\alpha)j}=g_{jk}~.
\end{equation}
Hence equation (\ref{2.34}) and (\ref{2.35}) gives expression for $g^{ij}$ and $g_{ij}$ respectively.\\\\
{\bf Solution 2.26:} By the property of determinant\,(see appendix II)
$$\frac{\partial g}{\partial x^k}=G^{ji}\frac{\partial}{\partial x^k}g_{ij}~~,~~G^{ji}=\mbox{cofactor~of}~g_{ij}~\mbox{in}~g.$$
But $$g^{ji}=\frac{G^{ji}}{g}~~~i.e,~~G^{ji}=g g^{ji}.$$
So, $$\frac{\partial g}{\partial x^k}=g g^{ji}\frac{\partial}{\partial x^k}g_{ij}$$
$$\mbox{or,}~~ \frac{1}{g} \frac{\partial g}{\partial x^k}=g^{ji}\frac{\partial}{\partial x^k}g_{ij}$$
$$\mbox{or,}~~ \frac{\partial}{\partial x^k}\log g=g^{ji}\frac{\partial}{\partial x^k}g_{ij}.$$
Again from the property of the reciprocal tensor
$$g^{ji} g_{ik}=\delta^j_k$$
$$i.e,~~ g^{ji} g_{ij}=\delta^j_j=n~~,~~n=\mbox{dimension of the space.}$$
$$i.e,~~ g^{ji} \frac{\partial}{\partial x^k} g_{ij}+g_{ij} \frac{\partial}{\partial x^k} g^{ji}=0$$
$$i.e,~~ g^{ji} \frac{\partial}{\partial x^k} g_{ij}=-g_{ij} \frac{\partial}{\partial x^k} g^{ji}$$
Thus, $$\frac{\partial}{\partial x^k}log g=-g_{ij} \frac{\partial}{\partial x^k} g^{ji}.$$
Hence the result.\\\\
{\bf Solution 2.27:} As $g_{ij}$ is a symmetric tensor of rank 2 so its transformation law gives
$${g'}_{ij}=\frac{\partial x^p}{\partial x^{'i}}\frac{\partial x^q}{\partial x^{'j}}g_{pq}~.$$
Taking determinant of both sides we have
$$\left|g'_{ij}\right|=\left|g_{\alpha \beta}\right|\left|\frac{\partial x^\alpha}{\partial x^{'i}}\right|\left|\frac{\partial x^\beta}{\partial x^{'j}}\right|$$
$$i.e,~~ g'=gJ^2~~,~~J=\left|\frac{\partial x^\alpha}{\partial x^{'i}}\right|~~\mbox{is the Jacobian of transformation.}$$
\begin{eqnarray}
\mbox{But}~~~~dx^1 dx^2 \ldots \ldots dx^n &=& \left|J\right|dx'^1 dx'^2 \ldots \ldots dx'^n \nonumber \\
&=& \sqrt{\frac{g'}{g}}dx'^1 dx'^2 \ldots \ldots dx'^n \nonumber \\
\Rightarrow \sqrt{g}dx^1 dx^2 \ldots \ldots dx^n &=& \sqrt{g'}dx'^1 dx'^2 \ldots \ldots dx'^n~. \nonumber
\end{eqnarray}
Hence~ $dV=\sqrt{g}dx^1 dx^2 \ldots \ldots dx^n$ is an invariant volume element.\\\\
{\bf Proof 2.28:} Let $A$ be any square matrix having determinant $\left|\det A\right|$. We consider a variation of the elements of the matrix.Then
\begin{eqnarray}
\delta \ln \left|\det A\right| &=& \ln \left|\det (A+\delta A)\right| - \ln \left|\det A\right| \nonumber \\
&=& \ln \left|\frac{\det (A+\delta A)}{\det A}\right|= \ln \left|\det A^{-1}(A+\delta A)\right| \nonumber \\
&=& \ln \left|\det (I+A^{-1}\delta A)\right|. \nonumber
\end{eqnarray}
Now, $\det (I+\delta)=1+\mbox{Tr}(\delta)+O(\delta^2)$ , $\delta $ is a small square metric of same order as $A$.\\
Thus
\begin{eqnarray}
\delta \ln \left|\det A\right| &=& \ln (1+\mbox{Tr}(A^{-1})\delta A) \nonumber \\
&=& \mbox{Tr}(A^{-1})\delta A \nonumber \\
\mbox{so,}~~ \frac{\partial}{\partial x^K} \ln \left|\det A\right| &=& Tr(A^{-1})\frac{\partial A}{\partial x^K}~. \nonumber
\end{eqnarray}
In particular, if we choose $A=g_{ij}$ , the metric tensor, then
$$\frac{\partial}{\partial x^K}\ln g=\frac{G^{ij}}{g}\frac{\partial g_{ij}}{\partial x^K}~,~~G^{ij}=~\mbox{Cofactor~of}~g_{ij}~\mbox{in}~g =g \cdot g^{ji}$$
$$\frac{1}{g}\frac{\partial g}{\partial x^K}=\frac{g.g^{ji}}{g}\frac{\partial g_{ij}}{\partial x^K}=g^{ji}\frac{\partial g_{ij}}{\partial x^K}~.$$\\
{\bf Solution 2.29:} Let us consider a co-ordinate transformations:
$$x^i\rightarrow\overline{x}^i\rightarrow\overline{\overline{x}}^i$$
$$i.e.,~~~~ \overline{x}^i=\overline{x}^i(x^k)~\mbox{and}~\overline{\overline{x}}^i=\overline{\overline{x}}^i(\overline{x}^k)$$
Let the Christoffel symbols in these co-ordinate systems be 
$\Gamma^i_{jk}~,~\overline{\Gamma}^i_{jk}~\mbox{and}~\overline{\overline{\Gamma}}^i_{jk}$ respectively.\\

For the first set of co-ordinate transformation $(i.e.,~~x^i\longrightarrow\overline{x}^i)$ the transformation of the Christoffel symbols are given by equation (\ref{2.53}) as
\begin{equation}\label{2.56}
\overline{\Gamma}^k_{ij}=\Gamma^c_{ab} \frac{\partial x^a}{\partial \overline{x}^i} \frac{\partial x^b}{\partial \overline{x}^j} \frac{\partial \overline{x}^k}{\partial x^c}+\frac{\partial^2 x^c}{\partial \overline{x}^i \partial \overline{x}^j} \frac{\partial \overline{x}^k}{\partial x^c}~.
\end{equation}
Similarly, corresponding to the second set of co-ordinate transformation 
\begin{equation}\label{2.57}
\overline{\overline{\Gamma}}^r_{pq}=\overline{\Gamma}^k_{ij} \frac{\partial \overline{x}^i}{\partial \overline{\overline{x}}^p} \frac{\partial \overline{x}^j} {\partial \overline{\overline{x}}^q} \frac{{\partial \overline{\overline{x}}^r}}{\partial \overline{x}^k}+\frac{\partial^2 \overline{x}^k}{\partial \overline{\overline{x}}^p \partial \overline{\overline{x}}^q}.\frac{\partial \overline{\overline{x}}^r}{\partial \overline{x}^k}~.
\end{equation}
Now combining equation (\ref{2.56}) and (\ref{2.57}) we have
\begin{eqnarray}\label{2.58}
\overline{\overline{\Gamma}}^r_{pq} &=& \Gamma^c_{ab} \frac{\partial x^a}{\partial \overline{x}^i} \frac{\partial x^b}{\partial \overline{x}^j} \frac{\partial \overline{x}^k}{\partial x^c} \frac{\partial \overline{x}^i}{\partial \overline{\overline{x}}^p} \frac{\partial \overline{x}^j} {\partial \overline{\overline{x}}^q} \frac{{\partial \overline{\overline{x}}^r}}{\partial \overline{x}^k}+\frac{\partial^2 x^c}{\partial \overline{x}^i \partial \overline{x}^j}\frac{\partial \overline{x}^i}{\partial \overline{\overline{x}}^p} \frac{\partial \overline{x}^j} {\partial \overline{\overline{x}}^q}\frac{\partial \overline{x}^k}{\partial x^c} \frac{\partial \overline{\overline{x}}^r}{\partial \overline{x}^k}+\frac{\partial^2 \overline{x}^k}{\partial \overline{\overline{x}}^p \partial \overline{\overline{x}}^q}\frac{\partial \overline{\overline{x}}^r}{\partial \overline{x}^k} \nonumber \\
&=& \Gamma^c_{ab} \frac{\partial x^a}{\partial \overline{\overline{x}}^p}\frac{\partial x^b}{\partial \overline{\overline{x}}^q}\frac{\partial \overline{\overline{x}}^r}{\partial x^c}+\frac{\partial^2 \overline{x}^k}{\partial \overline{\overline{x}}^p \partial \overline{\overline{x}}^q}\frac{\partial \overline{\overline{x}}^r}{\partial \overline{x}^k}+\frac{\partial^2 x^c}{\partial \overline{x}^i \partial \overline{x}^j}\frac{\partial \overline{\overline{x}}^r}{\partial \overline{x}^c}\frac{\partial \overline{x}^i}{\partial \overline{\overline{x}}^p} \frac{\partial \overline{x}^j} {\partial \overline{\overline{x}}^q}~.
\end{eqnarray}
From the chain rule of differentiation:
$$\frac{\partial x^c}{\partial \overline{x}^i} \cdot \frac{\partial \overline{x}^i}{\partial \overline{\overline{x}}^p}=\frac{\partial x^c}{\partial \overline{\overline{x}}^p}~,$$
differentiating both sides w.r.t. $\overline{\overline{x}}^q$, we get
$$\frac{\partial^2 x^c}{\partial \overline{x}^i \partial \overline{x}^j} \frac{\partial \overline{x}^i}{\partial \overline{\overline{x}}^p} \frac{\partial \overline{x}^j} {\partial \overline{\overline{x}}^q}+\frac{\partial x^c}{\partial\overline{x}^i}\frac{\partial^2 \overline{x}^i}{\partial \overline{\overline{x}}^p \partial \overline{\overline{x}}^q}=\frac{\partial^2 \overline{x}^c}{\partial \overline{\overline{x}}^p \partial \overline{\overline{x}}^q}$$
Now multiplying  both sides by $\frac{\partial \overline{\overline{x}}^r}{\partial x^c}$ we get
$$\frac{\partial^2 x^c}{\partial \overline{x}^i \partial \overline{x}^j}\frac{\partial \overline{x}^i}{\partial \overline{\overline{x}}^p} \frac{\partial \overline{x}^j} {\partial \overline{\overline{x}}^q}\frac{\partial \overline{\overline{x}}^r}{\partial \overline{x}^c}+\frac{\partial^2 \overline{x}^i}{\partial \overline{\overline{x}}^p \partial \overline{\overline{x}}^q}
\frac{\partial x^c}{\partial\overline{x}^i}\frac{\partial \overline{\overline{x}}^r}{\partial x^c}=\frac{\partial^2 x^c}{\partial \overline{\overline{x}}^p \partial \overline{\overline{x}}^q}\frac{\partial \overline{\overline{x}}^r}{\partial x^c}$$
$$\mbox{or,}~~~~ \frac{\partial^2 x^c}{\partial \overline{x}^i \partial \overline{x}^j}\frac{\partial \overline{x}^i}{\partial \overline{\overline{x}}^p} \frac{\partial \overline{x}^j} {\partial \overline{\overline{x}}^q}\frac{\partial \overline{\overline{x}}^r}{\partial x^c}+\frac{\partial^2 \overline{x}^k}{\partial \overline{\overline{x}}^p \partial \overline{\overline{x}}^q}
\frac{\partial \overline{\overline{x}}^r}{\partial \overline{x}^k}=\frac{\partial^2 x^c}{\partial \overline{\overline{x}}^p \partial \overline{\overline{x}}^q}\frac{\partial \overline{\overline{x}}^r}{\partial x^c}~.$$
Using this relation in equation (\ref{2.58}) we have
$$\overline{\overline{\Gamma}}^r_{pq} =\Gamma^c_{ab}\frac{\partial x^a}{\partial \overline{\overline{x}}^p}\frac{\partial x^b}{\partial\overline{\overline{x}}^q}\frac{\partial\overline{\overline{x}}^r}{\partial x^c}+\frac{\partial^2 x^c}{\partial \overline{\overline{x}}^p \partial \overline{\overline{x}}^q}\frac{\partial \overline{\overline{x}}^r}{\partial x^c}$$
Hence the transformation law for Christoffel symbols possesses transitive property.\\\\
{\bf Solution 2.30:} $$\overline{\nabla}_m \phi=\overline{\partial}_m \phi=\frac{\partial \phi}{\partial \overline{x}_m}=\frac{\partial \phi}{\partial x^k} \cdot \frac{\partial x^k}{\partial \overline{x}_m}=\frac{\partial x^k}{\partial \overline{x}^m}(\partial_k \phi)$$
$$~~~~~~~~~~=\frac{\partial x^k}{\partial \overline{x}^m}(\nabla _k \phi)~.$$
This shows that $\nabla _k \phi$ is a (0,1) tensor.\\\\
{\bf Solution 2.31:} $$\overline{A}_p=\frac{\partial x^i}{\partial \overline{x}^p} A_i~.$$
Then,
\begin{eqnarray}
\frac{\partial \overline{A}_p}{\partial \overline{x}^q} &=& \frac{\partial}{\partial \overline{x}^q}\left(\frac{\partial x^i}{\partial \overline{x}^p}A_i\right) \nonumber \\
&=& \frac{\partial^2 x^i}{\partial \overline{x}^q \partial \overline{x}^p}A_i+\frac{\partial x^i}{\partial \overline{x}^p}\frac{\partial A_i}{\partial \overline{x}^q} \nonumber
\end{eqnarray}
$$=\left(\overline{\Gamma}^s_{qp} \frac{\partial x^i}{\partial \overline{x}^s}-\frac{\partial x^i}{\partial \overline{x}^q}\frac{\partial x^k}{\partial \overline{x}^p}\Gamma^i_{jk}\right)A_i+\frac{\partial x^i}{\partial \overline{x}^p}\frac{\partial x^j}{\partial \overline{x}^q}\frac{\partial A_i}{\partial x^j}~~~~\mbox{(using~equation~(\ref{2.55}))}$$
\begin{eqnarray}
= \overline{\Gamma}^s_{qp} \overline{A}_s-\frac{\partial x^j}{\partial \overline{x}^q}\frac{\partial x^i}{\partial \overline{x}^p}\Gamma^s_{ji}A_s+\frac{\partial x^i}{\partial \overline{x}^p}\frac{\partial x^j}{\partial \overline{x}^q}\frac{\partial A_i}{\partial x^j} \nonumber
\end{eqnarray}
$$\mbox{or,}~~~~ \frac{\partial \overline{A}_p}{\partial \overline{x}^q} - \overline{\Gamma}^s_{qp} \overline{A}_s = \frac{\partial x^i}{\partial \overline{x}^p}\frac{\partial x^j}{\partial \overline{x}^q}\left(\frac{\partial A_i}{\partial x^j}-\Gamma^s_{ji}A_s\right)$$
$$\mbox{or,}~~~~ \overline{\nabla}_q \overline{A}_p=\frac{\partial x^i}{\partial \overline{x}^p}\frac{\partial x^j}{\partial \overline{x}^q} \nabla _j A_i~.$$
This shows that $\nabla _j A_i$ is a (0, 2) tensor.\\\\
In a similar way  it can be shown that $\nabla _k A^{i_1,i_2,\ldots , i_r}_{j_1,j_2,\ldots , j_s}$ is a $(r,~s+1)$ tensor.\\\\
\underline{\bf Note:} Covariant differentiation, increases the covariant order of a tensor by one.\\\\
{\bf Solution 2.32:} By the definition of Christoffel symbol
$$\Gamma_{mij}=\frac{1}{2}\left(\frac{\partial g_{ij}}{\partial x^m}+\frac{\partial g_{mj}}{\partial x^i}-\frac{\partial g_{mi}}{\partial x^j}\right)$$
and,
$$\Gamma_{mij}=\frac{1}{2}\left(\frac{\partial g_{ij}}{\partial x^m}+\frac{\partial g_{im}}{\partial x^j}-\frac{\partial g_{mj}}{\partial x^i}\right).$$
Adding and noting the symmetry of $g_{ij}$ we get
$$\partial_m g_{ij}=\Gamma_{mij}+\Gamma_{mji}~.$$
\\\\
{\bf Solution 2.33:} The angle between any two hypersurfaces is equal to the angle between their normals. Now $\underline{e}^i$ and $\underline{e}^j$ are normals $i$-th  and $j$-th co-ordinate hypersurfaces respectively, where $\lbrace \underline{e}^1,\underline{e}^2, \ldots \ldots , \underline{e}^n \rbrace$ is the reciprocal natural basis of the given co-ordinate system.
$$\cos \theta_{ij}=\frac{(\underline{e}^i \cdot \underline{e}^j)}{\sqrt{\underline{e}^i \cdot \underline{e}^i}\sqrt{\underline{e}^j \cdot \underline{e}^j}}~.$$
But we know that $\underline{e}^i \cdot \underline{e}^j=g^{ij}$ , hence
$$\cos \theta_{ij}=\frac{g^{ij}}{\sqrt{g^{ii}}\sqrt{g^{jj}}}.$$
\\\\
\underline{\bf Note:} The co-ordinate hypersurfaces are orthogonal to each other \textbf{iff} $g^{ij}=0$ whenever $i\neq j$.\\\\
{\bf Solution 2.34:} $$g= \det g_{ij}~.$$
Let $G^{ij}=$ cofactor of $g_{ji}$ in $g$ , then $g^{ij}=\dfrac{G^{ij}}{g}$ .\\
We know that
\begin{eqnarray}
\partial_m g &=& G^{ij}\partial_mg_{ji}=g g^{ij} \partial_m g_{ji} \nonumber \\
&=& gg^{ij}\left(\Gamma_{mji}+\Gamma_{mij}\right)=g\left(\Gamma^i_{im}+\Gamma^i_{mi}\right) \nonumber \\
&=& 2g\Gamma^s_{ms} \nonumber \\
\Gamma^s_{ms}=\frac{1}{2g}\partial_m g &=& \frac{1}{2}\partial_m \log g=\partial_m \log {\sqrt{g}}=\frac{1}{\sqrt{g}}\partial_m (\sqrt{g})~. \nonumber
\end{eqnarray}
\\
{\bf Solution 2.35:}
\begin{eqnarray}
\nabla _k \delta^i_j &=& \partial_k \delta^i_j+\Gamma^i_{ks}\delta^s_j-\Gamma^u_{kj}\delta^i_u \nonumber \\
&=& 0+\Gamma^i_{kj}- \Gamma^i_{kj}=0 \nonumber \\
\nabla _k g_{ij} &=& \partial_k g_{ij}-\Gamma^s_{ki}g_{sj}-\Gamma^u_{kj}g_{iu} \nonumber \\
&=& \partial_k g_{ij}-\left(\Gamma_{kij}+\Gamma_{kji}\right)=\partial_k g_{ij}-\partial_k g_{ij}=0. \nonumber
\end{eqnarray}
Also as $g^{ij} g_{jp}=\delta^i_p$ , so $\nabla _k\left(g^{ij} g_{jp}\right)=0$
$$\Rightarrow(\nabla _k g^{ij})g_{jp}+g^{ij}(\nabla _k g_{jp})=0~~~~\Rightarrow \nabla _k g^{ij}=0.$$\\
{\bf Solution 2.36:}~~~~We know that
$$\nabla_j A^p_q=\partial_j A^p_q+\Gamma^p_{sj} A^s_q-\Gamma^s_{qj} A^p_s$$
\begin{eqnarray}
\nabla_k \nabla_j A^p_q &=& \partial_k \left(\nabla_j A^p_q \right)-\Gamma^{\sigma}_{jk}\nabla_{\sigma}A^p_q+ \Gamma ^p_{\sigma k}\nabla _j A^{\sigma}_q-\Gamma^{\sigma}_{qk}\nabla_j A^p_{\sigma} \nonumber \\
&=& \partial_k \partial_j A^P_q + \left(\partial _k \Gamma ^p_{sj} \right)A^s_q + \Gamma ^p_{sj} \left(\partial _kA^s_q \right)- \left(\partial_k \Gamma ^s_{qj} \right) A^p_s \nonumber \\
&-& \Gamma^s_{qj}\partial_kA^p_s-\Gamma^{\sigma}_{jk}\left(\partial_{\sigma}A^p_q+\Gamma^p_{s \sigma}A^s_q-\Gamma^s_{q \sigma} A^p_s \right) \nonumber \\
&+& \Gamma^p_{\sigma k}\left(\partial_j A^{\sigma}_q+\Gamma^{\sigma}_{sj}A^s _q-\Gamma^s_{qj}A^{\sigma}_s \right) \nonumber \\
&-& \Gamma^{\sigma}_{qk} \left(\partial_j A^p_{\sigma}+\Gamma^p_{sj}A^s_{\sigma}-\Gamma^s_{\sigma j}A^p_s \right). \nonumber
\end{eqnarray}
Commuting on $k$ and $j$ we get
\begin{eqnarray}\label{2.67}
(\nabla_k \nabla_j-\nabla_j \nabla_k)A^p_q &=& \left(\partial_k \Gamma^p_{js}-\partial_j \Gamma^p_{ks}+\Gamma^p_{\sigma k}\Gamma^{\sigma}_{js}-\Gamma^p_{\sigma j}\Gamma^{\sigma}_{ks}\right)A^s_q \nonumber\\
&-& \left(\partial_k \Gamma^s_{jq}-\partial_j\Gamma^s_{kq}+\Gamma^s_{\sigma k}\Gamma^{\sigma}_{jq}-\Gamma^s_{\sigma j}\Gamma^{\sigma}_{kq}\right)A^p_s \nonumber\\
&=& -R^p_{sjk} A^s_q+R^s_{qjk}A^p_s.
\end{eqnarray}
In a similar way
\begin{equation}\label{2.68}
[\nabla_k,\nabla_j]A_{pq}=A_{sq}R^s_{pjk}+A_{ps}R^s_{qjk}
\end{equation}
and,
\begin{equation}\label{2.69}
[\nabla_k,\nabla_j]A^{pq}=-R^p_{sjk}A^{sq}-R^q_{sjk}A^{ps}.
\end{equation}\\
{\bf Solution 2.37:} Transvecting the given relation $R_{ij}=\lambda g_{ij}$ by $g^{ij}$ we get
$$R=\lambda n,~~i.e.,~~ \lambda=\frac{R}{n}$$
Hence, $$R_{ij}=\frac{R}{n} g_{ij}$$

A Riemannian space of dimension `$n$' in which the above equation holds is called an Einstein space of dimension `$n$'. In Einstein's general relativity the space-time world is a (pseudo)Riemannian 4--space of signature $(+, +, +, -)$ or $(+, -, -, -)$ which is an Einstein space.\\\\
{\bf \underline{Note:}} As $g^{ij} g_{ik}=\delta^i_k$ , so $g^{ij}~g_{ij}=\delta^i_{~i}=\delta^1_{~1}+\delta^2_{~2}+\ldots \ldots + \delta^n_{~n}=n$ .\\\\
{\bf Solution 2.38:} Let $n$ be the dimension of the space. Now transvecting the given relation by $g^{ij}$ we get
$$n \, R_{kl}-\delta^j_l R_{jk}+\delta^i_k~R_{il}-R \, g_{kl}=0$$
$$\mbox{or,}~~~~n \, R_{kl}-R_{lk}+R_{kl}-R \, g_{kl}=0$$
$$\mbox{or,}~~~~R_{kl}=\frac{R}{n} g_{kl}~.$$
Hence the space is Einstein.\\\\
{\bf Solution 2.39:} Let $V_n$ be a space of constant curvature of dimension `$n$'. Then we have
$$R_{hijk}=k\left(g_{hj}~g_{ik}-g_{hk}~g_{ij}\right)$$
Now multiplying by $g^{hk}$ and contracting on $h$ and $k$ we get
$$R_{ij}=k\left(\delta^k_{~j}g_{ik}-n g_{ij}\right)=k(1-n)g_{ij}~,$$
showing that the space is an Einstein space.\\\\
{\bf \underline{Note:}} Here $k(1-n)=\dfrac{R}{n} ~~~~~\Rightarrow R=kn(1-n)$.\\\\
{\bf Solution 2.40:} We have,~~$R^h_{~j}=g^{hi} R_{ij}$
\begin{eqnarray}
\nabla_hR^h_{~j}=g^{hi}\nabla_hR_{ij}=g^{hi}\nabla_hR^s_{~ijs} &=& g^{hi}\nabla_h\left(g^{sp}R_{pijs}\right) \nonumber \\
&=& g^{hi}g^{sp}\nabla_hR_{pijs}~. \nonumber
\end{eqnarray}
By Bianchi's 2nd identity we have
$$\nabla_hR_{pijs}+\nabla_jR_{pish}+\nabla_sR_{pihj}=0.$$
So we can write
\begin{eqnarray}
\nabla_h R^h_{~j} &=& -g^{hi}g^{sp}\left[\nabla_j R_{pish}+\nabla_s R_{pihj}\right] \nonumber \\
&=&-\nabla_j\left(g^{hi}g^{sp}R_{pish}\right)-\nabla_s\left(g^{hi}g^{sp}R_{pihj}\right) \nonumber \\
&=& \nabla_j \left(g^{hi}g^{sp}R_{pihs}\right)-\nabla_s\left(g^{sp}g^{hi}R_{pihj}\right) \nonumber \\
&=& \nabla_j\left(g^{hi} R_{ih}\right)-\nabla_s\left(g^{sp}R_{pj}\right) \nonumber \\
&=&\nabla_j R-\nabla_s R^s_{~j}=\frac{\partial R}{\partial x^j}-\nabla_h R^h_{~j} \nonumber \\
\Rightarrow \nabla_h R^h_{~j} &=& \frac{1}{2} \frac{\partial R}{\partial x^j}~. \nonumber
\end{eqnarray}
\\
{\bf Solution 2.41:} Einstein space of dimension `$n$' is defined by the relation 
$$R^k_{~j}=\frac{R}{n} \delta^k_{~j}$$
$$\nabla_k R^k_{~j}=\frac{1}{n}\delta^k_{~j} \nabla_k R=\frac{1}{n}\frac{\partial R}{\partial x^j}~.$$
But, $$\nabla_k R^k_{~j}=\frac{1}{2}\frac{\partial R}{\partial x^j}$$
so, we have~~~~~~~~ $(n-2)\dfrac{\partial R}{\partial x^j}=0$.\\\\
Thus if $n>2$ then scalar curvature $R$ is a constant.\\\\
{\bf Solution 2.42:}~~~~It is given that $g_{ij}=0~,~~\forall ~i,~j$ such that $i\neq j$ .\\

So~~~~~~ $g^{ij}=0~,~~\forall ~i,~j$ ~~s.t.~~ $i \neq j$ .\\

Also~~~~~~ $g^{ii}=\dfrac{1}{g_{ii}}~,~~\forall ~i$ .\\\\
From the antisymmetric property of curvature tensor
$$R_{hijk}=-R_{ihjk}~~~,~~~R_{hijk}=-R_{hikj}~.$$\\
Note that $i~,~j~,~h$ take values from 1 to 3 and they are all unequal. So for convenience we choose $i=1~,~j=2~,~h=3$.
\begin{eqnarray}
\mbox{Now,}~~~~~~~~~~R_{hj}&=& g^{pq}R_{phjq}=\sum^3_{p=1} g^{pp}R_{phjp}=\sum^3_{p=1} \frac{R_{phjp}}{g_{pp}} \nonumber \\
&=&\frac{R_{ihji}}{g_{ii}}+\frac{R_{jhjj}}{g_{jj}}+\frac{R_{hhjh}}{g_{hh}}=\frac{R_{ihji}}{g_{jj}}=\frac{R_{hiij}}{g_{ii}}~. \nonumber
\end{eqnarray}
\begin{eqnarray}
\mbox{Also,}~~~~~~~~~~R_{hh}&=&\sum^3_{p=1} \frac{R_{phhp}}{g_{pp}}=\frac{R_{ihhi}}{g_{ii}}+\frac{R_{jhhj}}{g_{jj}}+\frac{R_{hhhh}}{g_{hh}} \nonumber \\
&=& \frac{R_{hiih}}{g_{ii}}+\frac{R_{hjjh}}{g_{ij}}~. \nonumber
\end{eqnarray}\\\\


\chapter{Curves in a Riemannian Space}


\section{Parametric Representation}

~~~Let $x^i$ be a coordinate system defined in a coordinate neighbourhood of an $n$-dimensional Riemannian space $M$. A curve in $M$ is given by
\begin{equation}\label{3.1}
x^i=f^i(u)~,~~~i=1, 2, \ldots ,n,
\end{equation}
where $u$ is a real variable, called the parameter, defined in some interval $I$ of the real line $R$. If the functions $f^i$ ($i=1, 2, \ldots ,n$) are $C^{\infty}$, then the curve is also briefly called a $C^{\infty}$ curve. As the functions are single valued so for a single value of $u$, there corresponds a single point of the curve. If the converse is true then the parameter is also called a coordinate for the curve.\\

{\bf Example:} ~~~ $x^1=a \cos u$~,~~~$x^2=a \sin u$~,~~~$x^3=0~,~~\ldots~~, x^n=0$.\\

We have $P(0)=P(2\pi)$. So if we take the interval as $0 \leq u <2\pi$ then the correspondence between points and parametric values is one-one and the parameter become a coordinate.\\

{\bf Note:} If in the eq. (\ref{3.1}) of a curve $f^{i'}(u)$ are all identically zero then $f^i(u)$ are all constants and the locus of (\ref{3.1}) reduces to a point and then eq. (\ref{3.1}) is not called a curve. It may however happen that $f^{i'}(u)=(0, 0, \ldots ,0)$ at isolated points. The parameter will be said to be irregular at such points.\\

\section{Arc Length of a Curve} 

~~~If $s$ be the measure of the arc length of the curve (\ref{3.1}) then from\,(\ref{2.27}), \textit{i.e.},
$$ds^2=g_{ij}dx^i dx^j~,$$
we get
\begin{equation}\label{3.2}
\frac{ds}{du}=\sqrt{g_{ij}\frac{dx^i}{du}\frac{dx^j}{du}}~.
\end{equation}

On integration,
$$s=\int\limits _{u_0}^{u} \sqrt{g_{ij}\frac{dx^i}{du}\frac{dx^j}{du}}\,du~.$$

This gives the arc-length of the curve from a fixed point $A(u_0)$ to any point $P(u)$.\\

As $g_{ij}$ is positive definite, so it follows from (\ref{3.2}) that $\dfrac{ds}{du} \neq 0$. Hence from 
$$\frac{dx^i}{du}=\frac{dx^i}{ds}\frac{ds}{du}~,$$
it follows that $\dfrac{dx^i}{du} \neq 0$ which implies $\dfrac{dx^i}{ds} \neq 0$ and vice-versa. Also $\dfrac{ds}{du} \neq 0$ shows that the correspondence between $u$ and $s$ is one-one and so if $u$ is a coordinate for the curve then $s$ is also so. Further, the correspondence being one-one, $u$ is a function of $s$ and $s$ is a function of $u$ , {\it i.e.}, $u=\phi (s)$. Then the equation of the curve can be written as
$$x^i=f^i(\phi (s))=g^i(s)~~~~\mbox{(say)}.$$

Therefore, $s$ is also a parameter of the curve.\\

\section{Tangent and Normal to the Curve}

~~~The unit tangent vector $\textit{\textbf{t}}$ to the curve in $M$ at any point $P$ is given by the contravariant components
$$t^i=\frac{dx^i}{ds}~,~~~i=1, 2, \ldots ,n~.$$

{\bf Note:} For any general parameter $u$, the vector $\frac{dx^i}{du}$ is also called a tangent vector but it is not a unit vector. As $\textit{\textbf{t}}$ is a unit vector so
$$g_{ij}t^i t^j=1.$$

Taking intrinsic differentiation along the curve, we have
$$g_{ij}t^i \frac{\delta t^j}{ds}=0.$$

The above relation shows that (assuming $\dfrac{\delta \textit{\textbf{t}}}{ds}$ is non-zero) the vector $\dfrac{\delta \textit{\textbf{t}}}{ds}$ is normal to the tangent $\textit{\textbf{t}}$. The unit vector corresponding to this vector is called the principal normal or the first normal to the curve. The magnitude of the vector $\dfrac{\delta \textit{\textbf{t}}}{ds}$ is called the first curvature of the curve relative to $M$. It is denoted by $\kappa _1$, \textit{i.e.}, $\kappa _1=\left|\dfrac{\delta \textit{\textbf{t}}}{ds}\right|$.\\

{\bf Note:} The idea of curvature comes from the Euclidean space and we shall show it in the following corollary.\\

{\bf Corollary:} Show that $\left|\dfrac{d\textit{\textbf{t}}}{ds}\right|$, where $\textit{\textbf{t}}$ is the unit tangent vector to any curve $\gamma$ in $n$-dimensional space, under rectangular Cartesian coordinates gives the curvature of the curve (\textit{i.e.}, the arc-rate of turning of the tangent).\\

{\bf Proof:} Let $\textit{\textbf{t}}$ be the unit tangent vector at any point $P(s)$ on the curve and $\textit{\textbf{t}}+\Delta \textit{\textbf{t}}$ be the tangent vector at the neighbouring point $Q(s+\Delta s)$. Suppose $\Delta \theta$ be the angle between these tangents at $P$ and $Q$.Now through any point $O$, we draw $\textit{\textbf{OA}}=\textit{\textbf{t}}$ and $\textit{\textbf{OB}}=\textit{\textbf{t}}+\Delta \textit{\textbf{t}}$. As $OA=\left|\textit{\textbf{t}}\right|=1$ and $OB=\left|\textit{\textbf{t}}+\Delta \textit{\textbf{t}}\right|=1$ , so the perpendicular $ON$ bisects $AB$ as well as $\angle AOB$. From vector algebra, $\textit{\textbf{AB}}=\Delta \textit{\textbf{t}}$. As $\angle AOB=\Delta \theta$, so $\angle NOB=\dfrac{1}{2}\Delta \theta$ and $\textit{\textbf{NB}}=\dfrac{1}{2}\Delta \textit{\textbf{t}}$.\\

\begin{wrapfigure}[9]{r}{0.37\textwidth}
\includegraphics[height=5 cm , width=6.5 cm ]{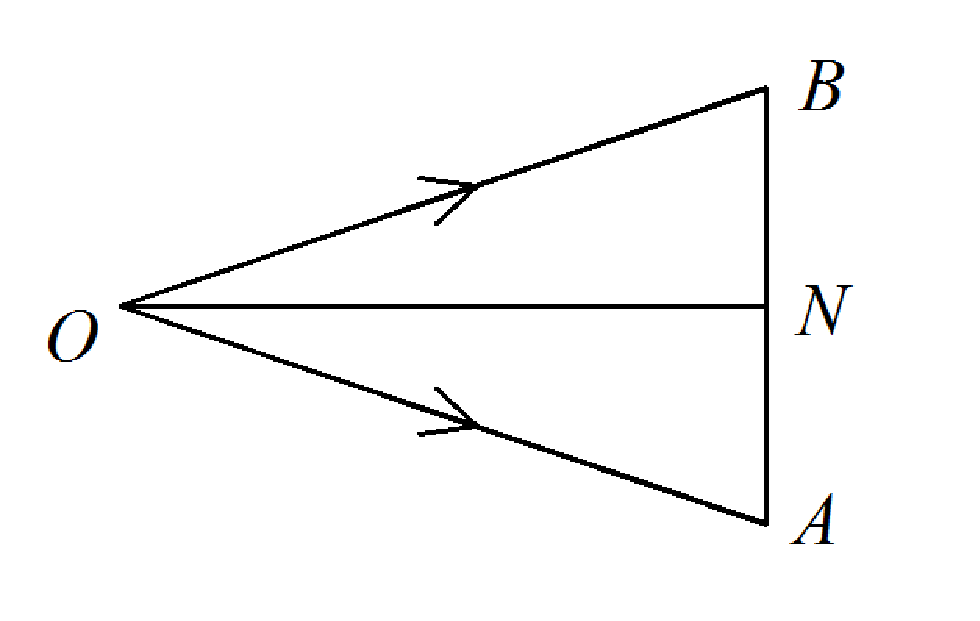}\vspace{-\intextsep}
\begin{center}
Fig. 3.1
\end{center}\vspace{-\intextsep}
\end{wrapfigure}
Now from the triangle $NBO$\,, we have
\begin{eqnarray} 
\sin \frac{1}{2}\Delta \theta &=& \frac{NB}{OB}=\frac{\left|\frac{1}{2}\Delta \textit{\textbf{t}}\right|}{1} \nonumber \\
i.e.~~~~\frac{\sin \frac{1}{2}\Delta \theta}{\frac{1}{2}\Delta \theta} &=& \frac{\left|\Delta \textit{\textbf{t}}\right|}{\left|\Delta \theta\right|} \nonumber
\end{eqnarray}

Now proceeding to the limit as $\Delta s \longrightarrow 0$ (\textit{i.e.}, $Q \longrightarrow P$), we have
\begin{eqnarray}
\left|\frac{d\textit{\textbf{t}}}{d\theta}\right| &=& 1 \nonumber \\
\mbox{Thus,}~~\left|\frac{d\textit{\textbf{t}}}{ds}\right| &=& \left|\frac{d\textit{\textbf{t}}}{d\theta}\right| \left|\frac{d\theta}{ds}\right|=\kappa \nonumber
\end{eqnarray}

{\bf Note:} The result $\left|\dfrac{d\textit{\textbf{t}}}{d\theta}\right|=1$ implies that if we consider any vector field of unit magnitude then the intrinsic derivative of that vector field will also be a unit vector if the intrinsic derivative is taken with respect to the angle $\theta$ in polar coordinates.\\

\section{Serret$-$Frenet formulae for a curve in a Riemannian space}

~~~Let $\textit{\textbf{t}}_{\bm 1}$ be the unit tangent vector to a curve $\gamma$ in $M$ at $P$ and `$s$' is the arc length along the curve $\gamma$ from a fixed point to $P$. The derived vector $\dfrac{\delta \textit{\textbf{t}}_{\bm 1}}{ds}$ is  identically zero throughout the curve if the curve is a geodesic. So we assume $\gamma$ to be a non-geodesic curve and hence $\dfrac{\delta \textit{\textbf{t}}_{\bm 1}}{ds} \neq 0$ in general.\\

As $\textit{\textbf{t}}_{\bm 1}$ is of constant magnitude (unit length) so $\dfrac{\delta \textit{\textbf{t}}_{\bm 1}}{ds}$ is normal to $\textit{\textbf{t}}_{\bm 1}$ (\textit{i.e.}, normal to the curve $\gamma$). Let us write
\begin{equation}\label{3.3}
\dfrac{\delta \textit{\textbf{t}}_{\bm 1}}{ds}=\kappa _1 \textit{\textbf{t}}_{\bm 2}
\end{equation}
where $\kappa _1$ is the magnitude of $\dfrac{\delta \textit{\textbf{t}}_{\bm 1}}{ds}$, called the first curvature and $\textit{\textbf{t}}_{\bm 2}$ is the corresponding unit vector. We call $\dfrac{\delta \textit{\textbf{t}}_{\bm 1}}{ds}$ the first curvature vector and $\textit{\textbf{t}}_{\bm 2}$ the first normal or principal normal or the second orthonormalized osculating vector.\\

Next we consider the derived vector $\dfrac{\delta \textit{\textbf{t}}_{\bm 2}}{ds}$ of $\textit{\textbf{t}}_{\bm 2}$ at $P$. Suppose $\dfrac{\delta \textit{\textbf{t}}_{\bm 2}}{ds} \neq 0$ and resolve it into two components, one in the plane of $\textit{\textbf{t}}_{\bm 1}$, $\textit{\textbf{t}}_{\bm 2}$ and the other normal to this plane. The latter is denoted by $\kappa _2 \textit{\textbf{t}}_{\bm 3}$, where $\kappa _2$ is its magnitude and $\textit{\textbf{t}}_{\bm 3}$ is the corresponding unit vector. Since $\textit{\textbf{t}}_{\bm 2}$ is of constant magnitude, so $\frac{\delta \textit{\textbf{t}}_{\bm 2}}{ds}$ is perpendicular to $\textit{\textbf{t}}_{\bm 2}$. Hence we write
$$\frac{\delta \textit{\textbf{t}}_{\bm 2}}{ds}=\sigma _1 \textit{\textbf{t}}_{\bm 1}+\kappa _2 \textit{\textbf{t}}_{\bm 3}~,$$
where the scalar $\sigma _1$ is given by $\dfrac{\delta \textit{\textbf{t}}_{\bm 2}}{ds} \cdot \textit{\textbf{t}}_{\bm 1}$ .\\

As $\textit{\textbf{t}}_{\bm 2} \cdot \textit{\textbf{t}}_{\bm 1}=0$ , so taking intrinsic derivative with respect to the arc length $s$, we have
$$\frac{\delta \textit{\textbf{t}}_{\bm 1}}{ds} \textit{\textbf{t}}_{\bm 2}+\textit{\textbf{t}}_{\bm 1}\frac{\delta \textit{\textbf{t}}_{\bm 2}}{ds}=0.$$

But using (\ref{3.3}), we have
$$\sigma _1=\frac{\delta \textit{\textbf{t}}_{\bm 2}}{ds} \cdot \textit{\textbf{t}}_{\bm 1}=-\textit{\textbf{t}}_{\bm 2} \cdot \frac{\delta \textit{\textbf{t}}_{\bm 1}}{ds}=-\kappa _1~.$$

Thus,
\begin{equation}\label{3.4}
\frac{\delta \textit{\textbf{t}}_{\bm 2}}{ds}=-\kappa _1 \textit{\textbf{t}}_{\bm 1}+\kappa _2 \textit{\textbf{t}}_{\bm 3}~.
\end{equation}

We call $\kappa _2$ the second curvature scalar and $\textit{\textbf{t}}_{\bm 3}$ the second normal or the third orthonormalized osculating vector.\\

We assume the derivative $\dfrac{\delta \textit{\textbf{t}}_{\bm 3}}{ds}$ of $\textit{\textbf{t}}_{\bm 3}$ to be a non-zero vector and resolve it into two components, one in the 3-plane of $\textit{\textbf{t}}_{\bm 1}$, $\textit{\textbf{t}}_{\bm 2}$ and $\textit{\textbf{t}}_{\bm 3}$ and the other normal to this plane which we denote by $\kappa _3 \textit{\textbf{t}}_{\bm 4}$. As $\textit{\textbf{t}}_{\bm 3}$ is a vector of constant magnitude so $\dfrac{\delta \textit{\textbf{t}}_{\bm 3}}{ds}$ will be orthogonal to $\textit{\textbf{t}}_{\bm 3}$ and hence the former component will be a linear combination of $\textit{\textbf{t}}_{\bm 1}$ and $\textit{\textbf{t}}_{\bm 2}$. So let us write
$$\frac{\delta \textit{\textbf{t}}_{\bm 3}}{ds}=\rho _1 \textit{\textbf{t}}_{\bm 1} +\rho _2 \textit{\textbf{t}}_{\bm 2} +\kappa _3 \textit{\textbf{t}}_{\bm 4}~,$$
where $\rho _1 =\dfrac{\delta \textit{\textbf{t}}_{\bm 3}}{ds} \cdot \textit{\textbf{t}}_{\bm 1}~~,~~\rho _2 =\dfrac{\delta \textit{\textbf{t}}_{\bm 3}}{ds} \cdot \textit{\textbf{t}}_{\bm 2}$ .\\

As $\textit{\textbf{t}}_{\bm 1} \cdot \textit{\textbf{t}}_{\bm 3}=0$ and $\textit{\textbf{t}}_{\bm 2} \cdot \textit{\textbf{t}}_{\bm 3}=0$, so by intrinsic differentiation 
$$\frac{\delta \textit{\textbf{t}}_{\bm 3}}{ds} \textit{\textbf{t}}_{\bm 1}+\textit{\textbf{t}}_{\bm 3} \frac{\delta \textit{\textbf{t}}_{\bm 1}}{ds}=0~~~~~~~~~~,~~~~~~~~~~\frac{\delta \textit{\textbf{t}}_{\bm 3}}{ds} \textit{\textbf{t}}_{\bm 2}+\textit{\textbf{t}}_{\bm 3}\frac{\delta \textit{\textbf{t}}_{\bm 2}}{ds}=0$$
$$~~~~~~~~~~~~i.e.,~~\rho _1+\textit{\textbf{t}}_{\bm 3}\kappa \textit{\textbf{t}}_{\bm 2}=0~~~~~~~~~~~~,~~~~~i.e.,~~\rho _2+\textit{\textbf{t}}_{\bm 3} \left(-\kappa _1 \textit{\textbf{t}}_{\bm 1}+\kappa _2\textit{\textbf{t}}_{\bm 3}\right)=0~~~~$$
$$i.e.,~~\rho _1=0~~~~~~~~~~~~~~~~~~~~~~~,~~~~~i.e.,~~\rho _2=-\kappa _2~.~~~~~~~~~~~~$$

Hence we get the relation
\begin{equation}\label{3.5}
\frac{\delta \textit{\textbf{t}}_{\bm 3}}{ds}=-\kappa _2 \textit{\textbf{t}}_{\bm 2}+\kappa _3 \textit{\textbf{t}}_{\bm 4}~.
\end{equation}

We continue the process until we obtain the relation
\begin{equation}\label{3.6}
\frac{\delta \textit{\textbf{t}}_{\bm {n-2}}}{ds}=-\kappa _{n-3}\textit{\textbf{t}}_{\bm {n-3}}+\kappa _{n-2}\textit{\textbf{t}}_{\bm {n-1}}~.
\end{equation}

Then we define $\textit{\textbf{t}}_{\bm n}$ as an unit vector perpendicular to $\textit{\textbf{t}}_{\bm 1}~,~\textit{\textbf{t}}_{\bm 2}~, \ldots ,~\textit{\textbf{t}}_{\bm {n-1}}$ and so directed as to make $\textit{\textbf{t}}_{\bm 1}~,~\textit{\textbf{t}}_{\bm 2}~, \ldots ,~\textit{\textbf{t}}_{\bm n}$ a right-handed basis of the tangent space to $M$ at $P$. So the next equation may be put in the form
\begin{equation}\label{3.7}
\frac{\delta \textit{\textbf{t}}_{\bm {n-1}}}{ds}=-\kappa _{n-2} \textit{\textbf{t}}_{\bm {n-2}}+\kappa _{n-1}\textit{\textbf{t}}_{\bm n}~.
\end{equation}

It should be noted that although $\kappa _1~,~\kappa _2~, \ldots ,~\kappa _{n-2}$ are all positive, $\kappa _{n-1}$ may be of any sign. Since there cannot be any vector orthogonal to $\textit{\textbf{t}}_{\bm 1}~,~\textit{\textbf{t}}_{\bm 2}~, \ldots ,~\textit{\textbf{t}}_{\bm n}$ , so the equation for $\frac{\delta \textit{\textbf{t}}_{\bm n}}{ds}$ will be
\begin{equation}\label{3.8}
\frac{\delta \textit{\textbf{t}}_{\bm n}}{ds}=-\kappa _{n-1} \textit{\textbf{t}}_{\bm {n-1}}~.
\end{equation}

The set of all these intrinsic derivative equations can be written in compact form as
\begin{eqnarray}\label{3.9}
\frac{\delta \textit{\textbf{t}}_{\bm 1}}{ds} &=& \kappa _1 \textit{\textbf{t}}_{\bm 2} \nonumber \\
\frac{\delta \textit{\textbf{t}}_{\bm r}}{ds} &=& -\kappa _{r-1} \textit{\textbf{t}}_{\bm {r-1}}+\kappa _{r}\textit{\textbf{t}}_{\bm {r+1}} ~~,~r=2, 3, \ldots , n-1 \nonumber \\
\frac{\delta \textit{\textbf{t}}_{\bm n}}{ds} &=& -\kappa _{n-1}\textit{\textbf{t}}_{\bm {n-1}} 
\end{eqnarray}
or in a more compact form
\begin{equation}\label{3.10}
\frac{\delta \textit{\textbf{t}}_{\bm l}}{ds}=-\kappa _{l-1} \textit{\textbf{t}}_{\bm {l-1}}+\kappa _{l}\textit{\textbf{t}}_{\bm {l+1}} ~~,~l=1, 2, \ldots , n
\end{equation}
with $\kappa _0=0=\kappa _n$ and $\textit{\textbf{t}}_{\bm {n+1}}=\overrightarrow{0}$.\\

These formulae are known as Serret-Frenet formulae or simply Frenet formulae for the curve $\gamma$ in $M$.\\\\
{\bf Corollary I:} For three dimensional Euclidean space $\textit{\textbf{t}}_{\bm 1}$, $\textit{\textbf{t}}_{\bm 2}$, $\textit{\textbf{t}}_{\bm 3}$ are respectively written as $\textit{\textbf{t}}$, $\textit{\textbf{n}}$, $\textit{\textbf{b}}$ and are called the tangent, the principal normal and the binormal vectors respectively. Also $\kappa _1$, $\kappa _2$ are generally written as $\kappa$, $\tau$ and are called the curvature and the torsion respectively. So the Frenet formulae under rectangular Cartesian coordinates take the form
\begin{eqnarray}\label{3.11}
\frac{d\textit{\textbf{t}}}{ds} &=& \kappa \textit{\textbf{n}} \nonumber \\
\frac{d\textit{\textbf{n}}}{ds} &=&-\kappa \textit{\textbf{t}}+\tau \textit{\textbf{b}} \nonumber \\
\frac{d\textit{\textbf{b}}}{ds} &=&-\tau \textit{\textbf{n}}
\end{eqnarray}

It may be noted that $\textit{\textbf{t}} \times \textit{\textbf{n}}=\textit{\textbf{b}}$ , $\textit{\textbf{n}} \times \textit{\textbf{b}}=\textit{\textbf{t}}$ and $\textit{\textbf{b}} \times \textit{\textbf{t}}=\textit{\textbf{n}}$.\\

At each point of the curve, the planes spanned by $\{\textit{\textbf{t}},\textit{\textbf{n}}\}$, $\{\textit{\textbf{t}},\textit{\textbf{b}}\}$ and  $\{\textit{\textbf{n}},\textit{\textbf{n}}\}$ are respectively known as osculating plane, rectifying plane and normal plane.\\

One can define a vector in the rectifying plane as $\textit{\textbf{d}}=\tau\textit{\textbf{t}}+\kappa\textit{\textbf{b}}$ and it is termed as Darboux vector. Here the term $\kappa\textit{\textbf{b}}$ represents the rate of turning about the binormal vector due to curvature while the term $\tau\textit{\textbf{t}}$ stands for the rate of turning about the tangent vector due to torsion. Further, one can write the above Serret-Frenet formulae (i.e., eq. (\ref{3.11})) compactly as
\begin{equation}
\frac{d\textit{\textbf{t}}}{ds}=\textit{\textbf{t}} \times \textit{\textbf{d}} ,~~\frac{d\textit{\textbf{n}}}{ds}= \textit{\textbf{n}} \times \textit{\textbf{d}} \mbox{~~and~~} \frac{d\textit{\textbf{b}}}{ds}=\textit{\textbf{b}} \times \textit{\textbf{d}}.
\end{equation}

If $\textit{\textbf{r}}=\textit{\textbf{r}(s)}$ be a curve in $E^3$, (assuming at least four continuous derivatives), then $$\textit{\textbf{t}}=\dfrac{d\textit{\textbf{r}}}{ds}$$ is the tangent vector of length unity.\\

If a particles moves along the curve such that one can identify time as the arc length then $$\textit{\textbf{v}}=\textit{\textbf{t}} \mbox{~and~} |\textit{\textbf{v}}|=1$$
then the curve is known as unit speed curve.\\

Now $$\frac{d\textit{\textbf{v}}}{d\tau}=\frac{d\textit{\textbf{t}}}{ds}=\kappa\textit{\textbf{n}}$$
where $\tau$ is identified as the time coordinate. So the acceleration of the particle is along the normal (as $|\textit{\textbf{v}}|=1$ so $\dfrac{d|\textit{\textbf{v}}|}{d\tau}=0$ and $\dfrac{v^2}{\rho}=\dfrac{1}{\rho}=\kappa$).\\

Thus if a particle moves along a unit speed curve then its acceleration is always along the principal normal of magnitude $\kappa$, the curvature of the curve at the point.\\

$\bullet$ We shall now determine the conditions for which the position vector (P.V.) of a particle moving in a space curve is always on the osculating plane\\

As the P.V. lies on the osculating plane, so we write
\begin{eqnarray*}
\textit{\textbf{r}}(s)&=&a(s)\textit{\textbf{t}}(s)+g(s)\textit{\textbf{n}}(s)\\
\mbox{i.e,~} \textit{\textbf{t}}=\frac{d\textit{\textbf{r}}}{ds}&=&a'\textit{\textbf{t}}+g'\textit{\textbf{n}}+a\kappa\textit{\textbf{n}}+g\left(-\kappa\textit{\textbf{t}}+\tau\textit{\textbf{b}}\right)\\
\Rightarrow
(1-a'+\kappa g)\textit{\textbf{t}}&=&(a\kappa+g')\textit{\textbf{n}}+g\tau\textit{\textbf{b}}
\end{eqnarray*}
$\Rightarrow\tau=0$ i.e. $\textit{\textbf{b}}=$constant vector and hence the curve is a plane curve. Also we have
$$\frac{1-a'}{g}=\frac{g'}{a},~~a\kappa+g'=0$$ 
$\Rightarrow a^2+g^2=2F(s),~a=f(s)$ with $f(s)=\dfrac{dF}{ds}$\\

$\therefore|\textit{\textbf{r}}|^2=2F(s)$ and $\kappa=\dfrac{\frac{d^2F}{ds^2}-1}{\sqrt{2F(s)-\left(\frac{dF}{ds}\right)^2}}$\\

Here $F(s)$ is an arbitrary differentiable (at least twice) +ve function of $s$.\\

$\bullet$ Conditions for P.V. to be on rectifying plane:\\

Let $~~~\textit{\textbf{r}}(s)=\lambda(s)\textit{\textbf{t}}(s)+\mu(s)\textit{\textbf{b}}(s)$\\
So $\textit{\textbf{t}}=\dfrac{d\textit{\textbf{r}}}{ds}=\lambda'\textit{\textbf{t}}+\mu'\textit{\textbf{b}}+\lambda\kappa\textit{\textbf{n}}-\mu\tau\textit{\textbf{n}}$\\
$\Rightarrow \lambda'(s)=1,~\lambda\kappa=\mu\tau,~\mu'(s)=0 $\\
i.e. $\lambda=s+\lambda_0$, $\mu=\mu_0$, a constant, $\dfrac{\tau}{\kappa}=\dfrac{s+\lambda_0}{\mu_0}$\\

Thus $l^2=\lambda^2+\mu^2=s^2+2\lambda_0s+\left(\lambda_0^2+\mu_0^2\right)$\\
i.e. distance function is a quadratic polynomial in arc length.\\

As $\dfrac{\tau}{\kappa}$ is not a constant so it is not a generalized helix, rather a twisted curve having ratio, a linear function of are length. This is known as rectifying curve.\\

Also $\textit{\textbf{r}}\cdot\textit{\textbf{n}}=0$ and $\textit{\textbf{r}}\cdot\textit{\textbf{b}}=\mu_0$, a constant; so the rectifying curve has constant normal component. Further $\textit{\textbf{r}}=\lambda \textit{\textbf{E}}+\mu ~\textit{\textbf{b}}~\alpha~(\tau \textit{\textbf{t}}+\kappa \textit{\textbf{b}})=\textit{\textbf{d}}$, the Darboux vector. So, the position vector of a rectifying curve is always along the direction of the Darboux vector.\\

$\bullet$ Conditions for P.V. to be on normal plane:\\

Let $~~~\textit{\textbf{r}}(s)=\alpha(s)\textit{\textbf{n}}(s)+\beta(s)\textit{\textbf{b}}(s)$\\
$\therefore\textit{\textbf{t}}=\dfrac{d\textit{\textbf{r}}}{ds}=\alpha'\textit{\textbf{n}}+\beta'\textit{\textbf{b}}+\alpha\left(-\kappa\textit{\textbf{t}}+\tau\textit{\textbf{b}}\right)-\beta\tau\textit{\textbf{n}}$\\
$\Rightarrow1+\alpha\kappa=0,~\alpha\tau+\beta'=0,~\alpha'-\beta\tau=0$\\
$\Rightarrow\alpha^2+\beta^2=$ constant $=l_0^2$\\
$\therefore|\textit{\textbf{r}}|^2=\alpha^2+\beta^2=l_0^2$\\
$\Rightarrow$ the curve is a spherical curve.\\

Suppose $\alpha=l_0\cos(m_0s)$, $\alpha=l_0\sin(m_0s)$\\
then $\tau=-m_0$, a constant and $\kappa=-\dfrac{1}{l_0\cos(m_0s)}$.\\

As $\kappa$ is not constant  so it is not a circular curve, rather a spherical curve of constant torsion.\\

We now write the Serret-Frenet formulae i.e. eq. (\ref{3.11}) in matrix form as
\begin{equation*}
\frac{d}{ds}\begin{bmatrix}
	\textit{\textbf{t}}\\\textit{\textbf{n}}\\\textit{\textbf{b}}
\end{bmatrix}=\begin{bmatrix}
0&\kappa&0\\-\kappa&0&\tau\\0&-\tau&0
\end{bmatrix}\begin{bmatrix}
\textit{\textbf{t}}\\\textit{\textbf{n}}\\\textit{\textbf{b}}
\end{bmatrix}
\end{equation*}

Here $S=\begin{bmatrix}
0&\kappa&0\\-\kappa&0&\tau\\0&-\tau&0
\end{bmatrix}$ is a skew-symmetric matrix and is termed as space matrix. The eigen values of this matrix are $0$, $\pm i\sqrt{\kappa^2+\tau^2}$ with eigen vectors
$\textit{\textbf{e}}_{\bm {0}}=\textit{\textbf{d}}$, the Darboux vector, and
 $\textit{\textbf{e}}_\pm= -\kappa\textit{\textbf{t}} \mp \textit{i} \sqrt{\kappa^2+\tau^2}\textit{\textbf{n}}+\tau\textit{\textbf{b}}$\\
 
 Note that $\textit{\textbf{e}}_{\bm {0}}$ and $\textit{\textbf{e}}_\pm$ are orthogonal to each other.\\
 
 In general, the P.V. of a point on a space curve can be written as
 $$\textit{\textbf{r}}(s)=u(s)\textit{\textbf{t}}+v(s)\textit{\textbf{n}}+w(s)\textit{\textbf{b}}$$
 
 Now \begin{eqnarray}
 \textit{\textbf{V}}&=&\frac{d\textit{\textbf{r}}}{dt}=\frac{\partial\textit{\textbf{r}}}{\partial t}  +\bm{\omega}\times\textit{\textbf{r}}\nonumber\\
 &=&\frac{\partial\textit{\textbf{r}}}{\partial s}  +\bm{\omega}\times\textit{\textbf{r}}\nonumber\\
 &=& (\dot{u}-\kappa v)\textit{\textbf{t}}+(\dot{v}+\kappa u-\tau w)\textit{\textbf{n}}+(\dot{w}+\tau v)\textit{\textbf{b}}\nonumber
 \end{eqnarray}
 \hfill{(an overdot denotes differentiation with respect to `$s$')}\\
 
 So \begin{equation}\label{eq3.13}
 \bm{\omega}\times\textit{\textbf{r}}
 = -\kappa v\textit{\textbf{t}}+(\kappa u-\tau w)\textit{\textbf{n}}+\tau v\textit{\textbf{b}}
 \end{equation}
 
 Suppose $~\bm{\omega}=\omega_1\textit{\textbf{t}}+\omega_2\textit{\textbf{n}}+\omega_3\textit{\textbf{b}}$, then
 \begin{eqnarray}\label{eq3.14}
 \bm{\omega}\times\textit{\textbf{r}}&=&
 \begin{vmatrix}
 \textit{\textbf{t}}&\textit{\textbf{n}}&\textit{\textbf{b}}\\\omega_1&\omega_2&\omega_3\\u&v&w
 \end{vmatrix}\nonumber\\
  &=&(w\omega_2-v\omega_3)\textit{\textbf{t}}+(u\omega_3-w\omega_1)\textit{\textbf{n}}+(v\omega_1-u\omega_2)\textit{\textbf{b}}
 \end{eqnarray}
 
 Thus comparing eqs. (\ref{eq3.13})  and (\ref{eq3.14}), one gets
 $w\omega_2-v\omega_3=-\kappa v,~u\omega_3-w\omega_1=\kappa u-\tau w$ and $v\omega_1-u\omega_2=\tau v$.\\\\
 Choosing $\omega_2=0,~\omega_1=\tau,\omega_3=\kappa$, the angular velocity $\bm{\omega}=\tau\textit{\textbf{t}}+\kappa\textit{\textbf{b}}=\textit{\textbf{d}}$ i.e. rotation is along the Darboux vector.
 $|\vec{\omega}|=\sqrt{\tau^{2}+\kappa^{2}}$ is the magnitude of the angular velocity. Thus Darboux vector is the instantaneous axis of rotation with magnitude of angular velocity $\sqrt{\tau^{2}+\kappa^{2}}$. Moreover, rectifying curves can be interpreted kinematically as those curves whose position vector field determines the axis of instantaneous rotation at each point of the curve.

{\bf Corollary II:} If $\kappa _{n-1}=0$ identically, then the equation for $\dfrac{\delta \textit{\textbf{t}}_{\bm {n-1}}}{ds}$ (\textit{i.e.}, eq. (\ref{3.7})) becomes
$$\frac{\delta \textit{\textbf{t}}_{\bm {n-1}}}{ds}=-\kappa _{n-2}\textit{\textbf{t}}_{\bm {n-2}}$$
and the vector $\textit{\textbf{t}}_{\bm n}$ is uncalled for. However, $\textit{\textbf{t}}_{\bm n}$ may be defined uniquely as a unit vector such that ($\textit{\textbf{t}}_{\bm 1}~,~\textit{\textbf{t}}_{\bm 2}~, \ldots ,~\textit{\textbf{t}}_{\bm n}$) form a right-handed orthonormal frame of the tangent space. Hence the Frenet frame is fully defined and the last equation (\textit{i.e.}, eq. (\ref{3.8})) of Frenet formulae becomes
$$\frac{\delta \textit{\textbf{t}}_{\bm n}}{ds}=0~.$$

Next suppose $\kappa _{r-1}=0$ identically for some $r<n$. Then the equation for $\dfrac{\delta \textit{\textbf{t}}_{\bm {r-1}}}{ds}$ becomes
$$\frac{\delta \textit{\textbf{t}}_{\bm {r-1}}}{ds}=-\kappa _{r-2}\textit{\textbf{t}}_{\bm {r-2}}$$ 
and the vector $\textit{\textbf{t}}_{\bm r}$ is undefined. Therefore, all subsequent $t$-vectors after $\textit{\textbf{t}}_{\bm {r-1}}$ are undefined. Also all curvatures after $\kappa _{r-2}$ are undefined and may be treated to be all equal to zero. The curve is than said to be $(r-2)$\,-curvatured.\\

However, for $n$-dimensional Euclidean space, it can be proved that the curve lies in a $(r-1)$\,-plane. We may define vectors $\textit{\textbf{t}}_{\bm r}~,~\textit{\textbf{t}}_{\bm {r+1}}~, \ldots ,~\textit{\textbf{t}}_{\bm n}$ as constant unit vectors which are mutually orthogonal and orthogonal to $\textit{\textbf{t}}_{\bm 1}~,~\textit{\textbf{t}}_{\bm 2}~, \ldots ,~\textit{\textbf{t}}_{\bm {r-1}}$ . In this case
$$\frac{\delta \textit{\textbf{t}}_{\bm r}}{ds}=0~,~\ldots~,~\frac{\delta \textit{\textbf{t}}_{\bm n}}{ds}=0$$
and $\kappa _{r-1}=0=\kappa _r = \cdots =\kappa _n$ identically. It may be mentioned that choice of the vectors $\textit{\textbf{t}}_{\bm r}~, \ldots ,~\textit{\textbf{t}}_{\bm n}$ are not at all unique.\\

\section{Equations of a geodesic}

\begin{wrapfigure}{r}{0.33\textwidth}
\includegraphics[height=4 cm , width=6 cm ]{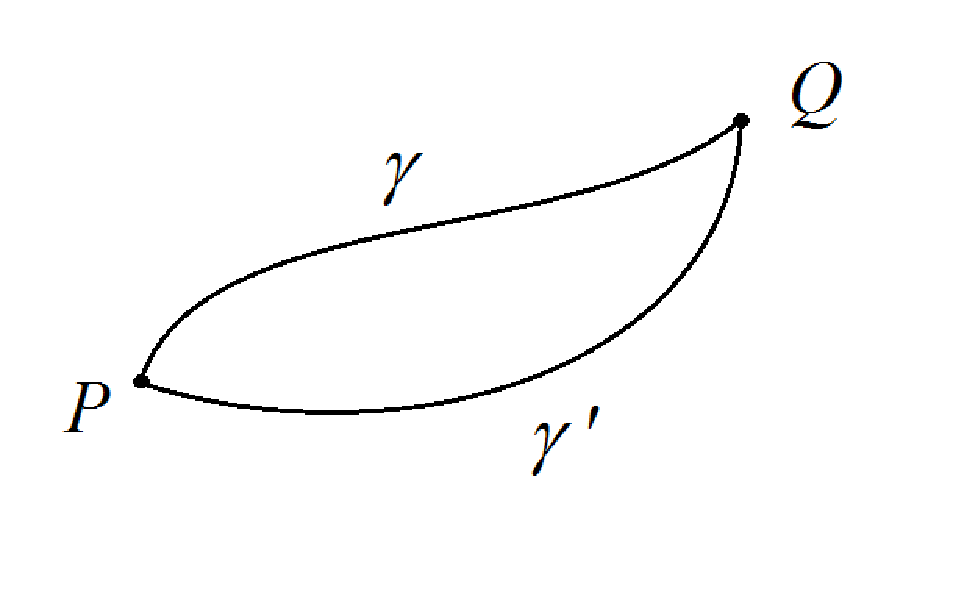}
\begin{center}
Fig. 3.2
\end{center}
\end{wrapfigure}
~~~Suppose
$\gamma : x^i=f^i(u)$
be a curve with parameter $u$ and let $P$, $Q$ be two points on it with parametric values $u_0$ and $u_1$ respectively. Let $\gamma '$ be a neighbouring curve which also passes through the points $P$ and $Q$ as shown in the figure.\\

So the equation of $\gamma '$ can be written as
\begin{equation}\label{3.13}
\overline{x}^i=x^i+\epsilon w^i~,
\end{equation}
where $\epsilon$ is a small scalar and $w^i$ are functions of $x^i$ (along the curve) such that
\begin{center}
$w^i=0$~~~for~ $u=u_0$~~and~~$u=u_1$~.
\end{center}

Now consider the integral
$$I= \int\limits _{P}^{Q} {\Phi(x^i, \dot{x}^i)du}~,~~i=1, 2, \ldots , n$$
where $\dot{x}^i=\dfrac{dx^i}{du}$ and $\Phi$ is an analytic function of the $2n$ arguments $x^i$ and $\dot{x}^i$ . Then
$$\delta I= \int\limits_{P}^{Q} {\left(\frac{\partial \Phi}{\partial x^i}\delta x^i+\frac{\partial \Phi}{\partial \dot{x}^i}\delta \dot{x}^i\right)du}+ \cdots \cdots ~,$$
where the dot terms are second and higher orders in the small quantities $\delta x^i$.\\

Now, 
\begin{eqnarray}
\int\limits_{P}^{Q} {\frac{\partial \Phi}{\partial \dot{x}^i}\delta \dot{x}^i du} &=& \frac{\partial \Phi}{\partial \dot{x}^i} \int\limits _{P}^{Q} {\frac{d}{du}\left(\delta x^i\right)du}-\int\limits _{P}^{Q}{\left[\left\lbrace \frac{d}{du}\left(\frac{\partial \Phi}{\partial \dot{x}^i}\right)\right\rbrace \int {\delta \dot{x}^i}du\right]du} \nonumber \\
&=& \left. \frac{\partial \Phi}{\partial \dot{x}^i}\delta x^i \right|_{P}^{Q}-\int\limits _{P}^{Q}{\frac{d}{du}\left(\frac{\partial \Phi}{\partial \dot{x}^i}\right)\delta x^i du} \nonumber \\
&=& -\int\limits _{P}^{Q}{\frac{d}{du}\left(\frac{\partial \Phi}{\partial \dot{x}^i}\right)\delta x^i du}~~~\left(\mbox{Since}~\delta x^i=\overline{x}^i-x^i=\epsilon w^i~,~\mbox{so}~\delta x^i=0~\mbox{at}~P~\mbox{and}~Q\right). \nonumber
\end{eqnarray}
Therefore,
\begin{equation}\label{3.14}
\delta I=\int\limits _{P}^{Q} {\left[\frac{\partial \Phi}{\partial x^i} -\frac{d}{du}\left(\frac{\partial \Phi}{\partial \dot{x}^i}\right) \right]\delta x^i du}~.
\end{equation}

The integral $\delta I$ is called the first variation of the integral $I$ and $I$ attains an extremal value on the curve $\gamma$ in its immediate neighbourhood if $\delta I=0$ for every set of functions $\delta x^i$ (or $w^i$) vanishing on $P$ and $Q$. A necessary and sufficient condition for this is
\begin{equation}\label{3.15}
\frac{\partial \Phi}{\partial x^i} -\frac{d}{du}\left(\frac{\partial \Phi}{\partial \dot{x}^i}\right)=0 ~,~~i=1, 2, \ldots , n~,
\end{equation}
which are known as Euler's equations on the condition of extremality.\\

Now, the length of the curve $\gamma$ from $P$ to $Q$ is
$$s=\int\limits _{u_0}^{u_1}{\sqrt{g_{\alpha \beta}\frac{dx^\alpha}{du}\frac{dx^\beta}{du}} du}=\int\limits _{u_0}^{u_1} {\sqrt{g_{\alpha \beta}\dot{x}^\alpha \dot{x}^\beta} du}~.$$

The arc length $s$ will be an extremal if (\ref{3.15}) holds with
\begin{equation}\label{3.16}
\phi =\sqrt{g_{\alpha \beta}\dot{x}^\alpha \dot{x}^\beta}=\frac{ds}{du}~.
\end{equation}

From geometric point of view, it is clear that $s$ is an extremal will mean that $s$ is a minimal.

Thus on differentiation,
\begin{eqnarray}
\frac{\partial \phi}{\partial \dot{x}^\mu} &=& \frac{1}{2\sqrt{g_{\alpha \beta}\dot{x}^\alpha \dot{x}^\beta}}\frac{\partial g_{\alpha \beta}}{\partial x^\mu} \dot{x}^\alpha \dot{x}^\beta \nonumber \\
&=& \frac{\frac{1}{2} \frac{\partial g_{\alpha \beta}}{\partial x^\mu} \dot{x}^\alpha \dot{x}^\beta}{\frac{ds}{du}} \nonumber
\end{eqnarray}
and
$$\frac{\partial \phi}{\partial \dot{x}^\mu} = \frac{1}{2\sqrt{g_{\alpha \beta}\dot{x}^\alpha \dot{x}^\beta}} 2g_{\mu \beta} \dot{x}^\beta =\frac{g_{\mu \beta} \dot{x}^\beta}{\frac{ds}{du}}~.$$

So
$$\frac{d}{du} \left(\frac{\partial \phi}{\partial \dot{x}^\mu}\right)=\frac{\left\lbrace \frac{ds}{du}\frac{d}{du}\left(g_{\mu \beta} \dot{x}^\beta \right)-g_{\mu \beta} \dot{x}^\beta \frac{d^2s}{du^2}\right\rbrace}{\left(\frac{ds}{du}\right)^2}~.$$

Hence from eq.\,(\ref{3.15}) ,
\begin{eqnarray}
&&\frac{1}{2}\frac{\partial g_{\alpha \beta}}{\partial x^\mu}\dot{x}^\alpha \dot{x}^\beta -\frac{d}{du}\left(g_{\mu \beta}\dot{x}^\beta\right)+g_{\mu \beta}\dot{x}^\beta \frac{\frac{d^2s}{du^2}}{\left(\frac{ds}{du}\right)} = 0 \nonumber \\
&i.e.,&~\frac{1}{2}\frac{\partial g_{\alpha \beta}}{\partial x^\mu}\dot{x}^\alpha \dot{x}^\beta -\frac{\partial g_{\mu \beta}}{\partial x^\alpha}\dot{x}^\alpha \dot{x}^\beta -g_{\mu \beta}\ddot{x}^\beta +g_{\mu \beta}\dot{x}^\beta \frac{\frac{d^2s}{du^2}}{\left(\frac{ds}{du}\right)} = 0 \nonumber \\
&i.e.,&~g_{\mu \beta}\ddot{x}^\beta +\left(\Gamma _{\alpha \mu \beta}+\Gamma _{\alpha \beta \mu}\right)\dot{x}^\alpha \dot{x}^\beta -\frac{1}{2}\left(\Gamma _{\mu \alpha \beta}+\Gamma _{\mu \beta \alpha}\right)\dot{x}^\alpha \dot{x}^\beta -g_{\mu \beta}\dot{x}^\beta \left(\frac{\ddot{s}}{\dot{s}}\right) = 0 \nonumber \\
&i.e.,&~g_{\mu \beta}\ddot{x}^\beta +\Gamma _{\alpha \beta \mu}\dot{x}^\alpha \dot{x}^\beta -g_{\mu \beta}\dot{x}^\beta \left(\frac{\ddot{s}}{\dot{s}}\right) = 0~. \nonumber
\end{eqnarray}

Transvecting by $g^{\delta \mu}$, we get
\begin{equation}\label{3.17}
\frac{d^2x^\delta}{du^2}+\Gamma _{\alpha \beta}^{\delta} \frac{dx^\alpha}{du}\frac{dx^\beta}{du}-\frac{dx^\delta}{du}\left(\frac{\ddot{s}}{\dot{s}}\right)=0.
\end{equation}

This is the differential equation for a geodesic in $V_n$ in terms of a general parameter $u$. However, if $u=s$ , the arc length then $\dfrac{ds}{du}=1$ and $\dfrac{d^2u}{ds^2}=0$. So the above geodesic equation simplifies to
\begin{eqnarray}\label{3.18}
&&\frac{d^2x^\delta}{ds^2}+\Gamma _{\alpha \beta}^{\delta} \frac{dx^\alpha}{ds}\frac{dx^\beta}{ds} = 0 \nonumber \\
&i.e.,&~\frac{\delta}{ds}\left(\frac{dx^\delta}{ds}\right) = 0~.\nonumber\\
&i.e.,&~ \frac{\delta t^{\delta}}{ds}= 0
\end{eqnarray}
\\

{\bf {\Large Some consequences of geodesic equations:}}\\

{\bf (a) Definition : Parallel vectors along a curve:} A vector field $V^i$ defined along a curve $\gamma : x^i=f^i (s)$ in $V_n$ is said to be parallel along the curve or parallely transported along the curve if
\begin{eqnarray}
&&\frac{\delta V^i}{ds}= 0 \nonumber \\
&i.e.,&~\frac{dV^i}{ds}+\Gamma _{kl}^{i} V^k \frac{dx^l}{ds} = 0. \nonumber
\end{eqnarray}

{\bf Note:} A curve is a geodesic if its tangent vector $\textit{\textbf{t}}$ is parallel along the curve. Hence a geodesic is also called an auto-parallel line.\\

{\bf (b) Theorem 3.1 : Any parallel vector field along a curve $\gamma$ is of constant magnitude.}\\

{\bf Proof:} As $\left|A\right|^2=g_{ij}A^i A^j$ , so
\begin{eqnarray}
\frac{d}{ds}\left|A\right|^2=\frac{d}{ds}(g_{ij}A^i A^j) &=& \frac{\delta}{ds}(g_{ij}A^i A^j) \nonumber \\
&=& \frac{\delta g_{ij}}{ds} A^i A^j+g_{ij}\frac{\delta A^i}{ds}A^j+g_{ij}A^i\frac{\delta A^j}{ds} \nonumber \\
&=& 0~, \nonumber
\end{eqnarray}
since $\dfrac{\delta g_{ij}}{ds}=0$, metric tensor is covariant constant and $\dfrac{\delta A^i}{ds}=0$ , $\textit{\textbf{A}}$ is constant along the curve $\gamma$.

Hence the vector field $\textit{\textbf{A}}$ is of constant magnitude.\\

{\bf Note:} As geodesic is an auto-parallel line, \textit{i.e.}, tangent vector $\textit{\textbf{t}}$ is parallel along the geodesic, so tangent vector to the geodesic is of constant magnitude.\\

As $t^i=\dfrac{dx^i}{ds}$ is the tangent vector to the geodesic, so from the geodesic equation (\ref{3.18}), we have $\dfrac{\delta t^i}{ds}=0$. Now,
\begin{eqnarray}
\left|\textit{\textbf{t}}\right|=g_{ij}\,t^i t^j=g_{ij}\frac{dx^i}{ds}\frac{dx^j}{ds} &=& \mbox{constant} \nonumber \\
i.e.,~~~~g_{ij}\frac{dx^i}{ds}\frac{dx^j}{ds} &=& \mbox{constant}~, \nonumber
\end{eqnarray}
along the geodesic. This is called the first integral of the geodesic equation.\\

{\bf (c) Theorem 3.2 : If $\textit{\textbf{A}}$ and $\textit{\textbf{B}}$ be two vector fields parallel along the curve $\gamma$ then $\textit{\textbf{A}}$ and $\textit{\textbf{B}}$ make a constant angle among them at every point on $\gamma$}.\\

{\bf Proof:}
$$\frac{d}{ds} (\textit{\textbf{A}}\textit{\textbf{B}})=\frac{d}{ds}(g_{ij}A^i B^j)=\frac{\delta}{ds}(g_{ij}A^i B^j)=0~~~(\mbox{Since}~\frac{\delta g_{ij}}{ds}=0=\frac{\delta A^i}{ds}=\frac{\delta B^i}{ds})~.$$

Hence the scalar product between the vectors $\textit{\textbf{A}}$ and $\textit{\textbf{B}}$ is a constant throughout the curve. Also the magnitudes of $\textit{\textbf{A}}$ and $\textit{\textbf{B}}$ are constants throughout the curve. So the angle $\theta$ between the vectors $\textit{\textbf{A}}$ and $\textit{\textbf{B}}$, \textit{i.e.},
$$\cos \theta =\frac{\textit{\textbf{A}} \cdot \textit{\textbf{B}}}{\left|\textit{\textbf{A}}\right| \left|\textit{\textbf{B}}\right|}$$
is also constant at every point of $\gamma$.\\

{\bf Note:} If $\textit{\textbf{v}}$ is a vector field parallel along a geodesic then $\textit{\textbf{v}}$ is a vector field of constant magnitude and is at constant angle with the tangent field of the geodesic, \textit{i.e.}, at constant angle with the geodesic.\\

{\bf (d) Definition : Geodesic coordinates:} Usually, Cartesian coordinates are one for which co-efficients of first fundamental form (\textit{i.e.} $g_{ij}$) are constants. In general Riemannian space such a coordinate system is not possible throughout the space. However, it is possible to have a coordinate system such that $g_{ij}$ are locally constants in the neighbourhood of a point $P \in M$, \textit{i.e.},
$$\frac{\delta g_{ij}}{\delta x^k}=0~~\mbox{at}~P~~~\mbox{and}~~~\frac{\delta g_{ij}}{\delta x^k} \neq 0~~\mbox{elsewhere}.$$
Then such a coordinate system is called a geodesic coordinate system with $P$ as the pole. So it is evident that Christoffel's symbols vanish at $P$ and consequently the covariant derivative reduces to partial derivative at $P$.\\

{\bf Note:} In geodesic coordinates, the geodesics at the pole $P$ becomes identical to those in Euclidean geometry.\\

{\bf Theorem 3.3 :} The necessary and sufficient conditions that a system of coordinates be geodesic with pole at $P$ are that their second order covariant derivative with respect to the metric of the space all vanish at the pole $P$.\\\\
{\bf Proof:} From the transformation law of Christoffel's symbols, we have (see eq.\,(\ref{2.54}))
\begin{eqnarray}
\bar{\Gamma}_{lm}^{q}\frac{\partial x^u}{\partial \bar{x}^q} &=& \frac{\partial x^i}{\partial \bar{x}^l} \frac{\partial x^k}{\partial \bar{x}^m}\Gamma _{ik}^{u}+\frac{\partial ^2 x^u}{\partial \bar{x}^l \partial \bar{x}^m} \nonumber \\
\mbox{or},~~-\Gamma _{ik}^{u} \frac{\partial x^i}{\partial \bar{x}^l} \frac{\partial x^k}{\partial \bar{x}^m} &=& \frac{\partial}{\partial \bar{x}^m} \left(\frac{\partial x^u}{\partial \bar{x}^l}\right) -\bar{\Gamma}_{lm}^{q}\frac{\partial x^u}{\partial \bar{x}^q} \nonumber.
\end{eqnarray}

Now, interchanging bar and unbar coordinate system, we have
\begin{equation}\label{3.19}
-\bar{\Gamma} _{ik}^{u} \frac{\partial \bar{x}^i}{\partial x^l} \frac{\partial \bar{x}^k}{\partial x^m} = \frac{\partial}{\partial x^m} \left(\frac{\partial \bar{x}^u}{\partial x^l}\right) -\Gamma _{lm}^{q}\frac{\partial \bar{x}^u}{\partial x^q}=\nabla _m{\left(\frac{\partial \bar{x}^u}{\partial x^l}\right)}=\bar{x}^u_{;lm}~,
\end{equation}
where we have assumed the bar coordinates as scalar functions of ${x^i}$\,'s. Now if we assume the bar coordinates as geodesic coordinates with $P$ as the pole then $\bar{\Gamma} _{ik}^{u}=0$ at $P$ and hence from (\ref{3.19}),
$$\bar{x}^u_{;lm}=0~~\mbox{at}~P~.$$
Conversely, if the second covariant derivatives vanish at $P$ then from (\ref{3.19}),
$$\bar{\Gamma} _{ik}^{u} \frac{\partial \bar{x}^i}{\partial x^l} \frac{\partial \bar{x}^k}{\partial x^m}=0~.$$
As $\dfrac{\partial \bar{x}^i}{\partial x^l} \neq 0$ at $P$, so $\bar{\Gamma} _{ik}^{u}=0$ at $P$, \textit{i.e.}, the bar coordinate system is a geodesic coordinate with $P$ as the pole.\\

{\bf Note:} At the pole of the geodesic coordinate system, the first order covariant derivatives are ordinary partial derivatives while the second order covariant derivatives of the coordinate system with respect to the metric of the space vanishes identically.\\

\section{Curves in three dimensions}

Let a curve $\gamma$ in three dimension is given by the parametric representation $x^i=f^i(u),~i=1, 2, 3$ and $s$ denotes the arc length along the curve $\gamma$. We shall start with the following theorem.\\

{\bf Theorem 3.4 :} If the parametric representation of a curve is regular and of class $C^k$ then the arc length $s$ is a regular parameter of class $C^k$.\\

{\bf Proof:} Let $P_0$ (a fixed point) and $P$ be two points on the curve having parameters $u_0$ and $u$. Then the arc length from $P_0$ to $P$ is
$$s=\int\limits _{u_0}^{u}{\sqrt{\sum _{i=1}^{3}{f^i}'(u){f^i}'(u)} du}=\psi (u)~~~(\mbox{say}).$$

If $f^1(u)$, $f^2(u)$ and $f^3(u)$ are functions of class $C^k$ then their derivatives are of class $C^{k-1}$. Hence when we take the integral we again get a function of class $C^k$. Thus $s=\psi (u)$ is a function of class $C^k$. Hence the theorem.\\

Again,
$$\frac{ds}{du}=\sqrt{{\left\lbrace {f^1}'(u)\right\rbrace}^2+{\left\lbrace {f^2}'(u)\right\rbrace}^2+{\left\lbrace {f^3}'(u)\right\rbrace}^2} \neq 0~.$$

Hence if $u$ is a regular parameter then so is $s$. Also the correspondence between $s$ and $u$ is one-one and therefore $\psi$ is invertible. So let us assume $u=\phi (s)$. Then $\phi (s)$ is also a one-one function of class $C^k$. The curve is then given by
$$x^i=f^i(\phi (s))=g^i(s)$$
which is also a function of class $C^k$.\\

{\bf Tangent Vector:} Let $x^i=x^i(u)$ be a given curve on which $P=x^i(u)$ and $Q=x^i(u+\Delta u)$ be two neighbouring points. Then the limiting position of the vector $\dfrac{\textit{\textbf{PQ}}}{\Delta u}$ as $Q \rightarrow P$, \textit{i.e.}, $\Delta u \rightarrow 0$ is a tangent vector to the curve at $P$. Now,
$$\lim_{\Delta u \rightarrow 0} \frac{\textit{\textbf{PQ}}}{\Delta u}=\lim_{\Delta u \rightarrow 0} \frac{x^i(u+\Delta u) -x^i(u))}{\Delta u}=\frac{dx^i}{du}~.$$

This is known as the parameter ruled forward tangent vector to the curve at $P$.\\

As
$$s=\int _{u_0}^{u}{\sqrt{\left(\frac{dx^1}{du}\right)^2+\left(\frac{dx^2}{du}\right)^2+\left(\frac{dx^3}{du}\right)^2} du}~,$$
so
$$\frac{ds}{du}=\sqrt{\left(\frac{dx^1}{du}\right)^2+\left(\frac{dx^2}{du}\right)^2+\left(\frac{dx^3}{du}\right)^2}~.$$

Thus putting $s=u$, we have
$$\left(\frac{dx^1}{ds}\right)^2+\left(\frac{dx^2}{ds}\right)^2+\left(\frac{dx^3}{ds}\right)^2=1~.$$
Hence $\dfrac{dx^i}{ds}$ is the unit forward tangent vector to the curve or simply the tangent vector.\\

{\bf Example:} The curve $x^i=(a\cos u, a\sin u, 0)$ represents a circle in the $x^1x^2$-plane. The tangent at the point $A(a,0,0)$ is $\left.\dfrac{dx^i}{du}\right|_{u=0}=\left.(-a\sin u, a\cos u, 0)\right|_{u=0}=(0,a,0)$.\\

\section{Curves in a plane}

{\bf Theorem 3.5 :} A necessary and sufficient condition for the curve $x^i=f^i(u)$ to lie on a plane is that
\begin{center}
$ \begin{vmatrix}
f^{1'} & f^{2'} & f^{3'} \\
f^{1''} & f^{2''} & f^{3''} \\
f^{1'''} & f^{2'''} & f^{3'''} \end{vmatrix}  =0~~~~\forall u$ . 
\end{center}

{\bf Proof:} First of all we shall prove that the condition is necessary. So we assume that $x^i=f^i(u)$ be a plane curve and we shall have to show that the above determinant to be zero. Suppose the equation of the plane be
\begin{equation}\label{3.20}
a_1 x^1+a_2 x^2+a_3 x^3+a_0=0.
\end{equation}

Then
$$a_1 f^1(u)+a_2 f^2(u)+a_3 f^3(u)+a_0=0~~~~\forall u~.$$

By successive differentiation with respect to $u$, we get
\begin{eqnarray}\label{3.21}
\left.
\begin{array}{c}
a_1 f^{1'}(u)+a_2 f^{2'}(u)+a_3 f^{3'}(u) = 0  \\
a_1 f^{1''}(u)+a_2 f^{2''}(u)+a_3 f^{3''}(u) = 0 \\
a_1 f^{1'''}(u)+a_2 f^{2'''}(u)+a_3 f^{3'''}(u) = 0~.
\end{array}
\right\}
\forall u
\end{eqnarray}

As $a_{i}'s$ are not all equal to zero, hence eliminating $a_{i}'s$ we get

$$\begin{vmatrix}
f^{1'} & f^{2'} & f^{3'} \\
f^{1''} & f^{2''} & f^{3''} \\
f^{1'''} & f^{2'''} & f^{3'''} \end{vmatrix}  =0~~~~\forall u ~. $$

So the condition is necessary.\\

To prove the condition sufficient, let us assume the condition. Then we know that for each value of $u$, numbers $b_1(u)$, $b_2(u)$ and $b_3(u)$ exist such that
\begin{eqnarray}
b_i f^{i'} ~\equiv& b_1 f^{1'}+b_2 f^{2'}+b_3 f^{3'}& = 0 \label{3.22} \\
b_i f^{i''} ~\equiv&  b_1 f^{1''}+b_2 f^{2''}+b_3 f^{3''}& =~ 0 \label{3.23} \\
b_i f^{i'''}~ \equiv&  b_1 f^{1'''}+b_2 f^{2'''}+b_3 f^{3'''}& = ~0~. \label{3.24}
\end{eqnarray}

(Note that for a fixed $u$, $b_1(u)$, $b_2(u)$ and $b_3(u)$ are non-zero solutions of $a_1$, $a_2$ and $a_3$ in eq. (\ref{3.20})).\\

Now, differentiating eq. (\ref{3.22}), we get
$$b_i f^{i''}+b_{i}'f^{i'}=0$$
which by virtue of eq. (\ref{3.23}) gives
\begin{eqnarray}\label{3.25}
&&b_{i}' f^{i'} = 0 \nonumber \\
&i.e.,&~b_{1}' f^{1'}+b_{2}' f^{2'}+b_{3}' f^{3'} = 0~.
\end{eqnarray}

Similarly, differentiating eq. (\ref{3.23}) and using eq. (\ref{3.24}), we get
\begin{eqnarray} \label{3.26}
&&b_{i}' f^{i''} = 0 \nonumber \\
&i.e.,&~b_{1}' f^{1''}+b_{2}' f^{2''}+b_{3}' f^{3''} = 0~. 
\end{eqnarray}

We now discuss the following two cases:\\

{\bf Case I:} Let the rank of $\begin{bmatrix}
f^{1'} & f^{2'} & f^{3'} \\
f^{1''} & f^{2''} & f^{3''} \end{bmatrix} $ be two.\\

Then the solution space of the equations
\begin{eqnarray}
x f^{1'}+y f^{2'}+z f^{3'} &=& 0 \nonumber \\
x f^{1''}+y f^{2''}+z f^{3''} &=& 0 \nonumber
\end{eqnarray}
in $x$, $y$, $z$ is of rank 1. But $b_i$ and $b_{i}'$ are both solutions and hence they must be proportional, \textit{i.e.},
$$\frac{b_{1}'}{b_1}=\frac{b_{2}'}{b_2}=\frac{b_{3}'}{b_3}=\phi '(u)~~~~(\mbox{say}).$$

Integrating with respect to $u$ gives
\begin{eqnarray}
\log b_i &=& \phi (u)+\log a_i \nonumber \\
i.e.,~~~b_i~~ &=& a_1 e^{\phi (u)}~~,~i=1, 2, 3.\nonumber
\end{eqnarray}

Hence from eq. (\ref{3.25}), we get (after cancelling the common factor $e^{\phi (u)}\phi '(u)$)
\begin{eqnarray}
a_1 f^{1'}+a_2 f^{2'}+a_3 f^{3'} &=& 0 \nonumber \\
i.e.,~~~a_1 f^{1}+a_2 f^{2}+a_3 f^{3}+a_0 &=& 0~~(\mbox{Integrating~once}) \nonumber . 
\end{eqnarray}

Hence the curve is a plane curve lying on the plane
$$a_1 x^1+a_2 x^2+a_3 x^3+a_0=0.$$

{\bf Case II:} Let the rank of $\begin{bmatrix}
f^{1'} & f^{2'} & f^{3'} \\
f^{1''} & f^{2''} & f^{3''} \end{bmatrix}$ be one.\\

Then $\dfrac{f^{1''}}{f^{1'}}=\dfrac{f^{2''}}{f^{2'}}=\dfrac{f^{3''}}{f^{3'}}=\xi '(u)$~~~(say).\\

So on integration
\begin{eqnarray}
\log f^{i'} &=& \xi (u)+\lambda _1~~~({\lambda _i}\mbox{'s ~are ~constants}) \nonumber \\
i.e.,~~~f^{i'} &=& a_i e^{\xi (u)}~~~(\mbox{writing} ~a_i = e^{\lambda _i}) \nonumber .
\end{eqnarray}

Integrating once more, we get
$$f^i=a_i \psi (u)+b_i~~,~~\psi (u)=\int {e^{\xi (u)} du}$$
with ${b_i}$'s as integration constants.\\

Hence we have
\begin{eqnarray}
\frac{f^1-b_1}{a_1} &=& \frac{f^2-b_2}{a_2} = \frac{f^3-b_3}{a_3} \nonumber \\
i.e.,~~~ \frac{x^1-b_1}{a_1} &=& \frac{x^2-b_2}{a_2} = \frac{x^3-b_3}{a_3} \nonumber .
\end{eqnarray}

This shows that the curve is a straight line and therefore it is a plane curve.\\

{\bf Definition:} A curve in $E_3$ which is not a plane curve is called a twisted curve.\\

{\bf Example:} A curve $x^1=au^1$, $x^2=bu^2$, $x^3=cu^3$ ($a$,$b$,$c\neq 0$) is called a twisted curve as
\begin{center}
$ \Delta =  \begin{vmatrix}
x^{1'} & x^{2'} & x^{3'} \\
x^{1''} & x^{2''} & x^{3''} \\
x^{1'''} & x^{2'''} & x^{3'''} \end{vmatrix} =12abc \neq 0$.
\end{center}

{\bf Osculating plane:} The osculating plane to a curve at a point $P$ on it is the plane having the highest order 
of contact with the curve at $P$.\\

Let $\gamma : x^i=x^i(u)$ be the given curve. Any plane passing through $x^i$ may be written as
$$\sum _i a_i(X^i-x^i)=0~,$$
\begin{equation}\label{3.27}
i.e.,~~~~a_1(X^1-x^1)+a_2(X^2-x^2)+a_3(X^3-x^3)=0
\end{equation}
where $X^i$ is a general point on the plane.\\

The point $x^i(u+\epsilon)$ will lie on the plane if
\begin{equation}\label{3.28}
\sum a_i \left\lbrace \epsilon x^{i'}+\frac{\epsilon ^2}{2!}x^{i''}+ \cdot \cdot \cdot +\frac{\epsilon ^{n-1}}{(n-1)!}x^{i^{(n-1)}}+\frac{\epsilon ^n}{n!}x^{i^{(n)}}(u+\epsilon \theta)\right\rbrace =0~~~(0<\theta <1).
\end{equation}

This gives at least one root of $\epsilon$ equal to zero implying that the plane passes through $P$. Let us choose ${a_i}^{'}$s such that
\begin{equation}\label{3.29}
\sum a_i x^{i'}=0.
\end{equation}

Then eq.\,(\ref{3.28}) gives at least two roots of $\epsilon$ equal to zero implying that the plane meets the curve at $P$ in at least two contiguous points (\textit{i.e.}, is of contact of order at least one). It may be noted that the plane now passes through the tangent at $P$.\\

Let $x^{i''}$ be not proportional to $x^{i'}$ at $P$. We choose ${a_i}$'s such that we also have
\begin{equation}\label{3.30}
\sum a_i x^{i''}=0.
\end{equation}

Now eliminating ${a_i}$'s from equations (\ref{3.27}), (\ref{3.29}) and (\ref{3.30}), we see that the equation of the plane is
\begin{equation}\label{3.31}
\begin{vmatrix}
X^1-x^1 & X^2-x^2 & X^3-x^3 \\
x^{1'} & x^{2'} & x^{3'} \\
x^{1''} & x^{2''} & x^{3''} \end{vmatrix} =0.
\end{equation}

For this plane, eq. (\ref{3.28}) gives at least three zero roots of $\epsilon$ and the plane meets the curve at $P$ in at least three contignous points and so the order of contact is at least two. The actual order of contact depends on the nature of the curve at $P$. If it happens that $x^{i(k)}$ are linear combination of $x^{i'}$'s and $x^{i''}$'s for $k=3, 4, \ldots ,p$ but not for $k=(p+1)$ then eq.\,(\ref{3.28}) gives $(p+1)$ zero roots of $\epsilon$ ($(p+1)$ contiguous points at $P$, \textit{i.e.}, order of contact $p$). In this case no plane can have a contact of order higher than $p$ because $a_i$ would be required to satisfy
$$\sum a_i x^{i'}=0~,~~~\sum a_i x^{i''}=0~,~~\cdots ,~\sum a_i x^{i^{(p+1)}}=0~.$$

But $x^{i'}~,~x^{i''}~, \ldots ,~x^{i^{(p+1)}}$ being independent vectors, implies $a_1=a_2=\cdots =0$ and the plane is undefined. Thus eq.\,(\ref{3.31}) gives the osculating plane for the curve.\\

\section{The moving trihedron\,(Frenet frame)}

~~~The osculating plane at a point $P$ is spanned by the vectors $\dfrac{d\textit{\textbf{r}}}{du}$ and $\dfrac{d^2\textit{\textbf{r}}}{du^2}$, where $u$ is the parameter of the curve (the two vectors are assumed to be independent). As this being true for any parameter so it is also true for arc length $s$. Let us write $\textit{\textbf{t}}=\dfrac{d\textit{\textbf{r}}}{ds}$, then $\dfrac{d^2\textit{\textbf{r}}}{ds^2}=\dfrac{d\textit{\textbf{t}}}{ds}$ and the osculating plane is spanned by these two independent vectors $\textit{\textbf{t}}$ and $\dfrac{d\textit{\textbf{t}}}{ds}$. Since $\textit{\textbf{t}}$ is the unit tangent vector so
$$\textit{\textbf{t}} \cdot \textit{\textbf{t}}=1~.$$

Thus on differentiation
$$\textit{\textbf{t}} \cdot \frac{d\textit{\textbf{t}}}{ds}=0~.$$

Thus $\textit{\textbf{t}}$ and $\dfrac{d\textit{\textbf{t}}}{ds}$ are independent, non-zero vectors perpendicular to each other and lie on the osculating plane. The unit vector $\textit{\textbf{n}}=\dfrac{\frac{d\textit{\textbf{t}}}{ds}}{\left|\frac{d\textit{\textbf{t}}}{ds}\right|}$ is called the principal normal vector. Suppose
$$\frac{d\textit{\textbf{t}}}{ds}=\kappa \textit{\textbf{n}}~~\mbox{with}~~\kappa =\left|\frac{d\textit{\textbf{t}}}{ds}\right|=\left|\frac{d^2\textit{\textbf{r}}}{ds^2}\right|~,$$
where $\kappa$ is called the first curvature or simply the curvature of the curve and is positive definite. We now define a vector $\textit{\textbf{b}}$ by the relation 
$$\textit{\textbf{b}}=\textit{\textbf{t}} \times \textit{\textbf{n}}~.$$

Then $\textit{\textbf{b}}$ is perpendicular to $\textit{\textbf{t}}$ and is thus a normal vector. It is called the binormal vector. The triad of vectors $\lbrace \textit{\textbf{t}}, \textit{\textbf{n}}, \textit{\textbf{b}}\rbrace$ forms a right handed orthonormal frame, called Frenet frame at the point $P$. As $P$ moves along the curve, we call the variable frame $\lbrace \textit{\textbf{t}}, \textit{\textbf{n}}, \textit{\textbf{b}}\rbrace$ the moving Frenet frame or moving trihedron.\\

{\bf Note:} Let $\textit{\textbf{v}}$ be a unit vector field defined along a curve, then $\left|\dfrac{d\textit{\textbf{v}}}{ds}\right|$ gives the arc rate of turning of the vector $\textit{\textbf{v}}$.\\

\subsection{Serret$-$Frenet Formulae}

~~~The formulae expressing $\textit{\textbf{t}}'$, $\textit{\textbf{n}}'$, $\textit{\textbf{b}}'$ ($' \equiv \dfrac{d}{ds}$) as linear combination of ($\textit{\textbf{t}}$, $\textit{\textbf{n}}$, $\textit{\textbf{b}}$) constitute what are called Serret-Frenet formulae or simply Frenet formulae. For convenience, let us consider any moving orthogonal frame $\lbrace \textit{\textbf{e}}_{\bm 1}, \textit{\textbf{e}}_{\bm 2}, \textit{\textbf{e}}_{\bm 3}\rbrace$ of vectors moving along the curve. We put
\begin{equation}\label{3.32}
\textit{\textbf{e}}'_{\bm i}=\sum _{j=1}^{3} a_{ij} \textit{\textbf{e}}_{\bm j}~.
\end{equation}

We show that the matrix $A=[a_{ij}]$ called the Cartan matrix of the frame $\lbrace \textit{\textbf{e}}_{\bm i}\rbrace$ is skew-symmetric. From eq. (\ref{3.32}), taking scalar product with $\textit{\textbf{e}}_{\bm k}$ , we get
$$\textit{\textbf{e}}'_{\bm i} \cdot \textit{\textbf{e}}_{\bm k}=\sum _{j} a_{ij}\textit{\textbf{e}}_{\bm j}\textit{\textbf{e}}_{\bm k}=\sum _{j} a_{ij} \delta _{jk}=a_{ik}~.$$

Again from the relation 
$$\textit{\textbf{e}}_{\bm i} \cdot \textit{\textbf{e}}_{\bm k}=\delta _{ik}~,$$
we get
$$\textit{\textbf{e}}'_{\bm i} \cdot \textit{\textbf{e}}_{\bm k}+\textit{\textbf{e}}_{\bm i} \cdot \textit{\textbf{e}}'_{\bm k}=0~,~~~i.e.,~~a_{ik}+a_{ki}=0.$$

Hence the Cartan matrix $[a_{ij}]$ is a skew-symmetric matrix. If we now consider vectors $\textit{\textbf{t}}$, $\textit{\textbf{n}}$, $\textit{\textbf{b}}$ for $\textit{\textbf{e}}_{\bm 1}$, $\textit{\textbf{e}}_{\bm 2}$, $\textit{\textbf{e}}_{\bm 3}$ respectively and noting that we already have
$$\textit{\textbf{t}}'=\kappa \textit{\textbf{n}}~,$$
we get $a_{12}=\kappa$ , $a_{13}=0$ , $a_{21}=-\kappa$, $a_{31}=0$ .\\

Also, if we put $a_{23}=\tau$ and call it the second curvature or the torsion of the curve then we have $a_{32}=-\tau$ and we obtain the Frenet formulae
$$\textit{\textbf{t}}'=\kappa \textit{\textbf{n}}~~,~~\textit{\textbf{n}}'=-\kappa \textit{\textbf{t}}+\tau \textit{\textbf{b}}~~,~~\textit{\textbf{b}}'=-\tau \textit{\textbf{n}}~.$$

From the third formula $\textit{\textbf{b}}'=-\tau \textit{\textbf{n}}$ , we get $\left|\textit{\textbf{b}}'\right|=\left|\tau\right|$. As $\textit{\textbf{b}}$ is a unit vector so $\textit{\textbf{b}}'$ gives the arc-rate of turning of the binormal. Its magnitude is equal to the magnitude of torsion. Thus within sign the torsion $\tau$ gives the arc-rate of turning of the binormal. $\tau$ may be positive, negative or zero.\\
A curve in $E^{3}$ is called a twisted curve if it has non-zero curvature and torsion.\\
{\bf Determination of $\textit{\textbf{t}}$, $\textit{\textbf{n}}$, $\textit{\textbf{b}}$, $\kappa$, $\tau$ :}\\

For any curve $\textit{\textbf{r}}=\textit{\textbf{r}}(s)$, we have
\begin{eqnarray}
\textit{\textbf{r}}'=\frac{d\textit{\textbf{r}}}{ds}&=&\textit{\textbf{t}}\label{3.33}\\
\textit{\textbf{r}}''=\frac{d^2\textit{\textbf{r}}}{ds^2}&=&\kappa \textit{\textbf{n}}\label{3.34}\\
\textit{\textbf{r}}'''=\frac{d^3\textit{\textbf{r}}}{ds^3} &=& \kappa '\textit{\textbf{n}}+\kappa \frac{d\textit{\textbf{n}}}{ds}=\kappa '\textit{\textbf{n}}+\kappa(-\kappa \textit{\textbf{t}}+\tau \textit{\textbf{b}}) \nonumber \\
&=& -\kappa ^2\textit{\textbf{t}}+\kappa '\textit{\textbf{n}}+\kappa \tau \textit{\textbf{b}}~.\label{3.35}
\end{eqnarray}

Eq.\,(\ref{3.33}) gives $\textit{\textbf{t}}=\textit{\textbf{r}}'$ .\\

Eq.\,(\ref{3.34}) gives $\kappa =\left|\textit{\textbf{r}}''\right|$ (Since $\kappa$ is non-negative).\\

Therefore,
$$\textit{\textbf{n}}=\frac{\textit{\textbf{r}}''}{\left|\textit{\textbf{r}}''\right|}~.$$

From equations (\ref{3.33}), (\ref{3.34}) and (\ref{3.35}), we get
\begin{eqnarray}
[\textit{\textbf{r}}', \textit{\textbf{r}}'', \textit{\textbf{r}}'''] &=& [\textit{\textbf{t}}, \kappa \textit{\textbf{n}}, -\kappa ^2\textit{\textbf{t}}+\kappa '\textit{\textbf{n}}+\kappa \tau \textit{\textbf{b}}] \nonumber \\
&=& \kappa ^2 \tau [\textit{\textbf{t}}, \textit{\textbf{n}}, \textit{\textbf{b}}]=\kappa ^2 \tau \nonumber
\end{eqnarray}

Therefore
\begin{equation}\label{3.36}
\tau =\frac{1}{\kappa ^2}[\textit{\textbf{r}}', \textit{\textbf{r}}'', \textit{\textbf{r}}''']=\frac{[\textit{\textbf{r}}', \textit{\textbf{r}}'', \textit{\textbf{r}}''']}{|\textit{\textbf{r}}''|^2}.
\end{equation}

Also 
\begin{equation}\label{3.37}
\textit{\textbf{b}}=\textit{\textbf{t}} \times \textit{\textbf{n}}=\textit{\textbf{r}}' \times \frac{\textit{\textbf{r}}''}{\left|\textit{\textbf{r}}''\right|}=\frac{\textit{\textbf{r}}' \times \textit{\textbf{r}}''}{\left|\textit{\textbf{r}}''\right|}~.
\end{equation}

Again from equations (\ref{3.33}) and (\ref{3.34}), we have
$$\textit{\textbf{r}}' \times \textit{\textbf{r}}'' =\textit{\textbf{t}} \times \kappa \textit{\textbf{n}}=\kappa \textit{\textbf{b}}$$
\begin{equation}\label{3.38}
i.e.,~~~~\textit{\textbf{b}}=\frac{\textit{\textbf{r}}' \times \textit{\textbf{r}}''}{\left|\textit{\textbf{r}}' \times \textit{\textbf{r}}''\right|}
\end{equation}
$$\mbox{and}~~~~\kappa =\left|\textit{\textbf{r}}' \times \textit{\textbf{r}}''\right|~.$$

Therefore, we can write
\begin{equation}\label{3.39}
\tau =\frac{\left[\textit{\textbf{r}}', \textit{\textbf{r}}'', \textit{\textbf{r}}'''\right]}{\left|\textit{\textbf{r}}' \times \textit{\textbf{r}}''\right|^2}~.
\end{equation}

However, if the equation of the curve is given in parametric form with parameter $u$ (say) (different from $s$), \textit{i.e.},
$$\textit{\textbf{r}}=\textit{\textbf{r}}(u)$$
then
\begin{eqnarray}
\frac{d\textit{\textbf{r}}}{du} &=& \frac{d\textit{\textbf{r}}}{ds}\frac{ds}{du} \nonumber \\
\frac{d^2\textit{\textbf{r}}}{du^2} &=& \frac{d^2\textit{\textbf{r}}}{ds^2}\left(\frac{ds}{du}\right)^2+\frac{d\textit{\textbf{r}}}{ds}\frac{d^2s}{du^2} \nonumber \\
\frac{d^3\textit{\textbf{r}}}{ds^3} &=& \frac{d^3\textit{\textbf{r}}}{ds^3} \left(\frac{ds}{du}\right)^3+3\frac{d^2\textit{\textbf{r}}}{ds^2}\frac{ds}{du}\frac{d^2s}{du^2}+\frac{d\textit{\textbf{r}}}{ds}\frac{d^3s}{du^3}, \nonumber
\end{eqnarray}
or in compact notation ($\cdot$ $\equiv$ $\dfrac{d}{du}$, $'$ $\equiv$ $\dfrac{d}{ds}$)
\begin{eqnarray}\label{3.40}
\dot{\textit{\textbf{r}}} &=& \textit{\textbf{r}}' \dot{s} \\
\ddot{\textit{\textbf{r}}} &=& \textit{\textbf{r}}'' \dot{s}^2+\textit{\textbf{r}}' \ddot{s} \nonumber \\
\ddot{\textit{\textbf{r}}} &=& \textit{\textbf{r}}''' \dot{s}^3+3\textit{\textbf{r}}''\dot{s} \ddot{s}+\textit{\textbf{r}}'\,\dddot{s}~. \nonumber
\end{eqnarray}

If we assume the arc length $s$ and the parameter $u$ in the same direction, \textit{i.e.}, $\dfrac{ds}{du}=\dot{s} >0$ then we have from eq.\,(\ref{3.40}),
\begin{equation}\label{3.41}
\dot{s}=\left|\dot{\textit{\textbf{r}}}\right|
\end{equation}
\begin{equation}\label{3.42}
\textit{\textbf{t}}=\frac{\dot{\textit{\textbf{r}}}}{\left|\dot{\textit{\textbf{r}}}\right|}~.
\end{equation}
Now
$$\dot{\textit{\textbf{r}}} \times \ddot{\textit{\textbf{r}}}=(\textit{\textbf{r}}' \times \textit{\textbf{r}}'') \dot{s}^3=(\textit{\textbf{t}} \times \kappa \textit{\textbf{n}})\dot{s}^3=\kappa \dot{s}^3 \textit{\textbf{b}}~.$$

As $\kappa$ and $\dot{s}$ are both positive so the unit vector $\textit{\textbf{b}}$ is given by
\begin{equation}\label{3.43}
\textit{\textbf{b}}=\frac{\dot{\textit{\textbf{r}}} \times \ddot{\textit{\textbf{r}}}}{\left|\dot{\textit{\textbf{r}}} \times \ddot{\textit{\textbf{r}}}\right|}~.
\end{equation}

Also
$$\kappa \dot{s}^3=\left|\dot{\textit{\textbf{r}}} \times \ddot{\textit{\textbf{r}}}\right|~.$$

Therefore
\begin{equation}\label{3.44}
\kappa =\frac{\left|\dot{\textit{\textbf{r}}} \times \ddot{\textit{\textbf{r}}}\right|}{\dot{s}^3}=\frac{\left|\dot{\textit{\textbf{r}}} \times \ddot{\textit{\textbf{r}}}\right|}{\left|\dot{\textit{\textbf{r}}}\right|^3}~.
\end{equation}

From equation (\ref{3.40}) , we get
$$\left[\dot{\textit{\textbf{r}}}, \ddot{\textit{\textbf{r}}}, \dddot{\textit{\textbf{r}}}\right]=\left[\textit{\textbf{r}}', \textit{\textbf{r}}'', \textit{\textbf{r}}'''\right] \dot{s}^6=\kappa ^2 \tau \dot{s}^6~.$$

Therefore,
\begin{eqnarray}\label{3.45}
\tau &=& \frac{1}{\kappa ^2 \dot{s}^6}\left[\dot{\textit{\textbf{r}}}, \ddot{\textit{\textbf{r}}}, \dddot{\textit{\textbf{r}}}\right] \nonumber \\
&=& \frac{\left|\dot{\textit{\textbf{r}}}\right|^6}{\left|\dot{\textit{\textbf{r}}} \times \ddot{\textit{\textbf{r}}}\right|^2} \frac{\left[\dot{\textit{\textbf{r}}}, \ddot{\textit{\textbf{r}}}, \dddot{\textit{\textbf{r}}}\right]}{\left|\dot{\textit{\textbf{r}}}\right|^6} \nonumber \\
&=& \frac{\left[\dot{\textit{\textbf{r}}}, \ddot{\textit{\textbf{r}}}, \ddot{\textit{\textbf{r}}}\right]}{\left|\dot{\textit{\textbf{r}}} \times \ddot{\textit{\textbf{r}}}\right|^2}~.
\end{eqnarray}

{\bf Problem 3.1.} Show that~ $\kappa =\dfrac{\sqrt{\left|\ddot{\textit{\textbf{r}}}\right|^2-\ddot{s}^2}}{\dot{s}^2}$.\\

\section{Cylindrical Helix}

~~~A cylindrical helix is a twisted curve lying on a cylinder and meeting all generators at a constant angle. The curve is often simply called a helix.\\

Let $x^3$-axis be parallel to the generators. We first take the equation of the helix in the parametric form as
$$x^1=f^1(u)~,~~~x^2=f^2(u)~,~~~x^3=f^3(u)~.$$

As it cuts the generators at a constant angle $\alpha$ so taking $\textit{\textbf{m}}=(0, 0, 1)$, we get
$$\cos \alpha =\frac{\textit{\textbf{m}} \cdot \frac{d\textit{\textbf{r}}}{du}}{\left|\textit{\textbf{m}}\right| \cdot \left|\frac{d\textit{\textbf{r}}}{du}\right|}=\frac{f^{3'}(u)}{\sqrt{{\left\lbrace f^{1'}(u)\right\rbrace}^2+{\left\lbrace f^{2'}(u)\right\rbrace}^2+{\left\lbrace f^{3'}(u)\right\rbrace}^2}}~.$$

\begin{wrapfigure}{r}{0.25\textwidth}\vspace{-\intextsep}
\includegraphics[height=6 cm , width=4 cm ]{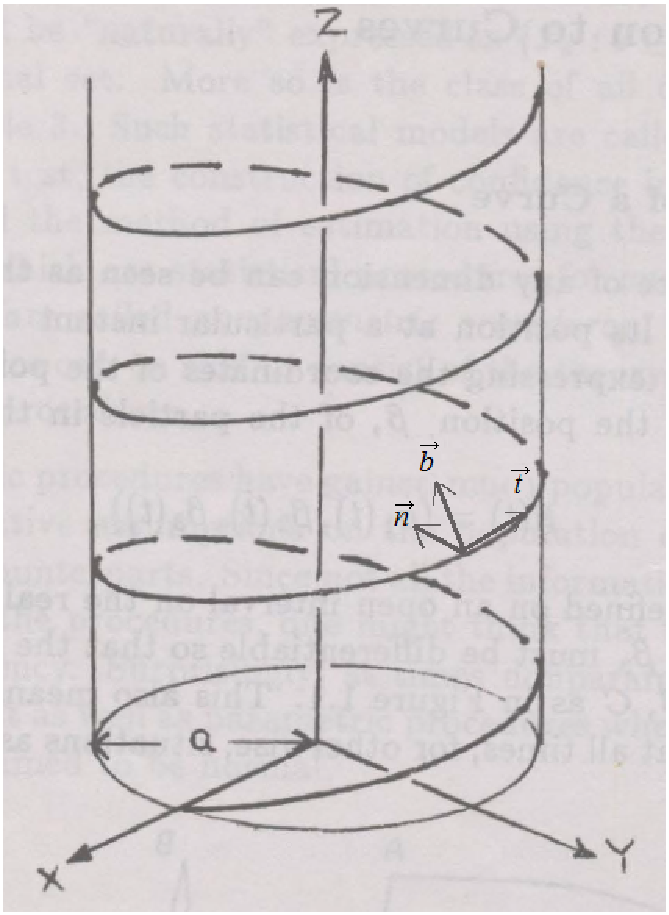}
\begin{center}\vspace{-\intextsep}
Fig. 3.3
\end{center}
\end{wrapfigure}
Hence
$$f^{3'}(u)=\cot \alpha \sqrt{\left(f^{1'}\right)^2+\left(f^{2'}\right)^2}~,$$
$$i.e.~~~~f^{3}(u)=\cot \alpha \int {\sqrt{\left(f^{1'}\right)^2+\left(f^{2'}\right)^2}} du~.$$

So the equation of the cylindrical helix can be written in the form
\begin{equation}\label{3.46}
x^1=f^1(u)~,~~~x^2=f^2(u)~,~~~x^3=\cot \alpha \int {\sqrt{\left(f^{1'}\right)^2+\left(f^{2'}\right)^2}} du~.
\end{equation}
{\bf Note\,:} $\alpha \neq \frac{\pi}{2}$ , otherwise the curve will be a plane curve in a normal section to the cylinder.\\

If the cylinder is a right circular cylinder then the helix is called a circular helix. So, if the circular cylinder is given by
$$x^1 = a \cos u~,~~~x^2 = a \sin u~,$$
then the circular helix has the equation
$$x^1 = a \cos u~,~~~x^2 = a \sin u~,~~~x^3 = a \cot \alpha u~,$$
where $a$ ($\neq 0$) , $\alpha \left(\neq \dfrac{\pi}{2}\right)$ are constants. Now putting $b = a \cot \alpha$, the equation of the circular helix may be written as
\begin{equation}\label{3.47}
x^1 = a \cos u~,~~~x^2 = a \sin u~,~~~x^3 = bu~.
\end{equation}

We shall now determine the tangent vector ($\textit{\textbf{t}}$) , the principal normal vector ($\textit{\textbf{n}}$) , the binormal vector ($\textit{\textbf{b}}$) , the curvature scalar ($\kappa$) and the torsion ($\tau$) for a circular helix :\\

For a circular helix,
$$\textit{\textbf{r}}=(a \cos u, a \sin u, bu)~.$$

Therefore,
\begin{eqnarray}
\frac{d\textit{\textbf{r}}}{du} &=& (-a \sin u, a \cos u, b)~, \nonumber \\
\textit{\textbf{t}} &=& \frac{\frac{d\textit{\textbf{r}}}{du}}{\left|\frac{d\textit{\textbf{r}}}{du}\right|}=\frac{1}{\sqrt{a^2+b^2}}(-a \sin u, a \cos u, b)~. \nonumber
\end{eqnarray}

Thus,
\begin{eqnarray}
\frac{ds}{du} &=& \bigg|\frac{d\textit{\textbf{r}}}{du}\bigg|=\sqrt{a^2+b^2} \nonumber \\
\frac{d\textit{\textbf{r}}}{ds} &=& \textit{\textbf{t}}=\frac{1}{\sqrt{a^2+b^2}}(-a \sin u, a \cos u, b) \nonumber \\
\frac{d^2\textit{\textbf{r}}}{ds^2} &=& \frac{1}{\sqrt{a^2+b^2}}(-a \cos u, -a \sin u, 0)\frac{1}{\sqrt{a^2+b^2}} \nonumber \\
&=& \frac{1}{a^2+b^2}(-a \cos u, -a \sin u, 0)~. \nonumber
\end{eqnarray}

Hence
\begin{eqnarray}
\kappa &=& \left|\frac{d^2\textit{\textbf{r}}}{ds^2}\right|= \frac{a}{a^2+b^2}~,~~a~\mbox{is positive} \nonumber \\
\textit{\textbf{n}} &=& \frac{\frac{d^2\textit{\textbf{r}}}{ds^2}}{\kappa}=(-\cos u, -\sin u, 0) \nonumber \\
\textit{\textbf{b}} &=& \textit{\textbf{t}} \times \textit{\textbf{n}}=(b \sin u, -b \cos u, a)\frac{1}{\sqrt{a^2+b^2}}~. \nonumber
\end{eqnarray}

Also
\begin{eqnarray}
\frac{d^2\textit{\textbf{r}}}{du^2} &=& (-a \cos u, -a \sin u, 0) \nonumber \\
\frac{d^3\textit{\textbf{r}}}{du^3} &=& (a \sin u, -a \cos u, 0) \nonumber
\end{eqnarray}

Therefore
\begin{eqnarray}
\tau &=& \frac{1}{\kappa ^2 s^6} \left[\frac{d\textit{\textbf{r}}}{du}~~~~\frac{d^2\textit{\textbf{r}}}{du^2}~~~~\frac{d^3\textit{\textbf{r}}}{du^3}\right] \nonumber \\
&=& \frac{\left(a^2+b^2\right)^2}{a^2}\frac{1}{\left(a^2+b^2\right)^3}  \begin{vmatrix}
-a\sin u & a\cos u & b \\
-a\cos u & -a\sin u & 0 \\
a\sin u & -a\cos u & 0 \end{vmatrix}  \nonumber \\
&=& \frac{b}{a^2+b^2}~. \nonumber
\end{eqnarray}

Thus $\kappa$ and $\tau$ are both constants for a circular helix.\\

{\bf Note:} In general, $\kappa$ and $\tau$ are given functions of the parameter of the curve. However, if $\kappa$ and $\tau$ are both constants then the curve is a circular helix.\\

{\bf Properties of a general cylindrical helix:}\\

Let $\textit{\textbf{m}}$ denotes the unit vector along the generator of the cylinder and $\alpha$ be the constant angle which the tangent vector makes with $\textit{\textbf{m}}$, \textit{i.e.}, the helix $H$ is characterized by the relation
\begin{equation}\label{3.48}
\textit{\textbf{t}} \cdot \textit{\textbf{m}}= \cos \alpha ~,~~\mbox{a constant.}
\end{equation}

Now differentiating with respect to the arc length $s$, we get
\begin{equation}\label{3.49}
\textit{\textbf{t}}' \cdot \textit{\textbf{m}}=0,~~~i.e.,~~\textit{\textbf{n}} \cdot \textit{\textbf{m}}=0~.
\end{equation}

Thus the principal normal is always perpendicular to the generators of the cylinder in which the helix lies. So the generator lies in the plane of $\textit{\textbf{t}}$ and $\textit{\textbf{b}}$. Further, as $\textit{\textbf{m}}$ makes an angle $\alpha$ with $\textit{\textbf{t}}$, so we can assume that it is so directed that it makes an angle $\dfrac{\pi}{2}-\alpha$ with $\textit{\textbf{b}}$. Then
\begin{equation}\label{3.50}
\textit{\textbf{b}} \cdot \textit{\textbf{m}}= \sin \alpha~.
\end{equation}

Again differentiating eq. (\ref{3.49}) with respect to $s$ , we have
\begin{eqnarray}\label{3.51}
\textit{\textbf{n}}' \cdot \textit{\textbf{m}}=0~,~~ i.e.,~~ (-\kappa \textit{\textbf{t}}+\tau \textit{\textbf{b}}) \cdot \textit{\textbf{m}} &=& 0 \nonumber \\
\mbox{or,}~~~ -\kappa \cos \alpha +\tau \sin \alpha &=& 0 \nonumber \\
i.e.,~~~~\frac{\kappa}{\tau} &=& \tan \alpha ~,~~\mbox{a constant.} 
\end{eqnarray}
Thus curvature and torsion are in a constant ratio.\\\\
Thus a general cylindrical helix has the following properties:\\

{\bf (a)} The tangent to the cylindrical helix makes a constant angle with a fixed direction, chosen as the generator of the cylinder.\\

{\bf (b)} The principal normal to the cylindrical helix is orthogonal to the fixed direction, \textit{i.e.}, perpendicular to the generator. In fact, the generator of the cylinder lies in the plane of $\textit{\textbf{t}}$ and $\textit{\textbf{b}}$.\\

{\bf (c)} The curvature and torsion bears a constant ratio at every point of the helix.\\
It is to be noted that any one of the above three conditions is sufficient for a twisted curve to be a cylindrical helix.
Now we shall show that any one of the above three conditions is sufficient for a twisted curve to be a cylindrical helix, \textit{i.e.}, eq.\,(\ref{3.48}) is satisfied.\\

Suppose for a twisted curve
\begin{eqnarray}
\textit{\textbf{n}} \cdot \textit{\textbf{m}} &=& 0 \nonumber \\
\Rightarrow \textit{\textbf{t}}' \cdot \textit{\textbf{m}} &=& 0 \nonumber \\
\Rightarrow \textit{\textbf{t}} \cdot \textit{\textbf{m}} &=& \mbox{constant}. \nonumber
\end{eqnarray}
Hence by eq. (\ref{3.48}) the curve is a cylindrical helix.\\

Suppose for a twisted curve
\begin{eqnarray}
\textit{\textbf{b}} \cdot \textit{\textbf{m}} &=& \mbox{constant} \nonumber \\
\Rightarrow \textit{\textbf{b}}' \cdot \textit{\textbf{m}} &=& 0 \nonumber \\
\Rightarrow \textit{\textbf{n}} \cdot \textit{\textbf{m}} &=& 0 \nonumber \\
\Rightarrow \textit{\textbf{t}} \cdot \textit{\textbf{m}} &=& \mbox{constant,} \nonumber
\end{eqnarray}
\textit{i.e.}, the curve is a cylindrical helix.\\

Suppose for a twisted curve
\begin{eqnarray}
&&\frac{\kappa}{\tau}=\rho ~,~~\mbox{a constant} \nonumber \\
&\Rightarrow& \kappa -\rho \tau = 0. \nonumber
\end{eqnarray}

By Frenet formulae, we have 
$$\textit{\textbf{t}}'=\kappa \textit{\textbf{n}}~~~\mbox{and}~~~\textit{\textbf{b}}'=-\tau \textit{\textbf{n}}~.$$

Therefore,
$$\textit{\textbf{t}}'+\rho \textit{\textbf{b}}'=0~~,~~i.e.,~~~\textit{\textbf{t}}+\rho \textit{\textbf{b}}=\textit{\textbf{l}}~,~~\mbox{a constant vector.}$$

Taking scalar product with $\textit{\textbf{t}}$, we get
$$\textit{\textbf{l}} \cdot  \textit{\textbf{t}}=1~.$$

Thus $\textit{\textbf{m}}=\dfrac{\textit{\textbf{l}}}{\left|\textit{\textbf{l}}\right|}$ is a constant unit vector such that
$$\textit{\textbf{m}} \cdot \textit{\textbf{t}}= \mbox{constant.}$$

Hence the twisted curve is a cylindrical helix.\\

{\bf Note:} $R=\dfrac{1}{\kappa}$ is called the radius of curvature and $T=\dfrac{1}{\tau}$ is called the radius of torsion.\\

\begin{center}
	\underline{-----------------------------------------------------------------------------------} 
\end{center}
\newpage
\vspace{5mm}
\begin{center}
	\underline{\bf Exercise} 
\end{center}
\vspace{3mm}
{\bf 3.1.} Show that~ $\kappa =\dfrac{\sqrt{\left|\ddot{\textit{\textbf{r}}}\right|^2-\ddot{s}^2}}{\dot{s}^2}$.\\\\
{\bf 3.2.} Show that for a circular helix, the principal normal is perpendicular to the generators of the cylinder.\\\\
{\bf 3.3.} Show that the arc length of a circular helix bears a constant ratio to that of its projection on the plane normal to the generator of the cylinder.\\\\
{\bf 3.4.} For the curve $\textit{\textbf{r}}=(3u, 3u^2, 2u^3)$, show that
$$R=T=\frac{3}{2}(1+2u^2)^2~.$$\\
{\bf 3.5.} For the curve $\textit{\textbf{r}}=\left(u, \dfrac{u^2}{2\alpha}, \dfrac{u^3}{6\alpha ^2}\right)$, show that
$$\kappa =\tau =\frac{4\alpha ^3}{(u^2+2\alpha ^2)^2}~.$$
{\bf 3.6.} Show that for the curve $\textit{\textbf{r}}=(u^3+3u, 3u^2, u^3-3u)$ , the curvature and the torsion are equal.\\\\
{\bf 3.7.}  Show that (a) $\tau \textit{\textbf{t}}'+\kappa \textit{\textbf{b}}'=0$~~;~~$\textit{\textbf{t}}' \cdot \textit{\textbf{b}}'=-\kappa \tau$ .\\
${}$~~~~~~~~~~~~~~~~~~~~~(b) $\textit{\textbf{b}}''=\tau (\kappa \textit{\textbf{t}}-\tau \textit{\textbf{b}})-\tau ' \textit{\textbf{n}}$ .\\
${}$~~~~~~~~~~~~~~~~~~~~~(c) $\textit{\textbf{n}}''=\tau ' \textit{\textbf{b}}-(\kappa ^2 +\tau ^2)\textit{\textbf{n}}-\kappa ' \textit{\textbf{t}}$ .\\\\
{\bf 3.8.} Show that~ $\dfrac{\left[\textit{\textbf{t}}', \textit{\textbf{t}}'', \textit{\textbf{t}}''' \right]}{\left[\textit{\textbf{b}}', \textit{\textbf{b}}'', \textit{\textbf{b}}''' \right]}=-\left(\dfrac{\kappa}{\tau}\right)^3$.\\\\
{\bf 3.9.} Show that $\left[T(R\textit{\textbf{r}}'')'\right]'+\left[\dfrac{T}{R}\textit{\textbf{r}}'\right]' +\dfrac{R}{T}\textit{\textbf{r}}'' =0$.\\\\
{\bf 3.10.} Find $\textit{\textbf{t}}$, $\textit{\textbf{n}}$, $\textit{\textbf{b}}$, $\kappa$, $\tau$ for the curve $\textit{\textbf{r}}=$($a\cos u$, $b\sin u$, $0$) , where $a$, $b$ are positive constants.\\\\
{\bf 3.11.} Show that a necessary and sufficient condition for a curve to be a straight line is $\kappa =0$ .\\\\
{\bf 3.12.} Show that a necessary and sufficient condition for a curve to be a plane curve is $\tau =0$.\\\\
{\bf 3.13.} Determine $f(u)$ so that the curve
$$x^1=a\cos u~,~~~x^2=a\sin u~,~~~x^3=f(u)$$
is a plane curve and find the nature of the curve.\\\\
{\bf 3.14.} Show that the torsion may be evaluated by the formula
$$\tau ^2=\frac{\left|\textit{\textbf{r}}'''\right|^2}{\kappa ^2}-\kappa ^2 -\left(\frac{\kappa '}{\kappa}\right)^2~.$$
{\bf 3.15.} Show that for a curve with non-vanishing curvature and torsion, the condition
$$\frac{R}{T}+(R'T)'=0$$
is equivalent to
$$\kappa ^3 \tau ^3 -\kappa ^2 \kappa '' \tau +2\kappa {\kappa '}^2\tau +\kappa ^2 \kappa ' \tau ' =0~.$$\\
{\bf 3.16.} Prove that the principal normal to the locus of the centre of spherical curvature is parallel to the principal normal to the original curve at the corresponding point.\\\\
{\bf 3.17.} Find the necessary and sufficient condition that a curve is a spherical curve.\\\\
{\bf 3.18.} If $M(\textit{\textbf{r}})$ describes a twisted curve of constant curvature and $M(\overline{\textit{\textbf{r}}})$ describes the locus of its centre of circular curvature then show that each curve is the locus of circular curvature of the other. Show also that they have the same curvature and this common curvature is the geometric mean of their torsions.\\\\
{\bf 3.19.} Show that if the principal normals of a curve be binormals of another then the curvature and torsion of the given curve must satisfy the relation
$$c(\kappa ^2+\tau ^2)=\kappa~~,~~\mbox{for some cosntant}~c~.$$
{\bf 3.20.} If all the osculating planes of a curve have a point in common then the curve is a plane curve.\\\\
{\bf 3.21.} If the $n^{th}$ derivative of $\textit{\textbf{r}}$ with respect to `$s$' is given by the recurrence relation:
$$\textit{\textbf{r}}^{\bm{(n)}}=a_n \textit{\textbf{t}}+b_n\textit{\textbf{n}}+c_n \textit{\textbf{b}}$$
then prove the following relations:\\
~~~~i) $a_{n+1}=a_n'-\kappa b_n$ ,\\
~~~~ii)$b_{n+1}=b_n'+\kappa a_n-\tau c_n$ ,\\
~~and~~iii) $c_{n+1}=c_n'+\tau b_n$ .\\\\
{\bf 3.22.} If the tangent and the binormal at a point of a curve make angles $\theta$ and $\phi$ respectively with a fixed direction then show that
$$\frac{\sin \theta \, d\theta}{\sin \phi \, d\phi}=-\frac{\kappa}{\tau}~.$$
{\bf 3.23.} Prove that principal normals at consecutive points do not intersect unless $\tau=0$ .\\\\
{\bf 3.24.} Prove that the position vector of the current point on a curve satisfies the differential equation:
$$\frac{d}{ds}\left\{T\frac{d}{ds}\left(R\frac{d^2 \textit{\textbf{r}}}{ds^2}\right)\right\}+\frac{d}{ds}\left(\frac{T}{R}\frac{d\textit{\textbf{r}}}{ds}\right)+\frac{R}{T}\frac{d^2 \textit{\textbf{r}}}{ds^2}=0~.$$
{\bf 3.25.} If $\bar{s}$ be the arc length of the locus of the centre of curvature, then show that
$$\left(\frac{d\bar{s}}{ds}\right)=\sqrt{(R')^2+\left(\frac{R}{T}\right)^2}~.$$\\

\begin{center}
	\underline{\bf Solution and Hints} 
\end{center}
\vspace{3mm}
{\bf Solution 3.1:}~~~~$\dot{\textit{\textbf{r}}} = \textit{\textbf{r}}' \dot{s}~,~~\ddot{\textit{\textbf{r}}}=\textit{\textbf{r}}'' \dot{s}^2 +\textit{\textbf{r}}' \ddot{s}$\\
Therefore
\begin{eqnarray}
\left|\ddot{\textit{\textbf{r}}}\right|^2 &=& \left|\textit{\textbf{r}}''\right|^2 \dot{s}^4+\left|\textit{\textbf{r}}'\right|^2 \ddot{s}^2 ~~(\mbox{since}~~\textit{\textbf{r}}' \cdot \textit{\textbf{r}}''=\textit{\textbf{t}} \cdot \kappa \textit{\textbf{n}}=0) \nonumber \\
&=&  |\textit{\textbf{r}}''|^2 \dot{s}^4+\ddot{s}^2~. \nonumber \\
&=& \kappa ^2 \dot{s}^4+\ddot{s}^2 \nonumber
\end{eqnarray}
Thus ~~~~~~~~~~~~~~~~~~~~$\kappa ^2 =\dfrac{\left|\ddot{\textit{\textbf{r}}}\right|^2-\ddot{s}^2}{\dot{s}^4}$\\\\
Hence ~~~~~~~~~~~~~~~~~~~$\kappa =\dfrac{\sqrt{\left|\ddot{\textit{\textbf{r}}}\right|^2-\ddot{s}^2}}{\dot{s}^2}$\\\\
{\bf Solution 3.2:} For a circular helix,
\begin{eqnarray}
\textit{\textbf{r}} &=& (a\cos u, a\sin u, bu) \nonumber \\
\Rightarrow \frac{d^2\textit{\textbf{r}}}{du^2} &=& (-a\cos u, -a\sin u, 0)~. \nonumber
\end{eqnarray}
So the principal normal $\textit{\textbf{n}}$ is along ($-\cos u, -\sin u, 0$). The generator of the cylinder is along (0, 0, 1). Hence $\textit{\textbf{n}}$ is orthogonal to the generators of the cylinder.\\\\
{\bf Solution 3.3:} Let the circular helix be given by
$$H:~\textit{\textbf{r}}=(a\cos u, a\sin u, bu)~.$$
So
$$\frac{d\textit{\textbf{r}}}{du}=(-a\sin u, a\cos u, b)~,~~i.e.,~~\left|\frac{d\textit{\textbf{r}}}{du}\right|=\sqrt{a^2+b^2}~.$$
Therefore,
\begin{eqnarray}
\textit{\textbf{t}} &=& \frac{d\textit{\textbf{r}}}{ds}=\frac{\frac{d\textit{\textbf{r}}}{du}}{|\frac{d\textit{\textbf{r}}}{du}|}=\frac{1}{\sqrt{a^2+b^2}}(-a\sin u, a\cos u, b) \nonumber \\
\Rightarrow \frac{ds}{du} &=& \sqrt{a^2+b^2}~. \nonumber
\end{eqnarray}
Now, a typical plane normal to the generator of the cylinder is the $x^1 x^2$-plane. So the projection of $H$ on $x^1 x^2$-plane is the curve
$$H^{*}: \textit{\textbf{r}}^{\bm *}=(a\cos u, a\sin u, 0)~.$$
If $s^{*}$ denote the arc length of this curve $H^{*}$ then from the above we have
\begin{eqnarray}
\frac{ds^{*}}{du} &=& a \nonumber \\
\Rightarrow \frac{ds^{*}}{ds} &=& \frac{\frac{ds^{*}}{du}}{\frac{ds}{du}}=\frac{a}{\sqrt{a^2+b^2}}~,~~\mbox{a constant.} \nonumber
\end{eqnarray}
Hence the result.\\\\
{\bf Solution 3.7:} (a) By Frenet's formula
$$\textit{\textbf{t}}=\kappa \textit{\textbf{n}}~~~\mbox{and}~~~\textit{\textbf{b}}'=-\tau \textit{\textbf{n}}~.$$
Therefore,
$$\tau \textit{\textbf{t}}'+\kappa \textit{\textbf{b}}'=0~.$$
Also,
$$\textit{\textbf{t}} ' \cdot \textit{\textbf{b}}'=\kappa \textit{\textbf{n}}(-\tau \textit{\textbf{n}})=-\kappa \tau~.$$
(b) \hspace*{5.1cm} $\textit{\textbf{b}}'=-\tau \textit{\textbf{n}}$
\begin{eqnarray}
\therefore ~~~~\textit{\textbf{b}}'' &=& -\tau \textit{\textbf{n}}'-\tau ' \textit{\textbf{n}} \nonumber \\
&=& -\tau (-\kappa \textit{\textbf{t}}+\tau \textit{\textbf{b}})-\tau ' \textit{\textbf{n}} \nonumber \\
&=& \tau (\kappa \textit{\textbf{t}}-\tau \textit{\textbf{b}})-\tau ' \textit{\textbf{n}}~. \nonumber 
\end{eqnarray}
(c) We have \hspace*{5.1cm} $\textit{\textbf{n}}'=-\kappa \textit{\textbf{t}}+\tau \textit{\textbf{b}}$ .
\begin{eqnarray}
\therefore ~~~~\textit{\textbf{n}}'' &=& -\kappa ' \textit{\textbf{t}}-\kappa \textit{\textbf{t}}'+\tau ' \textit{\textbf{b}}+\tau \textit{\textbf{b}}' \nonumber \\
&=& -\kappa ' \textit{\textbf{t}}-\kappa (\kappa \textit{\textbf{n}})+\tau ' \textit{\textbf{b}}+\tau (-\tau \textit{\textbf{n}}) \nonumber \\
&=& \tau ' \textit{\textbf{b}}-(\kappa ^2 +\tau ^2)\textit{\textbf{n}}-\kappa ' \textit{\textbf{b}}~. \nonumber
\end{eqnarray}
{\bf Solution 3.8:} 
\begin{eqnarray}
\textit{\textbf{t}}' &=& \kappa \textit{\textbf{n}} \nonumber \\
\textit{\textbf{t}}'' &=& \kappa ' \textit{\textbf{n}}+\kappa \textit{\textbf{n}}'=\kappa ' \textit{\textbf{n}}+\kappa (-\kappa \textit{\textbf{t}}+\tau \textit{\textbf{b}})=-\kappa ^2\textit{\textbf{t}}+\kappa '\textit{\textbf{n}}+\kappa \tau \textit{\textbf{b}} \nonumber \\
\textit{\textbf{t}}''' &=& -3\kappa \kappa ' \textit{\textbf{t}}-\left(\kappa ^3+\kappa \tau ^2 -\kappa ''\right)\textit{\textbf{n}}+(2\kappa ' \tau +\kappa \tau ')\textit{\textbf{b}}~. \nonumber
\end{eqnarray}
Then,
$$\left[\textit{\textbf{t}}', \textit{\textbf{t}}'', \textit{\textbf{t}}''' \right]=-\kappa ^3 (\kappa ' \tau -\kappa \tau ')~.$$
Similarly,
$$\left[\textit{\textbf{b}}', \textit{\textbf{b}}'', \textit{\textbf{b}}''' \right]=\tau ^3 (\kappa ' \tau -\kappa \tau ')~.$$
Thus,
$$\frac{\left[\textit{\textbf{t}}', \textit{\textbf{t}}'', \textit{\textbf{t}}''' \right]}{\left[\textit{\textbf{b}}', \textit{\textbf{b}}'', \textit{\textbf{b}}''' \right]}=-\left(\frac{\kappa}{\tau}\right)^3~,~~\mbox{provided}~~\kappa ' \tau -\kappa \tau ' \neq 0~.$$
{\bf Note:} If ~$\kappa ' \tau -\kappa \tau ' =0~$ then ~$\dfrac{\kappa}{\tau}=$\,constant and the curve is a helix. Both the scalar triple product vanish identically.\\\\
{\bf Solution 3.9:}
\begin{eqnarray}
LHS &=& \left[\frac{1}{\tau} \left\lbrace \frac{1}{\kappa} (\kappa \textit{\textbf{n}}) \right\rbrace ' \right]'+\left[\frac{\kappa}{\tau}\textit{\textbf{t}}\right]'+\frac{\tau}{\kappa}(\kappa \textit{\textbf{n}}) \nonumber \\
&=& \left[\frac{1}{\tau} \textit{\textbf{n'}}\right]'+\left[\frac{\kappa}{\tau}\textit{\textbf{t}}\right]'+\tau \textit{\textbf{n}} \nonumber \\
&=& \left[\frac{1}{\tau} (-\kappa \textit{\textbf{t}}+\tau \textit{\textbf{b}})\right]'+\left(\frac{\kappa}{\tau}\right)'\textit{\textbf{t}}+\frac{\kappa}{\tau}\textit{\textbf{t'}}+\tau \textit{\textbf{n}} \nonumber \\
&=& -\left(\frac{\kappa}{\tau}\textit{\textbf{t}}\right)' +\textit{\textbf{b'}} +\left(\frac{\kappa}{\tau}\right)'\textit{\textbf{t}}+\frac{\kappa ^2}{\tau}\textit{\textbf{n}}+\tau \textit{\textbf{n}} \nonumber \\
&=& -\left(\frac{\kappa}{\tau}\right)'\textit{\textbf{t}}-\frac{\kappa}{\tau}\kappa \textit{\textbf{n}}-\tau \textit{\textbf{n}}+\left(\frac{\kappa}{\tau}\right)'\textit{\textbf{t}}+\frac{\kappa ^2}{\tau}\textit{\textbf{n}}+\tau \textit{\textbf{n}} \nonumber \\
&=& 0~. \nonumber
\end{eqnarray}\\
{\bf Solution 3.11:} Let the curve be
$$x^1=x^1(u)~,~~~x^2=x^2(u)~,~~~x^3=x^3(u)~.$$
We know that
$$\kappa =\frac{\left|\textit{\textbf{r}}' \times \textit{\textbf{r}}''\right|}{\left|\textit{\textbf{r}}'\right|^3}~~~(\,'~\mbox{stands for differentiation with respect to}~u).$$
So, 
$$\kappa =0~~~ \Rightarrow ~~ \left|\textit{\textbf{r}}' \times \textit{\textbf{r}}''\right|=0 ~~~ \Rightarrow ~~ \textit{\textbf{r}}' \times \textit{\textbf{r}}'' =0$$
\begin{equation}\label{3.52}
\Rightarrow ~~ \frac{{x^1}''}{{x^1}'}=\frac{{x^2}''}{{x^2}'}=\frac{{x^3}''}{{x^3}'}=\xi '(u)~~~(\mbox{say}).
\end{equation}
We shall prove that eq.\,(\ref{3.52}) is also a necessary and sufficient condition for the curve to be a straight line.\\\\
Suppose equation (\ref{3.52}) holds. So on integration, we have
$${x^i}'=a^i e^{\xi (u)}~,~~~i=1, 2, 3~.$$
Integrating once more,
$$x^i=a^1 \psi (u)+b^i~,~~~\psi (u)=\int {e^{f(u)}du}~.$$
Thus we have
$$\frac{x^1-b^1}{a^1}=\frac{x^2-b^2}{a^2}=\frac{x^3-b^3}{a^3}~,$$
\textit{i.e.}, the curve is a straight line.\\\\
Conversely, let the curve be a straight line. Then its equation can be written as
$$\frac{x^1-b^1}{a^1}=\frac{x^2-b^2}{a^2}=\frac{x^3-b^3}{a^3}=\psi (u)~~(\mbox{say}),$$
\begin{eqnarray}
i.e.,~~~~ x^i &=& a^i \psi (u)+b^i~,~~~i=1, 2, 3 \nonumber \\
\Rightarrow ~~ {x^i}' &=& a^i \psi '(u)~,~~~{x^i}''=a^i \psi ''(u) \nonumber
\end{eqnarray}
Therefore,
$$\frac{{x^i}''}{{x^i}'}=\frac{\psi ''(u)}{\psi '(u)}=\xi (u)~~(\mbox{say}),$$
$$i.e.,~~~\frac{{x^1}''}{{x^1}'}=\frac{{x^2}''}{{x^2}'}=\frac{{x^3}''}{{x^3}'}~, $$
which is relation (\ref{3.52}).\\\\
Hence $\kappa =0$ is both necessary and sufficient condition for a curve to be a straight line.\\\\
{\bf Solution 3.12:} Let the curve be a plane curve. Then the osculating planes are constant, being the same as the plane of the curve. The binormal vector $\textit{\textbf{b}}$ is therefore a constant vector. Hence
$$\textit{\textbf{b}}'=0~,~~~i.e.,~~ -\tau \textit{\textbf{n}}=0~,~~~i.e.,~~ \tau =0~.$$
Conversely, let $\tau =0$, then $\textit{\textbf{b}}'=-\tau \textit{\textbf{n}}=0$. So $\textit{\textbf{b}}$ is a constant vector. Let
\begin{eqnarray}
b^i &=& \lambda ^i \nonumber \\
\Rightarrow ~~ \frac{d}{ds}\left(\sum b^i x^i\right) &=& \Sigma {b^i}'x^i+\Sigma b^i t^i = 0 \nonumber \\
\Rightarrow ~~~~~ \sum b^i x^i &=& \mbox{constant}=C \nonumber \\
\Rightarrow ~~ x^1 \lambda ^1+x^2 \lambda ^2+x^3 \lambda ^3 &=& C~,~~\mbox{a plane.} \nonumber
\end{eqnarray}
Hence the curve is a plane curve.\\\\
{\bf Solution 3.13:} The curve lies on the circular cylinder 
\begin{equation}\label{3.53}
x^1=a\cos u~,~~~x^2=a\sin u~.
\end{equation}
Since the projection of the curve in the $x^1x^2$-plane is the circle given by eq. (\ref{3.53}) so if the curve has to be a plane curve, it must be a plane section of the circular cylinder by a plane not parallel to the $x^3$-axis, \textit{i.e.}, the axis of the cylinder.\\\\
Now the necessary and sufficient condition that the curve is a plane curve is
\begin{eqnarray}
&&\left[\textit{\textbf{r}}', \textit{\textbf{r}}'', \textit{\textbf{r}}'''\right] = 0~, \nonumber \\
&i.e.,&~{\left| \begin{array}{ccc}
	-a\sin u & a\cos u & f'(u) \\
	-a\cos u & -a\sin u & f''(u) \\
	a\sin u & -a\cos u & f'''(u) \end{array} \right| } = 0~, \nonumber \\
&i.e.,&~f'''(u)+f'(u) = 0. \nonumber
\end{eqnarray}
Hence the solution for $f(u)$ is
$$f(u)=a_1 \cos u +b_1 \sin u +c_1~,$$
where $a_1$, $b_1$ and $c_1$ are arbitrary constants.\\\\
{\bf Solution 3.14:}
\begin{eqnarray}
\textit{\textbf{t}} &=& \frac{d\textit{\textbf{r}}}{ds}=\textit{\textbf{r}}' \nonumber \\
\Rightarrow ~~\textit{\textbf{r}}'' &=& \frac{d\textit{\textbf{t}}}{ds}=\textit{\textbf{t'}}=\kappa \textit{\textbf{n}} \nonumber \\
\Rightarrow ~~\textit{\textbf{r}}''' &=& \kappa '\textit{\textbf{n}}+\kappa \textit{\textbf{n'}}=\kappa ' \textit{\textbf{n}}+\kappa (-\kappa \textit{\textbf{t}}+\tau \textit{\textbf{b}}) \nonumber \\
\Rightarrow \left|\textit{\textbf{r}}'''\right|^2 &=& \kappa ^4+(\kappa ')^2+\kappa ^2 \tau ^2 \nonumber \\
\Rightarrow ~~\tau ^2 &=& \frac{\left|\textit{\textbf{r}}''\right|^2}{\kappa ^2}-\kappa ^2 -\left(\frac{\kappa '}{\kappa}\right)^2~. \nonumber
\end{eqnarray}\\
{\bf Solution 3.15:} The condition $\dfrac{R}{T}+(R'T)'=0$ is equivalent to $\dfrac{\tau}{\kappa}+\left(-\dfrac{\kappa '}{\kappa ^2 \tau}\right)'=0$ and this gives the result.\\\\
{\bf Solution 3.16:} Let $C(\textit{\textbf{r}})$ and $\overline{C}(\overline{\textit{\textbf{r}}})$ be the given curve and the locus of its centre of spherical curvature respectively with
$$\overline{\textit{\textbf{r}}}=\textit{\textbf{r}}+R \textit{\textbf{n}}+(R'T)\textit{\textbf{b}}~.$$
Now differentiating with respect to $s$, we get
\begin{eqnarray}
\overline{\textit{\textbf{t}}}\frac{d\overline{s}}{ds} &=& \textit{\textbf{t}}+R'\textit{\textbf{n}}+R\left(-\frac{1}{R}\textit{\textbf{t}}+\frac{1}{T}\textit{\textbf{b}}\right)+(R'T)'\textit{\textbf{b}}+(R'T)\left(-\frac{1}{T}\textit{\textbf{n}}\right) \nonumber \\
&=& \left[\frac{R}{T}+(R'T)'\right]\textit{\textbf{b}}~, \nonumber
\end{eqnarray}
where $\overline{s}$ is the arc length of the locus of the centre of spherical curvature. So we can write
$$\overline{\textit{\textbf{t}}}=\lambda \textit{\textbf{b}}~.$$
Again differentiating with respect to $s$, we get
$$(\overline{\kappa}\overline{\textit{\textbf{n}}})\frac{d\overline{s}}{ds}=-\lambda \tau \textit{\textbf{n}}~,$$
\textit{i.e.}, $\overline{\textit{\textbf{n}}}$ is parallel to $\textit{\textbf{n}}$ .\\\\
{\bf Solution 3.17:} A curve is a spherical curve if and only if its osculating sphere is a constant sphere. For the osculating sphere $S$, the centre $\textit{\textbf{w}}$ and radius $\sigma$ are given by
$$\textit{\textbf{w}}=\textit{\textbf{r}}+R\textit{\textbf{n}}+(R'T)\textit{\textbf{b}}~~~\mbox{and}~~~\sigma ^2=R^2+(R'T)^2~.$$
The osculating sphere is a constant sphere if its centre is fixed and radius is constant, \textit{i.e.},
$$\textit{\textbf{w}}'=0~~~\mbox{and}~~~\sigma '=0~.$$
Now,
\begin{eqnarray}
\textit{\textbf{w}}' &=& \frac{d\textit{\textbf{w}}}{ds}=\frac{d\textit{\textbf{r}}}{ds}+R'\textit{\textbf{n}}+R \frac{d\textit{\textbf{n}}}{ds}+(R'T)'\textit{\textbf{b}}+(R'T)\frac{d\textit{\textbf{b}}}{ds} \nonumber \\
&=& \textit{\textbf{t}}+R'\textit{\textbf{n}}+R(-\kappa \textit{\textbf{t}}+\tau \textit{\textbf{b}})+(R'T)'\textit{\textbf{b}}+(R'T)\left(-\frac{1}{T}\textit{\textbf{n}}\right) \nonumber \\
&=& \textit{\textbf{t}}+R'\textit{\textbf{n}}-\textit{\textbf{t}}+\frac{R}{T}\textit{\textbf{b}}+(R'T)'\textit{\textbf{b}}-R'\textit{\textbf{n}} \nonumber \\
&=& \left[\frac{R}{T}+(R'T)'\right]\textit{\textbf{b}} \nonumber
\end{eqnarray}
and
$$(\sigma ^2)'=2\sigma \sigma '=2RR'+2(R'T)(R'T)'=2(R'T)\left[\frac{R}{T}+(R'T)'\right]~.$$
Thus we see that the centre is fixed and the radius is constant if and only if 
$$\frac{R}{T}+(R'T)'=0~.$$
This is the necessary and sufficient condition for the curve to be a spherical curve.\\\\
{\bf Solution 3.18:} We have
$$\overline{\textit{\textbf{r}}}=\textit{\textbf{r}}+\frac{1}{\kappa}\textit{\textbf{n}}~~~(\kappa =\mbox{constant}).$$
Differentiating with respect to $s$, the arc length of the twisted curve $M$, we get
$$\overline{\textit{\textbf{t}}}\frac{d\overline{s}}{ds}=\overline{t}+\frac{1}{\kappa}(-\kappa \textit{\textbf{t}}+\tau \textit{\textbf{b}})=\frac{\tau}{\kappa}\textit{\textbf{b}}~.$$
As both $\textit{\textbf{t}}$ and $\textit{\textbf{b}}$ are unit vectors, so choosing the direction of increment of $\overline{s}$ properly, we have
$$\frac{d\overline{s}}{ds}=\frac{\tau}{\kappa}~~~\mbox{and}~~~\overline{\textit{\textbf{t}}}=\textit{\textbf{b}}~.$$
Now differentiating the above second relation with respect to $s$ , we have
$$\overline{\kappa}\overline{\textit{\textbf{n}}}\frac{d\overline{s}}{ds}=-\tau \textit{\textbf{n}}~,~~i.e.,~~ \overline{\kappa}\overline{\textit{\textbf{n}}}\left(\frac{\tau}{\kappa}\right)=-\tau \textit{\textbf{n}}~,~~i.e., \overline{\kappa}\overline{\textit{\textbf{n}}}=-\kappa\textit{\textbf{n}}
~~(\mbox{Since}~\tau \neq 0).$$
As both $\kappa$ and $\overline{\kappa}$ are positive, so we have
$$\overline{\kappa}=\kappa~~~\mbox{and}~~~\overline{\textit{\textbf{n}}}=
-\textit{\textbf{n}}~.$$
Again,
$$\overline{\textit{\textbf{b}}}=\overline{\textit{\textbf{t}}} \times \overline{\textit{\textbf{n}}}=\textit{\textbf{b}} \times (-\textit{\textbf{n}})=\textit{\textbf{t}}~.$$
Differentiating this relation, we have
\begin{eqnarray}
-\overline{\tau}\overline{\textit{\textbf{n}}}\frac{d\overline{s}}{ds} &=& \textit{\textbf{t}}'=\kappa \textit{\textbf{n}} \nonumber \\
i.e.,~-\overline{\tau}\overline{\textit{\textbf{n}}}\frac{\tau}{\kappa} &=& \kappa \textit{\textbf{n}} \nonumber \\
i.e.,~~~~~~~\kappa ^2~ &=& \tau \overline{\tau} \nonumber \\
i.e.,~~~~~~~\overline{\kappa}~~ &=& \kappa =\sqrt{\tau \overline{\tau}}~. \nonumber
\end{eqnarray}
Also the position vector $\overline{\overline{\textit{\textbf{r}}}}$ of the centre of circular curvature of the second curve is
$$\overline{\overline{\textit{\textbf{r}}}}=\overline{\textit{\textbf{r}}}+\frac{1}{\overline{\kappa}}\overline{\textit{\textbf{n}}}=\left(\textit{\textbf{r}}+\frac{1}{\kappa}\textit{\textbf{n}}\right)-\frac{1}{\kappa}\textit{\textbf{n}}=\textit{\textbf{r}}~.$$
Hence the result.\\\\
{\bf Solution 3.19:} Let the principal normal of a curve $C(\textit{\textbf{r}})$ be the binormal of another curve $\overline{C}(\overline{\textit{\textbf{r}}})$ . So we write
\begin{equation}\label{3.54}
\overline{\textit{\textbf{r}}}=\textbf{\textit{r}}+\lambda \textit{\textbf{n}}~,
\end{equation}
where $\lambda$ is some scalar function of the arc length $s$ and $\textit{\textbf{n}}$ at $\textit{\textbf{r}}$ and $\overline{\textit{\textbf{b}}}$ at $\overline{\textit{\textbf{r}}}$ are collinear. Now differentiating eq.\,(\ref{3.54}) with respect to the arc length $s$ of $C$, we get
\begin{eqnarray}\label{3.55}
\overline{\textit{\textbf{t}}} \frac{d\overline{s}}{ds} &=& \textit{\textbf{t}}+\lambda '\textit{\textbf{n}}+\lambda (-\kappa \textit{\textbf{t}}+\tau \textit{\textbf{b}}) \nonumber \\
&=& (1-\kappa \lambda)\textit{\textbf{t}}+\lambda '\textit{\textbf{n}}+\lambda \tau \textit{\textbf{b}}~.
\end{eqnarray}
As $\overline{\textit{\textbf{t}}}$ is orthogonal to $\overline{\textit{\textbf{b}}}$, which is parallel to $\textit{\textbf{n}}$, so $\overline{\textit{\textbf{t}}}$ is perpendicular to $\textit{\textbf{n}}$. So from the above eq. (\ref{3.55}), we have
$$\lambda '=0~,~~i.e.,~~ \lambda =~\mbox{a constant.}$$
Again differentiating (\ref{3.55}) with respect to $s$ , we have
\begin{eqnarray}
\overline{\textit{\textbf{t}}}\frac{d^2\overline{s}}{ds^2}+\overline{\kappa}\overline{\textit{\textbf{n}}} \left(\frac{d\overline{s}}{ds}\right)^2 &=& -\kappa ' \lambda \textit{\textbf{t}}+(1-\kappa \lambda)\kappa \textit{\textbf{n}}+\lambda \tau '\textit{\textbf{b}}+\lambda \tau (-\tau \textit{\textbf{n}}) \nonumber \\
&=& -\lambda \kappa '\textit{\textbf{t}}+(\kappa -\kappa ^2\lambda -\lambda \tau ^2)\textit{\textbf{n}}+\lambda \tau '\textit{\textbf{b}}~. \nonumber
\end{eqnarray}
Note that the L.H.S. is orthogonal to $\overline{\textit{\textbf{b}}}$, \textit{i.e.}, to $\textit{\textbf{n}}$, hence the component of $\textit{\textbf{n}}$ should be zero in the R.H.S.. So we have
\begin{eqnarray}
\kappa -\kappa ^2\lambda -\lambda \tau ^2 &=& 0 \nonumber \\
i.e.,~~~~~~\lambda (\kappa ^2+\tau ^2) &=& \kappa . \nonumber
\end{eqnarray}
Hence the result.\\\\
{\bf Solution 3.20:} The equation of the osculating plane at the point $P(\textit{\textbf{r}})$ of a curve may be written as
$$\left[\textit{\textbf{R}}-\textit{\textbf{r}},\textit{\textbf{t}},\textit{\textbf{n}}\right]=0~,$$
where $\textit{\textbf{R}}$ is the current point on the osculating plane. Let $A(\textit{\textbf{a}})$ be the common point of all the osculating planes. Then we have
\begin{equation}\label{3.56}
\left[\textit{\textbf{a}}-\textit{\textbf{r}},\textit{\textbf{t}},\textit{\textbf{n}}\right]=0~.
\end{equation}
Now differentiating the above equation with respect to `$s$' we get 
$$\left[-\textit{\textbf{t}},\textit{\textbf{t}},\textit{\textbf{n}}\right]+\left[\textit{\textbf{a}}-\textit{\textbf{r}}, \kappa \textit{\textbf{n}}, \textit{\textbf{n}}\right]+\left[\textit{\textbf{a}}-\textit{\textbf{r}}, \textit{\textbf{t}}, -\kappa \textit{\textbf{t}}+\tau \textit{\textbf{b}} \right]=0$$
$$i.e.,~~~~~~ \tau \left[\textit{\textbf{a}}-\textit{\textbf{r}}, \textit{\textbf{t}}, \textit{\textbf{b}}\right]=0~.$$
If $\tau =0$ then the curve is a plane curve, otherwise we have 
\begin{equation}\label{3.57}
\left[\textit{\textbf{a}}-\textit{\textbf{r}}, \textit{\textbf{t}}, \textit{\textbf{b}}\right]=0~.
\end{equation}
The relations (3.56) and (3.57) suggest that:
$\textit{\textbf{a}}-\textit{\textbf{r}}$ is orthogonal to $\textit{\textbf{t}} \times \textit{\textbf{n}}~~i.e.,~~ \textit{\textbf{a}}-\textit{\textbf{r}} \perp \textit{\textbf{b}}$ and $\textit{\textbf{a}}-\textit{\textbf{r}}$ is orthogonal to $\textit{\textbf{t}} \times \textit{\textbf{b}}$~~ \textit{i.e.}, $\textit{\textbf{a}}-\textit{\textbf{r}} \perp \textit{\textbf{n}}$.\\\\
Hence $\textit{\textbf{a}}-\textit{\textbf{r}}$ is parallel to $\textit{\textbf{t}}$ \textit{i.e.}, 
\begin{equation}\label{3.58}
\frac{dx^1}{x^1-a^1}=\frac{dx^2}{x^2-a^2}=\frac{dx^3}{x^3-a^3}=\lambda'~~(\mbox{say})
\end{equation} 
where $\textit{\textbf{a}}=(a^1,a^2,a^3)$ is a given vector.\\\\
Now integrating equation (\ref{3.58}) we get 
$$\log |x^i-a^i|=\lambda(s)+\log c^i$$
$$\mbox{or,}~~~~\frac{x^i-a^i}{c^i}=e^{\lambda(s)}$$
or equivalently, 
$$\frac{x^1-a^1}{c^1}=\frac{x^2-a^2}{c^2}=\frac{x^3-a^3}{c^3}$$
\textit{i.e.}, the curve is a plane curve.\\\\
{\bf Solution 3.21:}~~ We have
\begin{eqnarray}
\textit{\textbf{r}}^{(n)} &=& a_n \textit{\textbf{t}}+b_n\textit{\textbf{n}}+c_n \textit{\textbf{b}} \nonumber \\
\textit{\textbf{r}}^{(n+1)} &=& a_n' \textit{\textbf{t}}+a_n (\kappa \textit{\textbf{n}})+b_n' \textit{\textbf{n}}+b_n (-\kappa \textit{\textbf{t}}+\tau \textit{\textbf{b}})+c_n' \textit{\textbf{b}}+c_n (-\tau \textit{\textbf{n}}) \nonumber \\
&=& (a_n'-\kappa b_n)\textit{\textbf{t}}+(\kappa a_n+b_n'-\tau c_n)\textit{\textbf{n}}+(\tau b_n +c_n')\textit{\textbf{b}}~. \nonumber
\end{eqnarray}
But
$$\textit{\textbf{r}}^{\bm {n+1}}=a_{n+1} \textit{\textbf{t}} +b_{n+1} \textit{\textbf{n}} +c_{n+1}\textit{\textbf{b}}~.$$
Hence comparing the co-efficients we have 
$$a_{n+1}=a_n'-\kappa b_n~~,~~b_{n+1}=b_n'+\kappa a_n -\tau c_n~~,~~c_{n+1}=c_n'+\tau b_n~.$$\\
{\bf Solution 3.22:} ~~Let $\textit{\textbf{a}}$ be the unit vector along the given fixed direction. Then $\textit{\textbf{t}} \cdot \textit{\textbf{a}}=\cos \theta$ and $\textit{\textbf{b}} \cdot \textit{\textbf{a}}=\cos \phi$. Thus we have
$$\textit{\textbf{t}}' \cdot \textit{\textbf{a}}=-\sin \theta \, \frac{d\theta}{ds}~~\mbox{and}~~\textit{\textbf{b}}' \cdot \textit{\textbf{a}}=-\sin \phi \, \frac{d \phi }{ds}$$
$$\mbox{or}~~~~\kappa (\textit{\textbf{n}} \cdot \textit{\textbf{a}})=-\sin \theta \, \frac{d \theta}{ds}~~\mbox{and}~~ -\tau (\textit{\textbf{n}} \cdot \textit{\textbf{a}})=-\sin \phi \, \frac{d \phi }{ds}$$
$$\therefore ~~~~ \frac{\kappa}{\tau}=-\frac{\sin \theta \, d\theta}{\sin \phi \, d \phi}~.~~(\mbox{proved})$$\\
{\bf Solution 3.23:} ~~Let the consecutive points on the curve be $\textit{\textbf{r}}$ and $\textit{\textbf{r}}+d\textit{\textbf{r}}$ and the principal normals be $\textit{\textbf{n}}$ and $\textit{\textbf{n}}+d\textit{\textbf{n}}$ respectively. For intersection of the principal normals a necessary condition is that the 3 vectors $d\textit{\textbf{r}}~,~\textit{\textbf{n}}$ and $\textit{\textbf{n}}+d\textit{\textbf{n}}$ must be coplanar, \textit{i.e.}, 
$$\frac{d\textit{\textbf{r}}}{ds}~,~\textit{\textbf{n}}~,~\frac{d\textit{\textbf{n}}}{ds}~~\textrm{must be coplaner}.$$
$$i.e.,~~~~\left[\textit{\textbf{t}},~\textit{\textbf{n}}, -\kappa \textit{\textbf{t}}+\tau \textit{\textbf{b}}\right]=0~~~\Rightarrow ~~ \tau =0 $$\\
{\bf Solution 3.24:}
\begin{eqnarray}
LHS &=& \left\{T\left(R\textit{\textbf{t}}'\right)'\right\}'+\left(\frac{T}{R}\textit{\textbf{t}}\right)'+\frac{R}{T}\left(\textit{\textbf{t}}'\right) \nonumber \\
&=& \left\{T(R.\frac{1}{R}\textit{\textbf{n}})'\right\}'+\left(\frac{T}{R}\textit{\textbf{t}}\right)'+\frac{R}{T} \cdot \frac{1}{R}\textit{\textbf{n}} \nonumber \\
&=& \left\{T\left(-\frac{1}{R}\textit{\textbf{t}}+\frac{1}{T}\textit{\textbf{b}}\right)\right\}'+\left(\frac{T}{R}\textit{\textbf{t}}\right)'+\frac{1}{T}\textit{\textbf{n}} \nonumber \\
&=& -\left(\frac{T}{R}\textit{\textbf{t}}\right)'+\textit{\textbf{b}}'+\left(\frac{T}{R}\textit{\textbf{t}}\right)'-\textit{\textbf{b}}'~~=~0~~=RHS.
\end{eqnarray}\\
{\bf Solution 3.25:}
$$\bar{\textit{\textbf{r}}}=\textit{\textbf{r}}+R\textit{\textbf{n}}$$
So differentiating with respect to `$s$' we have 
$$\bar{\textit{\textbf{t}}}\frac{d\bar{s}}{ds}=\textit{\textbf{t}}+R'\textit{\textbf{n}}+R\left(-\frac{1}{R}\textit{\textbf{t}}+\frac{1}{T}\textit{\textbf{b}}\right)=R'\textit{\textbf{n}}+\frac{R}{T}\textit{\textbf{b}}$$
Squaring both sides we have
$$\left(\frac{d\bar{s}}{ds}\right)^2=R'^2+\left(\frac{R}{T}\right)^2~,~~\mbox{(proved).}$$\\\\


\chapter{Hypersurface in a Riemannian space}


\section{Basic Definition}

 A `$n$' dimensional hypersurface $V_n$ in an $(m+1)$-dimensional Riemannian space $M$ is given by the equations
$$y^\alpha=f^\alpha (x^1, x^2,\ldots ,x^n),~~\alpha=1,2,\ldots m+1$$
where $\{y^\alpha\}$ is a co-ordinate system in $M$ and $x^i$'s are $n$ real variables such that the Jacobian matrix
\[
J = \left[\frac{\partial y}{\partial x} \right]
=\left[
\begin{array}{cccc}
\frac{\partial y^1}{\partial x^1} & \frac{\partial y^1}{\partial x^2} & \ldots & \frac{\partial y^1}{\partial x^n}\\
\frac{\partial y^2}{\partial x^1} & \frac{\partial y^2}{\partial x^2} & \ldots & \frac{\partial y^2}{\partial x^n}\\
\vdots & \vdots & \ddots & \vdots \\
\frac{\partial y^{m+1}}{\partial x^1} & \frac{\partial y^{m+1}}{\partial x^2} & \ldots & \frac{\partial y^{m+1}}{\partial x^n}
\end{array}
\right]
\]
is of rank $n$.

In particular if $m>n$ then $V_n$ is called a subspace of $M$ or $M$ is called an enveloping space of $V_n$ . For $m=n$ , $V_n$ is called hypersurface of the enveloping space $M$.\\\\
{\bf Note:} The $n$ real variables $\{x^i\}$ is a co-ordinate system in $V_n$.\\\\
{\bf Induced metric in $V_n$:}

Let $a_{\alpha \beta}$ be the components of the metric tensor in $M$ under some $y-$co-ordinate system. Then the elementary distance `$dS$' between two neighbouring points in $V_n$ (which are therefore also in $M$) is given by
\begin{eqnarray}
dS^2=a_{\alpha \beta}dy^\alpha dy^\beta &=& a_{\alpha \beta}\left(\frac{\partial y^\alpha}{\partial x^i} dx^i\right) \left(\frac{\partial y^\beta}{\partial x^k} dx^k \right) \nonumber \\
&=& \left(a_{\alpha \beta}\frac{\partial y^\alpha}{\partial x^i}\frac{\partial y^\beta}{\partial x^k} \right)dx^i dx^k \nonumber \\
&=& g_{ik}dx^i dx^k~, \nonumber
\end{eqnarray}
where 
\begin{equation}\label{4.1}
g_{ik}=a_{\alpha \beta}\frac{\partial y^\alpha}{\partial x^i}\frac{\partial y^\beta}{\partial x^k}
\end{equation}
is the metric tensor in $V_n$.\\\\
{\bf Note:} Similar to $a_{\alpha \beta}$ , $g_{ik}$ is also symmetric in `$i$' and `$k$'.\\\\
{\bf Normal to the hypersurface:}\\

Let $N^\alpha$ be the contravariant components (in the $y-$co-ordinate system) of the unit normal $\textit{\textbf{N}}$ to $V_n \subset M$. For fixed $i~(i=1,2, \ldots ,n)$ the vector $\frac{\partial y^\alpha}{\partial x^i}$ is tangential to $V_n$ and hence orthogonal to the normal vector $\textit{\textbf{N}}$.

For another co-ordinate system $\bar{y}^\alpha$ in $M$ we write
$$\frac{\partial y^\alpha}{\partial x^i}=\frac{\partial y^\alpha}{\partial \bar{y}^\beta}\frac{\partial \bar{y}^\beta}{\partial x^i}~,$$
which shows that the tangential vector is a contravariant vector in $M$. Thus the orthogonality of normal vector $\textit{\textbf{N}}$ and the above tangent vector gives
\begin{equation}\label{4.2}
a_{\alpha \beta}\nabla _i y^\alpha N^\beta=0
\end{equation}
and the normalization of $\textit{\textbf{N}}$ gives
\begin{equation}\label{4.3}
a_{\alpha \beta}N^\alpha N^\beta=1~.
\end{equation}
with $\bigtriangledown_{_i}y^{\alpha}=\dfrac{\partial y^{\alpha}}{\partial x^i}$.

\section{Generalized Intrinsic and Covariant Differentiation\,: Differentiation on the hypersurface}

Let us start with the symbol convention : any Greek index stands for tensor character in $M$ while any Latin index denotes tensor character in the hypersurface $V_n$ . As we have seen $\nabla_i y^\alpha=\dfrac{\partial y^\alpha}{\partial x^i}$ , for fixed $i$ represents a contravariant vector in $y-$co-ordinates in $M$, so in a similar way, for fixed $\alpha$ , $\nabla_i y^\alpha$ is a covariant vector in the $x-$coordinates in $V_n$ .

Suppose `$s$' be the arc length along any curve $\gamma$ in $V_n$ and $A^\alpha_{\beta i}$ be an arbitrary tensor field along $\gamma$. According to the above symbol convention the tensor field $A^\alpha_{\beta i}$ is a (1, 1)-tensor in $y-$coordinates in $M$ and it is a (0, 1)-tensor in the $x-$coordinates in $V_n$. Let $u_\alpha$ , $v^\beta$ be the components in the $y-$coordinates of two unit vector fields parallel to $\gamma$ with respect to $M$ and $\omega^i$ be the components in the $x-$coordinates of a unit vector field parallel to $\gamma$ with respect to $V_n$ . So we have
\begin{eqnarray}
\frac{\delta u_\alpha}{ds} &=& 0~~~i.e.,~~ \frac{du_\alpha}{ds}-\Gamma^\theta_{\alpha \sigma} u_\theta \frac{dy^\sigma}{ds}=0 \nonumber \\
\frac{\delta v^\beta}{ds} &=& 0~~~i.e.,~~ \frac{dv^\beta}{ds}+\Gamma^\beta_{\theta \sigma} v^\theta \frac{dy^\sigma}{ds}=0 \nonumber \\
\textrm{and}~~~~\frac{\delta \omega^i}{ds} &=& 0~~~i.e.,~~\frac{d\omega^i}{ds}+\Gamma^i_{p m} \omega^p \frac{dx^m}{ds}=0~. \nonumber
\end{eqnarray}
Now we consider the intrinsic derivative of the scalar $A^\alpha_{\beta i}u_\alpha v^\beta \omega_i$ ~~\textit{i.e.}, we start with 
\begin{eqnarray}
\frac{\delta}{ds}(A^\alpha_{\beta i} u_\alpha v^\beta \omega^i) &=& \frac{d}{ds}(A^\alpha_{\beta i} u_\alpha v^\beta \omega^i) = \frac{d A^\alpha_{\beta i}}{ds}u_\alpha v^\beta \omega^i+A^\alpha_{\beta i}\frac{du_\alpha}{ds}v^\beta \omega^i+A^\alpha_{\beta i} u_\alpha \frac{d v^\beta}{ds}\omega^i+A^\alpha_{\beta i}u_\alpha v^\beta \frac{d \omega^i}{ds} \nonumber \\
&=& \frac{dA^\alpha_{\beta i}}{ds}u_\alpha v^\beta \omega^i+A^\alpha_{\beta i}v^\beta \omega^i \left(\frac{du_\alpha}{ds}-\Gamma^\theta_{\alpha \sigma} u_\theta \frac{dy^\sigma}{ds}\right)+A^\alpha_{\beta i} u_\alpha\omega^i\left(\frac{dv^\beta}{ds}+\Gamma^\beta_{\theta \sigma}v^\theta \frac{dy^\sigma}{ds}\right) \nonumber \\
&&+ A^\alpha_{\beta i}u_\alpha v^\beta \left(\frac{d \omega^i}{ds}+\Gamma^i_{pm} \omega^p \frac{dx^m}{ds}\right)+\Gamma^\theta_{\alpha \sigma}A^\alpha _{\beta i}v^\beta \omega^i u_\theta \frac{dy^\sigma}{ds} \nonumber \\
&&-\Gamma^\beta_{\theta \sigma}A^\alpha_{\beta i}u_\alpha \omega^i v^\theta \frac{dy^\sigma}{ds}-\Gamma^i_{pm}A^\alpha_{\beta i}u_\alpha v^\beta \omega^p \frac{dx^m}{ds} \nonumber \\
&=& \left\{\frac{dA^\alpha_{\beta i}}{ds}+\Gamma^\alpha_{\theta \sigma}A^\theta_{\beta i}\frac{dy^\sigma}{ds}-\Gamma^\theta_{\beta \sigma}A^\alpha_{\theta i}\frac{dy^\sigma}{ds}-\Gamma^p_{i m}A^\alpha_{\beta p}\frac{dx^m}{ds}\right\}u_\alpha v^\beta \omega^i \nonumber \\
&&+ A^\alpha_{\beta i} v^\beta \omega^i\frac{\delta u_\alpha}{ds}+A^\alpha_{\beta i}u_\alpha \omega^i\frac{\delta v^\beta}{ds}+A^\alpha_{\beta i}u_\alpha v^\beta \frac{\delta \omega^i}{ds}~. \nonumber
\end{eqnarray}
Using Leibnitz's rule to the left hand side we have for arbitrary $u_\alpha~,~v^\beta$ and $\omega^i$
$$\frac{\delta A^\alpha_{\beta i}}{ds}=\frac{d A^\alpha_{\beta i}}{ds}+\Gamma^\alpha_{\theta \sigma}A^\theta _{\beta i }\frac{dy^\sigma}{ds}-\Gamma^\theta_{\beta \sigma}A^\alpha_{\theta i} \frac{dy^\sigma}{ds}-\Gamma^p_{i m}A^\alpha_{\beta p}\frac{dx^m}{ds}$$
It is called the generalized intrinsic derivative of $A^\alpha_{\beta i}$ w.r.t `$s$' (\textit{i.e.}, along the curve $\gamma$). From the quotient law $\dfrac{\delta A^\alpha_{\beta i}}{ds}$ is of the same type as $A^\alpha_{\beta i}$. If the functions $A^\alpha_{\beta i}$ are defined throughout $V_n$ and $\gamma$ is an arbitrary curve in $V_n$ then we may write the R.H.S. of the above relation as 
$$\frac{\delta A^\alpha_{\beta i}}{ds}=\left[\frac{\partial A^\alpha_{\beta i}}{\partial x^m}+\Gamma^\alpha_{\theta \sigma}A^\theta_{\beta i}\nabla_my^\sigma-\Gamma^\theta_{\beta \sigma}A^\alpha_{\theta i}\nabla_my^\sigma-\Gamma^p_{i m}A^\alpha_{\beta p}\right]\frac{d x^m}{ds}$$
where we write $\nabla_m y^\sigma=\dfrac{\partial y^\sigma}{\partial x^m}$ for convenience. As $\dfrac{dx^m}{ds}$ is a contravariant vector in the x-coordinate system, so it follows that the expression within square bracket is a tensor of the type $A^\alpha_{\beta i m}$ \textit{i.e.}, a tensor of type (1, 1) in the y co-ordinate and of the kind (0, 2) in the $x$'s. We call it the generalized covariant $x^m$-derivative of $A^\alpha_{\beta i}$ and we write 
$$\nabla_m A^\alpha_{\beta i}=\frac{\partial A^\alpha_{\beta i}}{\partial x^m}+\Gamma^\alpha_{\theta \sigma}A^\theta_{\beta i}\nabla_my^\sigma-\Gamma^\theta_{\beta \sigma}A^\alpha_{\theta i}\nabla_my^\sigma-\Gamma^p_{i m}A^\alpha_{\beta p}$$
Now as covariant derivative of metric tensor is zero so we have
$$\nabla_m g_{ij}=0~;~\nabla_\sigma a_{\alpha \beta}=0$$
We shall now show that $\nabla_m a_{\alpha \beta}=0$.\\\\
By definition\\
\begin{eqnarray}
\nabla_m a_{\alpha \beta} &=& \frac{\partial a_{\alpha \beta}}{\partial x^m}-\Gamma^\theta _{\alpha \sigma}a_{\theta \beta}\frac{\partial y^\sigma}{\partial x^m}-\Gamma^\theta_{\beta \sigma} a_{\alpha \theta}\frac{\partial y^\sigma}{\partial x^m} \nonumber \\
&=& \left(\frac{\partial a_{\alpha \beta}}{\partial y^\sigma}-\Gamma^\theta_{\alpha \sigma}a_{\theta \beta}-\Gamma^\theta_{\beta \sigma}a_{\alpha \theta}\right)\frac{\partial y^\sigma}{\partial x^m} \nonumber \\
&=& \left(\nabla_\sigma a_{\alpha \beta}\right)\frac{\partial y^\sigma}{\partial x^m}=0. \nonumber
\end{eqnarray}
{\bf Note:} In general, we write
$$\nabla_m A^{\cdots \cdots}_{\cdots \cdots}=(\nabla_\sigma A^{\cdots \cdots}_{\cdots \cdots})\frac{\partial y^\sigma}{\partial x^m}~.$$

\section{Gauss's formula\,: Second Fundamental form}

We denote 
$$\frac{\partial y^\alpha}{\partial x^i }=\nabla_i y^\alpha$$
\begin{equation}\label{4.4}
 \nabla_j \nabla_i y^\alpha=\frac{\partial}{\partial x^j}\nabla_i y^\alpha-\Gamma^h_{i j}\nabla_h y^\alpha+\Gamma^\alpha_{\theta \sigma} \nabla_i y^\theta \frac{\partial y^\sigma}{\partial x^j}=\frac{\partial^2 y^\alpha}{\partial x^i \partial x^j}-\Gamma^h_{i j}\frac{\partial y^\alpha}{\partial x^h}+\Gamma^\alpha_{\theta \sigma}\frac{\partial y^\theta}{\partial x^i}\frac{\partial y^\sigma}{\partial x^j}
\end{equation}
Note that equation (\ref{4.4}) is symmetric in $(i,j)$.\\

As $g_{ij}=a_{\alpha \beta}\left(\nabla_i y^\alpha\right)\left(\nabla_j y^\beta\right)$ so taking generalized covariant derivative w.r.t. $x^k$ we have 
$$a_{\alpha \beta}\left(\nabla_k \nabla_i y^\alpha\right)\left(\nabla_j y^\beta\right)+a_{\alpha \beta}\left(\nabla_i y^\alpha\right)\left(\nabla_k \nabla_j y^\beta\right)=0~~~\left( \nabla_k g_{ij}=0~~\mbox{and}~~\nabla_k a_{\alpha \beta}=0\right).$$
Now rotating $(i,j,k)$ cyclically we get two more similar equations:
\begin{eqnarray}
a_{\alpha \beta}(\nabla_i \nabla_j y^\alpha)(\nabla_k y^\beta)+a_{\alpha \beta}(\nabla_j y^\alpha)(\nabla_i \nabla_k y^\beta)&=&0 \nonumber \\
\textrm{and}~~a_{\alpha \beta}(\nabla_j \nabla_k y^\alpha)(\nabla_i y^\beta)+a_{\alpha \beta}(\nabla_k y^\alpha)(\nabla_j \nabla_i y^\beta)&=&0. \nonumber
\end{eqnarray}
Now subtracting the first of these three equations from the sum of the last two and dividing by 2 and remembering that $\nabla_i \nabla_j y^\alpha$ is symmetric in i and j we get
\begin{equation}\label{4.5}
a_{\alpha \beta}(\nabla_i \nabla_j y^\alpha)(\nabla_k y^\beta)=0.
\end{equation}
Thus for any fixed i, j we see that $\nabla_i \nabla_j y^\alpha$ is a vector in $V_{n+1}$ and normal to $V_n$. So we write
\begin{equation}\label{4.6}
\nabla_i \nabla_j y^\alpha=b_{ij}N^\alpha~.
\end{equation}
Equations (\ref{4.4}) , (\ref{4.5}) and (\ref{4.6}) together are called Gauss's formula. Here $b_{ij}$ is a symmetric covariant hypersurface tensor of second order and $N^\alpha$ is a unit normal to the hypersurface $V_n$ . Also transvecting equation (\ref{4.6}) by $a_{\alpha \beta}N^\beta$ we get 
$$b_{ij}(a_{\alpha \beta} N^\alpha N^\beta)=a_{\alpha \beta}(\nabla_i \nabla_j y^\alpha)N^\beta$$
\begin{equation}\label{4.7}
i.e.,~~~~ b_{ij}=a_{\alpha \beta}(\nabla_i \nabla_j y^\alpha)N^\beta
\end{equation}
The elementary quadratic form $b_{ij}dx^idx^j$ is called the second fundamental form and $b_{ij}$ , the second fundamental tensor or the shape tensor.\\\\
{\bf Note:} If we choose $M=V_{n+1}=E_{n+1}$ , the $(n+1)$-dimensional Euclidean space then $y$ can be chosen to be rectangular Cartesian co-ordinates and equation (\ref{4.7}) simplifies to
$$b_{ij}=a_{\alpha \beta }\left(\partial_i \partial_j y^\alpha\right)N^\beta=(\partial_i \textit{\textbf{e}}_{\bm j}) \cdot \textit{\textbf{N}}~.$$

\section{Meusnier's Theorem and consequences}

\subsection*{Theorem 4.1 : Meusnier's Theorem\,:}

{\bf Statement:} If  for a curve $\gamma$ on a hypersurface $V_n$ of $M$\,($=V_{n+1}$) that passes through a point $P$ and have a given direction there at, the first normal relative to $V_{n+1}$ makes an angle $\theta$  with the normal to the hypersurface then the expression $\kappa_{1(a)}\cos \theta$ is an invariant for all such curves where $\kappa_{1(a)}$ is the first curvature relatively to $V_{n+1}$ .\\\\
{\bf Proof:} Suppose the curve $\gamma$ in $V_n$ has a given direction at $P$. Let $\textit{\textbf{t}}$ be the unit tangent to the curve at $P$ and let $t^\alpha$ and $t^i$ be the contravariant components of $\textit{\textbf{t}}$ in the $y$ co-ordinate system in $V_{n+1}$ and in the $x$-coordinate system in $V_n$ respectively.\\

Then
$$t^\alpha=(\nabla_i y^\alpha)t^i$$
Taking covariant derivative with respect to $x^j$ we get
\begin{eqnarray}
\nabla_j t^\alpha &=&(\nabla_j \nabla_i y^\alpha)t^i+(\nabla_i y^\alpha)(\nabla_j t^i) \nonumber \\
i.e.,~~~(\nabla_\sigma t^\alpha)\frac{\partial y^\sigma}{\partial x^j}&=&(b_{ij}N^\alpha)t^i+(\nabla_i y^\alpha)(\nabla_j t^i)~. \nonumber
\end{eqnarray}
Now transvecting both side with $\dfrac{dx^j}{ds}$ ($s$ is the arc length of the curve) and noting that $t^i=\dfrac{dx^i}{ds}$ , we get 
\begin{eqnarray}
(\nabla_\sigma t^\alpha)\frac{d y^\sigma}{ds}&=&(b_{ij}\frac{dx^i}{ds}\frac{dx^j}{ds})N^\alpha+(\nabla_i y^\alpha)(\nabla_j t^i)\frac{dx^j}{ds} \nonumber \\
\mbox{or,}~~~~~~~~\frac{\delta t^\alpha}{ds}&=&(b_{ij}\frac{dx^i}{ds}\frac{dx^j}{ds})N^\alpha+(\nabla_i y^\alpha)\frac{\delta t^i}{ds} \label{4.8} \\
\mbox{or,}~~\kappa_{1(a)}n_{1(a)}^\alpha &=& \kappa_n N^\alpha+\kappa_{1(g)} n_{1(g)}^i(\nabla_i y^\alpha) \label{4.9}
\end{eqnarray}
where $\kappa_{1(a)} \longrightarrow$ the first curvature scalar of the curve relative to $V_{n+1}$

$\textit{\textbf{n}}_{\bm {1(a)}}\longrightarrow$ the first normal vector to $\gamma$ relative to $V_{n+1}$ .

$\kappa_{1(g)}\longrightarrow$ the first curvature scalar of $\gamma$ relative to $V_n$

$\textit{\textbf{n}}_{\bm {1(g)}}\longrightarrow$ the first normal vector to $\gamma$ relative to $V_n$
\begin{equation}\label{4.10}
\mbox{and}~~~~ \kappa_n=b_{ij}\frac{dx^i}{ds}\frac{dx^j}{ds}~,
\end{equation}
is the normal curvature.\\\\
In vector notation, equation (\ref{4.9}) can be written as 
\begin{equation}\label{4.11}
\kappa_{1(a)}\textit{\textbf{n}}_{1(a)}=\kappa_n \textit{\textbf{N}}+\kappa_{1(g)}\textit{\textbf{n}}_{\bm {1(g)}}
\end{equation}
Now taking scalar product with $\textit{\textbf{N}}$ (the unit normal to the hypersurface $V_n$) we get
\begin{equation} \label{4.12}
\kappa_{1(a)} \cos \theta =\kappa_n~.
\end{equation}

From the expression (\ref{4.10}) we note that $\kappa_n$ is independent of the curve $\gamma$, it depends only on the direction of the tangent at the point $P$. Hence $\kappa_{1(a)} \cos \theta $ is an invariant for all curves in $V_n$ , passing through $P$ and having the given direction there at. Hence the theorem.\\\\

\subsection*{Theorem 4.2 : Darboux's Theorem\,:}

{\bf Statement:} For a curve $\gamma$ in a hypersurface $V_n$ in $V_{n+1}$ that passes through a given point $P$ and have a given direction at $P$, the projection of the first curvature vector relatively to $V_{n+1}$ upon the tangent space of $V_n$ at the point concerned is equal to the first curvature vector relatively to $V_n$ .\\\\
{\bf Proof:} In Meusnier's theorem, equation (\ref{4.11}) can be interpreted as follows:

The first curvature vector $\kappa_{1(a)}\textit{\textbf{n}}_{\bm {1(a)}}$ relatively to $V_{n+1}$ can be resolved into two orthogonal components:

i) the component $\kappa_n \textit{\textbf{N}}$ along the normal to the hypersurface at $P$.

ii) the components $\kappa_{1(g)}\textit{\textbf{n}}_{\bm {1(g)}}$ in the tangent space $T_P$ at $P$ to the hypersurface. Hence the theorem.\\\\
{\bf Note:} $\kappa_n$ is called the normal curvature of the curve at $P$ in the particular direction. $\kappa_{1(g)}\textit{\textbf{n}}_{\bm {1(g)}}$ is called the first curvature vector of the curve relatively to $V_n$. Also it is called the geodesic first curvature vector or Darboux vector. Its magnitude $\kappa_{1(g)}$ is the first curvature of the curve relatively to $V_n$ and is also called geodesic first curvature of the curve.\\\\

\subsection*{Theorem 4.3 : Another result from Meusnier's theorem\,:}

{\bf Statement:} A curve $\gamma$ on a hypersurface $V_n$ in $V_{n+1}$ is a geodesic in $V_n$ iff at every point of $\gamma$ the first curvature vector relatively to $V_{n+1}$ is normal to $V_n$ . Further, for a geodesic, its first curvature relatively to $V_{n+1}$ is equal to the normal curvature of the hypersurface in the direction of the geodesic.\\\\
{\bf Proof:} From equation (\ref{4.11}) we note that if for a curve through $P$ having the given direction there at, the first curvature vector relatively to $V_{n+1}$ has the direction of $\textit{\textbf{N}}$ then $$\kappa_{1(a)}\textit{\textbf{n}}_{\bm {1(a)}}=\kappa_n \textit{\textbf{N}}$$ and hence for such curve $\kappa_{1(a)}=\kappa_n$ . Hence the first part.\\
For the second part, we have again from equation (\ref{4.11}) , considering the magnitude,
$$\kappa_{1(a)}^2=\kappa_n^2+\kappa_{1(g)}^2$$
Thus at any point $\kappa_{1(a)}=\kappa_n$ implies $\kappa_{1(g)}=0$.

If this happen at every point of the curve, then 
$$\frac{\delta t^i}{ds}=0~~\textrm{(identically)}$$
and the curve is a geodesic in $V_n$ . Also in this case
$$\kappa_{1(a)}\textit{\textbf{n}}_{\bm {1(a)}}=\kappa_n \textit{\textbf{N}}.$$
Hence the theorem.\\\\
{\bf Note:} We often denote $\kappa_n $ by $\chi$.\\\\

\section{Principal curvatures and Principal directions}

At a point $P$ on the hypersurface $V_n$, a direction $\textit{\textbf{t}}$ in which the normal curvature $\chi$ attains an extreme value (local extreme) is called a principal direction and the extreme value of the normal curvature is called a principal curvature.\\

By definition,
$$\chi=b_{ij}t^i t^j~.$$
As $\textit{\textbf{t}}$ is a unit vector so $g_{ij}t^i t^j=1.$\\

Hence we have
\begin{equation}\label{4.13}
\chi=\frac{b_{ij}t^i t^j}{g_{ij}t^i t^j}~.
\end{equation}
Now, for variation of the direction $\textit{\textbf{t}}$, the extreme values\,(local extrema) of $\chi$ are given by
\begin{eqnarray} \label{4.14}
\frac{d \chi}{dt^i} &=& 0~~,~~~i=1,2,\ldots, n \nonumber \\
i.e.,~~~\left(g_{ij}t^i t^j\right)\left(b_{ij}t^j\right) &-& \left(b_{ij}t^i t^j\right)\left(g_{ij}t^j\right)=0 \nonumber \\
i.e.,~~~\left(b_{ij}-\chi g_{ij}\right)t^j &=& 0~.
\end{eqnarray}
Thus every direction at $P$ will be a principal direction if 
\begin{equation} \label{4.15}
b_{ij}-\chi g_{ij}=0
\end{equation}
In this case $b_{ij}$ are proportional to $g_{ij}$ and $\chi$ is independent of the direction $\textit{\textbf{t}}$ at such a point. This point is called an \underline{umbilic}.

Suppose that the point $P$ is not an umbilic. Then solutions will be obtained for values of $\chi$ given by the equation 
\begin{equation} \label{4.16}
\left| b_{ij}-\chi g_{ij} \right| =0
\end{equation}
This is called the characteristic equation for $b_{ij}$ in the metric of the hypersurface. Since $b_{ij}$ and $g_{ij}$ are both real symmetric and $g_{ij}$ is positive definite so the above characteristic will have n real roots $\chi _h~(h=1,2,\ldots n)$ with or without repetitions. These are called the $n$ principal curvatures. Any value of $t^i$ corresponding to any root is a principal direction. The principal directions corresponding to unequal roots of $\chi$ will be orthogonal to each other. On the other hand, if $\chi_P$ is a repeated root of multiplicity
`$r$' then the solution space of $\textit{\textbf{t}}_{\bm P}$ is of dimension `$r$' and we can choose in multiply infinite number of ways `$r$' mutually orthogonal directions for $\textit{\textbf{t}}$ and these principal directions will also be orthogonal to other principal directions corresponding to other principal curvatures. Thus there always exists `$n$' mutually orthogonal principal directions at any point $P$ in a hypersurface $V_n$ .\\\\
{\bf Note:} If every point of a hypersurface $V_n$ in $V_{n+1}$ is an umbilic then the hypersurface is said to be a totally umbilical hypersurface. This is the generalization of the notion of a sphere or a plane in $E_3$ or a hypersphere or a hyperplane in $E_{n+1}$ .\\\\
{\bf Theorem 4.4 :} Prove that any two distinct principal directions in the neighbourhood U of a point P of a hypersurface $V_n$ are mutually orthogonal.\\\\
{\bf Proof:} From equation\,(\ref{4.14})
$$b_{ij}t^j=\chi g_{ij}t^j$$
Let $\chi _1~, \chi _2$ be the principal normal curvatures and $t_1^i$ and $t_2^i$ be the corresponding two distinct principal directions. So we have
\begin{equation} \label{4.17}
b_{ij}t_1^j=\chi_1 g_{ij}t_1^j
\end{equation}
and
\begin{equation} \label{4.18}
b_{ij}t_2^j=\chi_2 g_{ij}t_2^j
\end{equation}
Now, (\ref{4.17})$\times ~t_2^i$ $-$ (\ref{4.18})$\times ~t_1^i$ ~~gives
$$(\chi_1-\chi_2)g_{ij}t_1^i t_2^j=0~,~~~\textrm{using symmetric property of}~ b_{ij} ~\mbox{and}~g_{ij}~.$$
As $$\chi_1 \neq \chi_2$$
so $$g_{ij}t_1^i t_2^j=0$$
\textit{i.e.}, $\textit{\textbf{t}}_{\bm 1}$ is orthogonal to $\textit{\textbf{t}}_{\bm 2}$.\\\\

\section{Mean curvatures of different orders and the total curvature}

Let $\chi_1, \chi_2, \ldots \chi_n$ be the `$n$' principal curvatures at $P$ of a hypersurface $V_n$ in $V_{n+1}$ . Then the sum of the products of the principal curvatures $\chi_1, \chi_2, \ldots \chi_n$  taken `$p$'\,($\leq n$) at a time is called the \underline{mean curvature} of order `$p$' or the $p$-th mean curvature and it will be denoted by $M_P$ . The first mean curvature $M_1=\chi_1+\chi_2+\ldots +\chi_n$ is denoted by $M$ and is called simply the mean curvature. The product $\chi_1 \cdot \chi_2 \cdots \chi_n$ of all the principal curvatures \textit{i.e.}, the $n$-th mean curvature $M_n$ is also called the total curvature or the Gaussian curvature and is denoted by $\mathcal{K}$.

The characteristic equation, 
$$\left|b_{ij}-\chi g_{ij}\right|=0$$
can be written as $$\left|g^{ki}\right| \left|b_{ij}-\chi g_{ij}\right|=0$$
\begin{center}
(as the metric tensor $g_{ij}$ is non-singular so is also $g^{ij}$)
\end{center}
So we have
$$\left|b^k_j-\chi \delta^k_j\right| =0$$
The above determinant in explicit form is given by 
$$\chi^n-b_1\chi^{n-1}+\cdots +(-1)^pb_p \chi^{n-p}+ \cdots +(-1)^nb_n=0$$
where $b_p$ is the sum of the principal minor of order `$p$' of the matrix $b^k_i$ and may be termed as the trace of order $p$ or the $p$-th trace of the matrix. So from the definition we have
$$b_p=M_p ~~,~ \forall p=1,2, \ldots n~.$$
In particular,
$$M=M_1=b^i_i=g^{ij}b_{ij}$$
and $$\mathcal{K}=M_n=\left|b^k_i\right|=\left|g^{ki}b_{ij}\right|=\left|g^{ki}\right|\left|b_{ij}\right|=\frac{b}{g}$$
If $M=0$ then the surface is called a minimal surface.\\\\

\section{Conjugate directions\,: Asymptotic line and Asymptotic direction}

Two vectors $\textit{\textbf{u}}$ and $\textit{\textbf{v}}$ at any point on the hypersurface are said to be \underline{conjugate} if $b_{ij}u^iv^j=0$. The directions, of the vectors $\textit{\textbf{u}}$ and $\textit{\textbf{v}}$ are said to be \underline{conjugate directions}.

A self conjugate direction at any point is called an \underline{asymptotic direction} at that point and a curve at any point of which the tangent direction is an asymptotic direction is called an \underline{asymptotic line}. It is clear that in a hypersurface there can be a real asymptotic line iff the second fundamental form is not definite (\textit{i.e.}, neither positive nor negative definite). The differential equation of an asymptotic line is
\begin{equation} \label{4.19}
b_{ij}dx^idx^j=0
\end{equation}
It follows that a curve in the hypersurface is an asymptotic line if $\kappa_n=0$ at every point of the curve in the direction of the curve there at.\\\\
{\bf Theorem 4.5 :} If two principal directions are orthogonal at a point of the hypersurface then they are not only orthogonal but are also conjugate.\\\\
{\bf Proof:} Let $\textit{\textbf{t}}_{\bm p}$ and $\textit{\textbf{t}}_{\bm q}$ be two principal directions at a point on the hypersurface and are orthogonal to each other. Suppose $\chi_p$ and $\chi_q$ are the corresponding principal curvatures. Then from equation\,(4.14) we have
$$(b_{ij}-\chi_p g_{ij})t^j_p=0$$
Now, multiply this equation by $t^i_q$ and summing over $i$ we get
$$b_{ij}t^j_p t^i_q=\chi_p g_{ij}t_q^i t^j_p$$
By condition, $\textit{\textbf{t}}_{\bm p}$ and $\textit{\textbf{t}}_{\bm q}$ are orthogonal so the R.H.S. vanishes. Hence $b_{ij}t_p^jt_q^i=0$ \textit{i.e.,} $\textit{\textbf{t}}_{\bm p}$ and $\textit{\textbf{t}}_{\bm q}$ are conjugate to each other.\\\\
{\bf Theorem 4.6 :} The normal curvature of the hypersurface $V_n$ for an asymptotic direction is zero.\\\\
{\bf Proof:} The normal curvature $\kappa_n$ of the hypersurface $V_n$ in the direction of a curve $\gamma$ is given by 
$$\kappa_n=b_{ij}\frac{dx^i}{ds}\frac{dx^j}{ds}$$
As $\gamma$ is an asymptotic line of $V_n$ (by condition) so we have $b_{ij}dx^idx^j=0~~i.e.,~~ \kappa_n=0$.\\\\

\subsection*{Theorem 4.7 : Euler's Theorem\,:}

{\bf Statement:} If $\chi_{_1}, \chi_{_2}, \ldots, \chi_{_n}$ are $n$ principal curvatures distinct or otherwise and $\textit{\textbf{t}}_{\bm {p_1}}~, \textit{\textbf{t}}_{\bm {p_2}}~, \ldots , \textit{\textbf{t}}_{\bm {p_n}}$ are mutually orthogonal principal directions corresponding to these principal curvatures, then the normal curvature $\chi$ in the direction $\textit{\textbf{l}}$ making an angle $\theta_i$ with the direction $\textit{\textbf{t}}_{\bm {p_i}}~~,~~(i=1,2, \ldots n)$ is given by
$$\chi=\sum _{i=1}^n \chi_i \cos ^2 \theta_i$$
{\bf Proof:} We have $\chi=b_{ij}l^il^j$

As $\textit{\textbf{l}}$ can be written as a linear combination of the principal directions so
$$\textit{\textbf{l}}=\lambda_1 \textit{\textbf{t}}_{\bm {p_1}}+\lambda_2 \textit{\textbf{t}}_{\bm {p_2}}+\ldots +\lambda_n \textit{\textbf{t}}_{\bm {p_n}}$$
Hence
\begin{eqnarray}
\textit{\textbf{l}} \cdot \textit{\textbf{t}}_{\bm {p_k}} &=& \lambda_k~~i.e.,~~ \lambda_k=\cos \theta _k \nonumber \\
\textit{\textbf{l}} &=& \sum _{k=1}^n\textit{\textbf{t}}_{\bm {p_k}} \cos \theta _k \nonumber \\
\mbox{and}~~~~~~\chi &=& b_{ij}\left(\sum _{k=1}^n\textit{\textbf{t}}_{\bm {{p_k}}}^i \cos \theta _k\right)\left(\sum _{r=1}^n\textit{\textbf{t}}_{\bm {{p_r}}}^j \cos \theta _r\right) \nonumber
\end{eqnarray}
As two orthogonal principal directions are also conjugate to each other so we have
$$b_{ij}t_{p_k}^it_{p_r}^j=0~~(k\neq r)~~~\mbox{and}~~~b_{ij}t_{p_k}^it_{p_k}^j=\chi_k~.$$
Thus the expression for $\chi$ gives
$$\chi =\sum _{k=1}^n\chi _k \cos ^2 \theta _k$$\\

\underline{\bf Note\,:} As the sum of principal curvatures is the sum of normal curvatures for $n$ mutually orthogonal directions in $V_n$  so the above sum may be the sum of the normal curvatures in any $n$ mutually orthogonal directions (Dupin's Theorem below)\\\\

\subsection*{Theorem 4.8 : Dupin's Theorem\,:}

{\bf Statement:} At any point of a hyper surface $V_n$ in $V_{n+1}$ , the sum of the normal curvatures in n mutually orthogonal directions is a constant, the mean curvature at that point.\\\\
{\bf Proof:} Let $\textit{\textbf{t}}_{\bm {p_1}}, \textit{\textbf{t}}_{\bm {p_2}}, \ldots , \textit{\textbf{t}}_{\bm {p_n}}$ be a set of $n$ orthogonal directions in $V_n$. Then sum of the normal curvatures of $V_n$ for these orthogonal directions is 
$$\sum _{p_k=1}^n b_{ij}t_{p_k}^it_{p_k}^j=b_{ij}g^{ij}=M,~~\mbox{a constant.}$$
Hence the theorem.\\\\
{\bf Theorem 4.9 :} A curve on a hypersurface $V_n$ in $V_{n+1}$ is a geodesic in $V_{n+1}$ , {\it iff} it is a geodesic as well as an asymptotic line in $V_n$.\\\\
{\bf Proof:} From Meusnier's theorem
$$\kappa_{1(a)}^2=\kappa_n^2+\kappa_{1(g)}^2$$
Now, $\kappa_{1(a)}=0$ implies $\kappa_n=0$ and $\kappa_{1(g)}=0$ and vice-versa. Hence it follows that the curve is a geodesic in $V_{n+1}$ iff it is a geodesic as well as an asymptotic line in $V_n$ .\\\\
\underline{\bf Totally geodesic hypersurface}\\

If all the geodesic of a hypersurface $V_n$ in $V_{n+1}$ are also geodesic of $V_{n+1}$ then the hypersurface is called a \underline{totally geodesic hypersurface} of the enveloping plane.\\\\
{\bf Theorem 4.10 :} A hypersurface $V_n$ in $V_{n+1}$ is totally geodesic in $V_{n+1}$ {\it iff} $b_{ij}=0$ identically.\\\\
{\bf Proof:} From Meusnier's theorem we have 
$$\kappa_{1(a)}^2=\kappa_{n}^2+\kappa_{1(g)}^2$$
Now whenever $\kappa_{1(g)}=0$ for any curve in $V_n$ in any direction at $P$ we must have $\kappa_{1(a)} =0$ at $P$ in that direction {\it iff} $\kappa_{n}=0$ at $P$ in every direction there at.\\

By definition, $$\kappa_n=b_{ij}\frac{dx^i}{ds}\frac{dx^j}{ds}$$
So $\kappa_n \equiv 0$ implies $b_{ij}\equiv 0$.\\\\
{\bf Note I :} This notion is a generalization of the notion of a plane in $E_3$ or hyperplane in $E_{n+1}$ .\\\\
{\bf Note II :} The above theorem can be generalized as follows :\\\\
{\bf Theorem 4.11 :} A totally geodesic hypersurface is a minimal hypersurface and its lines of curvature are indeterminate.\\\\
{\bf Proof:} We have seen that $b_{ij}=0$ identically for a totally geodesic hypersurface. So $M=b_{ij}g^{ij}=0$ , \textit{i.e.,} the hypersurface $V_n$ is a minimal hypersurface. Consequently, the equation 
$$b_{ij}=\frac{M}{n}g^{ij}$$
is identically satisfied. But it is the condition for indeterminant lines of curvature. Hence the theorem.\\\\

\section{Weingarten's Formula}

\begin{equation} \label{4.20}
\nabla_i N^{\alpha}=-b_i^k \nabla_ky^{\alpha}
\end{equation}
{\bf Proof:} We have 
$$a_{\alpha\beta}N^{\alpha}N^{\beta}=1.$$
Taking covariant derivative with respect to $x^i$ we have
$$a_{\alpha \beta}N^{\alpha}\,\nabla _{i}N^{\beta} = 0~.$$
This shows that the vectors $\nabla_1 N^{\beta},~\nabla_2 N^{\beta}, \ldots ,\nabla_n N^{\beta}$ are orthogonal to $\textit{\textbf{N}}$ and therefore are hypersurface vectors. Hence each of them is a linear combination of the vectors $\nabla_1 y^{\alpha},~\nabla_2 y^{\alpha}, \ldots,~\nabla_n y^{\alpha}$. Thus we write
\begin{equation} \label{4.21}
\nabla_i N^{\alpha}=t_i^k \nabla_ky^{\alpha}.
\end{equation}
Transvecting with $a_{\alpha\beta}\nabla_jy^{\beta}$ we get
$$a_{\alpha\beta}\left(\nabla_iN^{\alpha}\right)\left(\nabla_jy^{\beta}\right)=t_i^kg_{kj}~.$$
As $a_{\alpha\beta}N^{\alpha}\nabla_jy^{\beta}=0$ , so taking covariant derivative with respect to $x^i$ we have
\begin{eqnarray}
a_{\alpha\beta}\left(\nabla_iN^{\alpha}\right)\left(\nabla_jy^{\beta}\right) &+& a_{\alpha\beta}N^{\alpha}\left(\nabla_i\nabla_jy^{\beta}\right)=0 \nonumber \\
\mbox{or,}~~~~ a_{\alpha\beta}\left(\nabla_iN^{\alpha}\right)\left(\nabla_jy^{\beta}\right) &=& -a_{\alpha\beta}N^{\alpha}b_{ij}N^{\beta}=-\left(a_{\alpha\beta}N^{\alpha}N^{\beta}\right)b_{ij} \nonumber \\
\mbox{or,}~~~~ a_{\alpha\beta}t_i^k\left(\nabla_ky^{\alpha}\right)\left(\nabla_jy^{\beta}\right) &=& -b_{ij} \nonumber \\
\mbox{or,}~~~~~~~~~~~~ t_i^kg_{kj} &=& -b_{ij}~~~,~i.e.,~~~b_{ij}=-t_{ij} \nonumber
\end{eqnarray}
So from equation\,(\ref{4.21}), $\nabla _i N^{\alpha}=-b_i^k\nabla_ky^{\alpha}$ , the Weingarten's formula.\\\\
{\bf Note:} In Eucledian space, the Weingarten's formula takes the form : $\partial_iN^{\alpha}=-b_i^k\partial_ky^{\alpha}$.\\\\
{\bf Theorem 4.12 :} The derived vector of the unit normal with respect to $V_{n+1}$ , along a curve $\gamma$ in $V_{n}$ , will be tangential to the curve $\gamma$ provided $\gamma$ is a line of curvature of the hypersurface.\\\\
{\bf Proof:} Let $t^i$ be a unit tangent vector to a curve $\gamma$ in $V_n$ . Then the derived vector of $N^{\alpha}$ with respect to $V_{n+1}$ along $\gamma$ in $V_n$ is $\left(\nabla_iN^{\alpha}\right)t^i$ . But from Weingarten's formula
$$\nabla_iN^{\alpha}=-b_i^k\nabla_ky^{\alpha}$$
$$\mbox{or,}~~~~ \left(\nabla_iN^{\alpha}\right)t^i=-b_i^k\nabla_ky^{\alpha}t^i$$
By condition of the theorem $\left(\nabla_iN^{\alpha}\right)t^i$ will be along $t^i$ implies
\begin{eqnarray}
-b_i^k\nabla_ky^{\alpha}t^i &=& -Ky_{,i}^{\alpha}t^i~~~~,~~k~\mbox{is any constant} \nonumber \\
\mbox{or,}~~~~ b_{ij}g^{jk}y_{,k}^{\alpha}t^i\left(a_{\alpha\beta}y^{\beta}_{, h}\right) &=& Ky_{,i}^{\alpha}t^i\left(a_{\alpha\beta}y^{\beta}_{,h}\right) \nonumber \\
\mbox{or,}~~~~ b_{ij}g^{jk}t^ig_{kh} &=& K g_{ih}t^i \nonumber \\
\mbox{or,}~~~~ b_{ih}t^i-Kg_{ih}t^i &=& 0 \nonumber \\
\mbox{or,}~~~~ \left(b_{ih}-Kg_{ih}\right)t^i &=& 0. \nonumber
\end{eqnarray}
This implies that the directions $t^i$ is a principal direction for the symmetric tensor $b_{ij}$ , \textit{i.e.,} $t^i$ is a principal direction for the hypersurface. But $t^i$ is a unit tangent to the curve $\gamma$ in $V_n$ . Hence $\gamma$ is a line of curvature in $V_n$ .\\\\
{\bf Theorem 4.13 :} Prove that the normal to a totally geodesic hypersurface is parallel in the enveloping manifold.\\\\
{\bf Proof:} From Weingarten's formula (\ref{4.20})
\begin{eqnarray} \label{4.22}
\nabla_iN^{\alpha} &=& -b_i^k\nabla_ky^{\alpha} \nonumber \\
\mbox{or,}~~~~ \left(\nabla_iN^{\alpha}\right)e^i &=& -b_i^ke^i\nabla_ky^{\alpha}
\end{eqnarray}
where $e^i$ is any unit vector in the hypersurface $V_n$ .\\
By condition, $V_n$ is a totally geodesic hypersurface, \textit{i.e.}, each geodesic of $V_n$ is a geodesic in $V_{n+1}$ .\\\\
Hence, $\kappa_{(a)}=0=\kappa_{(g)}$ . But $\kappa_{(a)}^2=\kappa_{(n)}^2+\kappa_{(g)}^2$ . So $\kappa_{(n)}=0$ , \textit{i.e.},
$$b_{ij}\frac{dx^i}{ds}\frac{dx^j}{ds}=0.$$
But $\frac{dx^i}{ds}$ is arbitrary, so $b_{ij}=0$. Hence from (\ref{4.22}), $\nabla_iN^{\alpha}$ is orthogonal to $e^i$. This proves the theorem.\\\\
{\bf Rodrigues' Formula:} Along a line of curvature in $V_n$ embedded in $V_{n+1}$
$$d\textit{\textbf{N}}+\chi d\textit{\textbf{y}}=0~~,$$
where $\chi$ is the normal curvature of $V_n$ in $V_{n+1}$ in the direction of the line of curvature.\\\\
{\bf Proof:} By Weingarten's formula
$$\nabla_{i}N^{\alpha}=-b_i^k\nabla_ky^{\alpha}~.$$
Taking inner product with $\frac{dx^i}{ds}=t^i$ , we have
\begin{equation} \label{4.23}
\frac{\delta N^{\alpha}}{ds}=-\left(b_i^k t^i\right)\frac{\partial y^{\alpha}}{\partial x^k}~.
\end{equation}
Now, along a line of curvature we have,
\begin{eqnarray}
\left(b_{ij}-\chi g_{ij}\right)t^j &=& 0 \nonumber \\
i.e.,~~~\left(b_{j}^{k}-\chi \delta_j^k\right)t^j &=& 0 \nonumber \\
i.e.,~~~b_{j}^{k}t^j &=& \chi t^k \nonumber
\end{eqnarray}
Then from equation\,(\ref{4.23}) 
$$\frac{\delta N^{\alpha}}{ds}=-\chi t^k\frac{\partial y^{\alpha}}{\partial x^k}=
-\chi \frac{dy^{\alpha}}{ds}$$
$$i.e., ~~\delta \textit{\textbf{N}}+\chi d\textit{\textbf{y}}=0$$
Hence the theorem.\\\\
{\bf Note:} The above theorem (\textit{i.e.}, Rodrigues' formula) can be interpreted geometrically as follows :

Along a line of curvature, the normal to the hypersurface bends in the direction of the curve and the amount of bending is equal to $\chi$ -times the displacement along the curve in the opposite sense.\\\\
{\bf Theorem 4.14 :} The derived vector of the unit normal in an asymptotic direction of a hypersurface in a Riemannian space is orthogonal to that direction.\\\\
{\bf Proof:} Let $\textit{\textbf{l}}$ be an asymptotic direction at the current point $P$ of a hypersurface in a Riemannian space. Then
$$b_{ij}l^il^j=0.$$
Now the derived vector of the unit normal $N^{\alpha}$ to the hypersurface in the direction $\textit{\textbf{l}}$ is $l^i\nabla_iN^{\alpha}$.
The scalar product of this vector with the vector $\textit{\textbf{l}}$ is
\begin{eqnarray}
a_{\alpha\beta}\left(l^i\nabla_i N^{\alpha}\right)\left(l^j\nabla_j y^{\beta}\right) &=& a_{\alpha\beta}\left(\nabla_i N^{\alpha}\right)\left(\nabla_j y^{\beta}\right)l^il^j \nonumber \\
&=& -a_{\alpha\beta}\left(b_i^k\nabla_ky^{\alpha}\right)\left(\nabla_j y^{\beta}\right)l^il^j \nonumber \\
&=& -b_i^k\left\{a_{\alpha\beta}\left(\nabla_ky^{\alpha}\right)\left(\nabla_j y^{\beta}\right)\right\}l^il^j \nonumber \\
&=& -b_i^kg_{kj}l^il^j=-b_{ij}l^il^j=0  \nonumber
\end{eqnarray}
This shows that the derived vector $l^i\nabla_i N^{\alpha}$ of the unit normal $N^{\alpha}$ in the direction of $\textit{\textbf{l}}$ is orthogonal to $\textit{\textbf{l}}$. Hence the theorem.\\\\
{\bf Theorem 4.15 :} Prove that the normal to a totally geodesic hypersurface is parallel in the enveloping manifold.\\\\
{\bf Proof:} From Weingarten's formula :
$$\nabla_i N^{\alpha}=-b_i^k\nabla_ky^{\alpha}$$
$$\mbox{or,}~~~ \left(\nabla_i N^{\alpha}\right)e^i=-b_i^k\left(\nabla_ky^{\alpha}\right) \cdot e^i$$
where, $e^i$ is any unit vector of the hypersurface $V_n$.
As $V_n$ is a totally geodesic hypersurface, so each geodesic of $V_n$ is a geodesic of $V_{n+1}$, \textit{i.e.}, $\kappa_{(a)} =0 = \kappa_{(g)}$ . But we have $\kappa_{(a)}^2=\kappa_n^2+\kappa_{(g)}^2$. Hence $\kappa_n=0$ , \textit{i.e.}, $b_{ij}\dfrac{dx^i}{ds}\dfrac{dx^j}{ds}=0$.

As $\dfrac{dx^i}{ds}$ is arbitrary, so $b_{ij}=0$ , \textit{i.e.}, $\left(\nabla_iN^{\alpha}\right) \cdot e^i=0$

\textit{i.e.}, $\nabla_iN^{\alpha}$ is orthogonal to $e^i$

\textit{i.e.}, $\nabla_iN^{\alpha}$ is orthogonal to any direction $e^i$ of a totally geodesic hypersurface $V_n$ . Hence the result.\\\\

\section{Lines of Curvature\,: Differential Form}

A curve in a hypersurface $V_n$ in $V_{n+1}$ at every point of which the direction of tangent is a principal direction, is called a line of curvature.

We shall now determine the differential equation of the lines of curvature for 2-dimensional hypersurface $V_2$ in $E_3$ .\\\\
{\bf Note:} Through every point of $V_n$ there are $n$ mutually orthogonal lines of curvature.\\\\
From equation\,(\ref{4.14}) we have
$$b_{ij}t^j=\chi g_{ij}t^j~.$$
Now, putting $i=1,~2$, we get
$$b_{1j}t^j=\chi g_{1j}t^j$$
$$b_{2i}t^i=\chi g_{2i}t^i~.$$
So eliminating $\chi$ between these two equations we obtain
$$\left(g_{1j}b_{2i}-g_{2i}b_{1j}\right)t^it^j=0$$
$$\left(g_{1i}b_{2j}-g_{2i}b_{1j}\right)t^it^j=0~.$$
As $t^i=\frac{du^i}{ds}$ , the differential equation for the lines of curvature becomes
$$ \left(g_{1i}b_{2j}-g_{2i}b_{1j}\right)du^idu^j=0$$
or in explicit form :
\begin{equation} \label{4.24}
\left(g_{11}b_{12}-g_{12}b_{11}\right)\left(du^1\right)^2+\left(g_{11}b_{22}-g_{22}b_{11}\right)du^1du^2+\left(g_{12}b_{22}-g_{22}b_{12}\right)\left(du^2\right)^2=0
\end{equation}
This is the differential equation of the line of curvatures in two dimensional hypersurface $V_2$ in $E_3$ .\\\\
{\bf Minimal hypersurface:} A hypersurface $V_n$ in $V_{n+1}$ of vanishing mean curvature is called a minimal hypersurface.\\\\
{\bf Note:} The reason for this name is the fact that given a closed curve in $E_3$ , the surface of minimal area bounded by the curve is a surface of vanishing mean curvature.\\\\
\underline{\bf An useful formula for $M$ for a hypersurface $V_2$ in $E_3$}\\
$$M=g^{ij}b_{ij} =g^{11}b_{11}+2g^{12}b_{12}+g^{22}b_{22},$$
where $$g^{ij}=\mbox{cofactor of}~\frac{g_{ji}}{g}~~~,~~ g=\left|g_{ij}\right|=
 \left| \begin{array}{cc}
g_{11} & g_{12} \\
g_{21} & g_{22}
\end{array} \right|.$$
So,
$$g^{11}=\frac{g_{22}}{g}~~,~~g^{12}=-\frac{g_{12}}{g}~~,~~g^{22}=\frac{g_{11}}{g}~.$$
Hence,
\begin{equation} \label{4.25}
M=\frac{1}{g}\left(g_{11}b_{22}-2g_{12}b_{12}+g_{22}b_{11}\right) 
\end{equation}

{\bf Theorem 4.16 :} Prove that the mean curvature of a hypersurface is equal to the negative of the divergence of the unit normal.\\\\
{\bf Proof:} Suppose $N^{\alpha}$ be the unit normal vector to thr hypersurface $V_n$ in $V_{n+1}$ . Let $t^{\alpha}_h~~(h=1, 2, \ldots , n)$ be the `$n$' unit tangent vectors in $V_{n+1}$ to $n$ congruences $e_h~(h=1, 2, \ldots , n)$ of an orthogonal ennuple in $V_n$ . So $\textit{\textbf{t}}_{\bm h}$ is orthogonal to $\textit{\textbf{N}}$ {\it i.e,}
$$t^{\alpha}_h \cdot N_{\alpha}=0.$$

Taking co-variant derivative with respect to $y^{\beta}$ we have
$$t^{\alpha}_{h; \beta} N_{\alpha}+N_{\alpha;\beta} t^{\alpha}_h=0.$$

Multiplying both side by $t^{\beta}_h$ we get
$$\left(t^{\alpha}_{h; \beta} t^{\beta}_h\right)N_{\alpha}+\left(N_{\alpha;\beta} t^{\beta}_h\right)t^{\alpha}_h=0.$$

$\Longrightarrow$~~ Normal component of the 1st curvature of the curve\,(having tangent $\textit{\textbf{e}}_{\bm h}$) relative to\\

$V_{n+1}=-$(Tendency of $N^{\alpha}$ in the direction of the curve).\\

\textit{i.e.}, Normal curvature of $V_n$ in the direction of $\stackrel{\rightarrow}{e_h}$\\

$~~~~~~~~~~~~~~~~~~~~~~~~~~=-$Tendency of unit normal $N^{\alpha}$ in the direction of $\overrightarrow{\bm e}_{_{\bm h}}$ .\\

Now summing over $h$ from $1$ to $n$ we get
$$M=-\mbox{div}\, \textit{\textbf{N}}.$$
{\bf Theorem 4.17:} Prove that conjugate directions in a hypersurface are such that the derived vector of the unit normal in either direction is orthogonal to the other direction.\\\\
{\bf Proof:} From Weingarten's formula (equation\,(\ref{4.20}))
$$\nabla_i N^{\alpha}= -b^k_i \nabla_k y^{\alpha}$$ 
Let $a^i$ and $e^i$ be unit vectors in the hypersurface $V_n$ . The derived vector of $N^{\alpha}$ in the direction of $a^i$ is $\nabla_i N^{\alpha}a^i$. So projection of this vector along the direction $e^i$ is 
\begin{eqnarray}
\left(\nabla_i N^{\alpha}\right)a^i\left(a_{\alpha \beta}\nabla_h y^{\beta} e^h\right)&=&-b^k_i \nabla_k y^{\alpha} a^ia_{\alpha \beta}\nabla_h y^{\beta} e^h\nonumber \\
&=&-b^k_i\left(a_{\alpha \beta}\nabla_k y^{\alpha}\nabla_h y^{\beta}\right)a^i e^h  \nonumber \\
&=&-b^k_ia_{kh}a^ie^h \nonumber \\
&=&-b_{ih}a^ie^h \nonumber \\
&=& 0~~~ \mbox{if $a^i$ and $e^i$ are conjugate directions}. \nonumber
\end{eqnarray}
Hence the theorem.\\\\
{\bf Note:} If $a^i=e^i$ then
$$\nabla_i N^{\alpha} e^i\left(a_{\alpha \beta}\nabla_h y^{\beta} e^h\right)=-b_{ih}e^ie^h.$$
Suppose $e^i$ is tangent to a curve $\gamma$ in $V_n$ . Then $\gamma$ will be an asymptotic line of the hypersurface if $b_{ij}e^ie^j=0$.
 
Hence, $$\left(\nabla_i N^{\alpha}\right)e^i\left(a_{\alpha \beta} \nabla_h y^{\beta} e^h\right)=0$$

Thus we have the following result:\\

``The derived vector of the unit normal along a curve in the hypersurface will be orthogonal to the curve provided the curve is an asymptotic line in the hypersurface''.\\\\

\section{The Gauss and Codazzi Equations on a hypersurface}

{\bf Theorem 4.18 :} The Gauss and Codazzi equations on a hypersurface are given by
\begin{equation} \label{4.27}
R_{ijkl}=\left(b_{jl}b_{ki}-b_{jk}b_{il}\right)+\overline{R}_{\alpha \beta \gamma\delta}\left(\nabla_i y^{\alpha}\right)\left(\nabla_j y^{\beta}\right)\left(\nabla_k y^{\gamma}\right)\left(\nabla_l y^{\delta}\right)
\end{equation}
and
\begin{equation} \label{4.28}
\nabla_k b_{ij}-\nabla_j b_{ik}=\overline{R}_{\alpha \beta \gamma\delta}N^{\beta}\left(\nabla_i y^{\alpha}\right)\left(\nabla_j y^{\gamma}\right)\left(\nabla_k y^{\delta}\right).
\end{equation}
Here $\overline{R}_{\alpha \beta \gamma\delta}$ is the Riemannian curvature tensor on the enveloping space $V_{n+1}$ . The first equation is known as {\bf Gauss characteristic equations} and second one is called the {\bf Codazzi equations}.\\\\
{\bf Proof:} We have
\begin{equation} \label{4.29}
g_{ij}=a_{\alpha \beta}\left(\nabla_i y^{\alpha}\right)\left(\nabla_j y^{\beta}\right).
\end{equation}

As $N^{\alpha}$ is the unit normal vector so 
\begin{equation} \label{4.30}
a_{\alpha \beta}\left(\nabla_i y^{\alpha}\right)N^{\beta}=0.
\end{equation}

Also
\begin{equation} \label{4.31}
a_{\alpha \beta}N^{\alpha}N^{\beta}=1.
\end{equation}

Taking covariant derivative of equation\,(\ref{4.29}) with respect to $x^k$ we have
\begin{equation} \label{4.32}
\frac{\partial a_{\alpha \beta}}{\partial y^{\gamma}}\left(\nabla_k y^{\gamma}\right)\left(\nabla_i y^{\alpha}\right)\left(\nabla_j y^{\beta}\right) + a_{\alpha \beta}\left(\nabla_k \nabla_i y^{\alpha}\right)\left(\nabla_j y^{\beta}\right)+a_{\alpha \beta}\left(\nabla_i y^{\alpha}\right) \left(\nabla_k \nabla_j y^{\beta}\right)=0.
\end{equation}
Now rotating $(i,~j,~k)$ cyclically, we get two more equations. Sum of these two equations when substracted from equations\,(\ref{4.32}), we get
$$\left[\frac{\partial a_{\beta \gamma}}{\partial y^{\alpha}}+\frac{\partial a_{\gamma \alpha}}{\partial y^{\beta}}-\frac{\partial a_{\alpha \beta}}{\partial y^{\gamma}}\right]\left(\nabla_k y^{\gamma}\right)\left(\nabla_i y^{\alpha}\right)\left(\nabla_j y^{\beta}\right)+2a_{\alpha \beta}\left(\nabla_k y^{\alpha}\right)\left(\nabla_i \nabla_j y^{\beta}\right)=0$$
$$\mbox{or,}~~~~~~a_{\alpha \beta}\left(\nabla_i \nabla_j y^{\alpha}\right)\left(\nabla_k y^{\beta}\right)+\Gamma_{\alpha \beta \gamma}\left(\nabla_i y^{\alpha}\right)\left(\nabla_j y^{\beta}\right)\left(\nabla_k y^{\gamma}\right)=0$$
(Note that the Christoffel symbols of the 1st kind are formed with respect to $a_{\alpha \beta}$ evaluated at points of $V_n$)
$$\mbox{or,}~~~~~~a_{\alpha \beta}\left(\nabla_i \nabla_j y^{\alpha}\right)\left(\nabla_k y^{\beta}\right)+\Gamma^{\mu}_{\alpha \beta} a_{\mu \gamma}\left(\nabla_i y^{\alpha}\right)\left(\nabla_j y^{\beta}\right)\left(\nabla_k y^{\gamma}\right)=0$$
$$\mbox{or,}~~~~~~a_{\alpha \beta}\left(\nabla_i \nabla_j y^{\alpha}\right)\left(\nabla_k y^{\beta}\right)+\Gamma^{\alpha}_{\mu \nu} a_{\alpha \beta}\left(\nabla_i y^{\mu}\right)\left(\nabla_j y^{\nu}\right)\left(\nabla_k y^{\beta}\right)=0$$
$$\mbox{or,}~~~~~~a_{\alpha \beta}\left(\nabla_k y^{\beta}\right)\left[\left(\nabla_i \nabla_j y^{\alpha}\right)+\Gamma^{\alpha}_{\mu \nu}\left(\nabla_i y^{\mu}\right)\left(\nabla_j y^{\nu}\right)\right]=0.$$
Comparing this equation with equation (\ref{4.30}) we can write
\begin{equation} \label{4.33}
\left(\nabla_i \nabla_j y^{\alpha}\right)+\Gamma^{\alpha}_{\mu \nu}\left(\nabla_i y^{\mu}\right)\left(\nabla_j y^{\nu}\right)=b_{ij}N^{\alpha}.
\end{equation}
Now multiply both sides by $a_{\alpha \beta}N^{\beta}$ and summing for $\alpha$ we obtain
\begin{eqnarray} \label{4.34}
b_{ij} &=& a_{\alpha \beta}N^{\beta}\left(\nabla_i \nabla_j y^{\alpha}\right)+\Gamma^{\alpha}_{\mu \nu}a_{\alpha \beta}N^{\beta}\left(\nabla_i y^{\mu}\right)\left(\nabla_j y^{\nu}\right),\nonumber\\
&=&a_{\alpha \beta}N^{\beta}\left(\nabla_i \nabla_j y^{\alpha}\right)+\Gamma_{\mu \nu \beta}N^{\beta}\left(\nabla_i y^{\mu}\right)\left(\nabla_j y^{\nu}\right).
\end{eqnarray}
Taking covariant derivative of equation\,(\ref{4.30}) with respect to $x^j$ we get
\begin{eqnarray}
a_{\alpha \beta}\left(\nabla_i \nabla_j y^{\alpha}\right)N^{\beta}+a_{\alpha \beta}\left(\nabla_i y^{\alpha}\right)\left(\nabla_j N^{\beta}\right)&=&-\frac{\partial a_{\alpha}{\beta}}{\partial y^{\nu}}\left(\nabla_i y^{\alpha}\right)\left(\nabla_j y^{\nu}\right)N^{\beta},\nonumber\\
&=&-\left[\Gamma_{\alpha \nu \beta}+\Gamma_{\beta \nu \alpha}\right]\left(\nabla_i y^{\alpha}\right)\left(\nabla_j y^{\nu}\right)N^{\beta}, \nonumber
\end{eqnarray}
\begin{eqnarray}
\mbox{or,}~a_{\alpha \beta}\left(\nabla_i \nabla_j y^{\alpha}\right)N^{\beta}+\Gamma_{\alpha \nu \beta}\left(\nabla_i y^{\alpha}\right)\left(\nabla_j y^{\nu}\right)N^{\beta}=-\left[a_{\alpha \beta}\left(\nabla_i y^{\alpha}\right)\left(\nabla_j N^{\beta}\right)+\Gamma_{\beta \nu \alpha}\left(\nabla_i y^{\alpha}\right)\left(\nabla_j y^{\nu}\right)N^{\beta}\right], \nonumber
\end{eqnarray}
\begin{eqnarray} \label{4.35}
\mbox{or,}~~~a_{\alpha \beta}\left(\nabla_i \nabla_j y^{\alpha}\right)N^{\beta}&+&\Gamma_{\mu \nu \beta}\left(\nabla_i y^{\mu}\right)\left(\nabla_j y^{\nu}\right)N^{\beta} \nonumber \\
&=& -\left[a_{\alpha \beta}\left(\nabla_i y^{\alpha}\right)\left(\nabla_j N^{\beta}\right)+\Gamma^{\mu}_{\beta \nu}a_{\mu \alpha}\left(\nabla_i y^{\alpha}\right)\left(\nabla_j y^{\nu}\right)N^{\beta}\right] \nonumber \\
&=& -\left[a_{\alpha \beta}\left(\nabla_i y^{\alpha}\right)\left(\nabla_j N^{\beta}\right)+a_{\alpha \beta}\Gamma^{\beta}_{\mu \nu}\left(\nabla_i y^{\alpha}\right)\left(\nabla_j y^{\mu}\right)N^{\nu}\right].
\end{eqnarray}
Using equation\,(\ref{4.35}), equation\,(\ref{4.34}) simplifies to 
\begin{equation} \label{4.36}
b_{ij}=-a_{\alpha \beta}\left(\nabla_i y^{\alpha}\right)\left[\left(\nabla_j N^{\beta}\right)+\Gamma^{\beta}_{\mu \nu}\left(\nabla_j y^{\mu}\right)N^{\gamma}\right].
\end{equation}
Taking co-variant derivative of equation\,(\ref{4.31}) with respect to $x^j$ we have
$$a_{\alpha \beta}\left(\nabla_j N^{\alpha}\right)N^{\beta}+a_{\alpha \beta}\left(\nabla_j N^{\beta}\right)N^{\alpha}=-\frac{\partial a_{\alpha \beta}}{\partial y^{\nu}}\left(\nabla_j y^{\nu}\right)N^{\alpha}N^{\beta}.$$
\begin{eqnarray}
\mbox{or,}~~~~2a_{\alpha \beta}\left(\nabla_j N^{\beta}\right)N^{\alpha}&=& -\left[\Gamma_{\nu \alpha \beta}+\Gamma_{\nu \beta \alpha}\right]\left(\nabla_j y^{\nu}\right)\left(N^{\alpha}N^{\beta}\right) \nonumber \\
&=& -2\Gamma_{\nu \alpha \beta}\left(\nabla_j y^{\nu}\right)N^{\alpha}N^{\beta} \nonumber
\end{eqnarray}
\begin{eqnarray} \label{4.37}
\mbox{or,}~~~~a_{\alpha \beta}\left(\nabla_j N^{\beta}\right)N^{\alpha}+\Gamma^{\mu}_{\nu \alpha}a_{\mu \beta}\left(\nabla_j y^{\nu}\right)N^{\alpha}N^{\beta}&=&0 \nonumber \\
\mbox{or,}~~~~a_{\alpha \beta}N^{\alpha}\left(\nabla_j N^{\beta}\right)+\Gamma^{\alpha}_{\nu \mu}a_{\alpha \beta}\left(\nabla_j y^{\nu}\right)N^{\mu}N^{\beta}&=&0~~~(\mu\rightleftharpoons\alpha) \nonumber \\
\mbox{or,}~~~~a_{\alpha \beta}N^{\alpha}\left(\nabla_j N^{\beta}\right)+\Gamma^{\beta}_{\nu \mu}a_{\alpha \beta}\left(\nabla_j y^{\nu}\right)N^{\mu}N^{\alpha}&=&0~~~(\alpha \rightleftharpoons \beta) \nonumber \\
\mbox{or,}~~~~a_{\alpha \beta} N^\alpha \left[ \left(\nabla_i N^\beta \right)+\Gamma^\beta_{\nu \mu} \left(\nabla_j y^\nu \right)N^\mu \right]&=&0.
\end{eqnarray}
Now comparing equation\,(\ref{4.36}) with equation\,(\ref{4.30}) we write 
\begin{equation} \label{4.38}
\nabla_jN^\beta+\Gamma^\beta_{\mu \nu}(\nabla_j y^\mu)N^\nu=\Lambda^k_j\nabla_ky^\beta
\end{equation}
Using (\ref{4.37}) in equation\,(\ref{4.36}) we obtain 
$$b_{ij}=-a_{\alpha \beta}\left(\nabla_i y^\alpha \right)\left(\nabla_k y^\beta \right)\Lambda^k_j=-g_{ik}\Lambda^k_j=-\Lambda_{ij}~.$$
Hence equation\,(\ref{4.37}) can be written as 
\begin{equation} \label{4.39}
\nabla_jN^\beta+\Gamma^\beta_{\mu \nu} \left(\nabla_j y^\mu \right)N^\gamma=-b^k_j\nabla_k y^\beta .
\end{equation}
From the Ricci identity :
\begin{equation} \label{4.40}
y^\alpha_{,ijk} - y^\alpha_{,ikj}=y^\alpha_{,m}g^{mh}R_{hijk}~,
\end{equation}
where $R_{hijk}$ is the Riemann curvature tensor in $V_n$ with respect to the metric $g_{ij}$ . Now from equation\,(\ref{4.33})
\begin{equation} \label{4.41}
\nabla _i \nabla _j y^\alpha =b_{ij}N^\alpha -\Gamma ^\alpha_{\mu \nu}\left(\nabla _i y^\mu \right)\left(\nabla _j y^\nu \right)
\end{equation}
Therefore,
\begin{eqnarray}
\nabla_i \nabla_j \nabla_k y^\alpha&=&\left(\nabla_k b_{ij}\right)N^\alpha+b_{ij}\left(\nabla_k N^\alpha\right)-\frac{\partial}{\partial y^\lambda}\Gamma^\alpha_{\mu \nu}\left(\nabla_k y^\lambda\right)\left(\nabla_i y^\mu\right)\left(\nabla_j y^\nu\right) \nonumber \\
&-&\Gamma^\alpha_{\mu \nu}\left(\nabla_k \nabla_i y^\mu \right)\left(\nabla_jy^\nu \right) -\Gamma^\alpha_{\mu \nu} \left(\nabla_i y^\mu \right)\left(\nabla_k \nabla_j y^\nu \right) \nonumber
\end{eqnarray}
From (\ref{4.41}),
\begin{eqnarray}
\nabla_i \nabla_k y^\alpha &=& b_{ik}N^\alpha-\Gamma^\alpha_{\mu \nu}\left(\nabla_i y^\mu\right)\left(\nabla_k y^\nu\right) \nonumber \\
\nabla_i \nabla_k \nabla_j y^\alpha &=& \left(\nabla_jb_{ik}\right)N^\alpha+b_{ik}\left(\nabla_j N^\alpha\right)-\frac{\partial}{\partial y^\lambda}\Gamma^\alpha_{\mu \nu}\left(\nabla_jy^\lambda\right)\left(\nabla_iy^\mu\right)\left(\nabla_k y^\nu\right) \nonumber \\
&& -\Gamma^\alpha_{\mu \nu}\left(\nabla_j \nabla_i y^\mu\right)\left(\nabla_ky^\nu\right) -\Gamma^\alpha_{\mu \nu} \left(\nabla_i y^\mu\right)\left(\nabla_j \nabla_k y^\nu\right) \nonumber
\end{eqnarray}
Using (\ref{4.41}) {\it i.e.,}
$$\nabla_i \nabla_ky^\alpha =b_{ik}N^\alpha -\Gamma^\alpha_{\mu \nu}\left(\nabla_iy^\mu \right)\left(\nabla_ky^\nu \right)$$
we have,
\begin{eqnarray}
\nabla_i\nabla_k\nabla_jy^\alpha &=& \left(\nabla_jb_{ik}\right)N^\alpha + b_{ik}\left(\nabla_jN^\alpha \right)- \frac{\partial}{\partial y^\alpha}\Gamma^\alpha_{\mu \nu}\left(\nabla_iy^\mu\right)\left(\nabla_ky^\nu\right)\left(\nabla_j y^\lambda\right) \nonumber \\
&&- \Gamma^\alpha_{\mu \nu}\left(\nabla_j \nabla_i y^\mu\right)\left(\nabla_ky^\nu\right) -\Gamma^\alpha_{\mu \nu} \left(\nabla_i y^\mu\right)\left(\nabla_j \nabla_k y^\nu\right) \nonumber
\end{eqnarray}
Therefore,
\begin{eqnarray}
\left(\nabla_i \nabla_j \nabla_k-\nabla_i \nabla_k \nabla_j\right)y^\alpha &=& N^\alpha\left[\nabla_kb_{ij}-\nabla_jb_{ik}\right]+b_{ij}\left[-b^l_k\nabla_ly^\alpha-\Gamma^\alpha_{\mu \nu}\left(\nabla_ky^\mu\right)N^\nu\right] \nonumber \\
&&-b_{ik}\left[-b^l_j\nabla_ly^\alpha-\Gamma^\alpha_{\mu \nu}\left(\nabla_jy^\mu\right)N^\nu\right] \nonumber \\
&&+ \left[-\frac{\partial}{\partial y^\lambda}\Gamma^\alpha_{\mu \nu}+\frac{\partial}{\partial y^\nu}\Gamma^\alpha_{\mu \lambda}\right]\left(\nabla_iy^\mu\right)\left(\nabla_jy^\nu\right)
\left(\nabla_ky^\lambda\right) \nonumber \\
&&- \Gamma^\alpha_{\mu \sigma}\left(\nabla_jy^\sigma\right)\left[b_{ik}N^\mu-\Gamma^\mu_{\lambda \sigma}\left(\nabla_iy^\lambda\right)\left(\nabla_ky^\sigma\right)\right] \nonumber \\
&&+ \Gamma^\alpha_{\mu \nu}\left(\nabla_ky^\nu\right)\left[b_{ij}N^\mu-\Gamma^\mu_{\lambda \sigma}\left(\nabla_iy^\lambda\right)\left(\nabla_jy^\sigma\right)\right] \nonumber
\end{eqnarray}
$$\mbox{or,}~~~\nabla_my^\alpha g^{m h}R_{hijk} - N^\alpha\{\nabla_kb_{ij}-\nabla_jb_{ik}\} + b_{ij}b_{kh}g^{mh}\nabla_my^\alpha - b_{ik}b_{jh}g^{mh}\nabla_my^\alpha$$
$$= -b_{ij}\Gamma^\alpha_{\mu \nu}\left(\nabla_ky^\mu\right)N^\nu + b_{ik}\Gamma^\alpha_{\mu \nu}\left(\nabla_jy^\mu\right)N^\nu$$
$$~~~~~~~~~~~~~~~+ \left[\left(-\frac{\partial}{\partial y^\lambda}\Gamma^\alpha_{\mu \nu}+\frac{\partial}{\partial y^\nu}\Gamma^\alpha_{\mu \lambda}\right)\left(\nabla_iy^\mu\right)\left(\nabla_jy^\nu\right)
\left(\nabla_ky^\lambda\right)\right]$$
$$~~~~-b_{ik}\Gamma^\alpha_{\mu \nu}\left(\nabla_jy^\mu\right)N^\nu + b_{ij}\Gamma^\alpha_{\mu \nu}\left(\nabla_ky^\mu\right)N^\nu$$
$$~~~~~~~~~~~~~~~~~~~~~~~~~~~~~~+\Gamma^\alpha_{\mu \nu}\Gamma^\mu_{\lambda \sigma}\left(\nabla_iy^\lambda\right)\left(\nabla_jy^\nu\right)
\left(\nabla_ky^\sigma\right)-\Gamma^\alpha_{\mu \nu}\Gamma^\mu_{\lambda \sigma}\left(\nabla_iy^\lambda\right)\left(\nabla_jy^\sigma\right)
\left(\nabla_ky^\nu\right)$$
\\
$$\mbox{or,}~~~\left(\nabla_my^\alpha\right)g^{mh}\left[R_{hijk}-\left(b_{ik}b_{jh}-b_{ij}b_{kh}\right)\right]-N^\alpha\left(\nabla_kb_{ij}-\nabla_jb_{ki}\right)$$
$$= \left[-\frac{\partial}{\partial  y^\lambda}\Gamma^\alpha_{\mu \nu}+\frac{\partial}{\partial y^\nu}\Gamma^\alpha_{\mu \lambda}+\Gamma^\alpha_{\sigma \mu}\Gamma^\sigma_{\mu \lambda}-\Gamma^\alpha_{\sigma \lambda}\Gamma^\sigma_{\mu \nu} \right]\left(\nabla_iy^\mu\right)\left(\nabla_jy^\nu\right)
\left(\nabla_ky^\lambda\right)$$
\begin{equation} \label{4.42}
= \overline{R}^\alpha_{\mu \sigma \lambda}\left(\nabla_iy^\mu\right)\left(\nabla_j y^\sigma\right)\left(\nabla_k y^\lambda\right)
\end{equation}
Now multiplying equation (\ref{4.42}) by $a_{\alpha \beta}\nabla_ly^\beta$ and summing over $\alpha$ , we obtain,
\begin{eqnarray} \label{4.43}
R_{lijk}&=&\left(b_{ik}b_{jl}-b_{ij}b_{lk}\right)+\overline{R}_{\beta \mu \gamma \lambda}\left(\nabla_iy^\mu\right)\left(\nabla_jy^\gamma\right)
\left(\nabla_ky^\lambda\right)\left(\nabla_ly^\beta\right) \nonumber \\
i.e.,~~~~R_{ijkl}&=&\left(b_{jl}b_{ki}-b_{jk}b_{il}\right)+\overline{R}_{\alpha \beta \gamma \delta}\left(\nabla_i y^\alpha\right)\left(\nabla_j y^\beta\right)\left(\nabla_k y^\gamma\right)\left(\nabla_l y^\delta\right).
\end{eqnarray}
Again multiplying eq.\,(\ref{4.42}) by $a_{\alpha \beta}N^\beta$ we have
\begin{eqnarray} \label{4.44}
-\nabla_kb_{ij}+\nabla_jb_{ki}&=&\overline{R}_{\beta \mu \gamma \lambda}\left(\nabla_iy^\mu\right)\left(\nabla_jy^\gamma\right)
\left(\nabla_ky^\lambda\right)N^\beta \nonumber \\
i.e.,~~~~\nabla_k b_{ij}-\nabla_j b_{ik}&=&\overline{R}_{\alpha \beta \gamma \delta}(\nabla_i y^\alpha)(\nabla_j y^\gamma)(\nabla_k y^\delta)N^\beta~.
\end{eqnarray}
Here equations (\ref{4.43}) and (\ref{4.44}) are respectively called the Gauss equation and Codazzi equation on the hypersurface $V_n$ in $V_{n+1}$ .\\\\
{\bf Note:} The above formul\ae ~are also true for the hypersurface $V_n$ in $V_m~(m>n)$.\\\\

\section{Hypersurfaces in Euclidean Space\,: Spaces of constant curvature}

\subsection{Hyper-plane and Hyper-sphere}

$\bullet$ {\bf Hyper-plane}: Let $E_n$ be a flat space and $y^\alpha$ be the Euclidean co-ordinate in $E_n$ . Then the linear equation 
$$a_{\alpha} y^\alpha=d$$
represents a hyperplane in $E_n$ with $a_\alpha$'s and $d$ being constants.\\\\
As $$\nabla(a_{\alpha} y^\alpha)=a_\alpha$$
so the normal has the d.r. ($a_1, a_2, \ldots , a_n$). If $N^\alpha$ denotes the unit normal then 
$$N^\alpha=\frac{a_\alpha}{\sqrt{\sum ^n_{\alpha =1}a_\alpha^2}}~~~~ \forall \alpha $$
{\bf Note:} Angle between two hyperplanes mean angle between the corresponding normals. So hyperplanes are parallel if the corresponding normals are parallel.

The equation of a hyperplane passing through a given point $c^\alpha$ is given by $$a_\alpha \left(y^\alpha-c^\alpha\right)=0$$
In fact equation of a hyperplane through `$n$' points $c_h^\alpha$ $(h=1,2, \ldots ,n)$ can be written as
$$\left| \begin{array}{ccccc}
y^1&y^2&\ldots&y^n&1\\
c^1_1&c^2_1&\ldots&c_1^n&1\\
c^1_2&c^2_2&\ldots&c_2^n&1\\
..&..&..&..&..\\
c^1_n&c^2_n&\ldots&c_n^n&1
\end{array}\right|=0$$\\
$\bullet$ {\bf Hyper-sphere:} The locus of a point which is always at a constant distance\,$(r)$ from a fixed point $A(a^\alpha)$ is called a hypersphere. The equation of the hypersphere is
$$\sum^n_{\alpha=1}\left(y^\alpha-a^\alpha\right)^2=r^2$$
Here $A$ is the centre of the hyper-sphere and $r$ is the radius. From the equation of the sphere taking differential we have 
$$\sum^n_{\alpha=1}\left(y^\alpha-a^\alpha\right)dy^\alpha=0$$
As $dy^\alpha$ is along the tangent to the hyper-sphere so $(y^\alpha-a^\alpha)$ \textit{i.e.}, $\textit{\textbf{AP}}$ is along the normal.

The tangent hyper-plane at any point $B(b^\alpha)$ is given by 
$$\sum^n_{\alpha=1}\left(y^\alpha-a^\alpha\right)\left(b^\alpha-a^\alpha\right)=0$$
The equation of a hyper-sphere which has the points $A(a^\alpha)$ and $B(b^\alpha)$ as the two ends of a diameter is
$$\sum^n_{\alpha=1}\left(y^\alpha-a^\alpha\right)\left(y^\alpha-b^\alpha\right)=0.$$

\subsection{Central quadric hypersurfaces}

In flat space $E_n$ , the Riemannian co-ordinates of a point $P(y^\alpha)$ with respect to pole $O$ is defined as
\begin{equation} \label{4.45}
y^\alpha = st^\alpha
\end{equation}
where `$s$' is the arc length $OP$ in the direction of unit vector $t^\alpha$ at $P$. For any symmetric tensor $a_{\alpha \beta}$ the equation 
\begin{equation} \label{4.46}
a_{\alpha \beta}y^\alpha y^\beta =1
\end{equation}
represents a hypersurface, called central quadric hypersurface with centre at the pole $O$.

Using (\ref{4.45}) and (\ref{4.46}) we have
\begin{equation} \label{4.47}
\frac{1}{s^2}=a_{\alpha \beta} t^\alpha t^\beta
\end{equation}
So we have two equal and opposite values of s. Hence we have the following result:

``A straight line through the centre of a central quadric intersects the quadric in two points equidistant from the centre.''\\\\
The positive value of `$s$' given by (\ref{4.47}) is called the radius of the quadric along the direction $t^\alpha$.
Taking differential of equation\,(\ref{4.46}) we have
$$a_{\alpha \beta}y^{\alpha} dy^{\beta}=0~,$$
which shows that $y^{\alpha}$ is along the normal to the quadric. Thus equation of the tangent hyperplane at $(p^{\alpha})$ is\\
$$a_{\alpha \beta}p^{\alpha}\left(y^{\beta}-p^{\beta}\right)=0$$
$$a_{\alpha \beta}y^{\beta}p^{\alpha}=a_{\alpha \beta}p^{\alpha}p^{\beta}=1$$~~(as $p^{\alpha}$ is on the quadric).\\\\
{\bf Note:} $y^{\alpha}$ is any point on the tangent hyperplane.\\\\
So equation of the tangent hyperplane at $p^{\alpha}$ is 
$$a_{\alpha \beta}y^{\alpha}p^{\beta}=1.$$
{\bf Theorem 4.19 :} The sum of the inverse square of the radii of the quadric for $n$ mutually orthogonal directions at $0$ is invraiant and is equal to $a_{\alpha \beta}g^{\alpha \beta}$ , where $g_{\alpha \beta}$ is the metric tensor of flat space $E_n$ .\\\\
{\bf Proof:} Let $t^{\alpha}_h~(h=1,2,\ldots , n)$ be the $n$ unit tangent vectors to $n$ congruences $e_h~(h=1,2, \ldots , n)$ of an orthogonal ennuple in $E_n$ .  Now the radius $r_h$ corresponding to the direction $t^{\alpha}_h$ is given by
$$\frac{1}{r^2_h}=a_{\alpha \beta}t^{\alpha}_h t^{\beta}_h~.$$
Hence $$\sum^n_{h=1} \frac{1}{r^2_h}=\sum a_{\alpha \beta} t^{\alpha}_h t^{\beta}_h=a_{\alpha \beta}\sum_h t^{\alpha}_h t^{\beta}_h=a_{\alpha \beta} g^{\alpha \beta}.$$

\subsection{Evolute of hyperurface\,: Principal radii of normal curvature}

Let $V_n$ be a hypersurface of Euclidean space $E_{n+1}$ . Then from the Weingarten's formula\\
$$\nabla_i N^{\alpha}=-b^k_i \nabla_k y^{\alpha}$$
We write 
\begin{equation} \label{4.48}
N^{\alpha}_{,i}=-b^k_i y^{\alpha}_{,k}~~(\mbox{as}~ E_{n+1} ~\mbox{is Euclidean}).
\end{equation}
From Gauss formula\,(\ref{4.6}) in Euclidean space $E_{n+1}$
\begin{equation} \label{4.49}
y^{\alpha}_{,ij}=b_{ij} N^{\alpha}.
\end{equation}
Also the metric tensor in $V_n$ is given by 
\begin{equation} \label{4.50}
g_{ij}=\sum^{n+1}_{\alpha=1} y^{\alpha}_{,i}\,y^{\alpha}_{,j}.
\end{equation}
Let $\overline{p}(\overline{y})^{\alpha}$ be a point on the unit normal $N^{\alpha}$  at a distance $\rho$ from $p$ along the normal, then we write
\begin{equation} \label{4.51}
\overline{y}^{\alpha}=y^{\alpha}+\rho N^{\alpha}~.
\end{equation}
As $P$ moves on $V_n$ then the corresponding displacement of $\overline{p}$ is given by
$$d\overline{y}^{\alpha}=\left(y^{\alpha}_{,i}+\rho N^{\alpha}_{,i}\right)dx^i+N^{\alpha} d\rho.$$\\

The first term on the right hand side is tangential to $V_n$ while the second term is along the normal vector. So if we assume that $\overline{p}$ moves along the normal to the hypersurface then the first term vanishes, {\it i.e.,}
$$\left(y^{\alpha}_{,i}+\rho N^{\alpha}_{,i}\right)dx^i=0$$
$$i.e.,~~~\left(y^{\alpha}_{,i}-\rho b^k_i y^{\alpha}_{,k}\right)dx^i=0.~~~~\mbox{(using\,}(\ref{4.48}))$$\\
Now, multiply by $y^{\alpha}_{,l}$ and summing over `$\alpha$' we have
$$\left(g_{il}-\rho b^k_i g_{kl}\right)dx^i=0$$
\begin{equation} \label{4.52}
i.e.;~\left(g_{il}-\rho b_{il}\right)dx^i=0.
\end{equation}
The direction $dx^i$ given by (\ref{4.52}) are principal direction of the hypersurface and the roots $\rho$ of the equation
$$|g_{il}-\rho b_{il}|=0$$
are called principal radii of normal curvature. The locus of $\overline{p}$ satisfying equation\,(\ref{4.51}) is called the evolute of the hypersurface $V_n$ in $E_{n+1}$.\\\\
{\bf Note:} The evolute is also a hypersurface of $E_{n+1}$.\\\\
{\bf Theorem 4.20 :} A hyperplane and a hypersphere are the only hypersurfaces of an Euclidean space $E_{n+1}$ whose all points are umbilical points.\\\\
{\bf Proof:} As $E_{n+1}$ is Euclidean space so the metric $a_{\alpha \beta}=\delta_{\alpha \beta}$\\
$$i.e.,~~~ ds^2=a_{\alpha \beta}dy^{\alpha}dy^{\beta}=\sum^{n+1}_{\alpha=1}\left(dy^{\alpha}\right)^2.$$
So from the Weingarten's formula we have
$$N^{\alpha}_{,i}=-b^k_i y^{\alpha}_{,k}~.$$
As all points of the hypersurface $V_n$ in $E_{n+1}$ are umbilical points so we have
\begin{equation} \label{4.53}
b_{ij}=\mu g_{ij}~.
\end{equation}
Hence from (\ref{4.48}) we have
\begin{equation} \label{4.54}
N^{\alpha}_{,i}=-\mu \delta^k_i y^{\alpha}_{,k}= -\mu y^{\alpha}_{,i}~.
\end{equation}
$$i.e.,~~~ N^{\alpha}_{,i}+\mu y^{\alpha}_{,i}=0$$
Differentiating covariantly with respect to $x^j$ we have
$$N^{\alpha}_{,ij}+\mu_{,j}y^{\alpha}_{,i}+\mu y^{\alpha}_{,ij}=0.$$
Integrating $i~,~j$ we get
$$N^{\alpha}_{,ji}+\mu_{,i}y^{\alpha}_{,j}+\mu y^{\alpha}_{,ji}=0.$$
Subtracting these two equations we get
$$\mu_{,j}y^{\alpha}_{,i}-\mu_{,i}y^{\alpha}_{,j}=0.$$
Multiplying by $y^{\alpha}_{,l}$ and summing over $\alpha$ we get
$$\mu_{,j} g_{li}-\mu_{,i} g_{lj}=0~~\left(\mbox{as}~g_{ij}=\sum^{n+1}_{\alpha=1} y^{\alpha}_{,i}y^{\alpha}_{,j}\right).$$

Now multiplying by $g^{il}$ we obtain
$$n\mu_{,j}-\mu_{,j}=0$$
$$i.e;~(n-1) \mu_{,j}=0~~i.e;\frac{\partial \mu}{\partial x^j}=0.$$
So $\mu$ is constant throughout the hypersurface. Again integrating equation (\ref{4.54}) we have
\begin{equation} \label{4.55}
N^{\alpha}+\mu y^{\alpha}=c^{\alpha}
\end{equation}
where $c^{\alpha}$ is a constant vector.\\

{\bf Case: I}~~ ~~$\mu=0$\\

Then from (\ref{4.55})
\begin{eqnarray}
N^{\alpha}&=&c^{\alpha} \nonumber \\
i.e.,~~~N_{\alpha}y^{\alpha}_{,i}&=&c_{\alpha}y^{\alpha}_{,i} \nonumber \\
i.e.,~~~0&=&c_{\alpha}y^{\alpha}_{,i} \nonumber \\
i.e.,~~~c_{\alpha}y^{\alpha}&=&\lambda~,~~\mbox{a constant.} \nonumber
\end{eqnarray}

which is the equation of a hyperplane.\\

{\bf Case: II}~~ ~~ $\mu\neq 0$\\

Then from (\ref{4.55}),~~~~~~~~~~~$N^{\alpha}=c^{\alpha}-\mu y^{\alpha}$
$$\Rightarrow ~~~ \sum_{\alpha} N^{\alpha}N^{\alpha}=\sum_{\alpha} \left(c^{\alpha}-\mu y^{\alpha}\right)
\left(c^{\alpha}-\mu y^{\alpha}\right)$$
$i.e.,~~~\sum\left(y^{\alpha}-b^{\alpha}\right)^2=\frac{1}{\mu^2}=R^2$ , a hypersphere. Hence the theorem.\\\\

\section*{Appendix-I}

To prove\\
$$L_{\stackrel{\rightarrow}{V}}B=d\left[B(\stackrel{\rightarrow}{V})\right]+(dB)(\stackrel{\rightarrow}{V}),~~B~ \mbox{is a}~r~\mbox{form}.~~~~~~~~~~~~~~~~~~~~~~~~~~~~~~~~~~~~~~~~~(\mbox{I.1})$$\\
Note that the resulting $r-$form on the left hand side is expressed as the sum of two $r-$forms on the right hand side-- the first one is the exterior derivative of the contracted $(r-1)-$form $B(\stackrel{\rightarrow}{V})$ while the second one is the contraction of the $(r+1)$ form $dB$ with $\stackrel{\rightarrow}{V}$ {\it i.e.}, the first term corresponds to contraction followed by the exterior differentiation while in the second term exterior differentiation followed by contraction.\\

We shall prove the result by induction. For $r=0$ we choose $B=f$, a function. So\\
$$ L_{\stackrel{\rightarrow}{V}}f=\frac{df}{d \lambda}~,~~\mbox{if}~~\stackrel{\rightarrow}{V} =\frac{d}{d \lambda}~.$$
In this case contraction on $\stackrel{\rightarrow}{V}~~i.e.~~f(\stackrel{\rightarrow}{V})$ is zero by definition and
$$(df)(\stackrel{\rightarrow}{V})=\frac{df}{d \lambda}$$\\
Hence
$$ L_{\stackrel{\rightarrow}{V}}f=\frac{df}{d \lambda}=(df)(\stackrel{\rightarrow}{V})$$
So the result (I.1) is true for $r=0$.\\

Suppose $B=\underline{w}$, a one--form. Then in components
$$\underline{w}(\stackrel{\rightarrow}{V})=w_i V^i~~\mbox{ and hence}~~d[\underline{w}(\stackrel{\rightarrow}{V})]=d(w_i V^i)=(w_i V^i)_{,k}dx^k.$$
Also
\begin{eqnarray}
d\underline{w}=d(w_i dx^i)&=& dw_i \Lambda dx^i=w_{i, k}dx^k\Lambda dx^i\nonumber\\
&=& w_{i, k}(dx^k\otimes dx^i-dx^i\otimes dx^k)\nonumber.
\end{eqnarray}
\begin{eqnarray}
d\underline{w}(\stackrel{\rightarrow}{V})&=& w_{i,k}\left[dx^k(\stackrel{\rightarrow}{V})dx^i-dx^i(\stackrel{\rightarrow}{V})dx^k\right]\nonumber\\
&=& w_{i, k}\left[V^k dx^i-V^i dx^k\right]\nonumber.
\end{eqnarray}
\begin{eqnarray}
d[\underline{w}(\stackrel{\rightarrow}{V})]+(d\underline{w})(\stackrel{\rightarrow}{V})&=& w_{i, k} V^i dx^k+w_i V^i_{,k}dx^k+w_{i, k}V^k dx^i-w_{i,k}V^idx^k\nonumber\\
&=& w_{i,k}V^kdx^i+w_k V^k_{,i}dx^i\nonumber\\
&=& \left[w_{i, k}V^k+w_kV^k_{,i}\right]dx^i=L_{\stackrel{\rightarrow}{V}}\underline{w}~.\nonumber
\end{eqnarray}
Thus (I.1) is true for $r=1$. For higher forms we shall prove the result by induction. In general, an arbitrary $r-$form $B$ can be represented as a sum of functions times the wedge products of $r$ one-forms. So we can write
$$B=B_0 \alpha \wedge \beta.$$

where $B_0$ is a scalar function, $\alpha$ is a $s-$form $(s< r)$ and $\beta$ is a $(r-s)-$form.\\\\
Suppose the relations (I.1) is true for both $s-$form and $(r-s)-$forms, then
\begin{eqnarray}
L_{\stackrel{\rightarrow}{V}}B&=& \left(L_{\stackrel{\rightarrow}{V}}B_0\right)\alpha \wedge \beta+B_0\left(L_{\stackrel{\rightarrow}{V}}\alpha\right)\wedge \beta+B_0\alpha\wedge\left(L_{\stackrel{\rightarrow}{V}}\beta\right)\nonumber\\
&=& dB_0(\stackrel{\rightarrow}{V})\alpha \wedge \beta+B_0\left[d(\alpha(\stackrel{\rightarrow}{V}))+(d\alpha)(\stackrel{\rightarrow}{V})\right]\wedge \beta\nonumber\\
&+&B_0\alpha\wedge\left\{d\left[\beta(\stackrel{\rightarrow}{V})\right]+(d\beta)(\stackrel{\rightarrow}{V})\right\}. \nonumber
\end{eqnarray}
Now,
\begin{eqnarray}
d\left[\beta(\stackrel{\rightarrow}{V})\right]&=&d\left[B_0\alpha(\stackrel{\rightarrow}{V})\wedge\beta + (-1)^sB_0\alpha\wedge\beta(\stackrel{\rightarrow}{V})\right] \nonumber\\
&=&(dB_0)\left[\alpha(\stackrel{\rightarrow}{V})\wedge\beta + (-1)^s\alpha\wedge\beta(\stackrel{\rightarrow}{V})\right] \nonumber\\
&+& B_0\left\{d\left[\alpha(\stackrel{\rightarrow}{V})\right]\wedge\beta + (-1)^{s-1}\alpha(\stackrel{\rightarrow}{V})\wedge d\beta + (-1)^sd\alpha\wedge\beta(\stackrel{\rightarrow}{V})+\alpha\wedge\left[d\beta(\stackrel{\rightarrow}{V})\right]\right\} \nonumber
\end{eqnarray}
\begin{eqnarray}
(d\beta)(\stackrel{\rightarrow}{V})&=&\left[dB_0\wedge\alpha\wedge\beta + B_0d\alpha\wedge d\beta + (-1)^sB_0\alpha\wedge d\beta\right](\stackrel{\rightarrow}{V}) \nonumber\\
&=&dB_0(\stackrel{\rightarrow}{V})\alpha\wedge\beta - dB_0\wedge\left[(\alpha\wedge\beta)(\stackrel{\rightarrow}{V})\right] \nonumber \\
&+& B_0\left[d\alpha(\stackrel{\rightarrow}{V})\wedge\beta + (-1)^{s+1}d\alpha\wedge\beta(\stackrel{\rightarrow}{V}) + (-1)^s\alpha(\stackrel{\rightarrow}{V})\wedge d\beta + \alpha\wedge\beta(\stackrel{\rightarrow}{V})\right] \nonumber
\end{eqnarray}
Thus,
\begin{eqnarray}
d\left[\beta(\stackrel{\rightarrow}{V})\right] + (d\beta)(\stackrel{\rightarrow}{V})&=& dB_0(\stackrel{\rightarrow}{V})\alpha\wedge\beta + B_0\left\{d\left[\alpha(\stackrel{\rightarrow}{V})+(d\alpha)(\stackrel{\rightarrow}{V})\right]\wedge\beta\right. \nonumber \\
&+&\left. B_0\alpha\wedge\left\{d\left[\beta(\stackrel{\rightarrow}{V})\right] + (d\beta)(\stackrel{\rightarrow}{V})\right\}\right\} \nonumber \\
&=& L_{\stackrel{\rightarrow}{V}}B \nonumber
\end{eqnarray}
Hence by induction the relation (I.1) holds for any $B$.\\\\

\section*{Appendix-II : Differentiation of a determinant}

Let us consider a $n \times n$ determinant
$$\Delta = \left| \begin{array}{cccccc}
a_{11} & a_{12} & \ldots \ldots & a_{1n} \\
a_{21} & a_{22} & \ldots \ldots & a_{2n} \\
\ldots & \ldots & \ldots \ldots & \ldots \\
a_{r1} & a_{r2} & \ldots \ldots & a_{rn} \\
\ldots & \ldots & \ldots \ldots & \ldots \\
a_{n1} & a_{n2} & \ldots \ldots & a_{nn}
\end{array}\right|$$

Here the elements of $\Delta$ are functions of the variable $x$ (say).\\\\
So,
$$\frac{d\Delta}{dx} = \left| \begin{array}{cccccc}
\frac{da_{11}}{dx} & \frac{da_{12}}{dx} & \ldots \ldots & \frac{da_{1n}}{dx} \\
a_{21} & a_{22} & \ldots \ldots & a_{2n} \\
\ldots & \ldots & \ldots \ldots & \ldots \\
a_{r1} & a_{r2} & \ldots \ldots & a_{rn} \\
\ldots & \ldots & \ldots \ldots & \ldots \\
a_{n1} & a_{n2} & \ldots \ldots & a_{nn}
\end{array}\right| + \left| \begin{array}{cccccc}
a_{11} & a_{12} & \ldots \ldots & a_{1n} \\
\frac{da_{21}}{dx} & \frac{da_{22}}{dx} & \ldots \ldots & \frac{da_{2n}}{dx} \\
\ldots & \ldots & \ldots \ldots & \ldots \\
a_{r1} & a_{r2} & \ldots \ldots & a_{rn} \\
\ldots & \ldots & \ldots \ldots & \ldots \\
a_{n1} & a_{n2} & \ldots \ldots & a_{nn}
\end{array}\right| + \cdots \cdots + \left| \begin{array}{cccccc}
a_{11} & a_{12} & \ldots \ldots & a_{1n} \\
a_{21} & a_{22} & \ldots \ldots & a_{2n} \\
\ldots & \ldots & \ldots \ldots & \ldots \\
a_{r1} & a_{r2} & \ldots \ldots & a_{rn} \\
\ldots & \ldots & \ldots \ldots & \ldots \\
\frac{da_{n1}}{dx} & \frac{da_{n2}}{dx} & \ldots \ldots & \frac{da_{nn}}{dx}
\end{array}\right|.$$
If these $n$ determinants are denoted by $I_1~,~I_2~, \ldots ,~I_n$ then
\begin{eqnarray}
I_1 &=& \frac{da_{11}}{dx}A^{11} + \frac{da_{12}}{dx}A^{21} + \cdots \cdots + \frac{da_{1n}}{dx}A^{n1}~~~~\mbox{(expanding by row)} \nonumber\\
&=& \frac{da_{1j}}{dx}A^{j1} \nonumber
\end{eqnarray}
Similarly,
$$I_2 = \frac{da_{2j}}{dx}A^{j2}~~,~\ldots \ldots ~,~~I_r = \frac{da_{rj}}{dx}A^{jr}~~,~\ldots \ldots ~,~~I_n = \frac{da_{nj}}{dx}A^{jn}.$$\\
$$\mbox{Hence,}~~~~~~~~~ \frac{dA}{dx} = \frac{da_{ij}}{dx}A^{ji}.$$\\
\begin{center}
	\underline{-----------------------------------------------------------------------------------} 
\end{center}
\newpage
\vspace{5mm}
\begin{center}
	\underline{\bf Exercise} 
\end{center}
\vspace{3mm}
{\bf 4.1.} Show that $\sum\limits_{h=1}^n\kappa_h=M$ , where $\kappa_h$ are the principal curvatures.\\\\
{\bf 4.2.} Find the metric form and the asymptotic lines of the cylindroid : $x=u \cos v~,~y=u \sin v~,~z=m \sin 2v$ in $E_3$ where $(x,~y,~z)$ are rectangular cartesian coordinates.\\\\
{\bf 4.3.} Show that on the surface 
$$x=3u(1+v^2)-u^3~~,~~y=3v(1+u^2) -v^3~~,~~z=3(u^2-v^2)$$
the asymptotic lines are $u\neq v=$\,constants.\\\\
{\bf 4.4.} Show that the asymptotic lines on the paraboloid $2z=\dfrac{x^2}{a^2}-\dfrac{y^2}{b^2}$ lie on the planes $\dfrac{x}{a} \pm \dfrac{y}{b}=$\,constant.\\\\
{\bf 4.5.} If $b_{ij}\equiv 0$  for a hypersurface in $E_{n+1}$ , then show that the hypersurface is a hyperplane.\\\\
{\bf 4.6.}  If for a hypersurface in $E_{n+1}$ , if $b_{ij}=\lambda g_{ij}$ then $\lambda$ is a global constant and if this constant is different from zero then the hypersurface is a hypersphere.\\\\
{\bf 4.7.} On the right helicoid $\textit{\textbf{r}}=(u \cos v,~u \sin v,~cv)$, show that the principal curvatures are $\pm \dfrac{c}{u^2+c^2}$ and that the differential equation of the lines of curvature is 
$$du^2-\left(c^2+u^2\right)dv^2=0.$$\\
{\bf 4.8.} Show that the right helicoid $\textit{\textbf{r}}=(u \cos v,~u \sin v,~cv)$ is a minimal surface.\\\\
{\bf 4.9.} For the surface of revolution :
$$x=u \cos \phi ~,~y=u \sin \phi ~,~z=f(u)~,$$
prove that the parametric curves are the lines of curvature and find the principal curvatures.\\\\
{\bf 4.10.} Determine the orthogonal trajectories of the $u$-curves on the surface
$$\textit{\textbf{r}}=\left(u+v~,~1-uv~,~u-v\right)$$\\
{\bf 4.11.} Find the principal curvatures and the differential equation of the lines of curvature on the surface
$$x=a(u+v)~, ~y=b(u-v)~,~z=uv$$\\
{\bf 4.12.} Show that the lines of curvature of the paraboloid $xy=az$ lie on the surface $$\sinh ^{-1}\left(\frac{x}{a}\right)\pm \sinh ^{-1}\left(\frac{y}{a}\right)=~\mbox{constant}$$\\
{\bf 4.13} Find the asymptotic lines of the cylindroid $$x=u\cos v~,~y=u\sin v~,~z=m\sin 2v~.$$\\
{\bf 4.14.} Show that on the surface
$$x=3u(1+v^2)-u^3~,~y=3v(1+u^2)-v^3~,~z=3(u^2-v^2)~,$$
the asymptotic lines are $u\pm v= \mbox{constant}$.\\\\
{\bf 4.15.} Show that the asymptotic lines on the paraboloid
$$2z=\frac{x^2}{a^2}-\frac{y^2}{b^2}$$
lie on the planes $\dfrac{x}{a}\pm \dfrac{y}{b}= \mbox{constant}$.\\\\
{\bf 4.16.} Find the asymptotic lines on the right helicoid $$x=u\cos v~,~y=u\sin v~,~z=bv$$\\
{\bf 4.17.} Show that for a hypersurface in Euclidean space the Gauss and Codazzi equations reduce to $$R_{lijk}=b_{lj}b_{ik}-b_{lk}b_{ij}$$ and $$\nabla_k b_{ij}-\nabla_j b_{ik}=0.$$\\
{\bf 4.18.} For a hypersurface of a space of constant curvature $k$, the equations of Gauss and Codazzi reduce to 
$$R_{hijk}=(b_{hj}b_{ik}-b_{hk}b_{ij})+k(g_{hj}g_{ik}-g_{hk}g_{ij})$$ and $$\nabla_k b_{ij}-\nabla_j b_{ik}=0.$$\\
{\bf 4.19.} When the lines of curvature of a hypersurface of a space of constant curvature $\kappa_{n+1}$ are indeterminate, prove that the hypersurface has constant curvature $\kappa_n$, given by
$$\kappa_n=\frac{M^2}{n^2}+\kappa_{n+1}~.$$\\

\begin{center}
	\underline{\bf Solution and Hints} 
\end{center}
\vspace{3mm}
{\bf Solution 4.1:} Let $\textit{\textbf{t}}_{\bm 1},\textit{\textbf{t}}_{\bm 2}, \ldots, \textit{\textbf{t}}_{\bm n}$ be the unit vectors along the principal directions and hence they are orthogonal to each other.
$$\kappa_h=b_{ij}t_h^i t_h^j~~(h=1,2,\ldots n)$$
$$\sum _{h=1}^n\kappa_h=\sum _h b_{ij}t_h^i t_h^j=b_{ij} \sum _h t_h^i t_h^j=b_{ij} g^{ij}=M.$$\\
{\bf Solution 4.2:}
\begin{eqnarray}
\textit{\textbf{e}}_{\bm 1} &=& \left(\frac{\partial x}{\partial u}~,~\frac{\partial y}{\partial u}~,~\frac{\partial z}{\partial u}\right)=\left(\cos v~,~ \sin v~,~0\right) \nonumber \\
\textit{\textbf{e}}_{\bm 2} &=& \left(\frac{\partial x}{\partial u}~,~\frac{\partial y}{\partial u}~,~\frac{\partial z}{\partial u}\right)=\left(-u \sin v~,~u \cos v~,~2m \cos 2v\right) \nonumber \\
g_{11} &=& \textit{\textbf{e}}_{\bm 1} \cdot \textit{\textbf{e}}_{\bm 1}=\cos ^2v+ \sin ^2v=1 \nonumber
\end{eqnarray}
$$g_{12}=\textit{\textbf{e}}_{\bm 1} \cdot \textit{\textbf{e}}_{\bm 2}=0~~~,~~~g_{22}=\textit{\textbf{e}}_{\bm 2} \cdot \textit{\textbf{e}}_{\bm 2} =u^2+ 4m^2 \cos ^2 2v$$
Hence the metric form is
$$ds^2=du^2+\left(u^2+4m^2 \cos ^2 2v\right)dv^2$$
Now,
\begin{eqnarray}
\partial_1 \textit{\textbf{e}}_{\bm 1} &=& \frac{\partial \textit{\textbf{e}}_{\bm 1}}{\partial u}=(0,~0,~0) \nonumber \\
\partial_1 \textit{\textbf{e}}_{\bm 2} &=& \frac{\partial \textit{\textbf{e}}_{\bm 2}}{\partial u}=(-\sin v,~\cos v,~0) = \partial_2 \textit{\textbf{e}}_{\bm 1} \nonumber \\
\partial_2 \textit{\textbf{e}}_{\bm 2} &=& \frac{\partial \textit{\textbf{e}}_{\bm 2}}{\partial v}=(-u\cos v,-u\sin v,-4m\sin 2v) \nonumber
\end{eqnarray}
so
$$\textit{\textbf{e}}_{\bm 1} \times \textit{\textbf{e}}_{\bm 2}=(2m \sin v \cos 2v, -2m \cos v \cos 2v, ~u).$$
Hence,
$$\textit{\textbf{N}}=\frac{\textit{\textbf{e}}_{\bm 1} \times \textit{\textbf{e}}_{\bm 2}}{\left|\textit{\textbf{e}}_{\bm 1} \times \textit{\textbf{e}}_{\bm 2}\right|}=\lambda(2m \sin v \cos 2v, -2m \cos v \cos 2v, ~u)$$
where~~~ $\lambda=\frac{1}{\left|\textit{\textbf{e}}_{\bm 1} \times \textit{\textbf{e}}_{\bm 2}\right|}=\left(u^2+4m^2 \cos ^2 2v\right)^{-\frac{1}{2}}~.$\\
Thus,
$$b_{11}=\left(\partial_1 \textit{\textbf{e}}_{\bm 1}\right) \cdot \textit{\textbf{N}}=0$$
$$b_{12}=\left(\partial_1 \textit{\textbf{e}}_{\bm 2}\right) \cdot \textit{\textbf{N}}=\lambda(-2m \cos 2v)$$
$$b_{22}=\left(\partial_2 \textit{\textbf{e}}_{\bm 2}\right) \cdot \textit{\textbf{N}}=-4m u\lambda \sin 2v$$
The asymptotic directions are given by
$$b_{ij}dx^idx^j=0$$
$$i.e.,~~~ b_{11}du^2+2b_{12}dudv+b_{22}dv^2=0$$
which in the present case takes the form
$$-4m \cos 2v\,du dv - 4mu \sin 2v\,dv^2=0$$
$$i.e., ~~~\left(\cos 2v\,du + u \sin 2v\,dv\right)dv=0$$
So, we have two families of asymptotic lines :

(i) $dv=0$ , \textit{i.e.}, $v=$\,constant\,($u$-curves)

(ii)$\cos 2v\,du + u \sin 2v\,dv=0$ , \textit{i.e.}, $\frac{du}{u} + \tan 2v\,dv=0$ , \textit{i.e.}, $u=c\sqrt{\cos 2v}~.$\\\\
{\bf Hints 4.4:} Take $x=u,~y=v,~z=\dfrac{1}{2}\left(\dfrac{u^2}{a^2}-\dfrac{xv^2}{b^2}\right)$ and show that the asymptotic lines are $\dfrac{u}{a}\pm \dfrac{v}{b}=$ constant.\\\\
{\bf Solution 4.5:} Let us choose rectangular Cartesian coordinate system in $E_{n+1}$ . Then the Weingarten's formula takes the form :
$$\frac{\partial N^{\alpha}}{\partial x^i}=-b_i^k\frac{\partial y^{\alpha}}{\partial x^k}$$
where $b_i^k=g^{kp}b_{pi}$ . Now, if $b_{ij}\equiv 0$ then $b_i^k\equiv 0$ and we have $\frac{\partial N^{\alpha}}{\partial x^i}=0$, \textit{i.e.}, $N^{\alpha}=c^{\alpha}$, a constant.
For any tangent vector $\frac{dy^{\alpha}}{ds}$ to any curve in the hypersurface at any point in it we have 
$$\sum^{n+1}_{\alpha=1} N^{\alpha}\frac{dy^{\alpha}}{ds}=0~~~,~~i.e.,~~~ \sum^{n+1}_{\alpha=1} c^{\alpha}\frac{dy^{\alpha}}{ds}=0$$
so on integration,
$$\sum c^{\alpha}y^{\alpha}=\lambda~,~~~\mbox{a constant}.$$
It represents a hyperplane in $E_{n+1}$ .\\\\
{\bf Solution 4.6:} In rectangular Cartesian co-ordinate  the Weingarten's formula takes the form :
$$\frac{\partial N^{\alpha}}{\partial x^i}=-b_i^k\frac{\partial y^{\alpha}}{\partial x^k}$$
As $b_{ij}=\lambda g_{ij}~~~,~~i.e.,~~~b_i^k=\lambda \delta _i^k$ .
$$\mbox{so},~~\frac{\partial N^{\alpha}}{\partial x^i}=-\lambda \delta _i^k\frac{\partial y^{\alpha}}{\partial x^k}=-\lambda \frac{\partial y^{\alpha}}{\partial x^i}$$ 
Now, differentiating with respect to $x^j$ we have
$$\frac{\partial^2 N^{\alpha}}{\partial x^j\partial x^i}=-\frac{\partial \lambda}{\partial x^j} \frac{\partial y^{\alpha}}{\partial x^i} -\lambda \frac{\partial^2 y^{\alpha}}{\partial x^j\partial x^i}$$
commuting on $i$ and $j$ we get,
$$\frac{\partial \lambda}{\partial x^j}\frac{\partial y^{\alpha}}{\partial x^i}-\frac{\partial \lambda}{\partial x^i}\frac{\partial y^{\alpha}}{\partial x^j}=0.$$
As the vectors $\frac{\partial y^{\alpha}}{\partial x^1},~\frac{\partial y^{\alpha}}{\partial x^2}, \ldots ,~\frac{\partial y^{\alpha}}{\partial x^n}$ are independent, it follows that $\frac{\partial \lambda}{\partial x^i}=0~,~~\forall i=1,~2, \ldots , n$. Hence $\lambda$ is a global constant. Suppose $\lambda\neq 0$. Then we get
$$\frac{\partial N^{\alpha}}{\partial x^i}=-\lambda\frac{\partial y^{\alpha}}{\partial x^i}$$
Contracting with the tangent vector $\frac{dx^i}{ds}$ of any curve in the hypersurface, we get
$$\frac{dN^{\alpha}}{ds}=-\lambda\frac{dy^{\alpha}}{ds}$$
which on integration gives
\begin{eqnarray}
N^{\alpha} = -\lambda y^{\alpha}+\lambda a^{\alpha} \nonumber \\
\mbox{or,}~~~-N^{\alpha}= \lambda\left( y^{\alpha}- a^{\alpha}\right) \nonumber \\
i.e.,~~~ y^{\alpha}- a^{\alpha}=\rho N^{\alpha} \nonumber \\
i.e.,~~~ \sum_{\alpha=1}^{n}\left( y^{\alpha}- a^{\alpha}\right)\left( y^{\alpha}- a^{\alpha}\right)=\rho^2 \nonumber
\end{eqnarray}
This represents a hypersphere in $E_{n+1}$ .\\\\
{\bf Solution 4.7:} $$\textit{\textbf{e}}_{\bm 1}=\frac{\partial \textit{\textbf{r}}}{\partial u}=\left(\cos v, ~\sin v, ~0\right)$$
$$\textit{\textbf{e}}_{\bm 2}=\frac{\partial \textit{\textbf{r}}}{\partial v}=\left(-u \sin v,~ u \cos v, c\right)$$
$$g_{11}=\textit{\textbf{e}}_{\bm 1} \cdot \textit{\textbf{e}}_{\bm 1}=1~~,~~g_{12}=\textit{\textbf{e}}_{\bm 1} \cdot \textit{\textbf{e}}_{\bm 2}=0=g_{21}~~,~~g_{22}=\textit{\textbf{e}}_{\bm 2} \cdot \textit{\textbf{e}}_{\bm 2}=u^2+c^2$$
$$\textit{\textbf{e}}_{\bm 1} \times \textit{\textbf{e}}_{\bm 2}=(c \sin v,~-c \cos v, ~u)~~,~~\left|\textit{\textbf{e}}_{\bm 1} \times \textit{\textbf{e}}_{\bm 2}\right|=\sqrt{u^2+c^2}$$
So, $$\textit{\textbf{N}}=\frac{\textit{\textbf{e}}_{\bm 1}\times\textit{\textbf{e}}_{\bm 2}}{\left|\textit{\textbf{e}}_{\bm 1}\times\textit{\textbf{e}}_{\bm 2}\right|}=\frac{1}{\sqrt{u^2+c^2}}(c \sin v,~-c \cos v, ~u)$$
$$b_{11}=\left(\partial_1 \textit{\textbf{e}}_{\bm 1}\right) \cdot \textit{\textbf{N}}=0~~,~~ b_{22}=\left(\partial_2 \textit{\textbf{e}}_{\bm 2}\right) \cdot \textit{\textbf{N}}=0$$
$$b_{12}=\left(\partial_1 \textit{\textbf{e}}_{\bm 2}\right) \cdot \textit{\textbf{N}}=\frac{1}{\sqrt{u^2+c^2}}(-c \sin ^2 v-c \cos ^2v+0)=\frac{-c}{\sqrt{u^2+c^2}}~.$$
The differential equation for the lines of curvature is given by
$$\left(g_{11}b_{12}-g_{12}b_{11}\right)\left(du\right)^2+\left(g_{11}b_{22}-g_{22}b_{11}\right)dudv +\left(g_{12}b_{22}-g_{22}b_{12}\right)\left(dv\right)^2=0$$
$$i.e., ~~~~ \frac{-c}{\sqrt{u^2+c^2}}du^2+(u^2+c^2)\frac{c}{\sqrt{u^2+c^2}}dv^2=0$$
$$i.e., ~~~~ du^2 -(u^2+c^2)dv^2=0$$
This is the differential equation for the lines of curvature. The above differential equation can be factorized into
$$\left(du+\sqrt{u^2+c^2}\,dv\right)\left(du-\sqrt{u^2+c^2}\,dv\right)=0$$
Thus two principal directions are given by
$$(i)~~\frac{du}{\sqrt{u^2+c^2}}=\frac{dv}{-1}~~~\mbox{and}~~~(ii)~~\frac{du}{\sqrt{u^2+c^2}}=\frac{dv}{1}$$
The corresponding principal curvatures $\chi_1$ is
$$\chi_1=\frac{b_{11}(\sqrt{u^2+c^2})^2+2b_{12} \sqrt{u^2+c^2}(-1)+b_{22}(-1)^2}{g_{11}(\sqrt{u^2+c^2})^2+2g_{12} \sqrt{u^2+c^2}(-1)+g_{22}(-1)^2}=\frac{c}{u^2+c^2}$$
and Similarly, $$\chi_2=-\frac{c}{u^2+c^2}$$
Also integrating the two principal directions we get two families of lines of curvatures
$$\log \left(u+ \sqrt{u^2+c^2}\right)=-v+K_1~~~,~~i.e.,~~ \sinh ^{-1} \left(\frac{u}{c}\right)=-v+K_1$$
and 
$$\sinh ^{-1}\left(\frac{u}{c}\right)= v+K_2$$\\
{\bf Solution 4.8:} As in the preceding example we can obtain $g_{11}~,~g_{12}~,~g_{22}~,~b_{11}~,~b_{12}~,~b_{22}$. Hence the mean curvature
\begin{eqnarray}
M = g^{ij}b_{ij} &=& g^{11}b_{11}+g^{12}b_{12}+g^{21}b_{21}+g^{22}b_{22} \nonumber \\
&=& \frac{1}{g}\left[g_{22}b_{11}-2g_{12}b_{12}+g_{11}b_{22}\right] \nonumber \\
&=& \frac{1}{u^2+c^2}\left[\left(u^2+c^2\right).0-2.0\left(\frac{-c}{\sqrt{u^2+c^2}}\right)+1.0\right]=0 \nonumber
\end{eqnarray}
Hence the surface is a minimal surface.\\\\
{\bf Note:} Total curvature ${\cal K} =\chi_1\chi_2$ is given by
$${\cal K}=\frac{b_{11}b_{22}-b_{12}^2}{g_{11}g_{22}-g_{12}^2}=\frac{0-\frac{c^2}{u^2+c^2}}{1.(u^2+c^2)-0}=-\frac{c^2}{\left(u^2+c^2\right)^2}$$
Now $M=0$ means $\chi_1+\chi_2=0$.\\
So, $$\chi_1=\frac{c}{u^2+c^2}~~~,~~~\chi_2=-\frac{c}{u^2+c^2}~.$$\\
{\bf Solution 4.9:} $$\textit{\textbf{e}}_{\bm 1}=\left(\frac{\partial x}{\partial u}~,~\frac{\partial y}{\partial u}~,~\frac{\partial z}{\partial u}\right)=\left(\cos \phi~,~\sin \phi~,~f'(u)\right)$$
$$\textit{\textbf{e}}_{\bm 2}=\left(\frac{\partial x}{\partial \phi}~,~\frac{\partial y}{\partial \phi}~,~\frac{\partial z}{\partial \phi}\right)=\left(-u \sin \phi~,~u \cos \phi~,~0\right)$$
$$g_{11}=\textit{\textbf{e}}_{\bm 1} \cdot \textit{\textbf{e}}_{\bm 1}=1+\left\{f'(u)\right\}^2~~,~~g_{12}=\textit{\textbf{e}}_{\bm 1} \cdot \textit{\textbf{e}}_{\bm 2}=0~~,~~g_{22}=\textit{\textbf{e}}_{\bm 2} \cdot \textit{\textbf{e}}_{\bm 2}=u^2$$
$$\textit{\textbf{e}}_{\bm 1}\times\textit{\textbf{e}}_{\bm 2}=(-uf'(u) \cos \phi~,~-uf'(u) \sin \phi~,~u)$$
$$\left|\textit{\textbf{e}}_{\bm 1}\times\textit{\textbf{e}}_{\bm 2}\right|=u\sqrt{1+\left\{f'(u)\right\}^2}$$
$$\textit{\textbf{N}}=\frac{\textit{\textbf{e}}_{\bm 1}\times\textit{\textbf{e}}_{\bm 2}}{\left|\textit{\textbf{e}}_{\bm 1}\times\textit{\textbf{e}}_{\bm 2}\right|}=\frac{1}{\sqrt{1+\left\{f'(u)\right\}^2}}(-f'(u) \cos \phi~,~-f'(u) \sin \phi~, ~1)$$
$$b_{11}=\left(\partial_1\textit{\textbf{e}}_{\bm 1}\right) \cdot \textit{\textbf{N}}=\frac{f'(u)}{\sqrt{1+\{f'(u)\}^2}}$$
$$b_{12}=\left(\partial_1\textit{\textbf{e}}_{\bm 2}\right) \cdot \textit{\textbf{N}}=0$$
$$b_{22}=\left(\partial_2\textit{\textbf{e}}_{\bm 2}\right) \cdot \textit{\textbf{N}}=\frac{uf'(u)}{\sqrt{1+\{f'(u)\}^2}}~.$$
The differential equation for the lines of curvature is 
$$(g_{11}b_{12}-g_{12}b_{11})du^2+(g_{11}b_{22}-g_{22}b_{11})dud\phi+(g_{12}b_{22}-g_{22}b_{12})d\phi^2=0~~~, ~~i.e.,~~ du\,d\phi=0$$
Thus $du=0$, which are the $\phi$-curves and $d\phi =0$ which are the $u$-curves, are the lines of curvature. Hence the parametric curves are the lines of curvature. So the normal curvatures $\chi_1$ and $\chi_2$ along $u$-curve and $\phi$-curve respectively are the principal curvatures. These are given by 
$$\chi_1=\frac{b_{11}}{g_{11}}=\frac{f'(u)}{\left\{1+f'(u)^2\right\}^{\frac{3}{2}}}$$
$$\chi_2=\frac{b_{22}}{g_{22}}=\frac{f'(u)}{u\left\{1+f'(u)^2\right\}^{\frac{1}{2}}}~.$$\\
{\bf Solution 4.10:} Here $\frac{\partial \textit{\textbf{r}}}{\partial u}$ is a vector tangential to the $u$-curve and its contravariant components are $\delta^i_{(1)}$.
Let $du^i=(du^1,~du^2)=(du,~dv)$ gives the direction of the orthogonal trajectory to the $u$-curve. Then $$g_{ij}\delta^i_{(1)}du^j=0$$
$$i.e.,~~~g_{1j}du^j=0~~~,~~i.e.,~~~g_{11}du^1+g_{12} du^2=0 $$
\begin{equation} \label{4.26}
i.e., ~~~g_{11}du+g_{12} dv=0
\end{equation}
As $$\textit{\textbf{e}}_{\bm 1}=\frac{\partial \textit{\textbf{r}}}{\partial u}=(1,-v,1)~~~,~~~
\textit{\textbf{e}}_{\bm 2}=\frac{\partial \textit{\textbf{r}}}{\partial v}=(1,-u,-1)$$
$$g_{11}=\textit{\textbf{e}}_{\bm 1} \cdot \textit{\textbf{e}}_{\bm 1}=2+v^2~~~,~~~g_{12}=\textit{\textbf{e}}_{\bm 1} \cdot \textit{\textbf{e}}_{\bm 2}=uv$$
Hence equation\,(\ref{4.26}) becomes
$$(2+v^2)du+uvdv=0$$
$$i.e.,~~~\frac{du}{u}+\frac{vdv}{2+v^2}=0.$$
On integration,
$$\frac{1}{2}u(2+v^2)=\,\mbox{constant}$$
For different values of the constant the curve gives the orthogonal trajectories of the $u$-curves.\\\\
{\bf Solution 4.11:} We have 
$$\textit{\textbf{e}}_{\bm 1}=(a,~b,~v)~~,~~\textit{\textbf{e}}_{\bm 2}=(a,-b,~u)$$
So, $$g_{11}=\textit{\textbf{e}}_{\bm 1} \cdot \textit{\textbf{e}}_{\bm 1}=a^2+b^2+v^2~~~, ~~~g_{12}=\textit{\textbf{e}}_{\bm 1} \cdot \textit{\textbf{e}}_{\bm 2}=a^2-b^2+uv~~~,~~~g_{22}=\textit{\textbf{e}}_{\bm 2} \cdot \textit{\textbf{e}}_{\bm 2}=a^2+b^2+u^2$$
$$\textit{\textbf{e}}_{\bm 1}\times\textit{\textbf{e}}_{\bm 2}=\{b(u+v),-a(u-v),-2ab\}$$
$$\textit{\textbf{N}}=\frac{1}{\left|\textit{\textbf{e}}_{\bm 1}\times\textit{\textbf{e}}_{\bm 2}\right|}\{b(u+v),-a(u-v),-2ab\}$$
$$b_{11}=\left(\partial_1\textit{\textbf{e}}_{\bm 1}\right) \cdot \textit{\textbf{N}}=0~~,~~b_{12}=\left(\partial_1\textit{\textbf{e}}_{\bm 2}\right) \cdot \textit{\textbf{N}}=-\frac{2ab}{c}~~,~~b_{22}=\left(\partial_2\textit{\textbf{e}}_{\bm 2}\right) \cdot \textit{\textbf{N}}=0~~,~~c=\left|\textit{\textbf{e}}_{\bm 1}\times\textit{\textbf{e}}_{\bm 2}\right|~.$$
So the differential equation of the lines of curvature is 
$$(a^2+b^2+v^2)du^2-(a^2+b^2+u^2)dv^2=0$$
$$\left(\sqrt{(a^2+b^2+u^2)}\,du-\sqrt{(a^2+b^2+u^2)}\,dv\right)\left(\sqrt{(a^2+b^2+v^2)}\,du
+\sqrt{(a^2+b^2+u^2)}\,dv\right)=0.$$
Hence the principal directions are given by
$$\frac{du}{\sqrt{(a^2+b^2+u^2)}}=\pm\frac{dv}{\sqrt{(a^2+b^2+u^2)}}$$
The corresponding principal curvatures $\chi_1$ and $\chi_2$ are
$$\chi_1=\frac{b_{11}\left(\sqrt{(a^2+b^2+u^2)}\right)^2+b_{12}\sqrt{(a^2+b^2+u^2)}
	\sqrt{(a^2+b^2+v^2)}+b_{22}\left(\sqrt{(a^2+b^2+v^2)}\right)^2}{g_{11}\left(\sqrt{(a^2+b^2+u^2)}\right)^2+g_{12}\sqrt{(a^2+b^2+u^2)}
	\sqrt{(a^2+b^2+v^2)}+g_{22}\left(\sqrt{(a^2+b^2+v^2)}\right)^2}$$
$$=-\frac{2ab}{c}\left[\sqrt{(a^2+b^2+v^2)}\sqrt{(a^2+b^2+u^2)}+(a^2-b^2+uv)\right]^{-1}$$
and $$\chi_2=\frac{2ab}{c}\left[\sqrt{(a^2+b^2+v^2)}\sqrt{(a^2+b^2+u^2)}+(a^2-b^2+uv)\right]^{-1}$$\\
{\bf Solution 4.12:} The parametric form of the paraboloid $x=u~,~y=v~,~z=\frac{uv}{a}$ and then proceed as before.\\\\
{\bf Solution 4.17:} If $V_{n+1}$ is Euclidean space then $a_{\alpha \beta}=\delta_{\alpha \beta}$.\\\\
So~~ $\Gamma^{\gamma}_{\alpha \beta}=0$ and $\overline{R}_{\alpha \beta \gamma \delta}=0$.\\\\
Hence $R_{lijk}=b_{lj}b_{ik}-b_{lk}b_{ij}$ is the Gauss equation and the Codazzi equation becomes $$\nabla_k b_{ij}-\nabla_j b_{ik}=0.$$\\\\
{\bf Solution 4.18:} If $V_{n+1}$ is a space of constant curvature $k$ then
$$\overline{R}_{\alpha \beta \gamma \delta}=k\left(a_{\alpha \gamma}a_{\beta \delta}-a_{\beta \gamma}a_{\alpha \delta}\right)$$
So
$$\overline{R}_{\alpha \beta \gamma \delta}\left(\nabla_hy^\alpha\right)\left(\nabla_l y^\beta\right)\left(\nabla_j y^\gamma\right)\left(\nabla_k y^\delta\right)=k\left(a_{\alpha \gamma}a_{\beta \delta}-a_{\beta \gamma}a_{\alpha \delta}\right)\left(\nabla_hy^\alpha\right)\left(\nabla_l y^\beta\right)\left(\nabla_j y^\gamma\right)\left(\nabla_k y^\delta\right)$$
$$=k\left[\{a_{\beta \delta}\left(\nabla_ly^\beta\right)\left(\nabla_k y^\delta\right)\}\{a_{\alpha \gamma}\left(\nabla_h y^\alpha\right)\left(\nabla_j y^\gamma\right)\}-\{a_{\beta \gamma}\left(\nabla_ly^\beta\right)\left(\nabla_j y^\gamma\right)\}\{a_{\alpha \delta}\left(\nabla_h y^\alpha\right)\left(\nabla_k y^\delta\right)\}\right]$$
$$=k\left(g_{lk}g_{hj}-g_{lj}g_{hk}\right)$$
$$ R_{hijk}=\left(b_{hj}b_{ik}-b_{hk}b_{ij}\right)+k\left(g_{hj}g_{ik}-g_{hk}g_{ij}\right)$$
Again $$\overline{R}_{\alpha \beta \gamma \delta}\left(N^\alpha\right)\left(\nabla_l y^\beta\right)\left(\nabla_j y^\gamma\right)\left(\nabla_k y^\delta\right)=k\left(a_{\beta \delta}a_{\alpha \gamma}-a_{\beta \gamma}a_{\alpha \delta}\right)N^\alpha \left(\nabla_l y^\beta\right)\left(\nabla_j y^\gamma\right)\left(\nabla_k y^\delta\right)$$
$$=k\left[\{a_{\beta \delta}\left(\nabla_ly^\beta\right)\left(\nabla_k y^\delta\right)\}\{a_{\alpha \gamma}\left(\nabla_j y^\gamma\right)\left(N^\alpha\right)\}-\{a_{\beta \gamma}\left(\nabla_ly^\beta\right)\left(\nabla_j y^\gamma\right)\}\{a_{\alpha \delta}\left(N^\alpha\right)\left(\nabla_k y^\delta\right)\}\right]=0.$$\\
$$\mbox{Hence,}~~~~~~~~~~~~~ \nabla_k b_{ij}-\nabla_j b_{ik}=0.$$\\
{\bf Solution 4.19:} As $V_{n+1} $ is of constant curvature so $\overline{R}_{\alpha \beta \gamma \delta}=\kappa_{n+1}\left(a_{\beta \delta}a_{\alpha \gamma}-a_{\beta \gamma}a_{\alpha \delta}\right)$.\\
Also~~ $R_{hijk}=\kappa_n\left(g_{hj}g_{ik}-g_{hk}g_{ij}\right)$.\\\\
In the previous problem we have deduced that
$$\overline{R}_{\alpha \beta \gamma \delta}\left(\nabla_hy^\alpha\right)\left(\nabla_l y^\beta\right)\left(\nabla_j y^\gamma\right)\left(\nabla_k y^\delta\right)=\kappa_{n+1}\left(g_{lk}g_{hj}-g_{lj}g_{hk}\right)$$
So by Gauss equation,
$$R_{hijk}=\left(b_{hj}b_{ik}-b_{hk}b_{ij}\right)+\overline{R}_{\alpha \beta \gamma \delta}\left(\nabla_hy^\alpha\right)\left(\nabla_i y^\beta\right)\left(\nabla_j y^\gamma\right)\left(\nabla_k y^\delta\right)$$
$$\textrm{or,}~~~~\kappa_n\left(g_{hj}g_{ik}-g_{hk}g_{ij}\right)=\left(b_{hj}b_{ik}-b_{hk}b_{ij}\right)+\kappa_{n+1}\left(g_{lk}g_{hj}-g_{lj}g_{hk}\right)$$
Now, the lines of curvature of a hypersurface $V_n$ of $V_{n+1}$ will be indeterminate if 
$$b_{ij}=\frac{M}{n}g_{ij}$$
Hence, $$\kappa_n=\frac{M^2}{n^2}+\kappa_{n+1}~.$$\\\\


\chapter{Special Theory of Relativity: The Inside Geometry}


~~~In 1905, Einstein formulated the Special Theory of Relativity based on two postulates:\\

{\bf (i) \underline{The principle of Relativity:}}\\

All physical laws assume the same form in all inertial frames of references which are moving relative to each other with constant velocity.\\

{\bf (ii) \underline{Invariance of the speed of light:}}\\

The velocity of light does not depend on the relative motion of the source and the
 observer {\bf --} it is an invariant quantity.\\
 
\underline{{\bf Note:}} The second postulate is consistent with and suggested by the Michelson and Morley's experiment.\\

\section{Derivation of Lorentz Transformation From a geometric Point of view}

~~~According to Einstein, we should have four dimensional space-time as our physical world
 in which we have three space dimension and a time direction \textit{i.e.} the four dimensional
 space-time is characterized by co-ordinates $(t,x,y,z)$.\\

In Euclidean geometry, the distance between two points is invariant (Euclid's axiom). So
 if $(x,y,z)$ and $(x+ dx, y+ dy, z+ dz)$ are two neighbouring points in three dimension
 then
\begin{equation}\label{5.1}
dl^2 _3 = dx^2 + dy^2 + dz^2
\end{equation}
is invariant under co-ordinate transformation. So extending this axiom to Einstein's four
 dimensional space-time we have
\begin{equation}\label{5.2}
dl^2 _4 = dx^2 + dy^2 + dz^2 + \lambda ^{2}dt^2
\end{equation}
is an invariant quantity. Here $\lambda$ has the dimension of velocity and it is introduced
 on dimensional ground. Note that, $\lambda$ does not depend on co-ordinates as $dl^2 _4$ is
 invariant under co-ordinate transformation \textit{i.e.} $\lambda$ is an invariant notion of velocity.\\

Although, we are using space and time on the same footing in four dimensional space-time, but
 still there should be some separate identity for the time co-ordinate. The reason behind this separate identity are (i) for space co-ordinates, we can move both in forward and backward direction but time can move only in the forward direction (there is space reversibility but no time reversibility), (ii) for doing mechanics, time must have a separate identity from the space co-ordinates. To realize this
 identification we modify equation (\ref{5.2}) as
\begin{equation}\label{5.3}
ds^2 = dx^2 + dy^2 + dz^2 - \lambda ^{2}dt^2
\end{equation}
and assume that `$ds$' is an invariant quantity. Here `$ds$' is called space-time interval.
 Note that $ds^2$ is not positive definite (its consequences will be discussed later).\\
 
\begin{wrapfigure}{r}{0.43\textwidth}
\includegraphics[height=4 cm , width=7 cm ]{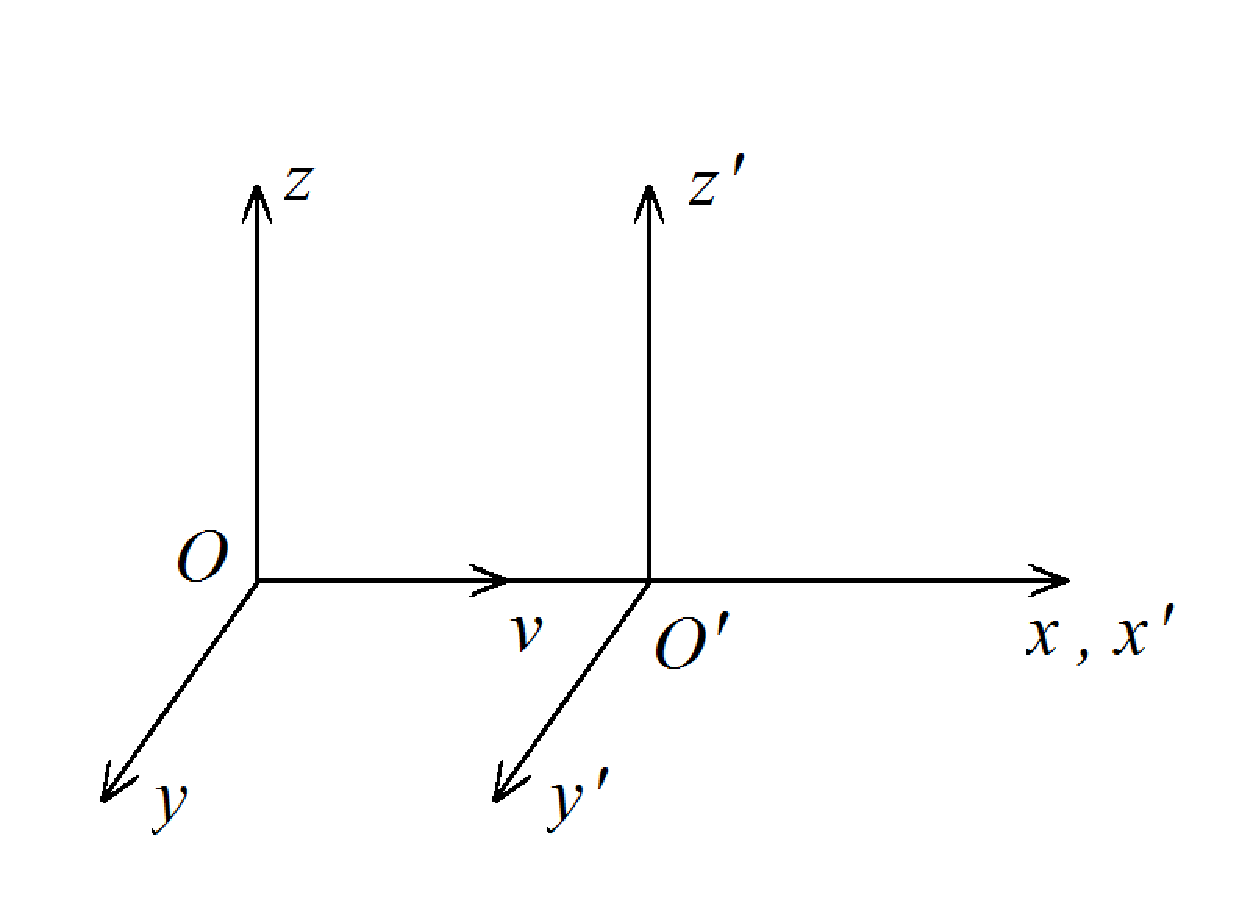}
\begin{center}
Fig. 5.1
\end{center}
\end{wrapfigure}

We now concentrate on those inertial co-ordinate systems for which equation (\ref{5.3}) is an
 invariant. Without any loss of generality, we choose two co-ordinate systems $(x,y,z,t)$
 (termed as $S$-frame) and $(x',y',z',t')$ (termed as $S'$-frame) in which $S'$-frame is
 moving relative to $S$-frame with constant velocity `$v$' along the common $x\textrm{-}x'$-axis.
 For invariance of the quadratic form (\ref{5.3}), the linear transformation equation for $y$(or
 $z$) co-ordinate can be written as
\begin{equation}\label{5.4}
y' = lx+ my+ nz+ kt ~,
\end{equation}
usually, the co-efficients $l,m,n$ and $k$ are constants or at most depend on the relative
 velocity `$v$'. As it is evident from the figure that the plane $y= 0$ is identical to
 $y' = 0$, so from (\ref{5.4}) we have $$lx+ nz+ kt= 0 ~~~, \forall~~x,z~~\textrm{and}~~t.$$

Hence we have,~~~ $l= 0= n= k$.\\

Thus we have~~~ $y' = m(v)y$\\

As for $y$ and $z$ coordinates it is immaterial whether $S'$ moves relative to $S$ along
 +ve or $-$ve direction of $x$-axis, so we must have $m(-v)= m(v)$ (\textit{i.e.} $m$ must be an even
 function of $v$). Further, as relative to $S'$-frame, $S$-frame moves with constant
 velocity $-v$ so we should write $y = m(-v)y'$. Hence $m(v)$ can have values $\pm 1$ and
 for convenience we choose $m(v)= +1$ \textit{i.e.} $y' = y$. Similarly we have $z' = z$. For the
 transformation of $x$-co-ordinate we write:
\begin{equation}\label{5.5}
x' = ax+ by+ cz+ dt
\end{equation}
where as usual the co-efficients $a,b,c$ and `$d$' are either constants or functions
 of `$v$'. Note that the plane $x' = 0$ in $S'$-frame is the plane $x= vt$ in $S$-frame.
 So from (\ref{5.5}) we get $$0= (av+ d)t+ by+ cz$$ and this holds for all $t,y$ and $z$. Thus
 we have $d= -av~,~b= 0= c$. As a result equation (\ref{5.5}) simplifies to
\begin{equation}\label{5.6}
x' = a(x-vt).
\end{equation}

Now for the time transformation we start with
\begin{equation}\label{5.7}
t' = \alpha x+ \beta y+ \gamma z+ \delta t
\end{equation}
with the co-efficients either constant or function of $v$ as usual.
 From the figure, it is evident that the plane $x= 0$ in $S$-frame is described by $x' = -vt'$
 in $S'$-frame. So from (\ref{5.6}) we have $$t' = at.$$ Using this value of $t'$ in (\ref{5.7})
 we have $$\beta y+ \gamma z+ (\delta -a)t= 0 ~~,~ \forall~~y,z~~\textrm{and}~~t.$$
 
This implies $$\beta = 0= \gamma~~~~\textrm{and}~~~~\delta = a.$$

Hence from (\ref{5.7}) we get
\begin{equation}\label{5.8}
t' = \alpha x+ at
\end{equation}

Thus, to obtain the complete transformation laws we shall have to determine the unknown
 co-efficients `$a$' and `$\alpha$'. Now, due to invariance of (\ref{5.3}) in $S$ and $S'$-frame
 we write
\begin{eqnarray}\label{5.9}
dx^{2}- \lambda ^{2}dt^{2}&=& (dx')^{2}- \lambda ^{2}(dt')^{2} \\
i.e.~~~~~~dx^{2}- \lambda ^{2}dt^{2}&=& a^{2}(dx- vdt)^{2}- \lambda ^{2}(\alpha dx+ adt)^{2} \nonumber \\
&=& (a^{2}- \lambda ^{2}\alpha ^{2})dx^{2}+ (a^{2}v^{2}- a^{2}\lambda ^{2})dt^{2}- 2dxdt(a^{2}v+
 \lambda ^{2}\alpha a) \nonumber
\end{eqnarray}

Now equating co-efficients of $dx^{2},~dt^{2}~\textrm{and}~dxdt$ , we have
\begin{eqnarray}
a^{2}- \alpha ^{2}\lambda ^{2}&=& 1,\label{5.10}\\
a^{2}(\lambda ^{2}- v^{2})&=& \lambda ^{2}\label{5.11}\\
\textrm{and}~~~~~~~a^{2}v+ \lambda ^{2}\alpha a&=& 0\label{5.12}
\end{eqnarray}

Solving these we obtain
\begin{equation}\label{5.13}
a= \frac{1}{\sqrt{1- \mu ^2}}~~,~~\alpha = -\frac{v/\lambda ^2}{\sqrt{1- \mu ^2}}~~,~~\mu = \frac{v}{\lambda}.
\end{equation}

So the transformation equations take the form
\begin{equation}\label{5.14}
x' = \frac{x- vt}{\sqrt{1- \mu ^2}}~~,~~y' = y~~,~~z' = z~~,~~t' = \frac{t- xv/\lambda ^2}{\sqrt{1- \mu ^2}}~,
\end{equation}
which is the Lorentz transformation with the absolute velocity $\lambda$ identified
 as the velocity of light.\\

Alternatively, we can derive the transformation laws as follows:\\

Suppose the linear transformation laws are chosen as
\begin{equation}\label{5.15}
x' = ax+ \delta t~~,~~t' = \alpha x+ \beta t.
\end{equation}

Then from the invariance relation (\ref{5.9}) we obtain $$dx^{2}- \lambda ^{2}dt^{2}= (adx+ \delta dt)^{2}
- \lambda ^{2}(\alpha dx+ \beta dt)^2.$$ 

So as before equating co-efficients of $dx^{2},~dt^{2}~\textrm{and}~dxdt$ we have
\begin{equation}\label{5.16}
a^{2}- \lambda ^{2}\alpha ^{2}= 1~~,~~\lambda ^{2}\beta ^{2}- \delta ^{2}= \lambda ^{2}~~~\textrm{and}~~~a\delta - \lambda ^{2}\alpha \beta = 0.
\end{equation}

Here we have three equations containing four unknown co-efficients $a,\delta ,\alpha~
\textrm{and}~\beta$. Hence for unique solution another relation among these co-efficients
 is specified from the geometry as:

 \begin{center}
 	``~The plane $x' = 0$ is equivalent to $x= vt$ in $S$-frame."
 \end{center}
 
So from equations (\ref{5.15}) we get
\begin{equation}\label{5.17}
\delta = -av .
\end{equation}

Solving these co-efficients we finally have the identical transformation equations --- the
 Lorentz transformation.\\

\section{Velocity Identity\,: Law of Composition of velocity}

~~~From the invariance of the space-time interval (\ref{5.3}) in $S$ and $S'$-frame we have
\begin{equation}\label{5.18}
dt\sqrt{1- \frac{u^2}{\lambda ^2}}= dt'\sqrt{1-\frac{(u')^2}{\lambda ^2}}
\end{equation}
where $u= \left\{\left(\dfrac{dx}{dt}\right)^{2}+ \left(\dfrac{dy}{dt}\right)^{2}+
 \left(\dfrac{dz}{dt}\right)^{2}\right\}^{1/2}$ is the speed of a particle in $S$-frame
 and that in $S'$-frame is $u'$. Also from the time transformation eq. (\ref{5.14}) we obtain
\begin{equation}\label{5.19}
dt' = dt\left(1- \frac{u_{x}v}{\lambda ^2}\right)/\sqrt{1- \mu ^2}.
\end{equation}

Now eliminating $dt'$ between equations (\ref{5.18}) and (\ref{5.19}) we have
\begin{equation}\label{5.20}
\sqrt{1- \frac{u^2}{\lambda ^2}}\sqrt{1- \frac{v^2}{\lambda ^2}}= \left(1- \frac{u_{x}v}{\lambda}\right)\left(1- \frac{u^{\prime 2}}{\lambda ^2}\right)^{1/2}
\end{equation}
where $u_x$ is the $x$-component of the velocity of the particle in $S$-frame. The relation
 (\ref{5.20}) is an identity connecting the speed of a particle in two frames of references. In
 particular, if the particle moves along the $x$-axis then the above identity becomes
\begin{equation}\label{5.21}
\sqrt{1- \frac{u^2}{\lambda ^2}}\sqrt{1- \frac{v^2}{\lambda ^2}}= \left(1- \frac{uv}{\lambda}\right)\left(1- \frac{u^{\prime 2}}{\lambda ^2}\right)^{1/2}
\end{equation}
which on simplification gives
\begin{equation}\label{5.22}
u= \frac{u' + v}{1+ \frac{u'v}{\lambda ^2}}~,
\end{equation}
the law of composition of velocity. We can also write the composition law by a new
 binary operation as
\begin{equation}\label{5.23}
u= u' \oplus v= \frac{u' + v}{1+ \frac{u'v}{\lambda ^2}}~.
\end{equation}

Further, the above law of composition of velocity can be written as
\begin{equation}\label{5.24}
u= \lambda \left[1- \frac{\left(1- \frac{u'}{\lambda}\right)\left(1- \frac{v}{\lambda}\right)}{1+ \frac{u'v}{\lambda ^2}}\right]
\end{equation}
which implies $$(i)~~~u_{max}= \lambda~~,~~(ii)~~~u' < \lambda~~,~~v< \lambda~~
\Rightarrow u< \lambda~,$$ \textit{i.e.} by composition of velocity it is not possible to have
 a velocity larger than the absolute velocity `$\lambda$'. Note that we always have
\begin{equation}\label{5.25}
u' \oplus \lambda = \lambda~~~;~~~\lambda \oplus \lambda = \lambda~.
\end{equation}

Moreover, in the non-relativistic limit : $\left|\dfrac{u}{c}\right| \ll 1~~,
~\left|\dfrac{v}{c}\right| \ll 1$ , the above identity (\ref{5.20}) simplifies to
 $$(u' _{x})^{2}+ (u' _{y})^{2}+ (u' _{z})^{2}= (u_{x}- v)^{2}+ u^{2} _{y}+ u^{2} _{z}$$
 which is identically satisfied by $$u' _{x}= u_{x}- v~~,~~u' _{y}= u _{y}~~,~~u' _{z}= u _{z}~,$$
 the law of composition of velocity in Newtonian theory.\\\\

\section{The invariance of the absolute velocity\,: The space-time Interval}

~~~In this section we shall show the following:\\

``~The invariance of the absolute velocity $\lambda$ implies the invariance of the
 space-time interval from one inertial frame to the other."\\

Let $A(x,y,z,t)$ and $B(x+ dx, y+ dy, z+ dz, t+ dt)$ be two neighbouring positions
 of a particle in an inertial frame $S$. So the space-time interval is given by
\begin{equation}\label{5.26}
ds^{2}= -\lambda ^{2}(dt)^{2}\left\{1- \frac{u^2}{\lambda ^2}\right\}
\end{equation}
where $u$ is the velocity of the particle.\\

Similarly, in $S'$-frame the space-time interval takes the form
\begin{equation}\label{5.27}
(ds')^{2}= -\lambda ^{2}(dt')^{2}\left\{1- \frac{u^{\prime 2}}{\lambda ^2}\right\}.
\end{equation}

Note that if $u' = \lambda$ then $ds' = 0$. As from the law of composition of velocity (\ref{5.25}) $u= \lambda$ so $ds= 0$. Thus $ds= 0= ds'$ if the particle moves with absolute velocity $\lambda$. However, if the particle moves with velocity less than the absolute velocity $\lambda$, then from the law of composition of velocity (\ref{5.22}), $ds^2$ can be considered as a function of $(ds') ^2$ \textit{i.e.} $ds^{2}= f\left\{(ds') ^2\right\}$. So by Taylor series expansion, in the non-relativistic limit we have $$ds^{2}= f_{0}+ f_{1}(ds') ^2,$$
  where $f_{0},f_{1}$ are either constants or at most depend on the relative speed between the two inertial frames $S$ and $S'$. As $ds= 0$ when $ds' = 0$ so $f_{0}= 0$ and we write $$ds^{2}= f_{1}(v)(ds') ^2.$$
 
  To determine an explicit form of $f_{1}$, we consider three inertial frame of references $S_{1}, S_{2}~\textrm{and}~S_{3}$ with relative velocities $v_{12}, v_{23}~~\textrm{and}~~v_{31}$ respectively, between the frames $(S_{1},S_{2}),(S_{2},
S_{3})~\textrm{and}~(S_{3},S_{1})$. Thus we write interrelation between the space-time intervals as $$ds^2 _{1}= f_{1}(v_{12})ds^2 _{2}~~,~~ds^2 _{1}= f_{1}(v_{31})ds^2 _{3}~~\textrm{and}~~ds^2 _{2}= f_{1}(v_{23})ds^2 _{3}~.$$ Hence for non-zero $ds^2 _{i}~(i= 1,2,3)$ we have $$f_{1}(v_{31})= f_{1}(v_{12})\cdot f_{1}(v_{23}).$$

 In general, $v_{31}$ depends not only on the magnitude of $v_{12}$ and $v_{23}$ but also on the angle between $v_{12}$ and $v_{23}$(even in the same direction for the relative velocities, the composition law (\ref{5.22}) gives $f_{1}\left(\frac{v_{12}+ v_{23}}{1+ \frac{v_{12}v_{23}}{\lambda ^2}}\right)=
 f_{1}(v_{12})f_{1}(v_{23})$). So the above relation is satisfied only for $f_{1}(v)= 1$ and we have $$ds^{2}= (ds') ^2.$$
 
  Hence space-time interval is an invariant quantity, does not depend on the inertial frame of reference under consideration, if the invariance of the absolute velocity is assumed.\\

\section{Consequences From Lorentz Transformation}

~~~The following are the results can be derived from the Lorentz transformation:\\

\textbf{I.} \textit{The set of all Lorentz transformations (having relative velocities in the same
 direction) forms a group. It is a commutative (abelian) group.}\\

If we denote the Lorentz transformation between two inertial frames $S$ and $S'$ as $L(v)$ \textit{i.e.}
$$L(v) :
		\left\{
				\begin{array}{ll}
								x' = (x- vt)/\sqrt{1- \mu ^2} \\
								y' = y \\
								z' = z \\
								t' = \frac{t- xv/\lambda ^{2}}{\sqrt{1- \mu ^2}}
				\end{array}
	  \right.$$
and that between $S'$ and $S''$ as $L(v')$ \textit{i.e.}
$$L(v') :
		\left\{
				\begin{array}{ll}
								x'' = (x' - v't')/\sqrt{1- (\mu ')^2} \\
								y'' = y' \\
								z'' = z' \\
								t'' = \frac{t' - x'v'/\lambda ^{2}}{\sqrt{1- (\mu ')^2}}
				\end{array}
	  \right.$$
then it can be shown that $$L(v)\circ L(v')= L(v'')~~~~\textrm{(closure property)}$$ where
 $v'' = \dfrac{v+ v'}{1+ \frac{vv'}{\lambda ^2}}$ .

It is easy to see that $L(0)$ gives the identity transformation and $$L(v)\circ L(-v)= L(0)$$
 \textit{i.e.} $L(-v)$ is the inverse of $L(v)$.\\

Thus set of all Lorentz transformations forms a group. Further, the symmetry of $v$ and $v'$
 in the expression for $v''$ shows that the group is commutative in nature.\\

\textbf{II.} \textit{There is no concept of absolute simultaneity -- it is a relative concept in
 special theory of relativity:}\\

We shall show that the concept of simultaneity is not absolute in nature according to
 Einstein's special theory of relativity.\\

As before, let $S$ and $S'$ be two inertial frames where $S'$ is moving relative to $S$
 along the common $x$-axis with constant velocity $v$. Suppose $B,A$ and $C$ (in order)
 be three points along the common $x$-axis with $AB= CA$, in $S'$-frame. Now two signals
 with speed $\lambda$ (the absolute velocity) start from $A$ in the directions of $B$ and
 $C$. As the points $B,A$ and $C$ are fixed in $S'$-frame so the two signals will reach
 $B$ and $C$ at the same instant. Hence we can say that relative to the observers at $B$
 and $C$ the two events are simultaneous in $S'$-frame. However, in $S$-frame the points
 are not fixed - $B$ approaches to $A$ while $C$ moves away from $A$. Hence the signal
 will reach $B$ earlier than at $C$, due to invariance of $\lambda$. So the two signals
 will not appear to be simultaneous in $S$-frame. Thus simultaneity is a relative concept.\\

Alternatively, if two events occur at $(x'_1, y'_1, z'_1, t')$ and $(x'_2, y'_2, z'_2, t')$ in $S'$-frame i.e., at same time but at different space points then by Lorentz transformation the time of the occurrence of these two events are given by $t_1=\dfrac{t'+x'_1\frac{v}{\lambda^2}}{\sqrt{1-\frac{v^2}{\lambda^2}}}$  and $t_2=\dfrac{t'+x'_2\frac{v}{\lambda^2}}{\sqrt{1-\frac{v^2}{\lambda^2}}}$. Hence the time difference in $S$-frame is $T=t_2-t_1=\frac{(x'_2-x'_1)\frac{v}{\lambda^2}}{\sqrt{1-\frac{v^2}{\lambda^2}}}$. Hence the two events will not appear to be simultaneous in $S$-frame.\\

In Newtonian theory, there is no absolute concept of velocity (due to absolute concept of
 time). Let $AB= CA= l$ and $\lambda$ be the velocity of the signals.\\

In $S'$-frame :~~~~~~~ $t= \dfrac{l}{\lambda}$ is the time taken by the signals to reach the
 points $B$ and $C$.\\

In $S$-frame :~~~~~~~ velocity of the signal along $AC= \lambda + v$ and that along $AB$ is
 $\lambda - v$.

~~~~~~~~~~~~~~~~~~~~~~~~  If $t_1$ and $t_2$ be the time taken by the signals to reach $B$ and
 $C$ then 
 \begin{eqnarray}t_{1}= \frac{l- vt}{\lambda - v}= \frac{(\lambda - v)t}{(\lambda - v)}= t\nonumber\\
 t_{2}= \frac{l+ vt}{\lambda + v}= \frac{(\lambda + v)t}{(\lambda + v)}= t.\nonumber\end{eqnarray}

Hence the concept of simultaneity is absolute in nature in Newtonian theory.\\

\textbf{III.} \textit{The rod appears to be contracted and moving clock goes slow in special
 theory of relativity.}\\

\textbf{IV.} \textit{The quadratic expression: $s^{2}= x^{2}+ y^{2}+ z^{2}- \lambda ^{2}t^{2}$
 is an invariant quantity in any inertial frame.}\\

\textbf{V.} \textit{The differential of the co-ordinates in an inertial frame transform as Lorentz
 transformation.}\\

\textbf{VI.} \textit{The Lorentz transformation can be viewed as a rotation of axes in $(x\,,\,it)$
-plane with an imaginary angle of rotation given by~ $\tanh \theta = v/\lambda$.}\\\\

\section{Universal Speed Limit}
~~~This section deals with a very well-known question in special theory of relativity namely ``Why is there a universal speed limit here?" Apparently, it seems quite arbitrary. However, a possible and probably unexpected answer to this question is wrong choice of the variable as a measure of speed. The correct variable  for velocity measure is called rapidity.\\

In the previous section (point VI) it has been shown that geometrically Lorentz transformation can be considered as a relation (a hyperbolic rotation) in a 2D plane. In particular in matrix notation one has
$$\begin{pmatrix}
x'\\T'
\end{pmatrix}=\begin{pmatrix}
\cosh\theta&-\sinh\theta\\-\sinh\theta&\cosh\theta
\end{pmatrix}\begin{pmatrix}
x\\T
\end{pmatrix}$$
with $T=\lambda t$ and $\beta=\dfrac{v}{\lambda}=\tanh\theta$\\

Further one can write the Lorentz transformation in $(x,t)$ plane as$$\begin{pmatrix}
x'\\T'
\end{pmatrix}=\gamma\begin{pmatrix}
1&-\beta\\-\beta&1
\end{pmatrix}\begin{pmatrix}
x\\T
\end{pmatrix}, ~~~~~\gamma=\sqrt{1-\beta^2}$$

Similarly, the Lorentz transformation between $(x',T')$ and $(x'',T'')$ can be written as$$\begin{pmatrix}
x''\\T''
\end{pmatrix}=\gamma'\begin{pmatrix}
1&-\beta'\\-\beta'&1
\end{pmatrix}\begin{pmatrix}
x'\\T'
\end{pmatrix}$$

Thus combining the two one gets the composite Lorentz transformation
\begin{eqnarray}\begin{pmatrix}
x''\\T''
\end{pmatrix}&=&\gamma\gamma'\begin{pmatrix}
1&-\beta'\\-\beta'&1
\end{pmatrix}\begin{pmatrix}
1&-\beta\\-\beta&1
\end{pmatrix}\begin{pmatrix}
x\\T
\end{pmatrix}\nonumber\\&=&\gamma''\begin{pmatrix}
1&-\beta''\\-\beta''&1
\end{pmatrix}\begin{pmatrix}
x\\T
\end{pmatrix}\nonumber
\end{eqnarray}
with $\beta''=\dfrac{\beta+\beta'}{1+\beta\beta'}$, the law of composition of velocity.\\

In Euclidean 2D plane a rotation with an angle $\theta$ is given by$$\begin{pmatrix}
x'\\y'
\end{pmatrix}=\begin{pmatrix}
\cos\theta&-\sin\theta\\\sin\theta&\cos\theta
\end{pmatrix}\begin{pmatrix}
x\\y
\end{pmatrix}=A\begin{pmatrix}
x\\y
\end{pmatrix}$$
with $\det A=1$.\\

This co-ordinate change by rotation only changes the direction of the co-ordinate lines (vectors) not their magnitude. Also for such two consecutive rotations as$$\begin{pmatrix}
\cos\theta&-\sin\theta\\\sin\theta&\cos\theta
\end{pmatrix}\begin{pmatrix}
\cos\phi&-\sin\phi\\\sin\phi&\cos\phi
\end{pmatrix}=\begin{pmatrix}
\cos(\theta+\phi)&-\sin(\theta+\phi)\\\sin(\theta+\phi)&\cos(\theta+\phi)
\end{pmatrix}$$
so the resulting rotation gives a rotation with angle $\theta+\phi$.\\

For Lorentz transformation, a rotation with hyperbolic angle, the corresponding determinant of the rotation matrix i.e.$$\det\begin{pmatrix}
	\cosh\theta&-\sinh\theta\\-\sinh\theta&\cosh\theta
\end{pmatrix}=1$$
and one has$$\begin{pmatrix}
\cosh\phi&-\sinh\phi\\-\sinh\phi&\cosh\phi
\end{pmatrix}\begin{pmatrix}
\cosh\theta&-\sinh\theta\\-\sinh\theta&\cosh\theta
\end{pmatrix}=\begin{pmatrix}
\cosh(\theta+\phi)&-\sinh(\theta+\phi)\\-\sinh(\theta+\phi)&\cosh(\theta+\phi)
\end{pmatrix}$$

Thus 2 successive Lorentz transformations with hyperbolic angle $\theta$ and $\phi$ results another Lorentz transformation with hyperbolic angle $\theta+\phi$.\\

Further, for the Lorentz transformation the hyperbolic angle $\theta$ is given by$$\sinh\theta=\gamma\beta,~\cosh\theta=\gamma,~\tanh\theta=\beta$$

From the property of the hyperbolic functions$$|\tanh\theta|\leq1,\mbox{~~~~for~any~real~}\theta$$

This implies $|\beta|\leq1$ i.e. $v\leq\lambda$.\\

So we have a barrier in speed limit. However, if $\theta$ measures the speed then there is no speed limit in special theory of relativity. Here the hyperbolic angle $\theta$ is called rapidity --- the measure of speed and it is unbounded as in Newtonian theory. Thus in Special theory of relativity, rapidity is the natural choice for speed measurement and velocity makes sense only in the non-relativistic limit $\theta\to0$. \\

\section{Curves and proper-time}
~~~Suppose $x^\mu(\lambda)$ be a curve (world line) parametrized by some real parameter $\lambda$. So for a fixed $\lambda$, $x^\mu(\lambda)$ represents a point $P$ on the manifold (the 4D Minkowskian space-time). The tangent vector to the curve is an element to the tangent space at $P$ and is defined as$$t^\mu=\dot{x}^\mu\equiv\frac{dx^\mu}{d\lambda}\,.$$

This tangent vector to the world line is a time like vector at every point i.e. $\|\dot{x}^\mu\|^2<0$, for all $\lambda$. So it is possible to define a notion of `time' as measured by a clock moving with the particle.\\

This time notion is clearly distinct from co-ordinate time and is termed as proper-time. The proper time is an observable while coordinate time is not an observable due to its dependence on its on the arbitrary choice of ordinates.\\

Mathematically, the differential of the proper time is defined as \begin{eqnarray}
d\tau^2&=&-ds^2=-g_{\mu\nu}dx^\mu dx^\nu\nonumber\\\mbox{i.e.~~} \left(\frac{d\tau}{d\lambda}\right)^2&=&-g_{\mu\nu}\dot{x}^\mu\dot{x}^\nu=-\|\dot{x}^\mu\|^2\nonumber
\end{eqnarray}

As for time like path $\|\dot{x}^\mu\|^2<0$, so $\tau$ has the integral form$$\tau=\int\sqrt{-\|\dot{x}^\mu\|^2}d\lambda\mbox{~~~~~i.e.~~~}\tau=\tau(\lambda)$$

Thus world line of a time-like particle can be parametrized by proper time. Further, $V^\mu=\dfrac{dx^\mu}{d\tau}$ is called the four velocity of the time-like particle along the world line with normalization $\|V^\mu\|^2=-1$.\\

On the other hand, for a mass less particle the world line is a null curve (having null tangent vector) i.e. $|\dot{x}^\mu(\lambda)|^2=0$, for any choice of $\lambda$ (tangent vector is not a zero vector). As a consequence$$\tau=\int\sqrt{-|\dot{x}^\mu(\lambda)|^2}d\lambda=0$$
i.e. $\tau=0$ between any two points on the world line. This can be interpreted as ``massless particles do not  experience the passage of time and hence they do not have a well-defined 4-velocity". Thus null paths do not have any preferred parameter and consequently, null geodesics passes a family of preferred parameters called affine parameters.\\\\

\section{Motion of time like, null and space like particles in Special Theory of Relativity}

\textbf{~~~(a) Time like particle}\\

We have seen in the last section that a massive particle moving along time like world line can be parametrized by proper time $\tau$ with normalization $\|\dot{x}^\mu\|^2=-1$ i.e.$$\|\dot{x}^\mu\|^2=-\dot{t}^2+\dot{x}^2+\dot{y}^2+\dot{z}^2=-1$$

In particular, if the particle is at rest then $\dot{x}^\mu=(1,0,0,0)$. This implies that the particle has no velocity along any of the spatial coordinates but it moves at the absolute speed $\lambda$ along the time coordinate. Also the above normalization implies $\dot{t}\neq0$ for a massive particle i.e. a massive particle must always move along time axis. (Note that $\dot{t}$ may be negative (past directed) or positive (future directed))\\

Suppose a massive particle is moving at constant spatial velocity $v$ along say $x$ axis, i.e. $$v=\frac{dx}{dt}$$

The corresponding 4-velocity will be$$\dot{x}^\mu=\gamma(1,v,0,0)$$
where $\gamma=\dfrac{dt}{d\tau}$ is termed as Lorentz factor.\\

In fact $\gamma$ measures the relation between coordinate time and proper time. Also $dt=\gamma d\tau$, indicates that the amount of time dialation is measured by $\gamma$. Also $\gamma$ can be estimated from the normalization $|\dot{x}^\mu|^2=-1$ as $\gamma=\dfrac{1}{\sqrt{1-v^2}}$, the Lorentz factor in STR,\\
i.e. $\gamma$ is equivalent to velocity normalization.\\

For a particle  moving with 3-velocity $v$ has energy (see the next section) $$E=m_0\gamma$$
As $v\to1$, $\gamma\to\infty$ so $E\to\infty$, i.e. a time like particle requires infinite amount of energy to accelerate the particle to the absolute speed.\\

\textbf{(b) Null particle:}\\

We have seen that for a particle with spatial 3-velocity $v$ has 4-velocity components $$\dot{x}^\mu=\gamma(1,v,0,0)$$
with $\|\dot{x}^\mu\|^2=\gamma^2(-1+v^2)$.\\

For a particle moving with absolute velocity (i.e. $v=1$), $\|\dot{x}^\mu\|^2=0$, so the path must be null. Hence, there is no need of choosing the normalization function $\gamma$ to be unity for convenience, i.e. $$x^\mu=(1,0,0,0) \mbox{~~~for~a~null~particle}$$

Now, due to norm invariance the null particle will always  move along a null path and the absolute speed is same in all inertial frames (i.e. for all observers), which is nothing but the 2nd postulate of STR. Finally, a particle moving at the absolute speed can never decelerate or accelerate to a different speed.\\

\textbf{(c) Space-like particle: Tachyon}\\

A particle moving with 3-velocity $v$ has 4-velocity vector $$\dot{x}^\mu=\gamma(1,v,0,0)$$ with $\|\dot{x}^\mu\|^2=\gamma^2(-1+v^2)$.\\

Now, if $v>1$ i.e. velocity is larger than the absolute speed then $\|\dot{x}^\mu\|^2>0$ so the particle moves along a space-like path. If the normalization is chosen as $\|\dot{x}^\mu\|^2=1$ then $\gamma=\dfrac{1}{\sqrt{v^2-1}}$.\\

So any particle moving faster than absolute velocity will move along space-like paths and are called tachyons. As before due to norm invariance tachyons will always move as space-like paths and they cannot be decelerate to the absolute speed or below. Note that as $v\rightarrow1$ (from above), $\gamma\rightarrow\infty$ so $E=m_0\gamma\rightarrow\infty$ as $v\rightarrow1$.\\

This means that infinite energy is required to decelerate a tachyon to absolute velocity.\\

Moreover, it is to be noted that $\gamma$ decreases as $v$ increases (beyond absolute velocity) and $\gamma\rightarrow0$ as $v\rightarrow\infty$. Thus the energy of a tachyon decreases as its velocity increases and finally, the tachyon has zero energy when its velocity is infinity. Then 
$$\dot{x}^\mu=\lim\limits_{v\rightarrow\infty}\gamma(1,v,0,0)=(0,1,0,0)$$
which may be considered as the rest position of the tachyon.\\

Therefore, a tachyon at rest moves only along a spatial direction while a normal massive particle at rest moves only along time direction.\\

\section{Time travel in STR} 
~~~ In this section, an interesting and fascinating issue namely the time travel will be discussed in the framework of STR. In the last section, it has been shown that theoretically within STR, a particle with velocity larger than absolute velocity (known as tachyon) follows a space-like path. An hypothetical experiment with tachyon motion will be described in the following and as a result, it is possible to move into past, leading to inconsistency.\\

In Minkowski space-time let $S_1$ and $S_2$ be two inertial frame of references moving relative to each other with constant velocity $u$ ($u<\lambda$). Let $A$ and $B$ two space stations in these two inertial frames. The coordinate systems $(t_1,x_1)$ and $(t_2,x_2)$ are the rest frames for $S_1$ and $S_2$ respectively. So world line for $A$ in $(t_1,x_1)$ frame is along $t_1$-axis and that for $B$ in $(t_2,x_2)$ frame is along $t_2$-axis. We now describe the hypothetical experiment with tachyon.\\

Suppose station $S_1$ at origin ($t_1=0,x_1=0$) sends a tachyon to a station $S_2$ with a velocity $v_1>\lambda$. Evidently, the tachyon reaches station $S_2$ at the instant its world line intersects $t_2$-axis. In spite of the space-like nature, the tachyon moves forward in time (see the last section) and hence it will be in future motion (no motion in the past). Now without any loss of generality one may consider this point of intersection as the origin of the $(t_2,x_2)$-coordinate system. Instantly at that instant $(t_2=0)$ station $S_2$ sends another tachyon back to station $A$ with speed $v_2>\lambda$. As before this tachyon  also moves forward in time i.e. the world line should be above the $x_2$-axis. Now if the relative velocity $u$ is sufficiently large then the $x_2$-axis intersects the $t_1$-axis below the origin i.e. at some negative $t_1$. This implies that if the tachyon is used as a carrier of some message then the message goes to the past of $A$ or the tachyon itself is detected in $S_1$-frame in the past. So this experiment may be considered as sending some message by an observer in $S_1$-station to his past -- a paradox. This type of paradox is well known in time machine. This paradox has similarity with grandfather paradox.\\

On the other hand, the above experiment can be interpreted in an alternative way. Suppose $A$ put a restriction on sending the tachyon: ``It sends a techyon at $t_1=0$ only if it did not receive a tachyon at any lime $t_1<0$". For $B$ station the restriction is that it sends a tachyon at time $t_2=0$ only if it received a tachyon exactly at that time $(t_2=0)$ i.e., station $B$ acts as a reflector (i.e. tachyon mirror) for the tachyon. Hence, assuming that $A$ did not receive any tachyon in earlier time (i.e. $t_1<0$) it sends a tachyon at $t_1=0$ and that tachyon is reflected back from station $B$ and it is received by A at time $t_1<0$ i.e., in past, which violate the restriction for station $A$. One can say it as ``A sends a tachyon at $t_1=0$ iff it does not send a tachyon at $t_1=0$!" An event can happen and not happen simultaneous -- a contradictory statement. This type of paradox is termed as consistency paradox in time machine.\\

\section{Relativistic Energy-momentum Relation\,: The Relativistic Mass}

~~~From the law of composition of velocity (\ref{5.22}) we write $$u' = \frac{u- v}{\left(1- \frac{uv}
{\lambda ^2}\right)}$$ \textit{i.e.} $$\frac{m_0}{\sqrt{1- \frac{u^{\prime 2}}{\lambda ^2}}}u' = \frac{m_{0}(u- v)}
{\left(1- \frac{uv}{\lambda ^2}\right)\sqrt{1- \frac{u^{\prime 2}}{\lambda ^2}}}= \frac{m_{0}(u- v)}
{\sqrt{1- \frac{u^2}{\lambda ^2}}\sqrt{1- \frac{v^2}{\lambda ^2}}}~~~~~~\textrm{(using the identity
 (\ref{5.21}))}$$ \textit{i.e.} $$m'u' = \frac{mu- mv}{\sqrt{1- \frac{v^2}{\lambda ^2}}}.$$

Here $m_0$ is constant having dimension of mass (known as rest mass), $m= \frac{m_0}{\sqrt{1-
 \frac{u^2}{\lambda ^2}}}~,~m' = \frac{m_0}{\sqrt{1- \frac{u^{\prime 2}}{\lambda ^2}}}$ are termed
 as relativistic mass of the moving particle. If we define the relativistic momentum as $$p_{x}
= mu_{x}$$ then from the above we write the transformation law for momentum along $x$-direction as
\begin{equation}\label{5.28}
p' _{x}= \frac{p_{x}- mv}{\sqrt{1- \frac{v^2}{\lambda ^2}}}
\end{equation}
$$\textrm{Now,}~~~~~~~~u' _{y}= \frac{dy'}{dt'}= \frac{\frac{dy'}{dt}}{\frac{dt'}{dt}}= \frac{\frac{dy}{dt}
\sqrt{1- \frac{v^2}{\lambda ^2}}}{\left(1- \frac{u_{x}v}{\lambda ^2}\right)}= \frac{u_{y}
\sqrt{1- \frac{v^2}{\lambda ^2}}}{\left(1- \frac{u_{x}v}{\lambda ^2}\right)}$$
\begin{equation}\label{5.29}
\textrm{So}~~~~~~~~p' _{y}= m'u' _{y}= \frac{m_0}{\sqrt{1- \frac{u^{\prime 2}}{\lambda ^2}}}\cdot \frac{u_{y}\sqrt{1- \frac{v^2}{\lambda ^2}}}{\left(1- \frac{u_{x}v}{\lambda ^2}\right)}= \frac{m_0}{\sqrt{1- \frac{u^2}{\lambda ^2}}}u_{y}= mu_{y}= p_y
\end{equation}
where in the third step (above) we have used the velocity identity (\ref{5.20}). Similarly, we have
\begin{equation}\label{5.30}
p' _{z}= p_z~.
\end{equation}

Also using the velocity invariant relation (\ref{5.20}) the relativistic mass transformation
 relation takes the form
\begin{equation}\label{5.31}
m' = \frac{m- \frac{p_{x}v}{\lambda ^2}}{\sqrt{1- \frac{v^2}{\lambda ^2}}}
\end{equation}

From equations (\ref{5.28}) to (\ref{5.31}) we see that $(p_{x}, p_{y}, p_{z}, m)$ transforms as Lorentz
 transformation of the space-time co-ordinate. So similar to the invariance of the quadratic
 form $$l^{2}= x^{2}+ y^{2}+ z^{2}- c^{2}t^{2}$$ we have $$p^2 _{x}+ p^2 _{y}+ p^2 _{z}- m^{2}
\lambda ^{2},$$ is invariant in any inertial frame. In a typical inertial frame in which the
 particle is at rest, the above quadratic form takes the value $-m^2 _{0}\lambda ^2$, hence
 we write
\begin{equation}\label{5.32}
p^{2}- m^{2}\lambda ^{2}= -m^2 _{0}\lambda ^2
\end{equation}

From dimensional analysis as momentum times velocity has the dimension of energy, so we write
\begin{equation}\label{5.33}
E^{2}= p^{2}\lambda ^{2}+ m^2 _{0}\lambda ^{4}= m^{2}\lambda ^4
\end{equation}

Hence we have the famous Einstein's energy-mass relation:
\begin{equation}\label{5.34}
E= m\lambda ^2
\end{equation}
and the energy-momentum conservation relation in special theory of relativity has the form
\begin{equation}\label{5.35}
E^{2}= p^{2}\lambda ^{2}+ m^2 _{0}\lambda ^4 .
\end{equation}

\section{Invariant Arc Length\,: Proper Time}

~~~Let $x^{\alpha}= x^{\alpha}(l)$ be the parametric form of a curve in 4D space-time with `$l$'
 as the parameter. Using the fact that the space-time interval is invariant, we can define in
 analogy an invariant arc length along the curve as
\begin{equation}\label{5.36}
L= \int\limits _{l_1}^{l_2} \sqrt{\left|ds^{2}\right|}= \int\limits _{l_1}^{l_2} \frac{\sqrt{\left|ds^{2}\right|}}{dl}dl= \int\limits_{l_1}^{l_2} \left|\left(\frac{d\textit{\textbf{r}}}{dl}\right)^{2}- \lambda ^{2}\left(\frac{dt}{dl}\right)^{2}\right|^{\frac{1}{2}}dl
\end{equation}

Such a trajectory in 4D is called a world line.\\

\underline{{\bf Note:}} In 3D such a trajectory is parametrized by time `$t$' \textit{i.e.} $x^{i}= x^{i}(t)
~,~i= 1,2,3$~~with $\textit{\textbf{v}}(t)= \dfrac{d\textit{\textbf{r}}}{dt}$ as the velocity. Also in 4D, we
 consider $l= vt$ as the parameter with $x^{\alpha}= (\lambda t, \textit{\textbf{r}})$.\\

We now consider the trajectory of a particle with respect to an inertial frame $S$. Suppose
 a clock is attached to the particle. Suppose during the time interval $(t, t+ dt)$ the particle
 (also clock) has moved through a distance $\left|d\textit{\textbf{r}}\right|$ relative to $S$-frame. Let $S'$ be another inertial frame, moving relative to $S$-frame with velocity same as the clock
 at the time instant `$t$'. Hence the clock (particle) is momentarily at rest with respect
 to $S'$-frame and we have $d\textit{\textbf{r}}'= 0$. So the lapse of time $= dt' = d\tau ~\textrm{(say)}$.
 Thus we obtain, $$ds^{2}= -\lambda ^{2}dt^{2}+ \left|d\textit{\textbf{r}}\right|^{2}= (ds')^{2}= -\lambda
 ^{2}(dt')^{2}= -\lambda ^{2}d\tau ^2$$ \textit{i.e.} $~~~~~~d\tau = \dfrac{\left|ds\right|}{\lambda}~,~~~~$
 denotes the time lapse in moving clock.
\begin{equation}\label{5.37}
i.e.~~~~~\tau = \int\limits _{t_1}^{t_2} \frac{\left|ds\right|}{\lambda}= \int\limits _{t_1}^{t_2} dt \sqrt{1- \frac{v^2}{\lambda ^2}}
\end{equation}

Here $\tau$ is called the proper-time along the trajectory of the clock between two events.
 The above expression shows that $\tau$ is invariant under Lorentz transformation. Also from
 the above equation (\ref{5.37}), one may note that the lapse of proper time is always smaller
 than the co-ordinate time interval $(t_{2}- t_{1})$ and hence one may conclude that moving
 clock always slows down.\\

\underline{{\bf Note:}} The world line of a particle is completely arbitrary, not necessarily moving
 with uniform velocity. So accelerated motion in some sense may be described by special
 theory of relativity. This will be elaborately described in subsequent section.\\

\section{General Lorentz Transformation\,: The Transformation Matrix}

~~~If the relative velocity $\textit{\textbf{v}}$ between two inertial frames $S$ and $S'$ is along any
 arbitrary direction then the position vector of any point can be written as $$\textit{\textbf{r}}=
 \textit{\textbf{r}}_{\bm {\parallel}}+ \textit{\textbf{r}}_{\bm {\perp}}~~~~~~\textrm{in}~S\textrm{-frame}$$ $$\textit{\textbf{r}}'= \textit{\textbf{r}}'_{\bm {\parallel}}+
 \textit{\textbf{r}}'_{\bm \perp}~~~~~~\textrm{in}~S'\textrm{-frame}$$ where $\textit{\textbf{r}}_{\bm {\parallel}}$ and $\textit{\textbf{r}}'_{\bm {\parallel}}$
 are parallel to $\textit{\textbf{v}}$ and $\textit{\textbf{r}}_{\bm \perp}$ and $\textit{\textbf{r}}'_{\bm \perp}$ are perpendicular to
 $\textit{\textbf{v}}$.\\

Clearly we write $\textit{\textbf{r}}_{\bm{\parallel}}= \dfrac{(\textit{\textbf{r}}\cdot \textit{\textbf{v}})\textit{\textbf{v}}}{\left|\textit{\textbf{v}}\right|^2}$\\

Previously, we have seen that there is no change of co-ordinate perpendicular to the relative
 velocity. So we have $\textit{\textbf{r}}'_{\bm \perp} = \textit{\textbf{r}}_{\bm \perp}$. Also the transformation parallel to
 the relative velocity is given by $$\textit{\textbf{r}}'_{\bm {\parallel}}= \frac{\textit{\textbf{r}}_{\bm {\parallel}}- \textit{\textbf{v}}t}{\sqrt{1-
 \frac{\left|\textit{\textbf{v}}\right|^2}{\lambda ^2}}}.$$ Thus
\begin{eqnarray}\label{5.38}
\textit{\textbf{r}}'&=& \textit{\textbf{r}}'_{\bm {\parallel}} + \textit{\textbf{r}}'_{\bm \perp} = \frac{\textit{\textbf{r}}_{\bm {\parallel}}- \textit{\textbf{v}}t}{\sqrt{1- \frac{\left|\textit{\textbf{v}}\right|^2}{\lambda ^2}}}+ \textit{\textbf{r}}_{\bm \bot} \nonumber\\
&=&\gamma (\textit{\textbf{r}}_{\bm {\parallel}}- \textit{\textbf{v}}t)+ (\textit{\textbf{r}}- \textit{\textbf{r}}_{\bm {\parallel}})  \nonumber\\
&=&\textit{\textbf{r}}+ (\gamma - 1)\textit{\textbf{r}}_{\bm {\parallel}}- \gamma \textit{\textbf{v}}t  \nonumber\\
&=&\textit{\textbf{r}}+ (\gamma - 1)\frac{(\textit{\textbf{r}}\cdot \textit{\textbf{v}})\textit{\textbf{v}}}{\left|\textit{\textbf{v}}\right| ^2}- \gamma \textit{\textbf{v}}t~~~~~,~~\gamma = \frac{1}{\sqrt{1- \frac{\left|\textit{\textbf{v}}\right|^2}{\lambda ^2}}}  \nonumber\\
&=&\textit{\textbf{r}}+ \frac{(\gamma - 1)}{\beta ^2}(\textit{\textbf{r}}\cdot {\bm \beta}){\bm \beta}- \gamma {\bm \beta}x^0~~~~,~~\beta = \left|\textit{\textbf{v}}\right|/\lambda~~,~x^{0}= \lambda t.
\end{eqnarray}
Also $$t' = \frac{t- \frac{(\textit{\textbf{r}}\cdot \textit{\textbf{v}})}{\lambda ^2}}{\sqrt{1- \frac{
\left|\textit{\textbf{v}}\right|^2}{\lambda ^2}}}= \gamma\left(t- \frac{\textit{\textbf{r}}\cdot {\bm \beta}}{\lambda}\right)$$
\begin{equation}\label{5.39}
i.e.~~~~~~\left(x^0\right)' = \gamma (x^{0}- \textit{\textbf{r}}\cdot {\bm \beta})= x^{0}+ (\gamma - 1)x^{0}- \gamma (\textit{\textbf{r}}\cdot {\bm \beta})
\end{equation}

Hence equations (\ref{5.38}) and (\ref{5.39}) together give the general Lorentz transformation.\\

The above equations of transformation may be considered as a linear transformation between
 two inertial frames $S$ and $S'$ so we write the above transformation equations in compact
 matrix form as
\begin{equation}\label{5.40}
x^{\prime \alpha}= \Lambda ^{\alpha} _{~\beta}x^{\beta}
\end{equation}
where $x^{\alpha}= (x^{0}, \textit{\textbf{r}})$ and $\textit{\textbf{n}}= \dfrac{\textit{\textbf{v}}}{\left|\textit{\textbf{v}}\right|}=
 \dfrac{\bm \beta}{\left|{\bm \beta}\right|}$ is the unit vector along the direction of 
 relative velocity. So the inverse transformation equations are
\begin{equation}\label{5.41}
x^{\alpha}= \Omega ^{\alpha} _{~\beta}(x')^{\beta}
\end{equation}
and the two $4\times 4$ matrices $\Lambda$ and $\Omega$ are related as
\begin{equation}\label{5.42}
\Lambda ^{\alpha} _{~\beta}\Omega ^{\beta} _{~\gamma}= \delta ^{\alpha} _{~\gamma}~~;~~\Lambda ^{\alpha} _{~\beta}\Omega ^{\delta} _{~\alpha}= \delta ^{~\delta} _{\beta}
\end{equation}
\textit{i.e.} the matrices $\Lambda$ and $\Omega$ are inverse of each other.\\

Further, one may note that $\Omega$ can be obtained from $\Lambda$ by changing $\beta$ to
 $-\beta$ \textit{i.e.} $$\Omega (\beta)= \Lambda (-\beta).$$
 
  In particular, the explicit form of
 the matrix components are
\begin{eqnarray}\label{5.43}
\Lambda ^0 _{~0}= \gamma = \Omega ^0 _{~0}~,~\Lambda ^0 _{~i}= -\gamma \left|{\bm \beta}\right|n^{i}= -\Omega ^0 _{~i}  \nonumber\\
\Lambda ^i _{~0}= -\gamma \left|{\bm \beta}\right|n^{i}= -\Omega ^i _{~0}. \\
\Lambda ^i _{~j}= \delta ^i _{~j}+ (\gamma -1)n^{i}n_{j}= \Omega ^i _{~j}. \nonumber
\end{eqnarray}

Also $~~~~\textrm{det}(\Omega ^{\alpha} _{~\beta})= \textrm{det}(\Lambda ^{\alpha} _{~\beta})= 1$.\\

We now examine the result of two successive Lorentz transformations (L.T.). At first for
 simplicity, we consider those L.T. whose relative velocities are along a particular co-
ordinate axis (say $x^1$-axis). Previously, we have shown that such a L.T. is equivalent
 to a rotation in $(ix^0$-$x^1)$-plane with an imaginary angle of rotation. So two
 successive L.T. with relative velocities along the same direction (\textit{i.e.} along the same
 co-ordinate axis $x^1$) is equivalent to another L.T. with equivalent angle of rotation
 equal to the sum of the previous angles of rotation \textit{i.e.} if $$\tanh \theta _{1}= \frac{v_1}
{\lambda}~,~\tanh \theta _{2}= \frac{v_2}{\lambda}~~~~\textrm{then}~~~\tanh (\theta _{1}+ \theta _{2})
= \frac{v_{12}}{\lambda}$$ with $v_{12}= \dfrac{v_{1}+ v_{2}}{1+ \frac{v_{1}v_{2}}{\lambda ^2}}$,
 the law of composition of velocities. So we write (as in section $\S$ 5.4) $$L(v_1)\circ
 L(v_2)= L(v_{12})= L(v_2)\circ L(v_1).$$

However, the situation completely changes if two successive L.T. are not in the same direction.
 Then the imaginary planes of rotation are not same for both the L.T. and hence the L.T. do not
 commute \textit{i.e.} $$L(v_{11})\circ L(v_{22})\neq L(v_{22})\circ L(v_{11})$$ where $v_{11}$ denotes
 the relative velocity along $x^1$-axis and $v_{22}$ that along $x^2$-axis. We shall now
 determine the measure of non-commutativity for two successive L.T. along any arbitrary 
directions. Suppose $$\textit{\textbf{v}}_{\bm 1}=\left|\textit{\textbf{v}}_{\bm 1}\right|\textit{\textbf{n}}_{\bm 1}~~,~~\textit{\textbf{v}}_{\bm 2}=\left|\textit{\textbf{v}}_{\bm 2}
\right|\textit{\textbf{n}}_{\bm 2}~~~~~(\textit{\textbf{n}}_{\bm 1}\neq \textit{\textbf{n}}_{\bm 2})$$ be the relative velocities of two successive
 L.T. So we write
\begin{equation}\label{5.44}
x^{\alpha} _{21}= \Lambda ^{\alpha} _{~\beta}(v_2)\Lambda ^{\beta} _{~\delta}(v_1)x^{\delta}
\end{equation}

Similarly, for the same two L.T. in reverse order we have
\begin{equation}\label{5.45}
x^{\alpha} _{12}= \Lambda ^{\alpha} _{~\beta}(v_1)\Lambda ^{\beta} _{~\delta}(v_2)x^{\delta}
\end{equation}

Thus the measure of non-commutativity is characterized by
\begin{equation}\label{5.46}
\delta x^{\alpha}= x^{\alpha} _{21}- x^{\alpha} _{12}
\end{equation}

Now for simplicity of calculation, we assume $\left|\textit{\textbf{v}}_{\bm {1,2}}\right| \ll \lambda$
 \textit{i.e.} $\left|{\bm \beta}_{\bm {1,2}}\right| \ll 1$ and retain terms in lowest power in 
$\left|{\bm \beta}_{\bm {1,2}}\right|$.\\

Using (\ref{5.43}) we have
\begin{equation}\label{5.47}
x^0 _{21}\simeq \left\{1+ \frac{1}{2}({\bm \beta}_{\bm 1}+ {\bm \beta}_{\bm 2})^2\right\}x^{0}- ({\bm \beta}_{\bm 1}+ {\bm \beta}_{\bm 2})\cdot \textit{\textbf{r}}= x^{0} _{12}
\end{equation}
$$\textit{\textbf{x}}_{\bm {21}}\simeq \textit{\textbf{r}}- ({\bm \beta}_{\bm 1}+ {\bm \beta}_{\bm 2})x^{0}+ ({\bm \beta}_{\bm 1}\cdot \textit{\textbf{r}}){\bm \beta}_{\bm 2}+ \frac{1}{2}\left\{({\bm \beta}_{\bm 2}\cdot \textit{\textbf{r}}){\bm \beta}_{\bm 2}+ ({\bm \beta}_{\bm 1}\cdot \textit{\textbf{r}}){\bm \beta}_{\bm 1}\right\}$$ and 
\begin{equation}\label{5.48}
\textit{\textbf{x}}_{\bm {12}}\simeq \textit{\textbf{r}}- ({\bm \beta}_{\bm 1}+ {\bm \beta}_{\bm 2})x^{0}+ ({\bm \beta}_{\bm 2}\cdot \textit{\textbf{r}}){\bm \beta}_{\bm 1}+ \frac{1}{2}\left\{({\bm \beta}_{\bm 1}\cdot \textit{\textbf{r}}){\bm \beta}_{\bm 1}+ ({\bm \beta}_{\bm 2}\cdot \textit{\textbf{r}}){\bm \beta}_{\bm 2}\right\}.
\end{equation}
Hence,
\begin{equation}\label{5.49}
\delta x^{\alpha}= ({\bm \beta}_{\bm 1}\cdot \textit{\textbf{r}}){\bm \beta}_{\bm 2}- ({\bm \beta}_{\bm 2}\cdot \textit{\textbf{r}}){\bm \beta}_{\bm 1}= \frac{1}{\lambda ^2}(\textit{\textbf{v}}_{\bm 1}\times \textit{\textbf{v}}_{\bm 2})\times \textit{\textbf{r}}
\end{equation}
\\

\underline{{\bf Note:}} The above result is upto the order $\beta ^2$.\\

The result in equation (\ref{5.49}) has an analogy in Newtonian theory where an infinitesimal
 change in the co-ordinates due to infinitesimal rotation of co-ordinate axes is given
 by $$\delta \textit{\textbf{r}}= {\bm \beta}\times \textit{\textbf{r}}~~,~~~\Omega \rightarrow \textrm{the
 angular velocity of rotation}.$$

Thus comparing with the above result we can say that the resultant effect of two L.T. is
 equivalent to a spatial rotation about the direction $\textit{\textbf{v}}_{\bm 1}\times \textit{\textbf{v}}_{\bm 2}$.\\

\underline{{\bf Note:}} The set of all Lorentz transformations do not form a group, as 
the combination of two infinitesimal L.T. in general involves a spatial rotation.\\

However, the set of all L.T. and rotations will form a group, called the Lorentz group. So
 each element of the Lorentz group corresponds either a Lorentz boost or a spatial rotation.\\\\

\section{Some aspects of Lorentz Group and its generators}

~~~An infinitesimal element of a Lorentz group will correspond to the transformation of space-time
 co-ordinates by $$x^{\prime a}= (\delta ^a _{~b}+ \omega ^a _{~b})x^b$$ where $\omega ^a _b$ are
 first order infinitesimal quantities. Then from the relation $$\Lambda ^a _{~b}\Lambda
 ^c _{~d}\eta _{ac}= \eta _{bd}$$ one can easily see that $\omega _{ab}$ is purely antisymmetric
 $$i.e.~~~\omega _{(ab)}= 0~~(\omega _{ab}= \omega _{[ab]}).$$

From the point of view of group representation, we can associate a square matrix with each element of the group such that product of any two such matrices (representing two elements of the group) will give the matrix corresponding to the element of the group obtained by group
 composition of the corresponding two elements. So if $D$ is the matrix representation of a group $G$ then $$D(g_1)\cdot D(g_2)= D(g_{1}\circ g_{2})$$ where `$\circ$' is the binary operation of the group and $g_1, g_2$ are any two elements of the group $G$. If $D$ is a $n\times n$ square matrix then it is said to be a `$n$' dimensional representation of the group.

Further, in analogy with quantum mechanics, the matrix representation corresponding to infinitesimal L.T. can be described as $$D= I+ \frac{1}{2}\omega ^{ab}\sigma _{ab}.$$ 
 
 Due to antisymmetric nature of $\omega ^{ab}$ the operator $\sigma _{ab}$ (corresponding to infinitesimal L.T.) is chosen to be antisymmetric so that out of the six independent components $\sigma _{0i}$ represents the Lorentz boosts while $\sigma _{ij}~~(i\neq j)$ corresponds to spatial rotations. In particular, the three
 operators $B_{i}\sigma _{0i}$ , generates the Lorentz boosts while spatial rotations are generated by the operators $R^{k}\alpha \epsilon ^{ijk}\sigma _{ij}$. Thus the above infinitesimal operator has the explicit form (with suitable normalization) $$D= I+ \frac{i}{2}B_{l}v^{l}+ \frac{i}{2}R_{k}\theta ^{k}.$$
 
  As we have mentioned earlier, the Lorentz boosts along the three spatial axes can be considered as rotation in $(x^{i}\textrm{-}t)$-plane with an imaginary angle so
 all the six operators $(B_{i}, R_{k})$ are related to the angular momentum operators in quantum mechanics. Thus we write, $$R^{i}= -2i\epsilon ^{ijk}x_{j}\partial _{k}~~~,~~~B_{k}= 2i(t\partial _{x_{k}}+ x_{k}\partial _{t})$$ with commutation relations $$[R_{i}, R_{k}]=
 i\epsilon _{ikl}R^{l}~~~;~~~[R_{i}, B_{l}]= i\epsilon _{ilm}B^{m}~~;~~[B_{i}, B_{l}]= -i\epsilon _{ilm}R^{m}.$$
 
  Here the first set of commutation relations are nothing but the usual commutation relations for angular momentum operator in quantum mechanics. The second set of commutation relations states that the boost operator behaves as a 3-vector under spatial rotation. Lastly,
 the commutation relation between two boosts is equivalent to a rotation (which  we have already shown). However, the commutation algebra can be closed by defining $$J_{i}= \frac{1}{2}(R_{i}+ iB_{i})~~,~~K_{i}= \frac{1}{2}(R_{i}- iB_{i})$$ so that $$[J_{i}, J_{j}]= i\epsilon _{ijk}J^{k}
~~~;~~~[K_{i}, K_{j}]= i\epsilon _{ijl}K^{l}~~~;~~~[J_{i}, K_{l}]= 0.$$ Therefore $J_i$ and $K_i$ are nothing but independent angular momentum operators in quantum mechanics.\\
In four dimension LG is a collection of $4 \times 4$ real matrices (denoted by $M(4,\mathbf{R})$) which preserve the Minkowskian metric and matrix multiplication as the group operation. Mathematically, the LG is denoted by $\mathbf{O}(3,1)$ and is defined as
\begin{equation}
	\mathbf{O}(3,1) \equiv \mathcal{L}\equiv \{ M \in M(4, \mathbf{R}): M^T \eta M= \eta \} \nonumber
\end{equation}
where $\eta = diag \{ -1, 1,1,1 \}$ is the Minkowski metric.\\

In a general `D' dimension the LG is denoted by $\mathcal{O}(D-1,1)$ and is defined as the set of $D \times D$ matrices:
\begin{equation}
	\mathcal{O}(D-1,1) \equiv \mathcal{L} \equiv \{ M \in M(D, \mathcal{R}): M^T \eta_D M= \eta_D \} \nonumber
\end{equation}
where $\eta_D = diag \{-1,1,1,...,1\}$

\textbf{Observations:}\\

(a) A $D$ dimensional LG is equivalent to a Lie group of real dimensions $\frac{D(D-1)}{2}$. Also it is analogous to the orthogonal group $\mathcal{O}(D)$.\\

(b) The rows and columns of a Lorentz matrix form a Lorentz basis of $\mathrm{R}^D$ i.e. a basis $\{e_0,e_1,...,e_{D-1} \}$ of $D$ vectors such that $e_\alpha^\mu \eta_{\mu \nu } e_\beta^\nu = \eta_{\alpha \beta}$. Similarly, the rows and columns of an orthogonal matrix form an orthonormal basis of $\mathcal{R}^D$.\\

(c) The LG corresponds to homogeneous linear transformation between two inertial space-times. So it may be considered as a subgroup of the Poincare group which consists of inhomogeneous transformations: $x'=\Lambda x+ a (x^\mu , a^\mu ~\mbox{are} ~4 \times 1 $ column vectors and $\Lambda^\mu_\nu $ are $4 \times 4$ square matrices) from one inertial (space-time) $ST$ to another inertial $ST$. Poincare group is also termed as inhomogeneous Lorentz group. The abstract structure of Poincare group is a semi-direct product $\mathcal{O}(3,1) \times \mathcal{R}^D$. ($\mathcal{R}^D $ being the group of translations) and the group operation is given by 
\begin{equation}
	(\Lambda, a). (\Lambda' , a')= (\Lambda.\Lambda' , a+\Lambda a') \nonumber
\end{equation}
The Poincare group is the group of all isometries of Minkowskian space-time, while LG is the group of isometries those leave the origin fixed.\\

\textbf{Lorentz transformation and Lorentz group:}

Suppose $\vec{X}= \begin{bmatrix}
	ct \\
	x \\
	y \\
	z
\end{bmatrix}$ , $\vec{X'}= \begin{bmatrix}
	ct' \\
	x' \\
	y' \\
	z'
\end{bmatrix}$ represent space-time co-ordinates (as column vectors) in two inertial frames S and $S'$. The space-time interval $S^2=x^2+y^2+z^2-c^2t^2$ can be written in matrix form as
\begin{equation}
	\vec{X}^T \eta \vec{X}
\end{equation}
The LT are the transformation of the ST which leaves the ST interval to be invariant i.e.
\begin{equation}
	\vec{X'}^T \eta \vec{X'}= \vec{X}^T \eta \vec{X}
\end{equation}
where $\vec{X'}= M \vec{X}$, $M$ being a $4 \times 4$ matrix.

The collection of all these $4 \times 4$ matrices $M$ which leave the above ST interval invariant is termed as LT with matrix multiplication. Using $\vec{X'}= M \vec{X}$ into the invariance of ST interval gives
\begin{eqnarray}
	\vec{X}^T M^T \eta M \vec{X}= \vec{X}^T \eta \vec{X} \nonumber \\
	\implies \eta = M^T \eta M \nonumber
\end{eqnarray}
i.e. matrices $M$ which preserve the Minkowski matrix form the LG.\\

\textbf{Properties:}

(i) By considering determinant of the above matrix condition for LG one gets, $det(M)=\pm 1$.

(ii) If the Minkowski metric $'\eta'$ can be written as a block matrix as
\begin{equation}
	\eta = \begin{bmatrix}
		-1 & \bar{0} \\
		\underbar {0} & \tilde{I}_{3 \times 3}
	\end{bmatrix} ~ \mbox{with} ~ \bar{0}=(0,0,0)~, ~  \underbar {0}=\begin{pmatrix}
		0 \\
		0 \\
		0
	\end{pmatrix}, I_{3 \times 3}, 3 \times 3 ~\mbox{identity matrix}.
\end{equation}
Then the matrix $M$ can be written as block matrix in most general form as 
\begin{equation}
	M= \begin{bmatrix}
		\Gamma & -\vec{a}^T \\
		- \vec{b} & \Lambda
	\end{bmatrix}
\end{equation}
where $\Gamma$ is a scalar, $\vec{a} ~,~ \vec{b} $ are vectors i.e. $\vec{a}=\begin{pmatrix}
	a_1 \\
	a_2 \\
	a_3
\end{pmatrix}$ and $\Lambda$ is a $3 \times 3$ matrix. By performing the block matrix multiplication the general restrictions on $\Gamma ~,~ \vec{a} ~,~ \vec{b}$ and $ \Lambda$ can be obtained as a result of invariance of the ST interval as

\begin{eqnarray}
	M^T \eta M &=& \begin{bmatrix}
		\Gamma & -\vec{b}T \\
		-\vec{a} & \Lambda^T
	\end{bmatrix} 
	\begin{bmatrix}
		-1 & \bar{0} \\
		\tilde{0} & I_{3 \times 3}
	\end{bmatrix}
	\begin{bmatrix}
		\Gamma & -\vec{a}T \\
		-\vec{b} & \Lambda
	\end{bmatrix} \nonumber \\
	&=& \begin{bmatrix}
		- \Gamma & -\vec{b}T \\
		\vec{a} & \Lambda^T
	\end{bmatrix} \begin{bmatrix}
		\Gamma & -\vec{a}^T \\
		-\vec{b} & \Lambda 
	\end{bmatrix} = \begin{bmatrix}
		-\Gamma^2 + \vec{b}^T \vec{b} & \Gamma \vec{a}^T-\vec{b}^T \Lambda \\
		\vec{a} \Gamma -\Lambda^T \vec{b} & -\vec{a} \vec{a}^T + \Lambda^T \Lambda
	\end{bmatrix} \nonumber \\
	&=& \begin{bmatrix}
		-1 & \bar{0}\\
		\tilde{0} & I_{3 \times 3}
	\end{bmatrix} \nonumber
\end{eqnarray}

\begin{eqnarray}
	\implies - \Gamma^2 +  |\vec{b}|^2 &=& -1 ~\mbox{i.e.}~ \Gamma^2= 1+ |\vec{b}|^2, \nonumber \\
	\Gamma \vec{a}^T - \vec{b}^T \Lambda &=& 0 ~\mbox{and}~ \Lambda^T \Lambda - \vec{a} \vec{a}^T = I_{3 \times 3}.\nonumber
\end{eqnarray}
Thus if the LT matrix is of the form $\begin{bmatrix}
	\Gamma & -\vec{a}^T \\
	-\vec{b} & \Lambda
\end{bmatrix}$ then one has the general following relations:

(i) $\Gamma^2 = 1+|\vec{b}|^2 $, (ii) $\Gamma \vec{a}^T = \vec{b}^T \Lambda$ and (iii) $\Lambda^T \Lambda - \vec{a} \vec{a}^T= I_{3 \times 3}$.

The first condition shows $\Gamma^2 \ge 1 ~\mbox{i.e.}~ \Gamma \ge 1 ~\mbox{or}~ \Gamma \le -1$.

Note that though $\Lambda \le -1$ is acceptable mathematically but from physical point of view, $\Gamma$ multiplies the time co-ordinate and has an effect on time symmetry.

For $\Gamma >0 ~\mbox{i.e.}~ \Gamma \ge 1 ~,~ \Gamma$ is known as Lorentz factor.

The LTs may be classified in the following 4 ways by the determinant of $M$ and the sign of $\Gamma$.

I. Proper LT : $det(M)=+1 (L_+)$

II. Improper LT : $det(M)=-1 (L_-)$

III. Antichronous LT : $\Gamma \le -1 (L^{\le})$

IV. Orthochronous LT : $ \Gamma \ge +1 (L^{\ge})$

Thus the full LG can be splitted into the union of the above four disjoint subsets:
\begin{equation}
	L= L^{\ge}_+ \cup L^{\le}_+ \cup L^{\ge}_- \cup L^{\le}_- \nonumber.
\end{equation}

In a group, a subgroup is closed under the same operation of the group (here matrix multiplication). This implies, if $M_1$ and $M_2$ are two LTs from a particular subgroup, the composite LT $M_1 M_2$ and $M_2 M_1$ must be in the same subgroup as $M_1$ and $M_2$. But one may note that the composition of two orthochronous and the composition of two improper LT is proper. Hence the sets $L_+^{\ge} ~,~L_+~,~\mbox{and}~ L^{\ge} ~\mbox{and}~L_0=L_+^{\ge} \cup L_-^{\le} $ form subgroups of $L$ while the set containing improper and for and / or antichronous transformation i.e. $L_+^{\le}~,~L_-^{\le} ~,~L_-^{\ge} $ do not form subgroups. \\

\textbf{Linear structure : The principle of inertia}

Suppose $S_1$ and $S_2$ are two inertial frames. Then according to principle of inertia of a particle moves along a straight line at constant velocity as seen by an observer $A$ in $S_1$ frame then the particle should move also along a straight line as seen by another observer $B$ in $S_2$-frame. Now, if  $x^\mu$ and $x'^\mu$ represent the space-time co-ordinates of the same particle w.r.t. above inertial frames $S_1$ and $S_2$, then the transformation between $x^\mu$ and $x'^\mu$ should be such that the straight line path in $S_1$ frame must be mapped to the straight line path in $S_2$ frame. The general transformation preserving this straight line nature is a prejective map as

\begin{equation}
	x'^\mu =\frac{a^\mu + \Lambda^\mu_\nu x^\nu}{b+c_\mu x^\mu}~,~ \mu =0,1,2,3 \nonumber
\end{equation}

Note that the repeated index $'\nu'$ in 2nd term of the numerator indicates a summation over $\nu=0,1,2,3.$ In the above transformation all the coefficients namely $a^\mu ~,~\Lambda^\mu \nu~,~b~ \mbox{and}~c_\mu$ are constants. If we impose that points having finite co-ordinates in $S_1$ should have finite co-ordinates in $S_2$ also then $c_\mu$ should vanish. Thus the above projective transformation reduces to

\begin{equation}
	x'^\mu = \Lambda^\mu_\nu x^\nu +a^\mu ~,~ \mu =0,1,2,3. \nonumber 
\end{equation}
This shows that the principle of Inertia results a linear structure of the ST. As $x^\mu$'s can also be obtained from $x'^\mu$ by inverting the above linear equation so the $4 \times 4$ matrix $\Lambda^\mu_{~\nu}$ should have an inverse.\\

\textbf{Mathematical properties:}

Usually the LG is the indefinite orthogonal group $\mathcal{O}(1,3)$, the proper LG is denoted by $\mathcal{SO}(1,3)$ and the restricted LG by $\mathcal{SO}^+(1,3)$. The LG is a Lie group of symmetries of the ST in STR. This group can be realized as a collection of matrices, linear transformations or unitary operators on some Hilbert space. This group is important because STR together with Quantum Mechanics are the two physical theories those are most throughly established and the conjunction of these two theories is the study of the infinite dimensional unitary representation of LG.

The LG is a 6 dimensional non-compact, non-abelian real Lie group that is not connected. The four connected components are not simply connected. The identity component (i.e. the component containing the identity element) of the LG is itself a group and is termed as the restricted LG and is denoted as $\mathcal{SO}^+(1,3)$. The restricted LG consists of those LTs which preserve both the orientation of space and the direction of time.

Due to Lie group nature of the LG $\mathcal{O}(1,3)$, it is not only a group but also has a topological description as a smooth manifold. As a manifold, it has four connected components i.e. it consists of four topologically separated spaces. These four connected components can be categorized by two transformation properties of its elements namely

(i) Some elements are reversed under time-inverting LTs, i.e. a future- pointing time-like vector would be inverted to a past-pointing vector.

(ii) Some elements have orientation reversed by improper LT.

Now LTs those preserve the direction of time are called orthochronous. This subgroup is denoted by $\mathcal{O}^+(1,3)$.

The transformations which preserve orientation are called proper and as linear transformations they have $det =\pm 1$ (note that the improper LTs have $det=-1$). The subgroup of proper LTs is denoted by $SO(1,3)$.

The subgroup of all LTs preserving both orientation and direction of time is called the proper orthochronous LG or restricted LG and is denoted by $SO^+(1,3)$.

Thus the set of the four connected components can be given by a group structure as the quotient group $\frac{\mathcal{O}(1,3)}{\mathcal{SO}^+(1,3)}$ and it is isomorphic to the Klein four group.

Every element of $\mathcal{O}(1,3)$ can be written as the semi director product of a proper orthochronous transformation and an element of the discrete group $\{1,P,T,PT \}$ where $P$ and $T$ are the parity and time reversal operators:
\begin{equation}
	P=diag (1, -1,-1,-1)~,~T= diag (-1,1,1,1) \nonumber.
\end{equation}

Thus an arbitrary LT can be specified as a proper orthochronous LT along with a further two bits of information, which pick out one of the four connected components - a pattern typical for finite dimension Lie groups.\\

\textbf{Restricted LG (RLG)}
The RLG consists of all LTs those can be connected to the identity by a continuous curve lying in the group. The RGL is a connected normal subgroup of the full LG with the same dimension (i.e. six).

The restricted LG is generated by ordinary spatial rotations and Lorentz boots (which are rotations in a hyperbolic space that includes a time-like direction). Every proper orthochronous LT can be written as a product of a rotation (specified by 3 real parameters) and a boost (also specified by three real parameters) i.e. 6 parameters to specify an arbitrary proper orthochronous LT.

The set of all rotations forms a Lie subgroup isomorphic to the ordinary rotation group $SO(3)$.

The set of all boosts however does not form a subgroup, as composition of two boosts does not in general result in another boost, rather a pair of non- colinear boosts is equivalent to a boost and a rotation, related to Thomas rotation. However, a boost in some direction or a rotation about some axis generates a one-parameter subgroup.\\

\textbf{Subgroups of LG:}
The defining property for a LG is $M^T \eta M=\eta$ . So each matrix $M$ in the LG has $det(M)=\pm 1$. Thus LG can be split into two disconnected subsets characterized by determinant $+1 ~\mbox{or}~ -1$. Lorentz matrices with $det +1$ span a subgroup, called the proper LG and is denoted by $\mathcal{SO}(1,3)$ or $\mathcal{L}_+$, the set of all LTs those preserve the orientation of space.

The set of all Lorentz matrices $M$ with $M^0_0 >0$ forms a subgroup of LG , called the orthochronous LG and is denoted by $\mathcal{O}(3,1)^+$ or $L^\uparrow
$. It preserves the direction of the arrow of time.

Then one has the proper orthochronous LG 
\begin{equation}
	\mathcal{SO}(3,1)^+ = L_+^\uparrow \equiv L_+ \cap L^\uparrow , \nonumber
\end{equation}
a subgroup of $L$ , the maximally connected subgroup of LG.

The group of orientation-preserving rotations of space $\mathcal{SO}(3)$, is a natural subgroup of $L_+^\uparrow$, consisting of matrices of the form $\begin{pmatrix}
	1 & 0 \\
	0 & R
\end{pmatrix}$ with $R \in \mathcal{SO}(3)$.

\textbf{Note:} $L_+$ can be generated by adding to $L_+^\uparrow$, the time -reversal matrix $T=\begin{pmatrix}
	-1 & 0 & 0 & 0 \\
	0 & 1 & 0 & 0 \\
	0 & 0 & 1 & 0 \\
	0 & 0 & 0 & 1
\end{pmatrix}$ . Similarly, $L^\uparrow$ can be obtained by adding to $L_+^\uparrow$, the parity matrix $P= \begin{pmatrix}
	1 & 0 & 0 & 0 \\
	0 & -1 & 0 & 0 \\
	0 & 0 & -1 & 0 \\
	0 & 0 & 0 & -1
\end{pmatrix}$ .

So in general the whole LG $'L'$ can be obtained by adding $T$ and $P$ to $L_+^\uparrow$. (Note that $T$ and $P$ do not commute with all the matrices in $L_+^\uparrow$).\\

\textbf{The notion of rapidity:}
In LT although the notion of velocity is the most intuitive one, but it is not the most practical one from mathematical view point. In particular, composition of two boosts with velocities $v$ and $w$ (in the same direction) does not yield a boost with velocity $v+w$. So for convenience, one has to determine a parameter for specifying boost so that one has an addition of the two parameters for combination of 2 boosts. This parameter is termed as rapidity and is defined as 
\begin{equation}
	\chi(v)= tanh^{-1}(v/c) \nonumber
\end{equation}
Thus the boost matrix can be written in terms of rapidity as
\begin{equation}
	M= \begin{pmatrix}
		cosh \chi & -sinh \chi & 0 & 0 \\
		-sinh \chi & cosh \chi & 0 & 0 \\
		0 & 0 & 1 & 0 \\
		0 & 0 & 0 & 1
	\end{pmatrix} \equiv L(\chi) \nonumber
\end{equation}
Then it can be verified that composition of two such boosts with rapidity $\chi_1$ and $\chi_2$ is a boost of the some form with rapidity $\chi_1+\chi_2$.

Hence rapidity is the additional parameter specifying Lorentz boosts. Note that boosts along a given axis form a non-compact, one parameter subgroup of $L_+^\uparrow$. It also readily provides a formula for the addition of velocities: the composition of two boosts with velocities $v$ and $w$ is a boost with rapidity $\chi(v)+\chi(w)$. So the velocity $V$ of the resulting boost is
\begin{eqnarray}
	\chi(V) &=& \chi(v)+\chi(w) \nonumber \\
	\mbox{i.e.} ~V &=& c. \{ tanh[\chi(v)+\chi(w) ]\} \nonumber \\
	&=& c. tanh [ arg~ tanh(v/c)+arg~ tanh(w/c)] \nonumber \\
	&=& \frac{(v+w)}{(1+\frac{vw}{c^2})} \nonumber.
\end{eqnarray}
Using this rapidity parameter any matrix $M \in L_+^\uparrow$ can be written as a product
\begin{equation}
	M = \tilde{R}_1~L(\chi)~ \tilde{R}_2 \nonumber
\end{equation}
where $R_1$, $R_2$ are rotations of the form 
\begin{equation}
	\tilde{R}_1 = \begin{pmatrix}
		1 & 0 \\
		0 & R_1
	\end{pmatrix} ~\mbox{and}~ \tilde{R}_2 = \begin{pmatrix}
		1 & 0 \\
		0 & R_2
	\end{pmatrix} \nonumber
\end{equation}
and $L(\chi)$ is a Lorentz boost of the form
\begin{equation}
	\begin{pmatrix}
		cosh \chi & -sinh \chi & 0 & 0 \\
		-sinh \chi & cosh \chi & 0 & 0 \\
		0 & 0 & 1 & 0 \\
		0 & 0 & 0 & 1
	\end{pmatrix}  \nonumber
\end{equation}
The above decomposition is a standard decomposition of a proper orthochronous LT. It is to be noted that the above decomposition is not unique.

\section{The Space-time geometry in Special Theory of Relativity\,: Minkowskian geometry and Null cone}

~~~We shall now discuss the intrinsic geometry of the four dimensional space-time in relativity theory. We have seen that the quadratic expression
\begin{equation}\label{5.50}
s^{2}= x^{2}+ y^{2}+ z^{2}- \lambda ^{2}t^2
\end{equation}
is an invariant quantity. Note that $s^2$ is indefinite in sign. We shall now discuss the three possibilities namely $s^{2}>, =, < 0$. For $s^{2}= 0$, we have $$x^{2}+ y^{2}+ z^{2}- \lambda ^{2}t^{2}= 0,$$ which represents the surface of a cone in four dimensional space-time having vertex at the origin $(0, 0, 0, 0)$ and axis along the time axis. The observer is situated at the vertex $O$. Now inside the cone we have $s^{2}< 0$ while $s^{2}> 0$ outside the cone.
 
\begin{wrapfigure}{r}{0.29\textwidth}\vspace{-\intextsep}
\includegraphics[height=6 cm , width=5 cm ]{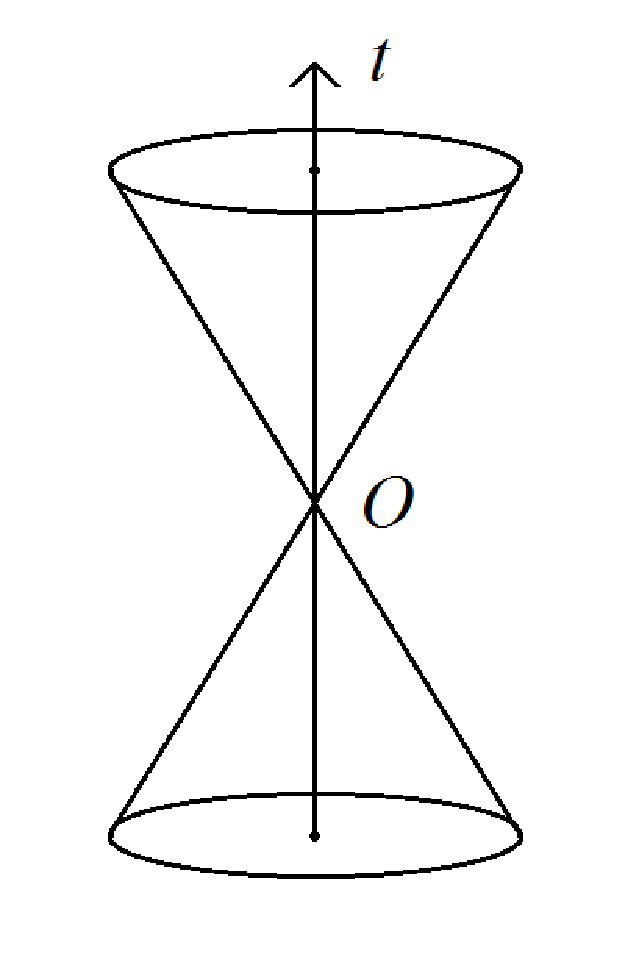}
\begin{center}\vspace{-\intextsep}
Fig. 5.2
\end{center}
\end{wrapfigure}

As $$s^{2}= \lambda ^{2}t^{2}\left(\frac{u^2}{\lambda ^2}- 1\right),$$ so particles inside the cone will have velocity less than the absolute velocity $\lambda$ while velocity
 will be greater than $\lambda$ outside the cone. Inside the cone is termed as time-like region and it is physically admissible. The region outside the cone is not physically acceptable as velocity exceeds the absolute velocity $\lambda$ and the region is termed as space-like region. On the surface of the cone the velocity coincides with the absolute velocity $\lambda$ and is termed as null surface or simply the null cone. The upper half of the cone is the future null cone and the lower half is termed as past null cone. All past events which are causally connected to the observer $O$ are confined to the past null cone while all future events which will be
 connected to the observer, will be confined to the future null cone. Thus the whole four dimensional space-time is divided into three regions --- the time-like and space-like regions are separated by the null cone. Thus in relativity theory, the space-time geometry is not
 Euclidean, rather it is pseudo-Euclidean or it is known as Minkowskian geometry. The first fundamental form of this 4D space-time can be written as
\begin{equation}\label{5.51}
ds^{2}= \eta _{\mu \nu}dx^{\mu}dx^{\nu}
\end{equation}
where $\eta _{\mu \nu}= \textrm{diag}\left\{-1, 1, 1, 1\right\}$ is the Minkowskian metric and
 $x^{0}= \lambda t~,~x^{1}= x~,~x^{2}= y~\textrm{and}~x^{3}= z$.

If $a^{\mu}$ be a four-vector in Minkowskian space then its length is defined as
\begin{equation}\label{5.52}
\left\|a^{\mu}\right\|^{2}= \eta _{\mu \nu}a^{\mu}a^{\nu}= (a^1)^{2}+ (a^2)^{2}+ (a^3)^{2}- (a^0)^{2} .
\end{equation}

Also the scalar product between two four-vectors is defined as
\begin{equation}\label{5.53}
(\utilde{a}, \utilde{b})= \eta _{\mu \nu}a^{\mu}b^{\nu}= a^{1}b^{1}+ a^{2}b^{2}+ a^{3}b^{3}- a^{0}b^{0}
\end{equation}

The indefiniteness in the sign of the norm classifies the four-vectors into three classes namely
 time-like vectors (having $\left\|a^{\mu}\right\|^{2}< 0$), space-like vectors (having $\left\|
a^{\mu}\right\|^{2}> 0$) and null vectors (for which $\left\|a^{\mu}\right\|= 0$). In pseudo-
Euclidean (or pseudo-Riemannian) geometry, a null vector is distinct from zero vector. A null
 vector may have all components to be non-zero but still its norm is zero. Thus a null vector
 is a zero vector but a zero vector is not a null vector.
 
 We shall now discuss few properties among the above three types of vectors. Suppose $a^{\mu}=(a^{0},a^{1},a^{2},a^{3})$ is a time-like or null vector i.e, $\left\|a^{\mu}\right\|^{2}\leq0$ i.e, $(a^1)^{2}+ (a^2)^{2}+ (a^3)^{2}- (a^0)^{2}\leq0$ i.e, 
 \begin{equation}
 	(a^1)^{2}+ (a^2)^{2}+ (a^3)^{2}\leq (a^0)^{2}\label{eq5.54}
 \end{equation}
 Suppose $b^{\mu}=(b^{0},b^{1},b^{2},b^{3})$ be any four vector orthogonal to $a^{\mu}$. The question is ``Can we infer about the nature of the vector $b^{\mu}$?" Due to orthogonality between the two vectors we have $a^{\mu}b_{\mu}=0$ i.e, 
 \begin{equation}
 	-a^{0}b_{0}+a^{1}b_{1}+a^{2}b_{2}+a^{3}b_{3}=0\nonumber
 \end{equation}i.e,
\begin{equation}
	a^{1}b_{1}+a^{2}b_{2}+a^{3}b_{3}=a^{0}b_{0}\label{eq5.55}
\end{equation}
 Now by Cauchy-Schwartz (CS) inequality we have 
 \begin{equation}
 	(a_{1}^{2}+a_{2}^{2}+a_{3}^{2})(b_{1}^{2}+b_{2}^{2}+b_{3}^{2})\geq (a_{1}b_{1}+a_{2}b_{2}+a_{3}b_{3})^{2}\label{eq5.56}
 \end{equation}
Using equations (\ref{eq5.54}) and (\ref{eq5.55}), the above inequality simplifies to
\begin{equation}
	a_{0}^{2}(b_{1}^{2}+b_{2}^{2}+b_{3}^{2})>a_{0}^{2}b_{0}^{2}\nonumber
\end{equation}
i.e, $b_{1}^{2}+b_{2}^{2}+b_{3}^{2}>b_{0}^{2}$ or equivalently, $-b_{0}^{2}+b_{1}^{2}+b_{2}^{2}+b_{3}^{2}\geq 0$. Hence, $b^{\mu}$ is a space-like vector. Thus, any vector orthogonal to a time-like or null vector must be a space-like vector. Thus any vector orthogonal to a time-like or null vector must be a space-like vector. On the other hand, if $a^{\mu}$ and $b^{\mu}$ are two null vectors, then is it possible that they are orthogonal to each other? Due to null nature of the vectors we have
\begin{equation}
	a_{1}^{2}+a_{2}^{2}+a_{3}^{2}=a_{0}^{2},~~b_{1}^{2}+b_{2}^{2}+b_{3}^{2}=b_{0}^{2}\label{eq5.57}
\end{equation}
Also, due to orthogonality of the two vectors we have the relation 
\begin{equation}
	a_{1}b_{1}+a_{2}b_{2}+a_{3}b_{3}=a_{0}b_{0}\label{eq5.58}
\end{equation}
So from the Cauchy-Schwartz inequality
\begin{equation}
	(	a_{1}^{2}+a_{2}^{2}+a_{3}^{2})(b_{1}^{2}+b_{2}^{2}+b_{3}^{2})\geq(a_{1}b_{1}+a_{2}b_{2}+a_{3}b_{3})^{2}
\end{equation}
Using the above two relations (\ref{eq5.57}) and (\ref{eq5.58}) we see that equality holds in the above Cauchy Schwartz inequality and we should have $\dfrac{a_{1}}{b_{1}}=\dfrac{a_{2}}{b_{2}}=\dfrac{a_{3}}{b_{3}}=\lambda$ (say). Thus, we also have $a_{0}=\lambda b_{0}$. Hence, $a^{\mu}$ and $b^{\mu}$ are parallel vectors. Therefore, we obtain a very peculiar result namely ``Two null vectors are simultaneously parallel and orthogonal to each other". Lastly, if $a^{\mu}$ is a space-like vector and $b^{\mu}$ is orthogonal to $a^{\mu}$ then as before one can use the Cauchy-Schwartz inequality but it is not possible to have any definite conclusion about the nature of the vector field $b^{\mu}$. Therefore, based on the above analysis we have the following results
\begin{itemize}
	\item A vector orthogonal to a time-like/ null vector must be a space-like vector.
	\item A vector orthogonal to a space-like vector can not have definite nature.
	\item A null vector can be simultaneously orthogonal and parallel to another null vector- a distinct feature in Minkowskian geometry.
\end{itemize}
\section{The Accelerated Motion in Special Theory of Relativity}

~~~In Minkowski space, let $x^\alpha(\tau)$ be the world line of a test particle. Its four velocity is given by
$$u^\mu=\dfrac{dx^\mu(\tau)}{d\tau}=(\dot{t},\dot{x},\dot{y},\dot{z})$$
with normalization $\|u^\mu\|^2=-1$. So one gets
$$\eta_{\mu \nu}u^\mu u^\nu=-1.$$

Now differentiating with respect to $\tau$ one has
$$\eta_{\mu \nu}a^\mu a^\nu=0.$$
where $a^\mu=\dfrac{du^\mu}{d\tau}$ is the four acceleration vector. Thus the four acceleration is always orthonormal to the four velocity vector. As $u^\mu$ is a time-like vector so $a^\mu$ is a space-like vector. In the inertial frame where the test particle is at rest i.e, $u^\mu=(1,0,0,0)$ then $a^\mu=(0,\textit{\textbf{a}})$. As $u^\mu$ is along the tangent to the world line so $a^\mu$ will be along the principal normal to the world line. Further, the magnitude of the 4-acceleration is related to the curvature of the world line.
%
%
%

\subsection{Null cone co-ordinates}

~~~If we make the transformation of co-ordinates $(t, x)\rightarrow (u, v)$ defined by $$u= 
\lambda t- x~~,~~v= \lambda t+ x$$ then the Minkowskian metric can be written as $$ds^{2}=
 -dudv~~~~\textrm{so that}~~~~g^{(n)} _{ab}=
\begin{pmatrix}
     0      & -\dfrac{1}{2} \\
-\dfrac{1}{2} &      0      \\
\end{pmatrix}
~~\textrm{with}~~a, b= 0, 1~~\textrm{and}~~x^{0}= u~,~x^{1}= v.$$

 The null co-ordinates
 $(u, v)$ are termed as null cone co-ordinates and $g^{(n)} _{ab}$ is termed as Minkowskian
 metric in null cone co-ordinates. Note that the scaling of the null co-ordinates as $$\widetilde{u}
= \mu u~~~,~~~\widetilde{v}= \frac{1}{\mu}v~~~,~~~\mu, a~~\textrm{constant}$$ preserve the above
 Minkowski metric and hence it can be considered as a Lorentz transformation.
\begin{eqnarray}
u' = \lambda t' - x' &=& \frac{\lambda t- \frac{v_{r}x}{\lambda}}{\sqrt{1- \frac{v^2 _r}{\lambda ^2}}}- \frac{x- v_{r}t}{\sqrt{1- \frac{v^2 _r}{\lambda ^2}}}= \frac{(\lambda t- x)+ \frac{v_r}{\lambda}(\lambda t-x)}{\sqrt{1- \frac{v^2 _r}{\lambda ^2}}}  \nonumber\\
&=&\frac{u\left(1+ \frac{v_r}{\lambda}\right)}{\sqrt{1- \frac{v^2 _r}{\lambda ^2}}}= u \sqrt{\frac{1+ \frac{v_r}{\lambda}}{1- \frac{v_r}{\lambda}}}  \nonumber
\end{eqnarray}

Similarly $$v' = v \sqrt{\frac{1- \frac{v_r}{\lambda}}{1+ \frac{v_r}{\lambda}}}~.$$
\begin{equation}\label{5.54}
\textrm{As}~~~~~u' = \mu u~~~,~~~v' = \frac{1}{\lambda}v~~~ \Rightarrow ~~~\mu = \sqrt{\frac{1+ \frac{v_r}{\lambda}}{1- \frac{v_r}{\lambda}}}~.
\end{equation}\\

\subsection{Trajectory of an Accelerated Observer}

~~~Let $x^{\alpha}(\tau)= (u(\tau), v(\tau))$ be the trajectory of a uniformly accelerated observer
 in the inertial frame with null cone co-ordinates. Due to normalization of velocity we have
 $$\dot{u}(\tau)\dot{v}(\tau)= -1$$ and $$\ddot{u}(\tau)\ddot{v}(\tau)= a^2.$$
 
 Now, as $~~~\dot{u}= -\dfrac{1}{\dot{v}}$~, so $~~~\ddot{u}= \dfrac{\ddot{v}}{\dot{v}^2}$. Hence we
 have $$\left(\frac{\ddot{v}}{\dot{v}}\right)^{2}= a^2,$$ which on integration (twice) gives $$v(\tau)
= \frac{A}{a}e^{a \tau}+ B.$$ $$\textrm{Thus}~~~~~~~~u(\tau)= - \frac{1}{Aa}e^{-a \tau}+ C.$$

Here $A, B$ and $C$ are integration constants. As $u \rightarrow \widetilde{u}= \mu u~~$ and $~~v
 \rightarrow \widetilde{v}= \frac{1}{\mu}v$ is a L.T., so we can choose $A= 1$. Further, one can
 choose $B= 0= C$ by shifting the origin of the corresponding inertial frame properly. Thus the
 trajectory in null co-ordinates take the parametric form
\begin{eqnarray}\label{5.55}
u(\tau)&= &- \frac{1}{a}e^{-a \tau}~~~~,~~~~v(\tau)= \frac{1}{a}e^{a \tau}\\
i.e.~~~~uv&=& - \frac{1}{a^2}.  \nonumber
\end{eqnarray}
\begin{wrapfigure}[15]{r}{0.5\textwidth}\vspace{-2\intextsep}
\includegraphics[height=8 cm , width=9 cm]{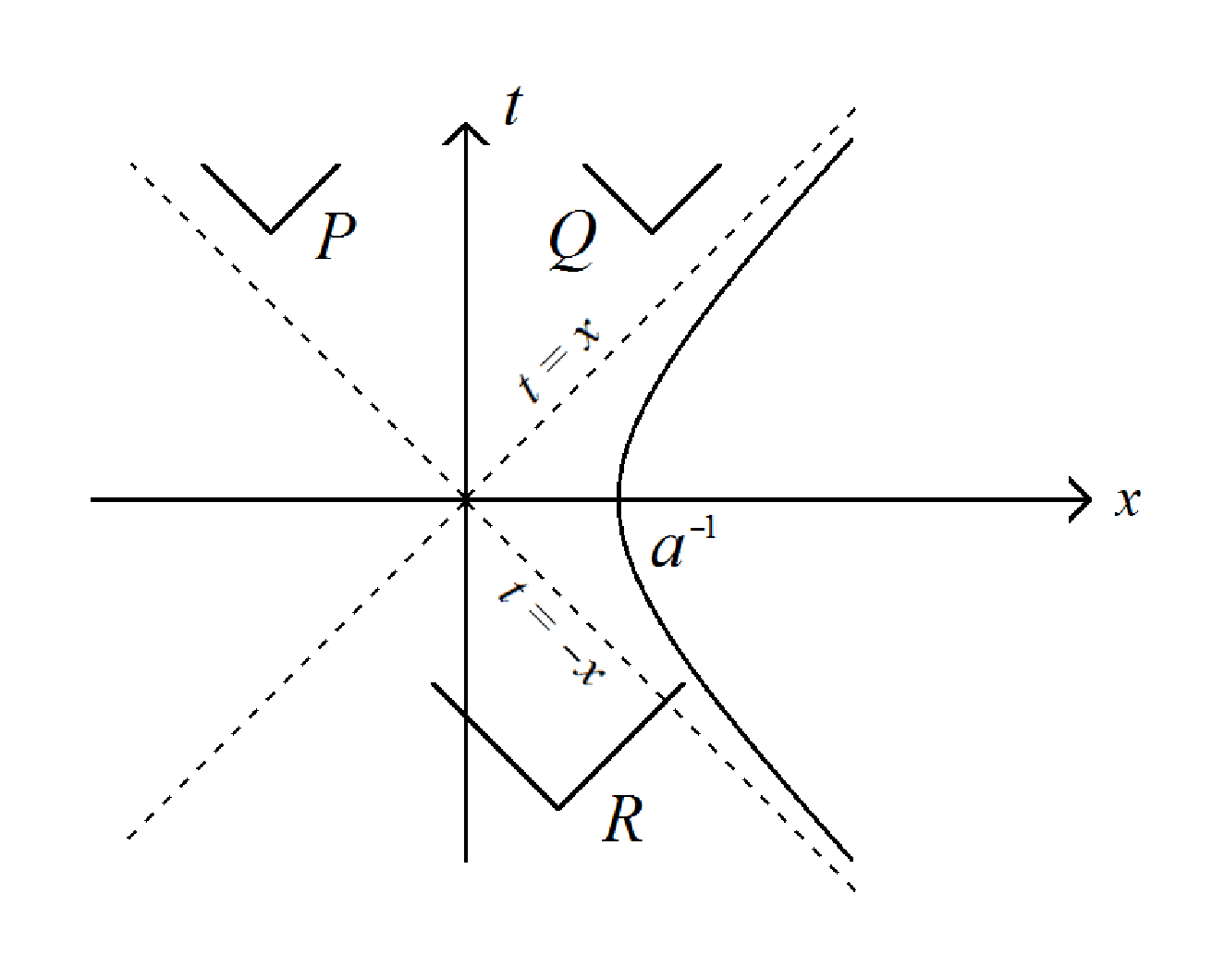}
\begin{center}\vspace{-1.1\intextsep}
Fig. 5.3
\end{center}
\end{wrapfigure}
Further, in terms of original Minkowskian co-ordinates we have the parametric form:\\

$x(\tau)= \dfrac{1}{a}\cosh (a \tau)~~,$\\

$t(\tau)= \dfrac{1}{a}\sinh (a \tau)$\\

$i.e.~~x^{2}- t^{2}= \dfrac{1}{a^2}.$\\

Hence the world line of the accelerated observer is a
 branch of the rectangular hyperbola in $(t, x)$-plane.\\

\underline{{\bf Note:}}  \textbf{I.} As $\left|t\right| \rightarrow \infty$, the world line approaches the
 null cone.\\

~~~\textbf{II.} The observer arrives from $x= + \infty$, decelerates and stops at
 $x= a^{-1}$, then accelerates back towards infinity.\\

The figure shows the world line of a uniformly accelerated observer (with proper acc.
 $\left|\textit{\textbf{a}}\right|$) in Minkowski space-time. The dash lines show the null cone. The
 observer cannot receive any signal from the events $P$ and $Q$ and cannot send signals to $R$.\\

\subsection{Comoving Frame of an Accelerated Observer\,: Rindler space-time}

~~~Let $(z^{0}, z^{1})$ be the comoving frame for an accelerated observer \textit{i.e.} a co-ordinate system
 in which the observer is at rest at $z^1 = 0$ and hence $z^0$ is the proper time along the observer's
 world line. We also choose the co-moving frame such that the metric is conformally flat \textit{i.e.}
\begin{equation}\label{5.56}
ds^{2}= -f(z^{0}, z^{1})\left\{(dz^{0})^{2}- (dz^{1})^{2}\right\}.
\end{equation}

\underline{{\bf Note:}} The conformally flat form of the metric simplifies quantization ({\it See Introduction to Quantum Effects in Gravity by Viatcheslav Mukhanov (Author), Sergei Winitzki (Author)}) of fields to a great extend.\\

We shall now address the questions namely (\textit{i}) whether such a co-ordinate transformation is possible or not and (\textit{ii}) if possible then find a relationship with Minkowski co-ordinates.\\

The above metric in terms of null cone co-ordinates of the co-moving frame takes the form:
\begin{equation}\label{5.57}
ds^{2}= -f_{0}(u_{z}, v_{z})du_{z}dv_{z}
\end{equation}
where $u_{z}= z^{0}- z^{1}$ and $v_{z}= z^{0}+ z^{1}$ are the null co-ordinates of the co-moving frame.\\

As along the world line of the observer we have $$z^{0}(\tau)= \tau~~~\textrm{and}~~~z^{1}
(\tau)= 0$$ so the null co-ordinates are given by $$v_{z}(\tau)= \tau = u_{z}(\tau).$$

 Hence at the observer's position we have $$f_{0}(u_{z}= \tau , v_{z}= \tau)= 1.$$
 
  Further, from the relation $$ds^{2}= -dudv= -f_{0}(u_{z}, v_{z})du_{z}dv_{z}$$ $$(u,v~~\textrm{are the usual null
 cone co-ordinates})$$ it is clear that $u$ and $v$ cannot be function of both the variables $u_z$ and $v_z$, each of them depends on only one of these two variables. So without loss of generality we assume $$u= u(u_z)~~~~\textrm{and}~~~~v= v(v_z).$$

 The explicit form of the functions will be determined by comparing observer's trajectory in these two co-ordinate systems.\\

We write, $$\frac{du(\tau)}{d \tau}= \frac{du(u_z)}{du_z}\cdot \frac{du_{z}(\tau)}{d \tau}$$

As the parametric form of the trajectory in Minkowskian space is given as $$u= - \frac{1}{a}
e^{-a \tau}~~~~,~~~~v= \frac{1}{a}e^{a \tau}~~,$$ which gives, $$\frac{du}{d \tau}= e^{-a \tau}=
 -au(\tau).$$
 
  As along the observer's world line  $u_{z}(\tau)= \tau = v_{z}(\tau)$\\

so $\dfrac{du_{z}(\tau)}{d \tau}= 1$ and hence $\dfrac{du}{du_z}= -au~~i.e.~~u= u_{0}e^{-au_z}$.\\

Similarly, $v= v_{0}e^{+au_z}$, where $u_{0}, v_{0}$ are integration constants.\\

Now, $f_{0}(u_{z}= \tau , v_{z}= \tau)= 1$ gives $a^{2}u_{0}v_{0}= -1$. So we choose $u=
 - \dfrac{1}{a}e^{-au_z}~~~,~~~v= \dfrac{1}{a}e^{av_z}$.\\

Thus we obtain $u$ as a function of $u_z$ alone and $v$ as a function of $v_z$ alone and consequently
 the metric becomes $ds^{2}= dudv= e^{a(v_{z}- u_{z})}du_{z}dv_{z}$.\\

Further going back to the space-time co-ordinates $(t, x)$ we have
\begin{eqnarray}
t(z^{0}, z^{1})= \frac{1}{a}e^{az^{1}}\sinh (az^0) \nonumber\\
x(z^{0}, z^{1})= \frac{1}{a}e^{az^{1}}\cosh (az^0) \nonumber
\end{eqnarray}
$i.e.~~~\dfrac{t}{x}= \tanh (az^0)$ and $x^{2}- t^{2}= \dfrac{1}{a^2}e^{2az^1}$.\\

Hence the metric in accelerated frame is
\begin{equation}\label{5.58}
ds^{2}= -e^{2az^1}\left[(dz^0)^{2}- (dz^1)^{2}\right]
\end{equation}
and is known as Rindler space-time. Clearly, the Rindler space-time is locally equivalent to
 Minkowski space-time. So it has zero curvature. The co-ordinate curves:  $z^{0}=\textrm{constant}$
 and $z^{1}=\textrm{constant}$ are family of straight lines through the origin and a family of
 rectangular hyperbolas respectively. The range of $(z^{0}, z^{1})$ is :  $- \infty < z^{0}, z^{1}
 < + \infty$. Note that the above Rindler metric covers only one quarter of the Minkowski space-time
 $\left(x> \left|t\right|\right)$.\\

Moreover in the 4D Rindler space-time $$ds^{2}= e^{2az^1}\left(-(dz^0)^{2}+ (dz^1)^{2}\right)+
 (dz^2)^{2}+ (dz^3)^{2}$$ if we make the transformation
\begin{equation}\label{5.59}
e^{az^1}= 1+ a \xi ^{1},
\end{equation}
then the above metric becomes
\begin{equation}\label{5.60}
ds^{2}= -(1+ a \xi ^{1})^{2}(dz^0)^{2}+ (d \xi ^1)^{2}+ (dz^2)^{2}+ (dz^3)^{2}.
\end{equation}

This metric has the familiar form of the weak field limit of Einstein gravity with $a \xi ^{1}= \phi$ (or $v$)
 as the Newtonian potential\,(see sect.\,6.4) and the acceleration experienced by the observer in the accelerated
 frame is `$-a$'. Further, the line element between two events in the space-time cannot have the
 Minkowskian form $ds^{2}= -dt^{2}+ dx^2$, rather, at least be modified to a form $$ds^{2}= -\left(1+
 \frac{2 \phi}{c^2}\right)c^{2}dt^{2}+ dx^2,$$ in the presence of a gravitational field.\\

This can be considered as the direct connection between the principle of equivalence and a geometrical
 description of a gravity.\\

Therefore, we conclude that\\

~~~~~~~\textbf{(i)} gravitational fields are locally indistinguishable from accelerated frames.\\

~and \textbf{(ii)} accelerated frames are described by a line element as (\ref{5.60}) and the gravitational
 field affects the rate of clocks in such a way that the clocks slow down in strong gravitational
 field as $$\Delta T= \Delta t\left(1+ \frac{\phi}{c^2}\right)$$ in the lowest order in $\dfrac{\phi}
{c^2}$. Here $\Delta t$ is the time interval measured by a clock in the absence of gravitational
 field while $\Delta T$ represents the corresponding interval measured by a clock located in the
 gravitational potential $\phi$.\\\\

\section{ The invariant notion of relative velocity in Special Theory of Relativity}

The relative velocity between two particles will be derived in the context of STR in an invariant way. Here all velocities (speeds) are measured in terms of the speed of light i.e. $c=1$ is chosen. Also we have the following conventions:\\

\begin{itemize}
	\item The inner product of two four vectors $x^\mu$ and $y^\mu$ is \\
	
	$x.y=x^\mu y_\mu = -x_0 y_0 +\Vec{x}.\Vec{y}$
	
	\item The four velocity $v^\mu= (\gamma, \gamma \Vec{v})$ with $\Vec{v}=\frac{d \Vec{x}}{dt}$, the 3 velocity, and $\gamma=\frac{1}{\sqrt{1-v^2}}$ the relativistic factor.
	
	\item The energy-momentum 4-vector of a free particle of mass $`m'$ is $p_\mu =(E, \Vec{p})$ with $p_\mu p^\mu= p^2 = \Vec{p}^2- E^2 =-m_0^2$, $m_0$ being the rest mass or proper mass of the particle.
	
	\item The trajectory (i.e. world line) of a particle in 4D Minkowski space is denoted by $z_\mu \equiv [ z^0(\tau), \Vec{z}(t)]$ with $\tau$ being the proper time.
	
\end{itemize}

\textbf{Relative velocity:} 
Let $\vec{v_1}$ and $\vec{v_2}$ be the 3 velocity of particles 1 and 2 in an inertial frame S. Then $\Vec{v_{21}}=\Vec{v_2}- \Vec{v_1}$, (a 3 vector) is called the relative velocity between two particles. In the following we shall introduce the relativistic definition of the relative velocity following\\

(a) relative momentum approach and (b) velocity addition approach.\\

\textbf{Relative momentum approach:} In STR, the energy of a free particle is the fourth component of the four vector $p^\mu \equiv (\Vec{p}, E)$. Note that the energy scalar is not an invariant scalar, rather it transform according to the fourth component in LT. However, the relative energy i.e. the energy of particle 2 as seen by particle 1 (i.e. by an observer on the rest frame of particle 1 ) can be defined in an invariant way as follows:\\

Suppose $E_2$, the energy of particle 2 when measured from a frame in which particle 1 is at rest is denoted by $E_2^{(1)}$, i.e. $E_2^1= E_{21}$.\\

Now the question is what is the relative energy $E_{21}$ as measured in any other frame?\\

As $p^\mu =( m_0 \gamma \Vec{v}, m_0 \gamma)= (\Vec{p},m)$

$\therefore || p^\mu ||^2= p^2-m^2 = -m_0^2$, (by energy- momentum conservation relation).

Thus we write,

$p_1^2 = - m_1^2~,~~ p_2^2 = -m_2^2$ \\

So $p_1^\mu p_{2\mu}= (\Vec{p_1}, m_1 \gamma_1). (\Vec{p_2}, m_2 \gamma_2)=\Vec{p_1} \Vec{p_2}-m_1 m_2 \gamma_1 \gamma_2 =\Vec{p_1} \Vec{p_2}-E_1 E_2$

As particle 1 is at rest so $\gamma_1=1~,~ \Vec{p_1}=0$, hence $p_1^\mu p_{2\mu}=-E_1E_2=-m_{10}E_{21}$ (as particle 1 is at rest). 

i.e. 
\begin{equation}
	E_{21}=- \frac{p_1^\mu . p_{2 \mu}}{m_{10}}, \label{e1}
\end{equation}
an invariant quantity. So in any other frame

$ E_{21}=-v_1^\mu . p_{2 \mu}$ ( $v_1^\mu \rightarrow$ 4 velocity of particle 1).

This may be termed as an invariant relative energy.As $p_2^{(1)}$ and $E_2^{~1}$ satisfy the usual energy-momentum relation, so we have
\begin{eqnarray}
	||p_2^\mu ||^2 & \equiv & (p_2^{(1)})^2- (E_2^{(1)})^2 = -m_{20}^2 \nonumber \\
	\mbox{i.e.} (p_2^{(1)})^2 &=& (E_2^{(1)})^2-m_{20}^2 \nonumber
\end{eqnarray}

\textbf{Note:} When $v_1^\mu = (1,0)$ then $E_{21}=E_2^{(1)}=E_{21}^{(1)}$

The relative momentum of 2 relative to 1 can be defined as (in any reference frame)
\begin{equation}
	|p_{21}^\mu|=(E_{21}^2-m_{20}^2)^{\frac{1}{2}}=\frac{1}{m_{10}}\left[(p_1^\mu . p_{2 \mu})^2- m_{10}^2 m_{20}^2 \right]^{\frac{1}{2}}\label{e2}
\end{equation}
So the corresponding magnitude of the relative velocity is then defined (in analogy to $v^\mu = p^\mu /E$) as
\begin{equation}
	|v_{21}^\mu|=\frac{|p_{21}^\mu|}{E_{21}}=\frac{\left[(p_1^\mu . p_{2 \mu})^2- m_{10}^2 m_{20}^2 \right]^{\frac{1}{2}}}{-(p_1^\mu p_{2 \mu })}=\left[1- \frac{m_{10}^2 m_{20}^2}{(p_1^\mu p_{2 \mu })^2} \right]^{\frac{1}{2}} \label{e3}
\end{equation}

\begin{equation}
	|v_{12}^\mu |= \left[1- \frac{1}{(v_1^\mu . v_{2 \mu})^2} \right]^{\frac{1}{2}}\label{e4}
\end{equation}
The above relation shows that the relative 4 velocity is completely symmetrical between 1 and 2.

Now $\gamma_{21}=(1-v_{21})^{-\frac{1}{2}}=|v_1^\mu v_{2\mu }|$ and $p_{21}^\mu = m_{20} \gamma_{21} |v_{21}^\mu|$, which shows that $p_{21}^\mu $ is not symmetrical between 1 and 2.

Now suppose $\vec{v}=tanh \alpha \hat{v}$, then $\gamma = cosh \alpha $ and $v^{\mu}=(cosh \alpha, sinh \alpha \hat{v})$.

So,
\begin{equation}
	(v_1^\mu . v_{2 \mu})^2= (cosh \alpha_1 cosh \alpha_2 + sinh \alpha_1 sinh \alpha_2)^2= cosh^2(\alpha_1- \alpha_2)\nonumber
\end{equation}
i.e. $ \gamma_{21}=cosh(\alpha_1- \alpha_2)>1$ and hence $v_{21}$ is always real.

Thus we have a relativistically invariant relative energy, momentum and velocity given by equations (\ref{e1}),(\ref{e2}) and (\ref{e3})/(\ref{e4}).\\

\textbf{The velocity addition approach:}

Suppose a frame $S'$ moves with velocity $\vec{u}$ relative to S and if $\vec{v}$ and $\vec{v'}$ are the velocities of a particle P relative to S and $S'$.

The general LT gives the transformation law
\begin{eqnarray}
	\vec{r'} &=& \vec{r}+\frac{(\gamma-1)}{\beta^2}(\vec{r}.\vec{\beta})\vec{\beta}-\gamma \vec{\beta} x^{\circ} ~,~ \vec{\beta}=\vec{u} \nonumber \\
	t' &=& \gamma (t- \frac{\vec{r}. \vec{\beta}}{c}),~~~~~~~ \gamma=\frac{1}{\sqrt{1-u^2}} \nonumber
\end{eqnarray}
On inversion (by changing $\vec{r} \rightleftharpoons \vec{r'}, t \rightleftharpoons t', \vec{\beta} \rightarrow - \vec{\beta}$)
\begin{equation}
	\vec{r}=\vec{r'}+ \frac{(\gamma-1)}{\beta^2}(\vec{r'.\vec{u}})\vec{u}+\gamma \vec{u} t'~,~t=\gamma (t'+ \vec{r'}. \vec{u}) \nonumber
\end{equation}
Now, \begin{eqnarray}
	\vec{v} &=& \frac{d \vec{r}}{dt}= \frac{d \vec{r}/dt'}{dt/ dt'}=\frac{\frac{d \vec{r}}{dt'}+\frac{(\gamma-1)}{u^2} (\frac{d \vec{r'}}{dt}.\vec{u})\vec{u}+\gamma \vec{u}}{\gamma (1+ \frac{d \vec{r}}{dt'}.\vec{u})} \nonumber \\
	\therefore ~ \vec{v} &=& \frac{\vec{v'}+ \gamma \vec{u}+\frac{(\gamma^2-1)}{(\gamma+1) u^2}(\vec{v'}. \vec{u})\vec{u}}{\gamma (1+ \vec{v'} \vec{u})} \nonumber
\end{eqnarray}

As, $\gamma^2-1= \frac{1}{1-u^2}-1= \frac{u^2}{1-u^2}=\gamma^2 u^2$, so
\begin{equation}
	\vec{v}=   \frac{\vec{v'}+\gamma \vec{u}(1+\frac{\gamma(\vec{u}. \vec{v'})}{\gamma+1})}{\gamma (1+ \vec{u}. \vec{v'})} \nonumber
\end{equation}

Now suppose a particle 1 is at rest in frame S. Then velocity of the particle 1 will be $\vec{v_1}=- \vec{u}$ as seen from S' frame. Now, if the above particle P is identified as particle 2, then its velocity relative to S' be $\vec{v_2}(= v')$. Thus the relative velocity $\vec{v_{21}}$ of particle 2 relative to particle 1 is given by

\begin{eqnarray}
	\vec{v_{21}} &=& \vec{v} \nonumber \\
	&=& \frac{\vec{v_2}-\gamma_1 v_1 (1- \gamma_1 \frac{(\vec{v_1}.\vec{v_2})}{(1+\gamma_1)})}{\gamma_1 (1-\vec{v_1}. \vec{v_2})} \nonumber \\
	\therefore |\vec{v_{21}}| &=& \frac{\left[ (\vec{v_1- \vec{v_2})^2+ (\vec{v_1}.\vec{v_2})^2}-v_1^2 v_2^2 \right]^{\frac{1}{2}}}{(1-\vec{v_1} \vec{v_2})} \label{e5}
\end{eqnarray}
(for detail derivation see problem 5.26)

We shall now show the equivalence between the formula (\ref{e4}) and (\ref{e5}) for the expression for relative velocity.\\

In (\ref{e4}) $v_1^\mu =(\gamma_1 , \gamma_1 \vec{v_1})~,~ v_2^\mu =(\gamma_2, \gamma_2 \vec{v_2})$
\begin{eqnarray}
	v_1^\mu v_{2 \mu} &=& \gamma_1 \gamma_2 -\gamma_1 \gamma_2 \vec{v_1} \vec{v_2} = \gamma_1 \gamma_2 (1- \vec{v_1}. \vec{v_2}) \nonumber \\
	\therefore && 1- \frac{1}{(v_1^\mu v_{2 \mu})^2}= 1- \frac{1}{\gamma_1^2 \gamma_2^2(1-\vec{v_1}.\vec{v_2})^2} \nonumber \\
	&=& 1- \frac{(1-v_1^2)(1-v_2^2)}{(1-\vec{v_1}.\vec{v_2})^2}= \frac{1-2 \vec{v_1}.\vec{v_2}+(\vec{v_1}.\vec{v_2})^2-1+v_1^2+v_2^2-v_1^2 v_2^2}{(1-\vec{v_1}.\vec{v_2})^2} \nonumber \\
	&=& \frac{(v_1^2+v_2^2-2 \vec{v_1}. \vec{v_2})+(\vec{v_1}.\vec{v_2})^2-v_1^2v_2^2}{(1-\vec{v_1}.\vec{v_2})^2} \nonumber \\
	&=& \frac{(\vec{v_1}-\vec{v_2})^2+ (\vec{v_1}. \vec{v_2})^2- v_1^2 v_2^2}{(1-\vec{v_1}.\vec{v_2})^2} \nonumber
\end{eqnarray}
Hence from (\ref{e5}) we have 
\begin{equation}
	|\vec{v_{21}}|= \{ 1- \frac{1}{(v_1. v_2)} \}^{\frac{1}{2}}. \nonumber
\end{equation}
Thus both from relative momentum approach and velocity addition approach we have the invariant expression for relative velocity.

\section{Non Commutativity of General Lorentz Transformation and Wigner Rotation: A Review }

The general Lorentz transformation (LT) between two inertial frames $S$ and $S'$ is given by (see section 5.11)
\begin{equation}
	\vec{r'}= \vec{r}+ \frac{(\gamma_u-1)(\vec{r}.\vec{u}).\vec{u}}{|\vec{u}|^2}-\gamma_u \vec{u}t \label{e1}
\end{equation}
and
\begin{equation}
	t' =\gamma_u(t-\frac{\vec{r}.\vec{\beta_u}}{c})~,~\beta_u=\frac{\vec{u}}{c}~,~ \gamma_u=\frac{1}{\sqrt{1-\frac{u^2}{c^2}}} \nonumber
\end{equation}
where $\vec{u}$ is the velocity of $S'$-frame w.r.t. $S$ -frame. Writing $T=ct$ the above transformation equations can be written as

\begin{equation}
	\left.\begin{array}{lll}
		\vec{r}&=& \vec{r}+(\gamma_u-1) \frac{(\vec{r}.\vec{\beta_u}).\vec{\beta_u}}{|\vec{\beta_u}|^2}-\gamma_u \vec{\beta_u}T  \\
		T' &=& \gamma_u (T-\vec{r}.\vec{\beta_u})
	\end{array}\right\} \label{e2}
\end{equation}
As 
\begin{eqnarray}
	\gamma_u^2-1 &=& \frac{1}{1- \frac{u^2}{c^2}}-1 = \frac{\frac{u^2}{c^2}}{1-\frac{u^2}{c^2}}=\gamma_u^2 \frac{u^2}{c^2} = \gamma_u^2 \beta_u^2 \nonumber \\
	\implies ~ \frac{\gamma_u-1}{\beta_u^2} &=& \frac{\gamma_u^2}{(\gamma_u+1)} \nonumber
\end{eqnarray}
So equation  $(\ref{e2})$ modifies to

\begin{equation}
	\left.\begin{array}{lll}
		\vec{r}&=& \vec{r}+ (\frac{\gamma_u^2}{\gamma_u +1}) \frac{(\vec{r}.\vec{\beta_u})\vec{\beta_u}}{|\vec{\beta_u}|^2}-\gamma_u \vec{\beta_u}T  \\
		T' &=& \gamma_u (T-\vec{r}.\vec{\beta_u})
	\end{array}\right\} \label{e3}
\end{equation}
Suppose a particle has velocity $\vec{V}$ w.r.t. $S$ frame and $\vec{V'}$ w.r.t. $S'$ frame. Then 
\begin{eqnarray}
	\vec{V'} &=& \frac{d \vec{r'}}{dt'}= \left[\frac{d \vec{r}}{dt}+ (\frac{\gamma_u^2}{1+\gamma_u})(\frac{d \vec{r}}{dt}.\vec{\beta_u}).\vec{\beta_u}-\gamma_u \vec{\beta_u}c \right] \frac{dt}{dt'} \nonumber \\
	&=& \left[ \vec{V}+(\frac{\gamma_u^2}{1+\gamma_u})(\vec{V}.\vec{\beta_u})\vec{\beta_u}-c \gamma_u \vec{\beta_u} \right](\frac{dt}{dt'}). \nonumber
\end{eqnarray}
Using $\frac{dt'}{dt}=\gamma_u (1-\frac{\vec{V}.\vec{u}}{c^2})$,
\begin{equation}
	\vec{V'} = \frac{\left[ \vec{V}+(\frac{\gamma_u^2}{1+\gamma_u})(\vec{V}.\vec{\beta_u})\vec{\beta_u}-c \gamma_u \vec{\beta_u} \right]}{\gamma_u (1-\frac{\vec{u}.\vec{V}}{c^2})} \nonumber
\end{equation}
or equivalently,

\begin{eqnarray}
	\vec{V} &=& \frac{\left[ \vec{V'}+(\frac{\gamma_u^2}{1+\gamma_u})(\vec{V'}.\vec{\beta_u})\vec{\beta_u}+c \gamma_u \vec{\beta_u} \right]}{\gamma_u (1+\frac{\vec{u}.\vec{V'}}{c^2})} \nonumber \\
	\mbox{i.e.,}~ \vec{V} &=& \vec{V'} \oplus \vec{u} .\label{e4}
\end{eqnarray}

The magnitude is given by (see problem $(5.26)$)
\begin{equation}
	|\vec{V}|^2= \frac{ \left[|\vec{u}+\vec{V'}|^2- \frac{1}{c^2}(\vec{u} \times \vec{V'})^2 \right]}{(1+ \frac{\vec{u}. \vec{V'}}{c^2})^2}. \label{e5}
\end{equation}
This is called the velocity transformation law.\\  \\

\textbf{Composition of two general LTs}\\
Suppose $\vec{u}$ is the relative velocity between two frames $S$ and $S'$ and let $\vec{v}$ be the relative velocity between two inertial frames $S'$ and $S''$. We shall now examine whether the composition of these two general LTs will be a LT or not. We start with time transformation.

\begin{eqnarray}
	T &=& \gamma_u (T' + \vec{r'}. \vec{\beta_u}) ~\mbox{and}~ T' = \gamma_v (T''+\vec{r''}.\vec{\beta_v}) \nonumber \\
	&=& \gamma_u \left[ \gamma_v (T''+\vec{r''}.\vec{\beta_v})+ \vec{r'}.\vec{\beta_u} \right] \nonumber \\
	&=& \gamma_u \left[ \gamma_v (T''+\vec{r''}.\vec{\beta_v})+ \vec{\beta_u} \left\{ \vec{r''}+(\gamma_v -1) \frac{(\vec{r''}. \vec{\beta_v})}{|\vec{\beta_v}|^2}\vec{\beta_v}+\gamma_v \vec{\beta_v} T'' \right \}\right] \nonumber \\
	&=& \gamma_u \gamma_v (1+ \vec{\beta_v }. \vec{\beta_u})T'' +\gamma_u \gamma_v \vec{\beta_v} \vec{r''} + \gamma_u \vec{\beta_u} \vec{r''}+ \frac{\gamma_u \gamma_v^2}{1+ \gamma_v}(\vec{r''}. \vec{\beta_v})(\vec{\beta_u}. \vec{\beta_v}) \nonumber \\
	&=& \gamma_u \gamma_v (1+\vec{\beta_u}. \vec{\beta_v}) \left[ T''+\frac{\vec{r''}}{\gamma_v (1+\vec{\beta_u}. \vec{\beta_v})} \left \{ \vec{\beta_u}+\gamma_v \vec{ \beta_v}+ (\frac{\gamma_v^2}{1+ \gamma_v})(\vec{\beta_u}.\vec{\beta_v}).\vec{\beta_v} \right \}\right] \nonumber \\
	&=& \gamma \left[ T''+ \vec{r''}. \frac{ \vec{w}}{c}\right]= \gamma \left[T'' + \vec{r''}. \vec{\beta_w} \right] \label{e6}
\end{eqnarray}
where
\begin{eqnarray}
	\vec{w} &=& \frac{\left\{ \vec{u}+ \gamma_v \vec{v}+ \frac{\gamma_v^2}{1+ \gamma_v}(\vec{\beta_u}. \vec{\beta_v})\vec{v} \right \}}{\gamma_v (1+ \vec{\beta_u}\vec{\beta_v})} \nonumber \\
	&=& \vec{u} \oplus \vec{v} \nonumber \\
	\mbox{and}~ \gamma & =& \gamma_u \gamma_v (1+\frac{\vec{u}. \vec{v}}{c^2}) \label{e7}
\end{eqnarray}
From equation $(\ref{e7})$ i.e.

\begin{eqnarray}
	\gamma &=&  \gamma_u \gamma_v (1+\frac{\vec{u}. \vec{v}}{c^2}) ,~ \mbox{one has} ~\nonumber \\
	\frac{1}{\sqrt{1- \frac{w^2}{c^2}}} &=& \frac{1+\frac{\vec{u}. \vec{v}}{c^2}}{\sqrt{1- \frac{u^2}{c^2}} \sqrt{1- \frac{v^2}{c^2}}} \nonumber \\
	\implies 1- \frac{w^2}{c^2} &=& \frac{(1-\frac{u^2}{c^2})(1-\frac{u^2}{c^2})}{(1+\frac{u^2. \vec{v}}{c^2})^2} \nonumber \\
	\mbox{i.e.}~ \frac{w^2}{c^2} &=& 1-  \frac{(1-\frac{u^2}{c^2})(1-\frac{u^2}{c^2})}{(1+\frac{u^2. \vec{v}}{c^2})^2} = \frac{\frac{1}{c^2}(\vec{u}+\vec{v})^2}{(1+\frac{u^2. \vec{v}}{c^2})^2} \nonumber \\
	\mbox{i.e.}~ w^2 &=& \frac{(\vec{u}+ \vec{v})^2}{(1+\frac{\vec{u}. \vec{v}}{c^2})^2} \nonumber
\end{eqnarray}
Note that though $\vec{w}$ is not symmetric in $\vec{u}$ and $\vec{v}$ but its magnitude is symmetric in  $\vec{u}$ and $\vec{v}$. i.e.
\begin{eqnarray}
	\vec{w} &=& \vec{u} \oplus \vec{v} \neq \vec{v} \oplus \vec{u} \nonumber \\
	\mbox{but}~  |\vec{w}|^2 &=& |\vec{u} \oplus \vec{v}|^2 = |\vec{v} \oplus \vec{u}|^2 \label{e8}
\end{eqnarray}
Thus an interchange of $\vec{u}$ and $\vec{v}$ implies a rotation of $\vec{w}$, keeping the magnitude same. This rotation is known as Thomas rotation/ Thomas- Wigner rotation or Wigner rotation.
\begin{equation}
	\mbox{As}~~ \gamma^2-1 =\gamma^2 \frac{w^2}{c^2} ~~\mbox{i.e.}~~w= \frac{c}{\gamma} \sqrt{\gamma^2-1} \label{e9}
\end{equation}
Suppose $\vec{u}$ be the velocity of an inertial frame $S$ w.r.t. an object $A$. Let $\vec{v}$ be the velocity of an object $B$ w.r.t. $S$ frame. It is assumed that $\vec{u}$ and $\vec{v}$ are not parallel. Then the velocity of $B$ as measured by $A$ is given by 

\begin{equation}
	\vec{V}_{AB} = \frac{1}{(1+\frac{\vec{u}.\vec{v}}{c^2})}\left[ \left\{ 1+\frac{\gamma_u}{1+\gamma_u}(1+\frac{\vec{u}.\vec{v}}{c^2}) \right\} \vec{u}+\frac{1}{\gamma_u}\vec{v} \right]= \vec{u}\oplus \vec{v} \label{e10}
\end{equation}
similarly, the velocity of $A$ as measured by $B$ is given by 

\begin{equation}
	\vec{V}_{BA} = \frac{1}{(1+\frac{\vec{u}.\vec{v}}{c^2})}\left[ \left\{ 1+\frac{\gamma_v}{1+\gamma_v}(1+\frac{\vec{u}.\vec{v}}{c^2}) \right\} \vec{v}+\frac{1}{\gamma_v}\vec{u} \right]= \vec{v}\oplus \vec{u}. \label{e11}
\end{equation}
Though $\vec{V}_{AB} \neq \vec{V}_{BA}$, but
\begin{equation}
	|\vec{V}_{AB}| =  |\vec{V}_{BA}|= \frac{1}{(1+\frac{\vec{u}.\vec{v}}{c^2})} \left[ |\vec{u}+\vec{v}|^2-\frac{1}{c^2}(\vec{u} \times \vec{v})^2 \right]^{\frac{1}{2}} \label{e12}
\end{equation}
and both have the same Lorentz factor
\begin{equation}
	\gamma= \gamma_{\vec{u} \oplus \vec{v}}= \gamma_{\vec{v} \oplus \vec{u}}= \gamma_u \gamma_v (1+\frac{\vec{u}. \vec{v}}{c^2})\label{e13}
\end{equation}

\subsection{Lorentz Transformation in (block) matrix form}
In formulating the general LT between two inertial frames $S$ and $S'$ we have assumed that the corresponding co-ordinate axes are parallel in the two frames. But it should be noted that though both the pairs $(S, S')$ and $(S', S'')$ have parallel coordinate axes but when viewed from $S$ the co-ordinate axes of $S$ and $S''$ are not parallel. Hence a complete description of the relation between the frames can not be provided by velocity addition, rather one has to formulate the complete description in terms of LT corresponding to the velocities.

A Lorentz boost with an arbitrary velocity $\vec{u}$ can be written symbolically as
\begin{eqnarray}
	\vec{X'}= L (\vec{u}) \vec{X} \label{e14}
\end{eqnarray}
with $\vec{X}= \begin{bmatrix}
	ct \\
	x \\
	y \\
	z
\end{bmatrix}, ~\vec{X'}= \begin{bmatrix}
	ct' \\
	x' \\
	y' \\
	z'
\end{bmatrix}$ i.e. $\vec{X}= \begin{bmatrix}
	ct \\
	\vec{r}
\end{bmatrix}$, $\vec{X'}= \begin{bmatrix}
	ct' \\
	\vec{r'}
\end{bmatrix}$
\begin{equation}
	L (\vec{u})= \begin{bmatrix}
		\gamma_u & -\gamma_u \beta_x &  -\gamma_u \beta_y & -\gamma_u \beta_z \\
		-\gamma_u \beta_x & 1+(\gamma_u-1)\frac{\beta_x^2}{\beta_u^2} & (\gamma_u-1)\frac{\beta_y \beta_x}{\beta_u^2} & (\gamma_u-1)\frac{\beta_z \beta_x}{\beta_u^2} \\
		-\gamma_u \beta_y &  (\gamma_u-1)\frac{\beta_x \beta_y}{\beta_u^2} & 1+(\gamma_u-1)\frac{\beta_y^2}{\beta_u^2} &(\gamma_u-1)\frac{\beta_y \beta_z}{\beta_u^2} \\
		-\gamma_u \beta_z & (\gamma_u-1)\frac{\beta_x \beta_z}{\beta_u^2} & (\gamma_u-1)\frac{\beta_z \beta_y}{\beta_u^2} & (\gamma_u-1)\frac{\beta_z^2}{\beta_u^2}
	\end{bmatrix} \label{e15}
\end{equation}
or equivalently it can be written in block matrix form as
\begin{equation}
	L_B(\vec{u})= \begin{bmatrix}
		\gamma_u & -\frac{\gamma_u}{c} \vec{u}^T \\
		-\frac{\gamma_u}{c} \vec{u} & I_{3 \times 3}+(\frac{\gamma_u^2}{1+\gamma_u})(\frac{\vec{u}.\vec{u}^T}{c^2})
	\end{bmatrix} .\label{e16}
\end{equation}
Note that $\vec{u}$, and $\vec{r}$ are $3 \times1$ column vectors while the transposes $\vec{u}^T$, $\vec{r}^T$ are row vectors. The boost matrix is a symmetric matrix having inverse $L(u)^{-1}= L(-u)$ i.e. $\vec{X}=L(-u)\vec{X'}$.

Also to each admissible velocity $\vec{u}$ there corresponds a pure Lorentz boost i.e. $u \leftrightarrow L(u)$.

The velocity addition $\vec{u} \oplus \vec{v}$ corresponds to the composition of boosts $L(\vec{v}) \circ L(\vec{u})$ i.e. $L(v)$ operates on $L(u) \vec{X}$ while the composition of boosts $L(\vec{u}) \circ L(\vec{v})$ is associated to the velocity addition $\vec{v}\oplus \vec{u}$. Thus one has

\begin{equation}
	\vec{X''}= L (\vec{v}) \vec{X'}~~,~~ \vec{X'} = L (\vec{u}) \vec{X} \nonumber
\end{equation}
i.e.
\begin{eqnarray}
	\vec{X''} &=& L (\vec{v}) ~L(\vec{u}) \vec{X} \nonumber \\
	&=& \begin{bmatrix}
		\gamma & - a^T \\
		-b & M
	\end{bmatrix} \vec{X} \nonumber
\end{eqnarray}
i.e. $\Lambda = \begin{bmatrix}
	\gamma & -a^T \\ 
	-b & M
\end{bmatrix}$ the matrix corresponding to composition of two non- parallel boosts. Here
\begin{equation}
	\gamma = \gamma_u \gamma_v (1+\frac{\vec{v}^T \vec{u}}{c^2}) \nonumber
\end{equation}
\begin{equation}
	\vec{a}= \frac{\gamma}{c}(\vec{u} \oplus \vec{v})~~,~~ \vec{b}= \frac{\gamma}{c} (\vec{v} \oplus \vec{u}) \label{e17}
\end{equation}
are column vectors and the $3 \times 3$ matrix $M$ has the expression

\begin{equation}
	M= \gamma_u \gamma_v (\frac{\vec{v} \vec{u}^T}{c^T})+   \left\{ I_{3 \times 3}+ \frac{\gamma_v^2}{1+\gamma}(\frac{\vec{u}. \vec{v}}{c^2}) \right\} \left\{ I_{3 \times 3} + \frac{\gamma_u^2}{1+\gamma_u}(\frac{\vec{u}.\vec{v}}{c^2}) \right\}\label{e18}
\end{equation}
The inverse LT from $S''$ to $S$ can be written as
\begin{equation}
	\Lambda^{-1}= \begin{bmatrix}
		\gamma & b^T \\
		a & M^T
	\end{bmatrix} = L (- \vec{u}) L (- \vec{v}) \nonumber
\end{equation}
Note that the matrix $\Lambda$ is not symmetric and hence it does not corresponds to a single boost i.e. 
\begin{equation}
	L (u \oplus v) \neq L(v) L(u). \nonumber
\end{equation}
This is known as incompleteness of velocity composition from the result of two boosts. 

Now for complete description one has to introduce a rotation before or after the boost. This rotation is also known as Thomas rotation. A rotation can be expressed in matrix form as

\begin{equation}
	\vec{X'}= \mu (\vec{\theta}) \vec{X}~~ \mbox{with}~~ \mu (\vec{\theta})= \begin{bmatrix}
		1 & 0 \\
		0 & R(\theta)
	\end{bmatrix} \nonumber
\end{equation}
where $R$ is a $3 \times 3$ rotation matrix, $\vec{\theta}= \theta. \vec{\epsilon}$ is termed as axis-angle vector with $\vec{\epsilon}$, a unit vector along the axis. Conventionally, the rotation is chosen to be $+ve$ in the anticlockwise direction. Precisely, the rotation matrix rotates any $3D$ vector about $\vec{\epsilon}$ axis through an angle $\theta$ (in the anticlockwise sense). Further, a boost followed or preceeded by a rotation is also a LT due to the invariance of the space-time interval. Let
\begin{equation}
	\Lambda (\vec{\theta},\vec{u})= \mu (\vec{\theta}) L(\vec{u}) ~\mbox{and }  \Lambda (\vec{v},\vec{\phi})= L(\vec{v}) \mu(\vec{\phi}) \nonumber
\end{equation}
corresponds to same Lorentz transformations then one has 

\begin{equation}
	L(\vec{u})= \mu (-\vec{\theta})L(\vec{v}) \mu (\vec{\phi}) \nonumber
\end{equation}
i.e. two Lorentz boosts are related by a matrix similarity transformation.\\

\subsection{Composition of two general Lorentz Transformation:}
Let us consider three inertial frame of references $S_1, S_2$ and $S_3$. Suppose $S_2$ is moving relative to $S_1$ frame with constant velocity $\vec{u}$ in an arbitrary direction. The axes of $S_2$ frame are assumed to be parallel relative to $S_1$ frame. Then the LT between $S_1, S_2$ frames can be written in block matrix form as

\begin{equation}
	\vec{X'}=L(\vec{u})\vec{X}~~i.e.~~ \begin{bmatrix}
		ct' \\
		x' \\
		y' \\
		z'
	\end{bmatrix}= \begin{bmatrix}
		\gamma_u & -a_u^T \\
		-b_u & M_u
	\end{bmatrix}\begin{bmatrix}
		ct \\
		x \\
		y \\
		z
	\end{bmatrix} \label{e19}
\end{equation}
similarly the LT between $S_2$ and $S_3$ frames reads as
\begin{equation}
	\vec{X''}=L(\vec{v})\vec{X'}~~i.e.~~ \begin{bmatrix}
		ct'' \\
		x'' \\
		y'' \\
		z''
	\end{bmatrix}= \begin{bmatrix}
		\gamma_v & -a_v^T \\
		-b_v & m_v
	\end{bmatrix}\begin{bmatrix}
		ct' \\
		x' \\
		y' \\
		z'
	\end{bmatrix} \label{e20}
\end{equation}
Now the relative velocity of frame 3 w.r.t. frame 1 is given by 
\begin{equation}
	\vec{V}_{31}= \frac{\left[ \{ 1+(\frac{\gamma_u}{1+\gamma_u})\frac{\vec{u}.\vec{v}}{c^2} \}\vec{u}+\frac{1}{\gamma_u}\vec{v} \right]}{(1+\frac{\vec{u}.\vec{v}}{c^2})}\label{e21}
\end{equation}
similarly, the relative velocity of frame 1 w.r.t. frame 3 is

\begin{equation}
	\vec{V}_{13}= \frac{\left[ \{ 1+(\frac{\gamma_v}{1+\gamma_v})\frac{\vec{u}.\vec{v}}{c^2} \}\vec{v}+\frac{1}{\gamma_v}\vec{u} \right]}{(1+\frac{\vec{u}.\vec{v}}{c^2})}\label{e22}
\end{equation}
(Note that $\vec{V}_{31} \neq \vec{V}_{13}$ but $|\vec{V}_{31}|= |\vec{V}_{13}|=\frac{c}{\gamma}\sqrt{\gamma^2-1}, \gamma=\gamma_u \gamma_v (1+\frac{\vec{u}.\vec{v}}{c^2})$).

Thus combining $(\ref{e19})$ and $(\ref{e20})$ one gets

\begin{equation}
	\vec{X''}= L(\vec{v}) \vec{X'}= L(\vec{v}) L(\vec{u}) \vec{X} \nonumber
\end{equation}
(Note that due to non-commutativity of the matrix product $L(\vec{v}) L(\vec{u}) \neq L(\vec{u}) L(\vec{v})$).\\

Now
\begin{eqnarray}
	L(\vec{v}) L(\vec{u})&=&\begin{bmatrix}
		\gamma_v & -\gamma_v \frac{\vec{v}^T}{c} \\
		-\gamma_v \frac{\vec{v}}{c} & I_3 +\frac{\gamma_v^2}{1+\gamma_v}(\frac{\vec{v}.\vec{v}^T}{c^2})
	\end{bmatrix} \begin{bmatrix}
		\gamma_u & -\gamma_u \frac{\vec{u}^T}{c} \\
		-\gamma_u \frac{\vec{u}}{c} & I_3 +\frac{\gamma_u^2}{1+\gamma_u}(\frac{\vec{u}.\vec{u}^T}{c^2})
	\end{bmatrix} \nonumber \\
	&=&\begin{bmatrix}
		\gamma_u \gamma_v (1+\frac{\vec{v}^T.\vec{u}}{c^2}) & -\gamma_u \gamma_v \frac{\vec{u}^T}{c}-\gamma_v \frac{\vec{v}^T}{c}- (\frac{\gamma_v \gamma_u^2}{1+\gamma_u})\frac{\vec{v}^T}{c} \frac{(\vec{u}.\vec{u}^T)}{c^2} \\
		-\gamma_u \gamma_v \frac{\vec{v}}{c}-\gamma_u \frac{\vec{u}}{c}-\frac{\gamma_u \gamma_v^2}{1+\gamma_v}(\frac{\vec{v}.\vec{v}^T}{c^2})\frac{\vec{u}}{c} & \gamma_u \gamma_v \frac{\vec{v}.\vec{u}^T}{c^2}+\{ I_3+(\frac{\gamma_v^2}{1+\gamma_v})(\frac{\vec{v}.\vec{v}^T}{c^2}) \}\times \{ I_3+(\frac{\gamma_u^2}{1+\gamma_u})(\frac{\vec{u}.\vec{u}^T}{c^2}) \}
	\end{bmatrix} \nonumber \\
	&=& \begin{bmatrix}
		\gamma & \frac{-\frac{\gamma_u \gamma_v}{c}(1+\frac{\vec{v}^T.\vec{u}}{c^2})\{ \vec{u}^T+\frac{\vec{v}^T}{\gamma_u}+(\frac{\gamma_u}{1+\gamma_u})\vec{v}^T(\frac{\vec{u}.\vec{u}^T}{c^2})\}}{(1+\frac{\vec{v}^T. \vec{u}}{c^2})} \\
		-\frac{\gamma_u \gamma_v}{c}(1+\frac{\vec{v}^T.\vec{u}}{c^2})\{ \vec{v}+\frac{\vec{u}}{\gamma_v}+(\frac{\gamma_v}{1+\gamma_v})(\frac{\vec{v}.\vec{v}^T}{c^2}) \vec{u}^\} & \gamma_u \gamma_v \frac{\vec{v}. \vec{u}^T}{c^2}+ \{ I_3+(\frac{\gamma_v^2}{1+\gamma_v})(\frac{\vec{v}.\vec{v}^T}{c^2}) \} \{ I_3+(\frac{\gamma_u^2}{1+\gamma_u})(\frac{\vec{u}.\vec{u}^T}{c^2}) \}
	\end{bmatrix} \nonumber \\
	&=& \begin{bmatrix}
		\gamma & -\vec{a}^T \\
		-\vec{b} & M
	\end{bmatrix} \nonumber
\end{eqnarray}
where 
\begin{eqnarray}
	\vec{a}^T &=& \frac{\gamma}{c} [\{ 1+(\frac{\gamma_u}{1+\gamma_u})(\frac{\vec{u}.\vec{v}}{c^2})\}\vec{u}^T+\frac{1}{\gamma_u}\vec{v}^T]=\frac{\gamma}{c}\vec{V}_{31}^T \nonumber \\
	\vec{b} &=& \frac{\gamma}{c}[ \{ 1+(\frac{\gamma_v}{1+\gamma_v})(\frac{\vec{u}.\vec{v}}{c^2})\}\vec{v}+\frac{1}{\gamma_v}\vec{u}]= \frac{\gamma}{c}\vec{V}_{13} \nonumber \\
	\mbox{and}~ M &=& \gamma_u \gamma_v \frac{\vec{v}. \vec{u}^T}{c^2}+ \{ I_3+(\frac{\gamma_v^2}{1+\gamma_v})(\frac{\vec{v}.\vec{v}^T}{c^2}) \} \{ I_3+(\frac{\gamma_u^2}{1+\gamma_u})(\frac{\vec{u}.\vec{u}^T}{c^2}) \} \nonumber
\end{eqnarray}
Thus if $\Lambda=L(\vec{v})\circ L(\vec{u})$, then $\Lambda^{-1}=L(-\vec{u})L(-\vec{v})$ and we have 
\begin{equation}
	\Lambda^{-1}= L(-\vec{u})L(-\vec{v})= \begin{bmatrix}
		\gamma & b^T \\
		a & M^T
	\end{bmatrix}. \nonumber
\end{equation}
i.e.
\begin{equation}
	\vec{X}= L(-\vec{u}) \vec{X'}= L(-\vec{u}) L(-\vec{v}) \vec{X''} = \Lambda^{-1} \vec{X''} \nonumber
\end{equation}
Note that though the Lorentz transformation is symmetric but the composition of two general LT does not give a symmetric matrix and hence it can not represent a single boost. Symbolically,
\begin{equation}
	L(\vec{u} \oplus \vec{v}) \neq L(v) L(u) ~\mbox{or}~ L(\vec{v} \oplus \vec{u}) \neq L(\vec{u}) L(\vec{v}). \nonumber
\end{equation}
We shall now show that the composition of two general boost corresponds to a boost along the composition of velocity followed (or preceeded) by a rotation i.e.
\begin{equation}
	L(\vec{v}) \circ L(\vec{u})= \Lambda(\vec{v},\vec{u})= R(\epsilon) L(\vec{A})= L(\vec{B}) R(\epsilon) \nonumber
\end{equation}
where $\vec{A}=\vec{u}\oplus \vec{v}$, $\vec{B}=\vec{v}\oplus \vec{u}$ and $R(\epsilon)$ is a $4 \times 4$ matrix of the form

\begin{equation}
	R(\epsilon) = \begin{bmatrix}
		1 & 0 \\
		0 & R(\vec{\epsilon})
	\end{bmatrix}\nonumber
\end{equation}
Here $R(\vec{\epsilon})$ is a $3 \times 3$ rotation matrix characterized by axis-angle representation with $\vec{\theta}=\theta . \vec{\epsilon}$. Here $\theta$ is the angle of rotation in the counter clockwise direction and $\vec{\epsilon}$ is the unit vector parallel to the axis of rotation. Thus

\begin{eqnarray}
	R(\epsilon) &=& \Lambda(\vec{v}, \vec{u}). L^{-1}(\vec{A}) \nonumber \\
	&=& \begin{bmatrix}
		\gamma & -\vec{a}^T \\
		-\vec{b} & M
	\end{bmatrix} \begin{bmatrix}
		\gamma & \frac{\gamma}{c}\vec{A}^T \\
		\frac{\gamma}{c}\vec{A} & I+\frac{\gamma^2}{1+\gamma}(\frac{\vec{A}.\vec{A}^T}{c^2})
	\end{bmatrix} ,~~~~~(\mbox{since} ~\vec{a}=\frac{\gamma}{c}\vec{A} ~\mbox{and}~ \vec{b}=\frac{\gamma}{c}\vec{B})\nonumber \\
	&=& \begin{bmatrix}
		\gamma^2-\frac{\gamma^2}{c^2}\vec{A}^T \vec{A} & \frac{\gamma^2}{c}\vec{A}^T-\frac{\gamma}{c}\vec{A}^T-\frac{\gamma^3}{1+\gamma}(\frac{\vec{A}^T}{c})(\frac{\vec{A}.\vec{A}^T}{c^2}) \\
		-\frac{\gamma^2}{c}\vec{B}+\frac{\gamma}{c}M\vec{A} & -\frac{\gamma^2}{c^2}\vec{B}.\vec{A}^T+M+\frac{M \gamma^2}{(1+\gamma)}(\frac{\vec{A}.\vec{A}^T}{c^2}) 
	\end{bmatrix} \nonumber \\
	&=& \begin{bmatrix}
		\gamma^2(1-\frac{\vec{A}^2}{c^2}) & \frac{\gamma(\gamma-1)}{c} \vec{A}^T-(\frac{\gamma^2}{1+\gamma})(\frac{\gamma \vec{A}^T}{c})(\frac{\vec{A}.\vec{A}^T}{c^2}) \\
		-\frac{\gamma}{c}(\gamma \vec{B}-M \vec{A}) & -\frac{\gamma^2}{c^2}\vec{B}\vec{A}^T+M \{ 1+\frac{\gamma^2}{c^2}(\frac{\vec{A}.\vec{A}^T}{1+\gamma}) \}
	\end{bmatrix} \nonumber \\
	&=& \begin{bmatrix}
		1 & \frac{\gamma(\gamma-1)}{c}\vec{A}^T-\frac{(\gamma-1)}{A^2/c^2}\frac{\gamma \vec{A}^T}{c}(\frac{\vec{A}^2}{c^2}) \\
		-\frac{\gamma}{c}(\gamma \vec{B}-M \vec{A}) & C_{\alpha \beta}
	\end{bmatrix} \nonumber \\
	&=& \begin{bmatrix}
		1 & 0 \\
		-\frac{\gamma}{c}(\gamma \vec{B}-M \vec{A}) & C_{\alpha \beta}
	\end{bmatrix}~~ C_{\alpha \beta}=M \{ 1+\frac{\gamma^2}{c^2}(\frac{\vec{A}.\vec{A}^T}{1+\gamma}) \} -\frac{\gamma^2}{c^2}\vec{B}\vec{A}^T \nonumber \\
	&=& \begin{bmatrix}
		1 & 0 \\
		0 & \vec{R}(\vec{\epsilon})
	\end{bmatrix}=R(\epsilon)~~~~~\mbox{(for detailed calculation to show, $M\vec{A}=\gamma \vec{B}$ see appendix A)}. \nonumber
\end{eqnarray}

Thus $\Lambda(\vec{v}, \vec{u})=R(\epsilon).L(\vec{A})$.\\

This shows that composition of two general boost is not a boost but rather a composition of a boost along the velocity composition together with a rotation $\theta$ given by 
\begin{equation}
	cos \theta=\frac{\vec{A}.\vec{B}}{|\vec{A}||\vec{B}|}\nonumber
\end{equation}
\textbf{Note:} Composition of two Lorentz boosts is not a Lorentz boost but rather a composition of a boost along the composition of the two velocities together with a rotation from $\vec{A}$ to $\vec{B}$. Further, as both the boost and rotation keep the space-time interval to be invariant so $\Lambda$ represents a LT but not a Lorentz boost.\\

The explicit form of $R(\epsilon)$ is given by (see appendix A)
\begin{equation}
	R(\epsilon)=M-\frac{ba^T}{(1+\gamma)}.\label{e23}
\end{equation}

As the two composite velocities are of equal magnitude but in different directions, so one must be a rotated copy of the other. So one may write (for derivation see appendix B)
\begin{equation}
	\vec{b}= R(\epsilon) \vec{a} \label{e24}
\end{equation}
i.e. the matrix $R$ rotates $\vec{a}$ in the anticlockwise direction  to give $\vec{b}$. Further, one can invert equation $(\ref{e24})$ to have $\vec{a}$ as
\begin{equation}
	\vec{a}=R^{-1}(\epsilon) \vec{b}\label{e25}
\end{equation}
However, a simplification of $R^T(\epsilon)\vec{b}$ gives $\vec{a}$ (see appendix C),i.e.
\begin{equation}
	R^T(\epsilon) \vec{b} = \vec{a} \label{e26}
\end{equation}
Thus we have
\begin{equation}
	R^T=R^{-1}\label{e27}
\end{equation}
i.e. $R$ is an orthogonal matrix and it justifies $R$ to be a rotation matrix. Thus,
\begin{eqnarray}
	\vec{a} &=&\frac{\gamma}{c}\vec{u} \oplus \vec{v}~~ \mbox{and}~~ \vec{b}=\frac{\gamma}{c}\vec{v} \oplus \vec{u} ~~ \mbox{are of some magnitude i.e.} \nonumber \\
	|\vec{a}| &=& \frac{\gamma}{c}.\frac{1}{(1+\frac{\vec{u}.\vec{v}}{c^2})}[(\vec{u}+\vec{v})^2-\frac{1}{c^2}(\vec{u}\times \vec{v})^2]^{\frac{1}{2}}= \vec{b} \nonumber \\
	&=& \sqrt{\gamma^2-1}\label{e28}
\end{eqnarray}
Also,
\begin{equation}
	\vec{a} \times \vec{b}= \frac{\gamma_u \gamma_v (\gamma^2-1)(1+\gamma+\gamma_u+\gamma_v)}{c^2(\gamma+1)(\gamma_u+1)(\gamma_v+1)}(\vec{u} \times \vec{v})\label{e29}
\end{equation}
(detailed calculation can be found in appendix D)\\

The above rotation shows that the axis of rotation is parallel to $\vec{u} \times \vec{v}$ and hence
\begin{equation}
	\hat{e}=\frac{\vec{u}\times \vec{v}}{|\vec{u}\times \vec{v}|}~~, \mbox{the unit vector along the axis of rotation.}
\end{equation}

Moreover, due to $R$ be the rotation matrix so $Tr (R)=1+2 cos \epsilon~~,\epsilon$ is the angle  rotation. Now taking trace of equation $(\ref{e23})$ and simplifying one gets (see appendix E)
\begin{equation}
	cos \epsilon =\frac{(1+\gamma+\gamma_u+\gamma_v)^2}{(1+\gamma)(1+\gamma_u)(1+\gamma_v)}-1 \nonumber
\end{equation}

\underline{\textbf{Appendix A:}} 

To show $M \vec{A}= \gamma \vec{B}$
\vspace{-0.5cm}
\begin{eqnarray}
	M &=& \gamma_u \gamma_v \frac{\vec{v}.\vec{u}^T}{c^2}+(I+(\frac{\gamma_v^2}{1+\gamma_v})(\frac{\vec{v}.\vec{v}^T}{c^2}))(I+(\frac{\gamma_u^2}{1+\gamma_u})(\frac{\vec{u}.\vec{u}^T}{c^2})) \nonumber \\
	M \vec{A} &=& \frac{\gamma_u \gamma_v}{1+\frac{\vec{u}.\vec{v}}{c^2}} \{ 1+(\frac{\gamma_u}{1+\gamma_u})(\frac{\vec{u}.\vec{v}}{c^2})\}(\frac{\vec{v}. \vec{u}^T}{c^2})\vec{u}+\frac{\gamma_v}{(1+\frac{\vec{u}.\vec{v}}{c^2})}(\frac{\vec{v}.\vec{u}^T}{c^2})\vec{v} + \\
	\nonumber
	&& \{I+(\frac{\gamma_v^2}{1+\gamma_v})(\frac{\vec{v}\vec{v}^T}{c^2})+(\frac{\gamma_u^2}{1+\gamma_u})(\frac{\vec{u}.\vec{u}^T}{c^2})+\frac{\gamma_u^2 \gamma_v^2}{(1+\gamma_u)(1+\gamma_v)}(\frac{\vec{v}.\vec{v}^T}{c^2}) (\frac{\vec{u}.\vec{u}^T}{c^2})\} \times \\ \nonumber
	&& \frac{\left[ \{ 1+(\frac{\gamma_u}{1+\gamma_u})(\frac{\vec{u}.\vec{v}}{c^2})\}\vec{u}+\frac{\vec{v}}{\gamma_u} \right]}{(1+\frac{\vec{u}.\vec{v}}{c^2})} \nonumber \\
	&=& \frac{\gamma_u \gamma_v}{(1+\frac{\vec{u}.\vec{v}}{c^2})} \left\{ 1+(\frac{\gamma_u}{1+\gamma_u})(\frac{\vec{u}.\vec{v}}{c^2}) \right \}(\frac{u^2}{c^2})\vec{v}+\frac{\gamma_v}{(1+\frac{\vec{u}.\vec{v}}{c^2})}(\frac{\vec{u}.\vec{v}}{c^2})\vec{v} \nonumber \\
	&& +\frac{\left \{ \vec{u}+(\frac{\gamma_u}{1+\gamma_u})(\frac{\vec{u}.\vec{v}}{c^2})\vec{u}+\frac{\vec{v}}{\gamma_u} \right \}}{(1+\frac{\vec{u}.\vec{v}}{c^2})}+ \frac{(\frac{\gamma_v^2}{1+\gamma_v})\{1+\frac{\gamma_u}{1+\gamma_u}(\frac{\vec{u}.\vec{v}}{c^2}) \}(\frac{\vec{u}.\vec{v}}{c^2})\vec{v}}{(1+\frac{\vec{u}.\vec{v}}{c^2})} \nonumber \\
	&& +\frac{\gamma_v^2 (\frac{v^2}{c^2})\vec{v}}{\gamma_u(1+\gamma_u)(1+\frac{\vec{u}.\vec{v}}{c^2})}+(\frac{\gamma_u^2}{1+\gamma_u}) \{ 1+\frac{(\frac{\gamma_u}{1+\gamma_u})(\frac{\vec{u}.\vec{v}}{c^2})} {(1+\frac{\vec{u}.\vec{v}}{c^2})}\} (\frac{u^2}{c^2})\vec{u}+\frac{(\frac{\gamma_u}{1+\gamma_u})(\frac{\vec{u}.\vec{v}}{c^2})\vec{u}}{(1+\frac{\vec{u}.\vec{v}}{c^2})} \nonumber \\
	&& + \frac{\gamma_u^2 \gamma_v^2 \{ 1+\frac{\gamma_u}{1+\gamma_u}(\frac{\vec{u}\vec{v}}{c^2})\}\frac{u^2}{c^2}.(\frac{\vec{u}\vec{v}}{c^2})\vec{v}}{(1+\gamma_u)(1+\gamma_v)(1+\frac{\vec{u}.\vec{v}}{c^2})}+\frac{\gamma_u \gamma_v^2 (\frac{\vec{u}.\vec{v}}{c^2})}{(1+\gamma_u)(1+\gamma_v)(1+\frac{\vec{u}.\vec{v}}{c^2})} \nonumber \\
	&=&  \frac{\{ 1+(\frac{\gamma_u}{1+\gamma_u})(\frac{\vec{u}.\vec{v}}{c^2}) \}}{(1+\frac{\vec{u}.\vec{v}}{c^2})} \left[ \gamma_u \gamma_v (\frac{u^2}{c^2})\vec{v}+\vec{u}+(\frac{\gamma_v^2}{1+\gamma_v})(\frac{\vec{u}.\vec{v}}{c^2})\vec{v}+(\frac{\gamma_u^2}{1+\gamma_u})\frac{u^2}{c^2}\vec{u}+\frac{\gamma_u^2.\gamma_v^2}{(1+\gamma_u)(1+\gamma_v^2)}(\frac{u^2}{c^2})(\frac{\vec{u}.\vec{v}}{c^2})\vec{v} \right] \nonumber \\
	&& +\frac{\vec{v}}{(1+\frac{\vec{u}.\vec{v}}{c^2})} \left[ \gamma_v (\frac{\vec{u}.\vec{v}}{c^2})+\frac{1}{\gamma_u}+\frac{\gamma_v^2}{\gamma_u(1+\gamma_u)}(\frac{v^2}{c^2})+\frac{\gamma_u.\gamma_v^2}{(1+\gamma_u)(1+\gamma_v)}(\frac{\vec{u}.\vec{v}}{c^2})^2 \right]+\frac{(\frac{\gamma_u}{1+\gamma_u})(\frac{\vec{u}.\vec{v}}{c^2})\vec{u}}{(1+\frac{\vec{u}.\vec{v}}{c^2})} \nonumber  
\end{eqnarray}

\vspace{-10cm}

$\mbox{As,}~~ \gamma_u^2= \frac{1}{1-\frac{u^2}{c^2}}~ \mbox{so}~~\frac{u^2}{c^2}=\frac{\gamma_u^2-1}{\gamma_u^2}~ \mbox{i.e.}~~ \frac{\gamma_u^2}{1+\gamma_u}= \frac{\gamma_u-1}{(\frac{u^2}{c^2})}$
\begin{eqnarray}
	&=& \frac{\{1+(\frac{\gamma_u}{1+\gamma_u})(\frac{\vec{u}.\vec{v}}{c^2}) \}}{1+\frac{\vec{u}.\vec{v}}{c^2}}\left[ \gamma_u \gamma_v \frac{\gamma_u^2-1}{\gamma_u^2}\vec{v}+\vec{u}+(\frac{\gamma_v^2}{1+\gamma_v})(\frac{\vec{u}.\vec{v}}{c^2})\vec{v}+(\gamma_u-1)\vec{u}+\frac{\gamma_v^2(\gamma_u-1)}{(1+\gamma_v)}(\frac{\vec{u}.\vec{v}}{c^2})\vec{v} \right] \nonumber \\
	&& +\frac{\vec{v}}{(1+\frac{\vec{u}.\vec{v}}{c^2})}\left[ \gamma_v (\frac{\vec{u}.\vec{v}}{c^2})+\frac{1}{\gamma_v}+\frac{\gamma_v-1}{\gamma_v}+\frac{\gamma_u \gamma_v^2 (\frac{\vec{u}.\vec{v}}{c^2})^2}{(1+\gamma_u)(1+\gamma_v)} \right]+(\frac{\gamma_u}{1+\gamma_u})\frac{(\frac{\vec{u}.\vec{v}}{c^2})\vec{u}}{(1+\frac{\vec{u}.\vec{v}}{c^2})} \nonumber \\
	&=& \frac{\{ 1+(\frac{\gamma_u}{1+\gamma_u})\frac{\vec{u}.\vec{v}}{c^2} \}}{(1+\frac{\vec{u}.\vec{v}}{c^2})} \left[\gamma_u \gamma_v(\frac{u^2}{c^2})\vec{v}+\gamma_u \vec{u}+\frac{\gamma_u \gamma_v^2}{(1+\gamma_v)(\frac{\vec{u}.\vec{v}}{c^2})\vec{v}} \right] \nonumber \\
	&& +\frac{\vec{v}}{(1+\frac{\vec{u}.\vec{v}}{c^2})} \left[ \gamma_v(\frac{\vec{u}.\vec{v}}{c^2})\{ 1+\frac{\gamma_u \gamma_v (\frac{\vec{u}.\vec{v}}{c^2})}{(1+\gamma_u)(1+\gamma_v)} \} +\frac{\gamma_v}{\gamma_u} \right]+(\frac{\gamma_u}{1+\gamma_u})\frac{(\frac{\vec{u}.\vec{v}}{c^2})\vec{u}}{(1+\frac{\vec{u}.\vec{v}}{c^2})} \nonumber \\
	&=& \frac{\{ 1+(\frac{\gamma_u}{1+\gamma_u})(\frac{\vec{u}.\vec{v}}{c^2})\}}{(1+\frac{\vec{u}.\vec{v}}{c^2})} \left[ \gamma_u.\vec{u}+\gamma_u \gamma_v (\frac{u^2}{c^2})\vec{v}+\frac{\gamma_u \gamma_v^2}{(1+\gamma_v)}(\frac{\vec{u}.\vec{v}}{c^2})\vec{v} \right]+ \nonumber \\ 
	&&\frac{\vec{v}}{(1+\frac{\vec{u}.\vec{v}}{c^2})} \left[ \gamma_v(\frac{\vec{u}.\vec{v}}{c^2}) \{ 1+\frac{\gamma_u \gamma_v (\frac{\vec{u}.\vec{v}}{c^2})}{(1+\gamma_u)(1+\gamma_v)} \}+\frac{\gamma_v}{\gamma_u} \right]+(\frac{\gamma_u}{1+\gamma_u})\frac{(\frac{\vec{u}.\vec{v}}{c^2})\vec{u}}{(1+\frac{\vec{u}.\vec{v}}{c^2})} \nonumber \\
	&=& \frac{\left[ \gamma_u \vec{u} +(\frac{\gamma_u^2}{1+\gamma_u})(\frac{\vec{u}.\vec{v}}{c^2})\vec{u}+\gamma_u \gamma_v (\frac{\gamma_u^2-1}{\gamma_u^2})\vec{v}+(\frac{\gamma_u}{1+\gamma_u})^2 \gamma_v (\frac{u^2}{c^2})\vec{v}(\frac{\vec{u}.\vec{v}}{c^2})+(\frac{\gamma_u \gamma_v^2}{1+\gamma_v})(\frac{\vec{u}.\vec{v}}{c^2})\vec{v} \right]  }{\big( 1+\frac{\vec{u}.\vec{v}}{c^2} \big)} \nonumber \\
	&& + \frac{\left[ \frac{\gamma_u^2 \gamma_v^2}{(1+\gamma_u)(1+\gamma_v)}(\frac{\vec{u}.\vec{v}}{c^2})^2 \vec{v}+\gamma_v (\frac{\vec{u}.\vec{v}}{c^2})\vec{v}+\frac{\gamma_u \gamma_v^2 (\frac{\vec{u}.\vec{v}}{c^2})^2}{(1+\gamma_u)(1+\gamma_v)}\vec{v}+\frac{\gamma_v}{\gamma_u}\vec{v}+(\frac{\gamma_u}{1+\gamma_u})(\frac{\vec{u}.\vec{v}}{c^2})\vec{u} \right]}{\big( 1+\frac{\vec{u}.\vec{v}}{c^2} \big)}  \nonumber \\
	&=& \frac{\gamma_u}{\big( 1+\frac{\vec{u}.\vec{v}}{c^2} \big)}\vec{u}+\frac{ \{(\frac{\gamma_u^2}{1+\gamma_u})(\frac{\vec{u}.\vec{v}}{c^2})\vec{u}+(\frac{\gamma_u}{1+\gamma_u})(\frac{\vec{u}.\vec{v}}{c^2})\vec{u} \}}{\big( 1+\frac{\vec{u}.\vec{v}}{c^2} \big)}+\frac{\gamma_u \gamma_v (\gamma_u^2-1)\vec{v}}{\gamma_u^2 (1+\frac{\vec{u}.\vec{v}}{c^2})}+ \frac{(\gamma_u-1)\gamma_v (\frac{\vec{u}.\vec{v}}{c^2})\vec{v}}{\big( 1+\frac{\vec{u}.\vec{v}}{c^2} \big)} \nonumber \\
	&& + \frac{\vec{v}}{\big( 1+\frac{\vec{u}.\vec{v}}{c^2} \big)} \left[ (\frac{\gamma_u \gamma_v^2}{1+\gamma_v})(\frac{\vec{u}.\vec{v}}{c^2})+ \frac{\gamma_u^2 \gamma_v^2 (\frac{\vec{u}.\vec{v}}{c^2})^2}{(1+\gamma_u)(1+\gamma_v)}+\gamma_v (\frac{\vec{u}.\vec{v}}{c^2})+ \frac{\gamma_u \gamma_v^2 (\frac{\vec{u}.\vec{v}}{c^2})^2}{(1+\gamma_u)(1+\gamma_v)} \frac{\gamma_v}{\gamma_u} \right] \nonumber \\
	&=& \frac{\gamma_u \vec{u}}{(1+\frac{\vec{u}.\vec{v}}{c^2})}\{1+\frac{\vec{u}.\vec{v}}{c^2} \}+ \frac{\vec{v}}{(1+\frac{\vec{u}.\vec{v}}{c^2})} \left[ \frac{\gamma_v}{\gamma_u}(\gamma_u^2-1)+\gamma_v (\gamma_u-1)(\frac{\vec{u}.\vec{v}}{c^2})+(\frac{\gamma_u \gamma_v^2}{1+\gamma_v})(\frac{\vec{u}.\vec{v}}{c^2}) \right] \nonumber \\
	&& + \frac{\vec{v}}{(1+\frac{\vec{u}.\vec{v}}{c^2})}\left[\frac{\gamma_u^2 \gamma_v^2}{(1+\gamma_u)(1+\gamma_v)}(\frac{\vec{u}.\vec{v}}{c^2})^2 + \gamma_v (\frac{\vec{u}.\vec{v}}{c^2})+ \frac{\gamma_u \gamma_v^2}{(1+\gamma_u)(1+\gamma_v)}(\frac{\vec{u}.\vec{v}}{c^2})^2 + \frac{\gamma_v}{\gamma_u} \right] \nonumber \\
	&=& \gamma_u \vec{u}+\frac{\vec{v}}{(1+\frac{\vec{u}.\vec{v}}{c^2})} \left[ \gamma_u \gamma_v +\gamma_u \gamma_v(\frac{\vec{u}.\vec{v}}{c^2})+(\frac{\gamma_u \gamma_v}{1+\gamma_v})(\frac{\vec{u}.\vec{v}}{c^2})+\frac{\gamma_u \gamma_v^2}{1+\gamma_v}(\frac{\vec{u}.\vec{v}}{c^2}) \right] \nonumber \\ 
	&=& \gamma_u \vec{u} + \frac{\vec{v}}{(1+\frac{\vec{u}.\vec{v}}{c^2})} \left[ \gamma_u \gamma_v (1+\frac{\vec{u}.\vec{v}}{c^2}) + \frac{\gamma_u \gamma_v^2}{(1+\gamma_v)} (\frac{\vec{u}.\vec{v}}{c^2})(1+\frac{\vec{u}.\vec{v}}{c^2}) \right] \nonumber \\
	&=& \gamma_u \vec{u} + \gamma_u \gamma_v \vec{v} \left[ 1+(\frac{\gamma_v}{1+\gamma_v})(\frac{\vec{u}.\vec{v}}{c^2}) \right] \nonumber \\
	&=& \gamma_u \gamma_v \left[ \{ 1+(\frac{\gamma_v}{1+\gamma_v}) \frac{\vec{u}.\vec{v}}{c^2}\} \vec{v}+ \frac{1}{\gamma_v} \vec{u} \right] \nonumber \\
	&=& \gamma . \vec{B} \nonumber \\
	&& \mbox{i.e.}~~ M \vec{A}= \gamma \vec{B} \nonumber
\end{eqnarray}

Now,
\vspace{-.9cm}
\begin{eqnarray}
	C_{\alpha \beta} &=& M \{I+\frac{\gamma^2}{c^2}\frac{(\vec{A}.\vec{A}^T)}{(1+\gamma)} \}-\frac{\gamma^2}{c^2} BA^T \nonumber \\
	&=& M + \frac{\gamma^2}{(1+\gamma)c^2}(M \vec{A})A^T-\frac{\gamma^2}{c^2}BA^T \nonumber \\
	&=& M+ \frac{\gamma^3}{(1+\gamma)c^2}BA^T-\frac{\gamma^2}{c^2}BA^T \nonumber \\
	&=& M+\frac{\gamma^2}{c^2(1+\gamma)}BA^T \{ \gamma-(1+\gamma)\} \nonumber \\
	&=& M- \frac{ba^T}{(1+\gamma)} = R (\epsilon), ~~(b=\frac{\gamma}{c}B, a=\frac{\gamma}{c}A) \nonumber
\end{eqnarray}\\

\underline{\textbf{Appendix D:}} \\

To show : $\vec{a} \times \vec{b}= \frac{\gamma_u \gamma_v (\gamma^2-1)(1+\gamma+\gamma_u+\gamma_v)}{c^2(\gamma+1)(\gamma_u+1)(\gamma_v+1)}(\vec{u} \times \vec{v})$

\begin{eqnarray}
	\vec{a} &=& \frac{\gamma}{c}(\alpha_u \vec{u}+\frac{1}{\gamma_u}\vec{v})~, ~~ \vec{b}= \frac{\gamma}{c}(\alpha_v \vec{v}+\frac{1}{\gamma_v}\vec{u}) \nonumber \\
	\alpha_u &=& 1+(\frac{\gamma_u}{1+\gamma_u})(\frac{\vec{u}.\vec{v}}{c^2})~,~~\alpha_v= 1+(\frac{\gamma_v}{1+\gamma_v})(\frac{\vec{u}.\vec{v}}{c^2}) \nonumber \\
	\therefore \vec{a} \times \vec{b} &=& \frac{\gamma^2}{c^2} (\alpha_u \alpha_v-\frac{1}{\gamma_u \gamma_v})(\vec{u} \times \vec{v}) \nonumber
\end{eqnarray}
Now, 
\begin{eqnarray}
	\alpha_u \alpha_v-\frac{1}{\gamma_u \gamma_v}&=& \{1+(\frac{\gamma_u}{1+\gamma_u})(\frac{\vec{u}.\vec{v}}{c^2}) \} \{1+(\frac{\gamma_v}{1+\gamma_v})(\frac{\vec{u}.\vec{v}}{c^2}) \} -\frac{1}{\gamma_u \gamma_v} \nonumber \\
	&=&  1+\frac{\gamma_u}{1+\gamma_u}(\frac{\vec{u}.\vec{v}}{c^2})+\frac{\gamma_v}{1+\gamma_v} (\frac{\vec{u}.\vec{v}}{c^2})+\frac{\gamma_u}{1+\gamma_u}\frac{\gamma_v}{1+\gamma_v}(\frac{\vec{u}.\vec{v}}{c^2})^2  -\frac{1}{\gamma_u \gamma_v} \nonumber \\
	&=& \frac{(1+\gamma_u)(1+\gamma_v)+ \gamma_u (1+\gamma_v)(\frac{\vec{u}.\vec{v}}{c^2})+\gamma_v (1+\gamma_u)(\frac{\vec{u}.\vec{v}}{c^2})+\gamma_u \gamma_v(\frac{\vec{u}.\vec{v}}{c^2})^2 }{(1+\gamma_u)(1+\gamma_v)} -\frac{1}{\gamma_u \gamma_v} \nonumber \\
	&=&\frac{ \gamma_u \gamma_v(1+\gamma_u)(1+\gamma_v)+ \gamma_u^2 \gamma_v (1+\gamma_v)(\frac{\vec{u}.\vec{v}}{c^2})+ \gamma_u \gamma_v^2 (1+\gamma_u)(\frac{\vec{u}.\vec{v}}{c^2})+\gamma_u^2 \gamma_v^2 (\frac{\vec{u}.\vec{v}}{c^2})^2 }{\gamma_u \gamma_v (1+\gamma_u)(1+\gamma_v)}-(1+\gamma_u) (1+ \gamma_v) \nonumber \\ 
	&=& \frac{\gamma_u \gamma_v (1+\gamma_v)+ \gamma_u^2 \gamma_v (1+\gamma_v)(1+\frac{\vec{u}.\vec{v}}{c^2})+\gamma_u \gamma_v^2 (\frac{\vec{u}.\vec{v}}{c^2})+\gamma_u^2 \gamma_v^2 (\frac{\vec{u}.\vec{v}}{c^2})(1+\frac{\vec{u}.\vec{v}}{c^2})-(1+\gamma_u)(1+\gamma_v) }{\gamma_u \gamma_v (1+\gamma_u)(1+\gamma_v)} \nonumber \\
	&=& \frac{\gamma_u \gamma_v (1+\gamma_v) +\gamma_u (1+\gamma_v) \gamma +\gamma_u \gamma_v^2 (\frac{\vec{u}.\vec{v}}{c^2})+\gamma_u \gamma_v (\frac{\vec{u}.\vec{v}}{c^2}) \gamma -(1+\gamma_u)(1+\gamma_v) }{\gamma_u \gamma_v (1+\gamma_u)(1+\gamma_v)} \nonumber \\
	&=& \frac{\gamma_u \gamma_v +\gamma_u \gamma_v^2 (1+\frac{\vec{u}.\vec{v}}{c^2})+\gamma \gamma_u + \gamma \gamma_u \gamma_v (1+\frac{\vec{u}.\vec{v}}{c^2})-(1+\gamma_u)(1+\gamma_v) }{\gamma_u \gamma_v (1+\gamma_u)(1+\gamma_v)} \nonumber \\
	&=& \frac{\gamma_u \gamma_v + \gamma \gamma_v +\gamma \gamma_u +\gamma^2 -1 - \gamma_u -\gamma_v -\gamma_u \gamma_v}{\gamma_u \gamma_v (1+\gamma_u)(1+\gamma_v)}  \nonumber \\
	&=& \frac{\{\gamma (\gamma_u +\gamma_v)-(\gamma_u+ \gamma_v)+\gamma^2-1 \}}{\gamma_u \gamma_v (1+\gamma_u)(1+\gamma_v)} \nonumber \\
	&=& \frac{(\gamma-1)(\gamma_u+\gamma_v+\gamma+1)}{\gamma_u \gamma_v (1+\gamma_u)(1+\gamma_v)} \nonumber 
\end{eqnarray}

\begin{eqnarray}
	\therefore ~ \vec{a}\times \vec{b} &=& \frac{\gamma^2}{c^2(1+\frac{\vec{u}.\vec{v}}{c^2})}(\alpha_u \alpha_v-\beta_u \beta_v)(\vec{u}\times \vec{v}) \nonumber \\
	&=&\frac{\gamma_u^2 \gamma_v^2 (1+\frac{\vec{u}.\vec{v}}{c^2})^2}{c^2(1+\frac{\vec{u}.\vec{v}}{c^2})^2}. \frac{(\gamma-1)(\gamma_u+\gamma_v+\gamma+1)}{\gamma_u \gamma_v (1+\gamma_u)(1+\gamma_v)} (\vec{u}\times \vec{v}) \nonumber \\
	&=&  \frac{ \gamma_u \gamma_v (\gamma^2-1)(\gamma_u+\gamma_v+\gamma+1)}{c^2(1+\gamma) (1+\gamma_u)(1+\gamma_v)}(\vec{u}\times \vec{v}) \nonumber
\end{eqnarray} \\

\underline{\textbf{Appendix B:}}\\

To prove $\vec{b}= R \vec{a}:$

\begin{eqnarray}
	R &=& M- \frac{1}{(\gamma-1)} \vec{b} \vec{a}^T \nonumber \\
	\mbox{Now ,}~~ \vec{a} &=& \frac{\gamma}{c(1+\frac{\vec{u}.\vec{v}}{c^2})} (\alpha_u \vec{u}+ \frac{1}{\gamma_u} \vec{v})= \gamma_0 (\alpha_u \vec{u}+ \frac{1}{\gamma_u} \vec{v}) \nonumber
\end{eqnarray}
where $ \gamma_0 = \frac{\gamma}{c(1+\frac{\vec{u}.\vec{v}}{c^2})}$

\begin{eqnarray}
	M \vec{a} = \gamma_u \gamma_v \frac{\vec{v}\vec{u}^T}{c^2} \vec{a} + \{I+ (\frac{\gamma_u^2}{1+\gamma_u})\frac{\vec{u}.\vec{u}^T}{c^2}+(\frac{\gamma_v^2}{1+\gamma_v})\frac{\vec{v}\vec{v}^T}{c^2} \nonumber \\
	+ \frac{\gamma_u^2 \gamma_v^2}{(1+\gamma_u)(1+\gamma_v)}(\frac{\vec{u}.\vec{v}}{c^2}).(\frac{\vec{v}.\vec{u}^T}{c^2}) \}\vec{a} \nonumber
\end{eqnarray}

\begin{eqnarray}
	\mbox{1st term} &=& \gamma_u \gamma_v \frac{\vec{v}
		\vec{u}^T}{c^2} \vec{a}= \gamma_u \gamma_v \gamma_0 (\frac{\vec{v}\vec{u}^T}{c^2})(\alpha_u \vec{u}+\frac{1}{\gamma_u}\vec{v}) \nonumber \\
	&=& \gamma_0 \gamma_u \gamma_v \left[ \alpha_u \frac{u^2}{c^2}+ \frac{1}{\gamma_u}(\frac{\vec{u}\vec{v}}{c^2}) \right] \vec{v} \nonumber \\
	&=& \gamma_0 \gamma_u \gamma_v \left[ \{ 1+(\frac{\gamma_u}{1+\gamma_u})(\frac{\vec{u}.\vec{v}}{c^2}) \} \frac{\gamma_u^2-1}{\gamma_u^2} + \frac{1}{\gamma_u}(\frac{\vec{u}.\vec{v}}{c^2}) \right] \vec{v} \nonumber \\
	&=& \gamma_0 \gamma_u \gamma_v \left[\frac{\gamma_u^2-1}{\gamma_u^2}+\frac{\gamma_u-1}{\gamma_u}(\frac{\vec{u}.\vec{v}}{c^2})+\frac{1}{\gamma_u}(\frac{\vec{u}.\vec{v}}{c^2})  \right] \vec{v} \nonumber \\
	&=& \gamma_0 \gamma_u \gamma_v \left[ \frac{\gamma_u^2-1}{\gamma_u^2}+(\frac{\vec{u}.\vec{v}}{c^2}) \right] \vec{v} ~~~~~~~~~~~~~~~~~~~~~~ .~.~.~.~.~(I) \nonumber 
\end{eqnarray}

\begin{eqnarray}
	\mbox{2nd term}  &=& (\frac{\gamma_u^2}{1+\gamma_u})(\frac{\vec{u}.\vec{u}^T}{c^2}) \vec{a} \nonumber \\
	&=& (\frac{\gamma_u^2}{1+\gamma_u})(\frac{\vec{u}.\vec{u}^T}{c^2}) \gamma_0 (\alpha_u \vec{u}+\frac{1}{\gamma_u}\vec{v}) \nonumber \\
	&=& \gamma_0 (\frac{\gamma_u^2}{1+\gamma_u}) \{ \alpha_u (\frac{u^2}{c^2}) \vec{u}+\frac{1}{\gamma_u}(\frac{\vec{u}.\vec{v}}{c^2})\vec{u} \} \nonumber \\
	&=& \gamma_0 (\frac{\gamma_u^2}{1+\gamma_u}) \left[ \{ 1+(\frac{\gamma_u}{1+\gamma_u})(\frac{\vec{u}.\vec{v}}{c^2}) \} (\frac{\gamma_u^2-1}{\gamma_u^2})+\frac{1}{\gamma_u}(\frac{\vec{u}.\vec{v}}{c^2}) \right] \vec{u} \nonumber \\
	&=& \frac{\gamma_0 \gamma_u^2}{(1+\gamma_u)} \left[ \frac{\gamma_u^2-1}{\gamma_u^2}+(\frac{\gamma_u-1}{\gamma_u})(\frac{\vec{u}.\vec{v}}{c^2})+\frac{1}{\gamma_u}(\frac{\vec{u}.\vec{v}}{c^2}) \right] \vec{u} \nonumber \\
	&=& \gamma_0 \left[ (\gamma_u-1) + (\frac{\gamma_u}{1+\gamma_u})(\frac{\vec{u}.\vec{v}}{c^2}) \right] \vec{u}~~~~~~~~~~~~~~~~~~~~~~ .~.~.~.~.~(II)  \nonumber
\end{eqnarray}

\begin{eqnarray}
	\mbox{3rd term} &=& (\frac{\gamma_v^2}{1+\gamma_v})(\frac{\vec{v}\vec{v}^T}{c^2}) \vec{a} \nonumber \\
	&=& (\frac{\gamma_v^2}{1+\gamma_v})(\frac{\vec{v}\vec{v}^T}{c^2}) \gamma_0 (\alpha_u \vec{u}+ \frac{1}{\gamma_u} \vec{v}) \nonumber \\
	&=& \gamma_0 (\frac{\gamma_v^2}{1+\gamma_v}) \left[ \alpha_u (\frac{\vec{u}.\vec{v}}{c^2})+ \frac{1}{\gamma_u}(\frac{v^2}{c^2}) \right] \vec{v} \nonumber \\ &=& \gamma_0 \frac{\gamma_v^2}{(1+\gamma_v)} \left[ \{1+(\frac{\gamma_u}{1+\gamma_u})(\frac{\vec{u}.\vec{v}}{c^2}) \}(\frac{\vec{u}.\vec{v}}{c^2})+ \frac{1}{\gamma_u} \frac{\gamma_v^2-1}{\gamma_v} \right] \vec{v} ~~~~~~~~~~~~~~~~~~~~~~ .~.~.~.~.~(III) \nonumber
\end{eqnarray}

\begin{eqnarray}
	\mbox{4th term} &=& \frac{\gamma_u^2 \gamma_v^2}{(1+\gamma_u)(1+\gamma_v)}(\frac{\vec{u}.\vec{v}}{c^2})(\frac{\vec{v}.\vec{u}^T}{c^2})\vec{a} \nonumber \\
	&=& \frac{ \gamma_0 \gamma_u^2 \gamma_v^2}{(1+\gamma_u)(1+\gamma_v)}(\frac{\vec{u}.\vec{v}}{c^2}) \{\alpha_u (\frac{u^2}{c^2})+\frac{1}{\gamma_u}(\frac{\vec{u}.\vec{v}}{c^2})\} \vec{v} \nonumber \\
	&=&  \frac{ \gamma_0 \gamma_u^2 \gamma_v^2}{(1+\gamma_u)(1+\gamma_v)}(\frac{\vec{u}.\vec{v}}{c^2}) \left[ \{1+(\frac{\gamma_u}{1+\gamma_u})(\frac{\vec{u}.\vec{v}}{c^2})\} \frac{\gamma_u^2-1}{\gamma_u^2}+ \frac{1}{\gamma_u}(\frac{\vec{u}.\vec{v}}{c^2}) \right]\vec{v} \nonumber \\
	&=& \frac{ \gamma_0 \gamma_u^2 \gamma_v^2}{(1+\gamma_u)(1+\gamma_v)}(\frac{\vec{u}.\vec{v}}{c^2}) \left[ \frac{\gamma_u^2-1}{\gamma_u^2}+\frac{\gamma_u-1}{\gamma_u}(\frac{\vec{u}.\vec{v}}{c^2})+\frac{1}{\gamma_u}(\frac{\vec{u}.\vec{v}}{c^2}) \right] \vec{v} \nonumber \\
	&=& \gamma_0 \left[ \frac{(\gamma_u-1)\gamma_v^2}{1+\gamma_v}+\frac{\gamma_u^2 \gamma_v^2}{(1+\gamma_u)(1+\gamma_v)}(\frac{\vec{u}.\vec{v}}{c^2}) \right] (\frac{\vec{u}.\vec{v}}{c^2})\vec{v} ~~~~~~~~~~~~~~~~~~~~~~ .~.~.~.~.~(IV)  \nonumber
\end{eqnarray}
Thus combining $(I)+ (II)+ (III)+(IV)$ we have

\begin{eqnarray}
	M \vec{a} &=& \gamma_0 \gamma_u \gamma_v \left[ \frac{\gamma_u^2-1}{\gamma_u^2}+(\frac{\vec{u}.\vec{v}}{c^2}) \right] \vec{v} +\gamma_0 \left[ \{1+(\frac{\gamma_u}{1+\gamma_u})(\frac{\vec{u}.\vec{v}}{c^2}) \}\vec{u}+\frac{1}{\gamma_u}\vec{v} \right] \nonumber \\
	&& + \gamma_0 \left[ (\gamma_u-1)+(\frac{\gamma_u^2}{1+\gamma_u})(\frac{\vec{u}.\vec{v}}{c^2}) \right]\vec{u}+\gamma_0 \left[ (\frac{\gamma_v^2}{1+\gamma_v}) \{ 1+(\frac{\gamma_u}{1+\gamma_u})(\frac{\vec{u}.\vec{v}}{c^2}) \}+\frac{\gamma_v-1}{\gamma_u} \right]\vec{v} \nonumber \\
	&& + \gamma_0 \left[ \frac{(\gamma_u-1)\gamma_v^2}{1+\gamma_v}+\frac{\gamma_u^2 \gamma_v^2}{(1+\gamma_u)(1+\gamma_v)}(\frac{\vec{u}.\vec{v}}{c^2}) \right] (\frac{\vec{u}.\vec{v}}{c^2}) \vec{v} \nonumber \\
	&=& \gamma_0 \left[ \frac{\gamma_v (\gamma_u^2-1)}{\gamma_u}+ \gamma_u \gamma_v (\frac{\vec{u}.\vec{v}}{c^2})+ \frac{1}{\gamma_u} +(\frac{\gamma_v^2}{1+\gamma_v})(\frac{\vec{u}.\vec{v}}{c^2})+ \frac{\gamma_u \gamma_v^2}{(1+\gamma_u)(1+\gamma_v)}(\frac{\vec{u}.\vec{v}}{c^2})^2 \right] \nonumber \\
	&& + \gamma_0 \left[ \frac{\gamma_v-1}{\gamma_u}+\frac{(\gamma_u-1)\gamma_v^2}{1+\gamma_v}(\frac{\vec{u}.\vec{v}}{c^2})+\frac{\gamma_u^2 \gamma_v^2}{(1+\gamma_u)(1+\gamma_v)}(\frac{\vec{u}.\vec{v}}{c^2})^2 \right] \vec{v} \nonumber \\
	&&+\gamma_0 \left[ \{1+(\frac{\gamma_u}{1+\gamma_u})(\frac{\vec{u}.\vec{v}}{c^2}) \}+(\gamma_u-1)+\frac{\gamma_u^2}{1+\gamma_u}(\frac{\vec{u}.\vec{v}}{c^2}) \right] \vec{u} \nonumber \\
	&=& \gamma_0 \left[ \gamma_u \gamma_v +(\gamma_u \gamma_v + \frac{\gamma_u \gamma_v^2}{1+\gamma_v})(\frac{\vec{u}.\vec{v}}{c^2}) +\frac{\gamma_u \gamma_v^2}{1+\gamma_v}(\frac{\vec{u}.\vec{v}}{c^2})^2 \right] \vec{v}+\gamma_0 \left[ \gamma_u \{ 1+\frac{\vec{u}.\vec{v}}{c^2} \} \right]\vec{u} \nonumber \\
	&=& \gamma_0 \left[ \gamma_u \gamma_v (1+ \frac{\vec{u}.\vec{v}}{c^2}) + (\frac{\gamma_u \gamma_v^2}{1+\gamma_v})(\frac{\vec{u}.\vec{v}}{c^2}) (1+ \frac{\vec{u}.\vec{v}}{c^2})\right] \vec{v} +\gamma_0 \left[\frac{1}{\gamma_v}\gamma_u \gamma_v \{ 1+\frac{\vec{u}.\vec{v}}{c^2} \} \right]\vec{u} \nonumber \\
	&=& \gamma_0 \left[ \gamma+ \frac{\gamma_v}{1+\gamma_v}(\frac{\vec{u}.\vec{v}}{c^2})\gamma \right]\vec{v} + \gamma_0 \frac{\gamma}{\gamma_v}\vec{u} \nonumber \\
	&=& \gamma_0 \left[ \{1+(\frac{\gamma_v}{1+\gamma_v})(\frac{\vec{u}.\vec{v}}{c^2}) \}\vec{v} +\frac{1}{\gamma_v} \vec{u} \right]\gamma \nonumber \\
	&=& \gamma ~ \frac{\gamma}{c(1+\frac{\vec{u}.\vec{v}}{c^2})}\left[ \{1+(\frac{\gamma_v}{1+\gamma_v})(\frac{\vec{u}.\vec{v}}{c^2}) \}\vec{v}+\frac{1}{\gamma_v} \vec{u} \right] \nonumber \\
	&=& \gamma . \vec{b} \nonumber
\end{eqnarray}

Now,
\begin{eqnarray}
	\frac{1}{(\gamma+1)}\vec{b} \vec{a}^T. \vec{a} &=& \frac{1}{(\gamma+1)}\vec{b} |\vec{a}|^2 = \frac{\gamma^2}{(\gamma+1)c^2}|\vec{u} \oplus \vec{v}|^2 \vec{b} \nonumber \\
	&=& (\gamma-1) \vec{b} \nonumber \\
	\therefore ~ R \vec{a} &=& M \vec{a}- \frac{1}{(\gamma+1)}\vec{b} \vec{a}^T. \vec{a} \nonumber \\
	&=& \gamma \vec{b}- (\gamma-1) \vec{b} = \vec{b} \nonumber \\
	\mbox{i.e.}~~ \vec{b}&=& R \vec{a}. \nonumber
\end{eqnarray} \\

\underline{\textbf{Appendix-C:}} \\

To show $R^T \vec{b}= \vec{a}$ 
\begin{eqnarray}
	R^T &=& M^T- \frac{1}{(\gamma+1)} \vec{a} \vec{b}^T \nonumber \\
	M^T &=& \gamma_u \gamma_v (\frac{\vec{u}\vec{v}^T}{c^2}) +I+ (\frac{\gamma_v^2}{1+\gamma_v})\frac{\vec{v}\vec{v}^T}{c^2}+\frac{\gamma_u^2}{1+\gamma}(\frac{\vec{u}.\vec{u}^T}{c^2}) +\frac{\gamma_u^2 \gamma_v^2} {(1+\gamma_u)(1+\gamma_v)}(\frac{\vec{u}.\vec{v}}{c^2})(\frac{\vec{u}.\vec{v}^T}{c^2}) \nonumber \\
	\therefore ~ M^Tb &=& M^T \gamma_0 (\alpha_v \vec{v}+\frac{1}{\gamma_v} \vec{u})~,~ \gamma_0= \frac{\gamma}{c(1+\frac{\vec{u}.\vec{v}}{c^2})},~ \alpha_v= 1+(\frac{\gamma_v}{1+\gamma_v})(\frac{\vec{u}.\vec{v}}{c^2}) \nonumber 
\end{eqnarray}

\begin{eqnarray}
	\mbox{1st term} &:& \gamma_u \gamma_v (\frac{\vec{u}.\vec{v}^T}{c^2}) \gamma_0 (\alpha_v \vec{v}+\frac{1}{\gamma_v} \vec{u}) \nonumber \\
	&=& \gamma_0 \gamma_u \gamma_v \left[ \alpha_v (\frac{v^2}{c^2})+ \frac{1}{\gamma_v}(\frac{\vec{u}.\vec{v}}{c^2}) \right] \vec{u} \nonumber \\
	&=&\gamma_0 \gamma_u \gamma_v \left[\{1+(\frac{\gamma_v}{1+\gamma_v})(\frac{\vec{u}.\vec{v}}{c^2}) \}\frac{\gamma_v^2-1}{\gamma_v^2}+\frac{1}{\gamma_v}(\frac{\vec{u}.\vec{v}}{c^2}) \right]\vec{u} \nonumber \\
	&=&\gamma_0 \gamma_u \gamma_v \left[ \frac{\gamma_v^2-1}{\gamma_v^2}+\frac{\gamma_v-1}{\gamma_v}(\frac{\vec{u}.\vec{v}}{c^2})+\frac{1}{\gamma_v}\frac{\vec{u}.\vec{v}}{c^2} \right]\vec{v} \nonumber \\
	&=& \gamma_0 \gamma_u \gamma_v \left[ \frac{\gamma_v^2-1}{\gamma_v^2}+ (\frac{\vec{u}.\vec{v}}{c^2}) \right]= \gamma_0 (\gamma-\frac{\gamma_u}{\gamma_v})\vec{u} \nonumber
\end{eqnarray} 

\begin{eqnarray}
	\mbox{2nd term} &:& (\frac{\gamma_v^2}{1+\gamma_v})(\frac{\vec{v}\vec{v}^T}{c^2})\vec{b} =(\frac{\gamma_0 \gamma_v^2}{1+\gamma_v})\left[ \alpha_v (\frac{v^2}{c^2})\vec{v}+\frac{1}{\gamma_v}(\frac{\vec{u}.\vec{v}}{c^2})\vec{v} \right] \nonumber \\
	&=& (\frac{\gamma_0 \gamma_v^2}{1+\gamma_v}) \left[ \{1+\frac{\gamma_v}{1+\gamma_v}(\frac{\vec{u}.\vec{v}}{c^2}) \} \frac{\gamma_v^2-1}{\gamma_v^2}+\frac{1}{\gamma_v}\frac{\vec{u}.\vec{v}}{c^2} \right] \vec{v}\nonumber \\
	&=& (\frac{\gamma_0 \gamma_v^2}{1+\gamma_v}) \left[ \frac{\gamma_v^2-1}{\gamma_v^2} +\frac{\vec{u}.\vec{v}}{c^2} \right]\vec{v}= \gamma_0 \left[ (\gamma_v-1)+ \frac{\gamma_v^2}{1+\gamma_v}(\frac{\vec{u}.\vec{v}}{c^2}) \right] \vec{v} \nonumber 
\end{eqnarray}

\begin{eqnarray}
	\mbox{3rd term} &:& (\frac{\gamma_u^2}{1+\gamma_u})(\frac{\vec{u}.\vec{u}^T}{c^2})\vec{b} = (\frac{\gamma_0 \gamma_u^2}{1+\gamma_u}) \left[ \alpha_v (\frac{\vec{u}.\vec{v}}{c^2})\vec{u} +\frac{1}{\gamma_v} (\frac{u^2}{c^2})\vec{u} \right] \nonumber \\
	&=& (\frac{\gamma_0 \gamma_u^2}{1+\gamma_u}) \left[ \{ 1+(\frac{\gamma_v}{1+\gamma_v})(\frac{\vec{u}.\vec{v}}{c^2}) \} (\frac{\vec{u}.\vec{v}}{c^2})+\frac{1}{\gamma_v} \frac{\gamma_u^2-1}{\gamma_u^2} \right]\vec{u} \nonumber \\
	&=& \gamma_0 \left[ \{ 1+(\frac{\gamma_v}{1+\gamma_v})(\frac{\vec{u}.\vec{v}}{c^2}) \} \frac{\gamma_u^2}{1+\gamma_u}(\frac{\vec{u}.\vec{v}}{c^2}) + \frac{\gamma_u-1}{\gamma_v} \right] \vec{u} \nonumber
\end{eqnarray}
\vspace{-1cm}

\begin{eqnarray}
	\mbox{4th term} &:& \frac{\gamma_u^2 \gamma_v^2}{(1+\gamma_u)(1+\gamma_v)} (\frac{\vec{u}.\vec{v}}{c^2}) \gamma_0 \left[ \alpha_v (\frac{v^2}{c^2})\vec{u}+ \frac{1}{\gamma_v}(\frac{\vec{u}.\vec{v}}{c^2}) \vec{u}\right] \nonumber \\
	&=&  \frac{\gamma_0 \gamma_u^2 \gamma_v^2}{(1+\gamma_u)(1+\gamma_v)} (\frac{\vec{u}.\vec{v}}{c^2}) \left[ \{ 1+(\frac{\gamma_v}{1+\gamma_v})(\frac{\vec{u}.\vec{v}}{c^2})  \} \frac{\gamma_v^2-1}{\gamma_v^2}+\frac{1}{\gamma_v} (\frac{\vec{u}.\vec{v}}{c^2}) \right]\vec{u} \nonumber \\
	&=&  \frac{\gamma_0 \gamma_u^2 \gamma_v^2}{(1+\gamma_u)(1+\gamma_v)} (\frac{\vec{u}.\vec{v}}{c^2}) \left[\frac{\gamma_v^2-1}{\gamma_v^2}+\frac{\vec{u}.\vec{v}}{c^2} \right]\vec{u} \nonumber
\end{eqnarray}

\begin{eqnarray}
	\frac{1}{\gamma+1} \vec{a}\vec{b}^T. \vec{b} &=& \frac{\vec{a}|\vec{b}|^2}{\gamma+1}= \frac{(\frac{\gamma^2}{c^2})}{(\gamma+1)}|\vec{v}\oplus \vec{u}|^2 \vec{a}=(\gamma-1)\vec{a}  \nonumber \\
	\therefore~ M^T \vec{b} &=& \gamma_0 (\gamma-\frac{\gamma_u}{\gamma_v})\vec{u} +\vec{b} +\gamma_0 \left[ (\gamma_v-1)+\frac{\gamma_v^2}{(1+\gamma_v)}(\frac{\vec{u}.\vec{v}}{c^2}) \right]\vec{v} \nonumber \\
	&&+ \gamma_0 \left[ \{ 1+(\frac{\gamma_v}{1+\gamma_v}) (\frac{\vec{u}.\vec{v}}{c^2})\}(\frac{\vec{u}.\vec{v}}{c^2})\frac{\gamma_u^2}{(1+\gamma_u)}+\frac{\gamma_u-1}{\gamma_u} \right]\vec{u } \nonumber \\
	&& +\frac{\gamma_0 \gamma_u^2 \gamma_v^2}{(1+\gamma_u)(1+\gamma_v)}(\frac{\vec{u}.\vec{v}}{c^2})\left[ \frac{\gamma_v^2-1}{\gamma_v^2}+\frac{\vec{u}.\vec{v}}{c^2} \right]\vec{u} \nonumber \\
	&=& \gamma_0 (\gamma-\frac{\gamma_u}{\gamma_v})\vec{u}+\gamma_0 \{ 1+(\frac{\gamma_v}{1+\gamma_v}) (\frac{\vec{u}.\vec{v}}{c^2}) \}\vec{v} +\frac{\gamma_0}{\gamma_v}\vec{u} +\gamma_0 \left[(\gamma_v-1)+(\frac{\gamma_v^2}{1+\gamma_v})(\frac{\vec{u}.\vec{v}}{c^2}) \right]\vec{v} \nonumber \\
	&&+\gamma_0 \left[ \{ 1+(\frac{\gamma_v}{1+\gamma_v})(\frac{\vec{u}.\vec{v}}{c^2}) \}(\frac{\gamma_u^2}{1+\gamma_u})(\frac{\vec{u}.\vec{v}}{c^2})+\frac{\gamma_u-1}{\gamma_v} \right]\vec{u} \nonumber \\
	&& +\frac{\gamma_0 \gamma_u^2 \gamma_v^2}{(1+\gamma_u)(1+\gamma_v)}(\frac{\vec{u}.\vec{v}}{c^2}) \left[ \frac{\gamma_v^2-1}{\gamma_v^2}+\frac{\vec{u}.\vec{v}}{c^2}\right] \vec{u} \nonumber \\
	&=& \gamma_0 \left[ \{\gamma_v +\gamma_v (\frac{\vec{u}.\vec{v}}{c^2}) \}\vec{v} + \{ \gamma- \frac{\gamma_u-1}{\gamma_v}+\frac{\gamma_u-1}{\gamma_v}+\frac{\gamma_u^2}{1+\gamma_u}\frac{\vec{u}.\vec{v}}{c^2} \}\vec{u} \right] \nonumber \\
	&& \gamma_0 \left[ \{ \frac{\gamma_u^2 \gamma_v}{(1+\gamma_u)(1+\gamma_v)}(\frac{\vec{u}.\vec{v}}{c^2})^2 + \frac{\gamma_u^2(\gamma_v-1)}{(1+\gamma_u)}(\frac{\vec{u}.\vec{v}}{c^2})+\frac{\gamma_u^2 \gamma_v^2}{(1+\gamma_u)(1+\gamma_v)}(\frac{\vec{u}.\vec{v}}{c^2})^2 \}\vec{u} \right] \nonumber \\
	&=& \gamma_0 \left[ \gamma_v \{ 1+\frac{\vec{u}.\vec{v}}{c^2}\}\vec{v}+ \{\gamma+\frac{\gamma_u^2 \gamma_v}{1+\gamma_u} \frac{\vec{u}.\vec{v}}{c^2}+ \frac{\gamma_u^2 \gamma_v}{1+\gamma_u}(\frac{\vec{u}.\vec{v}}{c^2})^2 \}\vec{u} \right] \nonumber \\
	&=& \gamma_0 \left[ \frac{\gamma}{\gamma_u}\vec{v}+ \{ \gamma +\frac{\gamma_u^2 \gamma_v}{1+\gamma_u}(\frac{\vec{u}.\vec{v}}{c^2}) \{ 1+\frac{\vec{u}.\vec{v}}{c^2}\} \}\vec{u} \right] \nonumber \\
	&=& \gamma_0 \left[ \frac{\gamma}{\gamma_u}\vec{v}+\{\gamma+\gamma\frac{\gamma_u}{1+\gamma_u}(\frac{\vec{u}.\vec{v}}{c^2}) \}\vec{u} \right] \nonumber \\
	&=& \gamma \gamma_0 \left[ \frac{1}{\gamma_u}\vec{v} + \{1+\frac{\gamma_u}{1+\gamma_u} (\frac{\vec{u}.\vec{v}}{c^2})\}\vec{u} \right] \nonumber \\
	&=& \gamma \vec{a} \nonumber \\
	& \therefore & R^T \vec{b}= \gamma \vec{a}-(\gamma-1)\vec{a} =\vec{a}. \nonumber
\end{eqnarray} \\

\underline{\textbf{Appendix-E:}}

To prove $cos \epsilon = \frac{(1+\gamma+\gamma_u+\gamma_v)^2}{(1+\gamma)(1+\gamma_u)(1+\gamma_v)}-1$ \\

As $R$ is an orthogonal matrix and represents the rotation matrix so $\mbox{Tr}(R)=1+2 \cos \epsilon$. But

\begin{eqnarray}
	R&=& M-\frac{1}{\gamma+1}b a^T \nonumber \\
	\therefore ~ \mbox{Tr}R &=& Tr M - Tr \{ \frac{ba^T}{\gamma+1}\} \nonumber
\end{eqnarray}
\begin{eqnarray}
	M &=& \gamma_u \gamma_v (\frac{\vec{v}.\vec{u}^T}{c^2})+I +(\frac{\gamma_u^2}{1+\gamma_u})(\frac{\vec{u}.\vec{u}^T}{c^2})+(\frac{\gamma_v^2}{1+\gamma_v})(\frac{\vec{v}.\vec{v}^T}{c^2})+\frac{\gamma_u^2 \gamma_v^2}{(1+\gamma_u)(1+\gamma_v)}(\frac{\vec{v}.\vec{v}^T}{c^2})(\frac{\vec{u}.\vec{u}^T}{c^2})\nonumber \\
	\therefore~ TrM&=& \gamma_u \gamma_v \frac{\vec{u}\vec{v}}{c^2}+3+ (\frac{\gamma_u^2}{1+\gamma_u})(\frac{u^2}{c^2})+\frac{\gamma_v^2}{1+\gamma_v} (\frac{v^2}{c^2})+\frac{\gamma_u^2 \gamma_v^2}{(1+\gamma_u)(1+\gamma_v)}(\frac{\vec{u}.\vec{v}}{c^2})^2 \nonumber \\
	&=& \gamma_u \gamma_v (\frac{\vec{u}.\vec{v}}{c^2}) \{ 1+\frac{\gamma_u \gamma_v}{(1+\gamma_u)(1+\gamma_v)}(\frac{\vec{u}.\vec{v}}{c^2}) \}+3+\frac{\gamma_u^2}{1+\gamma_u}.\frac{\gamma_u^2-1}{\gamma_u^2}+\frac{\gamma_v^2}{1+\gamma_v}.\frac{\gamma_v^2-1}{\gamma_v^2} \nonumber \\
	&=& \gamma_u \gamma_v (\frac{\vec{u}.\vec{v}}{c^2}) \{ 1+\frac{(\gamma-\gamma_u \gamma_v)}{(1+\gamma_u)(1+\gamma_v)} \}+3+\gamma_u-1+\gamma_v-1 \nonumber \\
	&=& (\gamma-\gamma_u \gamma_v) \frac{(1+\gamma+\gamma_u +\gamma_v)}{(1+\gamma_u)(1+\gamma_v)}+(1+\gamma+\gamma_u+\gamma_v)-\gamma \nonumber \\
	&=& (1+\gamma_u+ \gamma_v+\gamma) \{ \frac{(\gamma-\gamma_u \gamma_v)}{(1+\gamma_u)(1+\gamma_v)}+1 \} -\gamma \nonumber \\
	&=&(1+\gamma+\gamma_u+ \gamma_v) \frac{(1+\gamma+\gamma_u+ \gamma_v)}{(1+\gamma_u)(1+\gamma_v)}-\gamma \nonumber \\
	&=& \frac{(1+\gamma+\gamma_u+ \gamma_v)^2}{(1+\gamma_u)(1+\gamma_v)}-\gamma \nonumber
\end{eqnarray}

\underline{\textbf{Appendix-F:}}
\begin{eqnarray}
	\vec{b}\vec{a}^T &=& \frac{(\frac{\gamma^2}{c^2})}{(1+\frac{\vec{u}.\vec{v}}{c^2})^2} \left[ \{ 1+(\frac{\gamma_v}{1+\gamma_v})(\frac{\vec{u}.\vec{v}}{c^2}) \}\vec{v}+\frac{1}{\gamma_v}\vec{u} \right] \left[\{ 1+(\frac{\gamma_u}{1+\gamma_u})\frac{\vec{u}.\vec{v}}{c^2} \}u^T+\frac{1}{\gamma_u}\vec{v}^T \right] \nonumber \\
	&=& \frac{(\frac{\gamma^2}{c^2})}{(1+\frac{\vec{u}.\vec{v}}{c^2})^2} \left[\{ 1+(\frac{\gamma_u}{1+\gamma_u})(\frac{\vec{u}.\vec{v}}{c^2})\} \{ 1+(\frac{\gamma_v}{1+\gamma_v})\frac{\vec{u}.\vec{v}}{c^2} \}(\vec{v}\vec{u}^T)+\frac{1}{\gamma_u \gamma_v}\vec{u}\vec{v}^T \right] +\nonumber \\
	&& \frac{(\frac{\gamma^2}{c^2})}{(1+\frac{\vec{u}.\vec{v}}{c^2})^2} \left[\frac{1}{\gamma_u} \{1+(\frac{\gamma_v}{1+\gamma_v})(\frac{\vec{u}.\vec{v}}{c^2}) \}(\vec{v}\vec{v}^T)+\frac{1}{\gamma_v} \{ 1+(\frac{\gamma_u}{1+\gamma_u})(\frac{\vec{u}.\vec{v}}{c^2})\}(\vec{u}\vec{u}^T) \right] \nonumber \\
	\therefore~ Tr(\vec{b}\vec{a}^T) &=& \frac{\gamma^2}{(1+\frac{\vec{u}.\vec{v}}{c^2})^2} \left[ \{ 1+\frac{\gamma_u}{1+\gamma_u}(\frac{\vec{u}.\vec{v}}{c^2})+\frac{\gamma_v}{1+\gamma_v}(\frac{\vec{u}.\vec{v}}{c^2})+\frac{\gamma_u \gamma_v}{(1+\gamma_u)(1+\gamma_v)}(\frac{\vec{u}.\vec{v}}{c^2})^2 \}(\frac{\vec{u}.\vec{v}}{c^2}) \right] \nonumber \\
	&&+\frac{\gamma^2}{(1+\frac{\vec{u}.\vec{v}}{c^2})^2} \left[ \frac{1}{\gamma_u \gamma_v}(\frac{\vec{u}.\vec{v}}{c^2})+\frac{1}{\gamma_u} \{ 1+(\frac{\gamma_v}{1+\gamma_v})(\frac{\vec{u}.\vec{v}}{c^2}) \}\frac{v^2}{c^2}+\frac{1}{\gamma_v}\{1+\frac{\gamma_u}{1+\gamma_u}(\frac{\vec{u}.\vec{v}}{c^2}) \}\frac{u^2}{c^2}  \right]\nonumber \\
	&=& \gamma_u^2 \gamma_v^2 \left[\{ 1+\frac{1}{\gamma_u \gamma_v}+\frac{\gamma_v}{\gamma_u(1+\gamma_v)}.\frac{\gamma_v^2-1}{\gamma_v^2}+\frac{\gamma_u}{\gamma_v(1+\gamma_u)}.\frac{\gamma_u^2-1}{\gamma_u^2} \}(\frac{\vec{u}.\vec{v}}{c^2}) \right] \nonumber \\
	&&+ \gamma_u^2 \gamma_v^2 \left[ (\frac{\gamma_u}{1+\gamma_u}+\frac{\gamma_v}{1+\gamma_v})(\frac{\vec{u}.\vec{v}}{c^2})^2+\frac{\gamma_u \gamma_v}{(1+\gamma_u)(1+\gamma_v)}(\frac{\vec{u}.\vec{v}}{c^2})^3 +\frac{\gamma_v^2-1}{\gamma_u \gamma_v^2}+\frac{\gamma_u^2-1}{\gamma_u^2 \gamma_v} \right] \nonumber \\
	&=& \gamma_u^2 \gamma_v^2 \left[ \{ 1+\frac{1}{\gamma_u \gamma_v}+\frac{\gamma_v-1}{\gamma_u \gamma_v}+\frac{\gamma_u-1}{\gamma_u \gamma_v} \}(\frac{\vec{u}.\vec{v}}{c^2})+\frac{(\gamma_u+\gamma_v)(\gamma_u \gamma_v-1)}{\gamma_u^2 \gamma_v^2} \right] \nonumber \\
	&& + \gamma_u^2 \gamma_v^2 \left[ \frac{(\gamma_u+\gamma_v+2 \gamma_u \gamma_v)}{(1+\gamma_u)(1+\gamma_v)}(\frac{\vec{u}.\vec{v}}{c^2})^2+\frac{\gamma_u \gamma_v}{(1+\gamma_u)(1+\gamma_v)}(\frac{\vec{u}.\vec{v}}{c^2})^3 \right] \nonumber \\
	&=& (\gamma-\gamma_u \gamma_v)(\gamma_u\gamma_v+\gamma_u+\gamma_v-1)+(\gamma_u+\gamma_v)(\gamma_u \gamma_v-1) \nonumber\\
	&&+\frac{(\gamma_u+\gamma_v+2 \gamma_u \gamma_v)}{(1+\gamma_u)(1+\gamma_v)}(\gamma-\gamma_u \gamma_v)^2+\frac{(\gamma-\gamma_u \gamma_v)^3}{(1+\gamma_u)(1+\gamma_v)} \nonumber \\
	&=& (\gamma-\gamma_u \gamma_v)(\gamma_u+\gamma_v)+(\gamma-\gamma_u\gamma_v)(\gamma_u\gamma_v-1)+(\gamma_u+\gamma_v)(\gamma_u \gamma_v-1) \nonumber \\
	&&+ \frac{(\gamma-\gamma_u \gamma_v)^2}{(1+\gamma_u)(1+\gamma_v)} \{ \gamma_u+\gamma_v+2 \gamma_u \gamma_v +\gamma -\gamma_u \gamma_v \} \nonumber \\
	&=& (\gamma-1)(\gamma_u+\gamma_v)+\frac{(\gamma-\gamma_u \gamma_v)}{(1+\gamma_u)(1+\gamma_v)} \{ (1+\gamma_u+\gamma_v+\gamma_u \gamma_v)(\gamma_u \gamma_v-1)+ \nonumber \\
	&&(\gamma-\gamma_u \gamma_v)(\gamma+\gamma_u +\gamma_v+\gamma_u \gamma_v) \} \nonumber \\
	&=& (\gamma-1)(\gamma_u+\gamma_v)+\frac{(\gamma-\gamma_u \gamma_v)}{(1+\gamma_u)(1+\gamma_v)} \{ \gamma_u \gamma_v +\gamma_u^2 \gamma_v +\gamma_u \gamma_v^2+\gamma_u^2 \gamma_v^2-1-\gamma_u -\gamma_v \nonumber \\
	&&- \gamma_u \gamma_v +\gamma^2+\gamma \gamma_u +\gamma \gamma_v +\gamma \gamma_u \gamma_v -\gamma \gamma_u \gamma_v -\gamma_u^2 \gamma_v -\gamma_u \gamma_v^2-\gamma_u^2 \gamma_v^2 \} \nonumber  \\
	&=&(\gamma-1)(\gamma_u+\gamma_v)+\frac{(\gamma-\gamma_u \gamma_v)}{(1+\gamma_u)(1+\gamma_v)} \{ \gamma^2-1+\gamma(\gamma_u+\gamma_v)-(\gamma_u+\gamma_v) \}\nonumber \\
	&=& (\gamma-1)(\gamma_u+\gamma_v)+\frac{(\gamma-\gamma_u \gamma_v)(\gamma-1)(\gamma+1+\gamma_u+\gamma_v)}{(1+\gamma_u)(1+\gamma_v)} \nonumber \\
	Tr (\frac{ba^T}{\gamma+1}) &=& (\gamma-1) \left[ \frac{\gamma_u+\gamma_v}{1+\gamma}+1+\frac{(\gamma-\gamma_u \gamma_v)(\gamma+1+\gamma_u+\gamma_v)}{(1+\gamma)(1+\gamma_u)(1+\gamma_v)} \right]-(\gamma-1) \nonumber \\
	&=& \frac{(\gamma-1)(1+\gamma+\gamma_u+\gamma_v)}{(1+\gamma)} \left[1+\frac{(\gamma-\gamma_u)}{(1+\gamma_u)(1+\gamma_v)} \right]-(\gamma-1) \nonumber \\
	&=& \frac{(\gamma-1)(1+\gamma+\gamma_u+\gamma_v)^2}{(1+\gamma)(1+\gamma_u)(1+\gamma_v)}-(\gamma-1) \nonumber 
\end{eqnarray}

\begin{eqnarray}
	\therefore~ TrR &=& Tr M- \frac{Tr(\vec{b}\vec{a}^T)}{\gamma+1} \nonumber \\
	&=& \frac{(1+\gamma+\gamma_u+\gamma_v)^2}{(1+\gamma_u)(1+\gamma_v)}-\gamma-\frac{(\gamma-1)(1+\gamma+\gamma_u+\gamma_v)^2}{(1+\gamma)(1+\gamma_u)(1+\gamma_v)}+(\gamma-1) \nonumber \\
	&=& \frac{2(1+\gamma+\gamma_u+\gamma_v)^2}{(1+\gamma)(1+\gamma_u)(1+\gamma_v)}-1 = 1+ 2 cos \epsilon \nonumber \\
	\therefore ~ cos \epsilon &=& \frac{(1+\gamma+\gamma_u+\gamma_v)^2}{(1+\gamma)(1+\gamma_u)(1+\gamma_v)}-1
\end{eqnarray}

	 \section{ Curve in Minkowski geometry and proper time: Four velocity vector}
	
	Suppose $x^\mu= x^\mu (\lambda)$ describes the path of massive particle with $\lambda$ being some real parameter. The tangent vector to this curve is defined by
	\begin{equation}
		x'^\mu= \frac{dx^\mu}{d\lambda} \nonumber
	\end{equation}
	
	This tangent vector is time-like as it describes the world line of a massive particle. So one must have $|x'^\mu(\lambda)|^2 <0 , \forall ~\mu$. Now the proper time along the path is measured by a clock moving with the particle. Hence it is co-ordinate independent and can be considered  as an observable. Formally, the differential of the proper time is related to the line element as
	
	\begin{eqnarray}
		d\tau^2 &=&-dS^2 = - \eta_{\mu \nu} dx^\mu dx^\nu \label{e1} \\
		i.e., (\frac{d\tau}{d \lambda})^2 &=& -\eta_{\mu \nu}  x'^\mu  x'^\nu = -| x'(\lambda)|^2  \nonumber \\
		i.e., d\tau &=& \sqrt{-| x'(\lambda)|^2}  d\lambda  \nonumber \\
		\mbox{or equivalently,}~~ \tau &=& \int \sqrt{-|x'(\lambda)|^2}  d\lambda \nonumber
	\end{eqnarray}
	
	Thus proper time can be obtained as a function of the parameter $\lambda ~~i.e.~ \tau= \tau(\lambda)$. 
	
	Note that if the path of the particle is parameterized by proper time i.e. $x^\mu = x^\mu (\tau)$, then $v^\mu \equiv \dot x^\mu = \frac{dx^\mu}{d \tau}$ is termed as particle's four velocity. Now due to relation ($\ref{e1}$) we have 
	\begin{equation}
		||\dot x^\mu  ||^2=-1 \nonumber
	\end{equation}
	i.e. the 4-velocity is a time-like vector and is always normalized.\\
	
	\underline{\textbf{Note}:} We have seen that for massive particle, the path is a time-like curve and the proper time is well defined. However, for massless particle the worldlines is a null path having tangent vector a null vector i.e. $|x'(\lambda)|^2=0$, for any choice of the parameter $\lambda$. Hence $\tau=0$ between any two points on the null curve or equivalently, one may say that null particles do not experience the passage of time. Hence the null paths do not have any preferred parameter rather have a family of preferred parameters termed as affine parameters but a null particle does not have a well defined 4- velocity.\\
	
	\textbf{Question:} Suppose a particle is moving along a time-like / null geodesic. Is it possible for the particle to suddenly switch over to null/ time like geodesic? 
	
	We can answer this question using the mathematical result "the norm of the tangent vector to the geodesic is preserved due to parallel transport" i.e.
	\begin{equation}
		\dot x^\mu \nabla_\mu v^\mu =0, \nonumber ~~~~ \mbox{ ($v^\mu$ is the tangent vector being parallel transported)}
	\end{equation}
	Now,
	\begin{equation}
		\dot{x}^\mu \nabla_\mu (|v|^2)= \dot{x}^\mu (g_{\alpha \beta} v^\alpha v^\beta)= g_{\alpha \beta}(\dot{x}^\mu \nabla_\mu v^\alpha)v^\beta+v^\alpha g_{\alpha \beta} (\dot{x}^\mu \nabla_\mu v^\beta)=0 \nonumber
	\end{equation}
	Hence $|v|^2$ remains constant i.e. it is not possible to have an exchange of null or time-like geodesic in course of motion.\\
	
	\textbf{(a) Twin "Paradox" and possible solution}
	The problem is as follows:
	
	Suppose A and B are two twins. B has decided to have a to and fro journey to a nearby star with a speed $\lambda c$ (where $\lambda$ is very close to unity), while A remains on earth. When B returns to earth they have distinct observations due to time dilation. According to A, he is older than B while according to B, he is older than A- both of them are assuming that he is at rest and the other twin brother is moving relative to him and as a result there will be time dilation of the other. The solution of this paradox is as  follows:
	
	Twin A always remain in an inertial frame throughout the journey of B while B's motion was accelerated and hence he was no longer in inertial motion and the time of B experiences dilation. Therefore, A will be order than B after B's journey.
	
	From the point of view of GTR as A moves in an inertial frame so only force acting on it is gravity. Hence A moves in a time-like geodesic. On the otherhand, B moves in a time-like path as some non-gravitational forces are acting on him (for his accelerated motion). As the proper time is maximized along the time like geodesic as A has large proper time than B. Hence A is much older than B.\\

	\textbf{(b) Rapidity, the proper velocity parameter: A justification for a universal space limit}\\
	
	A question that we have in mind `` why is there a speed limit in relativistic theory?"
	
	The possible answer to this question is that due to bad choice of the velocity parameter we are obtaining such a universal speed limit. The appropriate parameter for the measurement of velocity is the rapidity parameter which varies over the entire real line. The justification is as follows:
	
	The LT can be interpreted as a type of rotation - a hyperbolic rotation in the $(x,t)$ plane. So the rotation in a plane where one dimension has a -ve signature in the metric and as a result the rotation is characterized by a hyperbolic angle. More explicitly, the LT in $(x,t)$-plane can be written in matrix form as
	
	\begin{equation}
		\begin{pmatrix}
			t' \\
			x'
		\end{pmatrix}= \begin{pmatrix}
			cosh \phi & -sinh\phi \\
			-sinh\phi & cosh\phi
		\end{pmatrix} \begin{pmatrix}
			t \\
			x
		\end{pmatrix} \nonumber
	\end{equation}
	with $sinh\phi=\gamma v~,~cosh\phi=\gamma$ and hence $\phi=tanh^{-1}v$. 
	
	Note that if we have another LT from $(x',t') \xrightarrow{} (x'',t'')$ as 
	
	\begin{equation}
		\begin{pmatrix}
			t'' \\
			x''
		\end{pmatrix}= \begin{pmatrix}
			cosh \phi' & -sinh\phi' \\
			-sinh\phi' & cosh\phi'
		\end{pmatrix} \begin{pmatrix}
			t' \\
			x'
		\end{pmatrix} \nonumber
	\end{equation}
	then combination of these two LTs gives
	\begin{equation}
		\begin{pmatrix}
			t'' \\
			x'' 
		\end{pmatrix}=\begin{pmatrix}
			cosh (\phi+\phi') & -sinh (\phi+\phi') \\
			-sinh (\phi+\phi') & cosh (\phi+\phi')
		\end{pmatrix}\begin{pmatrix}
			t \\
			x
		\end{pmatrix} \nonumber
	\end{equation}
	a LT between $(x,t)$ and $(x'',t'')$.
	
	As superposition of two rotations gives another rotation with additive angle of rotation so superposition of two LT can be interpreted as rotation with hyperbolic angle. Due to property of $tanhx$ we have $v$ restricted to $|v|<1$, while $\theta$ varies over the entire real line. Hence if the hyperbolic angle known as rapidity parameter is chosen as the velocity measurement parameter then there is no restriction on the velocity. Apparently, restriction appears due to bad choice of the velocity parameters.\\

	\textbf{(c) The notion of global and local velocity:}
	
	In STR, the line element is 
	\begin{equation}
		ds^2= -dt^2+dx^2+dy^2+dz^2 \nonumber
	\end{equation}
	If a massless particle moves along the $z-$direction then we have
	\begin{equation}
		0=-dt^2+dz^2~~\mbox{i.e.}~~\frac{dz}{dt}= \pm 1,
	\end{equation}
	which means particle moves $+ve$ or $-ve$ z-direction with speed of light. This is expected, as for a massless particle proper time can not be defined and hence the notion of four velocity is no longer there. Only locally (at the observer's point) it moves with the speed of light.
	
	On the otherhand for massive particle we have
	\begin{equation}
		ds^2=-dt^2+dz^2 <0 ~~ \mbox{i.e.}~~|\frac{dz}{dx}|<1 \nonumber
	\end{equation}
	This implies a massive particle locally moves with a speed less than the light speed.
	
	In GTR, the situation is little different. Here particles moves in curved space-time. For simplicity we choose the line element as 
	\begin{equation}
		ds^2=-V^2 dt^2+dz^2~,~V \in \mathbb{R}.
	\end{equation}
	For massless particle
	\begin{equation}
		\frac{dz}{dt}= \pm V
	\end{equation}
	This shows that massless particles has an arbitrary speed in a typical co-ordinate system. Here $V$ is the co-ordinate speed not the local speed. Now due to general covariance (i.e. diffeomorphism) nature of GTR this co-ordinate speed depends on the choice of co-ordinate system.
	
	However, it is easy to show that the local speed will be the speed of light if one uses the property that at a particular point P it is always possible to transform to locally inertial coordinates which have the property : (i)$g_{\mu \nu}(p)=\eta_{\mu \nu}$ and (ii) $\partial_\rho g_{\mu \nu}=0$. Then for the observer at p the ST is completely flat in his immediate vacinity and hence locally he will measure the speed of the massless particle to be the light speed.
	
	Now for the homogeneous and isotropic FLRW space-time the line element has the form
	\begin{equation}
		ds^2= -dt^2 +a^2(t) d\Sigma^2
	\end{equation}
	with $d\Sigma^2$, the line element for the 3D spatial hyper-surfaces of constant 't' having uniform curvature . Usually, we choose them to be flat i.e.
	\begin{equation}
		d \Sigma^2= dx^2+dy^2+dz^2= dr^2+r^2 d\Omega_2^2.
	\end{equation}
	Here $a(t)$ is called the scale factor and it simply scales the spatial distances measured within the spatial hypersurfaces. At present time $a(t)$ is defined to be unity so that the whole space-time metric is flat.
	
	The proper distance is the spatial distance measured with this metric and is modified by the scale factor. Due to expansion of the inverse $a(t)$ increases and thus if two galaxies are at rest, the proper distance between them still increases with time. However, the comoving distance which factors out the scale factor, is constant for those two galaxies. Hence a galaxy that is currently at a proper distance $D_0$ from us will be at a distance
	\begin{equation}
		D(t)=a(t) D_0, \nonumber
	\end{equation}
	from us at time $'t'$. So the recession velocity of the galaxy w.r.t. us is 
	\begin{equation}
		\dot{D}(t)=\dot a(t) D_0 = \frac{\dot a(t)}{a(t)} D(t)= H D(t) \nonumber
	\end{equation}
	where $H(t)=\frac{\dot a}{a}$ is called the Hubble parameter. This is Hubble's law. It states that the recession velocity of a galaxy from us is proportional to its distance from us. It is to be noted that the recession velocity is a global velocity due to expansion of the space itself. The galaxy's local velocity (known as peculiar velocity) in space relative to nearby galaxies is independent from $\dot D$ and always less than $'c'$, as it locally follows a time-like path. 
	
	Thus there is an ambiguity in the universal speed limit: local speed is within space and can not exceed the speed of light while the global velocity due to expansion of space itself is unbounded.\\

	\textbf{(d) The notion of velocity in time like path:}
	
	The time-like paths of massive particles are usually parametrized by the proper time ($\tau$) so that the norm of the tangent vector $\dot x^\mu \equiv \frac{dx^\mu}{d \tau}$ is normalised i.e. $||\dot x^\mu ||^2=-1$. Also this tangent vector gives the 4-velocity of the particle. In STR, this norm is given by
	\begin{equation}
		||\dot x^\mu||^2= - \dot t^2+ \dot x^2+ \dot y^2+\dot z^2 <0 \nonumber
	\end{equation}
	If the particle is at rest in the frame then $\dot x^\mu =(1,0,0,0)$. So it has no velocity along any of the spatial co-ordinate but it moves at the speed of light along the time co-ordinate. From the above equation due to negativity of the norm $\dot t \neq 0$ i.e. a massive particle must always move along time axis, but $\dot t $ may have $+ve$ or $-ve$ sign. The past directed or future directed particle is characterized by $\dot t <0$ or $\dot t >0$.
	
	Suppose a particle moving at constant spatial 3-velocity $v$ along z direction i.e. $v=\frac{dz}{dt}$. Then the corresponding 4-velocity will be
	
	\begin{equation}
		\dot x^\mu = (\gamma, 0, 0, \gamma v)= \gamma(1,0,0,v) \nonumber
	\end{equation}
	where $\gamma=\frac{dt}{d\tau}=\frac{1}{\sqrt{1-\frac{v^2}{c^2}}}$ is the Lorentz factor.
	
	Note that $\gamma$ measures the relation between co-ordinate time and proper time. Moreover, $'\gamma'$ measures the amount of time dilation as $dt=\gamma d \tau$. Also Lorentz factor can be considered as a normalization factor in STR. The energy of the particle moving with 3-velocity $v$ is $E=m_0 \gamma c^2$ so as $v \xrightarrow{}c, ~\gamma \xrightarrow{} \infty ~\mbox{and}~ E \xrightarrow{} \infty$.
	
	This implies that particle requires infinite energy to accelerate the particle to the speed of light. \\

	\textbf{(e) Null particles: The speed of light}

	A particle with spatial 3-velocity $v$ has 4 velocity 
	\begin{eqnarray}
		\dot x^\mu &=& \gamma (1,0,0,v) \nonumber \\
		\mbox{i.e.}~~ ||\dot x^\mu||^2 &=& \gamma^2 (-1+v^2) .\nonumber
	\end{eqnarray}
	Now for null particles $v=1$ (i.e. $c=1$) and hence $||\dot x^\mu||^2=0$. So there is no need of normalization factor. For convenience if we choose $\gamma=1$ then we have $\dot x^\mu = (1,0,0,1)$ for a null particle. Now due to norm invariance, the light -like particle will always move along a null path and the speed of light is the same in all inertial frames (i.e. for all observers). This is nothing but the 2nd postulate of STR. This implies that a particle moving at the speed of light can never decelerate or accelerate to a different speed.\\

	\textbf{(f) Particles moving with velocity $v>c$: Tachyons}
	
	For any particle moving along z direction with velocity $'v'$ we have
	\begin{equation}
		\dot x^\mu= \gamma(1,0,0,v) ~\mbox{i.e}~ ||\dot x^\mu||^2=\gamma^2 (-1+v^2). \nonumber
	\end{equation}
	Now if $v>1$ i.e. velocity is larger than the speed of light then $||\dot x^\mu||^2>0$ i.e. path is space-like in nature. If we normalize to $|\dot x^\mu|=1$ then $\gamma=\frac{1}{\sqrt{v^2-1}}$.
	
	Thus a particle moving faster than light, will travel along space-like paths and is called tachyon. Due to norm invariance, tachyons can not be decelerate to the speed of light or below and hence they always move along space-like paths.
	
	Note that as $v \xrightarrow{}1$ (from above), $\gamma \xrightarrow{} \infty$. So $E=m \gamma \xrightarrow{}\infty$ as $v \xrightarrow{}1$. Hence a tachyon requires infinite energy to decelerate to the speed of light. Further, as $\gamma$ decreases $v$ increases and $\gamma \xrightarrow{} 0$ as $v \xrightarrow{} \infty$. This implies that tachyon has less energy at higher velocities. In fact, tachyon is at rest when its velocity is infinity and its energy is minimum. Thus a tachyon at rest has $ \dot x^\mu=  \lim_{\gamma \to 0} \gamma (1,0,0,v)=(0,0,0,1) $. This shows that a tachyon at rest moves only along a space-like direction while a normal massive particle at rest moves only about time direction.\\

	\textbf{(g) Inconsistency due to motion of a tachyon in STR :}
	
	In STR, we have seen that a tachyon moves along a space-like path (locally) and has velocity faster than light. So it is natural to speculate that time travel or at least communication to the past is possible.
	
	In Minkowski space-time let S and $S'$ be two inertial frame of references (one may consider them to be two space stations) and $u<c$ be the relative velocity between them. Let $(t,x)$ be the co-ordinate system for rest frame of S and that of S' is $(t',x')$. So $t-$ axis is the world line for S station and that for S'- station is t'- axis. We now perform an hypothetical experiment with tachyons as follows:
	
	From the origin i.e. $(t,x)=(0,0)$ of the rest frame S, a tachyon is send to station S' with speed $v>c$. The tachyon arrives at station S' whenever its world line intersects the t'- axis. Although the tachyon is superluminal but still it is going forward in time so it will necessarily be in future (i.e. there is no concept of time travel). So without any loss of generality one may consider this point of contact as the origin of the $(t',x')$ co-ordinate system. 
	
	On the otherhand the station $S'$ (at time $t'=0$) sends another tachyon back to station S with speed $v'>c$. Then w.r.t $S'$, the emitted tachyon moves forward in time i.e. the world line should be above the $x'$-axis. However, the ST diagram of the above experiment shows that for sufficiently large u, the $x'$-axis  intersects the t-axis at a -ve value. If tachyon is used as a carrier of message then the message goes to past in S-frame or the tachyon is detected in S-frame in the past. Thus the experiment may be considered as a person at space station S uses tachyon to send a message to his past - a paradox. This type of paradox is well known in time machine i.e. time travel. For the present experiment the paradox may be formulated as follows:
	
	Suppose station S sends a tachyon at $t=0$ only if it did not receive a tachyon at any time $t<0$. Further, station S' sends a tachyon at time $t'=0$ only if it did receive a tachyon at that time i.e. it simply acts as a tachyon mirror. So if S sends a tachyon at $t=0$ implies it did not receive a tachyon at an earlier time but in that case the tachyon is reflected back from $S'$-station and it is received in $S'$-station at an earlier time (i.e. $t< 0$) i.e. in past. This means that station S could not have sent the initial tachyon at t=0. In otherwords, one can say that S station sends a tachyon at $t=0$ iff it does not send a tachyon at $t=0$  !! a paradox. An event can happen and not happen simultaneously- a contradictory statement.
	
\newpage
\vspace{5mm}

	\section{Problems with Solutions} 

\vspace{3mm}
%
{\bf 5.1.} For what value of $\beta \left(=\dfrac{v}{c}\right)$ will the relativistic mass of a particle exceeds its rest mass by a given fraction $f$ ?\\\\
{\bf 5.2.} If a body of mass `$m$' disintegrates while at rest into two parts of rest masses $m_1$ and $m_2$, show that the energies $E_1$ and $E_2$ of the parts are given by
$$E_1 = c^2\frac{(m^2+m_1^2-m_2^2)}{2m}~~~,~~E_2 = c^2\frac{(m^2-m_1^2+m_2^2)}{2m}\,.$$\\
{\bf 5.3.} Two particles of proper masses $m_1$ and $m_2$ move along the $x$-axis of an inertial frame with velocities $u_1$ and $u_2$ respectively. They collide and coalesce to form a single particle. Assuming the law of conservation of momentum and energy prove that the proper mass $m_3$ and velocity $u_3$ of the resulting single particle are given by
$$m_3^2=m_1^2+m_2^2+2m_1m_2\gamma _1\gamma _2\left(1-\frac{u_1u_2}{c^2}\right)$$
$$u_3=\frac{m_1\gamma _1u_1+m_2\gamma _2u_2}{m_1\gamma _1+m_2\gamma _2}$$
where $\gamma _i^{-2} = 1-\dfrac{u_i^2}{c^2}~~~,~~i=1,2.$\\\\
{\bf 5.4.} Two events are simultaneous though not coincident in some inertial frame $S$. Prove that there is no limit on time separation assigned to these events in different frames but the space separation varies from a minimum (which is the measurement is $S$-frame) to $\infty$\,.\\\\
{\bf 5.5.} Let $q$ and $q'$ are respectively the velocities of a particle in two inertial frames $S$ and $S'$ has a velocity $V$ relative to $S$ in the $x$-direction of $S$-frame. Show that
$$q^2 = \frac{(q')^2+V^2+2q'V\cos \theta' - \left(\frac{q'V\sin \theta'}{c}\right)^2}{\left(1+\frac{V}{c^2}q'\cos \theta'\right)^2}$$
where $\theta'$ is the angle which $q'$ makes with $x$-axis.\\\\
{\bf 5.6.} If $u, v$ are two velocities in the same direction and $V$ be their resultant velocity given by
$$\tanh^{-1}\frac{V}{c}=\tanh^{-1}\frac{u}{c} + \tanh^{-1}\frac{v}{c}$$
then find the law of composition of velocity.\\\\
{\bf 5.7.} A particle of proper mass $m_0$ moves on the $x$-axis of an inertial frame and attracted to the origin by a force $m_0k^2x$. If it performs oscillation
of amplitude `$a$' then show that the periodic time of this relativistic harmonic oscillator is
$$\tau = \frac{4}{c}\int\limits_0^a\frac{f\,dx}{\sqrt{f^2-1}}~~,~~~f=1+\frac{k^2}{2c^2}\left(a^2-x^2\right).$$
Also verify that as $c \rightarrow \infty$ then $\tau \rightarrow \dfrac{2\pi}{k}$ and show that if $\dfrac{ka}{c}$ is small then
$$\tau = \frac{2\pi}{k}\left(1+\frac{3}{16}\frac{k^2a^2}{c^2}\right)~(\mbox{approx}.)$$\\
{\bf 5.8.} Determine the relative speed $v$ for which the Galilean and Lorentz expressions for $x$ differ by 1\%.\\\\
{\bf 5.9.} A rod is moving with a speed is $0.4c$ along its length in the positive $x$-direction, and a particle is moving along the negative $x$-direction with a speed $0.8c$, both the speeds being measured in the same inertial frames and the length of the rod with respect to $S$ frame is 3.6 meters. Find the relative velocity of the rod in the rest frame of the particle. Find the time taken by the particle to cross the rod in the $S$ frame as well as in the rest frame
of the rod.\\\\
{\bf 5.10.} A man moving along the $x$-axis of some inertial frame $S$ at a velocity $V$ observes a body of proper volume $V_0$ moving at a velocity $u$
along the $x$-axis of frame $S$. Find the volume of the body as measured by the man.\\\\
{\bf 5.11.} Verify that in an inertial co-ordinate system $S(x,t)$ the general solution of the wave equation $\left(\dfrac{1}{c^2}\dfrac{\partial ^2}{\partial t^2} - \dfrac{\partial ^2}{\partial x^2}\right)\psi =0$ is of the form $\psi (x,t) = f(x-ct)+g(x+ct)$. Check that $\psi (x,t) = A\cos\left\{\dfrac{w}{c}(ct-x)\right\}$ satisfies the above wave equation. Suppose $S'(x',t')$ be another inertial frame
moving relative to $S$-frame with constant velocity $V$ along $x$-axis and the above solution becomes $\psi (x',t') = A\cos\left\{\dfrac{w'}{c}(ct'-x')\right\}$.
Find the relation between $w$ and $w'$.\\\\
{\bf 5.12.} A rod is of length $l_0$ in its rest frame $S_0$. In another inertial frame $S$ it is oriented in a direction of the unit vector $\textit{\textbf e}$ and is moving with a velocity $\textit{\textbf V}$. Show that the length of the rod in the second frame is
$$l=\frac{l_0\sqrt{c^2-V^2}}{\sqrt{c^2-V^2\sin^2\theta}}$$
where $V=\left|\textit{\textbf V}\,\right|$ and $\theta$ is the angle between $\textit{\textbf e}$ and $\textit{\textbf V}$.\\\\
{\bf 5.13.} The space and time co-ordinates of two events as measured in an inertial frame $S$ are as follows\,:

Event 1\,:~~$x_1=x_0~~,~y_1=z_1=0~~,~t_1=\dfrac{x_0}{c}$~~~and

Event 2\,:~~$x_2=3x_0~~,~y_2=z_2=0~~,~t_2=\dfrac{x_0}{4c}$.\\
$S'$ is another inertial frame (moving relative	to $S$) in which the above two events appear simultaneously. Find the relative velocity of $S'$ with respect to $S$ and also the time of occurrence of both the events in $S'$ frame.\\\\
{\bf 5.14.}  Calculate the orientation of a rod of length $l$ in an inertial frame that is moving with a velocity $\mu c ~ (\mu <1)$ in a direction making an angle $2 \theta$ with the rod. \\\\
{\bf 5.15.} A rod of rest length one meter is moving longitudinally on a smooth table with a velocity $0.8c$ relative to the table. A circular black spot of rest diameter half meter lies in its path. What is the diameter of the spot as seen by the rod? Explain with reasons, what will be the shape of the spot as seen by an insect sitting at the centre of the rod?\\\\
{\bf 5.16.} Let a constant force $\textit{\textbf F}$ be applied on an object with rest mass $m_0$ at a rest position. Prove that its velocity after a time $t$ is
$$v=\frac{cFt}{\sqrt{m_0^2c^2+F^2t^2}}.$$
Also show that the above result is in agreement with classical result. Further find $v$ after a very long time.\\\\
{\bf 5.17.} The space and time co-ordinates of two events as measured in a frame $S$ are as follows\,:

Event I\,:~~$x_1=x_0~~,~y_1=z_1=0~~,~t_1=\dfrac{x_0}{c}$

Event II\,:~~$x_2=2x_0~~,~y_2=z_2=0~~,~t_2=\dfrac{x_0}{2c}$.\\
There exists a frame $S'$ in which these two events occur at the same time. Find the relative velocity of this frame with respect to $S$. What is the time at which both the events occur in the frame $S'$\,?\\\\
{\bf 5.18.} If the position vectors of two points in 4D space-time are $\left(\dfrac{1}{c},0,0,0\right)$ and $\left(\dfrac{2}{c},2,1,1\right)$ then examine whether the two points are causally connected or not.\\\\
{\bf 5.19.} A particle of rest mass $m_0$ describes the trajectory $x=f(t),~y=g(t),~z=0$ in an inertial frame $S$. Find the four velocity components. Also show that the norm of the four velocity is $-c^2$.\\\\
{\bf 5.20.} Suppose a particle moves relative to the primed system $S'$ with a velocity $u'$ in the $x'y'$ plane so that its trajectory makes an angle $\theta '$ with the $x'$-axis. Show that the equations of motion in $S'$ frame are given by $x'=u't'\cos \theta ',~y'=u't'\sin \theta ',~z'=0$. If $S$ be another inertial frame that moves with respect to $S'$ frame with a velocity $v$ along the common $x$-$x'$ axis, then the magnitude and direction of its velocity in $S'$ is given by
$$u=\frac{\left[(u')^2+v^2-2u'v\cos \theta ' - \frac{u'^2v^2}{c^2}\sin ^2\theta '\right]^{1/2}}{\left(1-\frac{u'v}{c^2}\cos \theta '\right)}$$
$$\mbox{and}~~~\theta = \tan ^{-1}\left[\frac{u'\sin \theta ' \sqrt{1-\frac{v^2}{c^2}}}{u'\cos \theta ' - v}\right]\,.$$
{\bf 5.21.} Show that the set $\mathcal{F}$ of all linear transformations $L= 
\begin{bmatrix}
	l_{11} & l_{12} \\
	l_{21} & l_{22}
\end{bmatrix} $ from $\mathbb{R}^2$ to itself and characterized by the fact $l_{11}>0$, $\det L>0$ and $L^T
\begin{pmatrix}
	1 & 0 \\
	0 & -1
\end{pmatrix} L = 
\begin{pmatrix}
	1 & 0 \\
	0 & -1
\end{pmatrix} $ forms a group under composition of mappings. Using this result prove that the 2D Lorentz transformations form a group.\\\\
{\bf 5.22:} An astronaut wants to go to a star 5 light years away. The rocket accelerates quickly and then moves at a uniform velocity.  Calculate with what velocity the rocket must move relative to the earth if the astronaut is to reach there in 1 year, as measured by a clock being at rest inside the rocket.\\

{\bf 5.23:} A man moving along the x-axis of some inertial frame S at velocity $v$ observes a body of proper volume $V_0$ moving at a velocity $u$ along the x-axis of frame S. Find the volume of the body as measured by the man.\\

\textbf{5.24:} At what speed $v_0$ will the Galilean and the Lorentz expression for x differ by 1\% ?  \\

\textbf{5.25:} If $x^2+y^2 +2 z^2 =4$ be the equation of the boundary of a region in inertial frame S. What would the volume of the same region appear in another inertial frame $S'$ that moves along the common $Z-Z'$ axis with a uniform speed $v_0$ comparable to c ? For what value of $v_0$ the volume seems to be halved? \\

\textbf{5.26:} Using general Lorentz transformation find the expression for velocity $\Vec{u'}$ of a particle as observed in $S'$-frame if $\Vec{u}$ denotes the velocity of that particle in $S$-frame. Hence show that if $\mid \Vec{u} \mid$, $\mid \Vec{v} \mid$ be both less than c then $\mid u' \mid <c$. What will be the expression for $ \Vec{u'}$ if $\Vec{u}$ is parallel to $\Vec{v}$ ? \\

\textbf{5.27:} Show that $\psi(x,t)=A~ cos[ \frac{\omega}{c}(ct-x)]$ satisfies the wave equation $(\frac{1}{c^2}\frac{\partial^2}{\partial t^2}-\frac{\partial^2}{\partial x^2})\psi =0$ in an inertial frame $S(x,t)$. In another inertial frame $S'(x',t')$ the solution is
\begin{equation}
	\psi(x',t')=A~cos[ \frac{\omega'}{c}(ct'-x')]\nonumber
\end{equation}
for the wave equation in $s'$- frame. Find the relation between $\omega$ and $\omega'$.\\

\textbf{5.28:} What is the basic difference between $4D$ representations of Poincare and Minkowski as far as Lorentz transformation is concerned ?\\

\textbf{5.29:} Prove that in Minkowski's geometric representation angle between the space axes is same as that between the time axes. Why are the hyperbolas $c^2 t^2-x^2=\pm 1$ appearing in this representation are called calibration curves?\\

\textbf{5.30:} What do you mean by world line? \\

\textbf{5.31:} The car-garage paradox in STR.\\

\textbf{5.32:} Suppose an inertial frame $S'(x',y',z',t')$ is moving w.r.t. another inertial frame $S(x,y,z,t)$ with a relative velocity $`v$' along the common $x-x'$ axis and by keeping corresponding co-ordinate planes parallel. Prove that with the aid of 2 postulates of STR and any other assumptions ( to be stated) that $y'=y~,~ z'=z$. Also show that the transformation equations for $x$ and $t$ can be put in the form:
\begin{equation}
	x'=a_{11}(x-vt)~,~t'=a_{41}x +a_{44}t~,\nonumber
\end{equation}
where $a_{11}~,~a_{41}$ and $a_{44}$ are constants or functions of $v$. \\

\textbf{5.33:} Prove that a $2D$ LT `L' connecting two inertial frames $S$ and $S'$ satisfies the following conditions:

(i) L is +ve definite and  (ii) $L^T~gL= g = \begin{pmatrix}
	1 & 0 \\
	0 & -1
\end{pmatrix}$.

Hence or otherwise prove that these transformations form commutative group under usual matrix multiplication. Also prove that if a $2D$ linear transformation L connecting inertial frames $S$ and $S'$ satisfies conditions (i) and (ii) then it is a L.T.\\

\textbf{5.34:} In context of a $2D$ LT define rapidity $`\phi$' and hence show that it is an isomorphism from the $2D$ Lorentz group $(L~, ~\star)$ to $(R~,~+)$.\\

\textbf{5.35:} A Galilean transformation connecting inertial frames $S(x,y,z,t)~ \mbox{and}~ S'(x',y',z',t') $ is given by
\begin{equation}
	\begin{pmatrix}
		t \\
		x
	\end{pmatrix} = \begin{pmatrix}
		1 & \vec{O'} \\
		\vec{A} & A
	\end{pmatrix} \begin{pmatrix}
		t' \\
		\vec{r'}  
	\end{pmatrix} + \begin{pmatrix}
		C_0 \\
		\vec{C}
	\end{pmatrix} ~\mbox{with} ~A \in So(3) \nonumber
\end{equation}
Prove the following:
(i) $\vec{v}$ is the velocity of the inertial frame $S'$ w.r.t. $S$.

(ii) If $(t_1, x_1, y_1, z_1) ~\mbox{and}~ (t_2, x_2, y_2, z_2)$ be the space -time co-ordinates of two events in $S-$ frame, then prove that their time separation is invariant. Also prove that their spatial separation is invariant only if the events are simultaneous.\\ \\

\begin{center}
	\underline{\bf Solution and Hints} 
\end{center}
\vspace{3mm}
{\bf Solution 5.1:} ~~$f=\frac{m-m_0}{m_0}~~~~,~~m_0~~\rightarrow~~\mbox{rest mass},~~m~~\rightarrow~~\mbox{relativistic mass}$\\
$~~~~~~~~~~~~~~~~~~~~~~= \dfrac{1}{\sqrt{1-\beta ^2}}-1$

Hence~~~$\beta = \dfrac{\sqrt{f(2+f)}}{1+f}$\,.\\\\
{\bf Solution 5.2:} ~~Energy conservation :~~$E_1+E_2=$ Total energy $=mc^2~~~~~~~~~~~........(1)$

If $p_1, p_2$ be the momenta of the disintegrated
parts then momentum conservation gives
$p_1+p_2=0$.\\
The energy-momentum conservation relation : ~$E^2
= c^2p^2 + m^2c^4$
\begin{center}
	$~~~~~~~~~~~~~~~~~~~i.e.~~~E_1^2 = c^2(p_1^2 + m_1^2c^2)~,~~E_2^2 =
	c^2(p_1^2 + m_2^2c^2)~~~~~~~~~~~~~~........(2)$\\
\end{center}
Solve (1) and (2) for $E_1$ and $E_2$.\\\\
{\bf Solution 5.3:} ~~The conservation of momentum
and energy give
\begin{eqnarray}
	m_1\gamma _1u_1 + m_2\gamma _2u_2 &=&
	m_3\gamma _3u_3 \nonumber \\
	m_1\gamma _1c^2 + m_2\gamma
	_2c^2 &=& m_3\gamma _3c^2~,~~~\gamma
	_3=\left(1-\frac{u_3^2}{c^2}\right)^{-1/2} \nonumber
\end{eqnarray}
Solve for $m_3$ and $u_3$.\\

{\bf Solution 5.4:} ~~Simultaneity of two events means time separation between these events to be zero. So $(x_1,y_1,z_1,t)$ and $(x_2,y_2,z_2,t)$ be the space-time points where two events occur. So the spatial and temporal separation of these two events in $S'$-frame be
\begin{eqnarray}
	\Delta x' &=& \frac{(x_2-x_1)}{\sqrt{1-\frac{v^2}{c^2}}}~,~~~\Delta t' = \gamma \Delta t \frac{v}{c^2}~,~~~\gamma = \left(1-\frac{v^2}{c^2}\right)^{-1/2}, \nonumber \\
	&=& \gamma \Delta x \nonumber
\end{eqnarray}
As $|v|$ varies from zero to $c$ so $\Delta x'$ varies from $\Delta x$ to $\infty$ and $\Delta t'$ varies from $-\infty$ to $\infty$\,.\\\\
\begin{wrapfigure}{r}{0.4\textwidth}\vspace{-1cm}
	\includegraphics[height=4 cm , width=6 cm ]{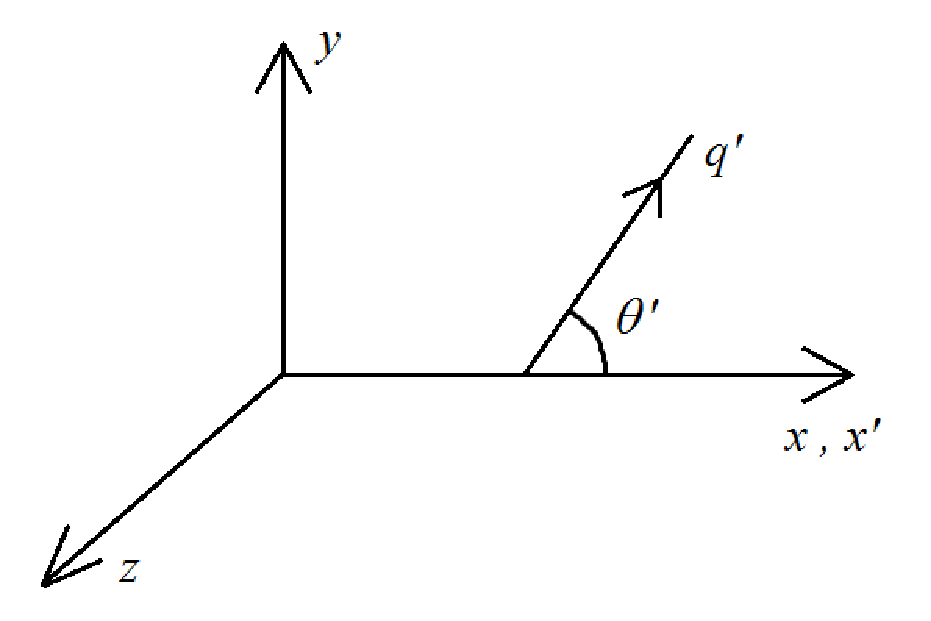}
\end{wrapfigure}
{\bf Solution 5.5:} ~~Suppose $q'\sin \theta'$ makes an angle $\phi '$ with $y'$-axis and $q\sin \theta$ makes an angle $\phi$ with $y$-axis where $\theta$ is the angle between $q$ and $x$-axis. The components of $q$ along the three axes are $(q\cos \theta, q\sin \theta \cos \phi, q\sin \theta \sin \phi)$ and those of $q'$ are $(q'\cos \theta ', q'\sin \theta '\cos \phi ', q'\sin \theta '\sin \phi ')$.

According to Lorentz transformation
\begin{eqnarray*}
	q\cos \theta &=& \dfrac{q'\cos \theta '+V}{1+\frac{q'\cos \theta '.V}{c^2}} \\
	q\sin \theta \cos \phi &=& \dfrac{q'\sin \theta '\cos \phi '\sqrt{1-\frac{V^2}{c^2}}}{1+\frac{q'\cos \theta '.V}{c^2}}  \\\\
	\mbox{and}~~q\sin \theta \sin \phi&=& \dfrac{q'\sin \theta '\sin \phi '\sqrt{1-\frac{V^2}{c^2}}}{1+\frac{q'\cos \theta '.V}{c^2}}
\end{eqnarray*}

Then obtain $q$.\\\\
\vspace{0.3cm}
{\bf Solution 5.6:} ~~Let $\alpha = \tanh^{-1}\dfrac{V}{c}~,~~\beta = \tanh^{-1}\dfrac{u}{c}~,~~\gamma = \tanh^{-1}\frac{v}{c}$

so we have ~~$\alpha = \beta + \gamma$.~~~~~~~~~~~~~~~~~~~~~~~~~~~~~~~~~~~~~~~~~~~~~~~~~~~~~~~~~~~~~~~~~~~~~........(1)

Now $\alpha = \tanh^{-1}\dfrac{V}{c}~~i.e.~~~\tanh \alpha = \dfrac{V}{c}$

$\Rightarrow ~~\dfrac{e^\alpha - e^{-\alpha}}{e^\alpha + e^{-\alpha}} = \dfrac{V}{c}$

$i.e.~~~~e^{2\alpha} = \dfrac{c+V}{c-V}$\,.\\
Similarly,~~$e^{2\beta} = \dfrac{c+u}{c-u}~,~~e^{2\gamma} = \dfrac{c+v}{c-v}$\\
So relation (1) gives
$$\ln \frac{c+V}{c+V} = \ln \frac{(c+u)(c+v)}{(c-u)(c-v)}\,.$$
This gives the relativistic law of composition of velocity.\\\\
{\bf Solution 5.7:} ~~The equation of motion of the particle is

\begin{wrapfigure}{r}{0.4\textwidth}
	\hspace{0.5cm}\includegraphics[height=1.5 cm , width=6 cm ]{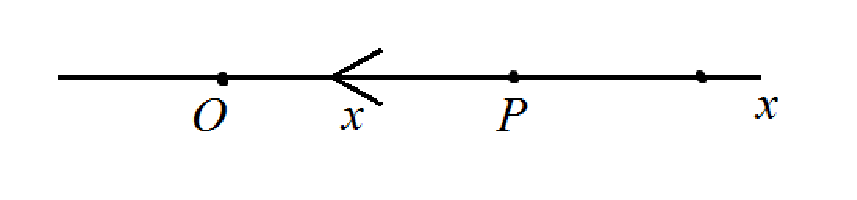}
\end{wrapfigure}
$$\frac{d}{dt}\left(\frac{m_0\dot{x}}{\sqrt{1-\left(\frac{\dot{x}}{c}\right)^2}}\right) = -m_0k^2x$$
$$i.e.~~~\int u\,d\left(\frac{u}{\sqrt{1-\frac{u^2}{c^2}}}\right) = A - \frac{k^2x^2}{2}~~,~~~~u=\dot{x}$$
$$i.e.~~~\frac{c^2}{\sqrt{1-\frac{u^2}{c^2}}} = A - \frac{k^2x^2}{2}$$
using initial condition :~~~$u=0$~~at $x=a$~,~~~$A= c^2 + \frac{k^2a^2}{2}$.\\
So solving for $u$ one obtains
$$u=\frac{c\sqrt{f^2-1}}{f}~,~~f=1+\frac{k^2}{2c^2}(a^2-x^2)$$
$$i.e.~~~t=\frac{1}{c}\int\limits ^a_0 \frac{f}{\sqrt{f^2-1}}dx$$
Hence time period :~~~$\tau = 4t = \dfrac{4}{c}\int\limits ^a_0 \dfrac{f}{\sqrt{f^2-1}}dx$\,.\\\\
\underline{2nd part:}~~ As $f = 1+\dfrac{k^2}{2c^2}(a^2-x^2)$,

so $f^2-1 = \dfrac{k^2}{2c^2}(a^2-x^2)\left[1+\dfrac{k^2}{2c^2}(a^2-x^2)\right]$\\
and hence~~ $\dfrac{1}{c}\dfrac{f}{\sqrt{f^2-1}} = \dfrac{1}{k\sqrt{a^2-x^2}}\left\{1+\dfrac{3}{8}\dfrac{k^2}{c^2}(a^2-x^2)+ \ldots \ldots \right\}$\,.\\
Thus as $c \rightarrow \infty$ , ~~$\dfrac{1}{c}\dfrac{f}{\sqrt{f^2-1}} \rightarrow \dfrac{1}{k\sqrt{a^2-x^2}}$\,.\\
Hence~~$\tau = \dfrac{4}{k}\int\limits ^a_0 \dfrac{dx}{\sqrt{a^2-x^2}} = \dfrac{2\pi}{k}$\,.\\\\
\underline{3rd part:}~~ If $\dfrac{ka}{c}$ is small and $x \leq a$ then $\dfrac{k^2}{c^2}(a^2-x^2)$ is a small quantity, and hence
$$\frac{1}{c}\frac{f}{\sqrt{f^2-1}} \simeq \frac{1}{k\sqrt{a^2-x^2}}\left\{1+\frac{3k^2}{8c^2}(a^2-x^2)\right\}\,.$$
As a result,
\begin{eqnarray}
	\tau &=& \frac{4}{k}\left[\int\limits ^a_0 \frac{dx}{\sqrt{a^2-x^2}} + \int\limits ^a_0 \frac{3k^2}{8c^2}(a^2-x^2)^{\frac{1}{2}}\,dx\right] \nonumber \\
	&=& \frac{2\pi}{k}\left[1+\frac{3}{16}\frac{k^2a^2}{c^2}\right] \mbox{(approx.)} \nonumber
\end{eqnarray}\\
{\bf Solution 5.8:} ~~$x_G'= x-vt$~,~~$x_L' = \dfrac{x-vt}{\sqrt{1-\frac{v^2}{c^2}}} = \dfrac{x_G'}{\sqrt{1-\frac{v^2}{c^2}}} > x_0'$\\
By condition,~~$\dfrac{x_L'-x_G'}{x_G'} = 0.01~~i.e.~~~1.01 = \dfrac{1}{\sqrt{1-\frac{v^2}{c^2}}}$\\
$\Rightarrow ~~v = 0.1401c$\,.\\\\
{\bf Solution 5.9:} The relative velocity of the rod in the rest frame of the particle is given by the law of addition of velocities in STR as
$$v_{rel.} = \frac{0.4c+0.8c}{1+\frac{(0.4c)(0.8c)}{c^2}} \simeq 0.909c\,.$$\\
\underline{2nd part:}~~ According to an observer in $S$ frame, the relative velocity between the particle and the rod is $(0.4c+0.8c) = 1.2c$. So the time taken by the particle to cross the rod according to the observer in $S$-frame $=\dfrac{3.6}{(1.2)\times 3 \times 10^8}$ sec.

As the rod is moving with velocity $0.4c$ relative to $S$-frame, so the length of the rod 3.6 meter (in $S$-frame) is not the proper length of the rod. If $l_0$ be the proper length of the rod then according to STR (length contraction)
$$3.6 = l_0\left[1-(0.4)^2\right]^{\frac{1}{2}}$$
$$i.e.~~~ l_0 \simeq 3.93~\mbox{meters}\,.$$
Now, in the rest frame of the rod, the particle moves with velocity $0,909c$ along the $-$ve $x$-axis. Hence in the rest frame of the rod the time taken by the particle to cross it will be
$$\frac{3.93}{(0.909)\times 3 \times 10^8}~\mbox{sec} \simeq 1.44 \times 10^{-8}~\mbox{sec.}\simeq 1.44 \times 10^{-8}\mbox{sec.} $$\\
{\bf Solution 5.10:} ~~Let $S'$ be the proper frame of the observer. So the relative velocity between $S$ and $S'$ is $V$ along the $+$ve $x$-axis. As $u$ be th velocity of the body relative to $S$ along $x$-axis so its velocity relative to $S'$ is $u'=\dfrac{u-v}{1-\frac{uv}{c^2}}$\,. Hence the volume of the body relative to the observer will be
$$V=V_0\sqrt{1-\left(\frac{u'}{c}\right)^2} = V_0\sqrt{1-\frac{1}{c^2}\left(\frac{u-v}{1-\frac{uv}{c^2}}\right)^2} = V_0\frac{\sqrt{(c^2-u^2)(c^2-v^2)}}{c^2-uv}\,.$$\\

{\bf Solution 5.11:} ~~It is easy to check that $\psi (x,t) = f(x-ct)+g(x+ct)$ satisfies the 2D wave equation. As a particular solution one may choose $g=0$ and hence $\psi (x,t) = A\cos\left\{\dfrac{w}{c}(ct-x)\right\}$ is a possible solution of the wave equation.
\begin{eqnarray}
	\mbox{Now,}~~~~\frac{w}{c}(ct-x) &=& \frac{w}{c}\left\{\gamma c\left(t' + \frac{vx'}{c^2}\right)-\gamma(x' + vt')\right\}~,~~~\gamma = \frac{1}{\sqrt{1-\frac{v^2}{c^2}}} \nonumber \\
	&=& \frac{w}{c}\left\{\gamma t'(c-t) -\gamma x'\left(1 - \frac{v}{c}\right)\right\} \nonumber \\
	&=& \frac{w}{c}\gamma\left(1 - \frac{v}{c}\right)(ct'-x') \nonumber
\end{eqnarray}
So on comparison, $w'=\dfrac{w}{\sqrt{1-\frac{v^2}{c^2}}}\left(1 - \dfrac{v}{c}\right)=w\sqrt{\dfrac{1 - \frac{v}{c}}{1 + \frac{v}{c}}}=w\sqrt{\dfrac{c-v}{c+v}}<w$\\

{\bf Hints 5.12:} ~~Standard length contraction problem.\\

{\bf Solution 5.13:} ~~Let the space-time co-ordinates of the two events in $S'$ frame be $(x'_1,y'_1,z'_1,t'_1)$ and $(x'_2,y'_2,z'_2,t')$ respectively.

According to Lorentz transformation\,:
$$t'_1 = \gamma \left(t_1 - \frac{v}{c^2}x_1\right)~,~~~t'_2 = \gamma \left(t_2 - \frac{v}{c^2}x_2\right)$$
For simultaneous occurrence of both the events in $S'$-frame
$$t'_1=t'_2~~\Rightarrow ~t_1 - \frac{v}{c^2}x_1 = t_2 - \frac{v}{c^2}x_2~~i.e.~~\frac{x_0}{c}-\frac{v}{c^2}x_0 = \frac{x_0}{4c}-\frac{v}{c^2}3x_0~~\Rightarrow ~v= -\frac{3c}{8}\,.$$
Also by L.T. ~~$t'_1 = \gamma \left(t_1 - \dfrac{vx_1}{c^2}\right) = \dfrac{1}{\sqrt{1-\left(\frac{3}{8}\right)^2}}\left\{\dfrac{x_0}{c}+\dfrac{3x_0}{8c}\right\} = \dfrac{11}{\sqrt{55}}\dfrac{x_0}{c}\,.$\\\\

{\bf Hints 5.14:}~Supposed the rod is at rest in S-frame and the rod makes an angle $2 \theta$ with x-axis. suppose $S'$ be another inertial frame moving relative to S along the common x-axis. Thus projection of the rod along the x-axis is $ l ~cos2 \theta$ and it is $ l ~sin2 \theta$ $\perp$ to x-axis. \\

Thus 
\begin{eqnarray}
	l'_{x} &=& \frac{l~ cos 2\theta}{\sqrt{1- \mu^2}}~,~ l'_{\perp} = l~ sin 2\theta ~, ~ \mu = \frac{v}{c} \nonumber \\
	\therefore ~~ (l')^2 &=& (l'_x)^2 + (l'_{\perp})^2 = \frac{l^2 ~ cos^2 2 \theta}{1- \mu^2} + l^2~ sin^2 2\theta = \frac{l^2- \mu^2~ l^2 sin^2 2\theta}{1- \mu^2} \nonumber \\
	\therefore ~~ l' &=& l \sqrt{\frac{l^2- \mu^2~ l^2 sin^2 2\theta}{1- \mu^2} } \nonumber
\end{eqnarray}\\

{\bf  Hints 5.15:} ~~Suppose the surface of the table is chosen as $xy$-plane of $S$-frame.\\

By Lorentz transformation\,: ~~~$x'_2 - x'_1 = \dfrac{(x_2-x_1) - v(t_2-t_1)}{\sqrt{1-\frac{v^2}{c^2}}}\,.$

Here $x'_2-x'_1 = \dfrac{1}{2}~,~~v=0.8c~,~~t_2-t_1$ (both ends of the diameter of the hole measured simultaneously).\\
Hence $x_2-x_1=0.3\,.$

Let $(x_0,y_0,0,t)$ and $(x,y,0,t)$ be the co-ordinates of the center of the hole and a point on the circumference of the hole relative to $S$-frame. The corresponding co-ordinates in $S'$-frame be $(x'_0,y'_0,0,t'_1)$ and $(x',y',0,t'_2)$ respectively so
\begin{eqnarray}
	x' - x'_0 &=& \frac{(x-vt) - (x_0-vt)}{\sqrt{1-\frac{v^2}{c^2}}} = \gamma (x-x_0)~~,~~~\gamma = \frac{1}{\sqrt{1-\frac{v^2}{c^2}}} \nonumber \\
	y' - y'_0 &=& y-y_0\,. \nonumber
\end{eqnarray}
Hence $\dfrac{(x'-x'_0)^2}{\gamma ^2} + (y'-y'_0)^2 = (x-x_0)^2 + (y-y_0)^2 = \dfrac{1}{4}$~~(as the rest radius of the hole is $=\dfrac{1}{2}$)

$\Rightarrow ~~\dfrac{16(x'-x'_0)^2}{\left({1-\frac{v^2}{c^2}}\right)^2} + 16(y'-y'_0)^2 = 1$ ,~~~an ellipse.\\

{\bf  Hints 5.16:} ~~Change in K.E. = Work done by the external force
$$\Rightarrow ~~\frac{1}{2}mv^2 - \frac{1}{2}m_0.0^2 = F.v.t$$
$$\Rightarrow  ~~\frac{v}{\sqrt{1-\frac{v^2}{c^2}}} = 2F.t~,~~\Rightarrow ~~v = \frac{2Fct}{\sqrt{m_0^2c^2+4F^2t^2}}$$
classical limit\,: ~~$c \rightarrow \infty$\\
velocity after long time $i.e.~t \rightarrow \infty$ ,~~$v \rightarrow c$\\

{\bf  Hints 5.18:} ~~The vector connecting these two points be
$$\textit{\textbf a} = \left(\frac{1}{c},2,1,1\right)$$
$$\left\|\textit{\textbf a}\right\|^2 = -\frac{1}{c^2}.c^2 + 4 + 1 + 1 = 5 > 0$$
so they are not causally connected.\\\\
\begin{wrapfigure}{r}{0.4\textwidth}
	\includegraphics[height=3.8 cm , width=8 cm ]{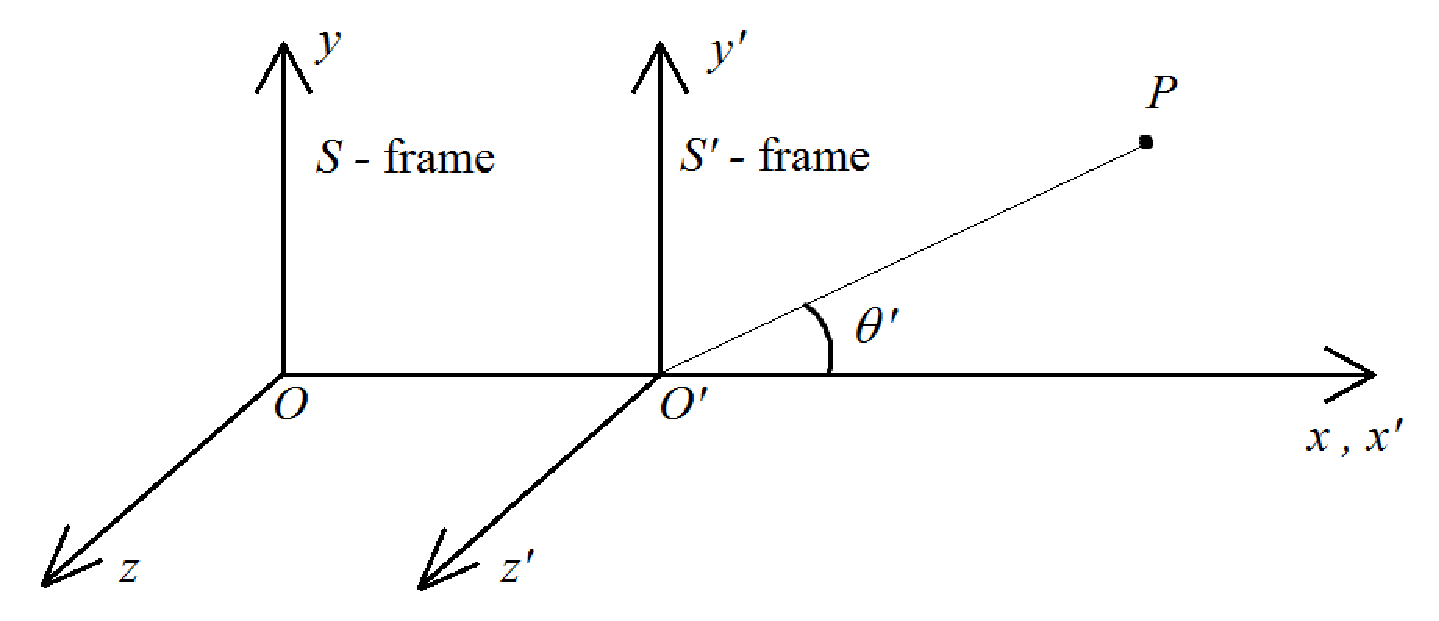}
\end{wrapfigure}
{\bf  Hints 5.20:} ~~$\dfrac{dx'}{dt'} = u'\cos \theta '~,~~\dfrac{dy'}{dt'} = u'\sin \theta '~,~~\dfrac{dz'}{dt'} = 0$\\
choosing initially ~$x'=0=y'=z'$ one gets: ~$x' = u't'\cos \theta '~,~~y' = u't'\sin \theta '~,~~z' = 0$\\
By Lorentz transformation for velocity components\,:
$$u_x = \frac{dx}{dt} = \frac{u'_x + v}{1+\frac{u'_xv}{c^2}} = \frac{u'\cos \theta ' + v}{1+\frac{u'\cos \theta '.v}{c^2}}$$
$$u_y = \frac{dy}{dt} = \frac{u'_y\sqrt{1-\frac{v^2}{c^2}}}{1+\frac{u'_xv}{c^2}} = \frac{u'\sin \theta '\sqrt{1-\frac{v^2}{c^2}}}{1+\frac{u'v\cos \theta '}{c^2}}$$
$$u^2 =u_x^2 + u_y^2\,,~~~\mbox{and}~~~~~\tan \theta = \frac{u_y}{u_x}$$\\
{\bf  Hints 5.21:} ~~Let $G=\left(
\begin{array}{rr}
	1 & 0 \\
	0 & -1
\end{array} \right)$ ~~then $L^TGL=G$

$i.e.~~G$ is self-conjugate under $L$\,.\\
Now, $\det G = \det (L^TGL)=(\det L)^2\det G$

$i.e.~~(\det L)^2 = 1~~~i.e.~~\det L = +1~~(\because \det L > 0~~\mbox{given})$\\
Now~~$(L_1L_2)^TG\,L_1L_2=L_2^T(L_1^TGL_1)L_2 = L_2^TGL_2 = G$\\
$\Rightarrow ~~L_1 \in \mathcal{F}~,~~L_2 \in \mathcal{F}$, then ~$L_1L_2 \in \mathcal{F}$

$I_2 = \left(
\begin{array}{cc}
	1 & 0 \\
	0 & 1
\end{array} \right)$ is clearly the identity.\\\\
Also $(L^{-1})^TG(L^{-1}) = (L^T)^{-1}GL^{-1} = (L^T)^{-1}(L^TGL)(L^{-1})$\\
$~~~~~~~~~~~=I_2GI_2 = G$\\
Hence $\mathcal{F}$ forms a group.\\\\
For 2D Lorentz transformation\,:
$$\left(
\begin{array}{c}
	ct' \\
	x'
\end{array} \right) = \left(
\begin{array}{cc}
	\gamma & -\frac{\gamma v}{c} \\
	-\frac{\gamma v}{c} & \gamma
\end{array} \right)\left(
\begin{array}{c}
	ct \\
	x
\end{array} \right)$$
$$i.e.~~~L=\left(
\begin{array}{cc}
	\gamma & -\frac{\gamma v}{c} \\
	-\frac{\gamma v}{c} & \gamma
\end{array} \right)~~~\Rightarrow ~~\det L = 1~,~~l_{11}=\gamma > 0\,.$$
Hence 2D Lorentz transformation forms a group.\\\\

{\bf  Hints 5.22:} Let $v$ be the velocity of the rocket. The time taken by the rocket to reach the star as measured by an observer on earth is $ \frac{5cY}{v}, ~Y= 365 \times 24 \times 60 \times 60 ~sec.$

By L T, 

\begin{eqnarray}
	t' &=& \frac{t- \frac{x~v}{c^2}}{\sqrt{1- \frac{v^2}{c^2}}} \implies Y = \frac{ \frac{5cY}{v}- \frac{5cYv}{c^2}}{\sqrt{1- \frac{v^2}{c^2}}} \nonumber \\
	\implies \sqrt{1-x^2} &=& 5 (\frac{1}{x}-x)~,~~ x= \frac{v}{c} \nonumber \\
	\implies x &=& \frac{5}{\sqrt{26}} ~~ i.e.~ v= \frac{5c}{\sqrt{26}} \nonumber
\end{eqnarray}\\

{\bf  Hints 5.23:} Let $S'$ be the proper frame of the observer. The relative velocity between $S$ and $S'$ is $v$ along the $+ve$ x- axis. As $u$ be the velocity of the body relative to $S$ along the x-axis so its velocity relative to $S'$ is

\begin{equation}
	u'= \frac{u-v}{(1-\frac{uv}{c^2})}\nonumber
\end{equation}
The volume of the body relative to the observer will be $V$ where 
\begin{equation}
	V=V_0 \sqrt{1-\frac{u'^2}{c^2}} = V_0 \sqrt{1-\frac{(u-v)^2}{c^2(1-\frac{uv}{c^2})^2}} =V_0 \frac{\sqrt{(c^2-u^2)(c^2-v^2)}}{(c^2-uv)} \nonumber
\end{equation}\\

{\bf  Hints 5.24:} 
\begin{equation}
	x'_G = x- v_0t
	~,~~ x'_L = \frac{x- v_0t}{\sqrt{1- \frac{v_0^2}{c^2}}} ~, ~ \frac{x'_G}{ \sqrt{1- \frac{v_0^2}{c^2}}} > x'_0 \nonumber
\end{equation}

As $x'_G$ and $x'_L$ differ by $1\%$ so $x'_L = 1.01 ~ x'_G$
\begin{equation}
	\implies (1.01)~x'_G = \frac{x'_0}{\sqrt{1-\frac{v_0^2}{c^2}}} \implies v_0 \approx 0.1401 ~c \nonumber
\end{equation}

{\bf  Hints 5.25:}

$V_0= \frac{4 \pi}{3}(2~.~2~.~\sqrt{2}) = \frac{16 \sqrt{2} \pi}{3}$

$ V' = V_0 \sqrt{1-\frac{v_0^2}{c^2}} = \frac{16 \sqrt{2}\pi}{3} \sqrt{1- \frac{v_0^2}{c^2}} $

Now if $V'= \frac{1}{2} V_0 \implies \frac{1}{2}= \sqrt{1- \frac{v_0^2}{c^2}} \implies v_0= \frac{\sqrt{3}}{2}~c$ \\

{\bf  Hints 5.26:} The general LT is
\begin{eqnarray}
	\Vec{r'} &=& \Vec{r} -\Vec{v}t + (\gamma -1) \frac{\Vec{v}}{v^2} ( \Vec{r}. \Vec{v}- t v^2) \nonumber \\
	\mbox{and}~~ t'&=& \gamma( t- \frac{\Vec{r} \Vec{v}}{c^2}) ~~,~ \gamma= \frac{1}{\sqrt{1-\frac{v^2}{c^2}}} \nonumber \\
	\therefore ~~\frac{d \Vec{r'}}{dt} &=& \Vec{u}- \Vec{v}+ \frac{(\gamma-1) \Vec{v}}{v^2} (\Vec{u}. \Vec{v}-v^2) \nonumber \\
	\frac{dt'}{dt} &=& \gamma(1- \frac{\Vec{u}. \Vec{v}}{c^2}) \nonumber
\end{eqnarray}
Now,

\begin{eqnarray}
	\Vec{u'} &=& \frac{d \Vec{r'}}{dt'} = \frac{ d\Vec{r'}/ dt}{dt'/ dt}= \frac{ (\Vec{u}-\Vec{v})+ \frac{(\gamma-1)\Vec{v}}{v^2} (\Vec{u}. \Vec{v}-v^2) }{\gamma(1- \frac{\Vec{u} .\Vec{v}}{c^2})} \nonumber \\
	\therefore~~ \mid \Vec{u'} \mid^2 &=&  \frac{ \mid (\Vec{u}-\Vec{v})+ (\gamma-1) \frac{\Vec{v}}{v^2} (\Vec{u}. \Vec{v}-v^2) \mid^2 }{\gamma^2 \left( 1- \frac{\Vec{u}. \Vec{v}}{c^2}\right)^2} \nonumber
\end{eqnarray}

\begin{eqnarray}
	\mbox{Numerator :} &=& \mid \Vec{u}- \Vec{v} \mid^2 + \mid (\gamma-1) \frac{\Vec{v}}{v^2} (\vec{u}. \vec{v}- \vec{v}. \vec{v})  \mid^2 + 2 (\vec{u}- \vec{v}) \frac{\frac{(\gamma-1) \vec{v}}{v^2} }{ ( \vec{u}-\vec{v}). \vec{v}} \nonumber \\
	&=& \mid \Vec{u}- \Vec{v} \mid^2 + \frac{(\gamma -1)^2}{v^2} \mid (\Vec{u}- \Vec{v}). \vec{v} \mid^2 + \frac{2 (\gamma -1)}{v^2} \{ (\Vec{u}- \Vec{v}). \vec{v} \} \times \{ (\Vec{u}- \Vec{v}). \vec{v} \} \nonumber \\
	&=& \mid \Vec{u}- \Vec{v} \mid^2 + \frac{(\gamma-1)^2}{v^2} \{\mid \Vec{u}- \Vec{v} \mid^2 \mid \vec{v} \mid^2 - \mid (\Vec{u}- \Vec{v}) \times \vec{v} \mid^2 \} + 2 \frac{(\gamma-1)}{v^2} \times \nonumber \\ 
	&& \{ \mid \Vec{u}- \Vec{v} \mid^2 v^2 - \mid (\Vec{u}- \Vec{v}) \times \vec{v} \mid^2 \}  ~~ \mbox{( using~ $\mid \Vec{a}~.~\Vec{b} \mid^2 = \mid \vec{a} \mid^2 ~ \mid \vec{b} \mid^2$ )} \nonumber \\
	&=& \mid \Vec{u}- \Vec{v} \mid^2 - \frac{(\gamma-1)^2}{v^2} \mid \vec{u} \times \vec{v} \mid^2 + \mid \Vec{u}- \Vec{v} \mid^2 (\gamma-1) (\gamma-1+2) +2 \frac{(\gamma-1)}{v^2} \mid \vec{u} \times \vec{v} \mid^2 \nonumber \\
	&=& \mid \Vec{u}- \Vec{v} \mid^2 - \frac{(\gamma^2-1)}{v^2} \mid \vec{u} \times \vec{v} \mid^2 - (\gamma^2-1) \mid \vec{u} - \vec{v} \mid^2 \nonumber \\
	&=& \gamma^2 \mid \Vec{u}- \Vec{v} \mid^2 -  \frac{(\gamma^2-1)}{v^2} \mid \vec{u} \times \vec{v} \mid^2 \nonumber \\
	\mbox{As}~~ \frac{(\gamma^2-1)}{v^2} &=& \frac{\frac{1}{1-\frac{v^2}{c^2}} -1}{v^2} = \frac{1}{c^2} \frac{1}{1- \frac{v^2}{c^2}} = \frac{\gamma^2}{c^2}.\nonumber
\end{eqnarray}

So the numerator becomes $\gamma^2 \left[ \mid \vec{u}-\vec{v} \mid^2 - \frac{1}{c^2} \mid \vec{u} \times \vec{v} \mid^2 \right]$

\begin{eqnarray}
	\therefore ~~ \mid \vec{u'} \mid^2 &=& \frac{ \mid \vec{u}-\vec{v} \mid^2 - \frac{1}{c^2} \mid \vec{u} \times \vec{v} \mid^2}{(1- \frac{\vec{u}.\vec{v}}{c^2})^2} \nonumber \\
	\therefore ~~ 1- \frac{\mid \vec{u'} \mid^2}{c^2} &=& 
	\frac{ (1- \frac{\vec{u}.\vec{v}}{c^2})^2 - \frac{1}{c^2}\mid \vec{u}-\vec{v} \mid^2 + \frac{1}{c^4} \mid \vec{u} \times \vec{v} \mid^2 }{(1- \frac{\vec{u}.\vec{v}}{c^2})^2} \nonumber \\
	&=& \frac{\left[ 1- 2 \frac{\vec{u}. \vec{v}}{c^2}+ (\frac{\vec{u}. \vec{v}}{c^2})^2 - \frac{1}{c^2} (u^2 + v^2 -2 \vec{u} \vec{v} )+ \frac{1}{c^4} \mid \vec{u} \times \vec{v} \mid^2 \right]}{(1- \frac{\vec{u}.\vec{v}}{c^2})^2} \nonumber \\
	&=& \frac{1+ \frac{\mid \vec{u} \mid^2}{c^2}.\frac{\mid \vec{v} \mid^2}{c^2} - \frac{ \mid \vec{u} \times \vec{v} \mid^2}{c^4} - \frac{u^2}{c^2}- \frac{v^2}{c^2} +\frac{1}{c^4}  \mid \vec{u} \times \vec{v}\mid^2 }{(1- \frac{\vec{u}.\vec{v}}{c^2})^2} \nonumber \\
	&=& \frac{(1-\frac{u^2}{c^2})(1-\frac{v^2}{c^2})}{(1-\frac{\vec{u}.\vec{v}}{c^2})^2} \nonumber
\end{eqnarray}
Now if $u<c~,~v<c$ then $u'<c$.

If $\vec{u}~||~\vec{v}$ then $\vec{u} \times \vec{v}=0$ and $\vec{u'}=\frac{\vec{u}-\vec{v}}{(1-\frac{\vec{u}.\vec{v}}{c^2})}$ \\

{\bf  Hints 5.27:} $x'=\gamma(x-vt),~t'=\gamma(t- \frac{xv}{c^2})$
Now
\begin{eqnarray}
	\frac{\omega'}{c}(ct'-x')&=& \frac{r \omega'}{c}[ct- \frac{xv}{c}-x+vt ]= \frac{\gamma \omega'}{c}[ (c+v)t- \frac{x}{c}(v+c)] \nonumber \\
	&=& \frac{\gamma \omega' (c+v)}{c^2}(ct-x) \nonumber
\end{eqnarray}
Comparing,
\begin{eqnarray}
	\frac{\omega}{c}&=& \gamma \frac{\omega' (c+v)}{c^2} \nonumber \\
	\mbox{i.e.~} \omega'&=& \frac{\omega c}{\gamma(c+v)} \nonumber \\
	\therefore ~ \omega' &=& \frac{\omega c}{c+v} \sqrt{1-\frac{v^2}{c^2}} = \omega \sqrt{\frac{c-v}{c+v}} \nonumber
\end{eqnarray}\\

{\bf  Hints 5.28:} In Poincare representation $x_4=ict$ (i.e. $x_4$ is purely imaginary) and here an orthogonal frame ($x_1~-~x_4$) is transformed to another orthogonal frame ($x'_1~-~x'_4$). On the otherhand, in case of Minkowski's representation $x_4=ct$ (i.e. $x_4$ is real) and an orthogonal frame $(x_1~-~x_4)$ is transformed to an oblique frame $(x_1'-x_4')$. So according to Poincare $2D$ LT (in $(x_1~-~x_4)$) can be thought of as a rotation of one orthogonal to another through an imaginary angle. Moreover, the units of length and time in the 1st system are not the same as those in the second one.\\

{\bf  Hints 5.29:} Let $\beta = \frac{v}{c},~ x_4=ct.$ Then the $2D$ L.T gives
\begin{equation}
	x_1'= \gamma(x_1- \beta x_4)~,~ x_4'=\gamma(x_4-\beta x_1)\nonumber
\end{equation}
Equation of $x_1'-$ axis (the space in $S'-$ frame) is $x_4'=0~i.e.~ x_4= \beta x_1$. So angle made by $x_1'-$ axis with $x_1$ axis is $tan^{-1} \beta$.

Similarly equation of $x_4'-$ axis (the time axis in $S'-$ frame) is given by $x_1'=0$ i.e. $x_1=\beta x_4$. So the angle made by $x_4'-$ axis with $x_4-$ axis is  $tan^{-1} \beta$. Hence they are equal.\\

The hyperbolas $c^2t^2-x^2= \pm1$ i.e. $x_4^2-x_1^2 = \pm 1$ are called calibration curves as their intersections the co-ordinate axes determine the units of length and units of time in the inertial frame $S$.\\

{\bf  Hints 5.30:} A world line is a curve in the $4D$ Minkowski space that represents a succession of events in the physical world. So it is taken to represent the history of a material point as it moves in time through the $3D$ physical space.\\

{\bf  Hints 5.31:} Consider a car and a garage both with proper length $l_0$. When at rest, car can be parked exactly inside the garage. Now a person driving the car towards the garage with speed $v$. For gateman, the car appears to be of length $l=l_0 \sqrt{1- \frac{v^2}{c^2}}<l_0$. So he realizes that the car smoothly enters the garage and the driver does not need to stop the car before the garage. On the otherhand, the driver realizes that the garage length is smaller than the car and he stops the car before the garage. This is called the car- garage paradox.\\

{\bf  Hints 5.32:} In addition to the postulates of STR one has to make assumptions of homogeneity and isotropy. Due to homogeneity the transformation equations are linear i.e. 

\begin{eqnarray}
	x' &=& a_{11}x + a_{12}y +a_{13}z +a_{14}t \nonumber \\
	y' &=& a_{21}x + a_{22}y +a_{23}z +a_{24}t \nonumber \\
	z' &=& a_{31}x + a_{32}y +a_{33}z +a_{34}t \nonumber \\
	t' &=& a_{41}x + a_{42}y +a_{43}z +a_{44}t \nonumber .
\end{eqnarray}
Here the coefficients $a_{\mu \nu}~ (\mu , \nu =1,2,3,4)$ may depend on relative velocity $v$.

As x axis coincides continuously with $x'-$ axis so one must have $y'=0=z'$ whenever $y=0=z$.

So $y'=a_{22}y+a_{23}z~,~z'=a_{32}y+a_{33}z$.

Similarly, the $xy$ plane $i.e.~z=0$ plane should transformed to $x'y'$ plane $i.e.~z'=0$ plane and similarly $zx$- plane i.e. $y=0$ goes to $z'x'-$ plane i.e. $y'=0$.

$~~~~~~~~~~~~~~~~~~~~~~~~~~\therefore  y'= a_{22}y~,~ z'= a_{33}z$

Now $a_{22}$ and $a_{33}$ can be determined from relativity postulate as $a_{22} = \frac{1}{a_{22}}$ and $a_{33} = \frac{1}{a_{33}}$ i.e. $y'=y~,~ z'=z$.

Further, $x'=0$ gives $x=vt$ so one has 
\begin{equation}
	x'= a_{11}(x-vt)\nonumber
\end{equation}
Due to isotropy condition $t'$ also does not depend on $y$ and $z$. Hence
\begin{equation}
	t'= a_{41}x +a_{44}t \nonumber
\end{equation}
Hence we have
\begin{equation}
	x'=a_{11}(x-vt)~,~y'=y~,~z'=z`,~t'= a_{41}x +a_{44}t. \nonumber
\end{equation}\\

{\bf  Hints 5.33:} The $2D$ L.T : $x'= \gamma(x-vt)~,~ t'= \gamma(t-\frac{xv}{c^2})$

\begin{equation}
	\therefore~~ \begin{pmatrix}
		x' \\
		ct'
	\end{pmatrix} =
	\begin{pmatrix}
		\gamma & -\gamma \frac{v}{c} \\
		-\gamma \frac{v}{c} & \gamma
	\end{pmatrix}
	\begin{pmatrix}
		x \\
		ct
	\end{pmatrix}~,~ \gamma= \frac{1}{\sqrt{1-\frac{v^2}{c^2}}}>1 \nonumber
\end{equation}
Let $\gamma= cosh \phi~,~ \beta= \frac{v}{c}= tanh \phi$ and we have 

\begin{equation}
	\begin{pmatrix}
		x' \\
		ct'
	\end{pmatrix} =
	\begin{pmatrix}
		cosh \phi & -sinh \phi \\
		-sinh \phi  & cosh \phi
	\end{pmatrix}
	\begin{pmatrix}
		x \\
		ct
	\end{pmatrix} \nonumber
\end{equation}

\begin{equation}
	\mbox{i.e.}~~ L=  \begin{pmatrix}
		x' \\
		ct'
	\end{pmatrix} =
	\begin{pmatrix}
		cosh \phi & -sinh \phi \\
		-sinh \phi  & cosh \phi
	\end{pmatrix}
	\begin{pmatrix}
		x \\
		ct
	\end{pmatrix} \nonumber
\end{equation}
$L_{11}= cosh \phi >0, ~ det(L)= cosh^2 \phi - sinh^2 \phi =1 >0$. So $L$ is $+ve$ definite.

Now,
\begin{eqnarray}
	L^TgL &=& \begin{pmatrix}
		cosh \phi & -sinh \phi \\
		-sinh \phi  & cosh \phi
	\end{pmatrix}
	\begin{pmatrix}
		1 & 0 \\
		0 & -1
	\end{pmatrix}
	\begin{pmatrix}
		cosh \phi & -sinh \phi \\
		-sinh \phi  & cosh \phi
	\end{pmatrix} \nonumber \\
	&=& 
	\begin{pmatrix}
		1 & 0 \\
		0 & -1
	\end{pmatrix} =g \nonumber
\end{eqnarray}
Let
\begin{equation}
	L_1 = \begin{pmatrix}
		cosh \phi & -sinh \phi \\
		-sinh \phi  & cosh \phi
	\end{pmatrix}~,~~ L_2 = \begin{pmatrix}
		cosh \psi & -sinh \psi \\
		-sinh \psi  & cosh \psi
	\end{pmatrix} \in I~\subset M(2, R) \nonumber
\end{equation}

\begin{equation}
	L_1 \circ L_2 = \begin{pmatrix}
		cosh (\phi + \psi) & -sinh (\phi + \psi) \\
		-sinh (\phi + \psi)  & cosh (\phi + \psi)
	\end{pmatrix} \in I \nonumber
\end{equation}
$\implies I$ is closed under matrix multiplication.
For $\phi=0$, $L_0 = \begin{pmatrix}
	1 & 0 \\
	0 & 1
\end{pmatrix} $, the identity element.

Associativity follows from matrix multiplication.

Putting $\psi = - \phi$
\begin{equation}
	L_{-\phi}= \begin{pmatrix}
		cosh \phi & sinh \phi \\
		sinh \phi  & cosh \phi
	\end{pmatrix} \nonumber
\end{equation}
and $L_\phi \circ L_{-\phi} = L_{- \phi} \circ L_{\phi} = I_{2 \times 2} \implies I$ is a commutative group.

Now let $S$ and $S'$ be two inertial frames connected by the linear transformation

\begin{equation}
	\begin{pmatrix}
		x' \\
		ct'
	\end{pmatrix} = \begin{pmatrix}
		L_{11} & L_{12} \\
		L_{21} & L_{22}
	\end{pmatrix} \begin{pmatrix}
		x \\
		ct
	\end{pmatrix} \nonumber
\end{equation}
As $L$ is $+ve$ definite so $L_{11}>0$ and $det L >0$. 

Also $ L^TgL =g~~, ~g = \begin{pmatrix}
	1 & 0 \\
	0 & -1
\end{pmatrix} $
\begin{eqnarray}
	\implies \begin{pmatrix}
		L_{11} & L_{12} \\
		L_{21} & L_{22}
	\end{pmatrix} \begin{pmatrix}
		1 & 0 \\
		0 & -1
	\end{pmatrix}  \begin{pmatrix}
		L_{11} & L_{12} \\
		L_{21} & L_{22}
	\end{pmatrix} = \begin{pmatrix}
		1 & 0 \\
		0 & -1
	\end{pmatrix} \nonumber \\
	\implies \begin{pmatrix}
		L_{11}^2 - L_{21}^2 & L_{11}L_{12}- L_{21}L_{22} \\
		L_{11}L_{12}- L_{21}L_{22} &  L_{12}^2 - L_{22}^2
	\end{pmatrix}= \begin{pmatrix}
		1 & 0 \\
		0 & -1
	\end{pmatrix} \nonumber
\end{eqnarray}
\begin{eqnarray}
	\implies ~    L_{11}^2 - L_{21}^2 &=& 1 \label{e1} \\
	L_{11}L_{12}- L_{21}L_{22} &=& 0 \label{e2} \\
	L_{12}^2 - L_{22}^2 &=& -1 \label{e3}
\end{eqnarray}

From equation $(\ref{e2})$
\begin{eqnarray}
	L_{11}L_{12} &=& L_{21}L_{22} \nonumber \\
	\implies \frac{L_{11}}{L_{21}} &=& \frac{L_{22}}{L_{12}} =\frac{1}{\lambda} (\mbox{say})~, \lambda >0 \\
	\implies L_{21}&=& \lambda L_{11}~,~ L_{12}= \lambda L_{22} \nonumber
\end{eqnarray}
From equation $(\ref{e1})$
\begin{eqnarray}
	\implies L_{11}^2 - L_{21}^2 &=& 1 \nonumber \\
	\implies L_{11}^2 - \lambda L_{11}^2 &=& 1 \nonumber \\
	\implies L_{11} &=& \frac{1}{\sqrt{1- \lambda^2}} ~~( \because L_{11} >0 ) \nonumber \\
	\therefore L_{21} &=& \frac{\lambda}{\sqrt{ 1- \lambda^2}} \nonumber
\end{eqnarray}

From equation $(\ref{e3})$
\begin{eqnarray}
	\lambda^2 L_{22}^2 - L_{22}^2 &=& -1 \nonumber \\
	L_{22}^2 &=& \frac{1}{1-\lambda^2} \nonumber\\
	\therefore L_{22} &=& \pm \frac{1}{\sqrt{1-\lambda^2}}~,~ L_{12}= \pm \frac{\lambda}{\sqrt{1- \lambda^2}} \nonumber
\end{eqnarray}
Now, if $L_{22}<0$ and $L_{12}<0$ then $det L<0$ which is not possible.

If $L_{22}<0 ~,~ L_{12}>0 ~,$ then $det L = -\frac{1}{1-\lambda^2}- \frac{\lambda^2}{1-\lambda^2}<0 $ not possible.

If $L_{22}>0 $ and $L_{12}<0 $ , then $det L = \frac{1}{1- \lambda^2}+\frac{\lambda^2}{1- \lambda^2} = \frac{1+ \lambda^2}{1- \lambda^2}>1~,$ not possible.

$\therefore L_{22}>0 ~\mbox{and}~ L_{12}>0 ~\mbox{i.e.}~ L_{22}=\frac{1}{\sqrt{1- \lambda^2}}~,~ L_{12}= \frac{\lambda}{\sqrt{1- \lambda^2}} ~\mbox{then}~ det L= \frac{1}{\sqrt{1- \lambda^2}} \begin{pmatrix}
	1 & \lambda \\
	\lambda & 0
\end{pmatrix} $.

Hence the above transformation is a LT with $ \gamma= \frac{1}{\sqrt{1-\lambda^2}}$ and $\lambda = \beta$.\\

\textbf{5.34:} For $2D$ LT : $S~ \xrightarrow[v]{} ~ S'$
\begin{equation}
	\begin{pmatrix}
		ct' \\
		x'
	\end{pmatrix} = \begin{pmatrix}
		\gamma & -\gamma \frac{v}{c} \\
		-\gamma \frac{v}{c} & \gamma
	\end{pmatrix} \begin{pmatrix}
		ct \\
		x
	\end{pmatrix} \nonumber
\end{equation}
if $\gamma=cosh \phi $ then $\beta= \frac{v}{c} =tanh \phi $ then

\begin{equation}
	L = \begin{pmatrix}
		\gamma & -\gamma \frac{v}{c} \\
		-\gamma \frac{v}{c} & \gamma
	\end{pmatrix} = \begin{pmatrix}
		cosh \phi & -sinh \phi \\
		-sinh \phi & cosh \phi
	\end{pmatrix} \nonumber
\end{equation}
here $\phi$ is termed as rapidity.

If $u$ be the relative velocity between $S$ and $S'$ then
\begin{equation}
	L_u = \begin{pmatrix}
		\gamma & -\gamma \frac{v}{c} \\
		-\gamma \frac{v}{c} & \gamma
	\end{pmatrix} = \phi(u)~,~ tanh \alpha_1 = \frac{u}{c} \nonumber
\end{equation}
Similarly $S' \xrightarrow[v]{}  S''~:~ L_v = \begin{pmatrix}
	\gamma'' & -\gamma'' \frac{v}{c} \\
	-\gamma'' \frac{v}{c} & \gamma''
\end{pmatrix} = \phi(v) ~,~ tanh\alpha_2 = \frac{v}{c} $

and $S ~\vec{w}~ S''~:~ L_w = \begin{pmatrix}
	\gamma'' & -\gamma'' \frac{w}{c} \\
	-\gamma'' \frac{w}{c} & \gamma''
\end{pmatrix} = \phi(w) ~,~ tanh\alpha_3 = \frac{w}{c} $.

As $w =\frac{u+v}{1+\frac{uv}{c^2}} \implies \phi(w) = \phi(u) + \phi (v) $ i.e. it is a homomorphism as $ tanh \alpha_3= tanh (\alpha_1 + \alpha_2) = \frac{tanh \alpha_1 + tanh \alpha_2}{1+ tanh \alpha_1 ~tanh \alpha_2} $.
\begin{eqnarray}
	L_1 &=& L_2 \nonumber \\
	\mbox{i.e.}~ \begin{pmatrix}
		cosh \phi & -sinh \phi \\
		-sinh \phi & cosh \phi
	\end{pmatrix} &=& \begin{pmatrix}
		cosh \psi & -sinh \psi \\
		-sinh \psi & cosh \psi
	\end{pmatrix}  \implies ~ \phi = \psi ~~ \implies \mbox{injective}. \nonumber 
\end{eqnarray}

Similarly, surjectivity can be proved since for every real no., one can get a corresponding $2D$ LT.

Hence the $2D$ LT is an isomorphism from the $2D$ Lorentz group to $(R~,~ +)$.\\

\textbf{5.35:}
The transformation has the explicit form
\begin{equation}
	t= t' + c_0~,~ \vec{r}= t' \vec{v} + A \vec{r'} + \vec{c} \nonumber
\end{equation}
Now,
\begin{eqnarray}
	\vec{u} &=& \frac{d \vec{r}}{dt} = \frac{dt'}{dt} \vec{v} + A \frac{d \vec{r'}}{dt} = \vec{v} + A \frac{d \vec{r'}}{dt'} \frac{dt'}{dt} \nonumber \\
	\implies ~ \vec{u} &=& \vec{v} + A \vec{u'}. \nonumber
\end{eqnarray}
Now, if $\vec{u'}=0$ then $ \vec{u}= \vec{v} \implies \vec{v} $ is the velocity of the inertial frame $S'$ relative to $S$.

Now, $t_2 - t_1 = (t_2' + c_0)- (t_1' + c_0)= t_2'- t_1'~ \implies$ if 2 events are simultaneous in $S-$ frame $i.e.~ t_1=t_2$ then they are also simultaneous in $S'-$ frame $i.e.~ t_1'= t_2'$.

Now, \begin{eqnarray}
	\vec{r_1} &=& A \vec{r_1'} + t_1' \vec{v} + \vec{c} \nonumber \\
	\vec{r_2} &=& A \vec{r_2'} + t_2' \vec{v} + \vec{c} \nonumber\\
	\therefore ~ \vec{r_1} - \vec{r_2} &=& A(\vec{r_1'} - \vec{r_2'}) + \vec{v} (t_1' - t_2') .\nonumber
\end{eqnarray}

Thus under the assumption of simultaneity
\begin{eqnarray}
	\vec{r_1} - \vec{r_2} &=& A(\vec{r_1'} - \vec{r_2'}) ~ \mbox{and}~ ( \vec{r_1} - \vec{r_2})^t = (\vec{r_1'} - \vec{r_2'})^t A^t \nonumber \\
	\therefore ~ \mid \vec{r_1} - \vec{r_2} \mid^2 &=& (\vec{r_1} - \vec{r_2})^t (\vec{r_1} - \vec{r_2}) = (\vec{r_1'} - \vec{r_2'})^t ~A^t A~ (\vec{r_1'} - \vec{r_2'}) \nonumber \\
	&=& \mid \vec{r'_1} - \vec{r'_2} \mid^2 ~~ \mbox{as}~ A^tA = I~, ~ \mbox{due to} ~ A \in So (3).
\end{eqnarray}
Hence spatial separation is also invariant.


\chapter[Einstein's GTR and Cosmology]{Einstein's General Theory of Relativity and Cosmology from Differential Geometric point of view}


\section{An introduction of differential geometric structure through the idea of equivalence principle}

~~~According to Einstein the gravitational field has only a relative existence similar to electric field generated by magneto electric induction. A freely falling observer does not experience any gravitational force in his surroundings. In fact, if he drops some object then it remains relative to him in a state of rest or of uniform motion. This is nothing but the equivalence principle. Based on this principle, Einstein formulated the general theory of relativity\,(which we shall discuss in the next sections).\\

In a static homogeneous gravitational field the particle's equation of motion can be described by Newton's second law as
$$m_i\frac{d^2\textit{\textbf{r}}}{dt^2} = m_g \cdot \textit{\textbf{g}} = \textit{\textbf{F}}_{\bm g}$$

Here $m_i$ and $m_g$ are termed as inertial and gravitational mass of the particle and $\textit{\textbf{g}}$ is the acceleration due to gravity, independent of the position of the particle in four dimensional space-time. In fact, one can interpret $m_i$ and $m_g$ as the measures of the body's resistance to the action of force and its capability of responding to the gravitational field respectively. However, E\"{o}tvos and collaborators showed experimentally that the above two masses are equal\,(another form of equivalence principle). So the above equation of motion simplifies to $\dfrac{d^2\textit{\textbf{r}}}{dt^2} = \textit{\textbf{g}}$ .\\

If we now switch over to a non-inertial frame described by
$$\textit{\textbf{r}}' = \textit{\textbf{r}} - \frac{1}{2}\textit{\textbf{g}}t^2$$
then the equation of motion becomes
$$\frac{d^2\textit{\textbf{r}}'}{dt^2} = 0$$
{\it i.e.} there is no effect of the gravitational field in the primed system. Here primed frame moves relative to the inertial frame with an acceleration $g$ and an observer will not experience any gravitational force there. This is another way of looking into the equivalence principle. Thus gravitational force can be taken into account when we switch over to non-inertial frame of reference.\\

The well known examples of non-inertial frames are {\it (i)} a frame rotating with respect to an inertial frame, {\it (ii)} a frame accelerated with respect to an inertial frame. Ideally, an inertial frame\,(in which Newtonian laws are valid) is specified as one in which a particle with no force on it appears to move with a uniform velocity in a straight line. One can imagine an inertial frame far away from any gravitating matter. However, in a gravitational field one can make it locally inertial\,(in a very small region).\\

We now examine how the metric tensor changes character in a non-inertial frame which is rotating about $z$\,-axis of an inertial frame. The transformation of co-ordinates gives
\begin{eqnarray}
x &=& x' \cos \omega t - y' \sin \omega t \nonumber \\
y &=& x' \sin \omega t + y' \cos \omega t \nonumber \\
z &=& z' \nonumber
\end{eqnarray}
where the constant $\omega$ is the angular velocity of rotation. Thus the Minkowski metric
$$ds^2 = -c^2dt^2 + dx^2 + dy^2 + dz^2$$
changes to
\begin{eqnarray}
ds^2 = -\left[c^2 - \omega ^2\left(x'^{~2} + y'^{~2}\right)\right]dt^2 &-& 2\omega dt \left(y'dx' - x'dy'\right) \nonumber \\
&+& \left(dx'^{~2} + dy'^{~2} + dz'^{~2}\right) \nonumber
\end{eqnarray}
{\it i.e.} the metric co-efficients are no longer constants rather they are functions of space-time co-ordinates. So in general for non-inertial co-ordinates one can write the line element as
\begin{equation} \label{6.1}
ds^2 = g_{\mu \nu}\left(x'\right)dx'^{\, \mu}~dx'^{\, \nu}.
\end{equation}

This is also true for accelerating frame {\it i.e.} when the new co-ordinates describe a frame accelerated with respect to an inertial frame.\\

In special theory of relativity, the equation of motion in Minkowski co-ordinates is given by
\begin{equation} \label{6.2}
\frac{d^2 x^{\alpha}}{d\tau ^2} = 0
\end{equation}
where $\{x^\alpha\}$ is an inertial co-ordinate system and $\tau$ denotes the proper time.\\

We now switch over to non-inertial frame of reference $\{y^\alpha\}$ so that
\begin{eqnarray} \label{6.3} 
\frac{dx^{\alpha}}{d\tau} &=& \left(\frac{\partial x^{\alpha}}{\partial y^{\mu}}\right)\left(\frac{dy^{\mu}}{d\tau}\right) \nonumber \\
\mbox{and}~~~~\frac{d^2 x^{\alpha}}{d\tau ^2} &=& \left(\frac{\partial x^{\alpha}}{\partial y^{\mu}}\right)\left(\frac{d^2 y^{\mu}}{d\tau ^2}\right) + \left(\frac{\partial ^2 x^{\alpha}}{\partial y^{\mu} \partial y^{\nu}}\right)\left(\frac{dy^{\mu}}{d\tau}\right)\left(\frac{dy^{\nu}}{d\tau}\right) = 0 \nonumber \\
i.e.~~~~\frac{d^2 y^{\lambda}}{d\tau ^2} &=& -\Gamma _{\mu \nu}^{\lambda}\left(\frac{dy^{\mu}}{d\tau}\right)\left(\frac{dy^{\nu}}{d\tau}\right)
\end{eqnarray}
\begin{equation} \label{6.4}
\mbox{where}~~~~~\Gamma _{\mu \nu}^{\lambda} = \left(\frac{\partial ^2 x^{\alpha}}{\partial y^{\mu} \partial y^{\nu}}\right)\left(\frac{\partial y^{\lambda}}{\partial x^{\alpha}}\right)
\end{equation}

Now due to invariance of $ds^2$ we have
$$ds^2 = \eta _{\alpha \beta}dx^{\alpha}dx^{\beta} = g_{\mu \nu}dy^{\mu}dy^{\nu}$$
\begin{equation} \label{6.5}
i.e.~~~g_{\mu \nu} = \eta _{\alpha \beta}\left(\frac{\partial x^{\alpha}}{\partial y^{\mu}}\right)\left(\frac{\partial x^{\beta}}{\partial y^{\nu}}\right)
\end{equation}
\begin{eqnarray} \label{6.6} 
\mbox{So}~~~~~\frac{\partial g_{\mu \nu}}{\partial y^{\lambda}} &=& \eta _{\alpha \beta}\left(\frac{\partial ^2 x^{\alpha}}{\partial y^{\mu} \partial y^{\lambda}}\right)\left(\frac{\partial x^{\beta}}{\partial y^{\nu}}\right) + \eta _{\alpha \beta}\left(\frac{\partial ^2 x^{\beta}}{\partial y^{\nu} \partial y^{\lambda}}\right)\left(\frac{\partial x^{\alpha}}{\partial y^{\mu}}\right) \nonumber \\
&=& g_{\nu \delta}\Gamma _{\mu \lambda}^{\delta} + g_{\mu \delta}\Gamma _{\lambda \nu}^{\delta}
\end{eqnarray}
which on simplification gives
\begin{equation} \label{6.7}
\Gamma _{\lambda \mu \nu} = \Gamma _{\lambda \mu}^{\delta}g_{\delta \nu} = \frac{1}{2}\left(\frac{\partial g_{\lambda \nu}}{\partial x^{\mu}} + \frac{\partial g_{\mu \nu}}{\partial x^{\lambda}} - \frac{\partial g_{\lambda \mu}}{\partial x^{\nu}}\right)
\end{equation}
and are termed as Christoffel symbols.\\

Equation\,(\ref{6.3}) is the geodesic equation in the non-inertial frame. In analogy with Newtonian theory $\Gamma _{\mu \nu}^{\lambda}$ can be interpreted as the force term and the metric tensor components $g_{\mu \nu}$ represent potential term\,(force is the gradient of the potential).\\

It is well known that partial derivative of any tensor is not a tensor. So to introduce a derivative operator that after differentiation will also be a tensor we proceed as follows\,:
\begin{eqnarray}
A_{\mu}& =& g_{\mu \nu}A^{\nu}\nonumber\\
\frac{\partial A_{\mu}}{\partial y^{\lambda}} &=& \frac{\partial g_{\mu \nu}}{\partial y^{\lambda}}A^{\nu} + g_{\mu \nu}\frac{\partial A^{\nu}}{\partial y^{\lambda}} \nonumber \\
&=& \left(\Gamma _{\mu \lambda}^{\delta}g_{\nu \delta} + \Gamma _{\lambda \nu}^{\delta}g_{\mu \delta}\right)A^{\nu} + g_{\mu \nu}\frac{\partial A^{\nu}}{\partial y^{\lambda}} \nonumber\\
i.e.~~~~\frac{\partial A_{\mu}}{\partial y^{\lambda}} - \Gamma _{\lambda \mu}^{\delta}g_{\nu \delta}A^{\nu}& =& g_{\mu \nu}\frac{\partial A^{\nu}}{\partial y^{\lambda}} + \Gamma _{\lambda \delta}^{\nu}g_{\mu \nu}A^{\delta}~~~~~~(\nu \rightleftharpoons \delta)\nonumber\\
i.e.~~~~\frac{\partial A_{\mu}}{\partial y^{\lambda}} - \Gamma _{\lambda \mu}^{\delta}A_{\delta}& =& g_{\mu \nu}\left(\frac{\partial A^{\nu}}{\partial y^{\lambda}} + \Gamma _{\lambda \delta}^{\nu}A^{\delta}\right)\nonumber
\end{eqnarray}

Thus if we define,~~~~~~~~$A_{\mu ;\lambda} = \dfrac{\partial A_{\mu}}{\partial y^{\lambda}} - \Gamma _{\lambda \mu}^{\delta}A_{\delta}$\\
and
\begin{equation} \label{6.8}
A_{~;\lambda}^{\nu} = \frac{\partial A^{\nu}}{\partial x^{\lambda}} + \Gamma _{\lambda \delta}^{\nu}A^{\delta}
\end{equation}
then we have
$$A_{\mu ;\lambda} = g_{\mu \nu}A_{~;\lambda}^{\nu}\,.$$

So by quotient law if $A_{~;\lambda}^{\nu}$ is a (1, 1)\,-tensor then $A_{\mu ;\lambda}$ is a (0, 2)\,-tensor and vice-versa.\\

The differentiation defined in (\ref{6.8}) is termed as covariant differentiation of contravariant vector $A^{\nu}$ and covariant vector $A_{\mu}$ respectively.\\

Further due to Leibnitz property for covariant differentiation one immediately gets $g_{\mu \nu ;\lambda} = 0$ {\it i.e.} the connection\,(Christoffel symbols) is metric compatible.\\

For partial derivatives, second order differentiation is commutative due to Schwarz but it is not true for covariant differentiation. This distinct feature of curved geometry has some interesting features in the geometric structure of the space-time. In fact, Riemann curvature tensor measures this non-commutativity as
\begin{equation} \label{6.9}
A_{~;\lambda \delta}^{\mu} - A_{~;\delta \lambda}^{\mu} = R_{~ \nu \delta \lambda}^{\mu}A^{\nu}
\end{equation}
with
\begin{equation} \label{6.10}
R_{~ \nu \delta \lambda}^{\mu} = \frac{\partial}{\partial x^{\delta}}\Gamma _{\lambda \nu}^{\mu} - \frac{\partial}{\partial x^{\lambda}}\Gamma _{\delta \nu}^{\mu} + \Gamma _{\lambda \nu}^{\alpha} \Gamma _{\alpha \delta}^{\mu} - \Gamma _{\delta \nu}^{\alpha} \Gamma _{\alpha \lambda}^{\mu}
\end{equation}

Thus by equivalence principle, it has been shown how gravity is introduced through accelerated frame of reference (i.e., non-inertial frame of reference) and space-time geometry changes to curved geometry.\\

\section{Concept of global and local velocity}

~~~In STR, it is found that a massive particle locally moves at a speed less than the velocity of light\,(the absolute velocity). Also it has a well defined four velocity with constant norm. On the other hand, a massless particle locally (i.e. at the same point as the observer) always moves at the speed of light. However, proper time cannot be defined for it and hence a null particle cannot have the idea of four velocity.\\

The situation is totally different in GTR. One of the major differences between STR and GTR is that in the former, inertial coordinate systems are globally defined while in the later, inertial coordinate systems, can only be defined locally, at a particular point of the space-time. Thus one has an arbitrary curved space-time in GTR. Let the line element be $$ds^2=-v^2dt^2+dx^2$$
with $v$ an arbitrary real number. So for massless particle (i.e. $ds=0$) the velocity is $$\frac{dx}{dt}=\pm v$$

Thus speed of a massless particle is arbitrary. This is termed as co-ordinate speed, not the local speed. Due to general covariance (i.e, diffeomorphism invariance) GTR holds in any co-ordinate system and the co-ordinate speed will naturally depend on the choice of the coordinate system.\\

However, the result ``a massless particle always locally moves at the speed of light" is a universal one and it holds in GTR also. This can be seen as follows: At any particular point $P$ (of the space-time) one can always construct a locally inertial coordinate system with properties\,: (i) $g_{\mu\nu}(P)=\eta_{\mu \nu}$ (Minkowski metric), (ii) $\partial_\alpha g_{\mu\nu}(P)=0$ and (iii) $\dfrac{\partial^2g_{\mu\nu(P)}}{\partial x^\alpha\partial x^\beta}\neq0$ for at least one of the choices for $\mu$, $\nu$, $\alpha$ and $\beta$. So an observer at $P$ will have flat Minkowski space-time in his neighbourhood. As a result the inertial observer will measure the speed of a massless particle to be the velocity of light. Thus in GTR the co-ordinate velocity is totally arbitrary (may even be larger than the velocity of light) while local velocity of a massless particle is universal.\\

Let us now consider the well known non-static but homogeneous and isotropic FLRW space-time (this space-time is the space-time of standard cosmology and it will be discussed in details in subsequent section) having line element $$ds^2=-dt^2+a^2(t)d\Sigma^2_3$$
where $d\Sigma^2_3$ is the line element for the 3D spatial $t$=constant hypersurfaces having uniform curvature (chosen to be flat). Hence $$d\Sigma^2_3=dx^2+dy^2+dz^2=dr^2+r^2(d\theta^2+\sin^2\theta d\phi^2).$$

In the above the function $a(t)$ is called the scale factor as it scales the spatial distances measured within the spatial hypersurfaces. Without any loss of generality, at present epoch  $a(t)=1$ is chosen to make the whole 4D space-time to be flat.\\

One can define the proper distance as the spatial distance measured with the metric (i.e. the co-ordinate distance multiplied by the scale factor). Due to expansion of our universe, $a(t)$ increases with time. As a consequence, if two galaxies are at rest, still the proper distance between them increases with time, while the co-moving distance between them remains constant. Thus if $D_0$ be the proper distance of a galaxy from us at present epoch its proper distance at a later time $t$ will be $$D(t)=a(t)D_0$$

Then the recession velocity of the galaxy with respect to us is given by $$\dot{D}(t)=\dot{a}(t)D_0=HD(t)$$
where, $H(t)=\dfrac{\dot{a}(t)}{a(t)}$ is called the Hubble parameter. At present epoch, $\dot{D}(t)=H_0D$ with $H_0=70$ Km/s/Mpc. This is Hubble's law. It states that the recession velocity of a galaxy from us is proportional to its distance from us.\\

Numerically, suppose there is a galaxy at a distance $1$ Mpc $\simeq$ $3$ Mly $\simeq$ $3\times10^{19}$ Km away from us, its recession velocity $\simeq$ $70$ Km/s. So a galaxy further away say $4.5$ Gpc $\simeq$ 14 Gly\,(approx.) is receding faster than light.\\

The recession velocity $\dot{D}$ is only a global velocity due to space itself expanding while the galaxy's local velocity (known as peculiar velocity) in space, relative to nearby galaxies is independent from $\dot{D}$ and always less than $c$ as it locally follows a time like path.\\

Thus there is an ambiguity in the universal speed limit: local speed is within space and is bounded  by the speed of light while the global velocity due to expansion of space itself is unbounded.\\

\section{Heuristic Derivation of Einstein's Equations for Gravity}

~~~There is a long of history how Einstein through continuous effort over ten years\,(1905$-$1915) was successful in moving from the formulation of the special theory of relativity\,(1905) to the theory of gravity\,(1915) -- the general theory of relativity. This theory shows a description of gravity and its action on matter in a pseudo-Riemannian manifold which is characterized by the metric tensor. The field equations show the source of the gravitational field determine the metric and vice-versa.\\

Einstein had in mind that the field equations should have some similarity with Newton's theory of gravity. The source of gravity in Newtonian theory is the mass density. So in relativistic arena the matter source should be a relativistic generalization of mass density {\bf --} the total energy which includes the rest mass. As $\rho$ is the energy density measured by a frame of reference so use of $\rho$ as the source of the field implies one class of observers is preferred than all others. This idea is at variance with the Einstein's idea of general covariance -- all co-ordinate systems on an equal footing. Further the equivalence of mass and energy from special relativity suggests that all forms of energy may be considered as sources for the gravitational field. Hence the whole of the stress-energy tensor $T_{\mu \nu}$\,($\rho$ is a component of it) is chosen as the source of the gravitational field.\\

The basic idea of Einstein's theory of gravitation consists of geometrizing the gravitational force {\it i.e.} mapping all properties of the gravitational force and its influence upon physical processes on to the properties of a (pseudo)\,Riemannian space. So considering (pseudo)\,Riemannian space as the geometry of space-time, Einstein derived logically the new fundamental physical law from the laws already known. This should show how the sources of the gravitational field determine the metric.\\

The logical arguments by which Einstein obtained the field equations for gravity are the following.\\

{\bf (a)} The space-time is a four dimensional pseudo-Riemannian manifold with a metric which can be put in the Minkowskian form $\eta _{\alpha \beta}$ at any point by an appropriate choice of co-ordinates (Locally inertial frame).\\

{\bf (b)} For a freely falling particle one can eliminate gravity locally and employ special relativity {\it i.e.} locally, one can not distinguish gravity from a uniformly accelerated inertial field and hence gravity can be considered as an inertial force\,(weak equivalence principle).\\

{\bf (c)} In special relativity, a free falling particle moves on time-like geodesic of the space-time. The effect of gravity through inertial force can be taken into account through metric connection of the four dimensional manifold.\\

{\bf (d)} To have an analogy with Newtonian theory the metric should play the role of the gravitational potential. As Poisson's equation describes Newtonian gravity so the field equations should be quasi linear second order partial differential equations in the metric.\\

{\bf (e)} Due to principle of general covariance the field equations must be tensorial in character.\\

{\bf (f)} From the point of view of non-local effects, gravity can be measured through the variation in the field which causes the test particle to travel on time-like geodesics. The convergence or divergence of these geodesics are described by geodesic deviation which is characterized by the Riemann curvature tensor.\\

{\bf (g)} As matter is described by stress-energy tensor $T_{\mu \nu}$ , a (0, 2)\,-tensor so gravity can be geometrized by a second rank tensor, obtained from Riemann curvature tensor through contraction.\\

{\bf (h)} The Ricci tensor $R_{\alpha \beta}$ is the natural (0, 2)\,-tensor obtained from Riemann curvature tensor through contraction. It is also symmetric as the stress-energy tensor. Further Ricci tensor contains second order partial derivatives of the metric\,(gravitational potential) and is quasi-linear in nature.\\

{\bf (i)} At first Einstein considered the equivalence of Ricci tensor and stress-energy tensor {\it i.e.} $R_{\alpha \beta} = \kappa T_{\alpha \beta}$ ($\kappa$ is the proportionality constant) as the field equations for gravity.\\

{\bf (j)} In Minkowski co-ordinates, the conversation equation for the energy-momentum tensor is
$$\partial _{\beta}T^{\alpha \beta} = 0.$$

Then due to the principle of minimal gravitational coupling, the general relativistic form of the conservation equation is written as
$$T_{~~;\beta}^{\alpha \beta} = 0$$

But
$$R_{~~;\beta}^{\alpha \beta} = \frac{1}{2}\frac{\partial R}{\partial x^{\alpha}}$$

Hence the field equations can not be chosen as the above form.\\

{\bf (k)} Einstein then tried to find a symmetric (0, 2)\,-tensor as a linear combination of the known (0, 2)\,-tensors namely $R_{\mu \nu}$ and $g_{\mu \nu}$ . So he considered
\begin{equation} \label{6.11}
R_{\mu \nu} + aRg_{\mu \nu} + \Lambda g_{\mu \nu} = G_{\mu \nu} + \Lambda g_{\mu \nu}
\end{equation}
as the desired (0, 2)\,-tensor. Note that $G_{\mu \nu}$ also contains second order derivatives of $g_{\mu \nu}$ and is quasi linear in nature. Now divergence of $G_{\mu \nu}$ gives\,(noting that covariant derivative of $g_{\mu \nu}$ vanishes)
\begin{eqnarray}
G_{~~;\nu}^{\mu \nu} &=& R_{~~;\nu}^{\mu \nu} + ag_{\mu \nu}\frac{\partial R}{\partial x^{\nu}} \nonumber \\
&=& \left(\frac{1}{2} + a\right)\frac{\partial R}{\partial x^{\mu}} \nonumber
\end{eqnarray}

Hence $G_{~~;\nu}^{\mu \nu} = 0$ gives $a = -\dfrac{1}{2}$ .\\

Thus $G^{\mu \nu} = R^{\mu \nu} - \dfrac{1}{2}Rg^{\mu \nu}$.\\

Therefore, the field equations for gravity take the form
$$G_{\mu \nu} + \Lambda g^{\mu \nu} = \kappa T_{\mu \nu}$$
\begin{equation} \label{6.12}
i.e.~~~~ R_{\mu \nu} - \frac{1}{2}Rg_{\mu \nu} + \Lambda g_{\mu \nu} = \kappa T_{\mu \nu}~.
\end{equation}

Here the constant $\Lambda$ is known as cosmological constant.\\

{\bf Note$-$I.} $\Lambda$ was introduced by Einstein for obtaining static model of the universe. However, when Hubble discovered that the universe is expanding then he discarded the $\Lambda$\,-term from the field equations.\\

{\bf II.} In four dimension, there are 10 field equations due to the symmetric nature of the tensors involved. However, due to Bianchi identities there are only six independent field equations containing 10 components of the metric tensor. This incompleteness in determination of the metric tensor is due to the invariance of the field equations under any general co-ordinate transformation {\it i.e.} co-ordinate freedom. The remaining four equations are known as constraints equations.\\

{\bf III.} The above field equations can also be written as follows\,:\\

Contracting the above field equation with the metric tensor we get
$$R - 2R + 4\Lambda = \kappa T$$
$$i.e.~~~R = 4\Lambda - \kappa T.$$

Substituting this value of $R$ in the field equation we obtain
$$R_{\mu \nu} = \Lambda g_{\mu \nu}+ \kappa\left(T_{\mu \nu} - \frac{1}{2}Tg_{\mu \nu}\right).$$

This is another form of the field equations in terms of Ricci tensor.\\

\section{Einstein's Equations from an Action Principle}

~~~In the same year, 1915, Hilbert and Einstein\,(within a gap of few weeks) derived Einstein's field equations for gravity from action principle.\\

For the gravitational equations, we need a scalar to use as the Lagrangian. Apart from a constant, the simplest scalar that we can think of is the Ricci scalar $R$. Note that $R$ is a function of $g_{ik}$ and its derivatives but $R$ also contains second derivatives of $g_{ik}$\,. During the process, we shall show that these second derivatives will not give any additional complications. Further, one may use other scalars {\it e.g.} $R_{ik}R^{ik}~,~R_{ijkl}R^{ijkl}$ etc. as Lagrangian but $R$ is the simplest choice. Also other choices lead to higher order field equations and/or modified gravity theories.\\

So we consider the variation of the action (known as Einstein-Hilbert action)
\begin{equation} \label{6.13}
\mathcal{A} = \int _{V} R\sqrt{-g}\,d^4 x~,
\end{equation}
defined over the space-time region $V$ with a bounding 3-surface $\Sigma$. An arbitrary small variation of the metric tensor gives
$$g_{ik}~~\longrightarrow~~g_{ik} + \delta g_{ik}~~~\mbox{and}~~~g^{ik}~~\longrightarrow~~g^{ik} + \delta g^{ik}~;$$
where $\delta g_{ik}$ and $\delta g_{ik,l}$ etc. vanish on $\Sigma$\,.\\

Now at any point $P$ in $V$ we have
\begin{equation} \label{6.14}
\delta\left(R\sqrt{-g}\right) = \delta\left(R_{ik}g^{ik}\sqrt{-g}\right) = \left(\delta R_{ik}\right)g^{ik}\sqrt{-g} + R_{ik}\delta\left(g^{ik}\sqrt{-g}\right).
\end{equation}

From the property of the reciprocal (2, 0) tensor $g^{ik}$ we have
$$g_{ik}\cdot g^{km} = \delta _{i}^{m}\,.$$
Taking variation we obtain
$$\Rightarrow \left(\delta g_{ik}\right)g^{km} + g_{ik}\left(\delta g^{km}\right) = 0$$
$$i.e.~~~\left(\delta g_{ik}\right)g^{km}\cdot g_{mn} = -g_{ik}g_{mn}\delta g^{km}$$
\begin{equation} \label{6.15}
i.e.~~~\delta g_{in} = -g_{ik}g_{mn}\delta g^{km}\,.
\end{equation}

From the differentiation of the determinant of the metric tensor we write\,(see Appendix II)\,:
$$\frac{\partial g}{\partial x^k} = G^{ji}\frac{\partial}{\partial x^k}g_{ij}$$
where $G^{ji}$ is the cofactor of $g_{ij}$ in $g$. So we write
$$G^{ji} = g\cdot g^{ji} = g\cdot g^{ij}\,.$$
$$\mbox{Thus,}~~~~~~~\frac{\partial g}{\partial x^k} = g\,g^{ij}\frac{\partial}{\partial x^k}g_{ij}$$
\begin{equation} \label{6.16}
i.e.~~~dg = g\cdot g^{ij}dg_{ij}\,.
\end{equation}

Hence,
$$\delta\left(\sqrt{-g}\right) = \frac{1}{2\sqrt{-g}}\delta\left(-g\right) = \frac{1}{2\sqrt{-g}}(-g)g^{ik}\delta g_{ik} = \frac{1}{2}\sqrt{-g}\,g^{ik}\delta g_{ik}\,.$$

Now,
\begin{eqnarray} \label{6.17}
R_{ik}\delta\left(g^{ik}\sqrt{-g}\right) &=& R_{ik}\left[\left(\delta g^{ik}\right)\sqrt{-g} + g^{ik}\delta\left(\sqrt{-g}\right)\right] \nonumber \\
&=& R_{ik}\left[\sqrt{-g}\,\delta g^{ik} - g^{ik}\frac{1}{2}\sqrt{-g}\,g^{pq}g_{pm}g_{ql}\delta g^{lm}\right] \nonumber \\
&=& R_{ik}\left[\sqrt{-g}\,\delta g^{ik} - g^{ik}\frac{1}{2}\sqrt{-g}\,g_{ml}\delta g^{lm}\right] \nonumber \\
&=& \left(R_{ik} - \frac{1}{2}Rg_{ik}\right)\sqrt{-g}\,\delta g^{ik}\,.
\end{eqnarray}

To evaluate $\delta R_{ik}$ , let us choose locally inertial co-ordinates at $P$. So for this choice of co-ordinates
$$R_{ik} = \Gamma _{il,k}^{l} - \Gamma _{ik,l}^{l}\,.$$
\begin{eqnarray} \label{6.18}
\mbox{Hence,}~~~~g^{ik}\delta R_{ik} &=& g^{ik}\delta\left(\Gamma _{il,k}^{l} - \Gamma _{ik,l}^{l}\right) \nonumber \\
&=& g^{ik}\delta \Gamma _{il,k}^{l} - g^{il}\delta \Gamma _{il,k}^{k} \nonumber \\
&=& \left(g^{ik}\delta \Gamma _{il}^{l} - g^{il}\delta \Gamma _{il}^{k}\right)_{,k}~~~~\mbox{(in a local inertial co-ordinates $g$'s are constant)} \nonumber \\
&=& W_{~,k}^k\,.
\end{eqnarray}

The transformation of the Christoffel symbols are given by
\begin{equation} \label{6.19}
\Gamma _{km}^{j} = \frac{\partial x^j}{\partial x'^{\,i}}\frac{\partial x'^{\,l}}{\partial x^k}\frac{\partial x'^{\,p}}{\partial x^m}\Gamma _{pl}^{\prime \,i} + \frac{\partial ^{2}x'^{\,i}}{\partial x^k \partial x^m}\frac{\partial x^j}{\partial x'^{\,i}}\,.
\end{equation}

Also $\Gamma _{km}^{j} + \delta \Gamma _{km}^{j}$ will also have the same transformation law, hence
$$\delta \Gamma _{km}^{j} = \frac{\partial x^j}{\partial x'^{\,i}}\frac{\partial x'^{\,p}}{\partial x^m}\frac{\partial x'^{\,l}}{\partial x^k}\delta \Gamma _{pl}^{\prime \,i}~~,$$
which shows that $\delta \Gamma _{km}^{j}$ is a (1, 2)\,-tensor. So from the quotient law $W^k$ is a vector. Thus we have
\begin{equation} \label{6.20}
g^{ik}\delta R_{ik} = W_{~,k}^k\,.
\end{equation}

As the left hand side is a scalar so the right hand side must be a scalar. This is possible if the ordinary derivative is replaced by covariant derivative {\it i.e.} we write
\begin{equation} \label{6.21}
g^{ik}\delta R_{ik} = W_{~;k}^k\,.
\end{equation}

Note that equation\,(\ref{6.20}) is nothing but eq.\,(\ref{6.21}) in the locally flat co-ordinate system at $P$. Since it is a scalar relation so it must hold in every co-ordinate system.\\

Thus,
$$\delta \mathcal{A} =  \int _{V} \left(R_{ik} - \frac{1}{2}Rg_{ik}\right)\sqrt{-g}\,\delta g^{ik}\,d^4 x + \int _{V} W_{~;k}^k\,\sqrt{-g}\,d^4 x~.$$

Now,
\begin{eqnarray}
W_{~;k}^k\sqrt{-g} &=& \frac{\partial W^k}{\partial x^k}\sqrt{-g} + \Gamma _{kl}^{k}\sqrt{-g}\,W^{l} \nonumber \\
&=& \frac{\partial W^k}{\partial x^k}\sqrt{-g} + \frac{\partial}{\partial x^l}\left\{\ln \sqrt{-g}\right\}\sqrt{-g}\,W^{l} \nonumber \\
&=& \frac{\partial W^k}{\partial x^k}\sqrt{-g} + \frac{\partial}{\partial x^l}\left(\sqrt{-g}\right)W^{l} = \frac{\partial}{\partial x^k}\left(W^{k}\sqrt{-g}\right) \nonumber\\
\therefore ~~~ \int _{V} W_{~,k}^k\sqrt{-g}\,d^4 x &= &\int _{\Sigma} W^{k}\sqrt{-g}\,d\Sigma = 0~~~~\left(\because \delta g_{ik} = 0~\mbox{on}~\Sigma ~,~\mbox{so}~\delta \Gamma _{jk}^{i} = 0~\mbox{on}~\Sigma\right)\nonumber
\end{eqnarray}
$$\therefore ~~~\delta \mathcal{A} = \int _{V} \left(R_{ik} - \frac{1}{2}Rg_{ik}\right)\sqrt{-g}\,\delta g^{ik}\,d^4 x\,.$$

So by variational principle~~~$\delta \mathcal{A} = 0~~~~\Rightarrow~~~G_{ik} = 0$ ,\\
which is nothing but the vacuum Einstein equations.\\

Further, if we also consider the matter Lagrangian into the action {\it i.e.}
$$\mathcal{A} = \frac{1}{2\kappa}\int _{V} R\sqrt{-g}\,d^4 x + \int _{V} \mathcal{L}_{\mbox{\tiny matter}}\sqrt{-g}\,d^4 x\,,$$
then
$$\delta \mathcal{A} = 0~~~~\Rightarrow~~~R_{ik} - \frac{1}{2}Rg_{ik} = -\kappa T_{ik}\,.$$

\section{Weak field approximation of Gravity\,: Linearization}
Every new physical theory should contain the old theory as a reduced or limiting case (the
correspondence principle). This is also true in Einstein’s theory of gravity. Here we shall obtain
Newtonian theory of gravitation as a limiting case of Einstein’s gravity and thereby the physical
meaning of $\kappa$ will be determined. \\

In Einstein's general theory of relativity, the field equations are highly non-linear in nature and it is very difficult to solve them. To understand the nature of these equations and their solutions, weak field approximation is employed.\\

In a weak gravitational field the metric tensor is characterized by $g_{\mu \nu} = \eta_{\mu \nu} + h_{\mu \nu }$ , where
$\eta_{\mu \nu } = diag (-1, +1, +1, +1)$ is the Minkowski metric and $ \mid h_{\mu \nu }\mid << 1$. So the space-time is
assumed to be nearly flat (or equivalently, the space-time is exactly flat and the tensor field $h_{\mu \nu}$
	is propagating in this flat space-time). In the linearized theory we retain terms that are linear
	in $h_{\mu \nu}$ or its derivatives. Thus the Christoffel symbols take the form

		\begin{eqnarray}
			\Gamma^{\alpha}_{\beta \gamma} &=& g^{\alpha \delta} \Gamma_{\beta \gamma \delta}=\frac{1}{2}g^{\alpha \delta} \left\{ \frac{\partial g_{\gamma \delta}}{\partial x^{\beta}}+\frac{\partial g_{\beta \delta}}{\partial x^{\gamma}}-\frac{\partial g_{\beta \gamma}}{\partial x^{\delta}} \right\} \nonumber \\
			&=& \frac{1}{2} \left(\eta^{\alpha \delta}+ h^{\alpha \delta} \right) \left\{ \frac{\partial h_{\gamma \delta}}{\partial x^{\beta}}+\frac{\partial h_{\beta \delta}}{\partial x^{\gamma}} -\frac{\partial h_{\beta \gamma}}{\partial x^{\delta}} \right\} \nonumber \\
			&\approx & \frac{1}{2} \eta^{\alpha \delta} \left \{ \frac{\partial h_{\gamma \delta}}{\partial x^{\beta}}+\frac{\partial h_{\beta \delta}}{\partial x^{\gamma}}-\frac{\partial h_{\beta \gamma}}{\partial x^{\delta}} \right\} \nonumber \\
			&=& \frac{1}{2} \left[ \partial_{\beta} h^{\alpha}_{\gamma}+\partial_{\gamma} h^{\alpha}_{\beta}-\partial^{\alpha} h_{\beta \gamma} \right] \label{e6.22}
		\end{eqnarray}
		
		In this linearized theory the lowering and raising of indices are done by Minkowskian metric  $\eta_{\mu \nu }$ and $\eta^{\mu \nu}$ respectively.\\
		
		The curvature tensor in this approximation takes the form 
		\begin{eqnarray}
			R^{\alpha}_{\beta \gamma \delta} &=& \partial_\gamma \Gamma^{\alpha}_{\beta \delta}- \partial_\delta \Gamma^{\alpha}_{\beta \gamma}+\Gamma^{\alpha}_{\gamma \rho} \Gamma^{\rho}_{\beta \delta}-\Gamma^{\alpha}_{\delta \rho} \Gamma^{\rho}_{\beta \gamma} \nonumber \\
			&\approx& \frac{1}{2} \left( \partial_\gamma ~\partial_\beta ~h_{\delta}^{\gamma}+\partial_\gamma ~\partial_\delta ~h_{\beta}^{\alpha} - \partial_\gamma ~\partial^\alpha ~h_{\beta \delta} \right) -\frac{1}{2} \left( \partial_\delta ~\partial_\beta ~h^{\alpha}_{\gamma}+\partial_\delta ~\partial_\gamma ~h_{\beta}^{\alpha} - \partial_\delta ~\partial^\alpha ~h_{\beta \gamma} \right) \nonumber \\
			&=& \frac{1}{2}\left[\partial_\gamma ~\partial_\beta ~h_{\delta}^{\gamma} - \partial_\gamma ~\partial^\alpha ~h_{\beta \delta}- \partial_\delta ~\partial_\beta ~h^{\alpha}_{\gamma} +\partial_\delta ~\partial^\alpha ~h_{\beta \gamma} \right] \label{e6.23}
		\end{eqnarray}
		
		Now contracting the indices $\alpha$ and $\delta$, we get the Ricci tensor as
		\begin{equation}
			R_{\beta \gamma} \equiv - R^{\alpha}_{\beta \gamma \alpha} = -\frac{1}{2} \left[ \partial_\gamma ~ \partial_\beta ~ h +\square ~ h_{\beta \gamma} -\partial_\gamma ~\partial_\mu~ h^{\mu}_{\beta}-\partial_{\mu
			}~ \partial_{\beta} ~ h^{\mu}_{\gamma} \right] \label{e6.24}
		\end{equation}
		and the Ricci scalar has the expression:
		
		\begin{equation}
			R=R^{\beta}_{~\beta} = \eta^{\beta \gamma} R_{\beta \gamma}= - \square ~h + \partial_\mu ~ \partial_\rho~ h^{\mu \rho} \label{e6.25}
		\end{equation}
		Thus the Einstein tensor is given by
		
		\begin{eqnarray}
			G_{\beta \gamma} &=& R_{\beta \gamma}-\frac{1}{2} R~g_{\beta \gamma} \nonumber \\
			&=& - \frac{1}{2} \left[ \partial_\gamma ~ \partial_\beta ~h + \square ~ h_{\beta \gamma}- \partial_\gamma \partial_\mu ~h^{\mu}_{\beta}- \partial_\mu ~ \partial_\beta ~h^{\mu}_\gamma \right] \nonumber \\
			&&-\frac{1}{2} \left( \eta_{\beta \gamma}+ h_{\beta \gamma} \right) \left(- \square~h +\partial_\mu ~\partial_\rho~ h^{\mu \rho} \right)\nonumber \\
			&=&- \frac{1}{2} \left[ \partial_\beta ~ \partial_\gamma ~h+ \square~ h_{\beta \gamma} - \partial_\gamma ~ \partial_\mu ~h^{\mu}_{\beta} - \partial_\mu ~ \partial_\beta ~h^{\mu}_{\gamma} - \eta_{\beta \gamma}~ \square~ h+\eta_{\beta \gamma} ~ \partial_\mu ~ \partial_\rho ~h^{\mu \rho}\right]\label{e6.26}
		\end{eqnarray}
		Hence the Einstein field equations in this linearized version can be written as
		
		\begin{equation}
			\partial_\beta ~ \partial_\gamma ~h +\square ~ h_{\beta \gamma}- \partial_\gamma~ \partial_\mu ~h^{\mu}_{\beta}- \partial_{\mu}~\partial_{\beta}~h^{\mu}_{\gamma}- \eta_{\beta \gamma}~\square~h+ \eta_{\beta \gamma}~\partial_\mu~\partial_\rho~h^{\mu \rho}=-16 \pi \kappa T_{\beta \gamma}\label{e6.27}
		\end{equation}

\subsection{Newtonian Limit}
Every new physical theory should contain the old theory as a reduced or limiting case ( the correspondence principle). This is also true in Einstein's theory of gravity. In the following we shall show that Newtonian theory of gravitation can be obtained as a limiting case of Einstein's gravity and thereby the physical meaning of $\kappa$ will be determined. Now to obtain the Newtonian limit of Einstein gravity we assume:\\

(i) The fields vary slowly so that derivatives with respect to $x^0$ are to be ignored.\\
(ii) In Newtonian gravity, the matter source is non-relativistic in nature and hence the dominant term in the energy momentum tensor will be $T_{00}=\rho c^2$ and other components are negligible compare to $T_{00}$.\\

Here we write down the Einstein field equations as 
\begin{equation}
	R_{\mu \nu}=\kappa \left( T_{\mu \nu}-\frac{1}{2}T g_{\mu \nu} \right) \nonumber
\end{equation}
The energy-momentum tensor for perfect fluid is given by
\begin{equation}
	T_{\mu \nu}= (\rho c^2 +p)u_\mu u_\nu +p g_{\mu \nu}. \nonumber
\end{equation}
Where $\rho$ is the energy density and $p$ is the thermodynamic pressure of the fluid and $u^\mu$ is the unit time-like vector.\\

Thus 
\begin{eqnarray}
	T &=& T_{\mu}^{\mu}=T_{\mu \nu}~g^{\mu \nu}= \left( \rho c^2+p \right)u_\mu u^{\mu}+4p=3p- \rho c^2 \approx -\rho c^2 ~~~~~~~~~\mbox{(as $p<< \rho c^2$,~ in Newtonian limit)}
	\nonumber \\
	T_{\mu}^{\nu} &=& \left(\rho c^2+p \right)u_{\mu}u^{\nu}+p \delta_{\mu}^{\nu} \nonumber \\
	\mbox{So}~ T_{0}^{0} &=& \left(\rho c^2 +p \right)u_0 u^0 +p \delta_{0}^{0}= -\left(\rho c^2 +p \right)+p = - \rho c^2 \nonumber \\
	\therefore ~ T_{00} &=& T_{0}^{0} \eta_{00}= \rho c^2 .\nonumber
\end{eqnarray}

Thus the $(00)$-component of the Einstein field equation becomes

\begin{eqnarray}
	R_{00} &=& \kappa \left( T_{00}-\frac{1}{2} \eta_{00}T \right) \nonumber \\
	& \approx& \kappa \left( \rho c^2- \frac{1}{2} (-1)(- \rho c^2) \right) \nonumber \\
	& \approx& \frac{1}{2} \kappa ~\rho c^2 \nonumber
\end{eqnarray}

\begin{eqnarray}
	\mbox{i.e.}~ -\frac{1}{2}~\square~h_{00} &= &\frac{1}{2}~\kappa~\rho c^2 \nonumber \\
	\mbox{i.e.}~ \nabla^2~ h_{00} &= & -\kappa ~ \rho c^2, \label{e6.28}
\end{eqnarray}
which is the well known Poisson equation.\\

Note that we have neglected terms containing time derivative in the expression for $R_{00}$ from equation $(\ref{e6.24})$. As Poisson equation appears in various physical context so every quantity which satisfies Poisson equation is not necessarily coincide with the Newtonian gravitational potential. Hence to show that $h_{00}$ is actually related to gravitational potential we shall examine the particle trajectories i.e. geodesics with the above approximations.\\

The geodesic equation in a curved space-time is given by 
\begin{equation}
	\frac{d^2x^\alpha}{d \tau^2}=-\Gamma^{\alpha}_{\beta \gamma} \frac{d x^{\beta}}{d \tau} \frac{dx^{\gamma}}{d \tau}\nonumber
\end{equation}
For non-relativistic particles (i.e. in Newtonian theory) proper time almost coincides with the co-ordinate time $t=\frac{x_0}{c}$ and the four velocity becomes
\begin{equation}
	\frac{dx^{\alpha}}{d \tau} \sim \left(c, \Vec{v} \right) ~\mbox{with} \mid\Vec{v}\mid<<c. \nonumber
\end{equation}

Thus for the above geodesic equation
\begin{equation}
	\frac{dx^{\alpha}}{d \tau^2}=-\Gamma^{\alpha}_{00}~c^2 = \frac{1}{2} \left( g_{00}, \beta \right) \eta^{\alpha \beta} c^2 = \frac{1}{2} \eta^{\alpha \beta} \left( h_{00}, \beta \right) c^2 = \frac{1}{2} \left( \nabla~ h_{00} \right)c^2 \nonumber
\end{equation}

If we now compare this equation of motion with that for a particle in the gravitational potential $U$, then
\begin{equation}
	\frac{d^2 \Vec{r}}{dt^2}=-\nabla U~ \mbox{and}~ \nabla^2 U=4 \pi G \rho \nonumber
\end{equation}
Thus comparing the two equations of motion we have

\begin{eqnarray}
	U &=& - \frac{1}{2} c^2~h_{00} ~ \mbox{i.e.}~ h_{00}=- \frac{2U}{c^2} \nonumber \\
	\mbox{So}~~ g_{00} &=& \eta_{00}+h_{00}=- \left(1+\frac{2U}{c^2} \right)\nonumber
\end{eqnarray}
Now,
\begin{eqnarray}
	\nabla^2 U &=& 4 \pi G \rho \nonumber \\
	\mbox{i.e.}~~ \nabla^2 \left(-\frac{1}{2}c^2~h_{00} \right) &=& 4 \pi G \rho \nonumber \\
	\mbox{i.e.}~~ \nabla^2 h_{00}=-\frac{8 \pi G \rho}{c^2} \nonumber
\end{eqnarray}

Thus comparing with Linearized Einstein equation $(\ref{e6.28})$, we have
\begin{equation}
	\kappa= \frac{8 \pi G}{c^4}=2.07 \times 10^{-48}~g^{-1}~cm^{-1}~s^2 \nonumber
\end{equation}

\subsection{Gravitational Waves as Linearized Einstein Gravity:}

For derivation of gravitational wave equation we start with the Einstein equation $(\ref{e6.27})$ in the Linearized version. We now introduce trace-reversal symmetric second rank tensor $\bar{h}_{\alpha \beta}$, defined as 

\begin{equation}
	\bar{h}_{\alpha \beta}=h_{\alpha \beta}-\frac{1}{2}\eta_{\alpha \beta}~h \label{e6.29}
\end{equation}

It is easy to see that $\bar{h}=-h$ and $\bar{h}_{\alpha \beta}-\frac{1}{2}\bar{h}~\eta_{\alpha \beta}=h_{\alpha \beta}$ and consequently $\bar{\bar{h}}=h$. Then the Einstein field equations in linearised form $(\ref{e6.27})$ simplifies to 
\begin{equation}
	\square ~ \bar{h}_{\alpha \beta}+\eta_{\alpha \beta}~\partial_{\delta}~\partial_{\rho}~\bar{h}^{\delta \rho}-\partial_{\delta}~\partial_{\alpha}~\bar{h}^{\delta}_\beta-\partial_{\delta}~\partial_{\beta}~\bar{h}^{\delta}_{\alpha}=-16 \pi \kappa T_{\alpha \beta}\label{e6.30}
\end{equation}
Using Fock coordinate conditions i.e. $\partial_{\alpha}~h^{\alpha \beta} -\frac{1}{2} \partial^{\beta}~h=0$ one gets $\partial_{\alpha}~\bar{h}^{\alpha \beta}=0 $ (gauge condition). As a consequence, the above Einstein field equations $(\ref{e6.30})$ simplify to

\begin{eqnarray}
	\square~ \bar{h}_{\alpha \beta}&=&-16~\pi \kappa T_{\alpha \beta} \label{e6.31} \\
	\mbox{i.e.}~~ \partial_\mu ~ \partial^{\mu}~\bar{h}_{\alpha \beta}&=&-16~\pi \kappa T_{\alpha \beta} \nonumber ,~~ \mbox{the wave equation for gravity. }
\end{eqnarray}\\

\textbf{\underline{Case-I:~ Vaccum: Homogeneous Wave Equation}} \\

The wave equation $(\ref{e6.31})$ now becomes
\begin{equation}
	\square ~ \bar{h}_{\alpha \beta}=0, \label{e6.32}
\end{equation}
wave equation far away from the gravitational source.\\

The solution can be written as 
\begin{equation}
	\bar{h}_{\alpha \beta}=A_{\alpha \beta} exp (i~l_\mu x^{\mu})\label{e6.33}
\end{equation}
We shall now use the coordinate conditions and gauge conditions to act on the constant tensor $A_{\alpha \beta}$ so that the gravitational wave (GW) solutions can be obtained with two polarization models.\\

The Fock coordinate conditions i.e. $\partial_{\alpha} \bar{h}^{\alpha \beta} =0$ gives
\begin{equation}
	l_{\alpha}A^{\alpha \beta}=0\label{e6.34}
\end{equation}

As $`\beta'$  is the only free index so the above constraint has four independent components.

We now consider the infinitesimal co-ordinate transformation:
\begin{eqnarray}
	{x'}^\mu &=& x^\mu + \xi^\mu (x) \label{e6.35} \\
	\mbox{i.e.}~~ \frac{\partial{x'}^\mu}{\partial x^{\rho}} &= &\delta^{\mu}_{\rho}+\partial_{\rho}~\xi^{\mu} \nonumber
\end{eqnarray}
The change in the metric tensor due to this co-ordinate transformation is given by

\begin{eqnarray}
	g'_{\alpha \beta} &=& \frac{\partial x^\mu}{\partial x'^{\alpha}} \frac{\partial x^\nu}{\partial x'^{\beta}} g_{\mu \nu} \nonumber \\
	&=& \left( \delta^{\mu}_{\alpha}-\partial_{\alpha}~ \xi^{\mu} \right)  \left( \delta^{\nu}_{\beta}-\partial_{\beta}~ \xi^{\nu} \right) g_{\mu \nu} \nonumber \\
	&=& g_{\alpha \beta}- \left( \partial_{\alpha} \xi^\mu \right) g_{\mu \beta} - \left( \partial_{\beta} \xi^\nu \right) g_{\alpha \nu} \nonumber ~ ~\mbox{(neglecting square and higher powers of $\xi$ or its derivatives)}\\
	&=& \eta_{\mu \nu}+h_{\mu \nu}-(\partial_{\alpha} \xi_{\beta})- (\partial_{\beta} \xi_{\alpha}) \nonumber
\end{eqnarray}
Further, if the weak field approximation is assumed to be valid even after co-ordinate transformation then 
\begin{equation}
	g'_{\alpha \beta}= \eta_{\alpha \beta}+h'_{\alpha \beta}(x')\nonumber
\end{equation}
Hence we have,

\begin{eqnarray}
	h'_{\alpha \beta}(x')&=&h_{\alpha \beta}-\partial_\alpha \xi_\beta -\partial_\beta \xi_\alpha \nonumber \\
	\mbox{i.e.}~~ h' &=& h-2 \left( \partial_\alpha ~\xi^\alpha \right)\nonumber
\end{eqnarray}
As before if we define $ \bar{h'}_{\alpha \beta}=h'_{\alpha \beta}-\frac{1}{2} \eta_{\alpha \beta} ~h'$, then 
\begin{equation}
	\bar{h'}_{\alpha \beta}= \bar{h}_{\alpha \beta}- \partial_{(\alpha}~\xi_{\beta)}+ \eta_{\alpha \beta}(\partial_\mu~\xi^{\mu})\label{e6.36}
\end{equation}
Then by Fock co-ordinate condition i.e.
\begin{equation}
	\partial^\alpha ~\bar{h'}_{\alpha \beta}= \partial^\alpha ~\bar{h}_{\alpha \beta}=0 \nonumber
\end{equation}
One has
\begin{eqnarray}
	\partial^\alpha (\partial_\alpha ~ \partial_\beta +\partial_\beta ~\partial_\alpha) &=& \eta_{\alpha \beta}~ \partial^\alpha (\partial_\mu ~ \xi^\mu)= \partial_\beta ~(\partial_\mu ~ \xi^\mu) \nonumber \\
	\mbox{i.e.}~~ \square ~\xi_\beta + \partial^{\alpha} \partial_\beta~ \xi_\alpha &=& \partial_\beta~\partial^{\mu}~\xi_\mu \nonumber \\
	\mbox{i.e.}~~ \square ~\xi_\beta &=&0~, ~~~ \mbox{i.e.}~ \xi_\beta = B_\beta~e^{il_\rho x^\rho}\nonumber
\end{eqnarray}
Using this solution for $`\xi'$ in equation $(\ref{e6.36})$ with equation $(\ref{e6.33})$ one gets
\begin{equation}
	A'_{\alpha \beta}= A_{\alpha \beta}- i~l_{(\alpha}~B_{\beta)}+i~ \eta_{\alpha \beta}~ l_\rho B^\rho \label{e6.37}
\end{equation}
Now the two gauge degrees of freedom can be chosen from the following two criteria :

\underline{\textbf{Choice-I:}}  $A_{\alpha}^{\alpha}=0 ~~\mbox{i.e.~ traceless}$.\\

From equation $(\ref{e6.37})$ one gets
\begin{equation}
	l_\mu ~ B^{\mu} = \frac{i}{2} A_{\alpha}^{\alpha} \nonumber
\end{equation}

\underline{\textbf{Choice-II:}} $ A'_{0 \alpha}=0 $\\

Putting $\alpha=0= \beta$ in equation $(\ref{e6.37})$ we have 
\begin{equation}
	B_0= \frac{1}{2il_0} (A_{00}+\frac{1}{2} A_{\alpha}^{~\alpha})\nonumber
\end{equation}
Similarly putting $\alpha=0,~ \beta=i$
\begin{equation}
	B_i = \frac{i}{2l_0^2} \left[ -2l_0~A_{0j}+l_j (A_{00}+\frac{1}{2} A^{\alpha}_\alpha ) \right] \nonumber
\end{equation}
Thus the above gauge conditions : $l_\alpha ~ A^{\alpha \beta} =0$ (by Fock co-ordinate condition) and $ A_{\alpha}^{~\alpha}=0$ and $A_{0 \alpha}=0$, obtained by the above gauge conditions are called the transverse traceless gauge condition of gravitational wave. Due to these restrictions the number of independent components of $A_{\alpha \beta}$ reduces to two. Further, due to transeverse traceless gauge this disturbance field ( i.e. the perturbed metric $h_{\alpha \beta}$) is characterized as 

\begin{equation}
	h^{TT}=0~, ~\bar{h}_{\alpha \beta}^{~TT}= h_{\alpha \beta}^{~TT}-\frac{1}{2} \eta_{\alpha \beta}~h^{TT}= h_{\alpha \beta}^{TT} \nonumber
\end{equation}
Now if it is assumed that the GW propagates along the z-direction i.e. $l^\mu= (\omega, 0,0,l^3)= (\omega, 0, 0, \omega)$ then 

\begin{eqnarray}
	l^\alpha ~ A_{\alpha \beta}= \omega~A_{0 \beta}+\omega~ A_{3b}&=&0 \nonumber \\
	\mbox{i.e.}~~ A_{3b}=0 \nonumber
\end{eqnarray}
Hence the coefficient tensor has the following matrix representation

\begin{equation}
	A_{\alpha \beta} =
	\begin{pmatrix}
		& 0 & 0 & 0 & 0 &\\ 
		& 0 & a_{11} & a_{12} & 0 & \\
		& 0 & a_{12} & -a_{11} & 0 &\\
		& 0 & 0 & 0 & 0 &
	\end{pmatrix} \nonumber
\end{equation}
Let us define $p_+ = a_{11}$ and $p_* =a_{12}$ and consequently, the plane wave solution takes the form

\begin{equation}
	h_{\alpha \beta}^{TT} =
	\begin{pmatrix}
		& 0 & 0 & 0 & 0 &\\ 
		& 0 & p_+ & p_* & 0 & \\
		& 0 & p_* & -p_+ & 0 &\\
		& 0 & 0 & 0 & 0 &
	\end{pmatrix}\nonumber e^{i`l_a x^a} =
	\begin{pmatrix}
		& 0 & 0 & 0 & 0 &\\ 
		& 0 & a_{11} & a_{12} & 0 & \\
		& 0 & a_{12} & -a_{11} & 0 &\\
		& 0 & 0 & 0 & 0 &
	\end{pmatrix} e^{i \frac{\omega (z-ct)}{c}} \nonumber
\end{equation}

Hence there are two polarization states of GW namely

\begin{equation}
	c^{+}_{\mu \nu} =h_+
	\begin{pmatrix}
		& 0 & 0 & 0 & 0 &\\ 
		& 0 & 1 & 0 & 0 & \\
		& 0 & 0 & -1 & 0 &\\
		& 0 & 0 & 0 & 0 &
	\end{pmatrix} ~ \mbox{and}~~c^{*}_{\mu \nu} = h_*
	\begin{pmatrix}
		& 0 & 0 & 0 & 0 &\\ 
		& 0 & 0 & 1 & 0 & \\
		& 0 & 1 & 0 & 0 &\\
		& 0 & 0 & 0 & 0 &
	\end{pmatrix} \nonumber
\end{equation}
Graphically, these polarizations can be described as \\ \\

\begin{figure}[h]
	\begin{minipage}{0.47\textwidth}
		\centering \includegraphics[height=3cm,width=5cm]{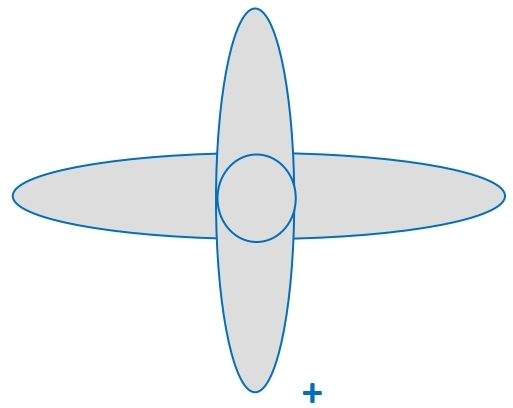}
	\end{minipage}\hfill
	\begin{minipage}{0.47\textwidth}
		\centering \includegraphics[height=3cm,width=5cm]{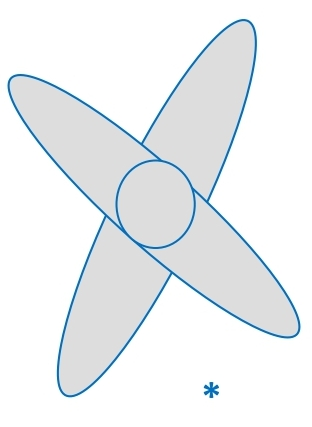}
	\end{minipage}

\end{figure}

\textbf{\underline{Case-2:} Inhomogeous Wave Equation}\\

The inhomogeneous wave equation $(\ref{e6.31})$ is presented in the flat space-time background so a general solution corresponds to the standard retarded boundary conditions as

\begin{equation}
	\bar{h}^{\alpha \beta} (\Vec{r},t)=4 \kappa \int T^{\alpha \beta} \frac{(t- \mid \overrightarrow{r-r_0}\mid, \Vec{r_0})}{ \mid \overrightarrow{r-r_0}\mid} d^3 r_0 \nonumber
\end{equation}

Thus in the linearization technique the gravitational influences (in the linear limit ) propagate at the speed of light.

There is nice analogy of GW with electromagnetic field as

(i) The relationship between $\bar h$ and $T$ is same as the relation between vector potential $A_\mu$ with current $J_\mu$ in electro magnetism.

(ii) A time dependent source will emit GW just as an accelerating charge will lead to electromagnetic radiation.

(iii) The curvature tensor in GR is analogous to the field tensor $F_{ab}$ in electromagnetism and both are gauge invariant.\\

\textbf{\underline{Observations:}}

(a) The linearized Einstein field equations $(\ref{e6.30})$ in the flat space-time background is identical to that of a spin - 2 field in flat space-time.

(b) The gauge transformation :
\begin{equation}
	h'_{\alpha \beta}= h_{\alpha \beta}- \partial_{\alpha}~ \xi_{\beta} - \partial_{\beta}~ \xi_{\alpha} \nonumber
\end{equation}
keep the curvature tensor as invariant.\\

\underline{Proof:}\\
\begin{eqnarray}
	R_{\alpha \beta \gamma \delta} &=& g_{\alpha \rho}~ R^{\rho}_{ \beta \gamma \delta} \nonumber \\
	&=& g_{\alpha \rho} \left[ \partial_{\gamma} \Gamma^{\rho}_{\beta \delta} - \partial_{\delta} \Gamma^{\rho}_{\beta \gamma} +\Gamma^{\rho}_{ \gamma \mu} \Gamma^{\mu}_{\beta \delta} - \Gamma^{\rho}_{\delta \mu} \Gamma^{\mu}_{\beta \gamma}  \right] \nonumber \\
	& \approx & g_{\alpha \rho} \left[ \partial_\gamma \Gamma^{\rho}_{\beta \delta} -  \partial_\delta \Gamma^{\rho}_{\beta \gamma} \right] \nonumber ~\mbox{( in the weak field approximation)} \\
	& \approx & \frac{\eta_{\alpha \rho}}{2} \left[ \partial_\gamma \partial_\beta h^{\rho}_{\delta} -  \partial_\delta \partial_\beta h^{\rho}_{\gamma} -  \partial_\gamma \partial^\rho h_{\beta \delta} + \partial_\delta \partial^\rho h_{\beta \gamma}  \right] \nonumber \\
	&=& \frac{1}{2} \left[ \partial_{\gamma} \partial_\beta h_{\alpha \delta} -  \partial_{\delta} \partial_\beta h_{\alpha \gamma}-  \partial_{\gamma} \partial_\alpha  h_{\beta \delta} +  \partial_{\delta} \partial_\alpha  h_{\beta \gamma}  \right] \nonumber
\end{eqnarray}

Due to the above gauge transformation

\begin{eqnarray}
	R_{\alpha \beta \gamma \delta} & \longrightarrow & R'_{\alpha \beta \gamma \delta} + \frac{1}{2} \left[ \partial_\gamma \partial_\beta ( \partial_\alpha \xi_\delta + \partial_\delta \xi_\alpha) - \partial_\delta \partial_\beta ( \partial_\alpha \xi_\gamma + \partial_\gamma \xi_\alpha) 
	- \partial_\gamma \partial_\alpha ( \partial_\beta \xi_\delta + \partial_\delta \xi_\beta) + \partial_\delta \partial_\alpha ( \partial_\beta \xi_\gamma + \partial_\gamma \xi_\beta) \right] \nonumber \\
	&=& R'_{\alpha \beta \gamma \delta} + \frac{1}{2} \left[ \xi_{\delta [, ~\beta ,~\gamma], \alpha} + \xi_{\alpha [, ~\beta  , ~\gamma ], \delta} - \xi_{\gamma [, ~\beta  , ~ \delta] , \alpha} - \xi_{\beta [, ~\gamma , ~\alpha],\delta} \right] \nonumber
\end{eqnarray}

As partial derivatives commute with each other so all the terms within the square bracket vanish and we have 
\begin{equation}
	R'_{\alpha \beta \gamma \delta} = R_{\alpha \beta \gamma \delta} \implies ~\mbox{ invariance ~ of ~ the ~ curvature ~ tensor.} \nonumber
\end{equation}

(c) Symmetry of the infinitesimal co-ordinate transformation :\\

Due to an infinitesimal co-ordinate transformation $x^i \longrightarrow x^i + \xi^i (x)$. The change in the metric tensor is given by

\begin{eqnarray}
	g'^{i k}(x'^a) &=& g^{lm} (x^a) \frac{\partial x'^i}{\partial x^l} \frac{\partial x'^k}{ \partial x^m} = g^{lm} ( \delta^{i}_{l} + \partial_l \xi^i ) ( \delta^{k}_{m}+ \partial_m \xi^k  ) \nonumber \\
	& \approx & g^{ik}(x^a) + g^{im} \partial_m \xi^k +g^{lk} \partial_l \xi^i \nonumber
\end{eqnarray}

Note that both $x'^a$ and $x^a$ identify the same physical event P but in two different co-ordinate systems. The last two terms in the R.H.S of the above equation denote the change in the components at a given point.

Now,

\begin{eqnarray}
	g'^{ik} (x'^{a}) &=& g'^{ik} ( x^a + \xi^a) = g'^{ik} (x^{a}) +  \xi^a ~\partial_a g^{i~k} \nonumber \\
	\implies g'^{ik} (x^a) &=&  g'^{ik} (x'^{a}) - \xi^a ~\partial_a g^{ik} \nonumber \\
	&=& g^{ik} (x^{a}) -  \xi^a ~\partial_a g^{ik} + g^{im} \partial_m \xi^k + g^{lk} \partial_l~ \xi^i \nonumber \\
	&=& g^{ik} (x^a) + \delta g^{ik} \nonumber
\end{eqnarray}

Now, $  \delta g^{ik} =g^{im} \partial_m  \xi^k + g^{lk}~ \partial_l \xi^i - \xi^a  \partial_a g^{ik} = \nabla^k  \xi^i + \nabla^i  \xi^k  $\\

similarly, $ \delta~ g_{ik} = - \nabla_i  \xi_k - \nabla_k  \xi_i $ with $ g'_{ik}= g_{ik} + \delta ~ g_{ik} $\\

Note that if $\xi^k$ is a Killing vector field then $\partial g_{ik} =0$ i.e. the functional form of the metric tensor does not change, a symmetry characterized by $\mathcal{L}_\xi ~ g_{ab}=0$. Here the four functions $\xi_k$ can be obtained by imposing 4 conditions on the tensor field $h_{m~n}$. By choosing the gauge condition : $\partial_r ~ \bar {h'}^{mr}=0~~,~ \xi^m $ can be determined as the solution of $ \square ~\xi^m = \partial_r  \bar{h}^{mr} $.

Note that the solution for $\xi^m$ is not unique as one may add to it any solution of $ \square ~ \xi^m =0$, the harmonic gauge.

\section{Einstein Equations on hypersurface}

\subsection{Normal Vector}

~~~We have already defined hypersurface and induced metric on it in section 4.1\,. We now introduce the notion of normal vector to the hypersurface $\Sigma$ :~~$f(x^{\alpha})=0$. As $f$ is constant along a hypersurface and it changes its value from one hypersurface to the other, so the vector $\dfrac{\partial f}{\partial x^{\alpha}}$ is directed along the normal to the hypersurface $\Sigma$\,. Thus, if $n_{\alpha}$ denotes the unit normal to the hypersurface\,(assuming it to be non-null) $i.e.~~n_{\alpha}n^{\alpha}=\epsilon$ ($\epsilon = +1$ for $\Sigma$ to be time-like, $\epsilon = -1$ for $\Sigma$ to be space-like) then we have
\begin{equation} \label{6.28}
n_{\alpha} = \epsilon \,f_{,\alpha} / \left\{\left|g^{\gamma \delta}f_{,\gamma}f_{,\delta}\right|\right\}^{1/2}
\end{equation}

It should be noted that $n^{\alpha}$ is directed along the increasing direction of $f~~~i.e.~~n^{\alpha}f_{,\alpha}>0$.\\

\subsection{Intrinsic Tensor to the hypersurface}

~~~Using the first fundamental form of the hypersurface we now introduce the notion of intrinsic tensor to the hypersurface.\\

Let us introduce the second rank tensor
\begin{equation} \label{6.29}
h^{\alpha \beta}= h^{ab}e_{a}^{\alpha}e_{b}^{\beta}= g^{\alpha \beta}- \varepsilon \, n^{\alpha}n^{\beta}
\end{equation}
where $e_{a}^{\alpha}= \dfrac{\partial y^\alpha}{\partial x^a}$ are basis vectors on $\Sigma$\,. Then we see that $h^{\alpha \beta}n_{\alpha}= h^{\alpha \beta}n_{\beta}= 0~~~i.e.~~h^{\alpha \beta}$ are defined only on $\Sigma$\,. Hence $h^{\alpha \beta}$ are purely tangent to the hypersurface. So any arbitrary tensor field $A^{\alpha \beta \cdots}$ of the manifold can be projected to the hypersurface so that only its tangential components survive as
\begin{equation} \label{6.30}
A^{\mu \nu \cdots}= A^{\alpha \beta \cdots}\cdot h_{\alpha}^{\mu}h_{\beta}^{\nu}\cdots
\end{equation}

Thus we can define
\begin{equation} \label{6.31}
A_{ab \cdots}= A_{\mu \nu \cdots}\, e_{a}^{\mu}e_{b}^{\nu}\cdots
\end{equation}
with $A^{mn \cdots}= h^{ma}h^{nb}\cdots \,A_{ab \cdots}$ as the intrinsic tensor field to the hypersurface associated with the tensor field $A^{\alpha \beta \cdots}$ to the manifold. It should be noted that an intrinsic tensor behaves as a tensor under a co-ordinate transformation $x^a~~\rightarrow~~x^{a\prime}$ to the hypersurface while it behaves as a scalar under a co-ordinate transformation to the manifold.\\

\subsection{Intrinsic covariant derivative and the relation to its partner in the manifold}

~~~We now try to define covariant derivative in $\Sigma$ in terms of a connection that is compatible with the induced metric $h_{ab}$ on $\Sigma$\,. We start with a tangent vector field $A^{\alpha}$ for which intrinsic vector field is $A^a~~i.e.$
\begin{equation} \label{6.32}
A^{\alpha}= A^{a}e_{a}^{\alpha}~~~,~~~A^{\alpha}n_{\alpha}= 0~~~,~~~A_{a}= A_{\alpha}e_{a}^{\alpha}\,.
\end{equation}

The intrinsic covariant derivative of an intrinsic vector field $A_a$ is defined as the projection of $A_{\alpha ; \beta}$ onto the hypersurface {\it i.e.}
\begin{eqnarray} \label{6.33}
A_{a|b} &=& A_{\alpha ; \beta}e_{a}^{\alpha}e_{b}^{\beta}= \left(A_{\alpha}e_{a}^{\alpha}\right)_{;\beta}e_{b}^{\beta}- A_{\alpha}e_{a;\beta}^{\alpha}e_{b}^{\beta} \nonumber \\
&=& A_{a,\beta}e_{b}^{\beta}- A_{\alpha}e_{b}^{\beta}\left(g^{\alpha \delta}e_{a\delta}\right)_{;\beta} \nonumber \\
&=& \frac{\partial A_a}{\partial y^{\beta}}\cdot \frac{\partial y^{\beta}}{\partial x^b}- \left(A_{\alpha}g^{\alpha \delta}\right)e_{b}^{\beta}e_{a\delta ; \beta} \nonumber \\
&=& \frac{\partial A_a}{\partial x^b}- A^{\delta}e_{b}^{\beta}e_{a\delta ; \beta} \nonumber \\
&=& \frac{\partial A_a}{\partial x^b}- A^{c}e_{c}^{\delta}e_{b}^{\beta}e_{a\delta ; \beta} \nonumber \\
&=& \frac{\partial A_a}{\partial x^b}- \Gamma _{cab}A^{c} \nonumber \\
&=& \frac{\partial A_a}{\partial x^b}- \Gamma _{ab}^{c}A_{c}
\end{eqnarray}
\begin{equation} \label{6.34}
\mbox{where}~~~~\Gamma _{cab}= e_{c}^{\delta}e_{b}^{\beta}e_{a\delta ; \beta}~~,~~\Gamma _{ab}^{d}= h^{dc}\Gamma _{cab}\,.
\end{equation}

We shall now show that the above intrinsic connection is compatible with the induced metric {\it i.e.}
$$\Gamma _{cab}= \frac{1}{2}\left(h_{ca,b}+ h_{cb,a}- h_{ab,c}\right)$$
or equivalently,
$$h_{ab|c}= 0.$$

By definition
\begin{eqnarray}
h_{ab|c} &=& h_{\alpha \beta ; \gamma}e_{a}^{\alpha}e_{b}^{\beta}e_{c}^{\gamma}= \left(g_{\alpha \beta} -\epsilon n_{\alpha}n_{\beta}\right)_{;\gamma}e_{a}^{\alpha}e_{b}^{\beta}e_{c}^{\gamma} \nonumber \\
&=& -\epsilon n_{\alpha ; \gamma}\left(n_{\beta}e_{b}^{\beta}\right)e_{a}^{\alpha}e_{c}^{\gamma}- \epsilon \left(n_{\alpha}e_{a}^{\alpha}\right)\left(n_{\beta ; \gamma}\right)e_{b}^{\beta}e_{c}^{\gamma}\,. \nonumber \\
&=& 0~~~~\left(\because ~~e_{b}^{\beta}n_{\beta} = 0 \right) \nonumber
\end{eqnarray}

We have seen that $A_{a|b}= A_{\alpha ; \beta}e_{a}^{\alpha}e_{b}^{\beta}$ are the tangential components of the vector field $A_{\alpha ; \beta}e_{b}^{\beta}$\,. So the natural question that arises what will be the normal component?\\

We write,
$$A_{\alpha ; \beta}e_{b}^{\beta}= g_{\alpha}^{~\delta}A_{\delta ; \beta}e_{b}^{\beta}.$$

Now by decomposing the metric $g_{\alpha}^{\delta}$ into tangential and normal components we obtain
\begin{eqnarray}
A_{\alpha ; \beta}e_{b}^{\beta} &=& \left(h_{pq}e_{\alpha}^{p}e^{\delta q}+ \epsilon n_{\alpha}n^{\delta}\right)A_{\delta ; \beta}e_{b}^{\beta} \nonumber \\
&=& h_{pq}\left(A_{\delta ; \beta}e_{b}^{\beta}e^{\delta q}\right)e_{\alpha}^{p}+ \epsilon\left(n^{\delta}A_{\delta ; \beta}e_{b}^{\beta}\right)n_{\alpha} \nonumber \\
&=& h_{pq}\left(A_{r|b}h^{rq}\right)e_{\alpha}^{p}- \epsilon\left(n_{~;\beta}^{\delta}A_{\delta}e_{b}^{\beta}\right)n_{\alpha} \nonumber \\
&=& h_{p}^{r}\left(A_{r|b}\right)e_{\alpha}^{p}- \epsilon A_{d}\left(n_{~;\beta}^{\delta}e_{\delta}^{d}e_{b}^{\beta}\right)n_{\alpha} \nonumber \\
&=& A_{r|b}e_{\alpha}^{r}- \epsilon A_{d}\left(g^{\delta \lambda}n_{\lambda ; \beta}e_{\delta}^{d}e_{b}^{\beta}\right)n_{\alpha} \nonumber \\
&=& A_{r|b}e_{\alpha}^{r}- \epsilon A_{d}\left(n_{\lambda ; \beta}e^{d\lambda}e_{b}^{\beta}\right)n_{\alpha} \nonumber \\
&=& A_{r|b}e_{\alpha}^{r}- \epsilon A^{d}\left(n_{\lambda ; \beta}e_{d}^{\lambda}e_{b}^{\beta}\right)n_{\alpha}\,. \nonumber
\end{eqnarray}

Let us define the intrinsic tensor
\begin{equation} \label{6.35}
K_{db}= n_{\lambda ; \beta}e_{d}^{\lambda}e_{b}^{\beta}
\end{equation}
as the extrinsic curvature or 2nd fundamental form of the hypersurface $\Sigma$\,. So we write
\begin{equation} \label{6.36}
A_{\alpha ; \beta}e_{b}^{\beta}= A_{a|b}\cdot e_{\alpha}^{a}- \epsilon A^{a}K_{ab}n_{\alpha}
\end{equation}
which shows that the vector field $A_{\alpha ; \beta}e_{b}^{\beta}$ of the manifold has tangential component $A_{a|b}$ while its normal component is $-\epsilon A^{a}K_{ab}$\,. Hence normal component vanishes iff extrinsic curvature vanishes.\\

Further multiplying the L.H.S. of eq.\,(\ref{6.36}) by $g^{\alpha \delta}$ and the 1st and 2nd term of the R.H.S. by the corresponding tangential and normal component we have
\begin{eqnarray} \label{6.37}
A_{~;\beta}^{\delta}e_{b}^{\beta} &=& A_{~|b}^{q}e_{q}^{\delta}- \epsilon A^{a}K_{ab}n^{\delta} \\
\Rightarrow ~~~A_{~;\beta}^{\delta}e_{b}^{\beta}e_{\delta}^{c} &=& A_{~|b}^{a}e_{\delta}^{a}e_{\delta}^{c}- \epsilon A^{a}K_{ab}\left(n^{\delta}\cdot e_{\delta}^{c}\right) \nonumber \\
\Rightarrow ~~~A_{~;\beta}^{\alpha}e_{b}^{\beta}e_{\alpha}^{c} &=& A_{~|b}^{c} \nonumber
\end{eqnarray}
which shows that the intrinsic covariant derivative of the intrinsic contravariant vector field $A^c$ is as before the tangential component of the vector field $A_{~;\beta}^{\alpha}e_{b}^{\beta}$. Also,
\begin{eqnarray} \label{6.38}
A_{~|b}^{c} &=& A_{~;\beta}^{\alpha}e_{b}^{\beta}e_{\alpha}^{c}= \left(A^{a}e_{a}^{\alpha}\right)_{;\beta}e_{b}^{\beta}e_{\alpha}^{c} \nonumber \\
&=& \left(A_{~,\beta}^{a}e_{b}^{\beta}\right)e_{a}^{\alpha}e_{\alpha}^{c}+ A^{a}e_{a;\beta}^{\alpha}e_{b}^{\beta}e_{\alpha}^{c} \nonumber \\
&=& \frac{\partial A^a}{\partial y^\beta}\cdot \frac{\partial y^\beta}{\partial x^b}\delta _{a}^{c}+ \left(g^{\alpha \delta}e_{a\delta}\right)_{;\beta}e_{b}^{\beta}e_{\alpha}^{c}\cdot A^a \nonumber \\
&=& \frac{\partial A^c}{\partial x^b}+ e_{a\delta ; \beta}e_{b}^{\beta}e^{c\delta}A^a \nonumber \\
&=& A_{~,b}^{c}+ \Gamma _{ab}^{c}A^{a}.
\end{eqnarray}

This is the explicit expression for intrinsic covariant derivative of an intrinsic contravariant vector field.\\

Moreover, writing $e_{a}^{\delta}$ for $A^{\delta}$ in equation\,(\ref{6.37}) we get
\begin{eqnarray} \label{6.39}
e_{a;\beta}^{\delta}\,e_{b}^{\beta} &=& \delta _{a~~|b}^{q}\,e_{q}^{\delta}- \epsilon \,\delta_{a}^{c}K_{cb}n^{\delta} \nonumber \\
&=& \Gamma _{db}^{q}\,\delta _{a}^{d}e_{q}^{\delta}- \epsilon \,K_{ab}n^{\delta} \nonumber \\
&=& \Gamma _{ab}^{c}e_{c}^{\delta}- \epsilon \,K_{ab}n^{\delta}.
\end{eqnarray}

This is known as Gauss-Weingarten equation. Now using the facts ({\it i}) the basis vectors $e_{a}^{\alpha}$ are orthogonal to the normal vector $n^\alpha$ {\it i.e.} $e_{a}^{\alpha}n_{\alpha}= 0$ , ({\it ii}) the basis vectors are Lie transported along one another {\it i.e.} $e_{a~;\beta}^{\alpha}e_{b}^{\beta} = e_{b~;\beta}^{\alpha}e_{a}^{\beta}$ .\\

One can easily see from the Gauss-Weingarten equation that the extrinsic curvature is symmetric in its two indices {\it i.e.} $K_{ab}= K_{ba}= \dfrac{1}{2}K_{(ab)}$\,.\\

Hence from (\ref{6.35}) we write
\begin{equation} \label{6.40}
K_{ab}= \frac{1}{2}n_{(\alpha ; \beta)}e_{a}^{\alpha}e_{b}^{\beta}= \frac{1}{2}\left(\mathcal{L}_{n}g_{\alpha \beta}\right)e_{a}^{\alpha}e_{b}^{\beta}\,.
\end{equation}

Hence one can say that extrinsic curvature is related to the normal derivative of the metric tensor of the manifold.\\

Now, the trace of the extrinsic curvature is given by
\begin{equation} \label{6.41}
K \equiv h_{ab}K^{ab}= n_{~;\alpha}^{\alpha}\,.
\end{equation}

The above result shows that if we have a congruence of\,(time-like or space-like)\,geodesics having tangent vector $n^\alpha$ {\it i.e.} the geodesics are hypersurface orthogonal then $K$ can be interpreted as the expansion of the congruence of geodesics. So one can say the hypersurface is convex or concave according as $K>~\mbox{or}~<0$ {\it i.e.} the congruence is diverging or converging.\\

{\bf Note:} A hypersurface $\Sigma$ of a manifold is completely characterized by the first and second fundamental forms {\it i.e.} by $h_{ab}$\,(the induced metric) and $K_{ab}$\,(the extrinsic curvature). $h_{ab}$ characterizes only the intrinsic properties of the hypersurface's geometry while $K_{ab}$ is related to the extrinsic properties {\it i.e.} how the hypersurface is embedded in the manifold.\\

\subsection{Relation between the intrinsic curvature and curvature of the manifold\,: Gauss-Codazzi equations}

~~~The intrinsic curvature to the hypersurface can be usually defined as the non-commutativity of the intrinsic covariant derivative as
\begin{equation} \label{6.42}
A_{~|pq}^{r}- A_{~|qp}^{r}= -R_{~spq}^{r}\,A^s
\end{equation}
where
$$R_{~spq}^{r}= \Gamma _{sq,p}^{r}- \Gamma _{sp,q}^{r}+ \Gamma _{mp}^{r}\Gamma _{sq}^{m}- \Gamma _{mq}^{r}\Gamma _{sp}^{m}.$$

We shall now try to relate this hypersurface curvature to the curvature of the manifold. One can consider the Gauss-Weingarten equation as a tensor equation on the manifold. So taking covariant derivative of both side of it and projecting it to the hypersurface we obtain
$$\left(e_{a;\beta}^{\alpha}e_{b}^{\beta}\right)_{;\gamma}e_{c}^{\gamma} = \left(\Gamma_{ab}^{d}e_{d}^{\alpha}- \epsilon \, K_{ab}n^{\alpha}\right)_{;\gamma}e_{c}^{\gamma}$$
$$\Rightarrow ~~~ e_{a;\beta \gamma}^{\alpha}e_{b}^{\beta}e_{c}^{\gamma}+ e_{a;\beta}^{\alpha}e_{b;\gamma}^{\beta}e_{c}^{\gamma} = \Gamma_{ab,c}^{d}e_{d}^{\alpha}+ \Gamma_{ab}^{d}e_{d;\gamma}^{\alpha}e_{c}^{\gamma}- \epsilon \, K_{ab,c}n^{\alpha}- \epsilon \, K_{ab}n_{~;\gamma}^{\alpha}e_{c}^{\gamma}$$
\begin{eqnarray}
\Rightarrow ~~~ e_{a;\beta \gamma}^{\alpha}e_{b}^{\beta}e_{c}^{\gamma}+ e_{a;\beta}^{\alpha}\left(\Gamma _{bc}^{d}e_{d}^{\beta}- \epsilon K_{bc}n^{\beta}\right) = \Gamma_{ab,c}^{d}e_{d}^{\alpha} &+& \Gamma_{ab}^{d}\left(\Gamma _{dc}^{e}e_{e}^{\alpha}- \epsilon K_{dc}n^{\alpha}\right) \nonumber \\
&-& \epsilon K_{ab,c}n^{\alpha}- \epsilon K_{ab}n_{~;\gamma}^{\alpha}e_{e}^{\gamma} \nonumber \\
\Rightarrow ~~~ e_{a;\beta \gamma}^{\alpha}e_{b}^{\beta}e_{c}^{\gamma}+ \Gamma _{bc}^{d}\left(\Gamma _{ad}^{e}e_{e}^{\alpha}- \epsilon K_{ad}n^{\alpha}\right)- \epsilon K_{bc}e_{a;\beta}^{\alpha}n^{\beta} &=& \Gamma_{ab,c}^{d}e_{d}^{\alpha}+ \Gamma_{ab}^{d}\left(\Gamma _{dc}^{e}e_{e}^{\alpha}- \epsilon K_{dc}n^{\alpha}\right) \nonumber \\
&-& \epsilon K_{ab,c}n^{\alpha}- \epsilon K_{ab}n_{~;\gamma}^{\alpha}e_{e}^{\gamma} \nonumber
\end{eqnarray}
\begin{eqnarray}
\Rightarrow ~~~ e_{a;\beta \gamma}^{\alpha}e_{b}^{\beta}e_{c}^{\gamma}= \Gamma_{ab,c}^{d}e_{d}^{\alpha}+ \left(\Gamma _{ab}^{d}\Gamma _{dc}^{e}- \Gamma _{bc}^{d}\Gamma _{ad}^{e}\right)e_{e}^{\alpha} &-& \epsilon \left(K_{dc}\Gamma _{ab}^{d}- K_{ad}\Gamma _{bc}^{d}\right)n^{\alpha} \nonumber \\
&-& \epsilon K_{ab,c}n^{\alpha}- \epsilon K_{ab}n_{~;\gamma}^{\alpha}e_{c}^{\gamma}+ \epsilon K_{bc}e_{a;\beta}^{\alpha}n^{\beta} \nonumber \\
\therefore ~~~ e_{a;\gamma \beta}^{\alpha}e_{b}^{\beta}e_{c}^{\gamma}= \Gamma_{ac,b}^{d}e_{d}^{\alpha}+ \left(\Gamma _{ac}^{d}\Gamma _{db}^{e}- \Gamma _{bc}^{d}\Gamma _{ad}^{e}\right)e_{e}^{\alpha} &-& \epsilon \left(K_{db}\Gamma _{ac}^{d}- K_{ad}\Gamma _{bc}^{d}\right)n^{\alpha} \nonumber \\
&-& \epsilon K_{ac,b}n^{\alpha}- \epsilon K_{ac}n_{~;\gamma}^{\alpha}e_{b}^{\gamma}+ \epsilon K_{cb}e_{a;\beta}^{\alpha}n^{\beta} \nonumber \\
\therefore ~~~ \left(e_{a;\gamma \beta}^{\alpha}- e_{a;\gamma \beta}^{\alpha}\right)e_{b}^{\beta}e_{c}^{\gamma}= \left(\Gamma_{ab,c}^{d}- \Gamma_{ac,b}^{d}\right)e_{d}^{\alpha} &+& \left(\Gamma _{ab}^{d}\Gamma _{dc}^{e}- \Gamma _{ac}^{d}\Gamma _{db}^{e}\right)e_{e}^{\alpha}- \epsilon \left(K_{dc}\Gamma _{ab}^{d}- K_{db}\Gamma _{ac}^{d}\right)n^{\alpha} \nonumber \\
&-& \epsilon \left(K_{ab,c}- K_{ac,b}\right)n^{\alpha}- \epsilon \left(K_{ab}e_{c}^{\gamma}- K_{ac}e_{b}^{\gamma}\right)n_{~;\gamma}^{\alpha} \nonumber
\end{eqnarray}
\begin{eqnarray} \label{6.43}
-R_{\mu \beta \gamma}^{\alpha}\,e_{a}^{\mu}e_{b}^{\beta}e_{c}^{\gamma} &=& \left[\left(\Gamma _{ab,c}^{e}- \Gamma _{ac,b}^{e}\right)+ \Gamma _{ab}^{d}\Gamma _{dc}^{e}- \Gamma _{ac}^{d}\Gamma _{db}^{e}\right]e_{e}^{\alpha} \nonumber \\
&-& \epsilon \left(K_{ab|c}- K_{ac|b}\right)n^{\alpha}- \epsilon \left(K_{ab}e_{c}^{\gamma}- K_{ac}e_{b}^{\gamma}\right)n_{~;\gamma}^{\alpha} \nonumber \\
\Rightarrow ~~~ R_{\mu \beta \gamma}^{\alpha}\,e_{a}^{\mu}e_{b}^{\beta}e_{c}^{\gamma} &=& R_{abc}^{e}e_{e}^{\alpha}+ \epsilon \left(K_{ab|c}- K_{ac|b}\right)n^{\alpha}+ \epsilon \left(K_{ab}e_{c}^{\gamma}- K_{ac}e_{b}^{\gamma}\right)n_{~;\gamma}^{\alpha}
\end{eqnarray}

Now proceeding along $e_{p\alpha}$ we get
\begin{eqnarray} \label{6.44}
R_{\delta \mu \beta \gamma}\,e_{a}^{\mu}e_{b}^{\beta}e_{c}^{\gamma}e_{p}^{\delta} &=& R_{dabc}\,g^{ed}e_{e}^{\alpha}e_{p\alpha}+ \epsilon \left(K_{ab|c}- K_{ac|b}\right)\left(n^{\alpha}e_{p\alpha}\right)+ \epsilon K_{ab}\left(e_{p\alpha}e_{c}^{\gamma}n_{~;\gamma}^{\alpha}\right)- \epsilon K_{ac}\left(e_{p\alpha}e_{b}^{\gamma}n_{~;\gamma}^{\alpha}\right) \nonumber \\
R_{\delta \alpha \beta \gamma}\,e_{a}^{\alpha}e_{b}^{\beta}e_{c}^{\gamma}e_{p}^{\delta} &=& R_{dabc}\,e^{\alpha d}e_{p\alpha}+ \epsilon K_{ab}K_{pc}- \epsilon K_{ac}K_{pb} \nonumber \\
R_{\delta \alpha \beta \gamma}\,e_{a}^{\alpha}e_{b}^{\beta}e_{c}^{\gamma}e_{p}^{\delta} &=& R_{pabc}+ \epsilon \left(K_{ab}K_{pc}- K_{ac}K_{pb}\right)
\end{eqnarray}

This is known as Gauss equation.\\

Again projecting equation\,(\ref{6.43}) along $n_{\alpha}$ we have
\begin{eqnarray} \label{6.45}
R_{\nu \mu \beta \gamma}\,g^{\nu \alpha}n_{\alpha}\,e_{a}^{\mu}e_{b}^{\beta}e_{c}^{\gamma} &=& R_{~abc}^{e}\left(e_{e}^{\alpha}n_{\alpha}\right)+ \epsilon \left(K_{ab|c}- K_{ac|b}\right)\left(n^{\alpha}n_{\alpha}\right)+ \epsilon \left(K_{ab}e_{c}^{\gamma}- K_{ac}e_{b}^{\gamma}\right)\left(n_{\alpha}\cdot n_{~;\gamma}^{\alpha}\right) \nonumber \\
\Rightarrow ~~~ R_{\nu \mu \beta \gamma}\,n^{\nu}e_{a}^{\mu}e_{b}^{\beta}e_{c}^{\gamma} &=& K_{ab|c}- K_{ac|b}
~~~\left(\because ~~ n_{\alpha}n_{~;\gamma}^{\alpha}= \frac{1}{2}\left(n^{\alpha}n_{\alpha}\right)_{;\gamma}= 0\right)
\end{eqnarray}

This is Codazzi equation.\\

{\bf Note:} Gauss-Codazzi equations express some components of the curvature tensor of the manifold in terms of the intrinsic and extrinsic curvatures of the hypersurface. However, there are other components of the manifold curvature tensor\,(for example $R_{\mu \alpha \beta \gamma}\,n^{\mu}e_{a}^{\alpha}n^{\nu}e_{b}^{\beta}$) which cannot be expressed only by the first and second fundamental forms of the hypersurface.\\

\subsection{Contraction of Gauss-Codazzi equations\,: Einstein equations on the hypersurface}

~~~The Ricci tensor and Ricci scalar of the manifold are given by
\begin{eqnarray} \label{6.46}
R_{\alpha \beta} &=& g^{\mu \nu}\,R_{\mu \alpha \nu \beta} \nonumber \\
&=& \left(\epsilon n^{\mu}n^{\nu}+ h^{ab}e_{a}^{~\mu}e_{b}^{~\nu}\right)R_{\mu \alpha \nu \beta} \nonumber \\
&=& h^{ab}\left(R_{\mu \alpha \nu \beta}\,e_{a}^{\mu}e_{b}^{\nu}\right)+ \epsilon\left(R_{\mu \alpha \nu \beta}n^{\mu}n^{\nu}\right)
\end{eqnarray}
\begin{eqnarray} \label{6.47}
R &=& g^{\alpha \beta}R_{\alpha \beta} = \left(\epsilon n^{\alpha}n^{\beta}+ h^{mn}e_{m}^{\alpha}e_{n}^{\beta}\right)\left[h^{ab}\left(R_{\mu \alpha \nu \beta}e_{a}^{\mu}e_{b}^{\nu}\right)+ \epsilon\left(R_{\mu \alpha \nu \beta}n^{\mu}n^{\nu}\right)\right] \nonumber \\
&=& 2\epsilon \, h^{ab}\left(R_{\mu \alpha \nu \beta}n^{\alpha}e_{a}^{\mu}n^{\beta}e_{b}^{\nu}\right)+ h^{ab}h^{mn}\left(R_{\mu \alpha \nu \beta}e_{a}^{\mu}e_{m}^{\alpha}e_{b}^{\nu}e_{n}^{\beta}\right)
\end{eqnarray}

(the other term vanishes due to the product of symmetric and anti-symmetric terms)\\

Thus the Einstein tensor of the manifold can be expressed as
\begin{eqnarray} \label{6.48}
G_{\rho \sigma} &=& R_{\rho \sigma}- \frac{1}{2}R\,g_{\rho \sigma} \nonumber \\
&=& h^{ab}\left(R_{\mu \rho \nu \sigma}e_{a}^{\mu}e_{b}^{\nu}\right)+ \epsilon\left(R_{\mu \rho \nu \sigma}n^{\mu}n^{\nu}\right) \nonumber \\
&-& \frac{1}{2}h^{ab}h^{mn}g_{\rho \sigma}\left(R_{\mu \alpha \nu \beta}e_{a}^{\mu}e_{m}^{\alpha}e_{b}^{\nu}e_{n}^{\beta}\right)- \epsilon \, h^{ab}g_{\rho \sigma}\left(R_{\mu \alpha \nu \beta}n^{\alpha}e_{a}^{\mu}n^{\beta}e_{b}^{\nu}\right)
\end{eqnarray}

Now,
\begin{eqnarray}
G_{\rho \sigma}\,n^{\rho}n^{\sigma} &=& h^{ab}\left(R_{\mu \alpha \nu \beta}e_{a}^{\mu}n^{\rho}e_{b}^{\nu}n^{\sigma}\right)+ \epsilon\left(R_{\mu \rho \nu \sigma}n^{\mu}n^{\nu}n^{\rho}n^{\sigma}\right) \nonumber \\
&-& \frac{1}{2}h^{ab}h^{mn}\left(g_{\rho \sigma}n^{\rho}n^{\sigma}\right)\left(R_{\mu \alpha \nu \beta}e_{a}^{\mu}e_{m}^{\alpha}e_{b}^{\nu}e_{n}^{\beta}\right)- \epsilon \, h^{ab}\left(g_{\rho \sigma}n^{\rho}n^{\sigma}\right)\left(R_{\mu \alpha \nu \beta}n^{\alpha}e_{a}^{\mu}n^{\beta}e_{b}^{\nu}\right) \nonumber \\
&=& \cancel{h^{ab}R_{\mu \rho \nu \sigma}e_{a}^{\mu}n^{\rho}e_{b}^{\nu}n^{\sigma}}- \frac{1}{2}\epsilon h^{ab}h^{mn}\left(R_{\mu \alpha \nu \beta}e_{a}^{\mu}e_{m}^{\alpha}e_{b}^{\nu}e_{n}^{\beta}\right) - \cancel{h^{ab}R_{\mu \alpha \nu \beta}n^{\alpha}e_{a}^{\mu}n^{\beta}e_{b}^{\nu}} \nonumber \\
\Rightarrow ~~ -2\epsilon G_{\rho \sigma}n^{\rho}n^{\sigma} &=& h^{ab}h^{mn}\left[R_{ambn}+ \epsilon\left(K_{mb}K_{an}- K_{mn}K_{ab}\right)\right] \nonumber \\
&=& R_{\Sigma}+ \epsilon\left(K_{ab}K^{ab}- K^{2}\right)~~~~~~~~~~~(\mbox{using}~(\ref{6.44})\,) \nonumber
\end{eqnarray}
\begin{equation} \label{6.49}
\therefore ~~~ -2\epsilon G_{\rho \sigma}n^{\rho}n^{\sigma}= R_{\Sigma}+ \epsilon\left(K_{ab}K^{ab}- K^{2}\right)
\end{equation}
where $R_{\Sigma}$ is the Ricci scalar of the hypersurface $\Sigma$\,.\\

Again from equation\,(\ref{6.48})
\begin{eqnarray}
G_{\rho \sigma}\,e_{~d}^{\rho}n^{\sigma} &=& h^{ab}\left(R_{\mu \alpha \nu \beta}e_{a}^{\mu}e_{d}^{\rho}e_{b}^{\nu}n^{\sigma}\right)+ \epsilon\left(R_{\mu \rho \nu \sigma}n^{\mu}n^{\nu}n^{\sigma}e_{d}^{\rho}\right) \nonumber \\
&-& \frac{1}{2}h^{ab}h^{mn}e_{d}^{\rho}\cdot\left(g_{\rho \sigma}n^{\sigma}\right)\left(R_{\mu \alpha \nu \beta}e_{a}^{\mu}e_{m}^{\alpha}e_{b}^{\nu}e_{n}^{\beta}\right)- \epsilon \, h^{ab}\left(g_{\rho \sigma}n^{\sigma}\right)\cdot e_{d}^{\rho}\left(R_{\mu \alpha \nu \beta}n^{\alpha}e_{a}^{\mu}n^{\beta}e_{b}^{\nu}\right) \nonumber \\
&=& h^{ab}\left[\left(R_{\sigma \nu \rho \mu}n^{\sigma}e_{b}^{\nu}e_{d}^{\rho}e_{a}^{\mu}\right)\right] \nonumber
\end{eqnarray}
(the second term vanishes due to product of symmetric and anti-symmetric product, 3rd and 4th terms vanish due to the fact $e_{d}^{\rho}n_{\rho}=0$\,)
\begin{eqnarray}
&=& h^{ab}\left[K_{bd|a}- K_{ba|d}\right]~~~~~~~~(\mbox{using}\,(\ref{6.45})\,) \nonumber \\
&=& K_{~d|a}^{a}- K_{|d}= K_{a|b}- K_{,a}\,. \nonumber
\end{eqnarray}
\begin{equation} \label{6.50}
\therefore ~~~ G_{\rho \sigma}e_{a}^{\rho}n^{\sigma}= K_{~a|b}- K_{,a}\,.
\end{equation}

{\bf Note\,:} $G_{\rho \sigma}e_{d}^{\rho}e_{l}^{\sigma}$ cannot be expressed only by the first and second fundamental form on the hypersurface.\\

We shall now simplify the R.H.S. of the equation\,(\ref{6.47}) so that the Ricci scalar of the manifold can be expressed in a more convenient form. We start with the first term on the R.H.S.\,:
\begin{eqnarray}
2\epsilon \, h^{ab}\,R_{\mu \alpha \nu \beta}n^{\alpha}e_{a}^{\mu}n^{\beta}e_{b}^{\nu} &=& 2\epsilon\left(h^{ab}e_{a}^{\mu}e_{b}^{\nu}\right)R_{\mu \alpha \nu \beta}n^{\alpha}n^{\beta} \nonumber \\
&=& 2\epsilon\left(h^{\mu \nu}\right)R_{\mu \alpha \nu \beta}n^{\alpha}n^{\beta} \nonumber \\
&=& 2\epsilon\left(g^{\mu \nu}- \epsilon \, n^{\mu}n^{\nu}\right)R_{\mu \alpha \nu \beta}n^{\alpha}n^{\beta} \nonumber \\
&=& 2\epsilon \, g^{\mu \nu}\, R_{\mu \alpha \nu \beta}n^{\alpha}n^{\beta}~~~~~~\mbox{(2nd term vanishes due to the product} \nonumber \\
&=& 2\epsilon \, R_{\alpha \beta}n^{\alpha}n^{\beta}~~~~~~~~~~~~~~\mbox{of symmetric and anti-symmetric part)} \nonumber \\
&=& 2\epsilon \, \left[R_{\alpha \beta}n^{\alpha}\right]n^{\beta} \nonumber \\
&=& 2\epsilon \, \left[n_{~;\beta \alpha}^{\alpha}- n_{~;\alpha \beta}^{\alpha}\right]n^{\beta} \nonumber \\
&=& 2\epsilon \left[\left(n_{~;\beta}^{\alpha}n^{\beta}\right)_{;\alpha}- n_{~;\beta}^{\alpha}n_{~;\alpha}^{\beta}- \left(n_{~;\alpha}^{\alpha}n^{\beta}\right)_{;\beta}+ n_{~;\alpha}^{\alpha}n_{~;\beta}^{\beta}\right] \nonumber \\
&=& 2\epsilon \left[\left(n_{~;\beta}^{\alpha}n^{\beta}- n_{~;\beta}^{\beta}n^{\alpha}\right)_{;\alpha}- n_{~;\beta}^{\alpha}n_{~;\alpha}^{\beta}+ K^2 \right] \nonumber
\end{eqnarray}
\begin{eqnarray}
\mbox{Now,}~~~n_{~;\beta}^{\alpha}n_{~;\alpha}^{\beta} &=& g^{\beta \mu}g^{\alpha \nu}n_{\nu ; \beta}n_{\mu ; \alpha}= g^{\beta \mu}g^{\alpha \nu}n_{\alpha ; \beta}n_{\mu ; \nu}~~~~~(\alpha \rightleftharpoons \nu) \nonumber \\
&=& \left(\epsilon \, n^{\beta}n^{\mu}+ h^{\beta \mu}\right)\left(\epsilon \, n^{\alpha}n^{\nu}+ h^{\alpha \nu}\right)\,n_{\alpha ; \beta}n_{\mu ; \nu} \nonumber \\
&=& \left(\epsilon \, n^{\beta}n^{\mu}+ h^{\beta \mu}\right)h^{\alpha \nu}\,n_{\alpha ; \beta}n_{\mu ; \nu}~~~~\left(\because ~~ n^{\alpha}n_{\alpha ; \beta}= \frac{1}{2}\left(n^{\alpha}n_{\alpha}\right)_{;\beta}= 0\right) \nonumber \\
&=& h^{\beta \mu}h^{\alpha \nu}\,n_{\alpha ; \beta}\,n_{\mu ; \nu} \nonumber \\
&=& \left(h^{bm}e_{b}^{\beta}e_{m}^{\mu}\right)\left(h^{an}e_{a}^{\alpha}e_{n}^{\nu}\right)\,n_{\alpha ; \beta}\,n_{\mu ; \nu} \nonumber \\
&=& h^{bm}h^{an}\left(n_{\alpha ; \beta}e_{a}^{\alpha}e_{b}^{\beta}\right)\left(n_{\mu ; \nu}e_{m}^{\mu}e_{n}^{\nu}\right) \nonumber \\
&=& h^{bm}h^{an}\,K_{ab}K_{mn}= K_{ab}K^{ab} \nonumber
\end{eqnarray}
$$\mbox{So,}~~2\epsilon \, h^{ab}R_{\mu \alpha \nu \beta}n^{\alpha}e_{a}^{\mu}n^{\beta}e_{b}^{\nu}= 2\epsilon\left[K^{2}- K_{ab}K^{ab}+ \left(n_{~;\beta}^{\alpha}n^{\beta}- n_{~;\beta}^{\beta}n^{\alpha}\right)_{;\alpha}\right].$$

Similarly, the second term on the R.H.S. of eq.\,(\ref{6.47})
\begin{eqnarray}
&=& h^{ab}h^{mn}\left(R_{\mu \alpha \nu \beta}e_{a}^{\mu}e_{m}^{\alpha}e_{b}^{\nu}e_{n}^{\beta}\right)= h^{ab}h^{mn}\left[R_{manb}+ \epsilon\left(K_{mb}K_{an}- K_{mn}K_{ab}\right)\right] \nonumber \\
&=& R_{\Sigma}+ \epsilon\left(K^{ab}K_{ab}- K^2\right) \nonumber
\end{eqnarray}
\begin{equation} \label{6.51}
\therefore ~~~ R= R_{\Sigma}+ \epsilon\left(K^{ab}K_{ab}- K^2\right)+ 2\epsilon\left(n_{~;\beta}^{\alpha}n^{\beta}- n^{\alpha}n_{~;\beta}^{\beta}\right)_{;\alpha}\,.
\end{equation}
which shows the expression of the Ricci scalar of the manifold evaluated on the hypersurface $\Sigma$\,.\\

Suppose our space-time is a $(n+1)$\,-dimensional manifold and the hypersurface $\Sigma$ is a $n$ -dimensional manifold and is space-like in nature. The Einstein field equations
$$G_{\mu \nu}= \kappa \, T_{\mu \nu}$$
on the manifold are $\dfrac{n(n+1)}{2}$ in number. Now when we express these field equations on the hypersurface $\Sigma$ then we have from equations (\ref{6.49}) and  (\ref{6.50}) (with $\epsilon = -1$ for space-like hypersurface).
\begin{equation} \label{6.52}
R_{\Sigma}+ \left(-K_{ab}K^{ab}+ K^{2}\right)= 2G_{\rho \sigma}n^{\rho}n^{\sigma}= 2\kappa \, T_{\rho \sigma}n^{\rho}n^{\sigma}= 2\kappa \, \rho
\end{equation}
and
\begin{equation} \label{6.53}
K_{~a|b}^{b}- K_{,a}= G_{\rho \sigma}e_{a}^{\rho}n^{\sigma}= \kappa \, T_{\rho \sigma}e_{a}^{\rho}n^{\sigma}= \kappa \, j_a\,
\end{equation}
when $n^{\mu}= (1,0,0, \ldots ,0)$ is the hypersurface orthogonal vector and  is along the time direction and $j_a$ is heat flow vector on the hypersurface $\Sigma$\,. Equations (\ref{6.52}) and (\ref{6.53}) are termed as constrain equations. Note that equation\,(\ref{6.52}) is a scalar equation and is known as scalar constrain equation, while equation\,(\ref{6.53}) is a vector equation and is known as vector constrain equation. Thus we have $(n+1)$ -constrain equations. The remaining Einstein equations\,: $G_{\mu \nu}e_{a}^{\mu}e_{b}^{\nu}= \kappa \, T_{\mu \nu}e_{a}^{\mu}e_{b}^{\nu}$ cannot be expressed solely in terms of hypersurface quantities\,$({\it i.e.} h_{ab}$ and $K_{ab})$ but they represent the evolution equations of $h_{ab}$ and $K_{ab}$ . Therefore we have\\

No. of field equations in $(n+1)$ -dimensional manifold $= \dfrac{(n+1)(n+2)}{2}$ .\\

No. of constrain equations in $n$ -dimensional hypersurface $= n+1$ .\\

No. of evolution equations in the hypersurface $= \dfrac{n(n+1)}{2}$ .\\

\begin{center}
\includegraphics[scale=0.30]{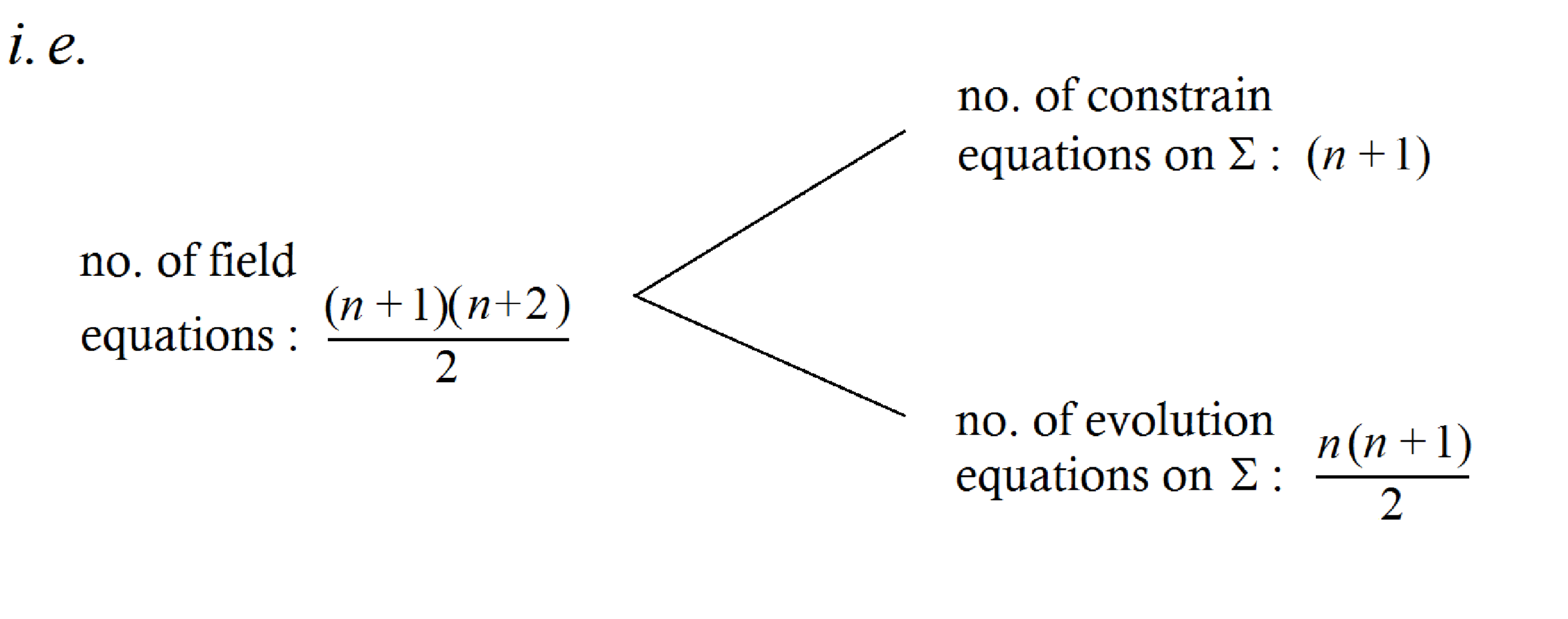}
\end{center}

Thus in usual four dimensional space-time we have 10 distinct field equations and in any space-like hypersurface\,({\it i.e.} in (3+1)\,-decomposition) there are four constrain equations and remaining six are the evolution equations.\\

\section{Geometrical characterization of different types of space-time models}

~~~In this section we shall discuss geometrical aspects of different space-time models namely stationary space-time, static space-time, and spherically symmetric space-time. In the previous section we have defined hypersurface orthogonal vector. At first we determine the condition for which the normal vector to be a Killing vector.\\

\subsection{Hypersurface orthogonal Killing vector field}

~~~Suppose a family of hypersurfaces are described by
\begin{equation} \label{6.54}
f(y^{\mu})= c
\end{equation}
\begin{wrapfigure}{r}{0.32\textwidth}\vspace{-\intextsep}
\includegraphics[height=3 cm , width=5.5 cm ]{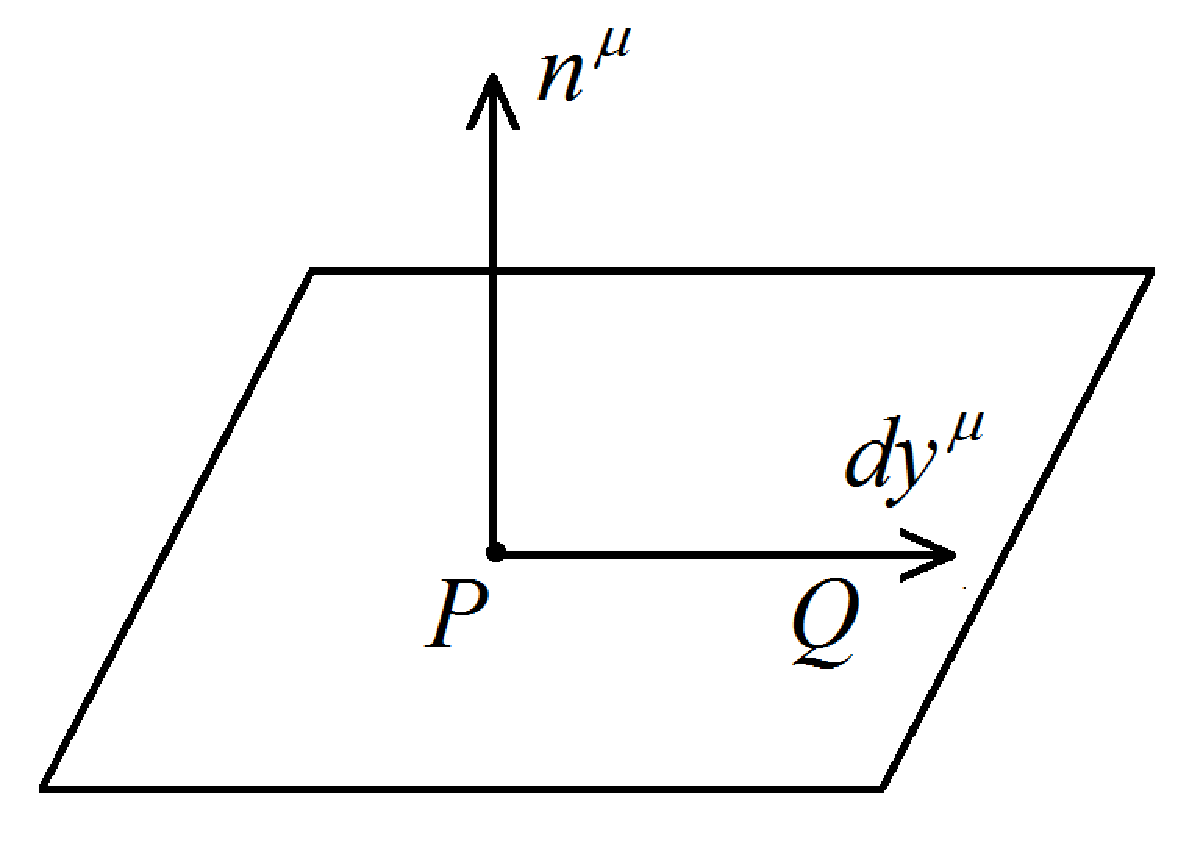}
\begin{center}\vspace{-\intextsep}
Fig. 6.1
\end{center}
\end{wrapfigure}
where different values of $c$ characterize different member of family. Let $P(y^{\mu})$ and $Q(y^{\mu}+ dy^{\mu})$ be two neighbouring points on the same hypersurface.\\

As~~~$f(y^{\mu}+ dy^{\mu})= c$\\
$i.e.~~f(y^{\mu})+ \dfrac{\partial f}{\partial y^{\mu}}dy^{\mu}= c~~(\mbox{at 1st order})$\\
\begin{equation} \label{6.55}
i.e.~~\frac{\partial f}{\partial y^{\mu}}dy^{\mu}= 0\, ,
\end{equation}
which shows that the covariant vector field $n_{\mu}= \dfrac{\partial f}{\partial y^{\mu}}$ is orthogonal to the hypersurface\,(as $dy^{\mu}$ lies on the hypersurface) at $P$. Thus if a vector field $N^{\mu}$ is proportional to $n^{\mu}$ at every point of the hypersurface then it is said to be hypersurface orthogonal. So we write
\begin{equation} \label{6.56}
N^{\mu}= K(y)\,n^{\mu}
\end{equation}
$$i.e.~~N_{\mu}= K(y)\,n_{\mu}= K(y)\,f_{,\mu}\,.$$

Note that, in general $K$ varies from point to point on the hypersurface.\\

Now,
\begin{equation} \label{6.57}
N_{\mu}\partial _{\nu}N_{\lambda}= Kf_{,\mu}K_{,\nu}f_{,\lambda}+ K^{2}f_{,\mu}f_{,\lambda \nu}\,.
\end{equation}

As the first term in the right hand side is symmetric in $\mu$ and $\lambda$ while the second term is symmetric in $\lambda$ and $\nu$ so the total anti-symmetrization of equation\,(\ref{6.57}) gives
\begin{equation} \label{6.58}
N_{[\mu}\partial _{\nu}N_{\lambda ]}= 0.
\end{equation}

As the Christoffel symbols are symmetric in its two lower indices so without any loss of generality we can replace the partial derivative in equation\,(\ref{6.58}) by the covariant derivative {\it i.e.} we have
\begin{equation} \label{6.59}
N_{[\mu}\nabla _{\nu}N_{\lambda ]}= 0.
\end{equation}

This is the condition for $N^{\mu}$ to the normal to the family of hypersurfaces\,(\ref{6.54})\,.

We shall now examine the situation if $N^{\mu}$ is a Killing vector field {\it i.e.}
\begin{equation} \label{6.60}
\mathcal{L}_{N}g_{\mu \nu}= 0~~~i.e.~~~\nabla _{\nu}N_{\mu}+ \nabla _{\mu}N_{\nu}= 0.
\end{equation}

Now using (\ref{6.60}) in (\ref{6.59}) we obtain
\begin{equation} \label{6.61}
N_{\mu}\nabla _{\nu}N_{\lambda}+ N_{\lambda}\nabla _{\mu}N_{\nu}+ N_{\nu}\nabla _{\lambda}N_{\mu}= 0.
\end{equation}

Contracting with $N^{\lambda}$ and writing $N^{2}= N^{\mu}N_{\mu}$ , we get
\begin{eqnarray}
N_{\mu}N^{\lambda}\nabla _{\nu}N_{\lambda} &+& N^{2}\nabla _{\mu}N_{\nu}+ N_{\nu}N^{\lambda}\nabla _{\lambda}N_{\mu}= 0 \nonumber \\
i.e.~~N_{\mu}\frac{1}{2}\nabla _{\nu}(N^2) &+& \frac{1}{2}N^{2}\left(\nabla _{\mu}N_{\nu}- \nabla _{\nu}N_{\mu}\right)- N_{\nu}N^{\lambda}\nabla _{\mu}N_{\lambda}= 0 ~~~~(\mbox{using (\ref{6.60}) in the second term}) \nonumber \\
i.e.~~N_{\mu}\nabla _{\nu}(N^2) &+& N^{2}\left(\nabla _{\mu}N_{\nu}- \nabla _{\nu}N_{\mu}\right)- N_{\nu}\nabla _{\mu}(N^2)= 0 \nonumber \\
i.e.~~N_{\mu}\partial _{\nu}N^{2} &-& N_{\nu}\partial _{\mu}N^{2}+ N^{2}\left(\partial _{\mu}N_{\nu}- \partial _{\nu}N_{\mu}\right)= 0 \nonumber \\
i.e.~~N^{2}\partial _{\mu}N_{\nu} &-& N_{\nu}\partial _{\mu}N^{2}= N^{2}\partial _{\nu}N_{\mu}- N_{\mu}\partial _{\nu}N^{2} \nonumber \\
i.e.~~\partial _{\mu}\left(\frac{N_{\nu}}{N^{2}}\right) &=& \partial _{\nu}\left(\frac{N_{\mu}}{N^{2}}\right).~~~~\left(N^{2}\neq 0~\mbox{as}~N^{\mu}~\mbox{is a non-null vector}\right) \nonumber
\end{eqnarray}

It shows that $\exists$ a scalar function $f$\,(say) such that
$$\frac{N_{\mu}}{N^{2}}= f_{,\mu}$$
\begin{equation} \label{6.62}
N_{\mu}= N^{2}\,f_{,\mu}~.
\end{equation}

This is the condition for $N^{\mu}$ to be hypersurface orthogonal Killing vector.\\

\subsection{Stationary and Static space-time}

~~~We shall define stationary and static space-times both in co-ordinate independent way or by imposing restriction on the metric of the space-time by a preferential co-ordinate system.\\

If there exists a typical co-ordinate system of the space-time such that all components of the metric tensor are time independent then the space-time is said to be stationary. Clearly, in an arbitrary co-ordinate system, the metric tensor depends on all the co-ordinates.\\

Now in the special co-ordinate system if we define a time-like vector field $K^{\mu}= \delta _{0}^{\mu}$ , then
\begin{eqnarray}
\mathcal{L}_{K}g_{\mu \nu} &=& K^{\lambda}g_{\mu \nu , \lambda}+ g_{\mu \lambda}K_{,\nu}^{\lambda}+  g_{\nu \lambda}K_{,\mu}^{\lambda} \nonumber \\
&=& \delta _{0}^{\lambda}g_{\mu \nu , \lambda}= \frac{\partial g_{\mu \nu}}{\partial x^0}= 0 \nonumber \\
i.e.~~~ \mathcal{L}_{K}g_{\mu \nu} &=& 0\,. \nonumber
\end{eqnarray}

As the last equation is a tensor equation so it holds in any other co-ordinate system and hence $K^{\mu}$ is Killing vector field of the space-time. Thus, a space-time is said to be stationary if $\exists$ a time-like Killing vector field.\\

{\bf Example:} The space-time described by the metric
$$ds^2 = dt^2 - e^{(2t/\alpha)}\left[dx^2 + dy^2 + dz^2\right]$$
is a stationary space-time\,(de Sitter space) because $\exists$ a preferential co-ordinate system $(\overline{t}, \overline{x}, \overline{y}, \overline{z})$ defined as
$$\overline{t}= t- \frac{1}{2}\alpha \ln \left[1- \frac{1}{\alpha ^2}\left(x^2 + y^2 + z^2\right)e^{(2t/\alpha)}\right]$$
$$\overline{x}= xe^{(t/\alpha)}~~,~~\overline{y}= ye^{(t/\alpha)}~~,~~\overline{z}= ze^{(t/\alpha)}$$
so that
$$ds^2 = \left[1- \frac{1}{\alpha ^2}\left(\overline{x}^2 + \overline{y}^2 + \overline{z}^2\right)\right]d\overline{t}^2 - d\overline{x}^2 - d\overline{y}^2 - d\overline{z}^2$$
{\it i.e.} all the metric co-efficients are time independent. Hence de-Sitter space-time is a stationary space-time.\\

For static space-time there are additional properties than the stationarity. If the line element of the stationary space-time in the preferential co-ordinate system has time reversibility then space-time is said to be static. This means that the line element in the preferential co-ordinate system should not contain any product term with `$dt$'\,. Thus in a static space-time $\exists$ a special co-ordinate system for which ({\it i}) all the metric co-efficients are time independent and ({\it ii}) the line element is invariant under time reversal {\it i.e.} $t \rightarrow -t$\,.\\

{\bf Remark:} A static space-time is always a stationary one but not the converse.\\

We shall now determine the extra condition on the time-like Killing vector field for stationary space-time so that the stationary space-time becomes static. In the special co-ordinate system the time-like Killing vector is $K^{\mu}= \delta _{0}^{\mu}$\,. So we have
$$K_{\mu}= g_{\mu \nu}K^{\nu}= g_{\mu \nu}\delta _{0}^{\nu}= g_{\mu 0}$$
So $$K^2 = K_{\mu}K^{\mu}= g_{\mu 0}\delta _{0}^{\mu}= g_{00}$$

So from the hypersurface orthogonality condition we have
$$g_{\mu 0}= g_{00}f_{,\mu}$$
$i.e.~~~f_{,0}= 1$ , which on integration gives
\begin{equation} \label{6.63}
f= x^0 + f_{0}(x^a)
\end{equation}
where $f_0$ is an arbitrary function of space co-ordinates only. We now consider a co-ordinate transformation, keeping the space co-ordinates unchanged as
\begin{equation} \label{6.64}
x_0 \rightarrow ~\overline{x}_{0}= x^0 + f_{0}(x^a)~~,~~x^a \rightarrow ~\overline{x}^{a}= x^a\,.
\end{equation}

In this new co-ordinate system the Killing vector and the metric components become
\begin{equation} \label{6.65}
\overline{K}^{\mu}= \delta _{0}^{\mu}~,~~\overline{g}_{\mu \nu ,0}= 0~,~~\overline{g}_{00}= g_{00}~,~~\overline{g}_{0a}= 0\,.
\end{equation}

Hence the line element in the new co-ordinate system does not contain any cross term and all metric co-efficients are time independent. So the space-time is static. Thus, a space-time will be static in nature if it possess a hypersurface orthogonal time-like Killing vector field.\\

{\bf Example:} The Schwarzschild space-time is described by the line element
$$ds^2 = +\left(1- \frac{2M}{r}\right)dt^2 - \frac{dr^2}{1- \frac{2M}{r}}- r^2 d\Omega _{2}^{2}~.$$

It is an example of static space-time.\\

\subsection{Spherically Symmetric space-time and line element}

~~~A space-time is said to be spherically symmetric if and only if it admits three linearly independent space-like Killing vector fields $K^l\,(l=1,2,3)$ having closed orbits\,({\it i.e.} topologically circles) and have closed commutation algebra as follows\,:
\begin{equation} \label{6.66}
\left[K^1 , K^2\right]= K^3~~,~~~\left[K^2 , K^3\right]= K^1~~,~~~\left[K^3 , K^1\right]= K^2\,.
\end{equation}

Further, in spherically symmetric space-time, $\exists$ a typical\,(cartesian) co-ordinate system in which the components of the Killing vector field $K^l$ are of the form\,:
\begin{equation} \label{6.67}
K^0 = 0~~,~~~K^a =w_{b}^{a}x^b~~,~~~w_{ab}= -w_{ba}
\end{equation}
where $w_{ab}$ depends on the Eulerian angles.\\

In spherically symmetric space-time, the general form of the line element can be written as
\begin{equation} \label{6.68}
ds^2 = A\,dt^2 - 2B\,dtdr - C\,dr^2 - D\,d\Omega _{2}^{2}
\end{equation}
where $A,B,C$ and $D$ are unknown functions of $t$ and $r$. We define a new radial co-ordinate $R= D^{\frac{1}{2}}$ so that the above line element becomes
\begin{eqnarray} \label{6.69}
ds^2 &=& A'(t,R)\,dt^2 + 2B'(t,R)\,dtdR - C'(t,R)\,dR^2 - R^2 \, d\Omega _{2}^{2} \nonumber \\
&=& \frac{1}{A'}\left(A'dt- B'dR\right)^2 - \left(\frac{B^{\prime \,2}}{A'}+ C'\right)dR^2 - R^2 \, d\Omega _{2}^{2}~.
\end{eqnarray}

In general, $A'dt- B'dR$ may not be perfect differential. However, by multiplying it by some function $\mu(R,t)$~, the above expression become exact {\it i.e.}
$$dT= \mu\left(A'dt- B'dR\right).$$

Hence line element\,(\ref{6.69}) becomes
$$ds^2 = \frac{1}{A'\mu ^2}dT^2 - \left(\frac{B^{\prime \,2}}{A'}+ C'\right)dR^2 - R^2 \, d\Omega _{2}^{2}~.$$

Now writing, $e^{\nu}= \left(A'\mu ^2\right)^{-1}~~,~~e^{\lambda}= \frac{B^{\prime \,2}}{A'}+ C'$ we have the line element.
\begin{equation} \label{6.70}
ds^2 = e^{\nu}dT^2 - e^{\lambda}dR^2 - R^2 \, d\Omega _{2}^{2}~.
\end{equation}

This is the general form of the spherically symmetric line element with $\nu = \nu(T,R)~,~\lambda = \lambda(t,R)$.\\

\section{Relativistic Cosmology}

~~~Relativistic cosmology has three main ingredients namely\\

({\it i}) The cosmological Principle,~~~({\it ii}) Weyl's postulate,~~~({\it iii}) General Relativity.\\

\subsection{The Cosmological Principle}

~~~This principle states that at each epoch, the universe presents the same aspect from every point except for local irregularities.\\

Mathematically, if we assume a cosmic time $t$ and construct space-like hypersurfaces\,:~~~$t=$\,
\\constant , then the above statement means that each slice has no privileged points {\it i.e.} it is homogeneous. Further, a space-like hypersurface is homogeneous if it admits a group of isometries which maps any point into any other point. Also, there exists three independent space-like Killing vectors at any point on each slice. Physically, it implies that at any instant each fundamental observer on the same hypersurface observes identical state of the universe around him. Thus the principle requires that not only should a slice\,(hypersurface) have no preferred points but it should have no preferred directions about any point. A manifold which has no privileged directions about a point is called isotropic and hence it must be spherically symmetric about that point. Thus according to cosmological principle space-time can be foliated into space-like hypersurfaces which are spherically symmetric about any point in them. Therefore, cosmological principle is a simplicity principle which states that the universe is both homogeneous and isotropic.\\

{\bf Note\,:} The homogeneity of the universe has same sense as the homogeneity of a gas. Further, homogeneity is applicable over the length scale of $10^8 - 10^9$ light years.\\

{\bf Observational evidence\,:}\\

The greatest support in favour of isotropy is the cosmic microwave background radiation\,(CMBR). According to CMBR the universe at present is pervaded by a bath of thermal radiation with a temperature of 2.7 K\,(anisotropy is observed to fractions of a percent). It is speculated that this radiation is a thermal remnant of the hot big bang. Further, the counts of galaxies and the linearity of the Hubble law can be considered as the observational support in favour of spatial homogeneity.\\

\subsection{Weyl's Postulate}

Weyl's Postulate states that particles of the substratum lie in space-time on a congruence of time-like geodesics diverging from a point in the finite or infinite past. Following Weyl's postulate, through each space-time point there is a unique geodesic of the family as geodesics can only intersect at singularity either at past\,(big-bang) or possibly at future\,(big-crunch). Hence substratum particles possess unique velocity at every space-time point. This characterizes the substratum to be perfect fluid -- the essence of Weyl's postulate.\\

{\bf Note\,:} Strictly speaking the galaxies do not have exactly this motion and the deviation from the general motion appear to be random in nature. However, the randomness is less than one-thousandth of the velocity of light. As the relative velocities of the galaxies due to general motion is of the order of the velocity of light so it is reasonable to neglect the random motion at least in the first order of approximation.\\

We shall now discuss the geometrical aspects of Weyl's postulate. According to this postulate, the geodesics of the substratum are orthogonal to a family of space-like hypersurfaces. So without any loss of generality we choose these hypersurfaces to be $t=$\,constant in a typical co-ordinate system $(t,x^1,x^2,x^3)$ so that the space co-ordinates $(x^1,x^2,x^3)$ are constant along the geodesics {\it i.e.} space-like co-ordinates of each substratum particle are constant along its geodesic. Such co-ordinate system is termed as co-moving system. In this co-ordinate system the line element of the space-time takes the form
\begin{equation} \label{6.71}
ds^2 = dt^2 - h_{ab}dx^{a}dx^{b}~~,~~~a,b=1,2,3
\end{equation}
where $h_{ab}= h_{ab}(t,x)$ and $t$ is identified as the cosmic time.\\

\subsection{The geometry of space-time as a consequence of Weyl postulate and cosmological principle}

In this section, we shall determine explicit geometry of space-time using both the cosmological principle and Weyl's postulate.\\

Let us consider a small triangle formed of three particles on the hypersurface $t= t_1$ . At a later instant $t= t_2\,(>t_1)$ these particles also form a triangle on the hypersurface $t= t_2$ . In general, there may not be any similarity between these two triangles. But due to cosmological principle, each hypersurface is homogeneous and isotropic {\it i.e.} no point and no direction on it will be preferential. Hence the second triangle\,(on the hypersurface $t= t_2$) must be similar to the first one\,(on the hypersurface $t= t_1$) and the ratio of the similar sides should be independent of the position of the triangle on the hypersurface. Thus the metric co-efficient $h_{ab}$ in equation\,(\ref{6.71}) must have the product form as follows\,:
\begin{equation} \label{6.72}
h_{ab}(t,x)= \left[R(t)\right]^{2}\cdot q_{ab}(x^a).
\end{equation}

As the magnification factor is the ratio of the values of $R(t)$ at the two hypersurfaces, hence $R(t)$ is called the scale factor. Note that $R(t)$ should be real otherwise a space-like interval at some instant may change to time-like interval at some other instant.\\

\begin{wrapfigure}{r}{0.37\textwidth}\vspace{-\intextsep}
	\includegraphics[height=7 cm , width=6.5 cm ]{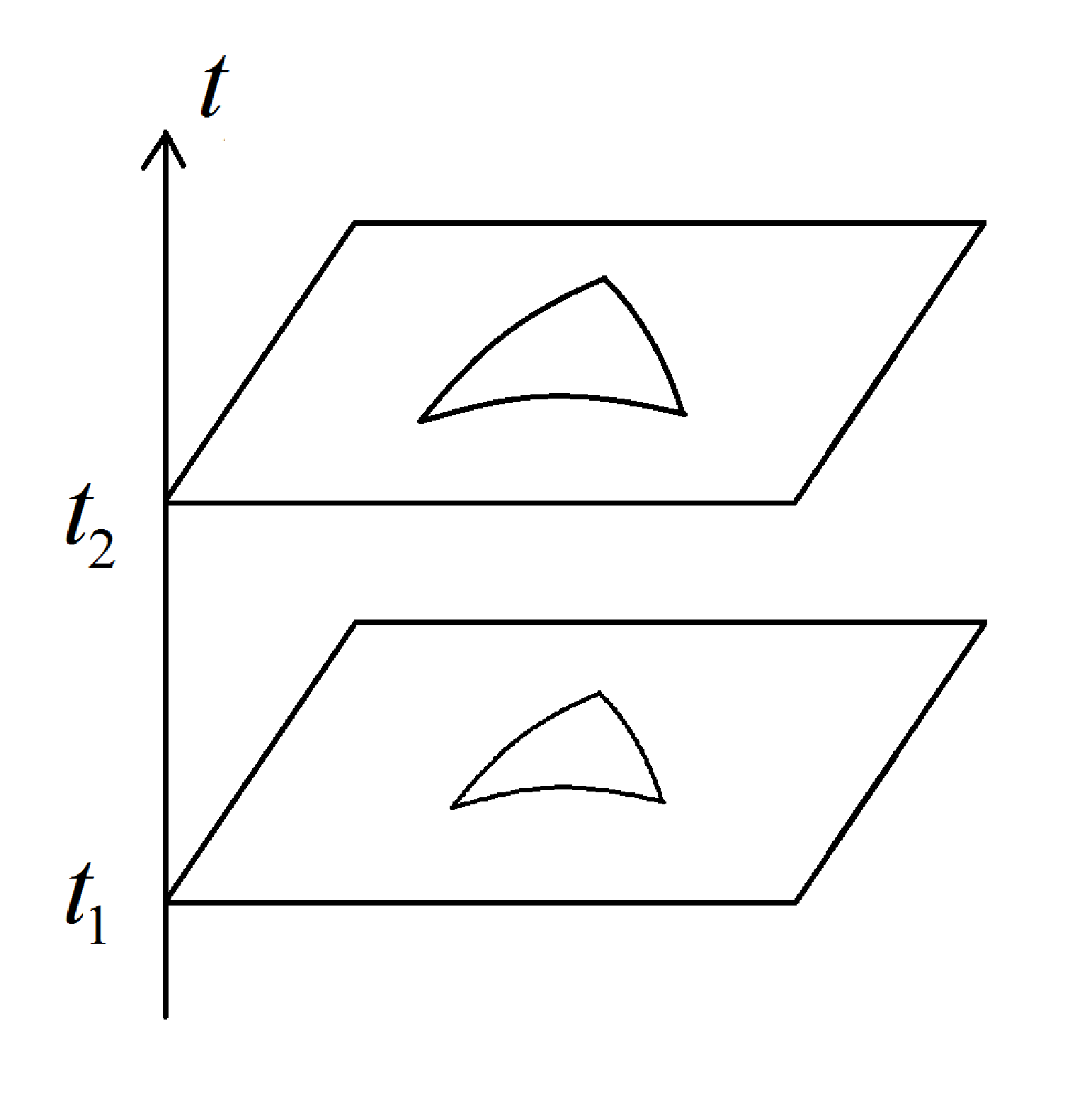}
	\begin{center}\vspace{-\intextsep}
		Fig. 6.2
	\end{center}
\end{wrapfigure}

Further, due to homogeneity and isotropic nature of the hypersurfaces the curvature at any point must be a constant, otherwise points on the hypersurface would not be geometrically identical. Hence it is a space of constant curvature. Mathematically, a space of constant curvature is identified by the following form of the curvature tensor
\begin{equation} \label{6.73}
R_{\alpha \beta \gamma \delta}= k\left(g_{\alpha \gamma}g_{\beta \delta}- g_{\alpha \delta}g_{\beta \gamma}\right)
\end{equation}
where the constant `$k$' is identified as the curvature and its sign will describe geometrically distinct spaces. So for the present 3-spaces\,(hypersurfaces) we write
\begin{equation} \label{6.74}
R_{abcd}= k\left(g_{ac}g_{bd}- g_{ad}g_{bc}\right).
\end{equation}

Contracting with $g^{ac}$ we obtain
\begin{equation} \label{6.75}
R_{bd}= R_{abcd}g^{ac}= 2k\,g_{bd}\,.
\end{equation}

Moreover, due to spherically symmetric nature of the hypersurfaces about every point, the line element on a hypersurface can be written as
\begin{equation} \label{6.76}
d\sigma ^2 = q_{ab}dx^{a}dx^{b}= e^{\lambda}dr^2 + r^2 d\Omega _{2}^{2}
\end{equation}
with $\lambda = \lambda (r)$.\\

For this line element the non-vanishing components of the Ricci tensor are
\begin{equation} \label{6.77}
R_{11}= \frac{\lambda ^1}{r}~,~~R_{22}= \mathrm{cosec}^{2}\theta \, R_{33}= 1+ \frac{r}{2}e^{-\lambda}\lambda ^{1}- e^{-\lambda}\,.
\end{equation}

Using (\ref{6.77}) in (\ref{6.75}) we get
\begin{equation} \label{6.78}
\frac{\lambda ^1}{r}= 2ke^{\lambda}~~~\mbox{and}~~~1+ \frac{r}{2}e^{-\lambda}\lambda ^{1}- e^{-\lambda}= 2kr^2
\end{equation}
which has the solution
\begin{equation} \label{6.79}
e^{-\lambda}= 1- kr^2
\end{equation}
and we have
\begin{equation} \label{6.80}
d\sigma ^2 = \frac{dr^2}{1- kr^2}+ r^{2}\,d\Omega _{2}^{2}\,.
\end{equation}

Using equations (\ref{6.80}) and (\ref{6.72}) in (\ref{6.71}), the four dimensional line element takes the form
\begin{equation} \label{6.81}
ds^2 = dt^2 - R^{2}(t)\left[\frac{dr^2}{1- kr^2}+ r^{2}\,d\Omega _{2}^{2}\right]
\end{equation}

Now choosing a new radial co-ordinate $\overline{r}$ as
\begin{equation} \label{6.82}
r= \frac{\overline{r}}{\left(1+ \frac{1}{4}k\overline{r}^2\right)}
\end{equation}
the line element (\ref{6.71}) takes the conformally flat form as
\begin{equation} \label{6.83}
ds^2 = dt^2 - \frac{R^{2}(t)}{\left(1+ \frac{1}{4}k\overline{r}^2\right)^2}\left[d\overline{r}^2 + \overline{r}^{2}\,d\Omega _{2}^{2}\right].
\end{equation}

Now to eliminate the arbitrariness in the magnitude of `$k$' we write
$$k= \left|k\right|\, \kappa ~~~~\mbox{for}~k \neq 0~,$$
where $\kappa = \pm 1$ and rescale the radial co-ordinate as
$$r^{\ast}= \left|k\right| ^{1/2}\,r~,$$
so that the line element (\ref{6.81}) becomes
\begin{equation} \label{6.84}
ds^2 = dt^2 - \frac{R^{2}(t)}{\left|k\right|}\left[\frac{dr^{\ast \,2}}{1- \kappa r^{\ast \,2}} + r^{\ast \,2}\,d\Omega _{2}^{2}\right].
\end{equation}

Defining,
\begin{eqnarray} \label{6.85}
\left. \begin{array}{rr}
S(t)= R(t)/\left|k\right| ^{1/2}&\mbox{for}~k \neq 0 \\
= R(t)~~~~~~~~~&\mbox{for}~k= 0
\end{array} \right.
\end{eqnarray}
and dropping the star symbol over the radial co-ordinate we obtain
\begin{equation} \label{6.86}
ds^2 = dt^2 - S^{2}(t)\left[\frac{dr^2}{1- \kappa r^2} + r^{2}\,d\Omega _{2}^{2}\right]
\end{equation}
or equivalently from (\ref{6.83})
\begin{equation} \label{6.87}
ds^2 = dt^2 - S^{2}(t)\left[\frac{d\overline{r}^{2}+ \overline{r}^{2}\,d\Omega _{2}^{2}}{\left(1+ \frac{\kappa}{4}\overline{r}^2\right)^2}\right]
\end{equation}
with $\kappa = 0, \pm 1$\,.\\

This is known as Friedmann--Leimatre-Robertson-Walker\,(FLRW) line element. The geometry of the hypersurface\,:~~$t= t_0$ is given by
\begin{equation} \label{6.88}
d\sigma ^2 = S_{0}^{2}\left[\frac{dr^2}{1- \kappa r^2} + r^{2}\,d\Omega _{2}^{2}\right]
\end{equation}
where $S_{0}= S(t_0)$ is the radius of the universe at the instant $t= t_0$\,.\\

\subsection{Geometry of 3-spaces\,(hypersurface) of constant curvature}

~~~In this section we shall discuss the geometry of the hypersurface for three different values of the curvature scalar $\kappa = 0, \pm 1$\,.\\

\textbf{a) \underline{$\kappa = +1$ :~~~closed model}}\\

For this choice of $\kappa$ we see from equation\,(\ref{6.88}) that the 3-space line element has singularity as $r \rightarrow 1$ (co-efficient of $dr^2$ become singular as $r \rightarrow 1$). To remove this co-ordinate singularity we introduce a new co-ordinate $\chi$ as $r= \sin \chi$ , so that line element becomes
\begin{equation} \label{6.89}
d\sigma ^2 = S_{0}^{2}\left[d\chi ^{2}+ \sin ^{2}\chi \, d\Omega _{2}^{2}\right].
\end{equation}

We now define a set of four variables $(x^1,x^2,x^3,x^4)$ as
\begin{equation} \label{6.90}
x^1 = S_{0}\cos \chi~,~~x^2 = S_{0}\sin \chi \sin \theta \cos \phi~,~~x^3 = S_{0}\sin \chi \sin \theta \sin \phi~,~~\mbox{and}~~x^4 = S_{0}\sin \chi \cos \theta~,
\end{equation}
so that
$$(x^1)^2 + (x^2)^2 + (x^3)^2 + (x^4)^2 = S_{0}^{2}~,$$
a hypersphere in four dimensional Euclidean space.\\

Also we have
\begin{equation} \label{6.91}
d\sigma ^2 = (dx^1)^2 + (dx^2)^2 + (dx^3)^2 + (dx^4)^2 = S_{0}^{2}\left[d\chi ^{2}+ \sin ^{2}\chi \, d\Omega _{2}^{2}\right].
\end{equation}

\begin{wrapfigure}[17]{r}{0.4\textwidth}\vspace{-\intextsep}
	\includegraphics[height=5.7 cm , width=7 cm ]{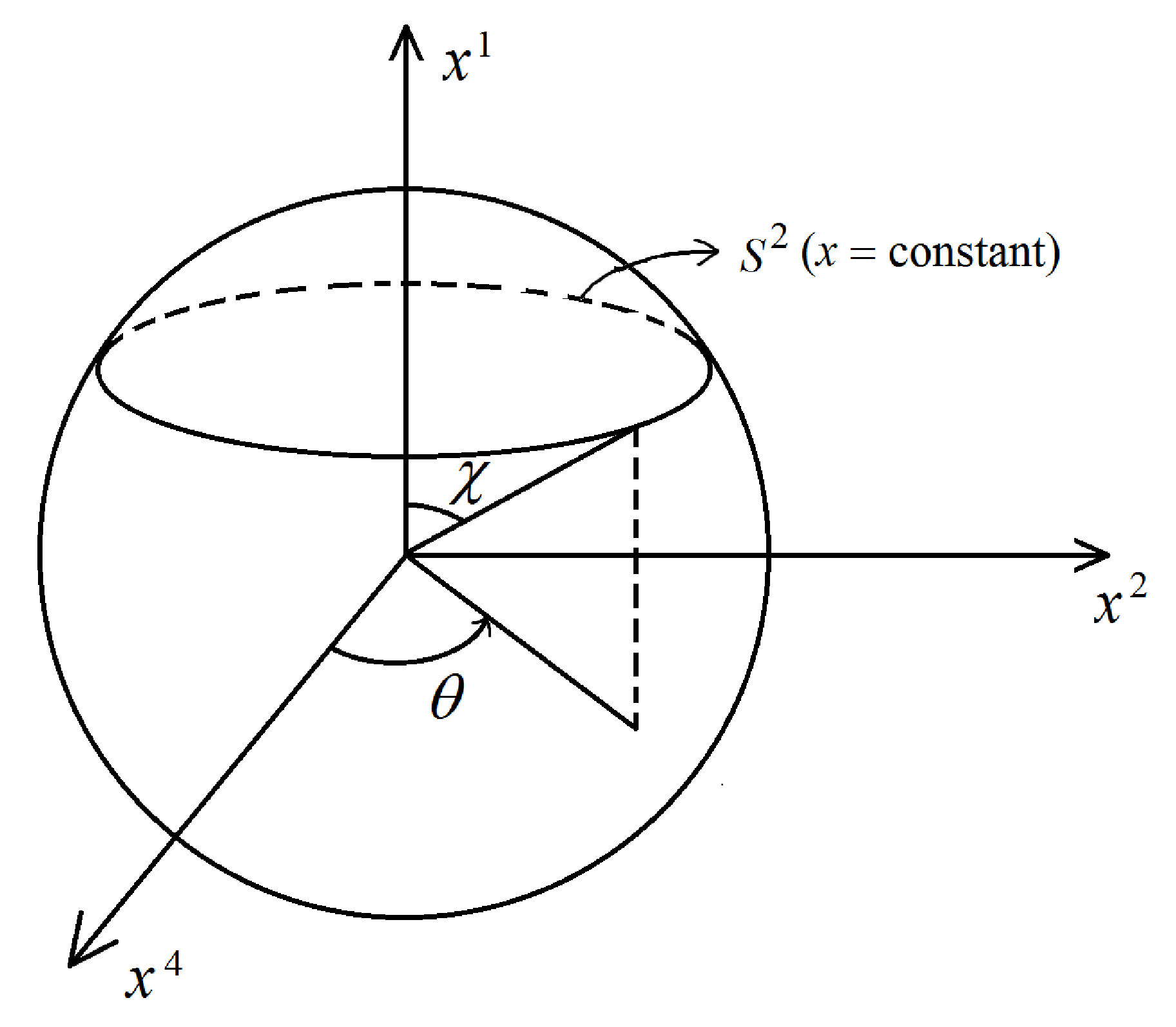}
	\begin{center}
		Fig. 6.3
	\end{center}
\end{wrapfigure}
Hence the hypersurface\,({\it i.e.} 3-space) is embedded in a four dimensional Euclidean space and in particular 3-space can be regarded as 3-sphere in four dimensional Euclidean space. Also over the hypersurface the three angular co-ordinates $(\chi , \theta , \phi)$ vary over the range\,: $0 \leq \chi \leq \pi$, $0 \leq \theta \leq \pi$, $0 \leq \phi < 2\pi$.\\

The figure shows the hypersurface where $x^3 = 0$ ({\it i.e.} $\phi = 0$) {\it i.e.} one dimension is suppressed. It is clear from the figure that the two surface $\chi =$\,constant appears as circles {\it i.e.} 2-spheres of surface area\,:
$$A_{\chi}= \int _{\theta = 0}^{\pi} \int _{\phi = 0}^{2\pi} \left(S_{0}\sin \chi \, d\theta \right)\left(S_{0}\sin \chi \sin \theta d\phi \right)$$
$$= 4\pi S_{0}^{2}\sin ^{2}\chi$$
and $(\theta , \phi)$ are the usual spherical co-ordinates on the 2-sphere. Note that the 2-sphere has vanishing area at the two poles and then gradually increases to a maximum at the equator. Further, the 3-volume bounded by the hypersurface is given by

\begin{wrapfigure}[10]{r}{0.2\textwidth}\vspace{-\intextsep}
\includegraphics[height=4 cm , width=3.5 cm ]{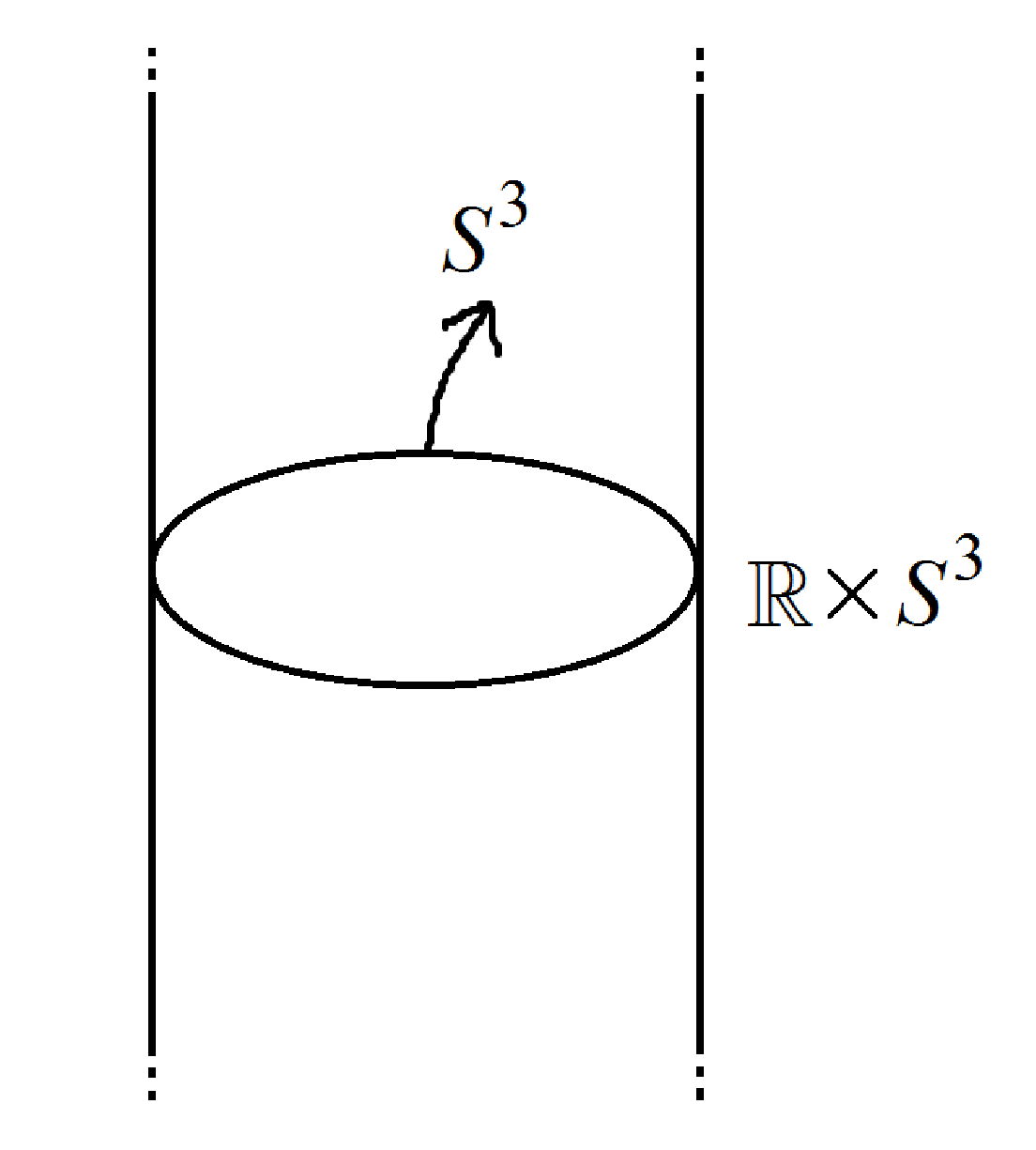}
\begin{center}
Fig. 6.4
\end{center}
\end{wrapfigure}

$$V= \int _{\chi = 0}^{\pi} \left(S_{0}d\chi\right) \int _{\theta = 0}^{\pi} \left(S_{0}\sin \chi \, d\theta \right) \int _{\phi = 0}^{2\pi} \left(S_{0}\sin \chi \sin \theta d\phi \right)= 2\pi ^{2}S_{0}^{3}~,$$
which justifies `$S_0$' as the radius of the universe. This 3-space is the generalization of an $S^2$ or 2-sphere as a three dimensional entity and is called as $S^3$ or 3-sphere. As it is the totality of everything that exists at any epoch so there are no physical points outside it nor does it have a boundary. Therefore, the topology of this 3-space is closed and bounded {\it i.e.} compact while that of the whole space-time is called cylindrical $\mathbb{R} \times S^3$ with cosmic time represented by $\mathbb{R}$.\\

\textbf{b) \underline{$\kappa = 0$ :}}\\

The transformation $(R,\theta , \phi) \rightarrow (x^1,x^2,x^3)$ as
$$x^1 = R\sin \theta \cos \phi~~,~~~x^2 = R\sin \theta \sin \phi~~,~~~x^3 = R\cos \theta$$
with $R= S_{0}\,r$ , simplifies the three dimensional line element
$$d\sigma ^2 = (dx^1)^2 + (dx^2)^2 + (dx^3)^2 = dR^{2}+ R^{2}\, d\Omega _{2}^{2}~.$$

Hence the hypersurface is a three dimensional Euclidean space which is covered by the usual spherical polar co-ordinates\,:\\
$$0 \leq R < \infty~~,~~0 \leq \theta \leq \pi~~,~~0 \leq \phi < 2\pi~.$$\\

The topology of the space-time is the four dimensional Euclidean space {\it i.e.} $\mathbb{R}^4$ and is open in nature.\\\\
\textbf{c) \underline{$\kappa = -1$ :}}\\

In this case by introducing a transformation of the radial co-ordinate\,: $r= \sinh \chi$ the 3D line element takes the form\,:
$$d\sigma ^2 = S_{0}^{2}\left[d\chi ^{2}+ \sinh ^{2}\chi \, d\Omega _{2}^{2}\right]$$
which clearly shows that the hypersurface can no longer be embedded in a 4D Euclidean space. However, similar to the transformation (\ref{6.90}) if we introduce
\begin{equation} \label{6.92}
p = S_{0}\cosh \chi~,~~q = S_{0}\sinh \chi \sin \theta \cos \phi~,~~u = S_{0}\sinh \chi \sin \theta \sin \phi~,~~\mbox{and}~~v = S_{0}\sinh \chi \cos \theta~,
\end{equation}
\begin{wrapfigure}{r}{0.45\textwidth}\vspace{-\intextsep}
	\includegraphics[height=7 cm , width=8 cm ]{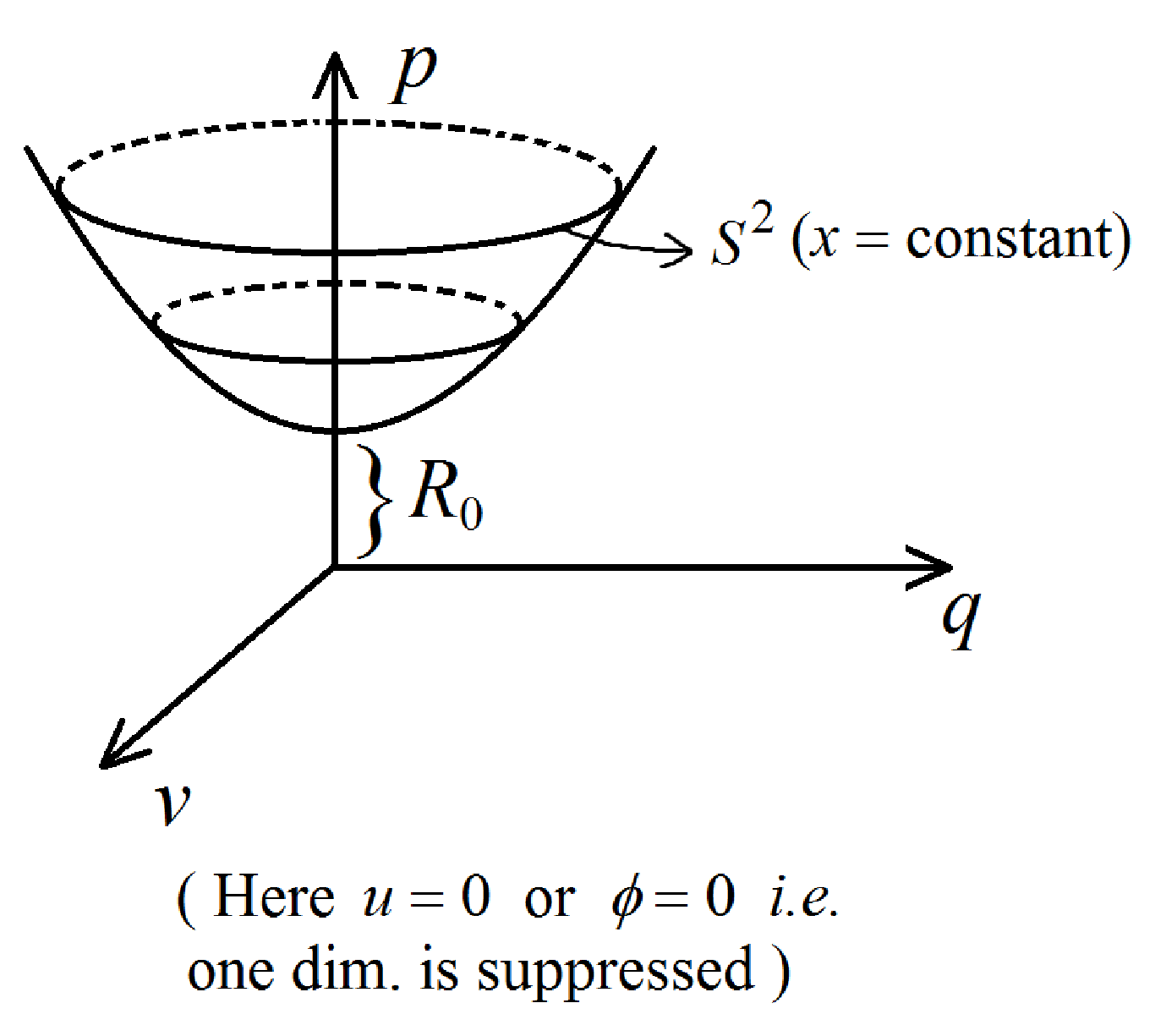}
	\begin{center}
		Fig. 6.5
	\end{center}
\end{wrapfigure}
then
$$d\sigma ^2 = -dp^2 + dq^2 + du^2 + dv^2$$
where
$$p^2 - q^2 - u^2 - v^2 = S_{0}^{2}\,.$$

Hence the 3-surface can be embedded in a flat Minkowskian space with signature $+2$\,. In particular, the hypersurface is a 3D hyperboloid in 4D Minkowski space as shown in the figure.\\

Here the co-ordinate range for the hypersurface is
$$0 \leq \chi < \infty~~,~~0 \leq \theta \leq \pi~~,~~0 \leq \phi < 2\pi~.$$

Also it is evident from the transformation that the 2-surface $\chi =$\,constant appears as circle {\it i.e.} 2-sphere
of surface area
$$A_{\chi}= 4\pi R_{0}^{2} \sinh ^{2}\chi~,$$
and $(\theta , \phi)$ are the standard spherical polar co-ordinates on these 2-spheres. As $\chi$ ranges from 0 to $\infty$ so the surface area of the successive 2-spheres increases from zero to infinite large value. In this case also the topology is $\mathbb{R}^4$ and open.\\

{\bf Note\,:} In the above three cases, we have only specified the simplest topology possible, however, it is possible to have other complicated topologies.\\

Thus following three ingredients of relativistic cosmology we have\,:\\

{\bf I.} The cosmological principle leads to the FLRW line element\,:
$$ds^2 = dt^2 - S^{2}(t)\left[\frac{dr^2}{1- \kappa r^2} + r^{2}\,d\Omega _{2}^{2}\right].$$

{\bf II.} Weyl's postulate requires that the substratum to be perfect fluid with energy momentum tensor
$$T_{\mu \nu}= (\rho + p)u_{\mu}u_{\nu} - pg_{\mu \nu}~.$$

{\bf III.} General Relativity gives the field equations
$$G_{\mu \nu}= \kappa \, T_{\mu \nu}~.$$

So in the preferred co-ordinate system\,(co-moving) $u^{a}= (1,0,0,0)$ and the explicit form of the field equations are
\begin{equation}\label{6.93}
	3\frac{\dot{S}^2}{S^2}+ 3\frac{\kappa}{S^2}= \kappa \, \rho
\end{equation}
and
\begin{equation}\label{6.94}
	2\frac{\ddot{S}}{S}+ 3\frac{\dot{S}^2}{S^2}+ \frac{\kappa}{S^2}= -\kappa \, p
\end{equation}	
with energy conservation relation\,:
\begin{equation}\label{6.95}
	\dot{\rho}+ 3\dfrac{\dot{S}}{S}(\rho + p)= 0\,.
\end{equation}

Note that equations (\ref{6.93}) - (\ref{6.95}) are not independent. One can be derived from the other two. Also the conservation equation can be written as
\begin{equation}
	\frac{dE}{dt}+p\frac{dV}{dt}=0
\end{equation}
with $E=\rho V$, $V=\dfrac{4}{3}\pi S^3$. This is nothing  but the 1st law of thermodynamics.\\

These field equations (\ref{6.93}) and (\ref{6.94}) are known as Friedmann equations and are fundamental equations in Relativistic cosmology.\\\\
\begin{center}
	\underline{-----------------------------------------------------------------------------------} 
\end{center}
\newpage
\vspace{5mm}
\begin{center}
	\underline{\bf Exercise} 
\end{center}
\vspace{3mm}
{\bf 6.1.} Show that Einstein field equations can be obtained from the action principle with action
$$\mathcal{A} = \int g^{\mu\nu}\left(\Gamma _{\mu\nu}^{\alpha}\Gamma _{\alpha\beta}^{\beta}-\Gamma _{\mu\alpha}^{\beta}\Gamma _{\beta\nu}^{\alpha}\right)\sqrt{-g}\,d^4x~.$$\\
{\bf 6.2.} If the Lagrangians $L(y,y',x)$ and $\overline{L}(y,y',x)$ differ by a divergence term $i.e.$
$$\overline{L} = L + \frac{dQ}{dx}(y,y',x)$$
the show that $L$ and $\overline{L}$ give rise to the same field equations.\\\\
{\bf 6.3.} Find the energy momentum tensor for a scalar field $\phi(t)$ for which the Lagrangian is given by
$$\mathcal{L} = \sqrt{-g}\left[\phi _{,a}\,\phi _{,b}\,g^{ab}+m_0^2\phi ^2\right]\,.$$\\
{\bf 6.4.} Show that the conservation equation for a perfect fluid can be written as
$$(\rho + p)u^\alpha \nabla _\alpha u^\delta + \left(u^\alpha u^\delta - g^{\alpha \delta}\right)\nabla _\alpha p=0\,.$$\\
{\bf 6.5.} Show that the Lagrangian
$$\mathcal{L} = \sqrt{-g}(R+2\Lambda) + \mathcal{L}_M$$
gives Einstein equations with cosmological term.

Also show that the above field equations can be obtained from the conservation of the energy momentum tensor $T_{ab}$ with
$$R_{ab} + \mathcal{L}Rg_{ab} - \Lambda g_{ab} = \kappa T_{ab}\,,$$
$\mathcal{L}, \Lambda$ and $\kappa$ being constants.\\\\
{\bf 6.6.} In the weak field approximation, the metric on the space-time manifold can be written as
$$g_{\alpha\beta}=h_{\alpha\beta}+\eta _{\alpha\beta}~~~~,~~\left|h_{\alpha\beta}\right|\ll 1\,.$$
Show that, under a background Lorentz transformation, $h_{\alpha\beta}$ transforms as if it is a tensor in special relativity.\\\\
{\bf 6.7.} Show that the Einstein tensor in weak field approximation can be written as
$$G^{\alpha\beta} = -\frac{1}{2}\square \,\overline{h}_{\alpha\beta}$$
with $\overline{h}_{\alpha\beta}= h_{\alpha\beta} -\dfrac{1}{2}h\,\eta ^{\alpha\beta}$ ~~(called trace-reverse of $h_{\alpha\beta}$).\\\\
{\bf 6.8.} Show that in the weak field approximation the line element\,:~~$ds^2 = g_{\mu\nu}dx^\mu dx^\nu$ simplifies to
$$ds^2= -(1+2\phi)dt^2 + (1-2\phi)\left(dx^2+dy^2+dz^2\right)\,.$$
Also interpret $\phi=-\dfrac{1}{2}h^{00}$ from Newtonian analogy.\\\\
{\bf 6.9.} Determine the components of the Riemannian tensor $R_{\alpha\beta\gamma\delta}$ for the weak field metric upto first order in $\phi$\,.\\\\
{\bf 6.10.} Show that in weak field approximation of Einstein gravity for empty space one gets the standard wave equation propagating with velocity $c$\,.

\chapter{Cosmological Solutions}
\section{Introduction}

~~~From the three basic pillars on which the modern cosmology is build up, namely the cosmological principle, Weyl postulate and Einstein's general theory of gravity, one gets the geometry of the space-time to be homogeneous and isotropic FLRW model described by the line element
\begin{equation}
	ds^2=-c^2dt^2+a^2(t)\left[\frac{dr^2}{1-kr^2}+r^2d\Omega^2_2\right],
\end{equation}
the cosmic fluid should be perfect fluid in nature having energy-momentum tensor
\begin{equation}
	T_{\mu\nu}=(\rho c^2+p)u_\mu u_\nu+pg_{\mu\nu}
\end{equation}
and the Einstein field equations for gravity show an equivalence between them. The explicit (non-vanishing) field equations are
\begin{equation}\label{7.3}
	\frac{\dot{a}^2}{a^2}+\frac{kc^2}{a^2}=\frac{8\pi G}{3}\rho+\frac{\lambda c^2}{3}
\end{equation}
and
\begin{equation}\label{7.4}
	2\frac{\ddot{a}}{a}+\frac{\dot{a}^2}{a^2}+\frac{kc^2}{a^2}=-\frac{8\pi G}{c^2}p+\lambda c^2
\end{equation}

The energy-momentum conservation equation $T^\nu_{\mu_{;\nu}}=0$ has the explicit form
\begin{equation}\label{7.5}
	\dot{\rho}+3H\left(\rho+\frac{p}{c^2}\right)=0
\end{equation}
with $H=\dfrac{\dot{a}}{a}$, the Hubble parameter.\\

Note that equations (\ref{7.3}) - (\ref{7.5}) are not independent; any one of them can be derived from the other two. Also combining equations (\ref{7.3}) and (\ref{7.4}) one gets
\begin{equation}\label{7.6}
	\frac{\ddot{a}}{a}=-\frac{4\pi G}{3}\left(\rho+\frac{3p}{c^2}\right)+\frac{\lambda c^2}{3}
\end{equation}

We shall now discuss cosmological solutions separately without and with cosmological constant.\\

\section{Cosmological Solutions without comological constant}
\subsection{Dust Cosmology: $p=0$}

~~~The conservation equation (\ref{7.5}) can be integrated to give
\begin{equation}\label{7.7}
	\rho=\rho_0\left(\frac{a_0}{a}\right)^3
\end{equation}
where $a_0$ and $\rho_0$ are the present values of the scale factor and energy density.\\

Now using (\ref{7.7}) in equations (\ref{7.3}) and (\ref{7.6}) give
\begin{equation}\label{7.8}
	\dot{a}^2=\frac{8\pi G\rho_0}{3}\frac{a_0^3}{a}-kc^2
\end{equation}
and
\begin{equation}\label{7.9}
	\ddot{a}=-\frac{4\pi G\rho_0}{3}\frac{a_0^3}{a^2}
\end{equation}

Suppose $\rho_0$ (i.e. $\rho$)=0 and $k=-1$, then equation (\ref{7.8}) gives $a=\pm ct$ (with $a=0$ at $t=0$).\\

The universe expands (or contracts) monotonically in a uniform fashion. This model is known as \textbf{Milne model}.\\

\begin{wrapfigure}[10]{r}{0.35\textwidth}\vspace{-1.9\intextsep}
	\includegraphics[height=6 cm , width=6 cm ]{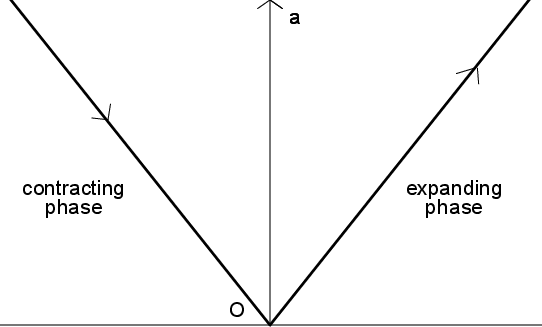}\vspace{-\intextsep}
	\begin{center}
		Fig. 7.1
	\end{center}\vspace{-\intextsep}
\end{wrapfigure}

$\mathbf{(i)~\rho_0\neq0,~k=0}$\\

In this case the evolution equation (\ref{7.8}) takes the form
\begin{eqnarray}
	\dot{a}^2&=&\frac{8\pi G\rho_0}{3}\frac{a_0^3}{a}\label{7.10}\\
	\mbox{i.e.~} H^2&=&\frac{8\pi G\rho_0}{3}\left(\frac{a_0}{a}\right)^3\nonumber
\end{eqnarray}

So at present epoch, $H_0^2=\dfrac{8\pi G\rho_0}{3}$;
i.e. $\rho_0=\dfrac{3H_0^2}{8\pi G}=\rho_c$\\

Here $\rho_0$, the energy density at the present epoch is also known as critical density and $m_0=\dfrac{3H_0^2}{8\pi Gc^2}$ is the present matter density.\\

The solution of equation (\ref{7.10}) gives
\begin{equation}\label{7.11}
	\left(\frac{a}{a_0}\right)^{\frac{3}{2}}=\frac{3}{2}H_0t~~~~~\mbox{(choosing the +ve sign)}
\end{equation}

So the present epoch is given by $$t_0=\frac{2}{3H_0}$$
and is called the age of the universe.\\

Also for the above solution 
\begin{eqnarray}
a=a_0\left(\frac{t}{t_0}\right)^{\frac{2}{3}},~~~H=\frac{2}{3t}>0,&&\\
\mbox{and~~~}q=-\left(1+\frac{\dot{H}}{H^2}\right)=\frac{1}{2}>0.~~~~&&\nonumber
\end{eqnarray}

Due to positivity of the Hubble parameter throughout the evolution starting from the big-bang singularity at $t=0$, it is an expanding model of the universe. Further, as $q>0$ so the expansion is in a decelerated manner. This model is known as Einstein-deSitter model.\\

On the other hand, if we choose the -ve sign in the solution of equation (\ref{7.10}) then the solution takes the form:
\begin{equation}
	a^{\frac{3}{2}}=a_0^{\frac{3}{2}}\left[1-\frac{3}{2}H_0(t-t_0)\right]
\end{equation}

This is a contracting model of the universe with $H=-H_0\left(\dfrac{a_0}{a}\right)^{\frac{3}{2}}$, $q=\dfrac{1}{2}$. Here $a=0$ at $t=t_0+\dfrac{2}{3H_0}$. This epoch is known as big chrunch singularity (future singularity).\\

\begin{wrapfigure}{r}{0.35\textwidth}\vspace{-1.9\intextsep}
	\includegraphics[height=6 cm , width=6 cm ]{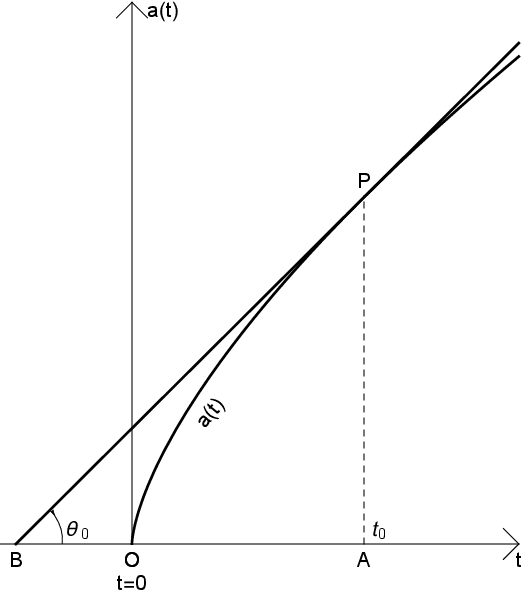}\vspace{-\intextsep}
	\begin{center}
		Fig. 7.2
	\end{center}\vspace{-\intextsep}
\end{wrapfigure}

The solution of the scale fctor given by equation (\ref{7.11}) is shown graphically by the curve $OP$. Here $A$ denotes the present epoch with $AP=a_0$. $PB$ is the tangent to the $a$-curve at present epoch and it meets the time-axis at $B$. Here $\tan\theta_0=\dot{a}\big|_{t=t_0}=H_0a_0.$.\\

So $AB=AP\cot\theta_0=\dfrac{a_0}{a_0H_0}=H_0^{-1}=\dfrac{3}{2}t_0$.\\

$\therefore \dfrac{AB}{OA}=\dfrac{3}{2}$ i.e. $OB=\dfrac{1}{2}AO$.\\

$\mathbf{(ii)~\rho_0\neq0,~k=-1:}$ \textbf{Open model}\\

In this case the evolution (\ref{7.8}) becomes
\begin{equation}\label{7.14}
		\dot{a}^2=\frac{8\pi G\rho_0}{3}\frac{a_0^3}{a}+c^2
\end{equation}

As $\dfrac{\ddot{a}}{a}=-q(t)H^2(t)$, so using (\ref{7.9}) one has
$$H^2(t)q(t)=\frac{4\pi G\rho_0}{3}\left(\frac{a}{a_0}\right)^3$$

At present epoch, $q_0=\dfrac{4\pi G\rho_0}{3H_0^2}>0$\\

i.e. $2q_0=\dfrac{\rho_0}{\frac{3H_0^2}{8\pi G}}=\Omega_0$ (the density parameter at the present epoch).\\

From equation (\ref{7.14}) at the present epoch
\begin{eqnarray}
	H_0^2&=&\frac{8\pi G\rho_0}{3}+\frac{c^2}{a_0^2}\nonumber\\&=&2q_0H_0^2+\frac{c^2}{a_0^2}\nonumber\\
	\mbox{i.e.~~}\frac{c^2}{a_0^2}&=&(1-2q_0)H_0^2
\end{eqnarray}

This implies, $0<q_0<\dfrac{1}{2}$ and hence $0<\Omega_0<1$. Again from equation (\ref{7.14}) one has
\begin{eqnarray}
	\dot{a}^2=c^2\left(\frac{\mu}{a}+1\right)&,&~~~ \mu=\frac{8\pi G\rho_0a_0^3}{3c^2}\nonumber\\
	\mbox{i.e.} \int\limits_0^a \frac{\sqrt{a}da}{\sqrt{\mu+a}}=\pm ct&&\nonumber
\end{eqnarray}

Note that the `-ve' sign correspond to contracting model. Hence for expanding model the solution can be written in parametric form
$$a=\frac{\mu}{2}(\cosh 2\theta-1)~,~~ct=\frac{\mu}{2}(\sinh2\theta-\theta).$$

So big bang singularity occurs at the parameter value $\theta=0$. As the scale factor `$a$' has no maximum, so starting from big bang singularity the universe expands infinitely as in $k=0$. So at present epoch (i.e. $\theta=\theta_0$)
$$\cosh2\theta_0=\frac{2a_0}{\mu}+1=\left(\frac{1-q_0}{q_0}\right),~\mbox{using~}\frac{c^2}{a_0^2}=(1-2q_o)H_0^2$$

So the present age of the universe is given by
\begin{eqnarray}
	t_0&=&\frac{\mu}{2c}\left[\sinh2\theta_0-2\theta_0\right]\nonumber\\&=&\frac{q_0}{(1-2q_0)^{\frac{3}{2}}}\left[\frac{\sqrt{1-2q_0}}{q_0}-\ln\bigg\{\frac{(1-q_0)+\sqrt{1-2q_0}}{q_0}\bigg\}\right]H_0^{-1}
\end{eqnarray}

It is easy to see that $t_0$ decreases as $q_0$ increases from $0$ to $\dfrac{1}{2}$. So maximum value of $t_0$ is $H_0^{-1}$ at $q_0=0$. But for $q_0=0$, one has $\mu=0$  so that $a=ct$. Then the line element becomes:
$$ds^2=-c^2dt^2+c^2t^2\left[\frac{dr^2}{1+r^2}+r^2d\Omega_2^2\right]$$
which can be written in Minkowski form
$$ds^2=-c^2d\tau^2+dR^2+R^2d\Omega_2^2$$
with $\tau=t\sqrt{1+r^2}$, $R=rct$.

This model is termed as Milne model as it has analogy with Milne's Kinematic relativity -- a cosmological theory without the notion of general relativity.\\

$\mathbf{(iii)~\rho_0\neq0,~k=+1:}$ \textbf{Closed model}\\

Here the evolution equation (\ref{7.8}) becomes
\begin{equation}\label{7.17}
	\dot{a}^2=\frac{8\pi G\rho_0}{3}\frac{a_0^3}{a}-c^2
\end{equation}

So as in $k=-1$ one has
\begin{equation}
	2q_0=\Omega_0\mbox{~~and~~}\frac{c^2}{a_0^2}=(2q_0-1)H_0^2
\end{equation}

Hence one has $q_0>\dfrac{1}{2}$ and $\Omega_0>1$. Thus closed model has energy density larger than the critical density.

The evolution equation (\ref{7.17}) takes the form
\begin{equation}\label{7.19}
	\dot{a}^2=c^2\left(\frac{l}{a}-1\right)\mbox{~~with~~}l=\frac{8\pi G\rho_0a_0^3}{3c^2}=\frac{2q_0a_0}{(2q_0-1)}
\end{equation}

So $\dot{a}=0$ when $a=l=a_{\max}$. Thus the universe expands till $a=a_{\max}$, and subsequently the universe contracts. Now solving (\ref{7.19}), the solution for $a$ can be written in parametric form as
\begin{equation}
	a=\frac{l}{2}(1-\cos 2\theta)~,~~ct=\frac{l}{2}(2\theta-\sin2\theta).
\end{equation}

Note that $\theta=0$ corresponds to big bang singularity while $\theta=\dfrac{\pi}{2}$ gives the maximum value $a_{\max}$ and $\theta=\pi$ represents the big chrunch singularity.

Now at the present epoch
$$	a_0=\frac{l}{2}(1-\cos 2\theta_0)\mbox{~and~~}ct_0=\frac{l}{2}(2\theta_0-\sin2\theta_0).$$

Hence $\cos2\theta_0=\dfrac{1-q_0}{q_0}$, i.e. $\dfrac{1}{2}<q_0<1$,
and the present age of the universe is given by
\begin{eqnarray}
	t_0=\frac{q_0}{(2q_0-1)^{\frac{3}{2}}}\left[\cos^{-1}\left(\frac{1-q_0}{q_0}\right)-\frac{\sqrt{2q_0-1}}{q_0}\right]H_0^{-1}
\end{eqnarray}

Also the time of reaching the maximum expansion $\left(\theta=\dfrac{\pi}{2}\right)$ is given by 
$$t_{\max}=\frac{\pi q_0H_0^{-1}}{(2q_0-1)^{\frac{3}{2}}}$$

Further if $T$ represents the life time of the universe (corresponding to $\theta=\pi$), then
$$T=\frac{2\pi q_0H_0^{-1}}{(2q_0-1)^{\frac{3}{2}}}=2t_{\max}$$

\begin{wrapfigure}{r}{0.35\textwidth}\vspace{-.5\intextsep}
	\includegraphics[height=6 cm , width=6 cm ]{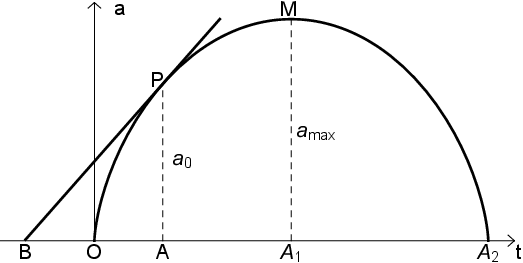}\vspace{-\intextsep}
	\begin{center}
		Fig. 7.3
	\end{center}\vspace{-\intextsep}
\end{wrapfigure}

In the figure, $OA=t_0$, $OA_1=t_{\max}$ and $OA_2=T$. Also $AP=a_0$, $A_1M=a_{\max}$ and $AM=H_0^{-1}$.\\

From the above solutions for $\rho_0\neq0$ with $k=0,\pm1$, one has the following observations:\\

(i) From the evolution equation (\ref{7.8}), one may note that at the very early phase of evolution (when $a$ is very small) the curvature term has no effect, so the universe evolutes like Einstein--de-Sitter model. However, at late time (when $a$ is large) the curvature term will characterize  the evolution.\\

(ii) Equation (\ref{7.9}) shows that $\ddot{a}<0$, $\forall a$ and for each choice of $k$. So the  path of evolution for the scale factor must be concave downwards and hence intersects the time axis at a finite point in the past (for expanding model). Hence all the above expanding models must have big-bang singularity (having infinite energy density).\\

\subsection{Perfect fluid solution}

~~~The Friedmann equations are
\begin{eqnarray}
	3\left(H^2+\frac{kc^2}{a^2}\right)&=&8\pi G\rho\label{7.22}\\
	2\left(\dot{H}-\frac{kc^2}{a^2}\right)&=&-8\pi G\left(\rho+\frac{p}{c^2}\right)\label{7.23}
\end{eqnarray}
and the energy conservation equation is
\begin{equation}\label{7.24}
	\dot{\rho}+3H\left(\rho+\frac{p}{c^2}\right)=0
\end{equation}

Assuming the perfect fluid to be of barotropic nature having equation of state: $\dfrac{p}{c^2}=\omega\rho$, $\omega$ a constant, equation (\ref{7.24}) can be integrated to give
\begin{equation}\label{7.25}
	\rho=\rho_0\left(\frac{a}{a_0}\right)^{-3(1+\omega)}
\end{equation}

Using (\ref{7.25}) in (\ref{7.22}) one has the evolution equation for the scale factor as
\begin{equation}\label{7.26}
	\dot{a}^2=\mu^2a^{-(1+3\omega)}-kc^2
\end{equation}
with $\mu^2=\dfrac{8\pi G\rho_0}{3}a_0^{3(1+\omega)}$.

Also elimination of $k$ between the two Friedmann equations (\ref{7.22}) and (\ref{7.23}) gives the acceleration as
\begin{equation}
	\frac{\ddot{a}}{a}=-\frac{4\pi G(1+3\omega)}{3}\rho
\end{equation}

Thus irrespective of the nature of the space-time (flat, open or closed) $\ddot{a}>=<0$ according as $\omega<-\dfrac{1}{3}$, $\omega=-\dfrac{1}{3}$ and $\omega>-\dfrac{1}{3}$ respectively.\\

\textbf{(i)} $\mathbf{k=0:}$ \textbf{flat model}\\

The evolution equation (\ref{7.26}) can be integrated to give
\begin{equation}
	a=\left\{\begin{array}{c c}
		\left[a_0^{\frac{3(1+\omega)}{2}}+\dfrac{3\mu}{2}(1+\omega)(t-t_0)\right]^{\frac{2}{3(1+\omega)}}&\mbox{,~if~}1+\omega>0\vspace{.2cm}\\a_0 e^{\mu(t-t_0)}&\mbox{,~if~}1+\omega=0\\\left[a_0^{-\frac{3|1+\omega|}{2}}+\dfrac{3\mu}{2}|1+\omega|(t_0-t)\right]^{-\frac{2}{3|1+\omega|}}&\mbox{,~if~}1+\omega<0
	\end{array}\right.
\end{equation}

where $a=a_0$ at the present epoch $t=t_0$.

Note that there is big bang singularity at $t=0$ for the choice $1+\omega>0$ while there is future singularity at $t=t_c$, given by
$$t_c=t_0+\frac{2}{3\mu|(1+\omega)|}a_0^{-\frac{3|1+\omega|}{2}}.$$

For the choice $\omega=-1$, there is neither any past nor future singularity (in finite time), only there is exponential expansion.

Further, for $1+\omega>0$, $H=\dfrac{2}{3(1+\omega)t}>0$, $q=\dfrac{1}{2}(1+3\omega)$ while for $1+\omega=0$, $H=\mu$, $q=-1$.\\

\textbf{(ii)} $\mathbf{k=-1:}$ \textbf{open model}\\

In this case the evolution equation (\ref{7.26}) takes the form
$$\dot{a}^2=c^2\left[\frac{\mu^2}{c^2}a^{-(1+3\omega)}+1
\right]$$

As $\dot{a}$ cannot vanish for any $a$ so the universe expands continuously to infinity. The evolution equation in integral form takes the form:
\begin{equation}
	ct=\left\{\begin{array}{c c}
		\int\limits_0^a\dfrac{a^{\frac{1+3\omega}{2}}da}{\sqrt{\mu_0^2+a^{(1+3\omega)}}}&\mbox{,~if~}1+3\omega>0\\\int\limits_0^a\dfrac{da}{\sqrt{\mu_0^2a^{-(1+3\omega)}+1}}&\mbox{,~if~}1+3\omega<0
	\end{array}\right.
\end{equation}
with $\mu_0=\dfrac{\mu}{c}$. For $\omega=-\dfrac{1}{3}$, we have Milne model of the universe. Or equivalently one has
\begin{equation}
	\int\limits_{\mu_0}^z \left[z^2-\mu_0^2\right]^{\frac{(1-3\omega)}{2(1+3\omega)}}dz=\frac{(1+3\omega)}{2}ct
\end{equation}
for $(1+3\omega)>0$ and $a=(z^2-\mu_0^2)^{\frac{1}{(1+3\omega)}}$, while for $(1+3\omega)<0$,
\begin{equation}
\int\limits_1^v (v^2-1)^{\frac{(2+3\omega)}{|1+3\omega|}}dv=\frac{|1+3\omega|}{2}\mu_0^{\frac{2}{|1+3\omega|}}ct	
\end{equation}with $a=\left(\frac{v^2-1}{\mu_0^2}\right)^{\frac{1}{|1+3\omega|}}$.

Alternatively, one can write in parametric form as
\begin{equation}
	\begin{array}{c}
		a=\mu_0^{\frac{2}{1+3\omega}}(\sinh\theta)^{\frac{2}{1+3\omega}}\\
		ct=\dfrac{2\mu_0^{\frac{2}{1+3\omega}}}{1+3\omega}\int\limits_0^\theta\sinh^{\frac{2}{1+3\omega}}\theta d\theta
	\end{array}
\end{equation}
for $1+3\omega>0$ and for $1+3\omega<0$, one has
\begin{equation}
	\begin{array}{c}
		a=\mu_0^{-\frac{2}{|1+3\omega|}}\sinh^{\frac{2}{|1+3\omega|}}\theta\\
		ct=\dfrac{2\mu_0^{-\frac{2}{|1+3\omega|}}}{|1+3\omega|}\int\limits_0^\theta \sinh^{\frac{2}{|1+3\omega|}-1}\theta d\theta
	\end{array}
\end{equation}

Note that $\int(x^2-b^2)^ndx$ is integrable for $n=\dfrac{m}{2}$, $m$ is an integer including zero. Hence the above integrals can be integrable for some suitable choices for $\omega$.\\

\textbf{(ii)} $\mathbf{k=+1:}$ \textbf{closed model}\\

Here the evolution equation becomes
\begin{equation}
	\dot{a}^2=c^2\left[\mu_0^2a^{-(1+3\omega)}-1\right]
\end{equation}

As $\dot{a}$ vanishes at $a_m=\mu_0^{\frac{2}{1+3\omega}}$ so the universe starting from big bang singularity expands till the scale factor has the maximum value $a_m$ and then there is a big-chrunch singularity, provided $1+3\omega>0$. On the other hand, for $1+3\omega<0$, the scale factor contracts from infinite value reaches a minimum at $a=a_m$ and then expands to infinity. So for $1+3\omega>0$, it is a cyclic model of the universe while for $1+3\omega<0$, it is a bouncing model of the universe. The solution in the integral form is similar to the case $k=-1$.\\

\section{$\lambda$-cosmology}

~~~From the Einstein field equations (\ref{7.3}) and (\ref{7.4}), one has the evolution equation equation (for dust)
\begin{equation}\label{7.35}
	\dot{a}^2=\frac{8\pi G\rho_0}{3}\frac{a_0^3}{a}-kc^2+\frac{\lambda c^2}{3}a^2
\end{equation}
and the measure of acceleration
\begin{equation}\label{7.36}
	\ddot{a}=\frac{4\pi G\rho_0}{3}\frac{a_0^3}{a^2}+\frac{\lambda c^2}{3}a
\end{equation}
where the solution of the conservation equation (\ref{7.5}) i.e. $\rho=\rho_0\left(\dfrac{a_0}{a}\right)^3$ has been used.

From the above equations it is easy to see that when $a$ is small, $\lambda$ term is insignificant compared to the first term on the r.h.s. So gravity is then attractive in nature and acceleration varies  as inverse square law as in Newtonian gravity. However, at large distance, $\lambda$-term dominates and it behaves as a repulsive force. Thus $\lambda$-term has no effect in the solar system or in the structure of our galaxy, rather it can influence on the scale of clusters of galaxies or larger.\\

\subsection{Einstein static universe}

~~~After formulating  the field equations for gravity Einstein was doubtful whether those coupled quasi-linear hyperbolic 2nd order partial differential equations will have any solution. But within few months Einstein was happy to see a very simple solution to his field equations due to Schwarzschild (known as Schwarzschild's vacuum solution or Schwarzschid's black hole solution). Then it was generally believed that by imposing symmetry to the space-time geometry the field equations may be solvable. At that time it was generally believed that universe is static and spherically symmetric in nature.

Einstein speculated that his equation for gravity should correctly describe the universe as a whole. For simplicity, he assumed the space-time to be homogeneous and isotropic in nature. As due to homogeneity, there should not be any pressure gradients in the universe. Hence he had chosen dust as the cosmic matter. Further, due to homogeneity and isotropy of the space-time geometry, the line element is given by FLRW model as
\begin{equation}
	ds^2=-c^2dt^2+a^2(t)\left[\frac{dr^2}{1-kr^2}+r^2d\Omega_2^2\right]
\end{equation}

Due to static nature, he had chosen $a(t)=a_0$, a constant and $k=+1$ for universe to be closed. So the Friedmann equations (i.e. equations (\ref{7.3}) and (\ref{7.4}) with $\lambda=0$) take the form
\begin{equation}\label{7.38}
	\frac{3c^2}{a_0^2}=8\pi G\rho_0 \mbox{~and~} \frac{c^2}{a_0^2}=0
\end{equation}
which has no finite realistic solution. Thus it is not possible to have any static homogeneous and isotropic model of the universe with dust as the cosmic matter in the frame work of Einstein gravity.

Einstein then argued that due to attractive nature of gravity the above static solution is not possible. However, stars are able to maintain a stationary shape as gravity is balanced by the outward pressure of the hot gas inside the star. So Einstein in 1917 cleverly introduced a term $\lambda$ in the r.h.s. of his field equation which effectively gives a negative pressure to balance gravity (this can be interpreted as the modification of Einstein gravity by addition if this term to the Ricci scalar in the Einstein Hilbert action). Thus the equations (\ref{7.38}) is now modified as
\begin{eqnarray}
	&&\frac{3c^2}{a_0^2}=8\pi G\rho_0+\lambda c^2 \mbox{~~~and~~~} \frac{c^2}{a_0^2}=\lambda c^2\label{7.39}\\
	\mbox{i.e.}&&a_0=\lambda^{-\frac{1}{2}},~~\rho_0=\frac{\lambda c^2}{4\pi G}
\end{eqnarray} 

This is known as Einstein static solution. The above solution shows that the scale factor (i.e. the radius) of the universe is inversely proportional to the square root of the matter density. Here $\lambda$ is termed by Einstein as cosmological constant.

As a rough estimate if $\rho_0$ is chosen as $\sim10^{-31}$ g.cm$^{-3}$ then $a_0\sim10^{29}$ cm and $\lambda\simeq10^{-58}$ cm$^{-2}$. Due to this very small value of $\lambda$ it cannot make any detectable difference from the prediction of standard general relativity (i.e. general relativity without $\lambda$ term). Hence no ambiguity will be there in solar system tests or any form of local tests of gravity.

However, this static model did not survive more than a decade when in 1929 Edwin Hubble observationally predict that the universe is not at all static, rather it is expanding. So at present this static model has only historical importance. On the other hand, the static model is of little interest in the context of singularity free model of the universe -- in emergent scenario the universe is assumed to be in the Einstein static phase as pre-inflationary era.

\subsection{de Sitter universe}

~~~Willem de Sitter in the same year 1917, obtained another solution to the Einstein field equations with cosmological term. Although similar to Einstein, he assumed  the space-time to be homogeneous and isotropic FLRW model (flat type) but he did not choose it to be static in nature. Also he has considered an empty model of the universe. Thus the Einstein field equations (\ref{7.3}) and $(\ref{7.4})$ become
\begin{equation}
	\begin{array}{c}
		3H^2=\lambda c^2 \mbox{~~and~~}\dot{H}=0\\
		\mbox{i.e.~~} H=\mbox{constant}=H_0=\sqrt{\dfrac{\lambda c^2}{3}}
	\end{array}
\end{equation}

Also $\dfrac{\dot{a}}{a}=H=H_0$ i.e. $a=a_0e^{H_0t}=a_0e^{\sqrt{\frac{\lambda c^2}{3}}t}$. So the FLRW line element becomes
\begin{equation}\label{7.42}
	ds^2=-dt^2+e^{2H_0t}[dr^2+r^2d\Omega_2^2]
\end{equation}

This is known as de Sitter solution. Using co-ordinate transformation:
\begin{equation}
	R=re^{H_0t},~ dT=dt+\frac{H_0R}{1-H_0^2R^2}dR
\end{equation}
the de Sitter metric can be written in stationary form as
\begin{equation}
	ds^2=-(1-H_0^2R^2)dT^2+\frac{dR^2}{1-H_0^2R^2}+R^2d\Omega_2^2
\end{equation}

Further, from the line element (\ref{7.42}) it is easy to see that with constant $\theta$ and $\phi$, any test particle follows time-like geodesics in the de Sitter space, with proper separation between any two particles increases with time as $e^{H_0t}$. As a consequence, these particles are all moving apart from one another, indicating the expanding nature of the universe.

However, these particles do not have any material status, nor they have any mass, so they do not influence the geometry of space-time. Thus the universe is empty in dynamic sense while kinematically it is expanding. Einstein nicely compared these two models stating de Sitter universe has motion without matter while Einstein static model has matter without motion.

Lastly, it is to be noted that the above empty solution of general relativity (due to de Sitter) does not satisfy Mach's criterion. According to Mach, there should be a background of distant matter due to which motion is measurable  -- without material background there is no meaning to say rest or motion. Einstein believed Mach conjecture and interestingly his static solution satisfies this conjecture -- it is a matter filled space i.e. a background of distant matter w.r.t. which a local observer can measure motion and formulate the laws of physics. Moreover, Einstein speculated that matter precisely characterise the geometry of space-time -- a unique feature of general relativity. Thus de Sitter solution is a counter example of Einstein's speculation.

\subsection{Generalized Einstein static model}

~~~At first Einstein static model will be introduced in a different way. For FLRW space-time model the Einstein field equations (with $\lambda=0$) (\ref{7.3}) and (\ref{7.4})  can be written as
\begin{eqnarray}
	&&\frac{\ddot{a}}{a}=-\frac{4\pi G}{3c^2}(\rho c^2+p)\label{7.45}\\
	\mbox{and}&&\frac{\ddot{a}}{a}+2\frac{\dot{a}^2}{a^2}+2\frac{kc^2}{a^2}=\frac{4\pi G}{c^2}(\rho c^2-p)\label{7.46}
\end{eqnarray}

Now for static model $\dot{a}=0=\ddot{a}$. Hence equation (\ref{7.45}) implies 
\begin{equation}
	\rho c^2+3p=0
\end{equation}
and (\ref{7.46}) implies,
\begin{eqnarray}
	&&\rho c^2-p=\frac{kc^4}{2\pi Ga^2}\\
	\mbox{i.e.}&&\rho c^2=-3p=\frac{3kc^4}{8\pi Ga^2} \label{7.49}
\end{eqnarray}

Thus for static model the fluid must have negative pressure (assuming $\rho$ to be positive) so that strong energy condition (SEC) is marginally violated. Thus the fluid is not physical (exotic in nature). Hence for realistic model Einstein introduced the cosmological constant term $\lambda$ so that effective energy density and pressure are given by
\begin{equation}\label{7.50}
	\bar{\rho}=\rho+\frac{\lambda c^2}{8\pi G}~,~~\bar{p}=p-\frac{\lambda c^4}{8\pi G}
\end{equation} 

Thus for this effective fluid static model condition (i.e. equation (\ref{7.49})) becomes
\begin{eqnarray}
	&&\bar{\rho} c^2=-3\bar{p}=\frac{3kc^4}{8\pi Ga^2}\label{7.51}\\
	\mbox{i.e.}&&\rho c^2=-3p+\frac{\lambda c^4}{4\pi G}\label{7.52}
\end{eqnarray}

Hence SEC is satisfied provided $\lambda$ should be positive. Also from equation ({\ref{7.49}}), $k$ should be $+1$. Thus static model is possible for normal fluid by introducing positive cosmological constant and FLRW space-time should have closed geometry.

Further, if the cosmic fluid is chosen as perfect fluid with barotropic equation of state $p=\omega\rho c^2$, then from equation (\ref{7.52}) one has
\begin{equation}\label{7.53}
	\rho=\frac{\lambda c^2}{4\pi G(1+3\omega)}
\end{equation} 

Using (\ref{7.51}) and (\ref{7.53}) in (\ref{7.50}) gives
\begin{equation}
	\frac{\lambda(1+\omega)}{(1+3\omega)}=\frac{1}{a^2}
\end{equation}

So Einstein static model gives
\begin{equation}
	\rho_0=\frac{\lambda_0 c^2}{4\pi G(1+3\omega)} \mbox{~~and~~} \frac{1}{a_0^2}=\frac{\lambda_0(1+\omega)}{(1+3\omega)}
\end{equation}

As energy density is constant so without any loss of generality we choose $\omega=0$ i.e. $p_0=0$. Hence, Einstein static model of the universe is a positive curvature finite size but unbounded in nature and the cosmic fluid is dust nature having constant energy density, depending on the fundamental constants $\lambda$ and $G$. Further, from the energy conservation equation (\ref{7.5}), for static model
\begin{equation}
	\begin{array}{c}
		\dot{\rho}+3H\rho=0\\
		\mbox{i.e.~} \rho a^3=\mbox{constant}=\rho_0 a_0^3=\dfrac{c^2}{4\pi G\sqrt{\lambda_0}}
	\end{array}
\end{equation} 

We shall now consider a generalization of the above Einstein static model. As a first step let us consider (choosing $c=1$)
\begin{equation}\label{7.57}
	\rho a^3=\frac{\mu}{4\pi G\sqrt{|\lambda|}}
\end{equation}
so that $\mu=1$ corresponds to Einstein static model. Now from the Friedmann equation
$$\frac{\dot{a}^2}{a^2}+\frac{k}{a^2}=\frac{8\pi G}{3}\bar{\rho}=\frac{8\pi G}{3}\rho+\frac{\lambda}{3}$$
one has (using equation (\ref{7.57}))
\begin{equation}\label{7.58}
	\dot{a^2}=-k+\frac{\lambda a^2}{3}+\frac{2\mu}{3\sqrt{|\lambda|}a}=F(a)\mbox{~~~(say)}
\end{equation}

Thus the qualitative behaviour of the scale factor $a(t)$ depends on the nature of zeros of $F(a)$ as well as on the maximum/minimum of $F(a)$ which is a cubic equation in $a$. We shall now consider the following 3 cases:\\

\textbf{(i)} $\mathbf{k=0,~\mu=0,~\lambda>0}$\\

Then the solution reads
$$a=a_0e^{H_0t}, ~~H_0=\sqrt{\frac{\lambda}{3}}$$
which is nothing but the de Sitter model. Further from the above evolution equation (\ref{7.58}) one may note that any model with positive cosmological constant will evidently be de Sitter asymptotically ($a\rightarrow\infty$). So for $\lambda>0,~k\leq0,~\dot{a}^2>0~\forall a$, leading to a monotonically expanding universe.\\

\textbf{(ii)} $\mathbf{k=+1,~\mu\neq0,~\lambda>0}$\\

Suppose $\mu>1$ i.e. the model has more matter than in a Einstein static model. It is to be noted that $F(a)$ has a minimum at $a=\dfrac{\sqrt[3]{\mu}}{\sqrt{\lambda}}$. At early stages of the evolution when $a$ is very small then 3rd term on the r.h.s. of equation (\ref{7.58}) dominates. So the model starts from big bang singularity and grows as $t^{\frac{2}{3}}$ as in Einstein - de Sitter model but gradually the expansion slows down till it reaches the minimum. Subsequently, the second term (i.e. $\lambda$ term) on the r.h.s. of equation (\ref{7.58}) slowly dominates over the others and consequently its expansion speeds up (i.e. accelerated) and ultimately approaches the de Sitter model asymptotically. This solution has the interesting feature namely around the minimum $a_m$ the model will stay for a while, known as ``coasting period" (or ``quasi-stationary phase"). In this era $a(t)$ remains very close to $a_m$ and the evolution equation (\ref{7.58}) can be approximated to
\begin{equation}\label{7.59}
	\dot{a}^2\simeq\left(\mu^{\frac{2}{3}}-1\right)+\left(a\sqrt{\lambda}-\mu^{\frac{1}{3}}\right)^2
\end{equation}
which has the solution
\begin{equation}
	a=\frac{\mu^{\frac{1}{3}}}{\sqrt{\lambda}}\left[1+\sqrt{1-\mu^{-\frac{2}{3}}}\sinh\left\{\sqrt{\lambda}(t-t_m)\right\}\right]
\end{equation}

Here $t_m$ is the time at which $\dot{a}$ reaches its minimum. Note that as $\mu\rightarrow1$, $a\rightarrow\dfrac{1}{\sqrt{\lambda}}$ i.e. Einstein static model. So if $\mu$ is very close to unity, then $a$ will remain close to Einstein static model.

This model is known as Lemaitre model. It is speculated that the quasi-stationary phase of this model would be favourable for the galaxy formation.\\

$\bullet$ \textbf{Eddington - Lemaitre model (EL model)}\\
 
The EL model is a limiting case of the Lemaitre model choosing $\mu\rightarrow1$. This model has an infinite ``coasting period", so that it may be considered as two distinct models: (a) starts from big bang singularity ($a=0$) at $t=0$ and then $a$ approaches asymptotically to the Einstein static value $\dfrac{1}{\sqrt{\lambda_0}}$ as $t\rightarrow\infty$ (EL1 model) (b) $a$ expands out gradually from the Einstein  static era at $t=-\infty$ and then $a$ grows monotonically to the de Sitter exponential expansion (EL2 model).

Due to this EL model it is easy to see that Einstein static model is unstable in nature -- if there is an infinitesimal expansion or contraction around $a=\dfrac{1}{\sqrt{\lambda}}$ then $a$ goes on expanding or contracting following the above two distinct models EL2 and EL1.

Observationally, it is found that there is a concentration of the redshifts of quasi-stellar objects (QSO) around $z\simeq2$. Lemaitre model has similar character around $a=\dfrac{\sqrt[3]{\mu}}{\sqrt{\lambda}}$, the `coasting' radius. By choosing $\mu$ close to unity it is possible to make the `coasting period' as long as we desire. Otherwise, the Lemaitre model does not give detail feature of the evolution of the QSO.

\begin{wrapfigure}{r}{0.35\textwidth}\vspace{-.5\intextsep}
	\includegraphics[height=6 cm , width=6 cm ]{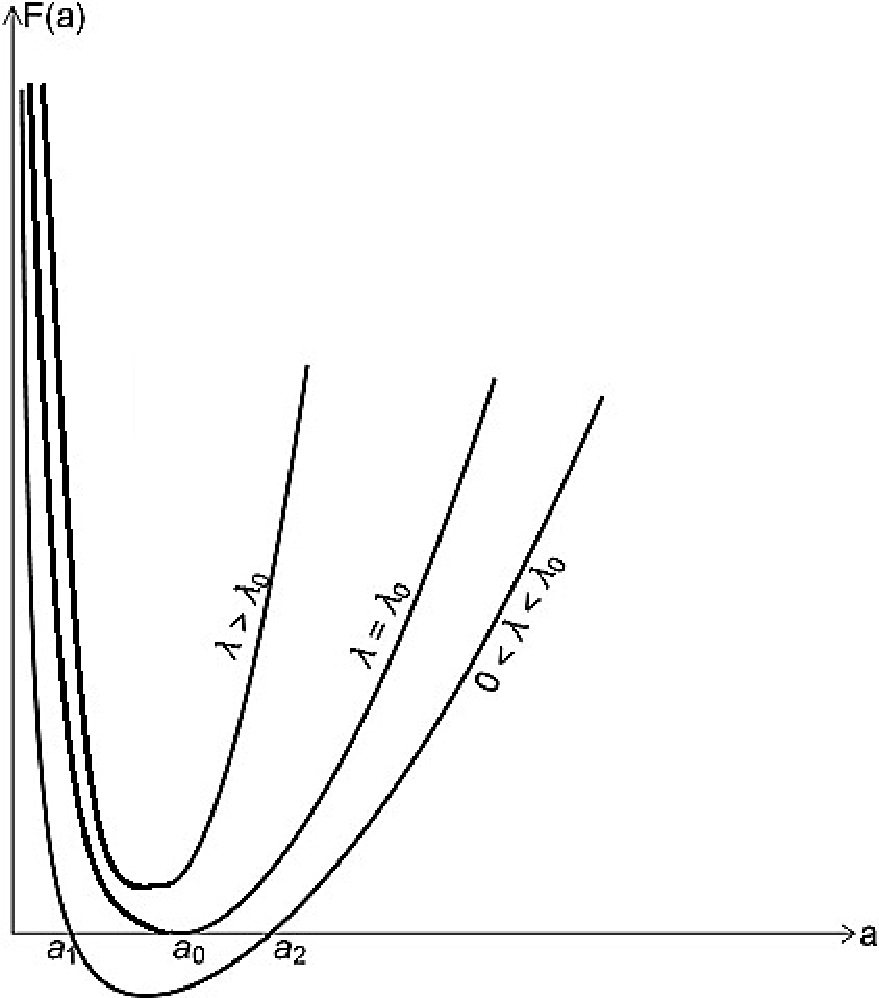}\vspace{-\intextsep}
	\begin{center}
		Fig. 7.4
	\end{center}\vspace{-\intextsep}
\end{wrapfigure}

Moreover, from (\ref{7.58}) one may note that $F(a)\rightarrow\infty$ as $a\rightarrow0$ or $\infty$. The graph of $F(a)$ in the figure shows that $F(a)$ has a minimum at $a=a_{\min}$, where $a_{\min}>0$ and $F(a_{\min})>0$ for $\lambda>\lambda_0$. For $\lambda=\lambda_0$, $a_{\min}=a_0$ and $F(a_0)=0$ and for $0<\lambda<\lambda_0$, $F(a_{\min})<0$.

Now for $\lambda>\lambda_0$, $F(a)>0$, $\forall a$, so that $\dot{a}>0$ $\forall a$, implying a monotonic expanding universe.

For $\lambda=\lambda_0$, the graph of $F(a)$ is the union of two parts: $A$ and $B$.

$A$: $F(a)$ starts from $\infty$ at $a=0$ to $0$ at $a=a_0$.

$B$: $F(a)$ starts from $0$ at $a=a_0$ to $\infty$ as $a\rightarrow\infty$.

For the first part (i.e. $A$) the universe starts from big bang singularity ($a=0$) expands monotonically to Einstein's static model asymptotically ($t\rightarrow\infty$). This is the model EL1 described above. Similarly, corresponding to part $B$ one has the model EL2.

However, if $\lambda=\lambda_0(1+\epsilon)$, $\epsilon\ll1$, the cosmic evolution is a combination of EL1 and EL2 and we have Lemaitre model as described above. It is to be noted that in the quasi-stationary phase the gravitational attraction is balanced by the cosmological repulsion due to $\lambda$. But subsequently, the repulsive force dominates and one has the EL2 model of expansion.\\

\textbf{(iii)} $\mathbf{\lambda<0}$\\ 

In this case from equation (\ref{7.58}) to keep $\dot{a}$ to be real $a$ should be finite so there is a zero (say $a=a_u$) of F(a), so that $\dot{a}=0$ but $\ddot{a}<0$ at $a=a_u$. So the scale factor increases from $0$ to $a_u$. So the scale factor increases from $0$ to $a_u$ and then $a$ decreases to zero again. This is true for any choice of $k$. Thus we have an oscillating model of the universe for $\lambda<0$ and for any choice of $k$.\\

\begin{figure}[h]
	\vspace{-.5\intextsep}
\begin{minipage}{.47\textwidth}\centering\includegraphics[height=4 cm , width=6 cm ]{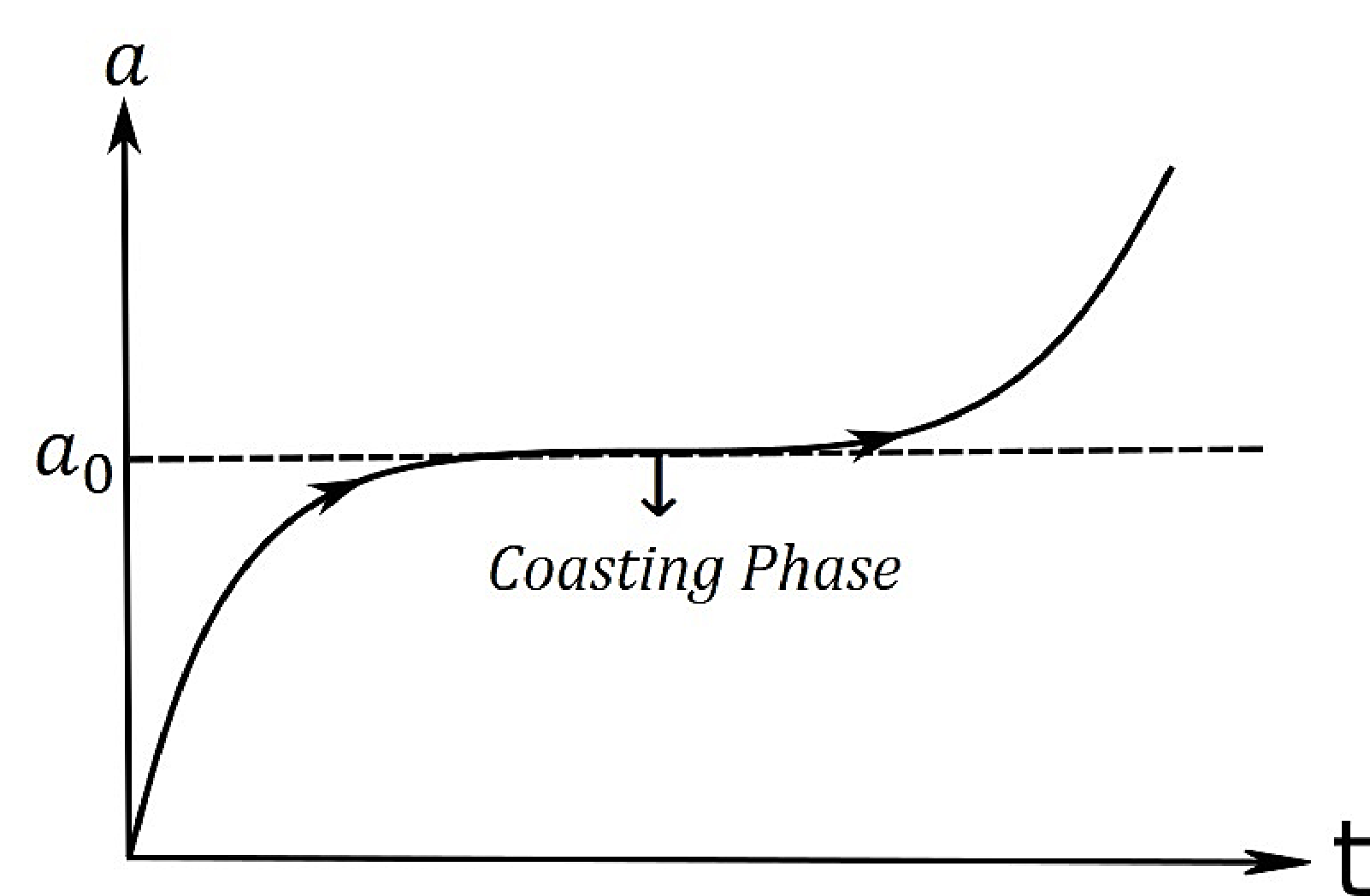}\vspace{-\intextsep}
\begin{center}
	Fig. 7.5: Lemaitre model
\end{center}\end{minipage}\begin{minipage}{.47\textwidth}
	\centering\includegraphics[height=4 cm , width=6 cm ]{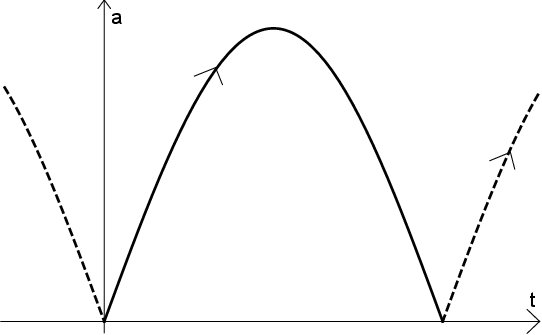}\vspace{-\intextsep}
	\begin{center}
		Fig. 7.6: Osculating model
	\end{center}\end{minipage}\vspace{-\intextsep}
\end{figure}

\begin{wrapfigure}{r}{0.35\textwidth}\vspace{-.5\intextsep}
	\includegraphics[height=6 cm , width=6 cm ]{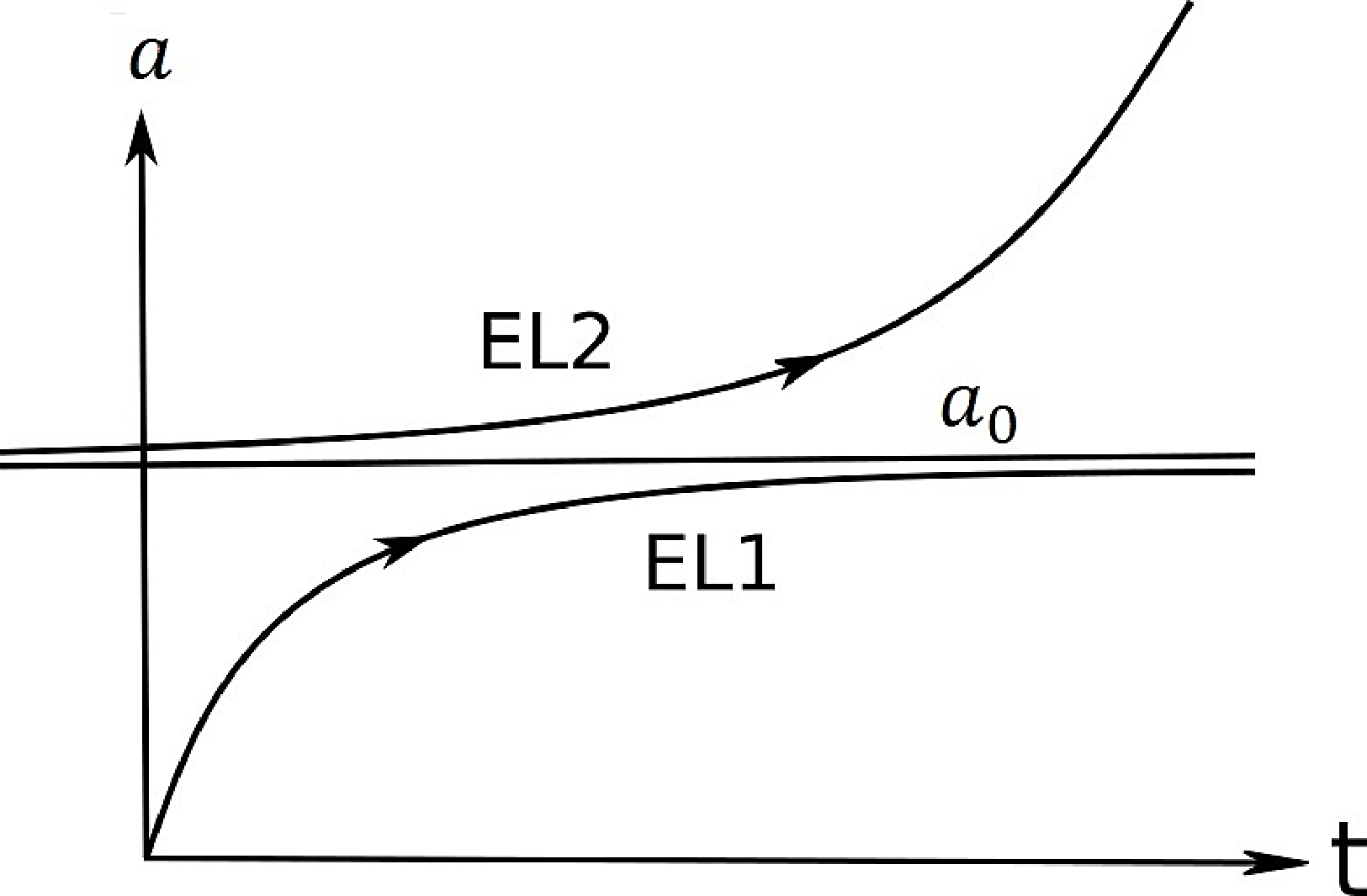}\vspace{-\intextsep}
	\begin{center}
		Fig. 7.7: Eddington Lemaitre model
	\end{center}\vspace{-\intextsep}
\end{wrapfigure}
\section{Cosmological parameters}

$$H(t)=\frac{\dot{a}(t)}{a(t)},~~q(t)=-\frac{a(t)\ddot{a}(t)}{\dot{a}^2(t)}$$
are termed as Hubble parameter and deceleration parameter. Also one can write $$\frac{\ddot{a}}{a}=-qH^2\mbox{~~i.e.~~}q=-\left(1+\frac{\dot{H}}{H^2}\right)$$

Note that $q$ is a dimensionless parameter while Hubble parameter has the dimension (time)$^{-1}$. There is another dimensionless parameter known as density parameter $\Omega(t)$ defined as 
$$\Omega(t)=\frac{8\pi G\rho(t)}{3H^2}$$

Also $\Lambda=\dfrac{\lambda}{3H^2}=\Omega_\lambda$ is dimensionless $\lambda$-parameter. From the Friedmann equation $$	\dfrac{\dot{a}^2}{a^2}+\dfrac{kc^2}{a^2}=\dfrac{8\pi G}{3}\rho$$ one has $$kc^2=H^2a^2(\Omega-1),$$
which implies $\Omega>1,~=1,~<1$ for $k=+1,~0$ and $-1$ respectively. For dust model
$$\frac{\ddot{a}}{a}=-\frac{4\pi G}{3}\rho \mbox{~~i.e.~~} qH^2=\frac{1}{2}\cdot\frac{8\pi G\rho}{3}\mbox{~~i.e.~~} 2q=\Omega$$

The only models in which $q$ (and hence $\omega$) is constant are Milne's model ($q=0=\Omega$) and Einstein de-Sitter model ($2q=\Omega=1$).\\

\section{Cosmological Constant}

~~~Einstein was not happy with his field equations for gravity as (in 1917) he was not able to obtain a static model of the universe. He thought that due to attractive nature of gravity, universe was initially at dynamic equilibrium and then contracted. To balance the gravity Einstein introduced the cosmological constant term and found a static model of the universe. But after about a decade Einstein withdraw his views when in 1929 Edwin Hubble observationally predicted that the universe is expanding. Also this observational fact was consistent with the cosmological solution to the Einstein field equations (without cosmological constant) by Mathematician Friedmann. Einstein termed it as his ``biggest blunder".

However, the static solution obtained by Einstein is not a stable one -- the equilibrium is unstable in a sense, if the universe expands slightly, then the expansion releases vacuum energy, causing more expansion. Similarly, if universe contracts a little bit will continue to do so.

Thus from 1930s till the late 1990s the cosmological constant was not an issue in cosmology and it is assumed to be zero. But dramatic changes occurred in 1998, again an observational data from a typeIa supernova indicates that our universe is expanding in an accelerated manner. Within the frame work of standard cosmology this observational fact can nicely be described by re-introducing a positive non-zero cosmological constant -- the simplest and promising candidate for Dark energy. Cosmology with this $\Lambda$ is termed as $\Lambda$CDM model.

In quantum field theory (QFT), an empty space is the vacuum state, consisting of quantum fields. The fluctuations of these quantum fields in their lowest energy state (i.e. ground state) are due to zero-point energy throughout the space. These vacuum fluctuations act as the cosmological constant. This theoretically calculated magnitude of the cosmological constant has extremely large value compared to the observed value from cosmology. In fact, theoretical prediction exceeds by 120 order of magnitude from observation. This huge discrepancy is termed as cosmological constant problem.

The $\lambda$-term is consistent with all the basic principles used in formulating Einstein's theory of gravity. Effectively $\lambda$ appears as a constant of integration. McCrea (1968) interpreted $\lambda$ as the energy density of vacuum. Alan Guth (1981) suggested a phase transition in the early universe from a state of very high vacuum energy density to the radiation era through a very small period of rapid exponential expansion (inflation). Due to this very short period of exponential expansion, the vacuum energy (i.e. $\lambda$ term) becomes vanshingly small.\\

\section{Cosmological constant as vacuum energy}

~~~The energy-momentum tensor $T^{\mu\nu}_V$ of the vacuum should be proportional to Minkowski metric $\eta^{\mu\nu}$ $\left(\eta^{00}=-1=\eta_{00},~\eta_{ii}=+1,~\eta_{i0}=0\right)$ in locally inertial co-ordinate system for the requirement of the Lorentz invariance. So in a general co-ordinate system  $T^{\mu\nu}_V$ must be proportional to $g^{\mu\nu}$. But in a general gravitational field, the energy-momentum tensor of a perfect fluid is given by
$$T^{\mu\nu}=pg^{\mu\nu}+(\rho c^2+p)u^\mu u^\nu$$
with $u^\mu$, a time-like vector field i.e. $g_{\mu\nu}u^\mu u^\nu=-1$. As the energy density $\rho$ and thermodynamic pressure $p$ are the coefficients in the energy-momentum tensor in a locally co-moving inertial co-ordinate system so they are scalars in nature. $u^\mu$ is defined by the requirement that it trans forms as a four vector under any general co-ordinate transformations. In particular $u^0=1$, $u^i=0$ in the locally co-moving carnelian inertial frame. Thus the above expression for $T^{\mu\nu}$ is generally covariant and it is true in locally inertial system. The conservation condition $T^{\mu\nu}_{;_{\nu}}=0$ gives
$$\dot{\rho}+3H\left(\rho+\frac{p}{c^2}\right)=0$$

As for vacuum energy $T^{\mu\nu}\propto g^{\mu\nu}$ so one should have $c^2\rho_V=-p_V$ and $T^{\mu\nu}_V=-c^2\rho_Vg^{\mu\nu}$. Also the conservation relation gives $\dot{\rho}_V=0$ i.e. $\rho_V=$ constant (i.e. independent of space-time coordinates). Further from the Friedmann equation
$$\frac{\dot{a}^2}{a^2}+\frac{k}{a^2}=\rho_V,$$
one must have $\rho_V>0$ if one takes $k=0$. Therefore cosmological constant $\lambda$ can be considered as the vacuum energy.\\

\section{Cosmological coincidence problem}

~~~In modern cosmology i.e. standard cosmology, $70\%$ of today's cosmic energy (with critical energy density $\rho_c\sim10^{-29}$ gm/cm$^3$) is the mysterious dark energy which is commonly accepted as cosmological constant. The remaining components (i.e. $30\%$) are matter due to dark matter and baryonic matter. There is an almost negligible amount of radiation (photons) ($\sim10^{-3}\%$). This is termed as $\Lambda$CDM model. 

As the densities of the components scale in different ways so accordingly, the cosmic history can be divided into three distinct epochs (early inflationary phase is not considered). In hot big bang model one has

(i) Initial State: A dense hot expanding fire ball. Here the dynamics of the universe was determined by radiation component, the dominant energy component at that epoch.

As $\rho_r\sim a^{-4}$ and $\rho_m\sim a^{-3}$ so with the expansion radiation energy decreases faster than the dark matter energy.

At redshift $z_{eq}=3400$, both the energy densities become of the same order and then we have,

(ii) The matter dominated era: Then dark matter dominates the expansion of the universe. During this era there is structure formation like stars, galaxies and galaxy clusters due to gravitational instability. 

Then with the expansion of the universe the energy density of DM gradually decreases and very recently at $z_{DE}\sim0.55$ the energy density of dark matter becomes the same order as dark energy and we have

(iii) Accelerated expansion era: Here gravity is no longer able to form super-galaxy clusters.

Although the nature of both dark matter and dark energy are unknown still cosmologist choose non-relativistic fluid as dark matter and cosmological constant $\Lambda$ as dark energy. Cosmologists speculate that we are living in a very special moment of cosmic history due to the remarkable fact that dark matter and dark energy densities of same order around the present time. As energy density of dark energy is constant while dark matter energy density varies as $a^{-3}$ so this coincidence implies a very fine-tuned initial conditions in the early universe. Both these energy densities were different by many order of magnitude in the early universe and it will be in the far future. So this strange coincidence in the order of magnitude of the dark matter and dark energy densities is termed as ``Cosmological coincidence problem".\\

\section*{Appendix}
\subsection*{Mathematical derivation of the instability of Einstein static model}

~~~Suppose the cosmological constant $\lambda$ changes from $\lambda_0$ by a small amount i.e. $\lambda=\lambda_0+\delta$ and consequently let $a=a_0+\epsilon$. Then from equation (\ref{7.35}) one gets

\begin{eqnarray}
	\dot{\epsilon}^2&=&-kc^2+\frac{8\pi G\rho}{3}(a_0+\epsilon)^2+\frac{c^2}{3}(\lambda_0+\delta)(a_0+\epsilon)^2\nonumber\\
	&\simeq&\left(-kc^2+\frac{8\pi G\rho_0}{3} a_0^2+\frac{c^2\lambda_0a_0^2}{3}\right)+\frac{2a_0\epsilon}{3}\left(8\pi G\rho_0+\lambda_0 c^2\right)+\frac{c^2a_0^2}{3}\delta\nonumber\\
	&&~~~~~~~~~~~~~~~~~~~~~~~~~~~~~~~~(\mbox{neglecting square and higher powers of~} \epsilon \mbox{~and~} \delta)\nonumber\\
&=& 2a_0\lambda_0c^2\epsilon+\frac{c^2a_0^2}{3}\delta \mbox{~~~~~~~~~~~~~~~(using equation (\ref{7.39}))}\nonumber
\end{eqnarray}

Thus if $\epsilon,\delta>0$ then $\dot{\epsilon}^2>0$ i.e. $\dot{\epsilon}$ cannot be zero anywhere and hence $a$ continues to increase or decreases for ever. Similarly, if $\epsilon,\delta<0$ then $\dot{\epsilon}^2<0$ which is impossible. This shows that Einstein's static model is unstable in nature.

Further fron equation (\ref{7.36}) one obtains
$$\frac{\ddot{\epsilon}}{a_0}=\frac{1}{3}(\lambda_0+\delta)c^2-\frac{4\pi G\rho_0}{3}=\frac{1}{2}\delta c^2$$
which clearly shows that $\epsilon$ is no longer oscillatory unless $\delta=-\epsilon$.

\subsection*{Derivation of equation (\ref{7.59})}

Expanding $F(a)$ in equation (\ref{7.58}) in Taylor series about $a=\dfrac{\sqrt[3]{\mu}}{\sqrt{\lambda}}$,
\begin{eqnarray}
	F(a)&=&F\left(\frac{\sqrt[3]{\mu}}{\sqrt{\lambda}}\right)+\left(a-\frac{\sqrt[3]{\mu}}{\sqrt{\lambda}}\right)\cancelto{0}{F'\left(\frac{\sqrt[3]{\mu}}{\sqrt{\lambda}}\right)}+\frac{\left(a-\frac{\sqrt[3]{\mu}}{\sqrt{\lambda}}\right)^2}{2}F''\left(\frac{\sqrt[3]{\mu}}{\sqrt{\lambda}}\right)\nonumber\\
	&&\mbox{~~~~~~~~~~~~~~~~~~~~~~~~~(neglecting third and higher order terms)}\nonumber\\
	&=& \left(\mu^{\frac{2}{3}}-1\right)+\frac{\left(a-\frac{\sqrt[3]{\mu}}{\sqrt{\lambda}}\right)^2}{2}\cdot2\lambda\nonumber\\
	&=&\left(\mu^{\frac{2}{3}}-1\right)+\left(a\sqrt{\lambda}-\mu^{\frac{1}{3}}\right)^2\nonumber
\end{eqnarray}


\end{document}